\documentclass[preprint2]{aastex}

\usepackage{lscape}
\usepackage{graphicx}
\usepackage{subfig}

\begin{document}

\title{Stellar spectropolarimetry with HiVIS: \\ Herbig AE/BE stars, circumstellar environments and optical pumping}

\author{David~M~Harrington}

\begin{abstract}

		The near-star environment around young stars is very dynamic with winds, disks, and outflows. These processes are involved in star and planet formation, and influence the formation and habitability of planets around host stars. Even for the closest young stars, this will not be imaged by the next generation of telescopes. Other proxies must be developed to probe the circumstellar environment. 

	The polarization of light across individual spectral lines is such a proxy that contains information about the circumstellar material on small spatial scales. Many models have been created to relate the circumstellar environment to observable polarization changes across spectral lines. However, measuring signals at the 0.1\% level requires a very careful control of systematic effects. We have recently built a high-resolution spectropolarimeter for the HiVIS spectrograph on the 3.67m AEOS telescope to address these issues.
		
	We have obtained a large number of high precision spectropolarimetric observations of Herbig Ae/Be, Classical Be and other emission-line stars collected on 117 nights of observations. Many stars showed systematic variations in the linear polarization amplitude and direction as a function of time and wavelength in the H$_\alpha$ line. The detected linear polarization varies from our typical detection threshold near $0.1\%$ up to $2\%$. Surprisingly, in several stars this polarization effect is {\it not} coincident with the H$_\alpha$ emission peak but is detected only in the absorptive part of the line profile and varies with the absorption. These detections are largely inconsistent with the traditional scattering models and inspired a new explanation of their polarization. 
	
	We developed a new spectropolarimetric model and argue that polarization in absorption is evidence of optical pumping. We argue that, while scattering theory fits many Be and emission-line star observations, this new theory has much more potential to explain polarization-in-absorption as seen in Herbig Ae/Be and other stellar systems. 

	This thesis presents a large spectropolarimetric study that combines new instrumentation, custom processing software, thorough calibrations, cross-instrument comparisons, a massive observing campaign on many targets larger than most studies to date combined, comparison of current theories on multiple objects and finally the creation of a new theory. 

\vspace{8mm}
	
\end{abstract}

\section{Introduction}

	Understanding how stars interact with their environment is a general problem important for many astrophysical systems. Many types of stars are surrounded by shells, envelopes or disks of material. Such circumstellar material often participates in accretion, polar outflows, winds and disk-star interactions. In young stars, these processes directly effect star and planet formation and evolution. Even for the closest young stars we study here, the stellar radius is an important spatial scale that is smaller than 0.1 milliarcseconds and it will not be imaged directly even by the next generation of telescopes. Other methods like interferometry can yield useful constraints, but the indirect techniques of spectroscopy and spectropolarimetry have much to offer. In particular, spectropolarimetric measurements can put unique constraints on the geometry of the system as well as thermodynamic and radiative environments of circumstellar material. 

	Many models show spectropolarimetry is a useful probe of circumstellar environments at small spatial scales. Circumstellar disks, rotationally distorted winds, magnetic fields, asymmetric radiation fields (optical pumping), and in general, any scattering asymmetry can produce a change in linear polarization across a spectral line such as H$_\alpha$ (cf. McLean 1979, Wood et al. 1993, Harries 2000, Ignace et al. 2004, Vink et al. 2005a, Kuhn et al. 2007). These signatures have the potential to directly constrain the density and geometry of the circumstellar material and probe the near-star environment. There have been several studies of young stars to date using low or medium resolution spectropolarimetry. Almost every study was focused on one class of objects and is typically based on only a few nights of observations. Though spectropolarimetric signatures have been detected in every class of star a many-epoch, high-resolution survey of any class of star has yet to be done.

	High-resolution linear spectropolarimetry measures the change in linear polarization across a spectral line. Typical spectropolarimetric signals are very small, often a few tenths of a percent change in polarization across a spectral line. Measuring these signals requires very high signal to noise observations and careful control of systematics to measure signals at the 0.1\% level. In order to address these issues, we built a dedicated spectropolarimeter for the HiVIS spectrograph on the 3.7m AEOS telescope and performed a telescope polarization calibration (Harrington et al. 2006, Harrington \& Kuhn 2008). In the process, we have upgraded much of the hardware, calibrations, and created a dedicated reduction package in IDL. This thesis provides a very careful and thorough characterization and analysis as well as a very large collection of data taken under a wide range of conditions in order to address these instrument-related issues. In addition to this large set of observations, a detailed comparison between instruments was performed. This data set was combined with recent and archival observations using another instrument as well as with previously published spectropolarimetry on three other instruments. The resulting 5-spectropolarimeter comparisons show that the HiVIS spectropolarimeter, as well as the results of the survey, are verified by all spectropolarimeters. 
			
	We collected a large number of stellar observations on over 100 nights from 2004 to 2008, monitoring roughly 30 Herbig Ae/Be stars and 30 Be and emission-line stars. These star types are thought to have very different circumstellar environments with various combinations of winds, disks, accretion, and jets (cf. Porter \& Rivinius 2003 and Waters 1998). However, there is still much debate over these processes. Scattering theory is the one common framework used to describe linear spectropolarimetric signatures. The scattering theory of spectropolarimetric line profiles was originally developed for Be stars but did not fit many of our Herbig Ae/Be observations. 

	In order to present this program in a coherent manner, there will be several independent parts. The differences between the star-types and geometries will be discussed in chapter 1. Then the instrument, observing procedures and the data reduction will be presented in chapters 2 and 3. The detailed polarization calibration of the instrument and telescope will be discussed in chapter 4. Chapter 5 is devoted to comparing results between instruments - measurements of several stars with many different instruments will be compared. Chapter 6 presents the Herbig Ae/Be observing campaign and the initial H$_\alpha$ line profiles, stellar variability, and some individual stellar properties. Before showing the detailed spectropolarimetric results in chapter 8, the current scattering paradigm will be outlined and some examples given in chapter 7. Since the Herbig Ae/Be observations do not conform to the current scattering theories, and a clear distinction between the HAe/Be stars and other stellar types must be made. Chapter 9 will present a survey of some bright Be and other emission-line stars which do seem to fit scattering theory much better. In chapter 10, a new theory based on optical pumping will be discussed. This theory has the potential to fit the observations significantly better than scattering theory and has a number of advantages. The entire thesis will be summarized in chapter 11.

\subsection{Stars, Circumstellar Disks and Stellar Winds}

	Herbig Ae/Be stars (HAeBe's) are intermediate-mass pre-main-sequence stars (see Waters 1998 for a review). They are young, typically $<$10Myr old, and of spectral type A or B with masses of 2-8 M$_\odot$ and photospheric temperatures of 7000-20000K. They typically are surrounded by an envelope of material and a circumstellar disk, a result of the star formation process. The modern criteria for a star to be classified as a HAeBe is that the A or B type star must have emission lines in it's spectrum and must emit an excess of infrared light from circumstellar material. The infrared excess requirement distinguishes the HAeBe stars from other A and B type stars, such as the classical Be stars, which are not young and have infrared excesses from free-free emission. Many HAeBe stars show evidence for active and dynamic circumstellar environments. Images taken in either the optical or near infra-red show large gaseous disks extending to nearly 1000AU from the star. These disks have holes in the central region. Mid- and Far-IR images as well as sub-mm measurements show emission from circumstellar dust from very near the star (IR) out to 100's of AU (sub-mm). Gaseous jets have been seen extending from the inner 10AU out to nearly 1000AU in some systems. Spectroscopic studies have suggested evidence for hot extended chromospheres, small amounts of $>$100,000K gas near the star (cf. Bouret et al. 1997), azimuthal structures such as wind-streamers and bullets, polar down-flows, accretion and cool opaque shells (cf. Bouret et al. 1997, Bouret \& Catala 1998, 2000, Bohm et al. 1996 Catala et al. 1999). In a typical description of an individual HAeBe star based on spectroscopy many, if not all of these phenomena will be evoked to explain the observations (cf. Beskrovnaya et al. 1994, 1995, 1998, 1999, Bohm \& Catala 1993, 1995, Bouret \& Catala 1998, 2000, Kozlova 2004, 2006, 2007, Pogodin 1994, 1997). A typical conclusion from a spectroscopic study of an individual stars will conclude that an extended chromosphere is present outside the photosphere. The presence of sporadic polar down-flows will be outlined based on red-shifted components of helium emission lines. A disk-wind or chromospheric-wind with complex and rapidly variable structure will be described based on blue-shifted components of hydrogen and sodium emission lines. Highly ionized species, such as nitrogen V, will be used as evidence for small clouds of high-temperature ionized gas near the star. A basic model for the circumstellar environment will be put forward that contains many of these components as well as discussions of the evidence. With so many varied effects involved that make specific statements about temperature, density, and geometry, a new and unique constraint on the circumstellar material would be most useful.
	
	To further develop spectropolarimetry as a technique and to compare results between different stellar systems, different star types must be observed over many epochs. A Be star is simply a B type star that shows emission lines, but they are not necessarily young pre-main-sequence stars (see Porter \& Rivinius 2003 for a review). Classical Be stars, as opposed to Herbig Be stars are main-sequence stars that rotate quickly and show emission lines from circumstellar material. This circumstellar material is not a result of a formation process, as with the young Herbig Be stars. The circumstellar material is a direct consequence of the stellar properties. The emission from the circumstellar material itself is transient. There are some models that show evidence for accretion and/or winds in these systems, but the geometry and density are thought to be quite different. For instance, the region where H$_\alpha$ emission forms is said to be a few stellar radii in HAeBe stars but up to 30 in Be stars. Comparing the spectroscopic modeling of AB Aurigae in Bouret \& Catala 1998 with the interferrometrically resolved H$_\alpha$ in $\zeta$ Tau from Quirrenbach et al. 1994 shows a striking contrast. Given that polarization arises from asymmetries in any system, a formation region that is ten times larger in size will have an inherently different asymmetry.

\subsection{Polarization}

   When talking about polarized radiation in most astronomical settings, the Stokes vector ({\bf I}) is usually the best description of the polarized light.  The Stokes Vector is an array of 4 numbers, IQUV, that gives the intensity of the the light in different polarization states.  I is the total intensity.  Q is the difference in intensity between 0$^\circ$ and 90$^\circ$ polarization states.  U is the difference between 45$^\circ$ and 135$^\circ$ polarization states.  V is the difference between the left and right handed circular polarization states.

\begin{small}
\begin{eqnarray}
  {\bf I} = \left( \begin{array} {c}  I \\ Q \\ U \\ V \end{array} \right) = 
            \left( \begin{array} {c}  Intensity \\ \updownarrow-\leftrightarrow \\ \nearrow-\searrow \\ \circlearrowleft-\circlearrowright \end{array} \right)
\end{eqnarray}
\end{small}

These parameters can be normalized to give the relative fraction of light in each polarization state.  I is always positive whereas QUV can be positive or negative with values from 0 to $\pm$I.  For example, if the Stokes Vector is [2,1,0,0] then the intensity is 2, the amount of light polarized at 0$^\circ$ (+Q) is 1 and there is no net polarization at 45$^\circ$, 135$^\circ$, left circular or right circular.  The fraction of light in the Q state is then q=Q/I=50\%.  In the stellar scattering context, once you have the Stokes vector of the scattered light and the stellar light, which is typically unpolarized, you can define the normalized Stokes parameters as the ratio of the flux in a particular polarization state and the total flux:

\begin{equation}
q=\frac{Q}{I}=\frac{Flux_Q}{Flux_{star}+Flux_{scat}}
\end{equation}
\begin{equation}
u=\frac{U}{I}=\frac{Flux_U}{Flux_{star}+Flux_{scat}}
\end{equation}

   The combined stellar and scattered flux, $Flux_{star}$ \& $Flux_{scat}$, are the total flux received from the source while the flux in any specific Stokes parameter, $Flux_Q$ or $Flux_U$, are the individual polarized fluxes. You can also define the degree of linear polarization, P, as:

\begin{equation}
P = \sqrt{q^2 + u^2}
\end{equation}

You can also project the primary vibrational direction of this polarization into your instrument's coordinate system as:

\begin{equation}
PA = \theta = \frac{1}{2}tan^{-1}(q/u)
\end{equation}

\subsection{Spectropolarimetry as a tool}

	Spectropolarimetry is a technique that measures the intensity as well as the polarization of light as a function of wavelength. This can be done many ways. Some instruments use broad-band filters and polarizers on an imaging instrument to measure the polarization in a few different colors (BVR). Others use low or high resolution spectrographs (R$\sim$100 to 300000) with polarizing optics to spectroscopically measure the polarization of light. Solar physicists have been using spectropolarimetric measurements of the sun for a century now and have observed a wealth of effects. Typically, high resolution (R$>$ 100,000) polarized spectra are taken of various regions of the sun (corona, sunspot umbra, photosphere, etc). This type of measurement is so common that the polarization spectrum is designated as the ``second solar spectrum".  

	Recently, people have begun to observe and model high resolution polarized spectra in a night-time astronomical setting. The star is unresolved and the polarization signatures are an average of all the stellar components. When you can resolve polarization changes across a single spectral line, significant effects are seen. For example, symbiotic binaries show 10\% polarization across Raman-scattered lines (cf. Harries \& Howarth 1996).   

	Recent spectropolarimetric observations of Herbig Ae/Be stars reveal numerous small linear polarization variations across H$_\alpha$ (cf. Vink et al. 2002, 2005b, Mottram et al. 2007, Harrington \& Kuhn 2007).  Our recent observations were the first to systematically search for and find spectropolarimetric variability with wavelength and epoch near  obscured H$_\alpha$ emission (Harrington \& Kuhn 2007). For most of our targets and a few in the literature, the polarization change is clearly $not$ coincident with the H$_\alpha$ emission peak. The polarization changes only in and around the absorptive part of the line profile, and the polarization near the emission peak is nearly identical to the continuum polarization. 
	
	Thomson scattering has been suggested to describe the earlier linear spectropolarimetric variations seen across emission lines in HAeBe stars.  Electron scattering produces differential linear polarization effects centered on the H$_\alpha$ line center (cf. Vink et al. 2005a) in this {\it scattering} paradigm.  A depolarization effect, where the polarized stellar continuum is diluted by unpolarized H$_\alpha$ emission, is invoked with O, B, and Wolf-Rayet (WR) stars (cf. McLean 1979, Oudmaijer \& Drew 1999, Harries Howarth \& Evans 2002) to account for these polarization variations.  An {\it apparent} polarization effect near the absorptive line profile is possible via this mechanism, but it must be accompanied by a corresponding signature (undetected in our survey) in the emissive component. Thus, scattering models can describe some classes of stars, but not the data we will describe. We note that magnetic effects may also contribute a linear polarization signal via the Hanle or Zeeman effects. However, observations of circular polarization in Herbig Ae/Be stars and many other star types have shown that the stellar magnetic fields are much too small to cause the observed amplitude of polarization seen in our survey (cf. Ignace et al. 2004, Wade et al. 2007, Catala et al. 2007).

\section{The HiVIS Spectropolarimeter}

\subsection{The AEOS Telescope \& HiVIS Spectrograph}

\begin{figure}
\centering
\subfloat[AEOS Telescope Schematic]{\label{fig:aeostel}
\includegraphics[width=0.4\textwidth]{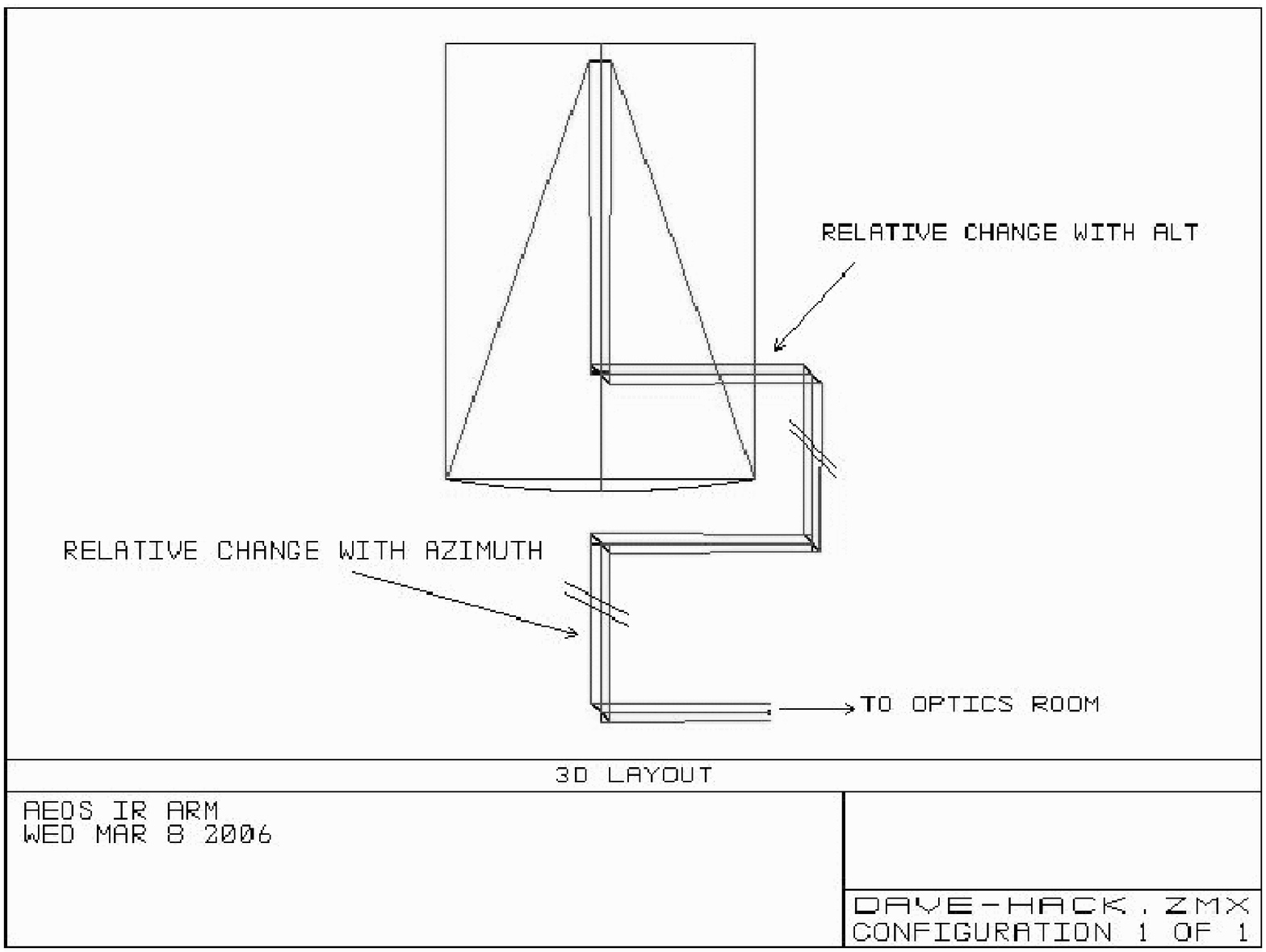}}
\quad
\subfloat[Optical Bench]{\label{fig:opben}
\includegraphics[width=0.4\textwidth]{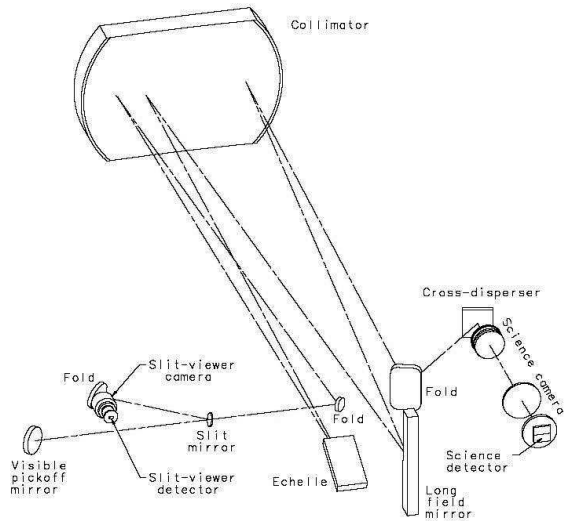}}
\vspace{5mm}
\quad
\subfloat[Coude Room Layout]{\label{fig:Coude_layout}
\includegraphics[width=0.9\textwidth, height=0.5\textheight]{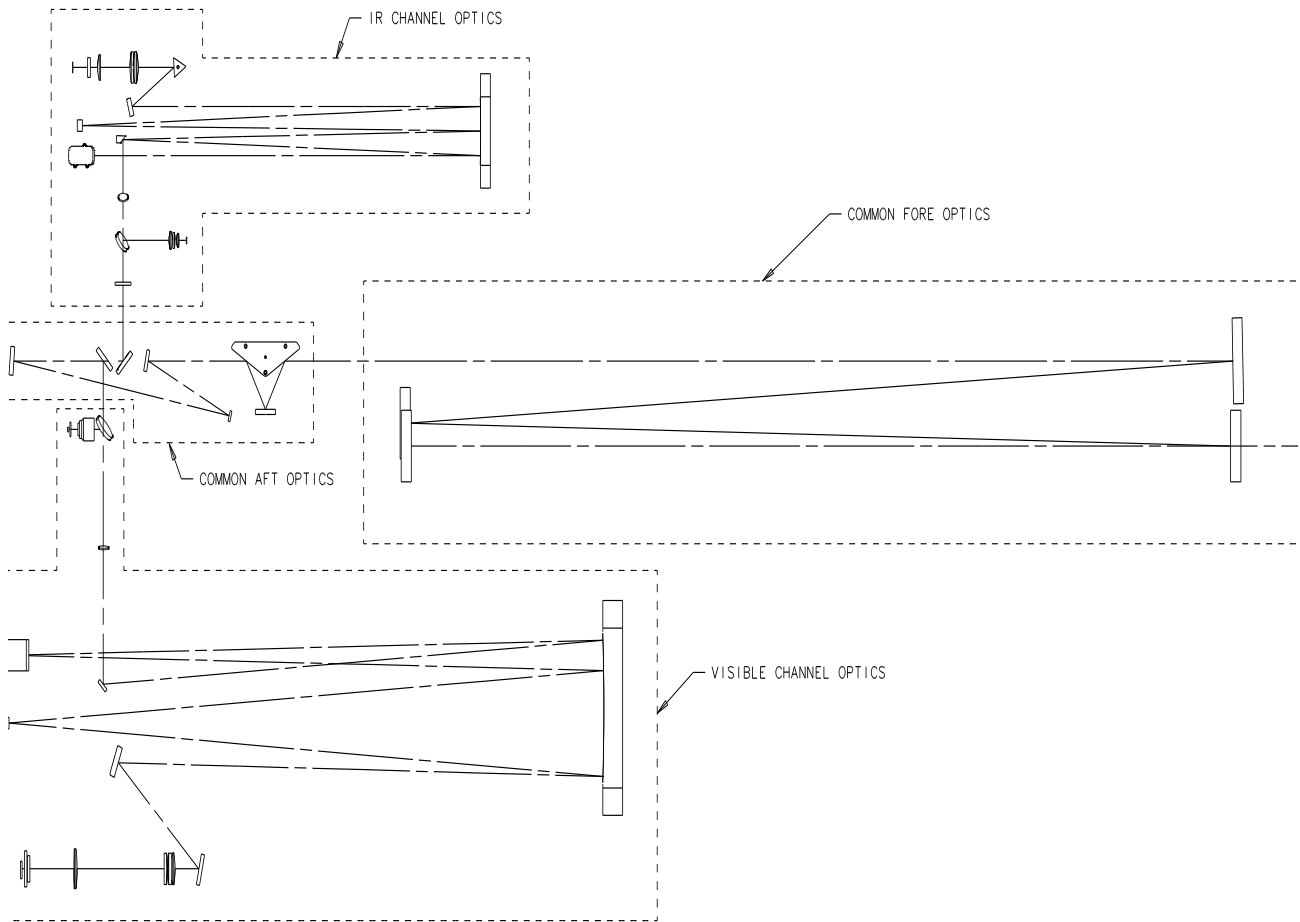}}
\vspace{3mm}
\caption[Optical and Mechanical Layouts]{ {\bf a)} Shows a Zemax schematic of the coud\'e  telescope shown pointing at the zenith, shortened for illustration purposes.  These are the 7 coud\'e reflections (m1 to m7) before the optics room.  {\bf b)}  Shows a view of the AEOS visible spectrograph optical bench - The spectropolarimetry module is mounted just downstream of the slit, near the bottom left corner of the figure.   {\bf c)}  Shows a view of the AEOS optics room.  The common fore-optics are in the center of the figure, with the visible (bottom) and IR (top) arms of the spectrograph on either side. The light enters the room from the center right after bouncing off the last coud\'e mirror, m7(not shown). }
\label{fig:room}
\end{figure}

	The Advanced Electro-Optical System telescope (AEOS) is a 3.67m, altitude-azimuth telescope, located on Haleakala mountain, Maui, Hawaii.  The spectrograph uses the f/200 coud\'e  optical path, with seven reflections (five at $\sim45^\circ$ incidence) before coming into the optics room.  This is illustrated in figure \ref{fig:aeostel}, taken from the Zemax (industry standard optical design software) telescope design.  The optics room, shown in figure \ref{fig:Coude_layout}, contains visible (0.5-1.0$\mu$m) and IR (1.0-2.5$\mu$m) cross-dispersed echelle spectrographs.  The visible spectrograph has 2 gratings which cover 500-770nm and 640-1000nm at resolutions of 12800 to 48900.  The common fore optics includes three fold mirrors, collimator, K-cell (image rotator), fold-mirror, tip-tilt correction mirror, and a reimaging spherical mirror.  A movable pickoff mirror allows for switching between the visible and IR channels (Thornton 2002, Thornton et al. 2003).
	
	The spectropolarimeter is designed for use with the visible arm of the AEOS spectrograph.  The visible arm is shown in figure \ref{fig:opben}.  From the pickoff mirror, shown in the bottom left corner of the figure, the beam passes through the slit in the slit-mirror, reflects off a fold mirror, the collimator, echelle, collimator, long mirror, collimator, fold mirror 2, and finally the cross disperser before being imaged on the CCD by a 5 lens camera.  There is a slit-viewer camera that is focused on the slit-mirror that allows real-time monitoring of the object.  The spectropolarimeter was mounted just downstream of the slit, near the image plane.  The plate scale is 0.16$''$ pixel$^{-1}$. The cross disperser's Red setting was used for all observations before Aug 28th 2007 and the Blue setting was used afterwards (to observe other wavelengths for other projects).

\subsection{The Spectropolarimeter}

	The spectropolarimetry module for the AEOS spectrograph consists of a rotating achromatic half-wave plate and a calcite Savart Plate.  The Savart plate is two crossed and bonded calcite crystals which separate the incoming, uncollimated beam into two parallel, orthogonally polarized beams displaced laterally by 4.5mm.  The half-wave plate allows the rotation of the linear polarization of the incoming light to allow complete measurement of the linear polarization of the incident light.  The beams are displaced in the spatial direction (along the long axis of the slit).  

	In a Savart plate, the two beams (extraordinary and ordinary rays) are swapped between the two crossed crystals so both displaced rays have the same optical path length through the calcite.  This design contrasts with other spectropolarimeters which use Fresnel rhomb retarders and a Wollaston prism to separate the polarization states (Donati et al. 1999, Kawabata 99 Goodrich 2003, Manset \& Donati 2003).
	
\begin{figure}
\centering
\subfloat[CCD Comparison]{\label{fig:chip}
\hspace{-4mm}
\includegraphics[width=0.4\textwidth]{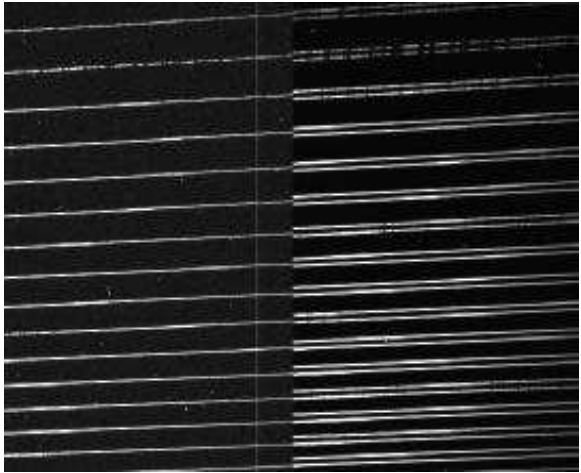}}
\hspace{-4mm}
\quad
\subfloat[Slit QU Geometry]{\label{fig:SlitQU}
\includegraphics[width=0.4\textwidth, height=0.33\textwidth]{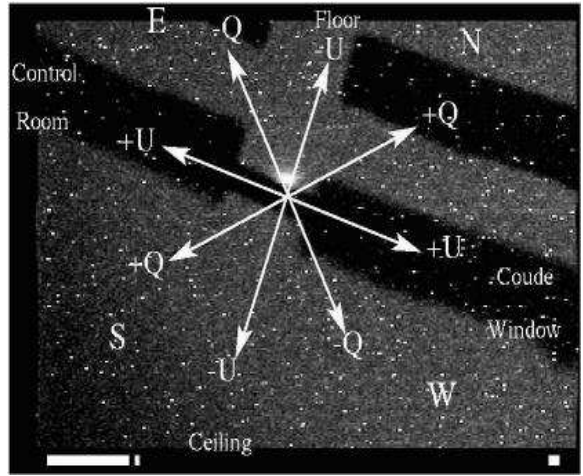}}
\vspace{5mm}
\quad
\subfloat[4 Waveplate Orientations]{\label{fig:4Ori}
\includegraphics[width=0.7\textwidth]{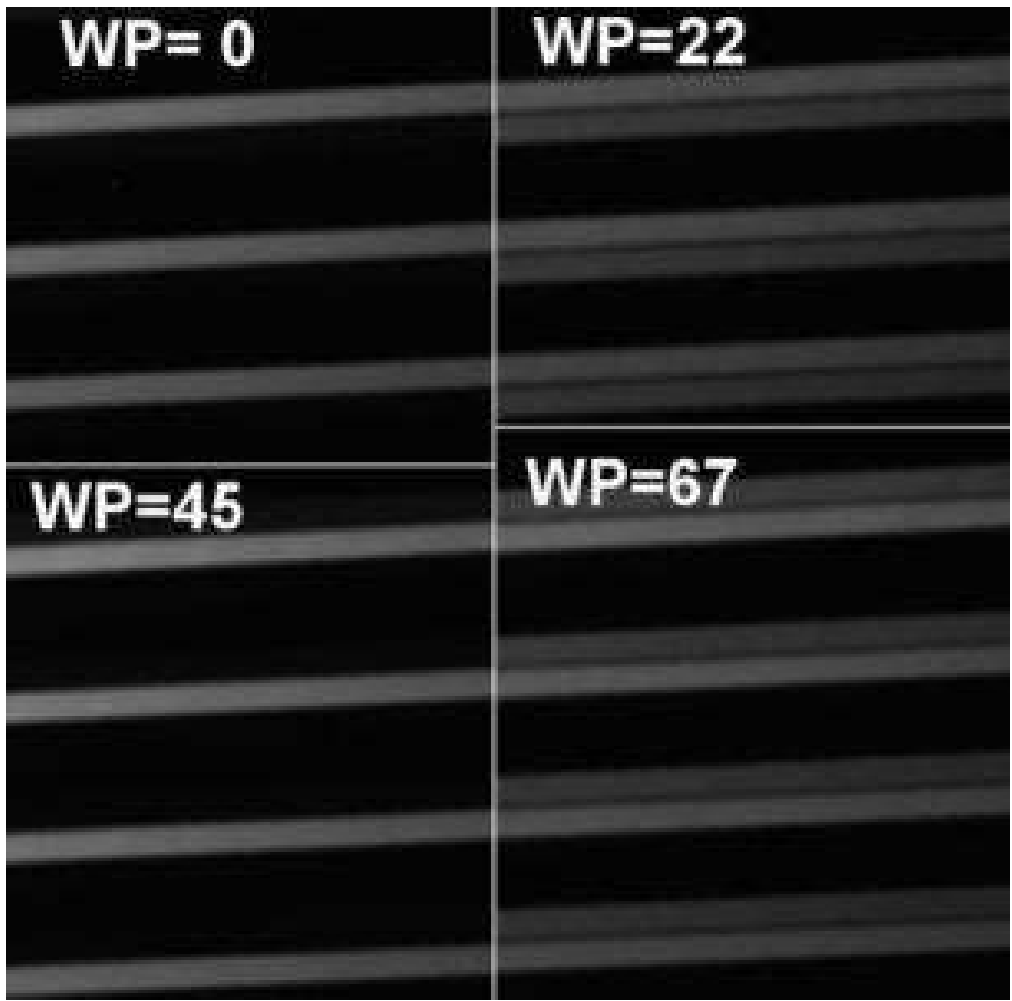}}
\caption[CCD Comparisons, Slit Geometry \& Observing Sequence]{ {\bf a)} Shows the spectropolarimetry module splits each order into two orthogonally polarized orders. The left image is a cropped raw image of a stellar spectra without the polarimeter (19 orders on the chip). The right image is same region of the image with the same star with the polarimeter in place, doubling each order (2x19 orders). {\bf b)} Shows the orientation of Q and U and the projection onto the sky as seen by the slitviewer camera at altitude 45$^\circ$ azimuth 225$^\circ$. The Stokes parameters are fixed to the instrument, whereas the projection on the sky changes with time when the image rotator is not used. {\bf c)} Shows a sequence with a polarizer mounted next to the slit to show how rotation of the waveplate swaps the order of polarized light. Successive waveplate rotations of 22.5$^\circ$ modulate the incoming polarization state. Since a 45$^\circ$ waveplate rotation swaps the polarized light from the top order to the bottom, instrumental systematic errors can be removed.}
\label{fig:basicdata}
\end{figure}

	The small deviation in the image plane splits each spectral order detected on the CCD into two orthogonally polarized orders separated by a few pixels, from which orthogonally polarized spectra are extracted separately.  For example, the red cross disperser setting originally shows 19 orders across the CCD, but after insertion of the Savart plate, it shows 38 (2 polarizations for 19 orders) as illustrated in figure \ref{fig:chip}.

	  These components were added to the Zemax optical ray trace model of the spectrograph to assess the aberrations and determine the specifications of the optics in order to produce the proper polarized-order separation on the CCD ($\sim$25 pixels).  The aberrations induced by the waveplate and Savart plate were predicted to be smaller than half a pixel leading to insignificant image degradation.  A deviation of 4.5mm at the slit was found to separate the polarized-orders by $\sim$24.5 pixels on the CCD.  A 60mm Savart Plate produces this separation (0.075$\times$60mm). A Halbo Optics Inc. Savart plate with a 15mm aperture was purchased and tested as well as a Bolder Vision Optik Inc. achromatic half-wave plate and a motorized, closed-loop rotation stage.  The half wave plate retardance was measured with a lab Ocean Optics spectrograph and two crossed polarizers.  The retardance varied roughly quadratically with wavelength at the 5\% level from 5000{\AA} to 8000{\AA}, the useful range of the polarizers) and was 0.49 waves retardance at H$_\alpha$.  The transmission of the Savart plate was above 95\% for these wavelengths. 

	After calibration and testing of the half-wave plate, Savart plate, and rotation stages, the instrument was assembled on the spectrograph on a movable plate which allows for the spectropolarimeter to be inserted or removed from the beam in under a minute.  A dekker was mounted upstream of the slit to reduce the slit length from 15$''$ to $\sim$7$''$ to minimize stray light introduced from the displacement.  This also assures polarized order separation on the CCD.   Flat fields taken after installation showed that the polarized orders were each $\sim$30 pixels across ($\sim$4$''$) with a separation of 7 pixels between each polarized beam.  

\clearpage

\onecolumn

\twocolumn
\clearpage

	To test the equipment on the telescope, a linearly polarizing filter was mounted just upstream of the the spectropolarimeter, in order to feed 100\% linearly polarzed light to the spectropolarimeter. HiVIS detected 95 to 98\% polarization over the useful range of the polarizer (5000{\AA}-8000{\AA}, orders 0-12).  This shows that HiVIS can detect linearly polarized light with the spectropolarimeter very efficiently.

\subsection{Hardware Improvements}

	There have been many hardware changes over the last few years. These were done to repair and improve the overall performance and reliability of HiVIS. The original CCD's failed and were replaced twice.  The polarimeter has been remounted with better alignment and less vignetting.  A new tip-tilt guiding corrector has been installed and calibrated.  A hard-mounted, adjustable dekkar and polarization calibration mount was installed to allow for quick, easy, and repeatable calibration and adjustment.  The failed CCD-ID20's were replaced with similar copies, but an unresolved hardware/RFI issue caused this repair to fail.  The original science camera failed and has been replaced by a new commercial 1k$^2$ PIXIS camera (Princeton Instruments) and is working wonderfully.  The calibration stage had lenses mounted that increased the flat-field flux by a factor of 6 and the ThAr lamp flux by a factor of 150.  These improvements greatly increase the ease and accuracy of spectral calibration.

\begin{figure}
\centering
\subfloat[ThAr Flux Increase]{\label{fig:thar-flux}
\includegraphics[ width=0.4\textwidth, angle=90]{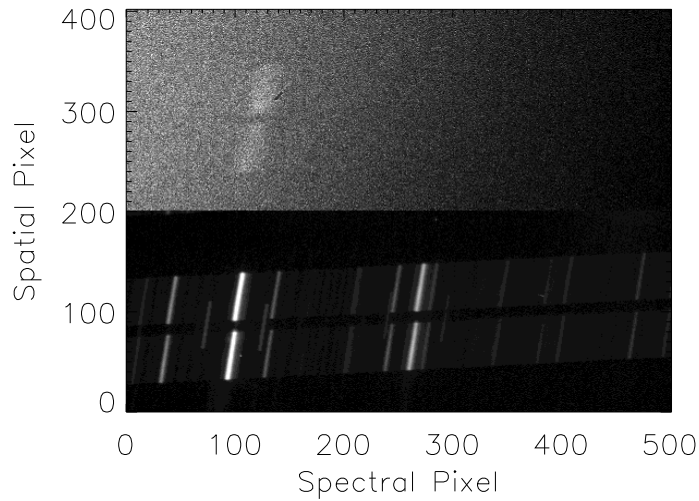}}
\quad
\subfloat[Savart Plate Leak]{\label{fig:savleak}
\includegraphics[ width=0.4\textwidth, angle=90]{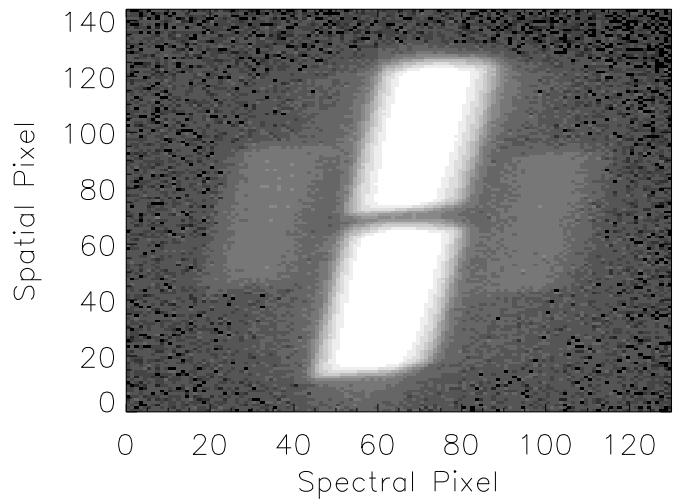}}
\caption[Th-Ar Flux Increases \& A Savart Plate Leak]{ {\bf a)}  A 20-minute Th-Ar spectra taken with the 1.5" slit before installing the lenses and mirrors in the calibration stage is on top.  A barely-detectable line is seen in this highly-stretched panel.  The brightest line in the H$_\alpha$ region is only 200 ADU above background whereas there are saturated lines in the new ThAr spectra even with a slit that is four times smaller.  {\bf b)}  The Savart plate has an inherent $\sim$1.3\% leak.  This leak should not affect the spectropolarimetry as it is stable with rotation of the waveplate - since the calculated polarization spectra are fractional differences, this stable pattern subtracts away.}
\label{fig:thar-sav}
\end{figure}

\subsection{CCD Replacements}

	The spectrograph was designed to use two 2K$\times$4K Lincoln Labs high resistivity CCD-ID20's mounted side by side with a 15 pixel gap to form a 4K$\times$4K array.  This setup was in use until September of 2006.  The CCD was set to rebin 2x2 so that the two output images are each 1024$\times$2048 pixels.

  	The original science camera CCD's were functional until October 2004 when one of the two intermittently failed.  After removing and testing the camera, it was determined that the bond-wires had come loose.  The one good CCD was replaced and the camera was remounted for a July 2005 run (Deep-Impact spectropolarimetry run, Harrington et al. 07).  Shortly thereafter, the CCD performance became intermittent.  The camera was then removed for repair.  In the summer of 2006, it was decided that both chips were irrepairable.  
  
  	A single spare CCD-ID20 was installed for the August 9, 21, 26, and Sept 12-14, 2006 observing runs.  This CCD was quite flat and free of cosmitic defects.  The camera worked well for some observations, but an intense, non-repeatable noise-signature was present on some nights, even after many attempted hardware fixes throughout the night.  The noise remained throughout the following months testing at the telescope, but was not present in any lab tests.  It is not certain what exactly is causing this error, but the non-repeatable nature of the noise as well as the complete saturation of the affected pixels, makes data with this CCD unusable. RFI is suspected because of the pulsed ringing nature of the noise which follows the readout direction, and the presence of this noise at the summit only.   

	On 9-18 2006, this spare CCD-ID20, as well as the entire dewar and camera mount, was taken off and replaced with a smaller commercial Pixis 1024BR 1k$^2$ array by Princeton Instruments on a custom-fabricated mount.  This camera, with slightly smaller pixels, performed quite well.  The peak QE is 95\% with $>$70\% QE past 900nm and 85\% QE at H$_\alpha$, making the performance of this camera almost identical to that of the HiVIS science camera, although with the loss of complete focal plane coverage.  The original CCD-ID20 chips had a plate scale of 0.16$''$ per pixel and a dispersion of 0.07{\AA} per pixel.  The original ccd's were rebinned 2$\times$2 onboard and the read time was about 1 minute. The Pixis camera had slightly smaller pixels (13$\mu m$) and thus a slightly greater plate scale and dispersion.  The Pixis camera is run in a unique way - with WinView software on a standard Windows-XP laptop.  The camera has GUI-set read-rate and gain parameters with a read rate options of 100 kHz (11se ) or 2 MHz (0.5sec) and gains of 1, 2, or 3.  Since the stellar-spectropolarimetry program required near-saturation photon counts, the Pixis camera, while slightly oversampling, was perfectly acceptable, especially with the hour-per-night of integration gained with the fast read time (0.5 s vs. 60 s).

\subsection{Calibration Stage Improvements}

  	On 11-17-2006, some lenses and a mirror were installed on the calibration stage to increase the flux of calibration light in the spectrograph.  The lens used with the flat field lamp focused the light onto the diffusion screen much more efficiently than the originally naked bulb, and caused a 6$\times$ increase in the flat-field flux.  The same lens setup when used with the ThAr lamp caused a similar increase in flux, but the improved ThAr exposure was still faint.  Originally, only 3 lines in the H$_\alpha$ order were seen at 100 ADU above background in a 20 minute exposure with the widest (1.5$''$) slit!  Even a factor of 6 increase wasn't enough.  Another mirror was installed on top of the diffusion screen to bypass the diffusion.  This increased the flux to a total of $\sim$150$\times$ the previous levels though introducing spatial structure to the ThAr exposures.  This improvement allowed for accurate line identification because many lines could be detected with the smallest slit at the highest resolution, allowing for a better comparison with the line lists and much more accurate wavelength solutions.

\subsection{Mounting \& Hardware Improvements}

  On 11-17-2006, a new dekkar was fabricated and mounted on the common-optics bench.  This was initiated because of the need to be able to move the dekker in and out of the beam quickly, easily, and repeatably by a single person.  The dekker is required to reduce the length of the slit from the original 15$''$ to $<$8$''$ to ensure complete polarized-order separation at the focal plane.  This dekkar has much more accurate mounting and an adjustable width, allowing for more precise alignment and control of the slit length, leading to proper order separation at the focal plane. The dekkar is not a true knife-edge and is mounted roughly 2cm away from the slit-mirror (and hence image plane) which leads to a $\sim$8 pixel wide edge to each order.  This is completely acceptable for the purpose of the survey. 

	A removable polarization calibrator unit was installed on the new dekkar mount.  A polarizing calibration optic (linear-polarizer, quarter-wave-plate, etc) can be mounted just up-stream of the polarimeter to allow the input of a known polarization state to the spectropolarimeter for calibration.  This type of mount allows for night-to-night calibrations as well as efficient trouble-shooting to be done easily and repeatably. 

	On 10-26-2006, the polarizing optics (Savart plate and wave-plate) were remounted closer to the slit with more robust and precise hardware.  The original polarimeter was mounted fixed on posts and had to be manually aligned.  The mount also required 3cm of optical path between the image plane (slit) and the entrance of the Savart plate.  Since the clear-aperture of the Savart plate was only 1cm and the beam diverges from the slit, the larger the distance from the image plane, the more difficult the alignment, and the greater the possibility of vignetting.  The Savart plate was remounted on a closer-fitting, remote-controllable xy-stage that put the Savart plate 2cm closer to the image plane, reducing the vignetting.  Both the Savart and the waveplate were then mounted on tip-tilt controllable rotation stages, allowing for a more precise alignment between the waveplate, Savart plate, and optical axis.  

	On 11-28-2006, a replacement tip-tilt corrector was installed.  The SBig STV unit takes an image of the slit-mirror, calculates the centroid position of the star on the slit, and sends a correction signal to a tip tilt mirror mounted on a piezo.  The tip-tilt piezo voltages and alignment mirrors were recentered and the unit was recalibrated over the following month.  The tip-tilt cannot centroid on a target star (V$\sim$6-8) faster than once per second or so, and thus can only correct slow guiding errors.  Since AEOS is specially designed for Air-force use, it does not have a normal sidereal-guiding tracker scope and thus is subject to slow pointing drift.  Depending on how recently a pointing model has been performed, the drift can be significant.  This is completely unacceptable and the tip-tilt corrector is essential to keep the star centered on the slit.

\onecolumn
\begin{figure}
\centering
\subfloat[RFI]{\label{fig:rfi}
\includegraphics[width=0.42\textwidth]{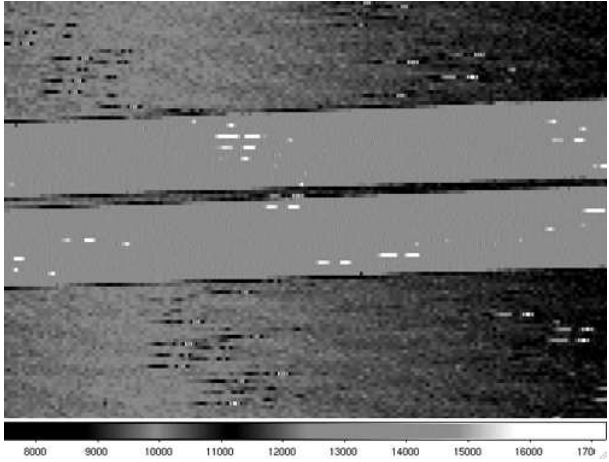}}
\quad
\subfloat[A New Dekkar]{\label{fig:dekkar}
\includegraphics[width=0.4\textwidth, height=0.3\textwidth]{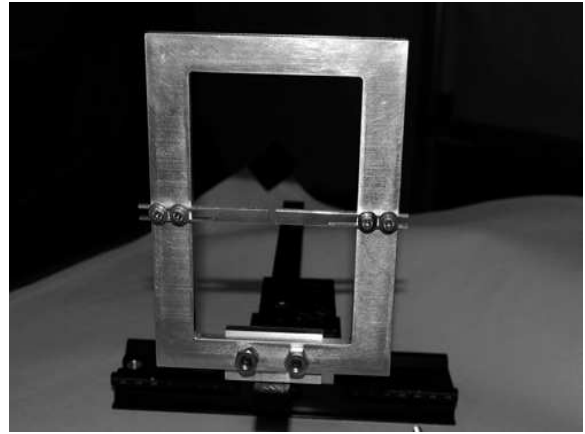}}
\vspace{10mm}
\hspace{-8mm}
\quad
\subfloat[Remounted Optics]{\label{fig:remount}
\includegraphics[width=0.45\textwidth]{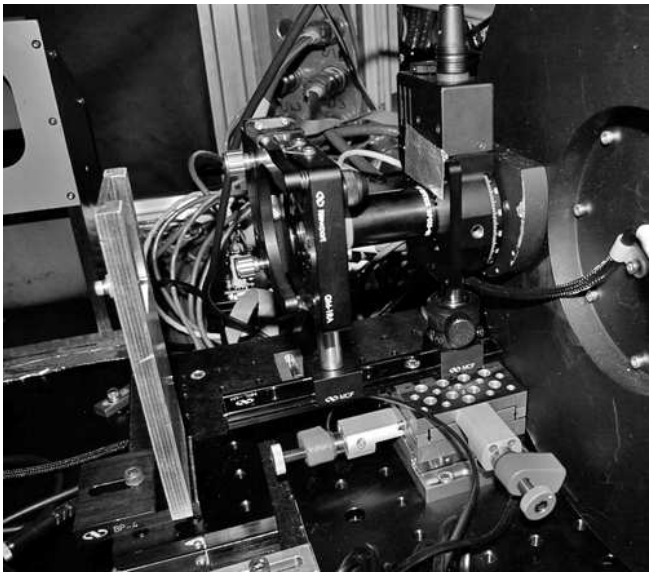}}
\quad
\subfloat[The Spare Chip]{\label{fig:sparechip}
\includegraphics[width=0.45\textwidth]{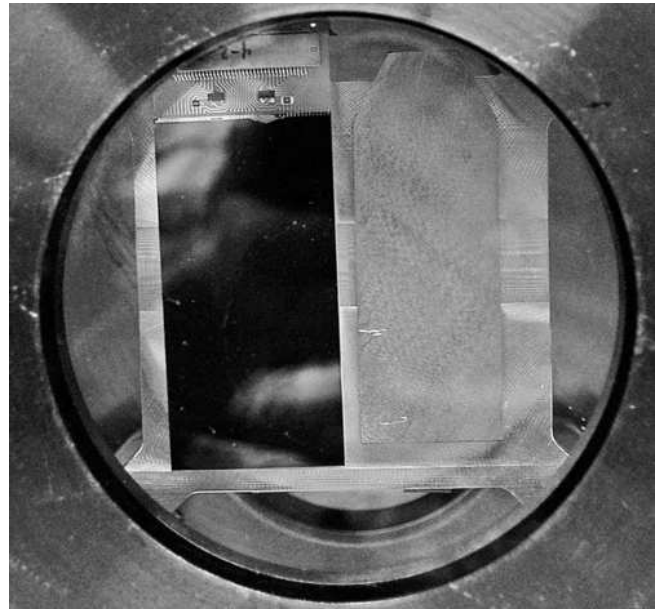}}
\vspace{5mm}
\caption[RFI, New Dekkar, Remounted Optics, Spare CCD]{ {\bf a)} The RFI present in the raw data in September 2006. An small region of a spectral order is shown with the flat-field lamp illuminating the instrument. This interference caused pulsed non-repeatable saturated spikes, seen as the small white and black streaks across the chip. The intermittent interference made this CCD repair useless. {\bf b)} A new dekkar for the slit. The movable arms allow for precise control of the slit-width reduction. {\bf c)} The polarizing optics were remounted in tip-tilt mounts with a smaller path length on a remote-controllable XY stage. {\bf d)} The spare CCD-ID20 mounted from August to October 2006 after attempted repairs on the science camera and original CCD's. }
\label{fig:instrum-improv}
\end{figure}
\twocolumn

\subsection{Instrument Summary}

	The polarimeter has been remounted with better alignment and less vignetting.  A new tip-tilt corrector has been installed and calibrated.  A hard-mounted, adjustable dekkar and polarization calibration mount have been installed and allow for quick, easy, and repeatable calibration and adjustment.  The failed CCD-ID20's were replaced with a similar copy, but an unresolved hardware/RFI issue caused this repair to fail.  A new commercial 1k$^2$ camera was mounted in place of the old science camera and is working wonderfully.  The calibration stage had lenses mounted that increased the flat-field flux by a factor of 6 and the ThAr lamp flux by a factor of 150.  These improvements greatly increase the ease and accuracy of spectral calibration.

\subsection{Observing With AEOS - Measuring Linear Polarization}

	The Stokes parameters I, Q and U fully specify the linear polarization state of light.  Q is the difference between horizontal and vertical polarizations.  U is the difference between +45$^\circ$ and -45$^\circ$ polarizations.  Typically, the fractional Stokes parameters, q=Q/I  and u=U/I, defined in chapter 1, are presented to quantify the fraction of the light which is polarized.  These are measured by taking the fractional difference between the orthogonally polarized spectra. 

	The observing sequence with the AEOS spectrograph is to take spectra at 0, 22.5, 45, and 67.5 degrees rotation of the half-wave plate with respect to the axis of the Savart Plate, hereafter called orientation 1, 2, 3, and 4.  Incoming light linearly polarized at some angle to the fast axis of the half wave plate will have its linear polarization states rotated by twice that angle.  For example, incoming light polarized at 22.5$^\circ$ with respect to the wave plate axis will exit polarized at 45$^\circ$ (Stokes U $\rightarrow$ Q).  Since the axis of the Savart plate remains fixed, the rotation of the wave plate by 45$^\circ$ will move linearly polarized light from one polarized beam to the other, swapping their location on the CCD and reversing their sign in the fractional polarization calculation (Q $\rightarrow$ -Q).  This allows for cancellation of systematic errors (derivative, misalignment, CCD response, etc.) in the Stokes Q and U calculations.  The waveplate angle can be controlled to 50$''$.
	
	The observing sequence is illustrated in figure \ref{fig:4Ori}.  A linear polarizer was mounted upstream of the spectropolarimeter aligned with the waveplate's axis and a normal data set (4 waveplate orientations) was obtained.  Notice how the bright order is swapped (top to bottom) with a 45$^\circ$ rotation of the waveplate and how the 22$^\circ$ and 67$^\circ$ orientations show uniform intensity for both spectra.  The rotation of the waveplate will allow the linearly polarized light to be sampled by moving it from one polarized order to the other in a systematic way.
 
	A frame from the slit-viewer camera is shown in figure \ref{fig:SlitQU} with the general geometry of the slit, dekker, and room as seen by the slit viewer camera.  The slit is parallel to the floor.  +U is parallel to the slit, and +Q is rotated $45^\circ$ counter clockwise, pointing to the upper right.  Since the image rotator was not used, the projection of the slit on the sky is a function of pointing and is not constant. 

 	The simultaneous imaging of orthogonal polarization states also allows for a greater efficiency observing sequence and for reduction of atmospheric and systematic effects.   Since both polarized spectra are imaged simultaneously with identical seeing, a single polarization measurement (Q or U) can be done with a single image, and consistency between images is easy to quantify.  
	
	The polarization angle is measured in the telescope frame and a projection of the slit's position angle onto the sky is done with the K-cell.  The AEOS image rotator was not used so that the absolute angle of the slit on the sky was not fixed.  However, the orientation of the K-cell mirrors in the optics room was fixed, making the polarization calibration much easier.

\section{The Data Reduction Package}

  The spectropolarimetric reduction pipeline can be broken into a few basic parts:  1) combining flats, darks, biases, 2) setting up the coordinate grids as x-pixel, y-pixel, slit-tilt and wavelength calibration,  3) calculating a first-pass spectrum by simply averaging all pixels at a given wavelength,  4) calculating more spectra based on optimal-extraction routines, with cosmic ray rejection and flux-weighting,  5) doing the polarization calculations, and 6) simulating and quantifying the systematic errors.

\subsection{Combining flats and setting up the coordinate grid}

  The first step is to produce an individual flat field, calculated as an average of many individual flats. Typically, an average of 15-40 individual exposures is used with a 3-$\sigma$ outlier rejection, a so-called clipped-mean, to produce an extremely accurate calibration.  Since this program depends on very high S/N measurements, a minimum of 20 flat field images of 10000 ADU each is required to get good S/N.  The dark and bias calibration is done in a less-than-traditional way because of a slowly-fluctuating, non-repeatable background level. This effect is actually quite common with most detectors.  There was an overscan region in the original science detector with the CCD-ID20's which allowed measurement of the changing dark levels throughout the night.  The replacement Pixis 1024BR array is significantly smaller than the focal plane and is always exposed to light.  Both the original Lincoln Labs CCD-ID20's and the Pixis detector showed a significant ($>$few\%) drift in the bias levels throughout a night.  The Pixis camera had no overscan region, but the median values of all the darks taken over many months did show variation in the overall background levels. 
  
\onecolumn
\begin{figure}
\centering
\subfloat[Reduction Coordinate System]{ \label{fig:coord} 
\includegraphics[width=0.35\textwidth, angle=90]{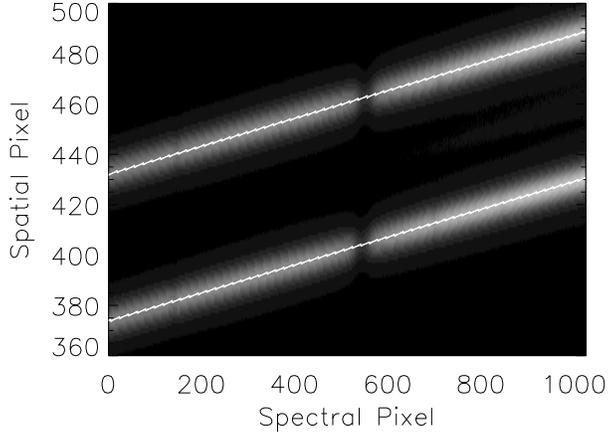}}
\vspace{-1mm}
\quad
\subfloat[Flat Field Spatial Cuts]{ \label{fig:flatslice} 
\includegraphics[width=0.35\textwidth, angle=90]{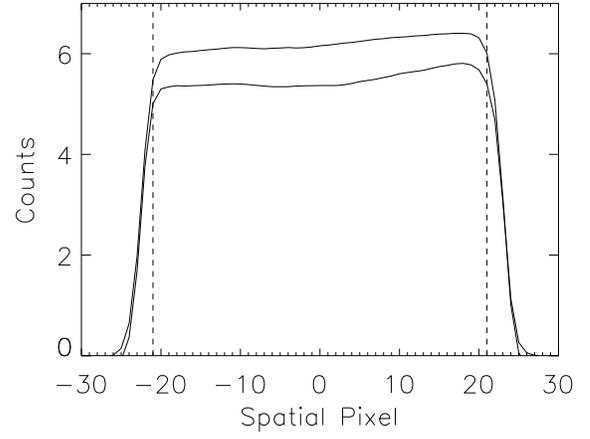}}
\quad
\subfloat[Calibrated Flat Field]{\label{fig:calibratedflat} 
\includegraphics[width=0.65\textwidth, angle=90]{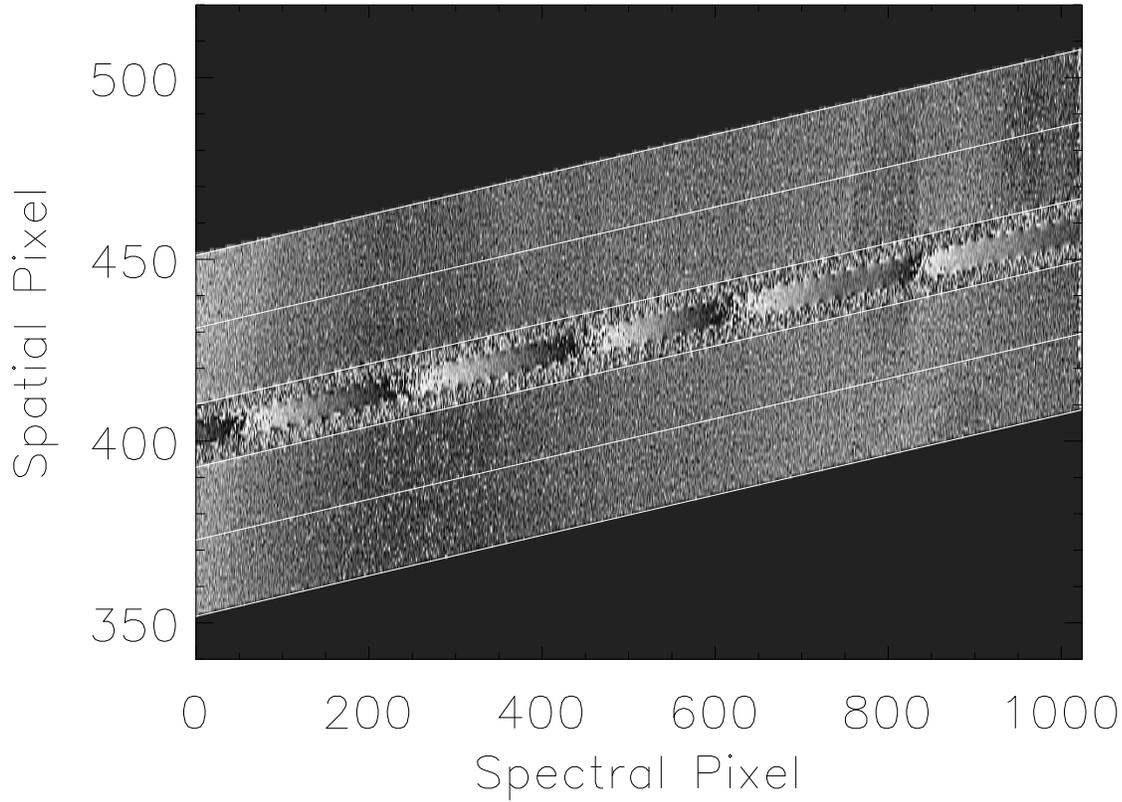}}
\vspace{0mm}
\caption[Coordinate Grid \& Flat fielding]{{\bf a)} The user-defined points serve as a starting point for the peak-finding routine that fits a 3$^{rd}$ order polynomial to the shape of the order, as specified by a bright stellar template. The gaussian-fit centers are plotted as the jagged line. The polynomial fits to the peak intensity follow the stellar brightness peak quite nicely. {\bf b)} The average spatial profile of the two polarized flat-field orders across the H$_\alpha$ region. The width of the orders is set by the dekkar width. The vertical lines mark the edge of the illuminated field selected for use. The small tilt of the spatial cuts traces the inhomogenity of the illumination field at the slit. The difference in height between the two cuts corresponds to the intrinsic continuum polarization of the instrument downstream of the calibration stage. {\bf c)} The calibrated flat field showing the center of each order and the width of each order. The region used for the background subtraction begins 10 pixels above and below the order.  The grey-scale runs from 0.95 to 1.05 showing the pixel-to-pixel structure. The region in the middle is excluded from the reduction.}  
\label{fig:flat-coord}
\end{figure}
\twocolumn  
  
  These variations with this instrumental setup and background levels required a background subtraction from each frame using the unexposed pixels outside the echelle orders as this directly tracked the background levels of the chip and reduced the systematic uncertainties.  For the short exposures of this program, the dark current was small, and is automatically corrected with the frame-by-frame background subtraction.

  Since the instrumental setup and the order locations on the CCD's has changed much over the last few years, and changed quite often during the instrument work done in the fall of 2006, setting up the coordinate grid requires more user-input than most dedicated reduction scripts.  The user must supply a guess at the polarized order locations, typically 5 points to fit a 3rd-order polynomial, and a bright stellar image taken that night to serve as a template for the curvature of the orders.  Figure \ref{fig:coord} shows a bright standard with the narrow H$_\alpha$ line near the center of the 1k$^2$ array.

\begin{figure}
\centering
\hspace{-2mm}
\subfloat[Thorium-Argon Spectra Comparison]{\label{fig:thar-fit} 
\includegraphics[width=0.35\textwidth, height=0.45\textwidth, angle=90]{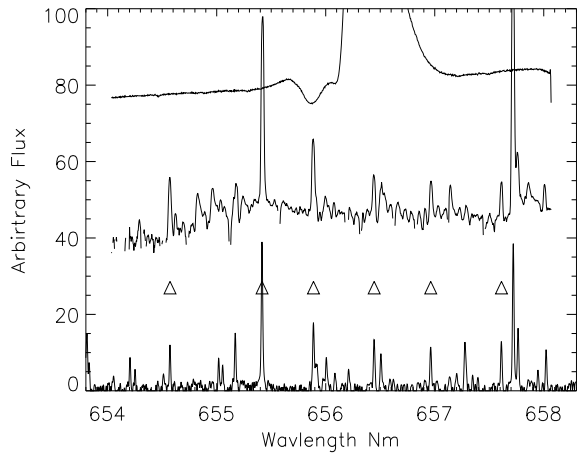}}
\quad
\vspace{2mm}
\subfloat[Thorium-Argon Tilt Alignment]{\label{fig:thar-align}  
\includegraphics[width=0.5\textwidth, height=0.45\textwidth, angle=90]{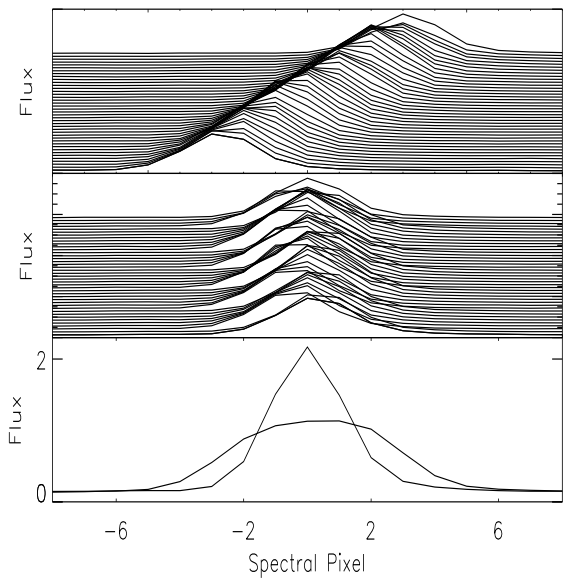}}
\vspace{4mm}
\caption[Wavelength Calibration \& Tilted Coordinates]{{\bf a)} The NOAO ThAr spectra is on the bottom. The calculated HiVIS 0.35" ThAr spectrum is in the middle. The triangle symbols mark the lines used in the polynomial fit. A sample unprocessed AB Aurigae spectra is shown as the top line to illustrate the wavelength coverage of a single spectral order. {\bf e)} The spectral shift of the Th-Ar lines along the slit is calculated and used to set up a tilted coordinate grid. Main result is a 70\% increase in resolution between reductions done with tilted and untilted coordinate grids (R=30000 to R=49000 for the 0.35" slit).}
\label{fig:lam}
\end{figure}

  The code sets up an initial polynomial fit to these user-input points, and then takes spatial (vertical) cuts at each spectral (horizontal) pixel. Each of these spatial cuts are then fit by a gaussian to determine a robust location of the peak intensity in the spatial direction across the CCD, reducing the influence of cosmic rays or chip defects.  These peak intensity positions allow for a robust fit to the order shape on the CCD with a slope of around 5\% (crossing a spatial pixel every 18 or 20 spectral pixels).  The orders are fit independently and the shapes are the same to the 0.1 pixel level ($\sim$0.2\% of an order width).  The order separation, caused by the Savart plate, is then calculated as the median difference between the two polynomial fits.  For the Pixis camera, this was 58-59 pixels.  The residuals between the calculated peak intensity and the polynomial fits are typically under half a pixel and show the expected oscillation with wavelength as spatial pixels are crossed.  
  
     The star observed for the template is never completely centered in the slit from small guiding errors.  Once these order shapes are calculated from the stellar observations, the polynomials are adjusted to reflect the true slit center.  This is done using the flat-field.  First an average spatial profile (slit illumination pattern) is calculated for the flat field, centered on the stellar template by simply averaging the flat field along the order template in the spectral direction.  These spatial profiles, shown in figure \ref{fig:flatslice} clearly show the slit boundaries, where the flat field intensity falls off sharply.  The flat-field spatial cuts are cross-correlated to find the offsets from the stellar order-shape polynomial fits to the center of the flat-field spatial cuts.     

  Once the offset from the stellar template to the flat-field region is calculated, the full coordinate grid can be computed with the order centers, edges, and background locations specified.  Once the coordinate grid is set, a calibrated flat field can be created, adjusting for the blaze function, and wavelength dependence of the flat-field lamp (blackbody). 

   A flat-field is supposed to correct for pixel-to-pixel gain and field-dependent efficiencies only.  The flat-field values should not introduce any blaze, lamp or  spatial illumination dependencies.  This means that the flat-field must be corrected for it's inherent spatial and spectral structure.  To correct the flat field spectral dependence (horizontal), a simple flat-field spectrum is made for each polarized order by averaging all pixels spatially at a given spectral position.  This spectrum has a bit of intrinsic noise in it so a low-order polynomial fit is made.  The flat-field image then has this spectral dependence removed.  

  The spatial structure is corrected by dividing each spatial cut along the flat-field spectrum by the average spatial profile, removing the spatial dependence of the flat (inhomogeneous slit illumination) at each wavelength.  This makes the flat-field much closer to a true pixel-to-pixel gain measurement, shown in figure \ref{fig:calibratedflat}.

\subsection{Wavelength Calibration}

  With the recent calibration-stage improvements that have implemented, it became possible to do very accurate wavelength calibration using low order polynomial fits to at least six clear, unblended lines, just over the H$_\alpha$ spectral region.  With a $>$600x increase in ThAr flux (15x without the diffuser-bypass mirror), the smallest slit (0.35$''$) at highest resolution (R=49000) can be used to clearly identify lines and compare the lines with standard line-lists obtained from the NOAO spectral atlas.  The new HiVIS detector has a wavelength scale of 0.037{\AA} per pixel (1.7 km/sec per pixel at H$_\alpha$).  Thus the rotation of the Earth can produce up to a $\frac{1}{3}$ pixel shift and seasonal variations can approach 40 pixels.  
  
  The wavelength calibration is done in a two-step process.  A Th-Ar spectrum taken with the 0.35" slit is used to identify the lines.  The Th-Ar spectra are calculated with the simple average over all illuminated pixels.  The unblended lines are identified and the peaks are fit by a gaussian to determine the line centers.  A 3$^{rd}$ order polyomial fit to the line centers is used to create a wavelength array.  The NOAO atlas spectrum, HiVIS spectrum, and a comparison AB-Auriga spectrum are shown in figure \ref{fig:thar-fit}.  The triangles mark the 6 lines used in the polynomial fits.  Since all stellar observations were done with the 1.5$''$ slit, a 4-orientation sequence of ThAr spectra are taken with this slit and then reduced with the simple-average method.  The resulting lower-resolution spectrum is cross-correlated with the higher-resolution spectra produced by the 0.35$''$ slit to determine any shifts between slits and waveplate orientations, the beam wander or wobble, which is a small fraction of a pixel.  

    The average slit tilt is then calculated to modify the coordinate-grid for the reduction program.  The tilt is a spectral-direction shift of $\sim$7 pixels between the bottom and top of each order (across the 50 pixel slit width).  This shift is calculated as a cross-correlation of a spectral cut beween each spatial position along the order.  This tilt is then implemented as a integer-pixel shift of the coordinate grid at each spatial point.  There is a very significant increase in resolution when the tilted-coordinates are used.  Figure \ref{fig:thar-align} shows the spectral cuts before and after alignment, as well as the intensity of the line calculated with both tilted and un-tilted coordinate grids.  The spectral resolution derived from the Th-Ar line fits goes from R=30000 to R=49000 when using tilted coordinates. 

\subsection{Calculating Raw Spectra}

  Once the tilted coordinate grid, gain-table, and wavelength calibration is done, the data can be processed.  Each exposure is divided by the gain-table (calibrated flat). A background spectrum is made by averaging the unexposed background region outside the polarized orders.  This background is noisy and so a linear fit to this background is subtracted from each order to remove the dark current and bias of the detector.  
  
  To create a simple, first-pass at the spectrum, every pixel at each wavelength was simply averaged to compute the intensity.  At the same time, an average spatial profile (PSF) is constructed by averaging all these spatial cuts spectrally.  While this method does not correct for cosmic rays, and does not weight the pixels by their fractional-stellar-flux or PSF, this simple average spectrum provides a starting point for more sophisticated reductions.
   
\subsection{Shift-n-Scale}
  
       In many parts of the reduction package, one curve must be shifted and scaled to another.  A least-squares routine that allows the calculation of the shift and scaling coefficients between two input curves was developed as part of the dedicated software package.  The routine assumes that the scale and shift from one curve to another can be written in terms of the input curve and it's derivative.  To illustrate the routine a shift-n-scale between the two polarized spectra, $I_1$ shifted and scaled to $I_2$, is calculated below:

	\begin{equation}
	I_2 = aI_1(x-b) 
	\end{equation}

	I is the intensity as a function of CCD pixel x, a is the relative gain between the two orthoganally polarized spectra, usually $\sim$1, and b is the spectral shift in pixels, usually $\leq$0.8 pixels.  The spectrum $I_1(x-b)$ can be expanded to first order as below:

	\begin{equation}
	 I_2=aI_1- ab(dI_1/dx)  
	\end{equation}

	Rewriting the constants $a=K_1$ and $-ab=K_2$ and bringing all terms to one side gives the difference between the spectra $I_2$ and $I_1$ scaled and shifted by a and b.  The least squares routine minimizes the function below for all pixels (x):

	\begin{equation}
	 \Sigma (I_2-K_1I_1-K_2dI_1/dx)^2
	\end{equation}

\begin{figure}
\centering
\subfloat[Shift-Scale Example]{ \label{fig:shfscl} 
\includegraphics[width=0.35\textwidth, angle=90]{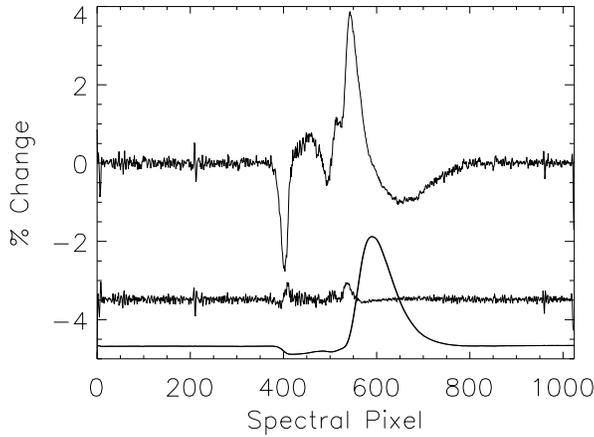}}
\quad
\subfloat[Cosmic Ray Removal]{ \label{fig:cosray} 
\includegraphics[width=0.35\textwidth, angle=90]{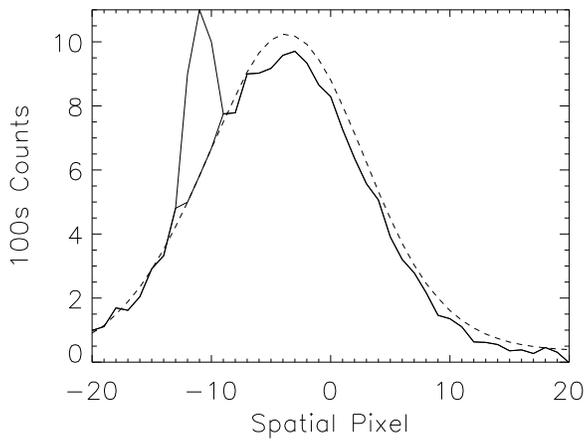}}
\vspace{0mm}
\caption[Shift-n-Scale \& Cosmic Ray Removal Routines]{  {\bf a)}  An example of the shift-n-scale algorithm correcting for optical misalignments.  A single spectrum, the bottom curve, was replicated and shifted spectrally by 1 pixel and then fed to the shift-n-scale routine.  The top curve is the resulting large-amplitude (5\% $\frac{a-b}{a+b}$) fractional differences.  This curve follows the derivative and must be corrected.  The shift-n-scale routine gave a scale of 0.999846 and a shift of 1.00742.  The shift-n-scale interpolation causes some residual noise, since the input curves were identical and should have subtracted completely.  This arises because of the smoothing done to calculate the derivative.  The middle, mostly flat curve shows this residual noise doubled for clarity.  The noise is at the level of 0.2\% with some residual systematic error.  {\bf b)}  This shows a cosmic-ray removal from a spatial cut.  The cosmic-ray removal routine computes the deviation of the spatial cut from the average line profile and identifies 4-$\sigma$ deviations in each spatial cut from the average spatial-profile (PSF) and replaces  the cosmic ray hits with the value for the least-squares scaled average spatial profile.  The smooth dashed curve is the computed least-squares scaled average spatial profile and the jagged curve is an individual spatial cut.  Three pixels were found to be $>4\sigma$ away from the average profile and were corrected.}
\label{fig:shfscl-cosray}
\end{figure}

	Expanding the square and collecting terms while implicitly assuming that each term is a sum over all pixels gives:

	\begin{equation}
	 I^2_2 + K_1^2 I^2_1 + K^2_2 \frac{dI_1}{dx}^2  - 2 K_1I_1 I_2 + 2 K_1 K_2 I_1 \frac{dI_1}{dx} - 2 K_2 I_2  \frac{dI_1}{dx}
	\end{equation}

	To solve for K$_1$ and K$_2$ two equations must simultaneously solved. Setting the derivative of this equation with respect to K$_1$ and with respect to K$_2$ equal to zero gives two equations that can be solved for the coefficients:
	
	\begin{equation}
	 I_1 I_2 = K_1 I^2_1 + K_2 I_1  \frac{dI_1}{dx}
	\end{equation}

	\begin{equation}
	 I_2  \frac{dI_1}{dx} = K_1 I_1 \frac{dI_1}{dx} + K_2 \frac{dI_1}{dx}^2
	\end{equation}
	
	Which gives the coupled equations that can be solved for the coefficients:
	
	\begin{eqnarray}
	  \left[ \begin{array} {c}  I_1 I_2 \\ I_2 \frac{dI_1}{dx} \end{array} \right] = 
         	  \left[ \begin{array} {ll}  I^2_1 &  I_1 \frac{dI_1}{dx} \\ I_1 \frac{dI_1}{dx} & \frac{dI_1}{dx}^2 \end{array} \right]
	  \left[ \begin{array} {c}  K_1 \\ K_2 \end{array} \right]
	\end{eqnarray}

	The derived coefficients $K_1$ and $K_2$ are then used to calculate the scaled and shifted functions.  An example of this routine applied to a spectrum of AB Aurigae is shown in figure \ref{fig:shfscl}.  A single spectrum was replicated and shifted by one pixel and corrected by the shift-n-scale routine.  The routine uses a smoothed profile to calculate the derivative with a variable amount of smoothing. For all the data presented here, a smoothing of 3 pixels was used.  This illustrates how a misalignment of the savart plate, simulated as a spectral shift of 1 pixel between the orthogonally polarized spectra, produces a fractional difference of order 5\% from this misalignment.  This signature is a systematic error that follows the derivative of the spectral line.  The scale and shift operation minimizes this derivative signal as shown (amplified) in the bottom line in the plot.  The residual noise is at the level of 0.2\% and has some residual systematic error in it from the smoothing used to calculate the derivative, but the derivative-based error has been attenuated by a factor of about 20.  

	The shift-n-scale routine also has some nonlinearity to it, and is best used for sub-pixel shifts when cross-correlation techniques fail. For example, when shift-n-scaling the original AB Aurigae spectrum to the simulated one in figure \ref{fig:shfscl}, the routine reproduced the shifts to better than 0.05\% for spectral shifts of 5 pixels, far greater than the range needed. However, when shift-n-scaling the simulated curves to the original curve, the calculated shifts had an error of $>$0.1 pixels for spectral shifts $>$0.8 pixels. This shows that for even strongly varying line-profiles, the shifts are reproduced very well and, when used with cross-correlation techniques for single-pixel alignment, can produce very accurate sub-pixel alignments.

\subsection{Optimal Spectral Extraction}

  This first-pass spectrum and average spatial profile are projected back on the original spatial cuts with the shift-n-scale routine.  The expected noise at each spatial pixel is found (using estimates of the gain and read-noise).  The standard deviation of each spatial pixel at each wavelength is then calculated and used to find any cosmic-ray hits as cosmic-rays (and chip defects) which will be strong devations ($>4\sigma$) from the spatial profile.  An example of this cosmic ray removal routine is shown in figure \ref{fig:cosray}.  The average spatial profile is the dashed line.  The spatial cut at this wavelength shows a few pixels deviating strongly from the average spatial profile from a cosmic ray hit.  These pixels are identified and replaced with the shifted-and-scaled spatial profile, greatly reducing the error in the spectrum.  Once this cosmic ray removal routine is done, a spectrum is calculated as the simple average of all the pixels.  The corrected spatial cuts are passed on to more optimal spectral extraction routines.
	
\begin{figure}[!h]
\includegraphics[width=0.4\textwidth, angle=90]{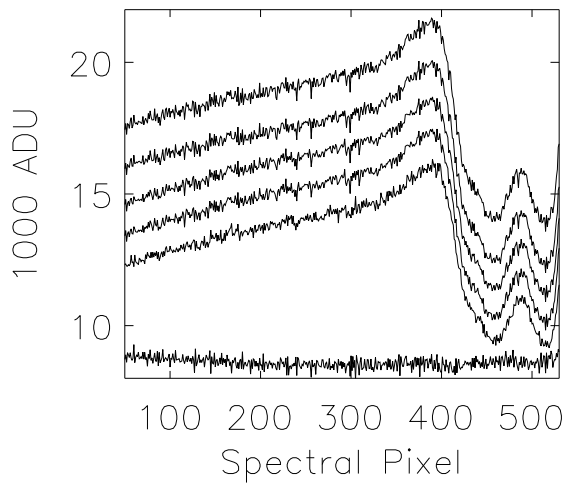}
\caption[Various Reduction Examples]{This shows a spectrum of AB Aurigae reduced with varying parameters to show the robustness of the reduction program.  Each curve represents a different reduction using order widths from 43 to 49, with or without flat fielding, and with or without tilted-coordinates.  From top to bottom 1) tilted, 49 width, no flat 2) tilted, 43 width, no flat 3) tilted, 43 width flattened 4) tilted 45 width flattened  5) untilted 49 width no flat.  The bottom curve shows the residuals from the unflattened untilted 49-width curve and the flattened tilted 43-width curve.}
\label{fig:difred} 
\end{figure}
  
    The first of the optimal extraction routines uses the shift-and-scale routine to fit the average spatial profile (PSF) to each of the individual spatial cuts at each wavelength.  This reduction process assumes that the spatial profile does not vary across the order, an assumption that all optimal reduction packages adopt (Horne 1986, Marsh 1989, Donati et al. 1997, Cushing et al. 2004).  The calculated error in the shifts between the optional fit and the calculated order location is dependent on the flux, but is typically less than 0.1 pixels.
        
    Another set of spectra are calculated as simple gaussian fits to the spatial cuts.  While the PSF (spatial cut) is not entirely gaussian, and certain seeing conditions or tracking-errors lead to not-quite-gaussian spatial cuts (PSF's), the fits are usually robust, and match the shift-n-scale and simple-average specta well.  The standard IDL gaussfit routine is applied to each spatial cut, with the initial estimates for the gaussian as:  peak intensity is twice the computed average-spectra, centered on the order, with a fwhm of one-quarter the order width and zero background.  The routine works well when the spatial cuts (PSF's) are roughly gaussian.
	
        Thus, there are three spectra with different systematic effects (simple-average, shift-n-scale, gaussian-fit) -  which are all included in the polarization analysis to probe the polarimetric reduction codes and systematic errors inherent in these reductions.  In practice, the exact choice of reduction parameters is unimportant.  Figure \ref{fig:difred} shows the simple-average method with and without a flat field or tilted coordinates, and varying order widths from 43 to 49 pixels (covering the edges of the illuminated region, see figure \ref{fig:flatslice}).  The tilted coordinates do add a significant increase in resolution, but for the high signal-to-noise spectropolarimetric observations the spectra are rebinned to lower resolution anyways.  The flat field does increase the signal-to-noise of the intensity calculation, but as shown below, the basic polarization calculations are independent of the flat field.  
        
        The AEOS telescope is designed for military applications and does not have a standard siderial guiding telescope.  There are significant tracking errors that can accumulate and there are offsets induced by manual telescope-operator corrections.  The operation requirements set out by the AFRL require that there is no control over the guiding by any outside equipemnt, and that the telescope operators completely control the pointing of the telescope.  There are a number of data set's where an operator's manual steps caused a non-gaussian spatial profile.  This inhomogeneity requires the reduction program to calculate the spectra using all three methods. The simple averaging and shift-n-scaling methods are the most robust. 

  	These three different reductions vary slightly between each other. The gaussian-fit and simple-average methods vary by only a few parts-per-million when the spatial cuts (psf's) are gaussian. However, the shift-n-scale method and the simple-average methods vary slightly across the line profile, giving an overall uncertainty in the shape of the intensity profile.  However, the spectropolarimetry is calculated as the difference between line profiles on the same image.  Since the same reduction method is applied to each polarized order in the same way for each polarization measurement, it is only any differential error that would cause systematic errors in the spectropolarimetry.  The reduction package uses all three calculation methods to assess the systematic errors that might be present in each method, or in the subsequent processing of the spectra.
   
\subsection{Continuum Subtraction}
  
  The focus in this work is on spectral and spectropolarimetric variability. With the many spurious polarization effects such as instrumental, telescope, interstellar, and intrinsic continuum polarizations, a continuum-polarization subtraction must be done to isolate the change across any individual spectral line. Since spectral variability is always referenced to continuum, no spectrophotometric or flux calibration is done, though standards are observed for future work.  
  
  The continuum is simply subtracted as a linear fit to the intensity on either side of the spectral line. This continuum subtraction normalizes the slope of the black-body of the star, removes the instrumental continuum polarization signals, and removes the instrumental blaze efficiency. Continuum subtraction is done by averaging 20 pixels sufficiently far enough away from chip edges.  There is a variable dark signal that curves the echelle orders near the edge of the chip, 10-20 pixels wide, in a non-repeatable way, so this region is avoided.  Cirrus, tracking errors, and wind-bounce all influence the throughput of the spectrograph and thus, even on a spectrophotometric night, a significant variation in the continuum flux is sometimes seen.
  
    The data taken before 10-27-2006 was imaged using an echelle order which placed the H$_\alpha$ line near the edge of the chip.  This placement was done before some of the instrumental corrections, and before the remounting of the Pixis camera to a different region of the focal plane.  In the more recent data, H$_\alpha$ is closer to the blaze-efficiency peak and is much closer to the center of the focal plane.  The continuum fits to the CCD-ID20 images were not as accurate because the red wings of the H$_\alpha$ line were near the order edge, giving rise to a small ($\sim$1\%) error in slope estimate.  However, this uncertainty is negligable for data taken on each night because the continuum is calculated in the same way for each instrumental setup on each night.  This only produces an uncertainty in the night-to-night variability of the line profiles, and in the overall slope of the calculated continuum and continuum polarization.

\subsection{Possibility of Spectro-astrometry}

\begin{figure} [!h]
\subfloat[Spectroastrometry MWC 166]{\label{fig:specast-mwc166} 
\includegraphics[width=0.34\textwidth, angle=90]{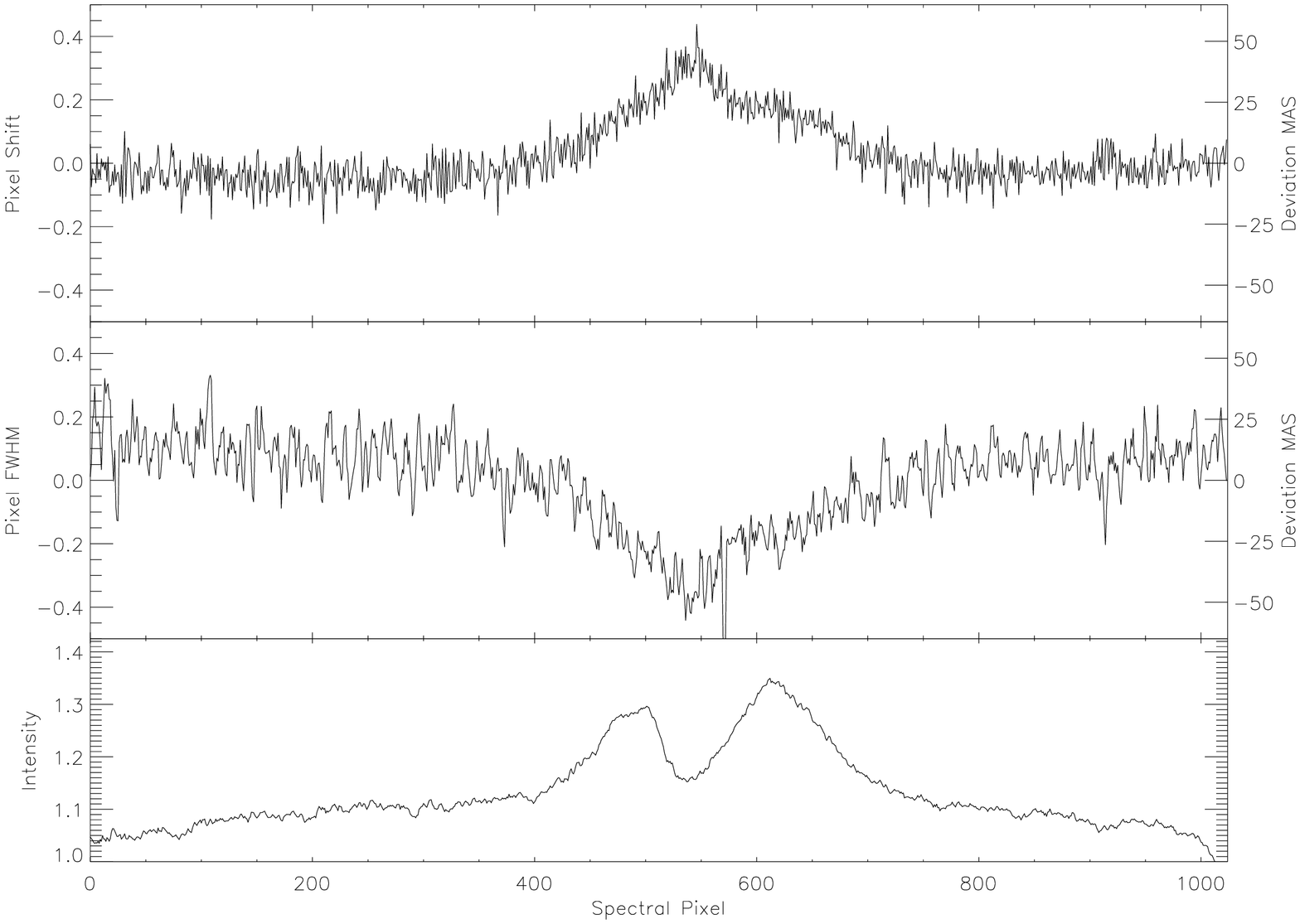}}
\quad
\subfloat[Spectroastrometry $\gamma$ Cas]{\label{fig:specast-gmcas}  
\includegraphics[width=0.34\textwidth, angle=90]{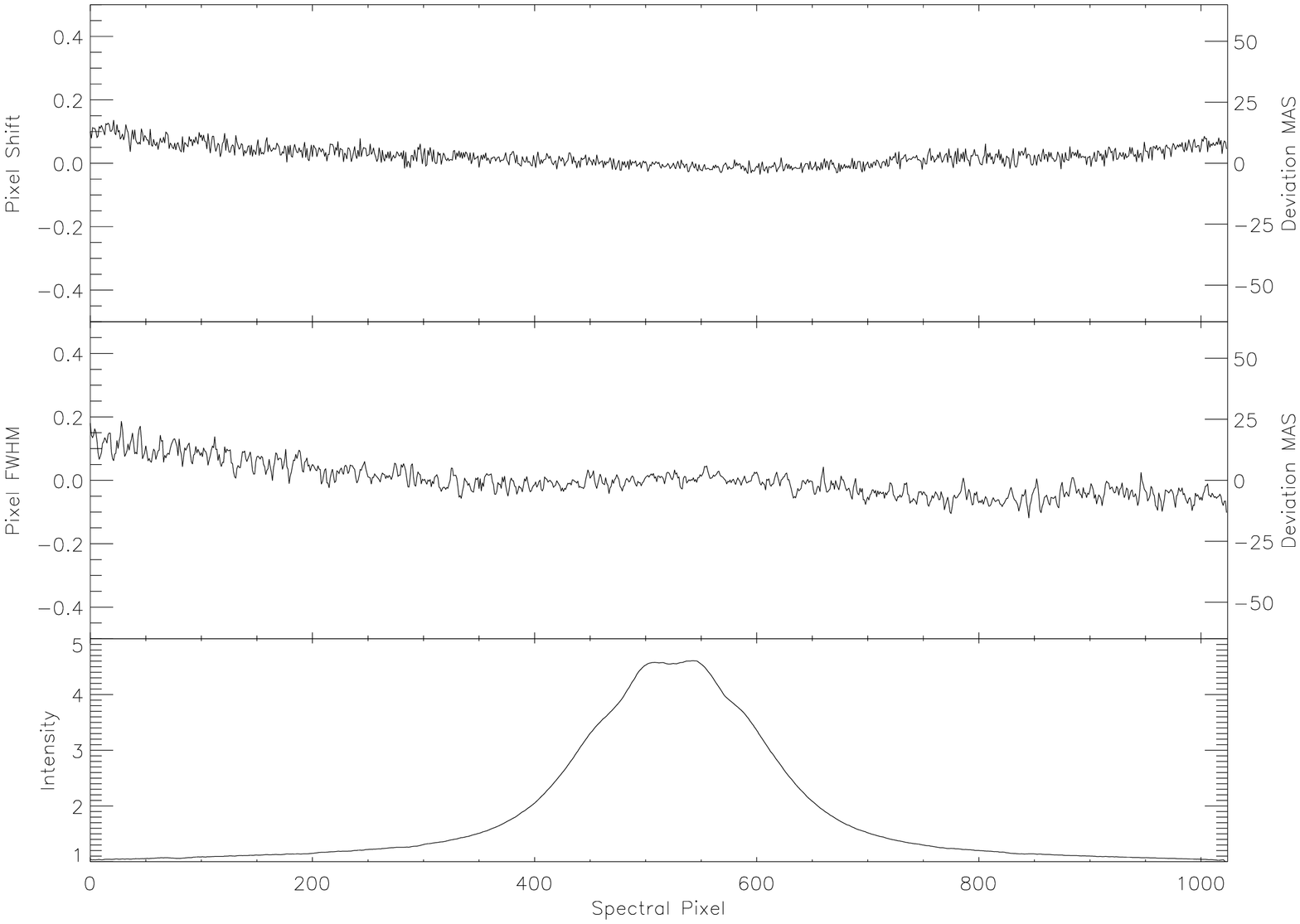}}
\caption[PSF Properties for MWC 166 \& $\gamma$ Cas]{  {\bf a)}  The spectroastrometry of MWC 166 and {\bf b)} $\gamma$ Cassiopeia. The top box shows the center of a Gaussian fit to the PSF at every spectral pixel. The middle box shows the calculated FWHM for each spectral pixel. The bottom box shows the resulting normalized intensity. A systematic effect is seen only across MWC 166.}
\label{fig:specast166-gmcas}
\end{figure}

 	At this point it is useful to demonstrate the coordinate grid and basic properties of the reduction package by demonstrating the wavelength dependence of the point-spread function properties. Several authors have argued that the change in centroid location (brightness peak) or full-width at half-max (fwhm) can be used as evidence for binarity in a set of spectroastrometric calculations (Beckers 1982, Bailey 1998, Baines et al. 2006). Spectroastrometry in principle is a technique that can detect unresolved binaries as a change in spatial profile width (psf) and location (centroid) across an emission or absorption line where the relative brightness of the two binary components significantly varies across the line. For example, if one star has an emission line and the other an absorption line, the centroid will shift towards the emitter in the core of the emission line. If the binary is significantly wide and bright, the fwhm would also decrease across the line as the relative contribution to the psf from the wide companion is reduced. This method makes use of the assumption that the psf across a line or spectral order does not change much, and that any significant change can be attributed to the source. Some have even developed routines to separate the fluxes from the individual binary components (Porter et al. 2004). HiVIS has been used to observe a number of young or emission-line stars as part of a large survey (Harrington \& Kuhn 2007) all of which have strong H$_\alpha$ lines. Given the reported detections and potential of this technique, the feasibility and capability of HiVIS to do similar work will be outlined.

\begin{figure} [!h]
\centering
\includegraphics[width=0.35\textwidth, angle=90]{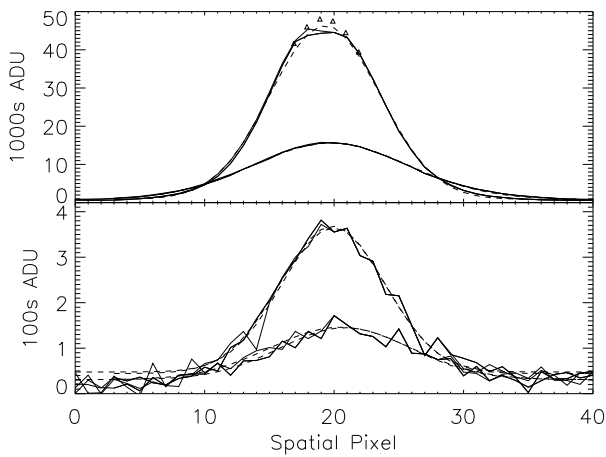}
\caption[PSF Properties for P-Cygni]{The spatial profile or psf of P-Cygni on two separate exposures and three different reductions.  One exposure was clipped (nearly saturated) to illustrate a systematic error in wavelength dependence of the psf.  The top panel shows the psf at the emission peak.  The triangles mark a gaussian fit to the non-clipped part of the psf.  The bottom panel is the psf in the absorptive trough.  The dashed lines in both panels show the Gaussian fits.  Notice how the wings of the Gaussian are significantly above the noise in the low-flux absorptive trough.}
\label{fig:psf-pcyg}
\end{figure}

	Binaries are common among pre-main-sequence stars and this technique can easily be applied to every data set taken with this instrument.  It also provides a good check on any systematic errors such as order mis-fitting or psf mis-fitting since any systematic error will show up with this measurement technique.  Since the centroid of a spatial profile on most spectrographs can be calculated to a fraction of a pixel accurately in high signal-to-noise data, this technique has the capability to detect binarity to less than 30mas separations or a third of a pixel when the pixels are 130mas (cf. Bailey 1998, Baines et al. 2006).  Normally, with a long-slit spectrograph, the slit of the instrument is rotated to a north-south line and an east-west line to fully sample the spatial changes.  This allows one to establish the position-angle of the binary system.  For polarimetric reasons expanded below, the image rotator is not used to keep this component fixed vertically during the observations.  This means that the position-angle rotates from exposure to exposure, but presenting the calculations is still useful to demonstrate the capability of the analysis software.

	For each spectrum, the optimal spectral extraction algorithms fit Gaussians and average spatial profiles (psf's) to the data at each wavelength, as was shown in figure \ref{fig:cosray}.  These spatial profiles can easily be compared with the nominal values for the average spatial profile's FWHM and centroid for each order.  At each wavelength, a deviation for the FWHM and the centroid can be calculated.  Because of small guiding errors, the centroid of each stellar observation does not coincide with the exact order center.  Because of the tilt of the orders on the detector, this will cause the centroid of the spectrum to cross spatial pixels at different spectral pixel locations.  In order to correct for this, the analysis package does a cross-correlation between the nominal order center location, as illustrated by the jig-saw shaped fit in figure \ref{fig:coord}, and the centroid location of the actual observed spectrum.  Once this small offset has been corrected, the centroid location and the FWHM across the order is calculated. Some sample calculations from data taken the 20$^{th}$ of September, 2007 will be outlined below.

	In Baines et al. 2006, the known binary MWC 166 was reported to have a centroid shift of roughly 40mas EW and 15mas NW with corresponding FWHM changes of 40mas and 5mas respectively.  This is tabulated in their paper as a total shift of 50mas at a PA of 287$^\circ$ where the known separation of the binary is 0.65$''$ at a PA of 298$^\circ$.  Figure \ref{fig:specast-mwc166} shows a HiVIS observation of this star.  The observations had a peak of 1500ADU and a continuum of 800-1300ADU across the order and a fwhm of 11.7 pixels (1.5$''$).  The deviation detected in FWHM and centroid offset are both around 40mas, both the correct magnitude for this system.  
	
	We also present observations of $\gamma$ Cassiopeia in figure \ref{fig:specast-gmcas} at the same scale as MWC 166 to show a non-detection across a strong line.  The peak intensity was near 26000ADU with continuum from 4000-6000ADU and a fwhm of 10.1 pixels (1.3$''$).  This non-detection shows that spurious signatures caused from the 1\% Savart plate leak, non-linearity in the chip, spatial dependence of the focus or any other effects are not present at an observable level across an emission line. 

	The star P-Cygni has a very strong H$_\alpha$ line.  This star was observed with two different exposure times separated by 90 minutes.  The reduction was performed with three separate settings: a width of 47 for tilted and untilted coordinates, as well as a width of 41 un-tilted.  This will be used to highlight changes in derived psf properties when including tilts and neglecting low-flux psf wings.  The altitude changed from 50$^\circ$ to 34$^\circ$ and the frame rotated by 28$^\circ$ between the observations.  The line has a normalized intensity of 0.194 in the absorption trough and 18.180 in the emission peak, or a flux ratio of 94.  This huge range in flux and the difference in seeing serves to highlight systematics.  One exposure is deliberately clipped (near saturated) to illustrate systematic errors that can occur with this technique.  The spatial profiles are shown in figure \ref{fig:psf-pcyg}.  The peak intensity of the saturated exposure was 44000ADU with continuum of 1800-3800ADU and only 400ADU in the absorptive component of the line. The calculated fwhm was 9.6 pixels (1.25$''$).  However, a Gaussian fit to the unsaturated region shows that the true peak should have been closer to 48000ADU, or a roughly 10\% clipping.  The unsaturatd exposure taken an hour and a half later and at higher airmass had considerably less flux, 16000ADU peak, continuum of 600-1100ADU, and only 130ADU in the absorptive component of the line.  The calculated fwhm was 14.1 pixels (1.83$''$).

\onecolumn
\begin{figure}
\centering
\includegraphics[width=0.85\textwidth, height=0.9\textwidth, angle=90]{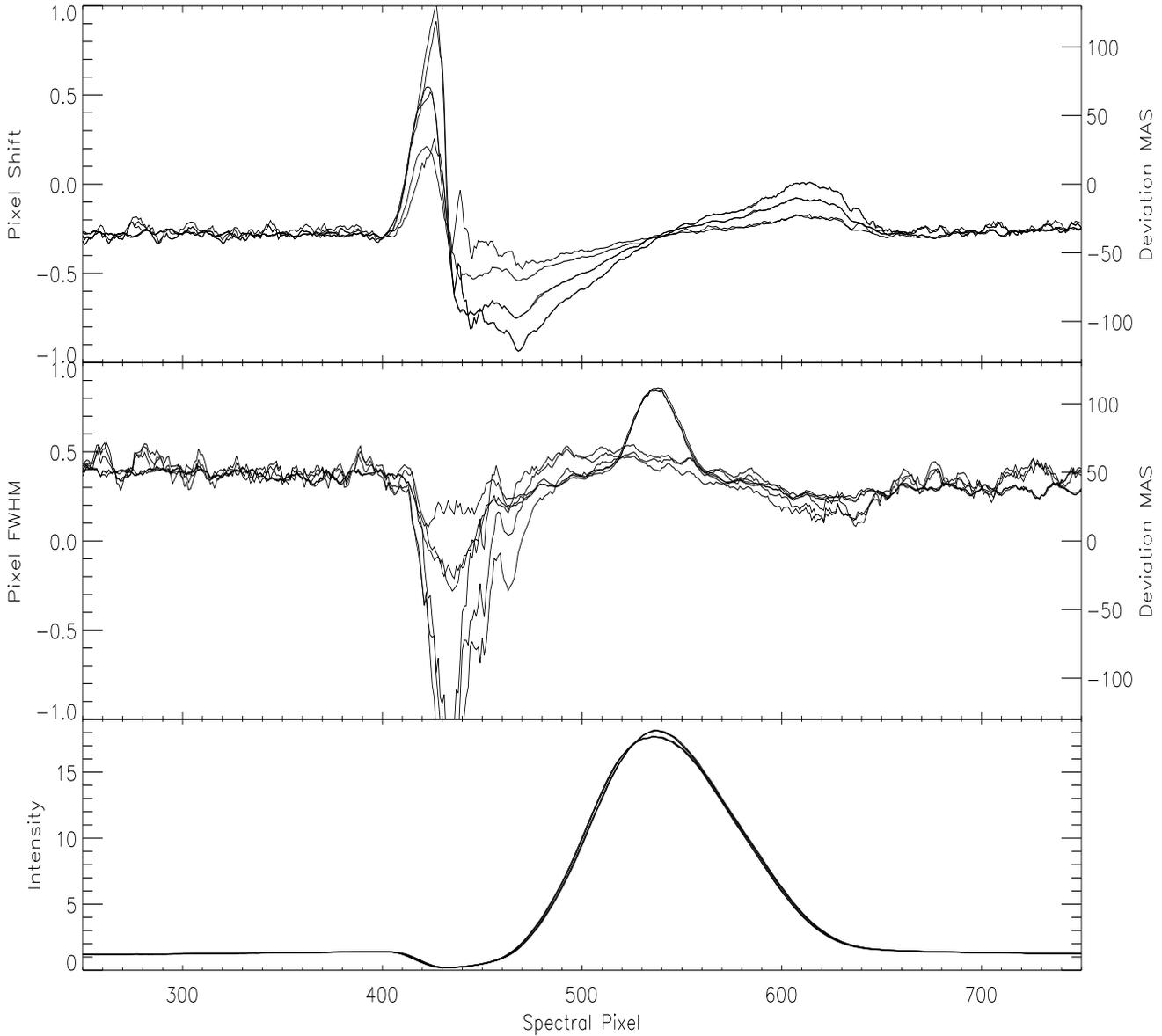}
\caption[Spectroastrometry for P-Cygni]{The wavelength dependent psf properties of P-Cygni for two separate exposures and three separate reductions.  The three different reductions are tilted and untilted coordinates with width of 47 pixels, and untilted coordinates with a width of 41 pixels. The top box shows the center of a Gaussian fit to the PSF at every spectral pixel. The saturated exposure shows a smaller deviation than the unsaturated exposure for every type of reduction meaning that more flux gives better fits. The tilted coordinates do not change the result. The middle box shows the calculated FWHM for each spectral pixel.  The saturated exposure shows a 50mas increase in fwhm in the clipped part of the line.  The unsaturated exposure shows a much greater 200mas decrease in fwhm in the absorptive component of the line. The tilted coordinates do not change the result, but using a smaller order width decreases the deviation, again suggesting that greater flux influences the fits. The bottom box shows the resulting normalized intensity. A systematic saturation effect is seen for all reductions across the emission peak of line where the saturated exposure is underestimated.}
\label{fig:specast-pcyg}
\end{figure}
\twocolumn

	The wavelength dependence of the psf properties shows striking changes across the line, shown in figure \ref{fig:specast-pcyg}.  Both exposures and all reductions show a similar change in centroid location, sharply changing by about 50-150mas in the blue side of the absorptive component, then rapidly dropping across the center of the absorptive component to -20mas to -80mas in the red side of the absorption.  The centroid then gradually rises over the emission to a peak of 5-15mas just before the emission line returns to continuum on the red side of the line.  The change is actually greater in magnitude for the unsaturated exposure with any reduction parameters.  There are actually six curves in the top panel of figure \ref{fig:specast-pcyg} but the difference between order widths of 41 and 47 is completely negligible.  The tilted coordinates show systematically lower deviations for both saturated and unsaturated exposures.  In essence, more flux gives less deviation presumably from better fits.  The tilted coordinates more accurately reflects the true wavelength dependence of the orders and reduces the systematic error.  
	
	The fwhm also shows strong differences between the different exposures and reduction parameters.  The unsaturated exposure with only 140ADU peak flux shows a 200mas decrease in fwhm in the absorptive component of the line in both tilted and untilted coordinates with a width of 47 pixels.  When the wings of the psf are neglected by using a width of 41, the calculated fwhm deviations are significantly smaller, near 75mas, presumably because the wings of the psf have a noise contribution.  This again shows that more flux leads to better fits and that large flux differences across a line lead to systematic effects.  The saturated exposure shows an extra 50mas increase in fwhm across the emission peak in the clipped region of the line for all reductions.

\begin{figure} [!h]
\subfloat[Spectroastrometry T Ori]{\label{fig:specast-tori} 
\includegraphics[width=0.32\textwidth, angle=90]{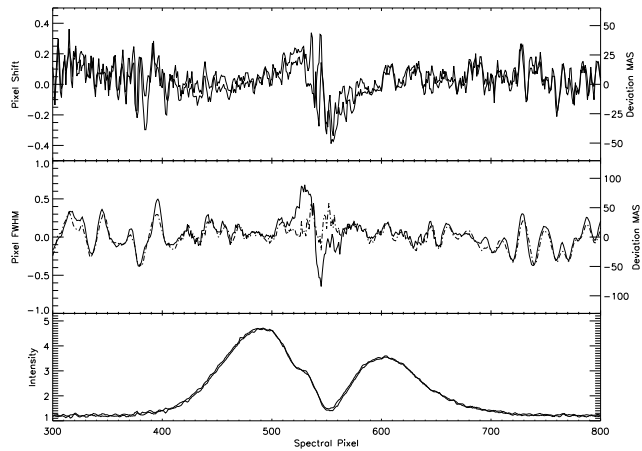}}
\quad
\subfloat[T Ori Resolved Emission]{\label{fig:tori} 
\includegraphics[width=0.35\textwidth, angle=90]{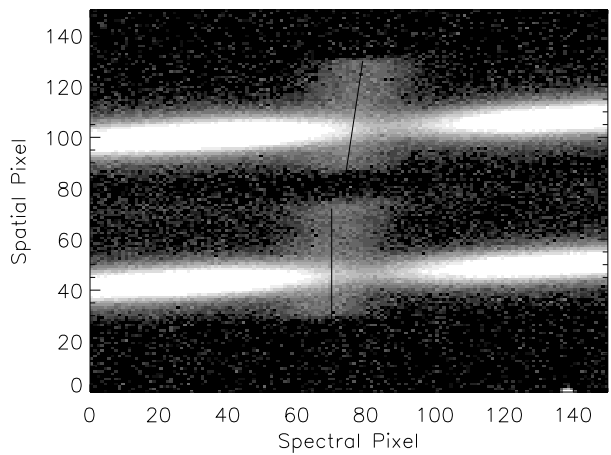}}
\caption[PSF Properties for T Ori with Resolved Emission]{  {\bf a)}  The wavelength dependence of the psf for T-Ori reduced with and without tilted coordinates.  The top box shows the center of a Gaussian fit to the PSF at every spectral pixel.  A systematic centroid shift effect is seen across the emission peak of line where there is extended, resolved H$_\alpha$ emission.  Using a tilted coordinate grid does not change the deviation.  The middle box shows the calculated FWHM for each spectral pixel, smoothed for clarity.  The tilted coordinates actually make the fwhm variation across the emission negligable, showing that this resolved emission does {\it not} contribute substantially to the fwhm.  The bottom box shows the resulting normalized intensity.  {\bf b)}  The raw exposure of T Ori showing the spatially resolved extended H$_\alpha$ emission near the absorptive component of the stellar line.  The lower beam has an untilted spectral slice shown, and the upper beam has a tilted coordinate grid overplotted.  This illustrates a systematic error that may result from extended resolved nebular emission around a source causing a change in centroid and fwhm.}
\label{fig:comp}
\end{figure}
	
	Another complication is extended emission.  A change in fwhm and centroid can be caused by significant resolved emission around an object, such as the observations of T Ori.   If there is no inspection of the raw data, this emission may be falsely reported as a binary detection.  Also - one cannot rule out the possibility of unresolved nebulosity as a significant contribution to the psf.  The observations of T Ori showed a peak signal of 600ADU with a continuum of 100-150ADU.  The fwhm was calculated as 10.1 pixels (1.3$''$).  The wavelength dependence of the psf for two reductions, using tilted and untilted coordinates, is shown in figure \ref{fig:specast-tori}.  The centroid shift is small, roughly 40mas and is independent of the order tilt.  The fwhm changes by roughly $\pm$80mas when using untilted coordinates, but is essentially zero when using tilted coordinates.  This can arise when a vertical order geometry intersects the fully-resolved emission near the edges of the order.  When using properly tilted coordinates, the fully resolved emission becomes an added background that does not influence the Gaussian fitting.  Figure \ref{fig:tori} shows the raw exposure and the extended emission with the tilted coordinates shown as the slanted black line across the extended emission.

	Further investigation of these effects is beyond the scope of this paper, but the analysis package is robust and subject to the same systematic errors as other analysis packages (Bailey 1998, Baines et al. 2006).  In summary, this analysis package can reproduce results of others, specifically MWC 166.  However, other systematic effects that always influence spectroscopic data must be considered as well.  This illustrates the robustness of the reduction geometry and coordinate grid as well as the psf fitting and analysis by the optimal spectral extraction routines.

\subsection{Polarization calculation}

	The simultaneous imaging of orthogonal polarization states also allows for a greater efficiency observing sequence and for reduction of atmospheric effects since polarization measurement can be done with a single image and both states are imaged simultaneously, with identical systematics.  Using the fractional polarizations measured as the difference between the orthogonally polarized spectra divided by the sum, q and u, the percent polarization, P, and the position angle, PA, can be calculated as follows:

\begin{equation}
q= \frac{Q}{I} = \frac{1}{2}(\frac{a-b}{a+b} - \frac{c-d}{c+d})= \frac{1}{2}(q_{0^\circ} + q_{45^\circ})
\end{equation}

\begin{equation}
P = \sqrt{q^2 + u^2}
\end{equation}

\begin{equation}
PA = \frac{1}{2}tan^{-1}\frac{q}{u}
\end{equation}

  It is important to note that many people use the ratio method defined as follows:

\begin{equation}
q= \frac{Q}{I} = ( \frac{ \sqrt{\frac{ad}{bc}}-1 }  { \sqrt{\frac{ad}{bc}}+1})
\end{equation}

  The Ukirt infrared polarimeter has yet another difference method published to calculate the polarimetry on their IRPol2 website.  
  
\begin{equation}
q= \frac{Q}{I} = (\frac{a-c-\frac{a+c}{b+d}(b-d)}{a+c+\frac{a+c}{b+d}(b+d)})
\end{equation}

  This method is a different version of the simple average method and can be reduced to a simple average where b and c switch places:
  
\begin{equation}
q= \frac{Q}{I} = \frac{1}{2}(\frac{a-c}{a+c} - \frac{b-d}{b+d})
\end{equation}

	When all three of these methods are applied to a data set, the difference between the calculated polarized spectra is entirely negligable.  Since no method was significantly different than any other, the easiest to implement was selected - a simple average applied to each individual exposure set.  

	The subtraction of the two terms allows for cancellation of systematic errors resulting from any misalignment of the polarized orders since each term is an independent measurement of Stokes q.  Measurement of Stokes U is the same formula, but with the waveplate at 22.5$^\circ$ and 67.5$^\circ$.  Since this is a fractional quantity, any correction applied equally to each order and to each chip, such as spectrophotometric calibrations, vignetting corrections or skyline subtractions, will not change the polarimetry in any significant way.  That is to say that no standard calibrations effect the spectropolarimetry.  
	
	We wish to note that no instrumental polalarization ripple of the kind mentioned in Aitken \& Hough 2001 or seen in ISIS (Harries et al. 1996a) has been seen in the data.  This ripple is thought to arise from Fabry-Perot type fringes in the wave-plate.  Though the instrument uses a standard achromatic waveplate, a thin layer of polymer sandwiched between two glass substrates, no ripple near the 0.1\% level has been seen.

\subsection{Aligning the spectra}

  The Savart plate axis can never be perfectly aligned with the slit.  This produces a small shift in the dispersion direction between the two polarized orders. This shift was calculated by cross-correlating the ThAr spectra from the two polarized orders.  The derived shifts were 2 pixels in 2004, 12 pixels in 2005 before the Savart plate reinstallation, and 7 pixels after the 2006 remounting.  
  
\begin{figure} [!h]
\subfloat[Stokes q Nine Ways]{\label{fig:q9ways} 
\includegraphics[width=0.35\textwidth, angle=90]{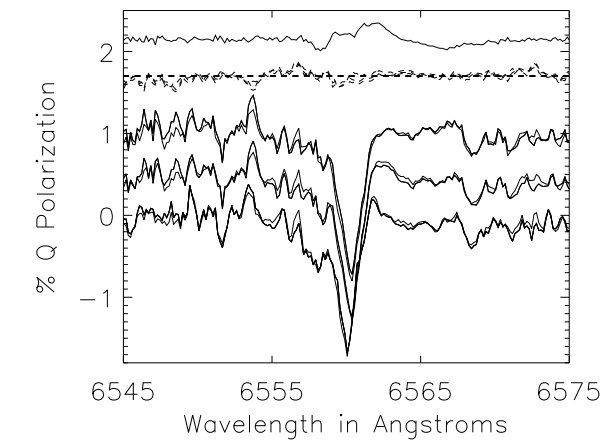}}
\quad
\subfloat[Waveplate Wobble]{\label{fig:wobble-exp} 
\includegraphics[width=0.35\textwidth, angle=90]{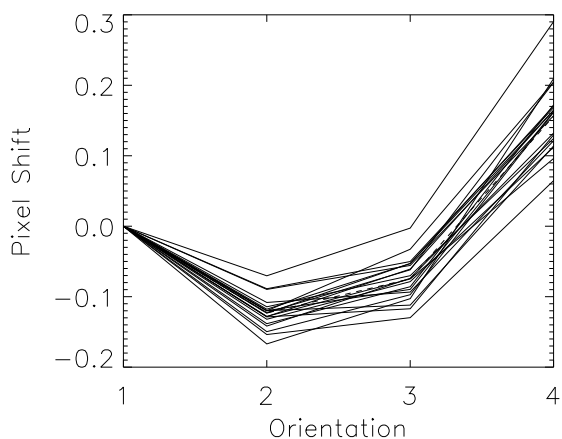}}
\caption[Nine Polarization Calculations and Beam Wobble Measurements]{  {\bf a)}  Stokes q calculated 9 different ways and the differences between them- from the bottom, no alignment, integer-pixel alignment, subpixel-alignment, and the differences between all of them and the first one (simple average, no alignment).  They're all quite similar in average shape.  The changing S/N can be seen across the trough and peak of the PCygni line.  The noise increases when any spectral alignment procedure is applied.  {\bf b)}  The beam wobbles (shows a small wavelength shift) as the waveplate rotates.  The wobble is calculated by applying the shift-n-scale routine independently to many lines in the Th-Ar lamp spectra.  The wobble shows a sinusoidal dependence on waveplate angle with an amplitude of 1 pixel.  The 67.5$^\circ$ region of operation is near the crest of the sine-curve and hence has an amplitude of 0.4 pixels.}
\label{fig:comp}
\end{figure}

  To characterize any derivative noise, there are a number of ways to check the wavelength alignment of the spectra and calculate the polarization.  The three ways are 1) no alignment 2) cross-correlation an a simple integer-pixel shift (which preserves noise properties) and 3) integer-pixel shift followed by the shift-n-scale sub-pixel alignment.  When no alignment is done, the spectra and polarization are simply calculated where the are on the detector.  In the integer-pixel-shift alignment method, a cross-correlation is done on the spectra to ensure that the spectra aligned to a single pixel, as a double-check on the wavelength calibration.  Since any interpolation can effect and corrupt the noise statistics of the spectra, as shown by the shift-n-scale example in figure \ref{fig:shfscl}, subpixel alignment must minimize the derivative signal more than it introduces extra noise.  To evaluate all three alignments, the polarization analysis is done on the three different spectra using the three different alignment methods.  These nine polarization spectra are similar in their overall structure.  A sample of this is shown in figure \ref{fig:q9ways}.  The polarization signature is around 1.5\% and the differences between the different methods are very small.

  The spectral shift (beam wobble) measured from Th-Ar lamp spectra taken at many waveplate orientations shows that there is a one pixel shift induced over a complete rotation of the waveplate with a roughly sinusoidal shape.  The alignment of the waveplate with the savart plate causes a rotational offset between the rotation-stage zero point and the waveplate zero point of 85$^\circ$.  Thus the operation of the waveplate is from rotation angles of 85$^\circ$ to 152.5$^\circ$ near the crest of the sine-curve.  

  Measurements of ThAr spectra taken each night, shown in figure \ref{fig:wobble-exp}, show that the beam wobble is roughly 0.3 pixels between the four waveplate orientations.  The derivative signal caused by a 0.3 pixel shift for strong emission lines with contrast of 5-20 is roughly 1-3\%.  It will be shown later that the derivative signal subtracts away in the dual-beam polarization analysis and that a beam wobble of 0.3 pixels produces a negligible error.  This beam-wobble signal follows regions of curvature and is proportional to the initial derivative error, as it is a derivative-of-derivative error (acceleration).  The sub-pixel shifts between the three different intensity calculation methods is calculated using the shift-n-scale routine.  The shifts were found to be a small fraction of a pixel.

	The wavelength of each pixel for each horizontal order is assigned by a 3$^{rd}$ order polynomial fit to thorium-argon calibration lamp images.  This analysis is done independently on each order for the two orthogonal polarization states.  Typically, reduction packages, such as Libre Esprit for the ESPaDOnS spectrograph on CFHT (Donati et al. 1997) use the wavelength fits as the alignment method between the polarized orders.  The simple average without alignment method was found to work best because the flat-field cancels in the polarization calculation, leaving the measurement independent of the pixel-to-pixel gain.

\subsection{Rotating Frames}

  Since HiVIS uses the alt-az coud\'{e} focus, the spectropolarimeter's Q-U reference frame rotates on the sky as the telescope tracks.  There are two main rotations: one with altitude, between the tertiary mirror and m4, as the altitude changes, the other with azimuth, between m6 and the coud\'{e} pickoff mirror.  Normally, coud\'{e} spectrographs use a 3-mirror image rotator somewhere downstream (in an pupil plane) to compensate for the telescope's motion and keep the orientation of the slit on the sky fixed.  HiVIS has an image rotator, but since the rotator is 3 oblique reflections, it is not used to increase the accuracy of the polarimetric calibration of the instrument.  Changing the mirror orientations would change the polarization induced by the mirrors.  The orientation of the instrument on the sky is allowed to rotate and exposures are kept short so that the rotation is small.  
  
  The analysis software takes the date and time of the exposure and calculates the altitude and azimuth of the telescope during the exposure.  Knowing the orientation of the telescope mirrors, the projection of the spectropolarimeter's Q and U axes on the sky can be reconstructed, and all measurements rotated to a common frame.  A common convention is to set +Q to North on the sky. 
  
  The rotation angle between the 4 individual exposures in each set is calculated as Position-angle + Altitude - Azimuth and is checked to quantify the inter-set smear.  Very large rotation angles occured when an object is tracked near transit (high azimuth rates).  For measurements with small rotations, the polarization measurements are assumed to be accurate and are rotated to an absolute frame (of the first exposure) on the sky.

  For every data set, rotation in an individual exposure causes each exposure to become a mix of Stokes Q and U, reducing the efficiency of the measurement and adding some uncertainty to the Stokes parameters.  Imagine an incident +Q beam.  A rotation of the instrument of 22.5$^\circ$ during an exposure would cause the projected instrument frame to go from +Q to +U and the resulting measurement would only show 50\% polarization, when 100\% was incident. 
  
    Rotation between sets also induces systematic error.  As the instrument rotates on the sky, different amounts of time will be spent sampling the polarization states.  As an extreme example, 22.5$^\circ$ rotation of the instrument between each individual exposure would cause all images to be oriented on the sky with +Q whereas the reduction package would assume the sequence was +Q, +U, -Q, -U.  The systematic error with this high rotation rate causes complete loss of information about U while doubling the time spent measuring Q.  It also causes the benefits of beam-swaping to be lost entirely and it causes the reduction methods to fail.

\subsection{Rebin By Flux}

  Since the expected spectropolarimetric signals are subtle differences in the polarized line profile shapes of less than one percent, and the few reported detections show polarization of at most 1\%, the lines must be calculated with at least S/N$ >$ 300 in all spectral-resolution-elements.  

    Most young emission-line stars have large line/continuum ratio's (5-10).  Some, such as AB Auriga, have PCygni profiles with emission-peak to absorption-trough ratio's of over 20.  The star P-Cygni itself has a ratio of nearly 100.  Even when saturating the peak of the emission line on the detector, a S/N of only 100 or so was achieved in the absorption trough of AB Auriga.  A flux-dependent averaging procedure has been developed to bin pixels spectrally by flux, to ensure that the S/N per spectral-pixel is roughly constant over the line at the expense of regular wavelength coverage.   

\begin{figure} [!h]
\subfloat[Rebin by Flux]{\label{fig:rb} 
\includegraphics[width=0.3\textwidth, angle=90]{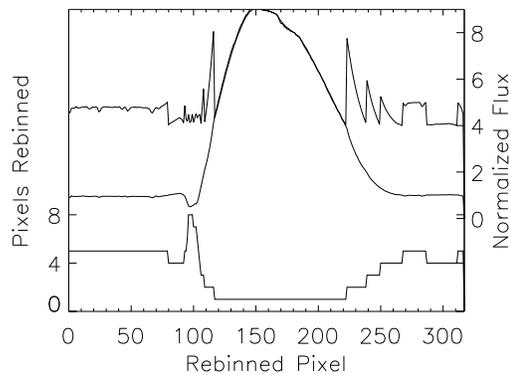}}
\quad
\subfloat[Stokes q Noise Reduction]{\label{fig:qnoise} 
\includegraphics[width=0.3\textwidth, angle=90]{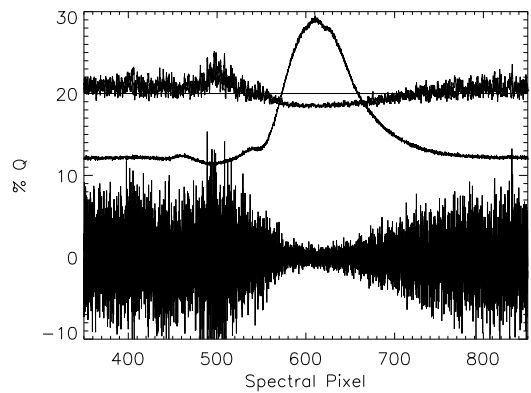}}
\caption[Rebin by Flux and Noise Reduction]{  {\bf a)}  The rebin-by-flux routine adaptively averages pixels to equalize the S/N per spectral pixel.  The number of physical pixels averaged per rebinned spectral pixel in the adaptive rebinning with a threshold of 4 is the bottom curve.  In the P-Cygni absorption trough, 8 pixels are binned, but in the emission peak, there is no binning. The middle curve is the rebinned spectra which shows a highly compressed P-Cygni trough because of the high binning.  The top curve is the flux per rebinned spectral pixel, with a minimum of 4 (the threshold) and a maximum of the emission peak.  {\bf b)}  The systematic error in a simulated data set caused only by varying noise.  Random noise proportional to $\sqrt{N}$ was added to eight identical copies of the H$_\alpha$ line of AB Aurigae, plotted above.  The resulting Stokes parameters were calculated and show a noise of roughly 5\% at continuum and 2\% in the emission line peak.. The raw Stokes q and u are the bottom curves, centered about zero, with the noise obviously varying across the line profile.  The spectropolarimetric signal, $\sqrt{q^2+u^2}$, is the top curve and it varies by over 2\% across the line profile.  The variation is caused only by the varying noise, rising with higher noise.}
\label{fig:comp}
\end{figure}

	  Another systematic error occurs when you calculate the degree of polarization for a spectrum, a squared quantity, with noise amplitudes that vary across the line.  Since the average absolute values of noisy data will be higher than clean data, and there will be significant noise variations across strong emission lines, varying noise can mascarade as a systematic spectropolarimetric error.  Figure \ref{fig:qnoise} shows a simulated polarization measurement where flux-dependent noise was added to eight copies of a single line-profile.  The noise varied from roughly 5\% at continuum to roughly 2\% in the center of the emission line.  The resulting Stokes q and u parameters were centered about zero, but with a varying noise amplitude.  When these measurements are converted to a degree of polarization via $\sqrt{q^2+u^2}$, this varying noise shows up as a systematic error in the degree of polarization.

\begin{figure} [!h]
\subfloat[Rebin Stokes q \& u]{\label{fig:rbqu}
\includegraphics[width=0.35\textwidth, angle=90]{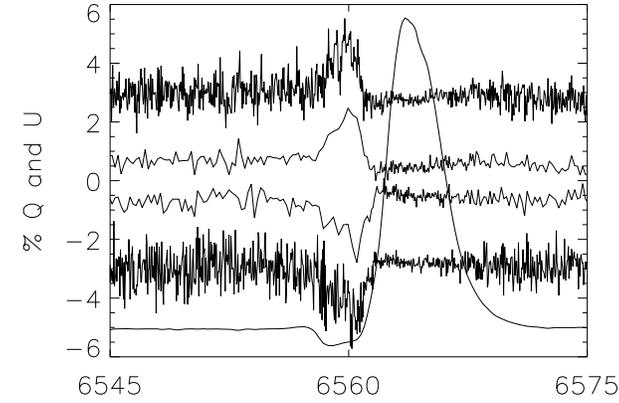}}
\quad
\subfloat[Derivative Simulation]{\label{fig:derivsim} 
\includegraphics[width=0.35\textwidth, angle=90]{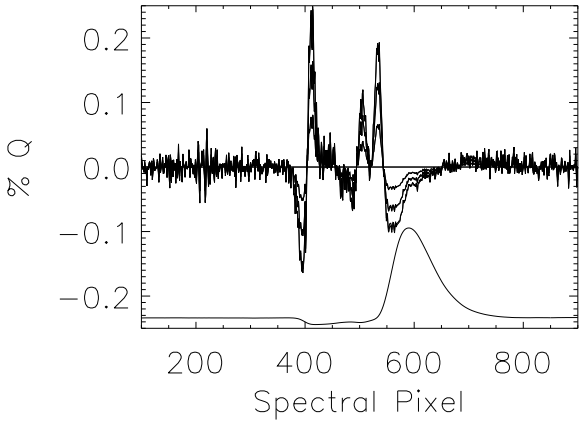}}
\caption[Rebin and Beam Wobble Stokes q \& u  Examples]{  {\bf a)}  An example of the rebin-by-flux routines applied to a data set.  The noisy curves show Stokes q (top) and Stokes u (bottom) with a uniform spectral resolution, but an uneven S/N.  The smoother curves near the middle have a more uniform noise per pixel and the wavelength dependence of the polarization is much more clearly seen.  The average H$_\alpha$ line is overplotted.  {\bf b)}  A beam-wobble polarization simulation - modeling the systematic error in polarization caused by a spectral shift of the spectra induced by beam wobble or optical misalignments.  As the waveplate rotates, the spectra of one orientation shift by a small fraction of a pixel to those of another orientation.  This simulation greatly exaggerates the shift to show what a strong systematic error would look like.  The curves show the simulated Stokes Q spectra for beam wobble amplitudes of 0, 0.5, 1.0, and 1.5 pixels.  The largest shift shows a peak-to-peak amplitude of 0.4\%.  The systematic effect follows the curvature since beam wobble is a derivative of derivative.}
\label{fig:comp}
\end{figure}
  
 To rebin-by-flux, the average line profile for each data set (after alignment) is calculated.  This provides the average flux and shape of all 8 lines.  A user-defined threshold, typically 1-5, sets the minimum flux in each rebinned spectral pixel.  The number of physical pixels to rebin at each wavelength is calculated and each individual spectrum is then rebinned accordingly.  A rebinned wavelength array for each set is also calculated since the number of pixels binned in the spectral direction is different for each star and data-set.  The polarization analysis is then performed on these rebinned data sets as specified above and the pixel-to-pixel variation in the noise is greatly attenuated.  Figure \ref{fig:rbqu} shows a comparision of raw and rebinned polarization spectra.  The rebinned spectra only show a few resolution elements over the absorption trough with many covering the emission peak.  The polarization signature in the absorptive part of the line stands out very clearly.

\subsection{Systematics - Alignment Error}

 The calculated polarization is sensitive to systematic errors in the alignment of the polarized spectra.  There are two main alignments of the spectra - between the polarized spectra in each individual image, and between the specra from each individual image. 

 The reduction code automatically makes derivative-error simulations for every data set to show the form of the systematic errors caused by misalignments.   The misalignment between polarization states causes a derivative signal in each fractional polarization measurement  $\frac{(a-b)}{(a+b)}$ that can be at the few \% level.  As long as the shift between polarization states is constant in each image, then the derivative signal is subtracted away when the second polarization measurement, $\frac{(c-d)}{(c+d)}$, is subtracted.  The polarization state shift between images is very stable.  There is never more than a 0.05 pixel shift difference between the calculated wavelength solutions of each polarized order as determined by ThAr line fits.  

\begin{figure} [!h]
\includegraphics[width=0.35\textwidth, angle=90]{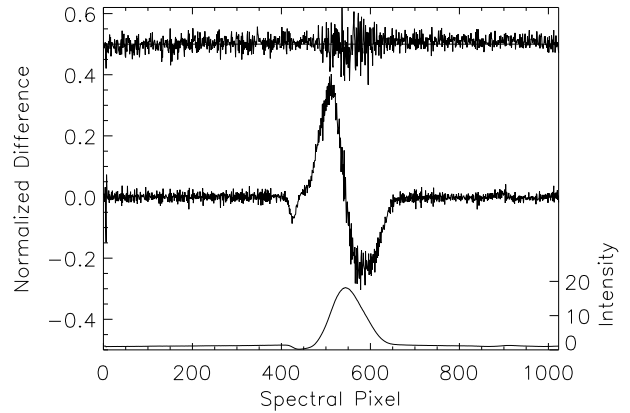}
\caption[Polarized Flux - P-Cygni]{\label{fig:pfx-pcyg}
An example of the polarized flux calculations for P Cygni.  The top curve is the polarized flux after all the wavelength alignment systematic errors have been removed.  The polarized flux is essentially zero and this data set shows no polarization signature.  The middle curve shows a 1-pixel misalignment to illustrate the magnitude of the alignment systematic errors.  The bottom curve shows the line profile, the normalized Stokes I, with the scale on the right.}
\end{figure}

\onecolumn
\begin{figure}
\subfloat[Systematic 1/I Error - P-Cygni]{\label{fig:sysq-pcyg}
\includegraphics[width=0.35\linewidth, angle=90]{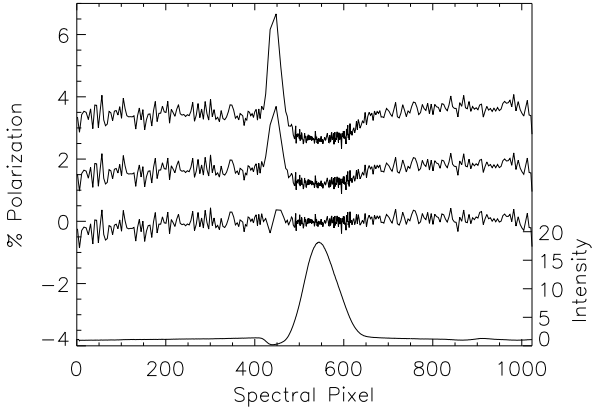}}
\quad
\subfloat[Systematic 1/I Error - AB Auriagae]{\label{fig:sysq-ab}
\includegraphics[width=0.35\linewidth, angle=90]{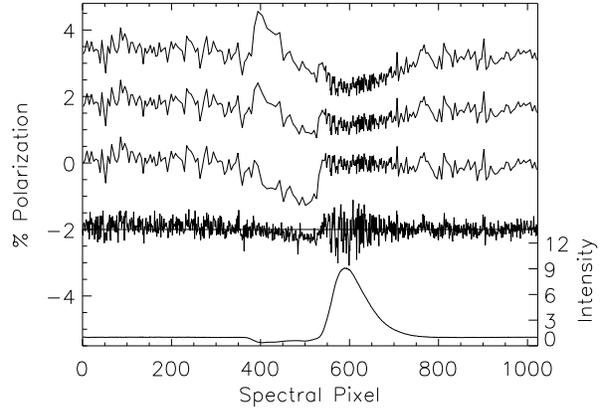}}
\caption[1/I Systematic Error P-Cygni \& AB Aurigae]{An illustrartion of the continuum misfit systematic error in Stokes q for {\bf a)}  P Cygni observed on 9-20-2007 and {\bf b)} AB Aurigae observed on 12-28-2006. The bottom curve in each plot shows the normalized H$_\alpha$ line profile with the scale on the right. The three top curves above show increasing levels of continuum misfitting - 0\%, 2\% and 4\%, in accord with eqn \ref{eq:conterr}. The Stokes q spectra have been vertically offset for clarity. In each, the \% polarization in the emissive part of the line profile is 0\% without any continuum error. In P-Cygni, the 4\% continuum error gives rise to a false polarization spectrum showing 2\% polarization change in the absorption trough and a nearly 1\% change across the emission peak. For AB Aurigae, the magnitude of polarization error is roughly half this since the line is about half as intense. The AB Aurigae plot has an extra line which shows the polarized flux (Q=q*I) that shows a small but significant deviation in the P-Cygni absorption. This gives rise to the detected q signature that is always present in the absorption, regardless of continuum error.}
\label{fig:1onIsys}
\end{figure}

\begin{figure}
\centering
\subfloat[Systematic 1/I Error - MWC 158]{\label{fig:sysq-158}
\includegraphics[width=0.35\linewidth, angle=90]{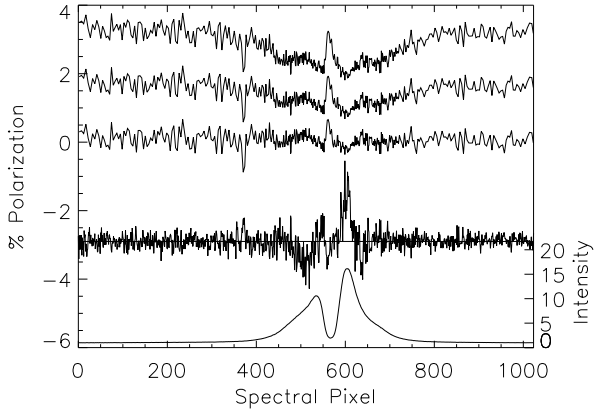}}
\quad
\subfloat[Systematic 1/I Error MWC 143]{\label{fig:sysq-143}
\includegraphics[width=0.35\linewidth, angle=90]{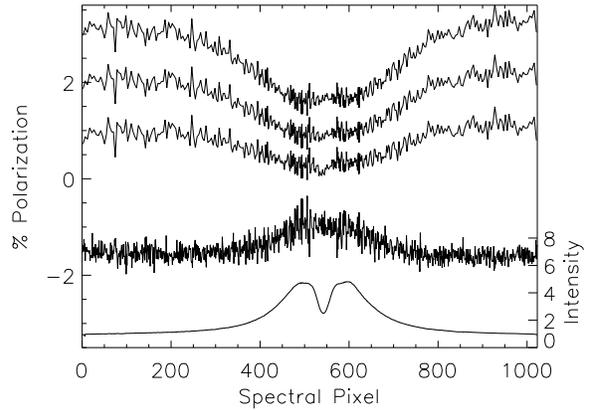}}
\caption[1/I Systematic Error MWC 158 \& MWC 143]{An illustrartion of the continuum misfit systematic error in Stokes q for  {\bf a)}   MWC 158 observed on 12-28-2006 and {\bf b)}  MWC 143 on 9-20-2007. The bottom curve shows the normalized H$_\alpha$ line profile with the scale on the right. The top three curves above show increasing levels of continuum misfitting - 0\%, 2\% and 4\%, in accord with eqn \ref{eq:conterr}. The Stokes q spectra have been vertically offset for clarity. The curve just above the H$_\alpha$ line is the polarized flux, (Q=q*I) multiplied by 10 to fit the left hand scale. In MWC 158, the \% polarization at the maximum emission part of the line profile is nearly 0\% but the 4\% continuum error gives rise to a false polarization spectrum showing 2\% polarization change in the absorption trough and a nearly 1\% change across the emission peak. In MWC 143, increasing error changes the magnitude of the polarization at the line center, but doesn't change the shape as the real effect has the same overall shape as the H$_\alpha$ line.}
\label{fig:1onIsys2}
\end{figure}
\twocolumn

  There is also a second order effect - beam wobble - that can cause a systematic polarization error.  If the spectra are misaligned between images, from the rotating waveplate or guiding errors in un-tilted coordinate systems, then a derivative-of-derivative effect can show up when the two polarization measurements are differenced.  This effect is simulated in figure \ref{fig:derivsim} by applying a orientation-dependent wavelength shift between the exposures that increases with waveplate angle.  This simulates any beam-wobble introduced by misalignment of the waveplate.  The induced error traces the curvature of the emission line, with polarization errors showing up where curvature is greatest.  Using ThAr spectra, this wavelength shift was measured to be $\sim$0.1 pixels, depending on waveplate orientation, shown in \ref{fig:wobble-exp}.  This small wobble has a very small influence on the polarization, as it takes a misalignment of several pixels to produce a noticable acceleration signal.  
  
  The derivative simulations in this code show that the simple derivative error subtracts away completely if the polarized-order-separation is constant (which it is to 0.05 pixels).  The code also shows that it takes a wobble of at least 2 full pixels between each orientation, for a total of a 6-pixels misalignment to produce a 1\% systematic error in a Stokes parameter for strong lines.  Using the shift-n-scale routine, as well as the ThAr measurements, this wobble was measured to be around 0.1 pixels and thus present only well below the 0.1\% level proportional to the curvature (derivative of derivative).

\subsection{1/I Systematic Errors }
	
	In Harrington \& Kuhn 2007, several systematic errors common in spectral reduction packages were explored. There are two main sources for error when calculating fractional differences: the wavelength alignment and the continuum subtraction of the 8 spectra.  The major source of error when calculating polarization with strong lines is typically the wavelength alignment. This was thoroughly explored in Harrington \& Kuhn 2007. However, in the course of the observations some additional continuum subtraction errors were noticed in some extremely strong lines. This continuum effect will be explored in this section.

	The continuum subtraction in many packages, such as Libre Esprit, is done by fitting a low order polynomial over a spectral order. In the single-line reduction package, a simple linear fit to the continuum on either side of the line is used, just like in the standard IRAF package. Each of the 8 polarized spectra that are used to calculate q and u have the continuum normalized independently by dividing each spectrum by the linear continuum fit. Depending on the width of the line and the number of pixels selected, these fits are typically robust to 1\% or better.

	This continuum normalization eliminates the telescope induced continuum polarization, as well as any intrinsic polarization.  Also, since this subtraction centers all normalized continuum values to 1, the polarized flux calculation no longer gives the actual flux of polarized light. It gives the deviation from continuum which can be a positive or negative value. Figure \ref{fig:pfx-pcyg} shows an example of the polarized flux calculated for a single observation on 9-20-2007. A 1-pixel derivative error, the middle curve in the figure, shows the amplitude of the wavelength errors that are subtracted by the reduction package, leaving the essentially featureless polarized flux as the top curve. 

	The systematic error arises when calculating the degree of polarization, a 1/I normalized quantity, when this polarized flux is not properly centered at zero.  This featureless curve, when normalized by a strongly varying line profile can give rise to a 1/I type of error.  When dividing a featureless line by the P Cygni profile, the absorptive trough with it's small numbers, amplifies the continuum error and falsely gives a high degree of polarization while the emission peak reduces the error.
	
	In order to illustrate this effect, a simulated continuum error, $\epsilon$, is implemented as a constant offset to a spectrum with this formula:
	
\begin{equation}
 \label{eq:conterr}
 q= \frac{1}{2}(\frac{a-b-\epsilon}{a+b+\epsilon} - \frac{c-d}{c+d})
\end{equation}
	
	We find that this error can cause significant polarization effects across strong emission lines when $\epsilon$ is near 0.01.  In order to be certain that a polarization effect is real, an examination of the polarized flux must be done to ensure that the continuum fitting has been done properly.  Since the widths of the emission lines in our targets varies greatly, there are atmospheric water vapor lines near H$_\alpha$, and there are other emission lines in the area, this is best done by visual inspection.

\section{Polarization Calibration}

	The next step in the instrument development was calibrating the instrumental polarization response.  Absolute polarization at all wavelengths requires very careful calibration of the telescope.  Internal optics can induce or reduce polarization (U,Q$\rightarrow$I or I$\rightarrow$Q,U) in the beam, cause cross-talk between polarization states (QU$\rightarrow$V, V$\rightarrow$QU), and also rotate the plane of polarization (Q$\rightarrow$U, U$\rightarrow$Q).  For an instrument at the coud\'{e} focus of an alt-az telescope, all of these effects are functions of wavelength, altitude and azimuth.  For AEOS, there are five 45$^\circ$ reflections before the coud\'e room, two of which change their relative orientation.  This is particularly severe, and only a few put a spectropolarimeter after so many mirrors.  The calibration is difficult, but some have tried this for many different telescope designs (cf. S\'{a}nchez Almeida et al. 1991, Kuhn et al, 1994, Giro et al. 2003) 
	
	Since there are no spectropolarimetric standards one usually observes polarimetric standards that have a precise polarization value averaged over some bandpass such as Hsu \& Breger 1982, Gil-Hutton \& Benavidez 2003, Schmidt \& Elston 1992, Fossati et al. 2007.  The calibration and creation of a telescope model is done by measuring unpolarized standard stars and polarized sources at many pointings to calibrate the effects of the moving mirrors.  
	
	In general, the telescope will have different responses to linearly polarized light since the absorption coefficients of the optical components vary with wavelength and the differential absorption is a function of the angle of the incident polarization with respect to the mirror/component axes.  Zeemax models of the telescope were used to compute the assumed polarization properties of the coud\'{e} focus given various optical constants for aluminum.  These show the expected altitude-azimuth dependence of the mueller-matrix terms, though the amplitudes of the computed effects are significantly different than measured.  While the models can calculate the full Mueller matrix as functions of altitude, azimuth, and wavelength, all of these terms cannot be practically measured.  However, certain parts of the Mueller matrix can be measured given proper sources (polarized standards, unpolarized standards, and scattered sunlight).  

	Unpolarized standard stars are the first tool used to calibrate the telescope.  Observations of these standards at as many pointings as possible allows one to construct a model of the telescopes response to unpolarized light and to quantify the telescope-induced polarization (but not the depolarization, cross-talk, or plane rotation which requires a linearly polarized standard).

	Polarized sources were the last calibration source used.  Using a polarized source allows one to measure the depolarization and rotation of the plane of polarization, all of which are, in principle, orientation dependent.  Polarized standard stars are polarized in their continuum light, but the constant value quoted in standard references is an average polarization in a relatively line-free, wide bandpass image.  The degree of polarization is usually not more than a few percent making instrument calibration difficult with these sources.  Another calibration method which will be explored below, is to use twilight (scattered sunlight) as a highly polarized source with a fairly well known degree and orientation of polarization.  The polarization is certainly not constant and varies with atmospheric properties, time, and pointing, but it is a very bright and highly polarized source that has given valuable insight into the polarization properties of the mirrors.

\subsection{Flat Field Polarization: Fixed Optics Polarization}
 
\begin{figure} [!h]
\includegraphics[width=0.4\textwidth, height=0.5\textwidth, angle=90]{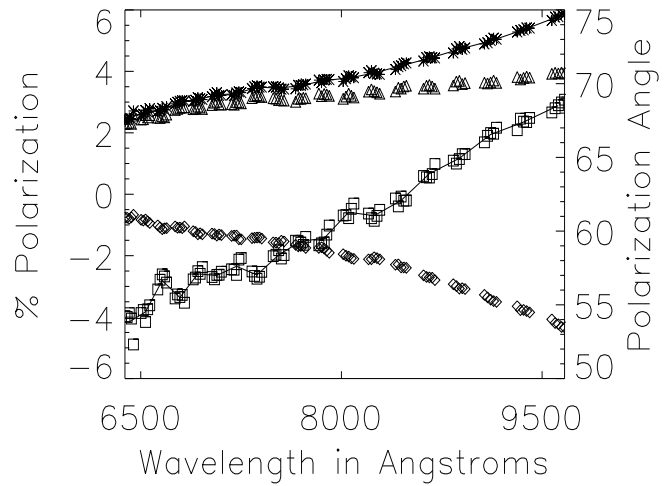}
\caption[Flat Field Polarization]{\label{fig:flatpol}
Flat field polarization analysis, averaged 200:1 for clarity. Stokes q is shown as diamonds , u is shown as triangles, and the degree of polarization is the top curve, shown as + symbols. The squares represent the position-angle of polarization (PA) with the y-axis scale on the right. The polarization (top curve) begins at 650nm being dominated by u (triangles), but begins to rise as -q (diamons) gets larger. The PA rotates by about 25$^\circ$, going from nearly 55$^\circ$ or mostly +u to being nearly 70$^\circ$ or a mix of -q and +u near (-3,4).}
\end{figure}

	All of the moving mirrors in the AEOS telescope are located upstream of the polarizing optics (analyzer). As an illustration of the polarization induced by the reflections between the halogen flat field lamp ($T_{eff} \sim 2900K$) and the spectropolarimeter, figure \ref{fig:flatpol} shows the polarization reduction applied to a set of flat field frames with the 4 waveplate orientations taken in 2004.   The flat field lamp is located just after the coud\'{e} room entrance port, immediately after the coud\'{e} pickoff mirror (m7) on the far right in figure \ref{fig:Coude_layout}.  The lamp is reflected off a diffuser screen and is assumed to be unpolarized.
	
	The flat fields show a significant apparent polarization which changes with wavelength. At 6500{\AA}, the induced polarization is dominated by the +U term, but the U and -Q terms become equal at 90. The position angle varies almost linearly across all wavelengths while the degree of polarization rises from 2\% at 650nm to 6\% at 950nm.  Since the remounting of the optics and a movement of the image-rotator, the values of the flat field polarization have changed, but this illustrates the magnitude and wavelength dependence of polarization induced by the 11 fixed reflections in the optics room. 

	We also have tested the polarization of the common-fore-optics.  A linear polarizer was mounted (with a mask) at various points along the optical path, from the calibration stage just after the coud\'{e} port to just in front of the slit.  A linear polarizer was measured to have 98.2\% polarization at the slit, 98.3\% polarization after the k-cell, 97.4\% before the k-cell, and 97.7\% after the flat field screen.  This shows that the linear polarization does not vary much between the beginning of the spectrograph optics and the slit.  Another polarizer mounted at the slit was measured to have 98.4\%.  This polarizer was put at the calibration-stage and was measured to have 98.1\% 97.7\% 97.8\% and 98.0\% polarization for angles -03.2, 44.0, 87.5, and 134.2 degrees in the slit's QU frame when manually rotated to 0, 45, 90, and 135 degrees.  The measured angles match completely within the manual uncertainty.  This shows that the degree of linear polarization is preserved through the spectrograph optics.
	
	The flat field also shows a difference in beam intensities (vignetting or beam transmission difference), in addition to it's intrinsic polarization.  The polarization in a single frame, computed as $\frac{a-b}{a+b}$ shows +q at  -4.55\% -q at 1.36\% +u at  -3.63\% and -u at 0.45\%.  These values are not symmetric about zero, as would be the case if there was only intrinsic polarization.  To make the polarization values symmetric about zero, a constant difference of 1.59\% between the two beams is applied.  This was calculated as 1.588 and 1.591 for q and u respectively as the null polarization spectra (sometimes called the check spectra: +q + -q).  Restated, this means the $\pm$q $\pm$u spectra are symmetric about 1.59\%, ie that the rotation of the waveplate swaps polarization around a constant difference in beam intensity of 1.59\%.  Subtracting the 1.59\% difference in beam intensities from all single-frame polarizations $\frac{a-b}{a+b}$ makes these polarization spectra symmetric about 0\% reflecting the true polarization of the beams.  
	
	We note that the old flat field polarization measurements showed a polarization of 2.78\% at a PA of 55.0$^\circ$ (-0.93\% 2.57\% qu) for 6567{\AA}.  This is not what is measured with the new setup.  This is expected since the image rotator was moved, the flat-field lenses were installed, and the new dekkar is slightly smaller than the old one, changing the illumination pattern.  The new flat field polarization is just as repeatable and robust as under the old configuration.

\begin{table}
\begin{center}
\caption[Flat Field Polarization]{Flat field polarizations \label{flatpol}}
\begin{tabular}{lllll}
\hline
{\bf Name:}          & {\bf +q}           & {\bf -q}           & {\bf +u}            & {\bf -u}     \\
Pol \%:                 & -4.546           & 1.364             &   -3.629           & 0.452     \\
Pol-1.59\%:        & -2.956            & 2.954             &  -2.039             & 2.042      \\
\hline
\end{tabular}
\end{center}
\end{table}

	The calculated flat field polarization after background correction is -2.95\% for q and -2.0\% for u calculated with the average method as 0.5*( +q - -q ) giving a degree of polarization of 3.59\% at an angle of 287.3 degrees. This test was performed on a sequence of 40 flat field frames and each individual exposure set gives the same answer to 0.01\%
	
	We did a test of the dichroic for the LWIS mounted inbetween m4 and m5.  The dichroic is roughly 12" square and causes an estimated 12\% reduction in throughput.  The twilight polarization was measured without, with, and again without the dichroic mounted.  The insertion and removal takes less than one minute so the twilight polarization and PA can be assumed constant throughout the $\sim$5 minute observation.  The measured polarization values were nearly identical with and without the dichroic.

\subsection{Unpolarized Standards: Telescope Induced Polarization}

	Observations of unpolarized standard stars allow the measurement of the variation of the telescope-induced polarization as a function of pointing.  Since the incident light is unpolarized, any observed polarization signature directly traces the polarization induced by the telescope - the IQ IU mueller matrix terms.  Many unpolarized standards have been observed in November 2004, June \& July 2005, Sept 2006 - Jan 2007, and June 2007 totaling 224 data sets.  Since September 2006, the image rotator was moved, the camera was changed in September 2006, and a better wavelength solution was obtained.  The observations from 2004-2005 will be presented in one section dealing mostly with the wavelength dependence of the polarization.  Then, the 126 observations after September 2006 will be presented as the telescope properties present during the bulk of the observations.  The calculated degree of polarization for each unpolarized standard observed is averaged to a single measurement over the entire H$_\alpha$ region.  Note that all the unpolarized standards showed no significant line polarization, and had a statistical error of less that 0.05\%, typically 0.02\%.

\subsection{2004-2005 Multi-wavelength observations}

\begin{figure}
\subfloat[All Unpolarized Standards q]{\label{fig:all_unpol_q}
\includegraphics[ width=0.35\textwidth, angle=90]{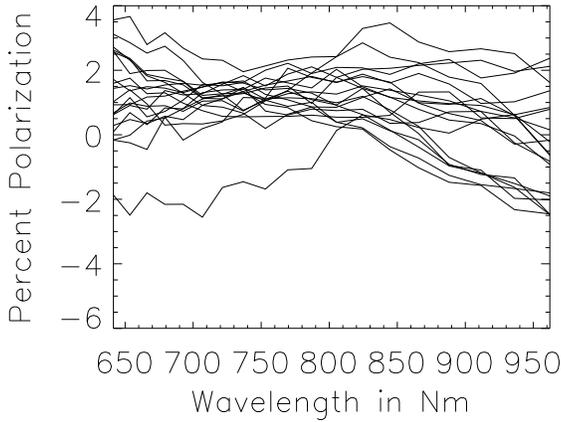}}
\quad
\subfloat[All Unpolarized Standards u]{\label{fig:all_unpol_u}
\includegraphics[ width=0.35\textwidth, angle=90]{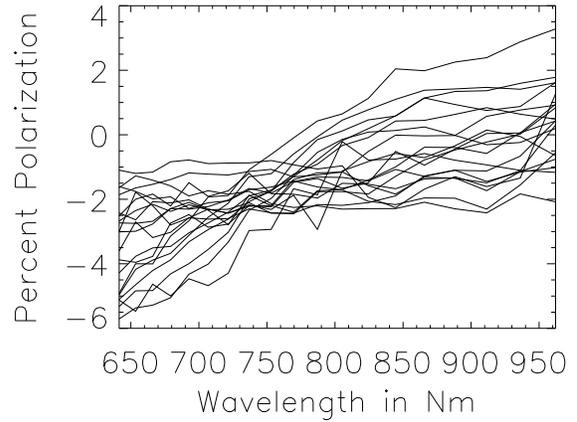}}
\caption[2005 Unpolarized Standard Star Spectropolarimetry]{ {\bf a)} Stokes q, rebinned 1000:1, for the unpolarized standard star observations in table \ref{upobs}.  Most observations show similar trends with wavelength (q falling).  {\bf b)} Stokes u, rebinned 1000:1, for the unpolarized standard star observations in table \ref{upobs}.  They all show similar trends with wavelength (u rising).}
\label{fig:comp}
\end{figure}

	Many unpolarized standards have been observed in November of 2004 and in July of 2005 totaling 19 data sets (152 spectra).   A plot of q and u for all the unpolarized standard stars, where each 1000-pixel order has been rebinned to 1 data point,  is shown in figures \ref{fig:all_unpol_q} and \ref{fig:all_unpol_u}. They show the induced polarization as a function of pointing and wavelength.  There is a fairly consistent trend between all stars where q goes from 0-3\% at 650nm falling to between -2\% to +2\% at 950nm, and u starts between 0\% and -6\% at 650nm and then rises to between -2\% to +3\%.

   Calibration of the induced polarization to any future observations is done by observing the unpolarized standard stars at different pointings and creating a map of the telescopes response by projecting these measurements in alt-az space.  This map can then be interpolated to any pointing one wishes to calibrate.  The ultimate aim being to have a high resolution map of the telescope's polarization response.

\subsection{Examples of Pointing Dependence}

	The telescope-induced polarization for the first three of the standards in table \ref{upobs} are given in figures \ref{fig:125pc} through \ref{fig:142pc}.  Each star shows different polarization and illustrates how the induced polarization varies with wavelength and pointing since each star is unpolarized. These stars cover declinations of -7, +27, and +42, allowing the measurement the induced polarization at many different azimuths and altitude, from far north to far south.  

	The spectropolarimetry for HD125184 is shown in figure \ref{fig:125pc}.  It shows moderate change with wavelength ($\sim$1.5\% increase in the blue and decrease in the red).  The change in altitude is about 15$^\circ$ (63-49) and the azimuth changes by 55$^\circ$ (175-230).

\onecolumn

\begin{figure}
\subfloat[2005 Induced Polarization Map 6400{\AA}]{\label{fig:unsurf0}
\includegraphics[ width=0.35\textwidth, angle=90]{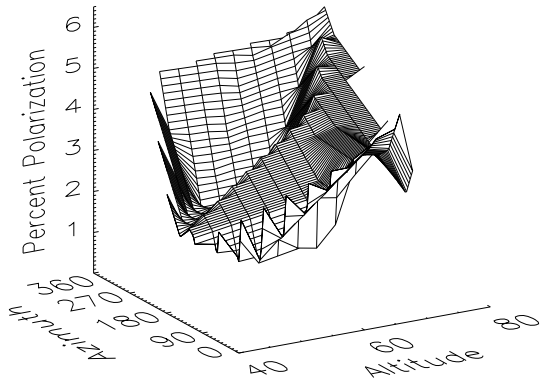}}
\quad
\subfloat[2005 Induced Polarization Map 9600{\AA}]{\label{fig:unsurf18}
\includegraphics[ width=0.35\textwidth, angle=90]{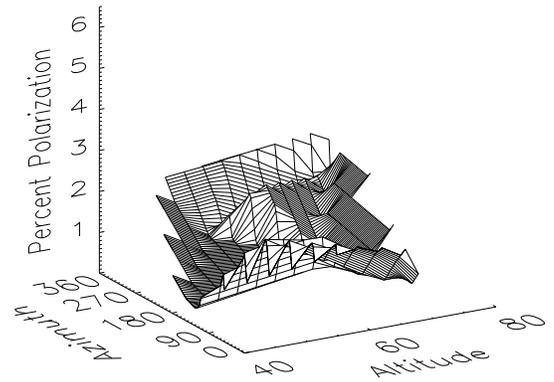}}
\caption[Unpolarized Standard Star All-Sky Maps 6400{\AA} \& 9600{\AA}]{ {\bf a)}  The measured polarization of unpolarized standard stars for order 0 (640nm) in the alt-az plane. The peak is at 6\%.  The left axis is azimuth and the horizontal axis is altitude.  This is a triangulation of 19 independent observations.  The wavelength dependence for a few of the points in this plot are shown in figures \ref{fig:125pc} through \ref{fig:142pc}.  {\bf b)}  The measured polarization of unpolarized standard stars for order 18 (960nm) in the alt-az plane. The peak is near 4\%.  The left axis is azimuth and the horizontal axis is altitude.  This is a triangulation of 19 independent observations.  The wavelength dependence for a few of the points in this plot are shown in figures \ref{fig:125pc} through \ref{fig:142pc}. }
\label{fig:unsurf}
\end{figure}

\begin{table}
\begin{center}
\caption{{Unpolarized Standard Star Observation}\label{upobs}}
\begin{tabular}{lcccccrr}
\tableline\tableline
Star & RA (hm) & Dec (dm) &  Date(UT) & Time(UT) & Exp(s) & Alt & Azi \\
\tableline
\tableline
HD114710 & 13 12 & +27 53 & 702 & 5:40 & 10 & 76 & 309  \\
HD114710 & 12 12 & +27 53 & 701 & 5:40 & 10 & 76 & 309  \\
HD114710 & 13 12 & +27 53 & 630 & 5:55 & 20 & 73 & 302  \\
HD114710 & 13 12 & +27 53 & 630 & 7:00 & 20 & 60 & 292  \\
HD114710 & 13 12 & +27 53 & 702 & 7:15 & 10 & 56 & 291  \\
HD114710 & 13 12 & +27 53 & 701 & 8:10 & 10 & 44 & 290  \\
\tableline
HD125184 & 14 15 & -07 19 & 701 & 5:45 & 30 & 63 & 175  \\
HD125184 & 14 15 & -07 19 & 703 & 5:55 & 30 & 61 & 180  \\
HD125184 & 14 15 & -07 19 & 702 & 7:20 & 30 & 56 & 218  \\
HD125184 & 14 15 & -07 19 & 701 & 8:00 & 30 & 49 & 230  \\ 
\tableline
HD142373 & 15 51 & +42 35 & 702 & 5:45 & 20 & 57 & 040  \\
HD142373 & 15 51 & +42 35 & 630 & 7:20 & 10 & 67 & 006  \\
HD142373 & 15 51 & +42 35 & 701 & 7:50 & 10 & 67 & 350  \\
HD142373 & 15 51 & +42 35 & 701 & 8:30 & 10 & 64 & 335  \\
\tableline
HD154345 & 17 01 & +47 08 & 702 & 5:50 & 30 & 45 & 043  \\
HD154345 & 17 01 & +47 08 & 701 & 8:40 & 30 & 63 & 002  \\
HD154345 & 17 01 & +47 08 & 704 & 8:50 & 15 & 63 & 035  \\
\tableline 
HD165908 & 18 05 & +30 33 & 703 & 5:40 & 30 & 57 & 065  \\
HD165908 & 18 05 & +30 33 & 702 & 7:30 & 30 & 34 & 066  \\
\tableline
\tableline
\end{tabular}
\end{center}
\end{table}

\begin{figure}
\subfloat[HD 125184 Polarization]{\label{fig:125pc}
\includegraphics[ width=0.22\textwidth, angle=90]{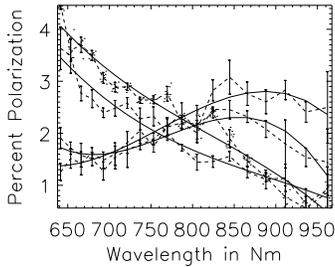}}
\quad
\subfloat[HD 114710 Polarization]{\label{fig:114pc}
\includegraphics[ width=0.22\textwidth, angle=90]{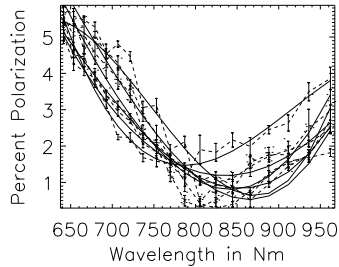}}
\quad
\subfloat[HD 142373 Polarization]{\label{fig:142pc}
\includegraphics[ width=0.22\textwidth, angle=90]{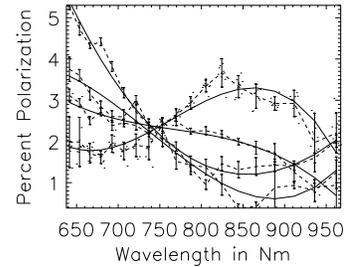}}
\caption[2005 Unpolarized Standard Star Spectropolarimetry]{  {\bf a)}  The degree polarization in percent for the unpolarized standard star HD125184. The progression in time at 650nm from the bottom curve to the top curve is 7:20, 8:00, 5:55, 5:45UT.  {\bf b)}  The degree of polarization in percent for the unpolarized standard star HD114710.  The progression in time is not very significant.  The curve that comes above the rest from 800nm to 950nm is at time 7:00UT.  {\bf c)}  The degree of polarization in percent for the unpolarized standard star HD142373.  The progression in time at 650nm from the bottom curve to the top curve is 5:45, 7:20, 8:30, 7:50UT.  }
\label{fig:comp}
\end{figure}

\twocolumn

	HD114710 spectropolarimetry for many different pointings is shown in figure \ref{fig:114pc}.  The spectropolarimetry does not show much change with pointing, but it does show a strong change with wavelength at all pointings, varying from 1\% to 6\% in a roughly quadratic way.  The change in azimuth was 20$^\circ$ (76-44) and the altitude changed by 30$^\circ$ (309-290).

	Spectropolarimetry of HD142373, shown in figure \ref{fig:142pc} shows one of the strongest changes of polarization with pointing seen, but was in the north.  During the observations, the star's elevation does not change much ($\sim8^\circ$, 57-64) , but the stars azimuth changes by almost 100$^\circ$ (40-335) and the star transits in the middle of the data set, between 7:20 and 7:50UT.  The blue polarization (650nm) drops by 3\% while the 850nm polarization goes from near zero to 3\% and the 950nm rises by only 1\%.  

	Using these types of measurements for unpolarized standard stars at many different pointings, one can construct a response map for the telescope.

\subsection{The Induced Polarization All-Sky Map}

	The observations of the degree of polarization for all the unpolarized standards observed, averaged to a single measurement per order are plotted on the alt-az plane to illustrate the induced polarization for a single order (wavelength) as a function of pointing.   Figure \ref{fig:unsurf0} shows the measurements for a single order (640nm).  There are 19 of these sky-map surfaces (19 orders in alt-az) that constitute the telescope response used for calibrating the instrument.  Note how the induced polarization reaches a maximum of $\sim$6\% at high altitude in the North and has a double-valley structure falling to around 2\% in the south (180$^\circ$) at lower elevations ($30^\circ$).  This order has the highest induced polarization meaning that the induced polarization is strongest at shorter wavelengths for this instrument. The longest wavelength order is shown in figure \ref{fig:unsurf18}.  The peak still occurs at high elevation in the north, but has a much lower magnitude of 3.5\% and falls to 1\% at low elevations in the south.  These maps illustrate the wavelength dependence of the polarization as a function of position on the sky.  As seen in figures 13-15, the errors are 0.1-1.0\% with a dependence on wavelength and pointing, making the calibration uncertain at this level.  
	
	To calibrate future spectropolarimetry, an interpolation of all 19 surfaces (wavelengths) to the pointing of the telescope in the middle of each data set can be used as the induced polarization calibration applied to the measurements.  The error in this calculation is then calculated as the combined error in the nearest observations used in the interpolation. It should be noted that for high-resolution line polarimetry, where one is only interested in the polarization changes across a spectral line (e.g. H$_\alpha$), this calibration serves as the baseline.  Since the mirror-induced polarization effects do not change significantly over a single spectral line, the calibration process is simplified by removing the wavelength dependence of all the corrections.

\subsection{2006-2008 Unpolarized Standard Star Observations}

\begin{figure}
\subfloat[Unpolarized Standard Stokes q \& u]{\label{fig:unpol-qu}
\hspace{-5mm}
\includegraphics[ width=0.35\textwidth, angle=90]{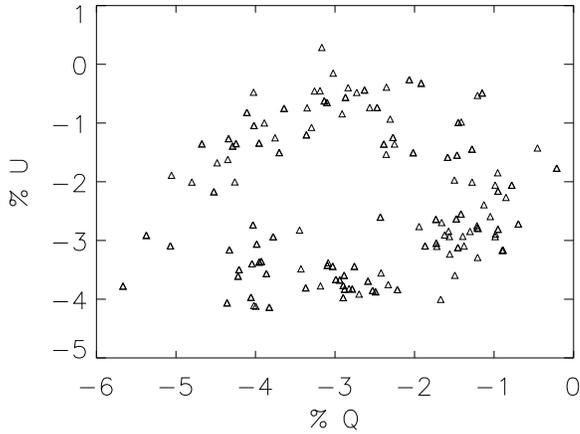}}
\quad
\subfloat[Alt-Azimuth Coverage]{\label{fig:alaz-unpol}
\includegraphics[width=0.35\textwidth, angle=90]{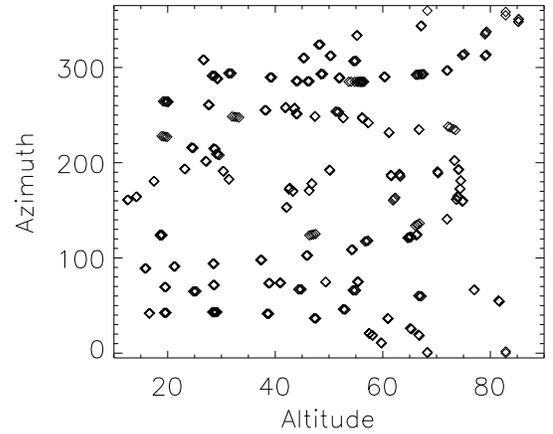}}
\quad
\subfloat[Induced Polarization Surface]{\label{fig:unsurf0}
\includegraphics[ width=0.4\textwidth, angle=90]{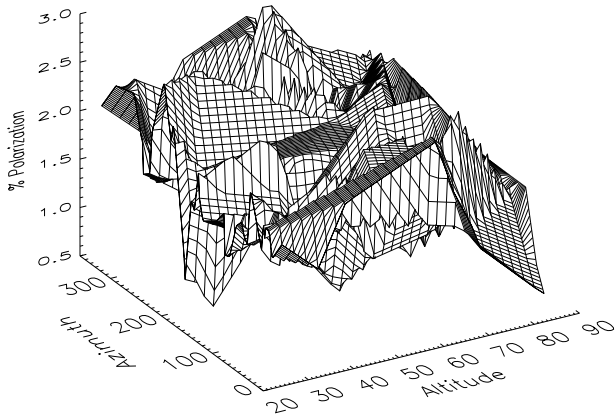}}
\quad
\subfloat[Focus Dependence]{\label{fig:unpol-foc}
\includegraphics[ width=0.3\textwidth, angle=90]{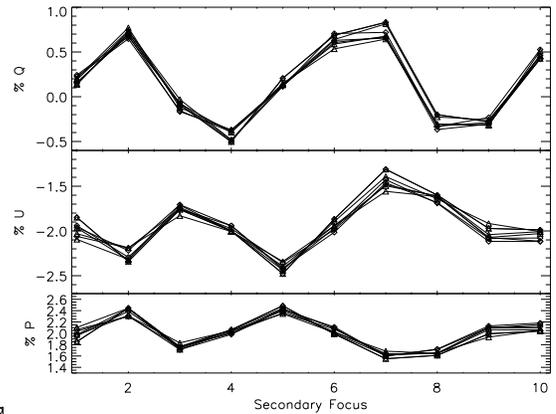}}
\caption[Unpolarized Standard Star Coordinates and Meausrements]{  {\bf a)} The measured polarization of unpolarized standard stars for the H$\alpha$ region plotted in the QU plane. The average value is -2.74\% q  -2.38\% u or an average of 3.8\% at 295$^\circ$, which is essentially the flat-field polarization.  There was no obvious correlation with these points and either altitude or azimuth.  {\bf b)}  The altitude-azimuth coverage for all the unpolarized standard stars used in making the telescope-induced polarization maps.  {\bf c)}  The measured polarization of unpolarized standard stars for the H$\alpha$ region in the alt-az plane. The range of polarization is 0.5\% to 3\% for 130 independent observations.  {\bf d)}  The measured polarization of an unpolarized standard star - HR8430 - for the H$\alpha$ region as a function of the secondary mirror focus. There is a strong dependence on focus value.  The best focus value was 5 and the width of the psf more than doubled for focus values of 1 and 10.}
\label{fig:unpolcoord}
\end{figure}

	After the camera was changed in September 2006, the image rotator was moved, and a better wavelength solution was obtained, another set of unpolarized standard star measurements was performed.  Only the 126 observations after September 2006 were used to produce the maps in this section.  The calculated degree of polarization for each unpolarized standard observed is averaged to a single measurement over the entire H$\alpha$ region.  Note that all the unpolarized standards showed no significant line polarization, and had a statistical error of less that 0.05\%, typically 0.02\%.  These polarization measurements are shown on the q-u plane in figure \ref{fig:unpol-qu}.  The measurements are centered around the flat-field value of -2.95\%q -2.04\%u.  When adjusted for the flat field polarization, they show a range of polarizations from 0.5\% to 3.0\%

	 The observations can be plotted on an altitude-azimuth plane to make an all-sky induced-polarization map.  The altitude-azimuth coverage, shown in figure \ref{fig:alaz-unpol}, was fairly uniform from altitudes of 20 to 80 degrees.  There is a slight gap low in the North that comes from a lack of unpolarized stars in this region - the highest declination star is at +50 (HR7469).  These observations are interpolated to a regular alt-az grid.  Figure \ref{fig:unsurf0} shows the all-sky map of the telescope-induced polarization.  This surface does not show a significant alt-az structure.  To understand the apparent variability in the map, a test of an unpolarized standard star, HR8430, as a function of the secondary mirror focus was performed.  There is a significant variability, or order 1\% as the focus changes shown in figure \ref{fig:unpol-foc}.  This magnitude of variability with secondary focus, combined with the induced-polarization map suggest that the telescope induced polarization map has a very small alt-az dependence, of the same order as the focus dependence.    	

\subsection{Spectropolarimetry of Scattered Sunlight }	

	At twilight, scattered sunlight is a linearly polarized source with a roughly known position angle and degree of polarization at all pointings.   This assumes singly-scattering of incident solar radiation.  Singly-scattered light is polarized orthogonal to the scattering plane with the degree of polarization.  Since the scattering geometry changes with pointing, so does the degree of polarization of the scattered sunlight (which reaches a maximum 90$^\circ$ from the sun).  Observing scattered sunlight at many different altitudes and azimuths allows the measurement of the instrument's response to linearly polarized light.

\onecolumn
\twocolumn

\subsection{Rayleigh Sky Polarization Models}

	Scattered sunlight is reasonably described by a simple single-scattering Rayleigh model (Coulson 1988).  Singly-scattered light is polarized orthogonal to the scattering plane with the degree of polarization proportional to $sin^2 \theta$.  The twilight polarization is highest at a scattering angle of 90$^\circ$ which reaches a maximum 90$^\circ$ from the sun on the North-Zenith-South great circle when the sun is in the West.  The degree of polarization in a Rayleigh atmosphere can be simply described with this equation:
	
\begin{equation}
d=d_{max} \frac{\sin{g}^2}{1+\cos{g}^2}
\end{equation}

\begin{equation}
\cos{g}=\sin{t_z}\sin{t}\cos{p}+\cos{t_z}\cos{t}
\end{equation}

Where g is the angular distance between the pointing and the sun, $t_z$ is the solar elevation, t is the pointing-to-zenith angle, and p is the pointing-to-solar meridian distance.  Contained the variable g is the altitude-azimuth dependence for a given location of the sun.  At sunset, $t_z\sim$0, simplifying the model.  The observations were done with the sun lower than about 15$^\circ$, typically from 45 minutes before sunset to just after sunset.  Using this simplification, the $\cos{g}$ term can be rewritten as $\sin{t}\sin{p}$ to simplify the degree of polarization as:

\begin{equation}
d=d_{max} \frac{\sin{g}^2}{1+(\sin{t}\sin{p})^2}
\end{equation}

\begin{figure} [!h]
\includegraphics[ width=0.35\textwidth, angle=90]{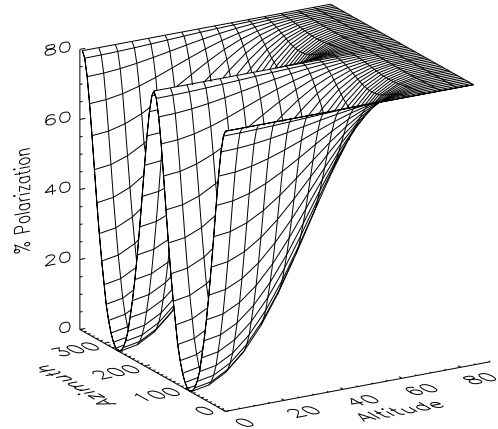}
\caption[Rayleigh Sky Image]{\label{fig:skyim}
The Rayleigh model - the degree of polarization of scattered sunlight with the sun setting in the west projected onto the alt-azi plane.  The polarization is highest where the image is white.  The maximum polarization was set to 80\% }   
\end{figure}

This gives a simple square-sinuosoid in angular distance from the sun (g) modulated by the pointing-to-zenith and meridian distances.  The North-Zenith-South meridian is 90$^\circ$ from the sun and represents the maximum polarization $d_{max}$.  The plane of polarization would always be parallel to the altitude axis (vertical).  In the absence of optical depth effects (which depolarize the light near the horizon), the polarization would always be $d_{max}$ on this meridian.  On the meridian from the East-Zenith-West, the degree of polarization would vary as sin$^2$, being zero in the East and West and $d_{max}$ at the zenith.  The plane of polarization would always be perpindicular to the EZW meridian and aligned with the azimuth axis of the telescope (horizontal).  Figure \ref{fig:skyim} shows this simple Rayleigh model for the twilight polarization projected onto the altitude-azimuth plane.

	At the high-elevation (10000ft) the dry, clear atmospheric conditions are particularly good for using a Rayleigh model because of the low optical-depth and high degree of polarization.  Near sunset at normal telescope pointings, scattered sunlight has a known position angle and a degree of polarization of typically 30-80\%.  Observing scattered sunlight at many different altitudes and azimuths allows a check of the instrument's response to linearly polarized light as a function of mirror angles.  

	This scattered light is not a perfect source however.  The degree of polarization is a strong function of the scattering properties of the air.  In particular, aerosol content (salt, dust, water, etc), optical depth (multiple scattering changing with altitude and horizon distance) and reflections from the Earth's surface (land and ocean) can significantly change the observed linear polarization (Lee 1998, Liu \& Voss 1997, Suhai  \& Horv\'{a}th 2004, Cronin et al. 2005).  The site is surrounded by a reflecting ocean surface, and it is unknown what effect this has on the twilight polarization at the site.  Since the sun was up during most of the observations, the polarized light reflected off the ocean surface has potential to complicate the measurements.  However, this effect is expected to be small because the maximum polarization at zenith obsered over many nights was relatively constant, being $\sim 75\%$, while the cloud covering the ocean below changed significantly.  The broad trends in the degree of polarization measured are in agreement with the singly-scattered sky polarization model, but the linear polarization observed by many researchers has been shown to be functions of time, atmospheric opacity, aerosol size distributions, and composition, all of which vary significantly from day to day (Coulson 1988, Liu \& Voss 1997, Suhai \& Horv\'{a}th 2004).  There are no accurate, quantitative simultaneous observations of the twilight polarization to compare with the HiVIS observations, but there are still many useful things to learn from measurements of a highly polarized source.

\onecolumn
\begin{figure}
\subfloat[Sky Polarization 75$^\circ$ Altitute]{\label{fig:sky75}
\includegraphics[ width=0.35\textwidth, angle=90]{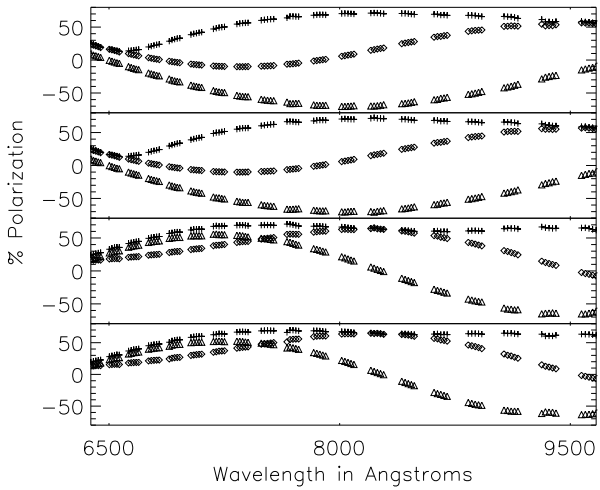}}
\quad
\subfloat[Twilight Polarization 65]{\label{fig:skys65}
\includegraphics[ width=0.35\textwidth, angle=90]{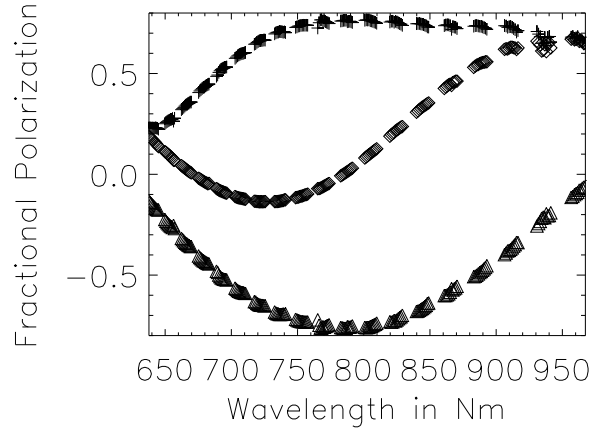}}
\quad
\subfloat[Twilight Polarization]{\label{fig:skyp}
\includegraphics[ width=0.35\textwidth, angle=90]{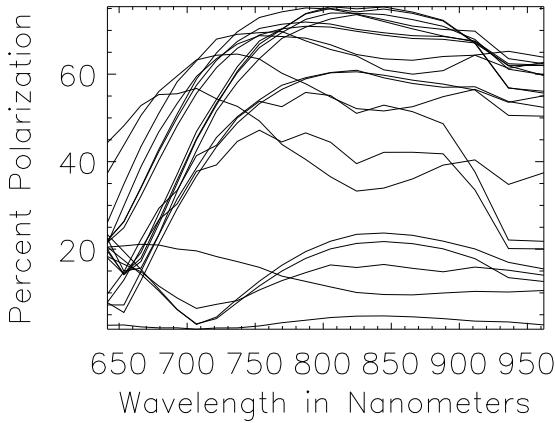}}
\quad
\subfloat[Sky Polarization Angle]{\label{fig:skytheta}
\includegraphics[ width=0.35\textwidth, angle=90]{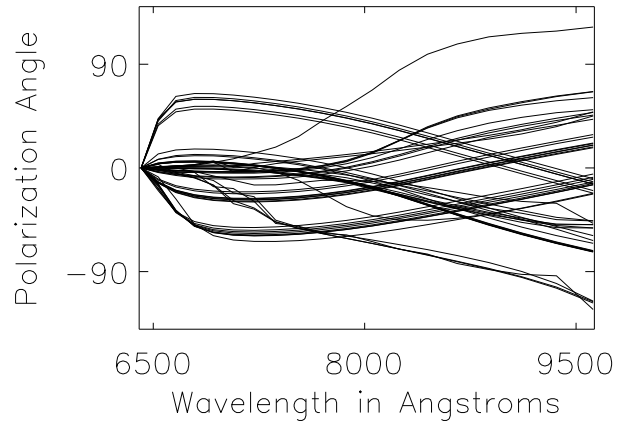}}
\caption[Unpolarized Standard Star Coordinates and Meausrements]{  {\bf a)}  Scattered sunlight polarization at an elevation of 75$^\circ$ for azimuths of NSEW from top to bottom.  Polarization of scattered sunlight is maximum at 90$^\circ$ scattering angle and at sunset, this is the arc from North to South through the Zenith.  The polarization spectra have been averaged 200:1 for ease of plotting.  The symbols are: q=$\Diamond$,  u=$\triangle$,  P=$\sqrt{Q^2+U^2}$=+.  P is the top curve with Stokes q and u are the more sinuosoidal curves with a strong rotation of the plane of polarization ($\frac{1}{2}tan^{-1}\frac{q}{u}$).  {\bf b)}  Scattered sunlight polarization at azimuth of 180 and elevation 65.  Polarization of scattered sunlight is maximum at 90$^\circ$ scattering angle and at sunset, this is the arc from North to South through the Zenith.  The polarization spectra have been averaged 200:1 for ease of plotting.  The symbols are: q=$\Diamond$,  u=$\triangle$,  P=$\sqrt{Q^2+U^2}$=+.  P is the top curve which begins at 25\% at 650nm, rises to 70\% by 750nm and stays flat until 950nm.  Stokes q and u are the more sinuosoidal with a strong rotation of the plane of polarization ($\frac{1}{2}tan^{-1}\frac{q}{u}$).  {\bf c)}  The fractional degree of polarization of scattered sunlight at combinations of altitudes [30,50,65,75,90] and azimuths [N,E,S,W].  The degree of polarization of the scattered sunlight is intrinsically a function of wavelength and pointing since the scattering geometry changes, making quantatative interpretation of the HiVIS degree of polarization difficult, but there is good correlation with theoretical skylight polarization models.  {\bf d)}  The calculated angle of polarization ($\frac{1}{2} tan^{-1}\frac{Q}{U}$) for scattered sunlight at altitudes [30,50,65,75] and azimuths [N,E,S,W] with the sun .  The 52 observations were taken June 26th to July 7th 2005 with somewhat irregular sampling on this alt-az grid.}
\label{fig:unpolcoord}
\end{figure}

\begin{figure}
\subfloat[Sky Polarization Surface]{\label{fig:sky-ps}
\includegraphics[ width=0.35\textwidth, angle=90]{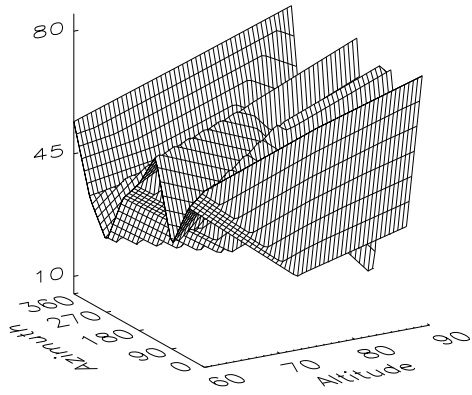}}
\quad
\subfloat[Sky Polarization with Altitude]{\label{fig:sky-pva}
\includegraphics[ width=0.35\textwidth, angle=90]{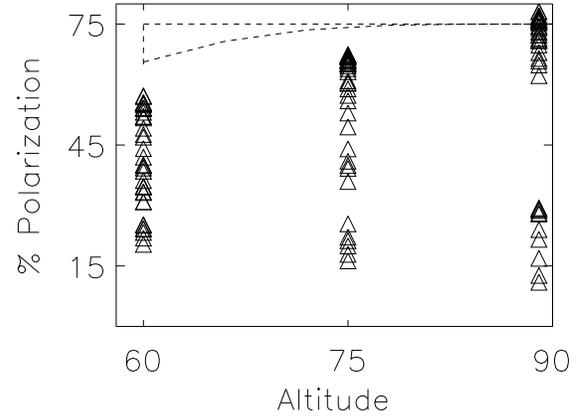}}
\caption[2007 Twilight Polarization Surface \& Altitude Dependence]{ {\bf a)} The average measured polarization of scattered sunlight as a function of altitude and azimuth during sunset for all observations on the 20th-22nd.  The twilight polarization is assumed to follow a simple rayleigh-scattering model with an azimuth-independent maximum at zenith.  The saw-tooth form at high altitudes is entirely the telescope.  {\bf b)} The measured polarization of scattered sunlight as a function of altitude during sunset.  The polarization should rise with altitude, but one rising trend (along cardinal NESW azimuths) and one falling trend (along intermediate azimuths SW, SE, NW, NE) is distinctly seen. The dashed line represents the region all the points would occupy for a simple Rayleigh model with a maximum degree of polarization of 75\%.}
\label{fig:skysurf-and-pva}
\end{figure}

\begin{figure}
\subfloat[Sky Zenith Polarization with Azimuth]{\label{fig:sky89}
\includegraphics[ width=0.23\textwidth, angle=90]{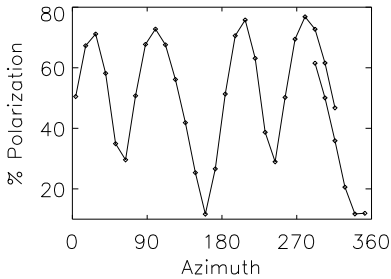}}
\quad
\subfloat[Sky Zenith PA]{\label{fig:sky89ang}
\includegraphics[ width=0.23\textwidth, angle=90]{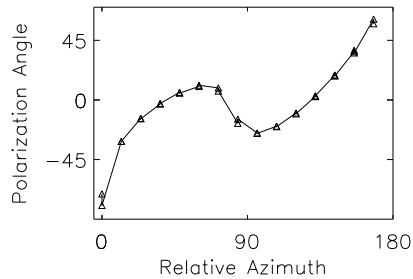}}
\quad
\subfloat[Sky Zenith Polarization]{\label{fig:sky89-depol}
\includegraphics[ width=0.23\textwidth, angle=90]{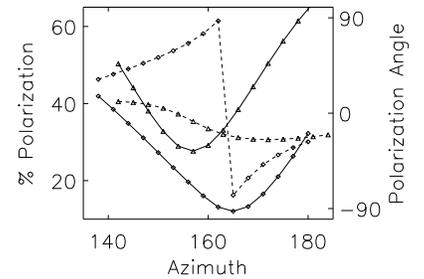}}
\caption[2007 Twilight Zenith Polarization Properties]{ {\bf a)}  The measured polarization of scattered sunlight at the zenith during sunset as the telescope rotates in azimuth.  The first measurement is at 292$^\circ$ at 4:16UT, wrapping around and finishing at 316$^\circ$ at 4:56UT.  The polarization should be constant with all azimuths, but a sinosoidal variation is seen.  This signature is caused by the crossing and uncrossing of the m6-coud\'{e} pickoff mirror pairs.  The minima at 160$^\circ$ and 340$^\circ$ are around 10\% whereas the minima at 70$^\circ$ and 250$^\circ$ are around 30\%.  The maxima are always 65-75\% with a systematic, nearly linear increase with time.  {\bf b)}  The measured position-angle of polarization of scattered sunlight at the zenith during sunset in the instrument frame as the telescope rotates in azimuth.  Since there is a 180$^\circ$ ambiguity in the Stokes parameters, the data have been folded by 180$^\circ$ with the measured angles matching almost perfectly.  The polarization should be rotate linearly with azimuth, but an extra variation of up to 30$^\circ$is seen.  This signature is caused by the crossing and uncrossing of the m6-coud\'{e} pickoff mirror pairs.  The relative azimuth is set to 160$^\circ$ to match the minimum polarization seen in the twilight measurements.  {\bf c)}  The measured degree of polarization and position-angle of polarization of scattered sunlight at the zenith during sunset near the maximum depolarization azimuths.  The solid lines show the degree of polarization with the scale on the left.  The dashed lines show the instrumental position angle of polarization with the scale on the right.  The polarization measurements at azimuths of 232-274$^\circ$, shown in triangles, have been shifted by 90$^\circ$ to fit on this plot.}
\label{fig:zenpol}
\end{figure}

\twocolumn

\subsection{Twilight Spectropolarimetry}

	In 2005, 52 polarization measurements (416 spectra) of scattered sunlight were obtained at altitudes and azimuths of [30$^\circ$, 50$^\circ$, 65$^\circ$, 75$^\circ$, 90$^\circ$] and [0$^\circ$, 90$^\circ$, 180$^\circ$, 270$^\circ$] to create a map of the degree of polarization and position angle for the sky.  These spectra show varied wavelength dependences.  A plot of the degree of polarization for some of the 52 sky polarization measurements is shown in figure \ref{fig:skyp}, and the position angle is shown in figure \ref{fig:skytheta}.  The rotation of the plane of polarization was a very strong function of wavelength and pointing, with over 90$^\circ$ rotation from 6500$\AA $ to 9500{\AA}.   Due to time constraints and the much longer read-time of the old camera, it was impossible to obtain more than 5 measurements on any one evening, and there is significant variability at single pointings of over 5\% night-to-night. Then, these measurements were compared to expected sky polarizations (Coulson 1988, Horv\'{a}th et al. 2002, Pomozi et al. 2001) and found significant telescope effects.

	These spectra showed varied wavelength dependences of polarization and position-angle.  A sample of polarization spectra in the north, south, east, and west at 75$^\circ$ elevation are shown in figure \ref{fig:sky75}.  They have been binned to 200 times lower spectral resolution, five points per spectral order, for ease of plotting.  The figure shows the north-south and east-west pairs are almost identical even though the north and west spectra were taken one night after the south and east spectra.  Most of the 52 spectra showed stronger polarization changes in degree and angle at shorter wavelengths with the degree of polarization becoming more constant at longer wavelengths.  However, the rotation of the plane of polarization was a very strong function of wavelength and pointing, with sometimes 90$^\circ$ of rotation from 6500{\AA} 9500{\AA}).  The strong rotation of the plane of polarization for all 2005 twilight measurements is shown in figure \ref{fig:skytheta}.  These curves do not sample the alt-az grid evenly, but they all show strong, relatively smooth rotation with wavelength.  Some show a very strong dependence between 6500{\AA} and 7000{\AA}.  Even if the degree of polarization becomes more constant, the rotation of the plane of polarization continues with wavelength.  It is hard to explain this as anything but the telescope.   
		
	Much more extensive twilight observations were done with the new hardware in June 20th-22nd 2007.  The AEOS dome walls were raised, but the aperture was open, allowing observations only at altitudes above 60$^\circ$. 109 polarization measurements were obtained over the evenings of June 20th, 21st, and 22nd, typically spending an hour to completely map the sky.  On all three nights, the coverage included at least one complete map and multiple overlapping observations.  Since the sun's azimuth at sunset was 296$^\circ$ during these three nights, the azimuth coverage was offset to reflect this azimuth as the west zero-point.  The observations spanned 4:46-5:19UT on 6-20, 4:11 to 5:01UT on 6-21 and 4:23 to 5:01UT on 6-22.  Over this time frame, the sun's altitude dropped by about 10$^\circ$ and the azimuth changed by 5$^\circ$.  The night of the 20th, one complete grid was obtained at altitudes of 60,75, and 89 degrees elevation at azimuths of 26, 116, 206, and 296 degrees, the equivalent of NESW in the simple rayleigh model.  On the subsequent two nights, the azimuth coverage was extended to include azimuths 71, 161, 251, and 341, or the secondary coordinates (NE, SW, etc).  Nearly three full patterns were obtained on the 21st and nearly two full patterns on the 22nd.

	As with the unpolarized standards, the scattered sunlight polarization measurements can be plotted on the altitude-azimuth plane.  Figure \ref{fig:sky-ps} shows the measured degree of polarization projected on the sky for the June 2007 measurements around the H$\alpha$ line.  There is a striking 4-ridge structure in the degree of polarization that immediately stands out at the zenith.  The oscillations are strongest at high altitude and decrease in amplitude as the altitude falls, becoming double peaked by an altitude of 60$^\circ$.  The position-angle also shows strong variation with pointing.  

	Figure \ref{fig:sky-pva} shows all the measurements plotted versus altitude.  The polarization at the zenith is at its maximum in the simple rayleigh model and should not vary at all with changing azimuth.  The spread in the values at each azimuth shows the effect of the telescope on polarized light.  On all three nights there were observations at a pointing repeated an hour after each other.  The change in polarization in that hour was as much as 10\% (rising by 10\% at zenith as the sun set).  However, the change in polarization from the telescope was roughly 5 times this amount in azimuth.  In an ideal setting, the points would all converge to a maximum polarization ($\sim$75\%) at the zenith while maintaining a spread from low to high at lower altitudes.  The dashed line in the figure shows the simple Rayleigh model range for a maximum polarization of 75\%.  This is obviously not the case, and two things stand out. The first is that there are no points at maximum polarization at lower altitudes.  In the simple rayleigh model, observations to the north and south should be just as strongly polarized as the zenith.  The second is that the polarizations measured at the zenith range from 10\% to 80\%.  Since the measurements at the zenith represent a 2$^\circ$ patch of sky, and the time variability is of order 10\%, the telescope must cause an effective depolarization from 75\% to 15\% for some pointings.

	In order to further examine these surfaces, more twilight measurements were performed at the zenith on July 27th and 28th 2007.  Both nights were relatively clear but with slightly higher humidity.   On the 27th, the telescope was pointed at the zenith and spun the azimuth axis over a number of points, covering 360$^\circ$ in 12$^\circ$ steps with three extra points overlapping.  Figure \ref{fig:sky89} shows the computed degree of polarization and figure \ref{fig:sky89ang} shows the position angle of polarization in the instrument frame.  The first measurement is at 292$^\circ$ at 4:16UT, wrapping around and finishing at 316$^\circ$ at 4:56UT.  Since the telescope never deviated from the zenith, the sinuosoidal oscillation in the degree of polarization is entirely caused by the telescope.  In the hour it took to make these measurements, the sun had dropped several degrees in altitude, increasing the overall degree of polarization at the zenith.  This is causing the points in the right side of the figure to not close the sine-curve.  The plot of position angle has been folded back on itself by 180$^\circ$ since the Stokes parameters are inherently 180$^\circ$ ambiguous. The position angle should go through two complete 90$^\circ$ rotations.  The position angle does this rotation, however it does not do it linearly!  The deviation is at least 30$^\circ$ from linear and the deviation is nearly identical over the hour of the measurement, in contrast to the 10\% change seen in the degree of polarization over the same time period.  Since the rotation of the plane of polarization matches so well over the entire hour of observations, this rotation is entirely caused the telescope.  Also, during the hour this experiment was performed, the overall intensity of the sunlight decreased linearly by 50\% with no dependence on azimuth, showing that there is minimal absorption of different polarization states.  
	
	Another thing to notice is that the minimum polarization at 160$^\circ$ and 340$^\circ$ is close to 10\%, whereas the minimum at 70$^\circ$ and 250$^\circ$ is more like 30\%.  On the 28th, a more in-depth study of the polarization minima was done in order to more fully quantify the minima.  Azimuths of 138-180$^\circ$ and 232-274$^\circ$ in 3$^\circ$ were observed steps spending 13 minutes at each group at 4:26UT and 4:40UT respectively.  Figure \ref{fig:sky89-depol} shows the degree of polarization and measured instrumental position angle for these observations.  The azimuth's have been shifted by 90$^\circ$ to center them.  The deep minimum shows a very strong PA flip through the minimum.  The shallow minimum shows a much weaker angle dependence.  
	
	In summary, the twilight observations show that the telescope can cause a 75\% linearly polarized source to depolarize to 15\%.  The telescope can also rotate the plane of polarization by at least 30$^\circ$ with pointing.  The telescope also causes the rotation of the plane of polarization with wavelength by up to 90$^\circ$ from 6500{\AA} to 9500{\AA}.  All of these effects are strong functions of pointing.  The 2005 measurements suggest that the depolarizing effect greatly diminishes at longer wavelengths, but that the rotation of the plane of polarization continues.  These are very strong telescope polarization effects that must be considered when discussing results with this spectropolarimeter.

\subsection{Zemax Models and Mueller Matricies}

	We constructed a model of the telescope's polarization response using Zemax to compare the predicted polarization effects with the observations.  Zemax software allows one to propagate any number of arbitrarily polarized rays of light through any optical design.  Each material or surface can be given any desired material property (thickness and complex index of refraction).  Programs were written to compute the Mueller matrix of the telescope for any altitude or azimuth by tracking polarized light through the telescope with a range of mirror angles.  The polarization properties of the telescope were traced from the pupil to the first coud\'e focus, just after the coud\'{e} pickoff mirror (mirror 7).  This gives a qualitative idea of the altitude-azimuth dependence of the polarization effects induced by the relative rotation of the altitude and azimuth mirror pairs.

	We started by using the optical constants for aluminum from the Handbook of Optical Constants of Solids II (Palik 1991) and ended up testing a range of refractive indicies from 1.5-2.5 for the real part and 6-10 for the imaginary part.  These values are near the range of values measured for aluminum in the green to near infrared regions.  For instance, in the Handbook of optics the index goes from n+ik of (0.77, 6.1) to (1.47, 7.8) to (2.80, 8.5) for wavelengths of 500nm to 650nnm to 800nm respectively (Bass 1995).  Also, Giro et al. (2003) find an index of (0.64, 5.0) in a V filter (530nm) using a model of the Nasmyth focus of the alt-azimuth TNG telescope to fit observations of unpolarized standard stars.  This is entirely within the reasonable range.  

	To compute the Mueller matrix, six pure polarization states $\pm$Q, $\pm$U, and $\pm$V were input at 400 pupil points each, propagate the light through the telescope at a given pointing, and average the resulting polarized light at the image plane.  Each reflection in the Zeemax models are given the same complex index of refraction.  The six results enable you to calculate the 16-element Mueller matrix for the telescope at a given pointing as the transfer from each input polarization to each output polarization.  These six calculations were computed for each pointing in 1$^\circ$ steps in altitude and azimuth mirror angles to project the Mueller matrix elements on to an all-sky map.  These maps were computed for a number of refractive indicies to determine the dependence of the telescope polarization on the assumed real and complex index components.  From these sky maps the expected amplitude and pointing-dependence of the telescopes polarization effects can be seen.  For notational simplicity, the Mueller matrix will be discussed as in this equation:

\begin{equation}
{\bf M} =
  \left ( \begin{array}{rrrr}
  II   &   QI   &  UI   &  VI          \\
  IQ &  QQ  &  UQ &  VQ     \\
  IU &  QU  & UU  & VU         \\
  IV &  QV  &  UV  &  VV     \\ 
  \end{array} \right ) 
\end{equation}

	The complex component of the refractive index was the dominant term - the higher the complex index, the shallower the wave can penetrate the aluminum and hence the less significant the polarization effects.  A model was run with a complex index of 100, meaning essentially no aluminum penetration, and the models predicted perfect polarization response by the telescope.  The real component of the index had a much smaller effect, with higher indicies meaning more perfect polarization response, as expected.

\onecolumn
\begin{figure}
\subfloat[Zeemax Induced Polarization]{\label{fig:zee-IP}
\includegraphics[ width=0.35\textwidth, angle=90]{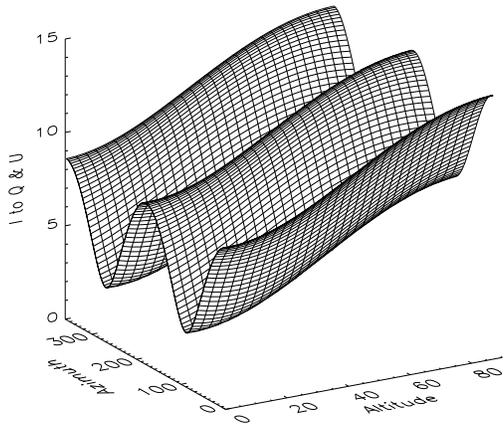}}
\quad
\subfloat[Zeemax QQ]{\label{fig:zee-QQ}
\includegraphics[ width=0.35\textwidth, angle=90]{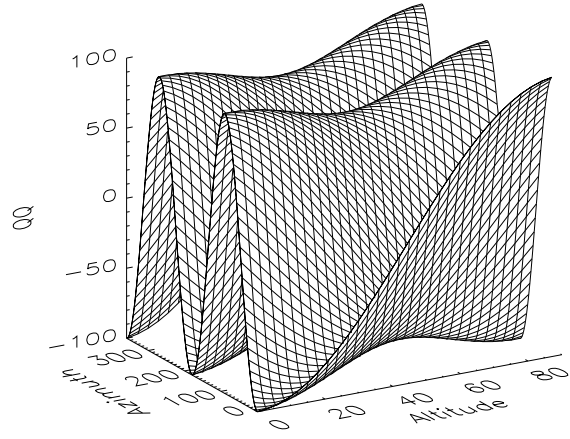}}
\quad
\subfloat[Zeemax Rotation Angle]{\label{fig:zeerot}
\includegraphics[ width=0.35\textwidth, angle=90]{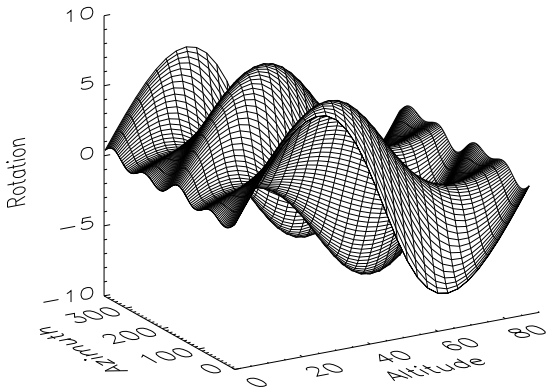}}
\quad
\subfloat[Zeemax VV]{\label{fig:zee-VV}
\includegraphics[ width=0.35\textwidth, angle=90]{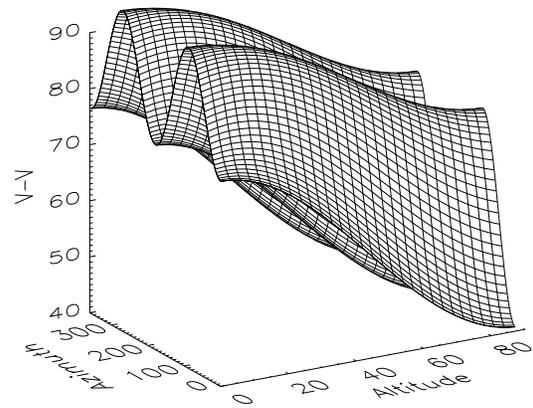}}
\caption[Unpolarized Standard Star Coordinates and Meausrements]{  
{\bf a)} The Zeemax calculated induced polarization, calculated as 100*(IQ$^2$+IU$^2$)$^{\frac{1}{2}}$ at the coud\'{e} focus for an index of refraction n=1.5-8i.  This shows the expected continuum polarization is of order 2\% to 15\% as a function of pointing.  A higher complex index reduces the degree of polarization, but does not significantly change the pointing dependence - all indicies give a double-sine with azimuth, rising with altitude.  {\bf b)}  The Zeemax QQ Mueller matrix term at the coud\'{e} focus for an index of refraction n=1.5-8i.  The matrix element has been multiplied by 100 for axis clarity.  This term shows how the overall reference frame of the instrument rotates with azimuth - a double-sine curve with azimuth is expected for a telescope with no polarization effects.  The slight bow seen at mid-altitudes reflects rotation of the plane of polarization by the telescope, causing small deviations from a perfect double-sine curve.  {\bf c)}  The Zeemax calculated rotation angle in degrees at the coud\'{e} focus for an index of refraction n=1.5-8i.  The angle is calculated as the difference between the ideal telescope (with complex index -$\inf$) and the simulated telescope.  The angle is calculated as half the inverse tangent of QQ on QU.  {\bf d)}  The Zeemax calculated VV Mueller matrix term at the coud\'{e} focus for an index of refraction n=1.5-8i.  The matrix element has been multiplied by 100 for axis clarity.  This term shows a significant loss of circular polarization (V) to other polarization states at certain pointings.}
\label{fig:unpolcoord}
\end{figure}
\twocolumn

	The telescope induced polarization from unpolarized incident light was computed as the IQ and IU Mueller matrix terms added in quadrature, 100$\times$(IQ$^2$+IU$^2$)$^{\frac{1}{2}}$, shown in figure \ref{fig:zee-IP}.  The induced polarization is expected to be minimal when the mirrors are crossed and maximal when they are aligned.  The rotation of the plane of polarization can be calculated as the deviation of the frame rotation from the ideal frame rotation.  The ideal frame rotation is shown in figure \ref{fig:zee-QQ}.  This rotation angle is computed as $\frac{1}{2}$tan$^{-1}$(QQ/QU) with the large complex refractive index.  Figure \ref{fig:zeerot} shows an example of the telescopes rotation of the plane of polarization.  The rotation angles should be maximal at intermediate angles.  Since the azimuth axis has two angles where the mirrors are aligned (parallel and antiparallel), a double-ridge structure is expected in azimuth.

  	These models proved ultimately to be quantitatively inaccurate.  Table \ref{zee} shows the range of Mueller matrix elements for a range of optical constants as well as the range of polarization-plane rotation angles and induced polarizations.  The predicted telescope-induced polarization (IQ and IU) were much greater, 2-28\% depending on pointing and index, than the measured 0.5\%-3.0\% polarization in unpolarized standards.  Even the double-sine shaped pointing dependence predicted by these induced-polarization models is masked by other effects.  
	
	The rotation of the plane of polarization was predicted to be quite small, 3$^\circ$ for a complex index of 10 to 8$^\circ$ for a complex index of 6 with an altitude-azimuth dependence illustrated in figure \ref{fig:zeerot}.  However, very strong rotations were observed, certainly at least 30$^\circ$, as previously seen in figures \ref{fig:skytheta} and \ref{fig:sky89ang}.  The model predicted the depolarization terms, QI UI and VI terms to be nearly identical to the induced polarization terms, IQ, IU and IV.  This predicted moderate to low depolarization, which clearly contradicts the twilight observations in figure \ref{fig:sky89}.  
	
	The rotation of the plane of polarization was also predicted to be only mildly dependent on wavelength since the complex index rises with wavelength from 6 at 500nm to near 11 at 1000nm to over 20 at 2000nm.  The deviation of the QQ, UU, QU and UQ terms from an ideal rotation matrix was small, at most 8$^\circ$ with a complex index of 6, and the rotation decreases with a rising complex index.  The indicies of refraction for aluminum do not change by much over this wavelength range.  In the Handbook of optics, the index of refraction goes from (1.4, 7.8) at 6500{\AA} to (2.8, 8.5) at 8000{\AA} and then to (1.2, 11.2) at 11300{\AA} (Bass 1995).  The complex index of refraction is always 8 or higher, meaning the model will predict less than 4$^\circ$ rotation of the plane of polarization.  This also has been measured to be completely false by the twilight observations since the plane of polarization rotates by over 90$^\circ$ for some pointings, shown previously in figure \ref{fig:skytheta}.  

\begin{table}
\begin{center}
\begin{scriptsize}
\caption[Zeemax Mueller Matrix Terms]{Zeemax Mueller Matrix Terms \label{zee}}
\begin{tabular}{lcccc}
\hline
{\bf Mueller}              & {\bf Index}                   & {\bf Index}                    & {\bf Index}                       & {\bf Index}            \\                
{\bf Term  }               &{ \bf (1.5, 6)}                 & {\bf (1.5, 8)}                 & {\bf (1.5, 10)}                  & {\bf (2.5, 8)}              \\
\hline
IQ, QI                        & -0.17 to 0.28               & -0.10 to 0.17              &  -0.06 to 0.11                & -0.16 to 0.26           \\
IU, UI                        & -0.22 to 0.22               & -0.13 to 0.13              &  -0.09 to 0.09                & -0.20 to 0.20           \\
IV, VI                        & -0.04 to 0.08               & -0.02 to 0.03              &  -0.01 to 0.02                 & -0.03 to 0.05           \\
VQ, QV                    & -0.22 to 0.77               & -0.17 to 0.63              &  -0.14 to 0.53                 & -0.16 to 0.60           \\
VU, UV                    & -0.87 to 0.62               & -0.75 to 0.50              &  -0.64 to 0.41                 & -0.71 to 0.47           \\
VV                            & 0.41 to 1.00                & 0.64 to 1.00               &  0.76 to 1.00                  & 0.66 to 1.00             \\
Rot ($^\circ$)         & -8.1 to 7.6                   & -4.4 to 4.3                  &  -2.8 to 2.7                     & -4.2 to 3.7                 \\
\% Pol                     & 5.8 to 28.2                  & 3.4-16.4                     & 2.2 to 10.1                     & 5.2 to 25.5                 \\
\hline
\end{tabular}
\end{scriptsize}
\end{center}
\end{table}

	We are unable to measure circular polarization with the current instrument configuration, but the circular-induced term (IV and VI) were predicted to be very small, less than 0.1 for the lowest complex index and less than 0.02 for the highest.  The circular cross-talk terms (VQ, VU, UV, QV) were all predicted to be very significant, running up to nearly 1 as a function of pointing and index.  This also implies that significant V sensitivity will be lost at certain pointings, as illustrated by the VV term being less than 50\% for some pointings in figure \ref{fig:zee-VV}.  
	
	There is also further evidence that the index of refraction is significantly different for this telescope at longer wavelengths.  Circular cross talk has been measured from other instruments using the AEOS telescope.  The Lyot project uses a near-infrared (JHK) coronographic imaging polarimeter mounted in a different coud\'{e} room (Oppenheimer et al. 2003).  This room changes the zero-point of the azimuth mirror, but does not change the optical path.  They also utilize the adaptive-optics system for the telescope, which adds several reflections just before the coud\'{e} pickoff and outputs a collimated beam instead of a converging beam.  The instrument first has to change the beam diameter by a number of oblique folds.  The polarization analysis is then performed in the collimated beam just after the coronograph with liquid-crystal waveplates and a Wollaston prism.  Despite the differences between the optics and wavelengths, they also found very significant Q, U, V crosstalk that was pointing dependent in H band (1600nm) images (Oppenheimer et al. 2008).  The total polarization $\sqrt{q^2+u^2+v^2}$ for AB Aurigae was constant over a range of pointings though the individual terms varied.  Since the complex index is predicted to be over 16 at this wavelength the Zeemax models would produce very small cross-talk even with several added mirrors.   

  	The aluminuum oxide coating that forms on every mirror is a major source of uncertainty.  Since it is impossible to precisely model the effect of aluminum oxide on the mirror surfaces, which always form, are significant, and possibly variable in time, the models are not expected to be quantitative.  The handbook of optics states that the oxide layers typically reduce the optical constants by 25\% in the IR, 10-15\% in the visible, and not much in the UV (Bass 1995).  This reduction is small compared to the range of input parameters used for this analysis, however, with so many mirrors, the contribution is suspected to be significant. Furthermore, the model does not include any of the spectrographs common fore optics.

\onecolumn
\begin{figure}
\centering
\subfloat[Example QU Loop]{\label{fig:iqu}
\includegraphics[ width=0.6\textwidth, angle=90]{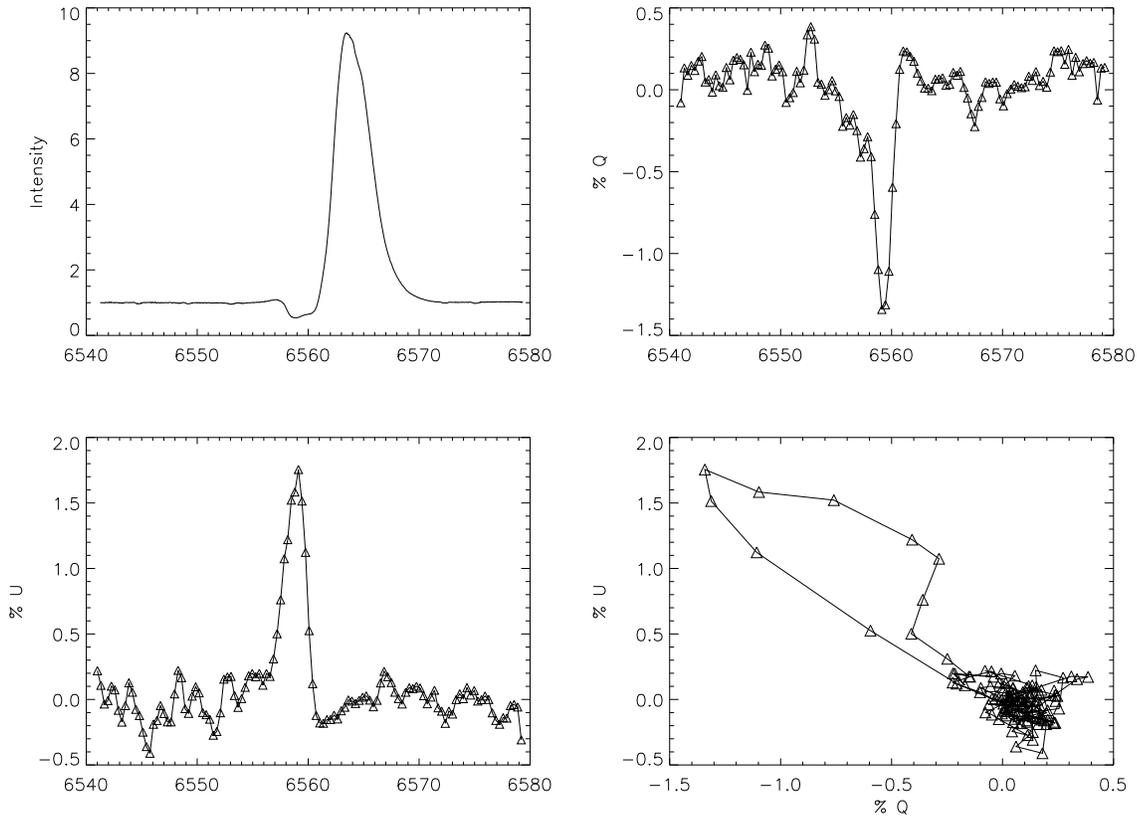}}
\quad
\subfloat[Telscope Polarization Effects]{\label{fig:telpol}
\includegraphics[ width=0.45\textwidth, angle=90]{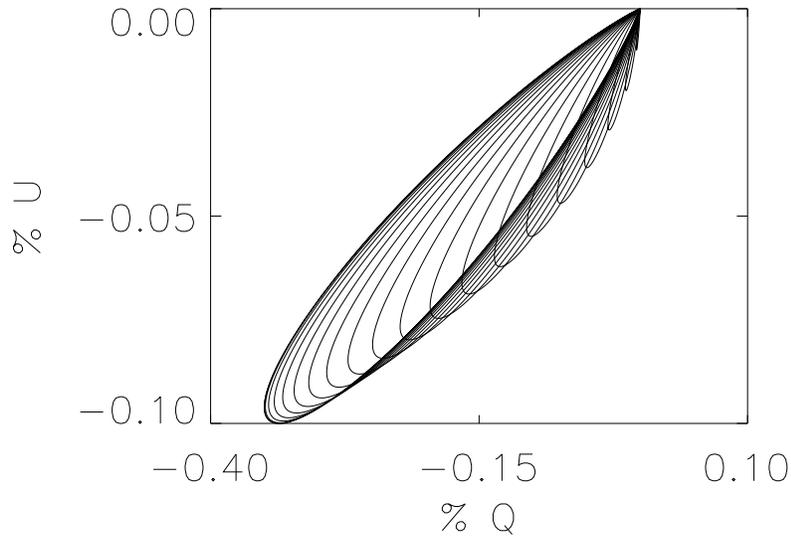}}
\caption[QU Loop Example with Telescope Effects]{ {\bf a)} A sample intensity, q, and u spectra of AB Aurigae, averaged heavily and plotted with wavelength and in the qu plane.  The top left box shows the intensity.  The top right and lower left show q and u respectively.  The bottom right shows a plot of q vs u, illustrating a qu-loop.  The continuum polarization has been set to zero, so a cluster of points representing this continuum for a knot around zero.  The loop is caused by the change of q and u across the absorptive component of the line.  {\bf b)} A simulation showing how the telescope nullifies a qu-loop as it becomes an increasingly strong polarizer.  The telescope attenuates polarization at some direction until one polarization state is completely extinguished.  If the qu-loop is aligned with the position-angle of the polarizing telescope, the qu-loop is completel nullified.  However, to extinguish a qu-loop, the telescope must be nearly 100\% polarizing.}
\label{fig:queff}
\end{figure}
\twocolumn

	The polarization measurements of these common optics described above showed that the degree of polarization remains nearly identical to the input polarization and that there are no orientation-dependent effects.  Complete linear polarization at the coud\'{e} focus (via a linear polarizer at the calibration stage) is completely reproduced by the polarimeter.  This suggests that these common foreoptics are not that significant in their overall effect on the degree of polarization for any input angle.  However the measurement does not constrain an orientation-independent rotation (a PA offset).  The flat field measurements shows 3.6\% polarization, showing that the polarization induced by the common fore optics is also small, even though there are 8 reflections including the three highly-oblique image-rotators reflections between the calibration stage and the slit.  This also suggests that most of the severe polarization effects are caused by the telescope mirrors.  These Zeemax models are only an illustrative guide of what polarization the relative rotation of the mirrors can induce and what the pointing dependence may look like.     
	
	In Harrington et al. 2006 it was claimed that the Zeemax model gives justification for neglecting circular polarization.  The circular polarization terms IV and VI were predicted to be quite small, typically less than 0.05.  The VQ, QV, UV and VU terms, which describe incident circular polarization becoming measured linear polarization were much larger, typically 0-0.7.  Since most astronomical sources have nearly zero circular polarization and the measured V signatures across H$_\alpha$ in Herbig Ae/Be stars is very small it was concluded that neglecting circular polarization was justified (cf. Catala et al. 2007 or Wade et al. 2007).  Now it has been shown that these Zeemax models are quantitatively inaccurate and there is uncertainty about the telescope-induced circular polarization.  However, even though these effects are expected to be severe, since this survey focuses on observing line polarization, even severe circular crosstalk will not impart more than a continuum-offset to the polarization measurements since they have negligible wavelength dependence across a single spectral line.  The linear polarization detections reported are typically an order of magnitude greater than the circular signatures reported elsewhere (cf. Harrington \& Kuhn 2007 and Wade et al. 2007) and these cannot be the result of cross-talk.

\subsection{Polarization Calibration Summary}

	The telescope polarization effects can be quite severe and complicated.  The effect on stellar spectropolarimetry is significant, but useful observations can still be performed with this instrument.  The telescope can not induce any polarization effects across a spectral line because the wavelength dependence of the induced polarization is quite weak at these wavelength scales.  This means that any observed spectropolarimetric signature at some wavelength in a spectral line must be real, and at least at the observed magnitude.  The measured signature could have been depolarized, rotated, translated, or be a cross-talk signature from the stars circular polarization, but the wavelength of the change in the line cannot be affected and the amplitude serves as a lower limit on the true spectropolarimetric effect.
	
	The effect of telescope polarization is best illustrated in the so-called qu-loops.  Polarization spectra are typically presented as either q and u spectra in i-q-u plots, or as degree of polarization and position angle spectra in the i-p-t plots.  Some authors augment these with plots of the polarization in qu-space, ie plotting q vs u (cf. Vink et al. 2005b or Mottram et al, 2007).  When a polarization spectrum is plotted in qu space, where each wavelength becomes a point in q and u, the telescope's polarization effects can be simply represented as rotations, translations, shears, and amplitude-attenuations.  Figure \ref{fig:iqu} shows an example of a q and u spectrum of AB Aurigae plotted in the qu plane.  The spectra have been smoothed heavily to remove noise and clarify the example.  The overall intensity is plotted in the top left panel.  The q and u spectra, shown in the top right and bottom left respectively, show strong spectropolarimetric effects in the absorptive part of the line profile.  In the qu plane, when plotting q vs u, these spectra show a loop.  The polarization of the continuum outside the line creates the so-called continuum knot at zero.  Since the continuum polarization has been removed from these spectra, each point on the graph from outside the spectral line plots near zero forming a cluster of points.  The spectra show a change of -1.4\% and 1.7\% for q and u respectively, which occurs near the red side of the absorption trough.  These changes send the qu-loop off to at a PA of 135$^\circ$ to a maximum amplitude point (-1.4, 1.7).

\section{Instrument Comparisons: ESPaDOnS, WWS, HPOL and ISIS}
     	
	Spectropolarimetric measurements are some of the more difficult measurements to make in an astronomical setting. Measuring physically interesting signals requires accurately measuring 0.1\% differences in the shape of polarized spectra. There has not been a systematic cross-instrument comparison to verify the magnitude and stability of these signatures at high spectral resolution. With this in mind, a number of observations that can be used to compare different instruments was compiled.
	
	There are a few other medium and high resolution spectropolarimeters with published stellar spectropolarimetry we can use for comparisons. There is the ESPaDOnS spectropolarimeter on Mauna Kea's 3.6m Canada-France-Hawaii telescope (CFHT), the ISIS spectropolarmeter on the 4m William Herschel Telescope (WHT), the HPOL spectropolarimeter at the 0.9m Pine-Bluff observatory and there are commissioning observations with William-Wehlau spectropolarimeter (WWS) mounted on the 1.6m Observatoire du mont M\'{e}gantic (OMM) telescope.
	
	ESPaDOnS is a fiber-fed spectropolarimeter with a resolution of 68000, roughly 5 times higher than HiVIS in it's typical polarimetric mode (Donati et al. 1999, Manset \& Donati 2003). The spectropolarimeter has a completely different design than HiVIS. The polarizing optics are mounted after the Cassegrain focus in a triplet-lens collimated beam. The instrument uses a dual-beam design with three fresnel-rhomb retarders and a Wollaston prism which produces the two orthogonally polarized beams. The two polarized beams are then focused onto two optical fibers (with two additional for sky that are unused in polarimetric mode) that feed the spectrograph. The typical exposure sequence is 8 exposures at different Frenel-rhomb orientations, 4 each for Q and U. The telescope is equatorial and, with the polarization analysis performed at the Cass focus, does not have any frame-rotation issues as the Wollaston and fibers are always at a fixed orientation on the sky. This makes the instrument very useful for long integrations. The fiber entrances do however introduce a strong instrumental continuum polarization, requiring a polynomial subtraction similar to HiVIS, and making continuum polarization studies impossible with this instrument. This instrument has a very different setup from HiVIS and as such is a useful cross-check.  
	
	The WHT is an alt-az telescope and ISIS is a long-slit spectropolarimeter. The slit is aligned on the sky by physically rotating the entire spectrograph, so there are no highly oblique reflections off moving mirrors. The polarizing optics are almost identical to the HiVIS design - a Savart plate just after the slit creates the orthogonal polarizations with a rotating quarter- and half-wave retarder mounted in front of the slit (not behind like HiVIS). A polarization observation is the same as our instrument - four sets of four-wave-plate rotations. The slit however has only the primary and secondary mirrors upstream, and those near-normal reflections do not induce significant instrumental polarization (Harries et al. 1996). The telescope-induced polarization of this telescope is insignificant. However, the waveplate does produce a very significant 0.2\% amplitude ripple that must be subtracted from every high-resolution data set.  
	
	The HPOL spectropolarimeter (Wolf, Nordsieck \& Nook 1996) also has a different design and is useful for comparison. The spectropolarimeter is a dedicated instrument on the University of Wisconsin's 0.9m Pine-Bluff observatory (PBO) and is also a visitor instrument on the 4-m WIYN telescope. As with the ISIS spectropolarimeter, the slit is at the Cassegrain focus with a 4-layer achromatic waveplate mounted just behind the slit. The beam is collimated before being passed through a Wollaston prism. The beam is then dispersed by a grating and focused onto the detector.
	
	The WWS has a design very similar to ESPaDOnS, but using two achromatic quarter-wave plates instead of two half-wave and one quarter-wave Fresnel rhombs (Eversberg et al. 1998). The polarization optics are also at the Cassegrain focus and feed optical fibers.
	
	Comparisons between observations with all of these instruments will show that HiVIS, though having severe telescope-induced polarization, can provide good quality measurements.

\subsection{Nearly-Simultaneous ESPaDOnS-HiVIS Observations}

	During the beginning of our major observing campaign in 2006, we observed MWC361 with the ESPaDOnS spectropolarimeter to compare with our HiVIS observations. We observed on August 1$^{st}$ and 3$^{rd}$ of 2006 with ESPaDONS and with HiVIS on November 7$^{th}$. These measurements were quite similar, but the three month time difference left open any questions about variability. 
	
	We then took a back-to-back observation set of MWC361 and HD163296 using HiVIS on June 18-24th 2007 and ESPaDOnS June 26th and 27th. All data sets were taken in fair weather and reduced with the dedicated, automated Libre-Esprit data reduction package provided by CFHT (Donati et al. 1999). As a further test of this package, we adapted our HiVIS reduction scripts to process the ESPaDOnS data in the H$_\alpha$ region and found the results matched very closely.  
	
\begin{figure}
\centering
\subfloat[CFHT MWC 361 Spectropolarimetry]{\label{fig:cfh361}
\includegraphics[width=0.5\textwidth, height=0.8\textwidth, angle=90]{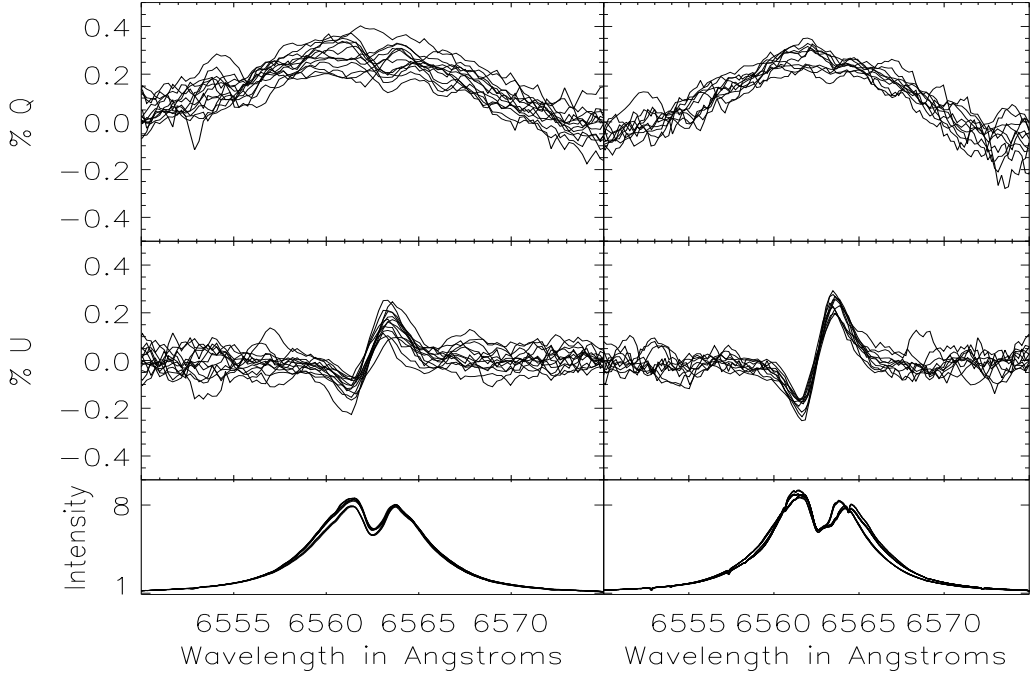}}
\quad
\subfloat[CFHT HD 163296 Spectropolarimetry]{\label{fig:cfh163}
\includegraphics[width=0.5\textwidth, height=0.8\textwidth, angle=90]{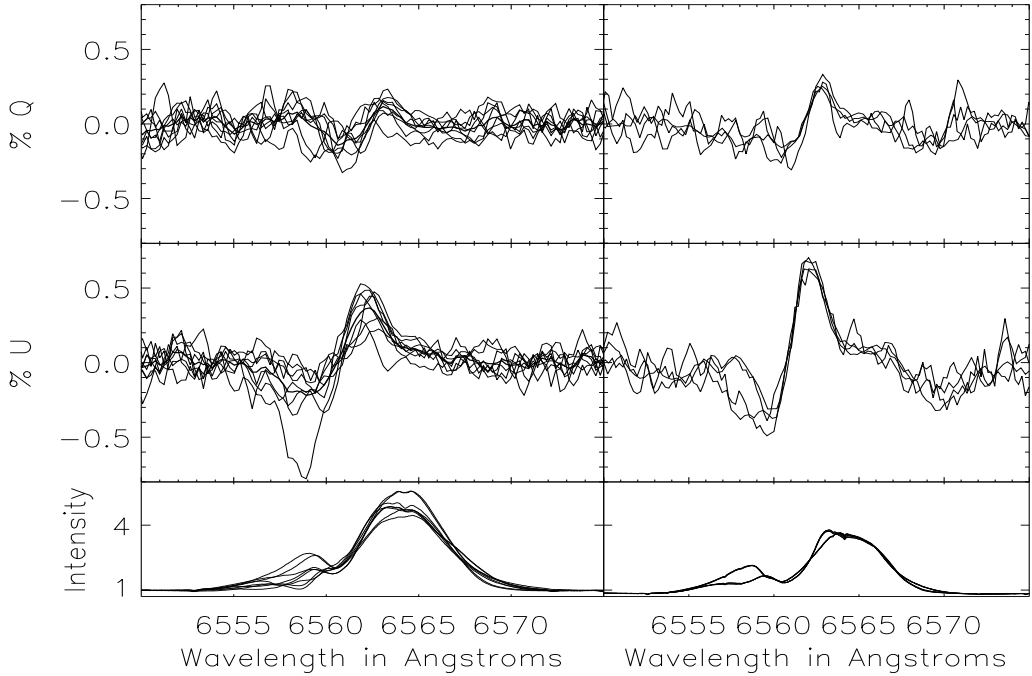}}
\caption[CFHT \& HiVIS Polarization of MWC 361 \& HD 163296]{  {\bf a)}  The spectropolarimerty of MWC361 from HiVIS on the nights of June 18-21st 2007 is shown on the left and from ESPaDOnS on the nights of June 23 \& 24th 2007 on the right.  The HiVIS data has been rotated to align the HiVIS qu-loops with those from ESPaDOnS. {\bf b)} The spectropolarimerty of HD163296 from HiVIS on the nights of June 18-21st 2007 is shown on the left and from ESPaDOnS on the nights of June 23 \&24th 2007 on the right. No rotation has been applied to the HiVIS data. The HiVIS data spans a range of pointings and exposure times. There is a resemblance between some HiVIS and ESPaDOnS observations, but significant telescope polarization effects cause the amplitude of the HiVIS signature to be smaller. All observations have been rebinned to lower spectral resolution for clarity.}
\label{fig:comp}
\end{figure}

	The ESPaDOnS data shows that over a period of three nights in 2006 (Aug 01 to Aug 03) and again over 2 nights in 2007 (June 23rd and 24th), the spectropolarimetry from MWC361 is completely stationary - there is no detectable variability in the polarization spectra. The spectropolarimetry nearly a year apart matches even though the intensity profiles changed slightly. Our HiVIS observations, while at lower spectral resolution, show a similar polarized spectrum. A side-by-side comparison of the MWC361 data is shown in figure \ref{fig:cfh361}. Each data set from HiVIS has been rotated by an arbitrary angle to take out any telescope frame rotation and plane of polarization rotations. These by-eye rotations align the polarization spectra quite well, showing that HiVIS can reproduce the ESPaDOnS measurements quite well.   	

	The ESPaDOnS observations of HD163296 show a large amplitude polarization change, almost 1\% peak-to-peak, near the absorptive component of the emission line. Observations with HiVIS also show a similar effect at these same wavelengths, but the magnitude of the change in the HiVIS data is smaller. Figure \ref{fig:cfh163} shows the side-by-side comparsions of HiVIS and ESPaDOnS data. The HiVIS data has also been de-rotated but the fits are not as good for many of the data sets.  This target had a different and wider range of pointings and could have the polarization amplitude modulated by the telescopes polarization properties.  Still, one can clearly see the proper qu shape and even similar values in the HiVIS observations.  

\onecolumn		

\begin{figure}
\centering
\subfloat[ESPaDOnS MWC 480]{\label{fig:cfht-480comp}
\includegraphics[width=0.35\textwidth, angle=90]{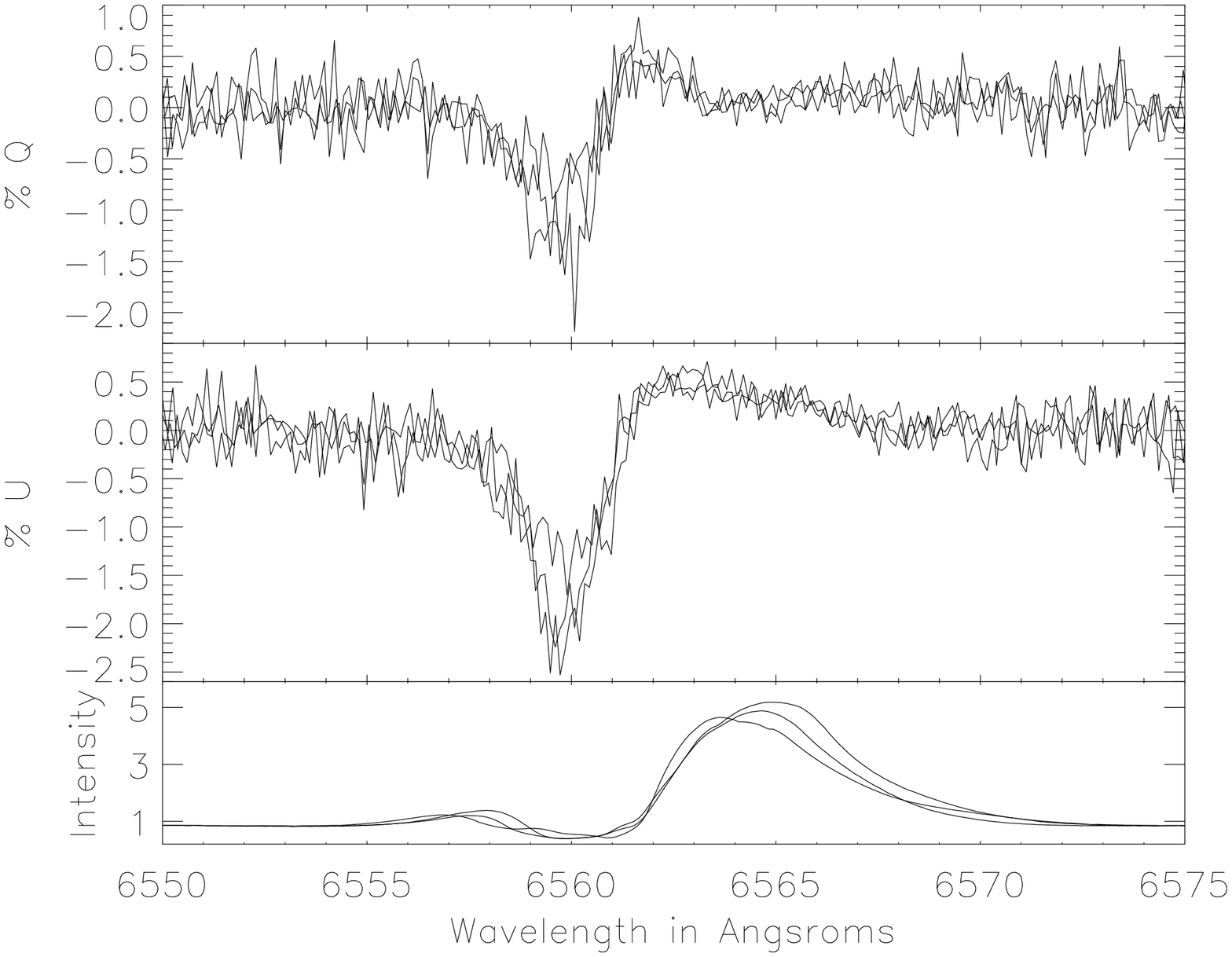}}
\quad
\subfloat[HiVIS MWC 480]{\label{fig:hivis-480comp}
\includegraphics[width=0.35\textwidth, angle=90]{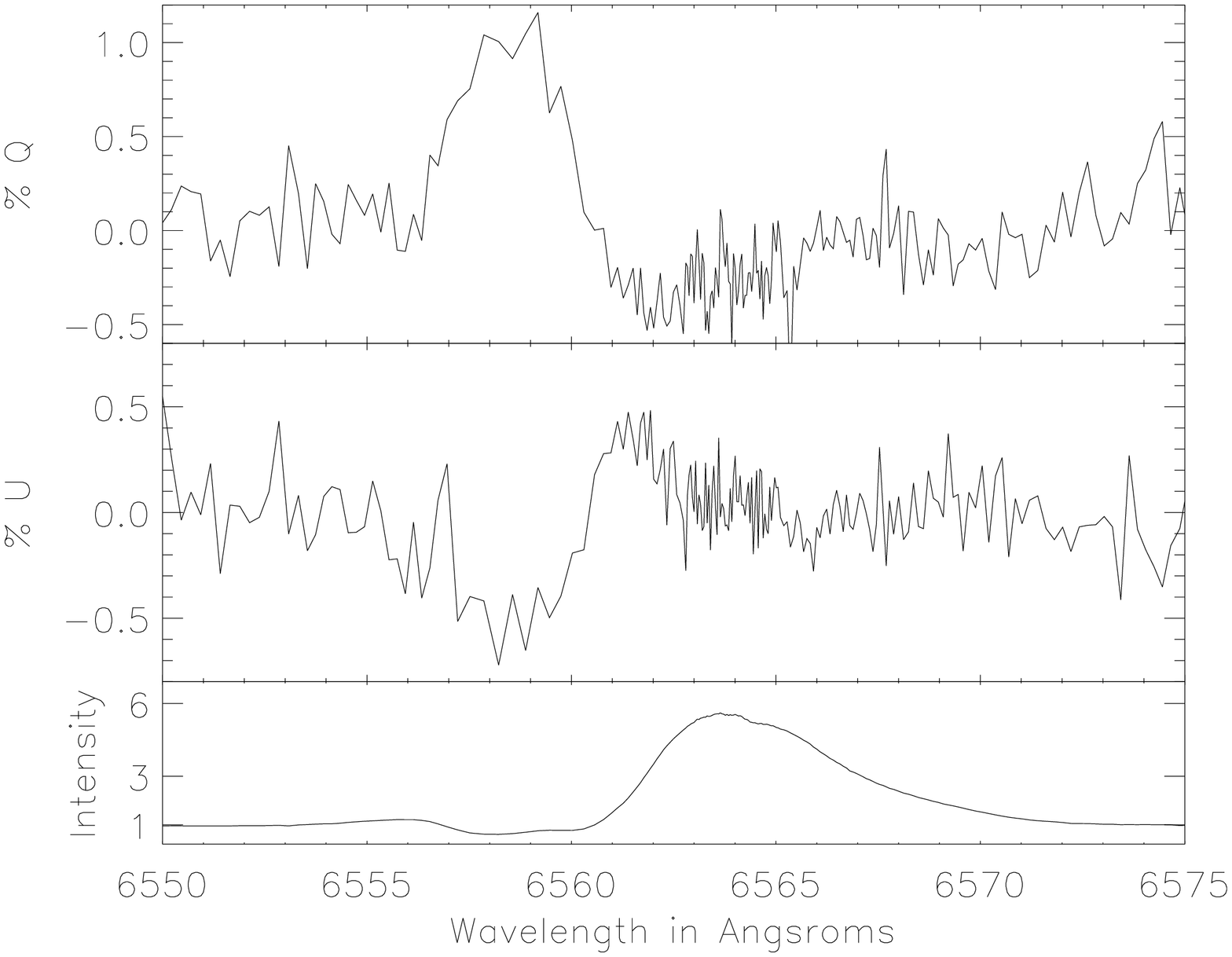}}
\quad
\subfloat[ESPaDOnS MWC 158]{\label{fig:cfht-158comp}
\includegraphics[width=0.35\textwidth, angle=90]{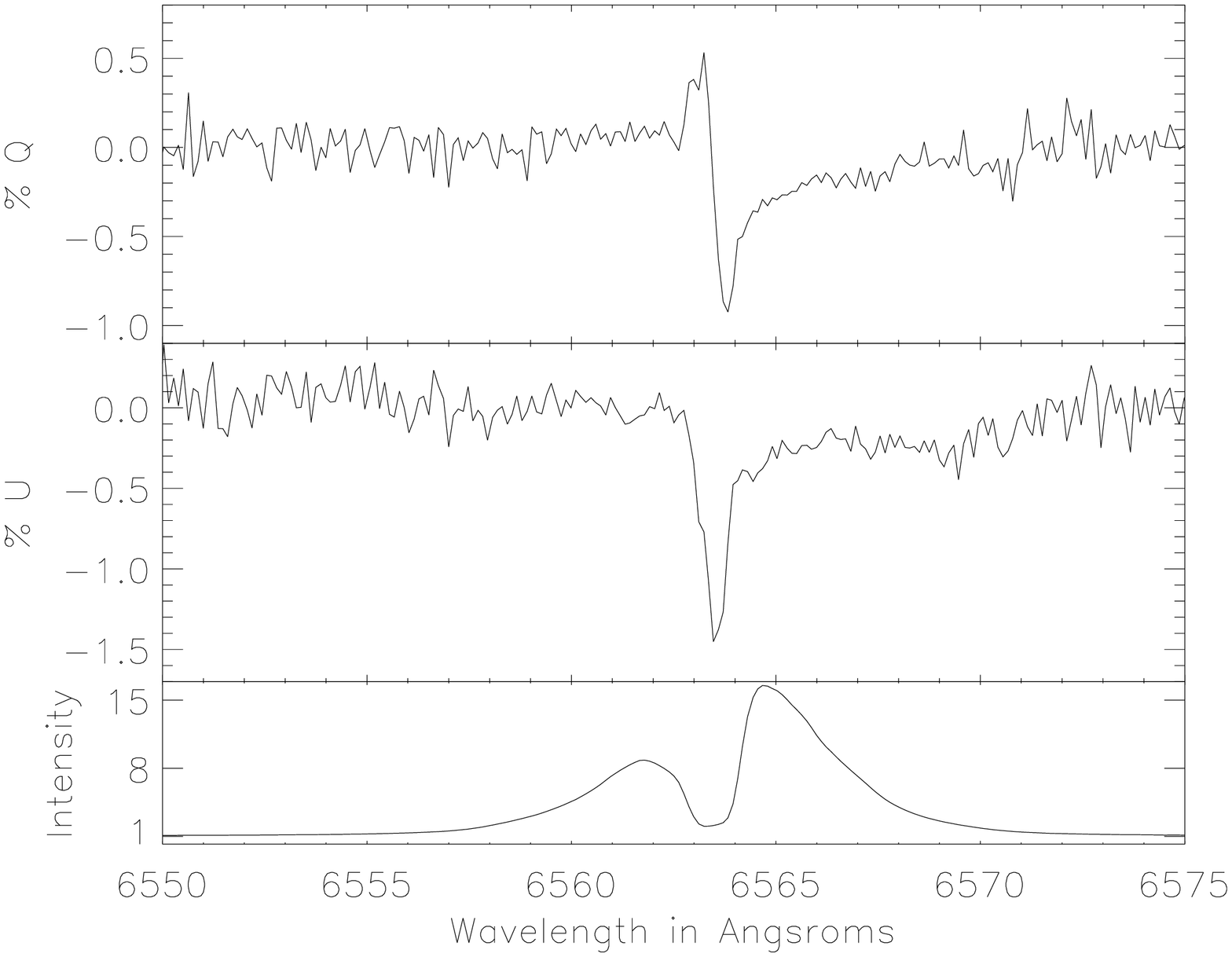}}
\quad
\subfloat[HiVIS MWC 158]{\label{fig:hivis-158comp}
\includegraphics[width=0.35\textwidth, angle=90]{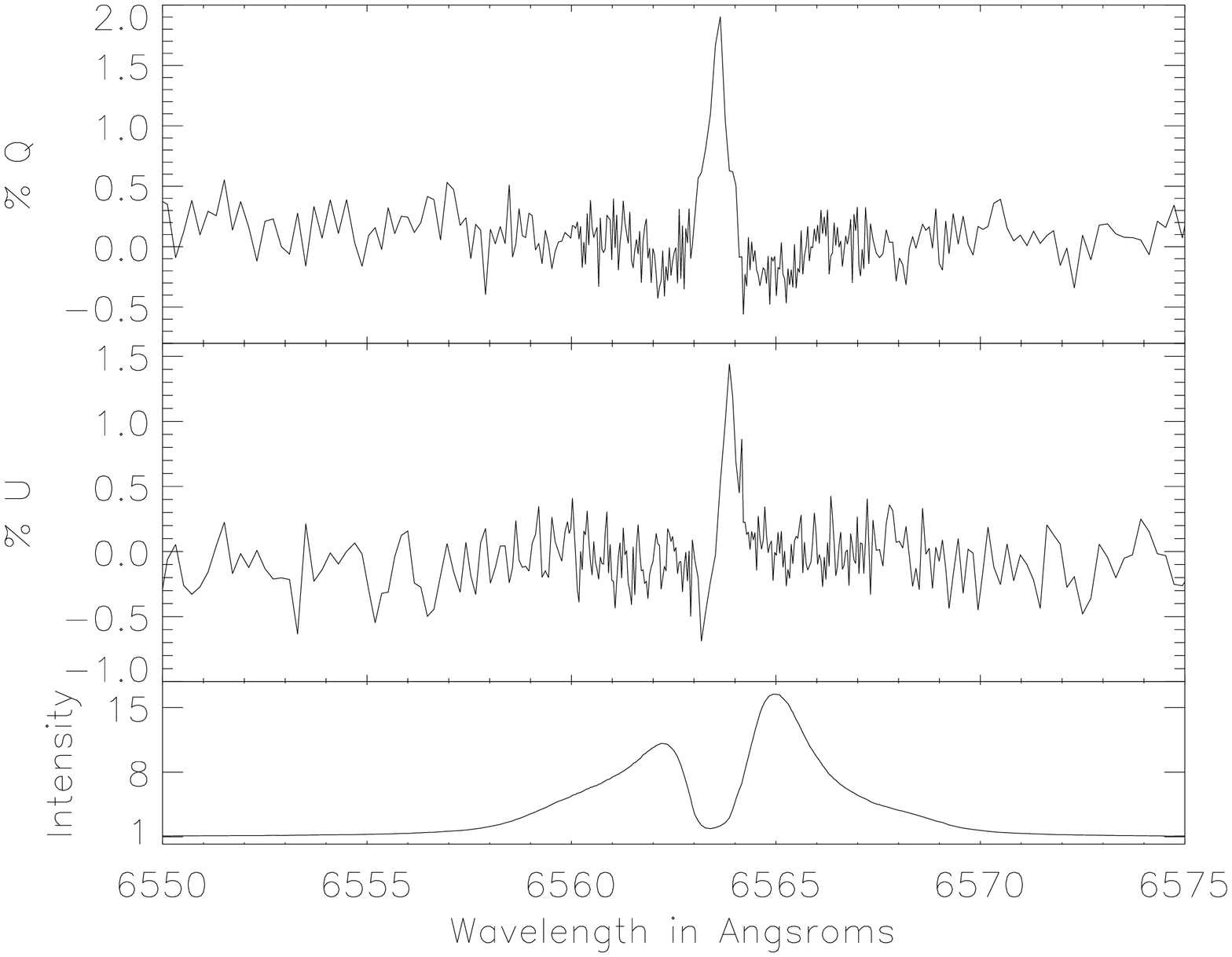}}
\quad
\subfloat[ESPaDOnS MWC 120]{\label{fig:cfht-120comp}
\includegraphics[width=0.35\textwidth, angle=90]{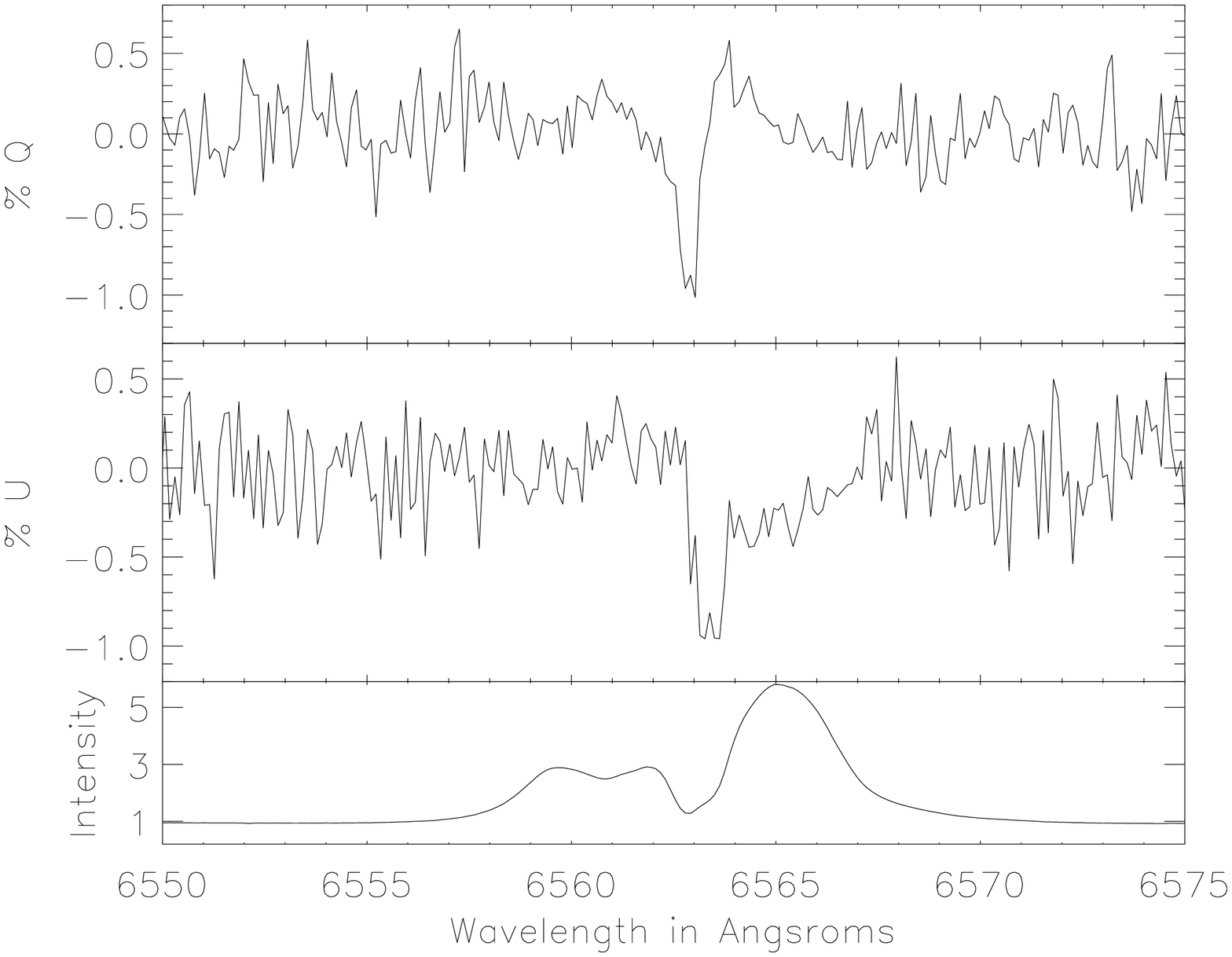}}
\quad
\subfloat[HiVIS MWC 120]{\label{fig:hivis-120comp}
\includegraphics[width=0.35\textwidth, angle=90]{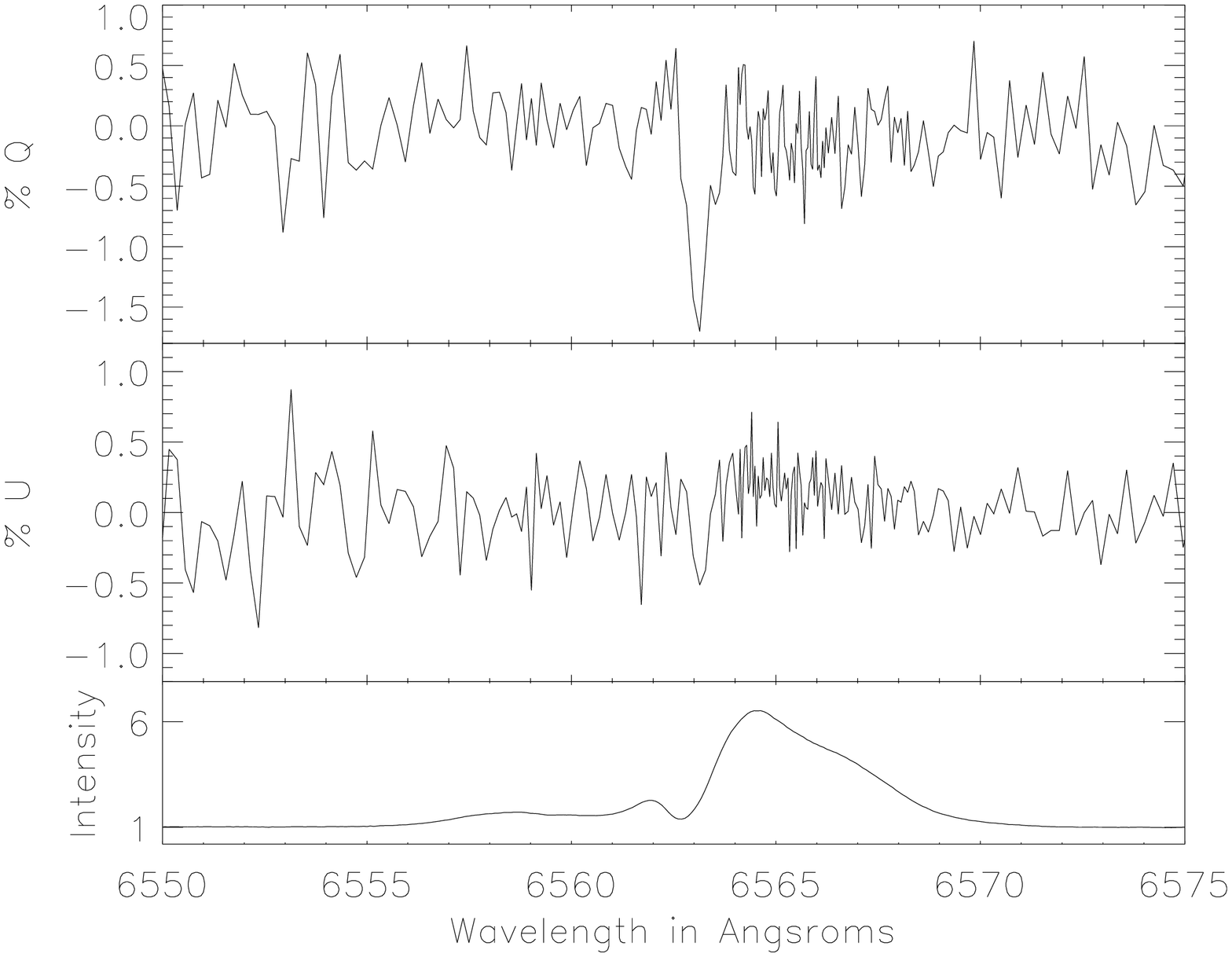}}
\caption[ESPaDOnS vs. HiVIS - MWC 480, MWC 158 \& MWC 120]{  {\bf a)}  The spectropolarimetry of MWC 480 from the CFHT archive and {\bf b)}  HiVIS. {\bf c)}  The spectropolarimetry of MWC 158 from the CFHT archive and {\bf d)}  HiVIS.  {\bf e)}  The spectropolarimetry of MWC 120 from the CFHT archive and {\bf f)}  HiVIS. No rotation has been applied to the HiVIS data, so the exact form of the HiVIS qu spectra is subject to rotation, but the overall magnitude and location of the polarization effects will remain unchanged. The observations match in magnitude and wavelength quite well.}
\label{fig:esp-hiv-mwc}
\end{figure}
\twocolumn

	This illustrates the type of conclusions one can draw with HiVIS data. MWC361 has a high declination, +68, and the range of pointings for all observations were quite small. When above an airmass of 2, the star is always low and in the north, from altitudes of 30-42 and azimuths of 340 to 20. HiVIS was quite successful at reproducing the spectropolarimetric measurements between ESPaDOnS and ISIS. HD163296 however has a declination of -21 and spans a much greater range of pointings - altitudes from 30-48 and azimuths of 132-226.  The signature is still detected at the correct wavelengths, but at a reduced amplitude.

\subsection{ESPaDOnS Archive Comparisons}

	We took archival ESPaDOnS data from February and August 2006 to compare with our October to December 2006 observations. Many individual examples will be discussed in later chapters, but a few examples now can illustrate some of the telescope effects. Figure \ref{fig:esp-hiv-mwc} shows a side-by-side comparison of MWC 480, MWC 158 and MWC 120 spectropolarimetry with ESPaDOnS and HiVIS. The observations were taken at different times, 6-months to more than a year apart. Though the H$_\alpha$ line types are the same, the line profiles have changed significantly between the observations making detailed comparisons difficult. There has been no rotation applied to the HiVIS data. We did not attempt to align the polarization measurements so that the style of comparison used in later chapters can be illustrated. However, even with these complications, the form and magnitude of the polarization measurements matches quite well. 
	
	MWC 480 shows a clear P-Cygni H$_\alpha$ line in both ESPaDOnS and HiVIS observations. The polarization change is over 1\% in each, with the change centered on the P-Cygni absorption. The ESPaDOnS observations show a large magnitude, 2\% in u and 1.5\% in q on two occasions and a smaller 0.5\% q, 1\% u on another. The polarization at the peak of the emission at most 0.3\% away from continuum while the polarization on red side of the line, $>$6567$\AA$, is identical to continuum. The HiVIS observations show 1\% changes in q and 0.5\% in u with a very similar morphology. The polarization changes are largest in the absorptive component with a small sign change on the blue side of the emission, 6563$\AA$. The emission peak is roughly 0.2\% different from continuum, with the return to continuum polarization happening on the red side of the line, near 6566$\AA$. By observing that the operation q $\rightarrow$ u and u $\rightarrow$ -q matches the ESPaDOnS data with the HiVIS data, a crude estimate of 135$^\circ$ rotation between the two observations can be inferred, assuming the star is not variable. Though we can't disentangle telescope polarization effects from actual stellar variability, the form and magnitude are similar. 
	
	MWC 158 also shows a very strong polarization effect in the central absorption feature. The ESPaDOnS observations show an antisymmetric change of 0.5\% in q with a narrow 1.5\% spike in u. HiVIS shows a 1.5\% spike in q with a 0.5\% drop then 1.5\% rise in u over the same narrow wavelength range. In this case the HiVIS amplitudes are greater than the ESPaDOnS data. Since the AEOS telescope can only reduce the magnitude of a spectropolarimetric effect, this star must have a variable spectropolarimetric signature. 
	
	The MWC 120 line profiles changed significantly - the ESPaDOnS observations show a much weaker blue-shifted absorption compared to the HiVIS observations. The ESPaDOnS observations, as in MWC 158, showed a strong change in the central absorptive component. Both q and u show decreases of roughly 1\%. HiVIS shows a 1.5\% drop at the same wavelengths in q only. There is also a smaller but significant broad change on the red side of the line profile in the ESPaDOnS observations. This can also be seen as a small 0.3\% rise in the HiVIS u spectrum. This comparison with the archival data shows that the spectropolarimetry from HiVIS, though subject to magnitude-reduction and rotation from the telescope polarization, can still provide a good measurement of the wavelength and relative magnitude of the polarization changes across a line. Both the narrow large-amplitude change in the absorption and the broad low-amplitude change across the line are reproduced.

\begin{figure} [!h]
\centering
\subfloat[Comparision with Vink]{\label{fig:cfht-vink}
\includegraphics[width=0.3\textwidth, angle=90]{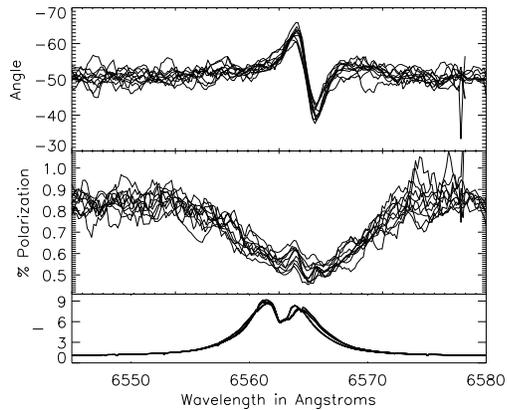}}
\quad
\subfloat[CFHT MWC 361 QU loops]{\label{fig:cfht-loop}
\includegraphics[width=0.3\textwidth, angle=90]{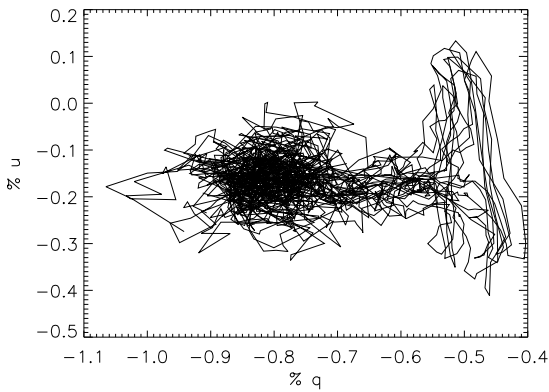}}
\caption[CFHT \% Polarization and PA for MWC 361]{  {\bf a)}  The spectropolarimerty of MWC361 from CFHT on Aug 01 and 03 2006 plotted as degree-of-polarization and position angle.  The continuum polarization has been shifted to match the Mottram et al. 2007 polarization of 0.82\% at 95.6$^\circ$ or q-u continuum of (-0.80\%, -0.16\%) before calculating these polarization spectra.  {\bf b)}  The qu-loops from MWC361 from CFHT on Aug 01 and 03 2006 shown with the Mottram et al. continuum of -0.80\%, -0.16\%.  Stokes q decreases broadly over the line and u provides the small vertical deviations at the qu-loop's maximum amplitude.}
\label{fig:vinkcomp}
\vspace{-3mm}
\end{figure}

\subsection{ISIS Comparison}

	We can also compare our observations to those of MWC361 taken with the ISIS spectropolarimeter, presented in the Vink et al. 2002, 2005b and Mottram et al. 2007 papers. These observations had a resolution of 35km/s or R$\sim$8500, whereas the 1.5" slit on the AEOS spectrograph has a resolution of R$\sim$13000 in the typical polarimetric mode. All data sets are then binned by flux to a constant polarimetric error. Vink et al. 2002, hereafter V02, present data from December 1999, Vink et al. 2005b, hereafter V05, presents data from December 2001 and Mottram et al. (2007) present data from late September 2004. All three data sets show a similar qu-loop and polarization spectrum, though the Vink et al. 2002 data has much higher noise. The continuum polarization was reported as 0.80\% to 0.82\% at a PA of 94$^\circ$ to 96$^\circ$ in V02 and V05 respectively. We can take our CFHT and HiVIS data and scale them to these continuum values (q=-0.8\% u=-0.15\%) and calculate the degree of polarization and PA. We find a very similar polarization spectrum, though with a significantly different PA spectrum, shown in figure \ref{fig:cfht-vink}. This changing PA is caused almost entirely by the u spectrum changing sharply in the center of the line while q is positive over the whole broad region around the line. In the V02 and V05 measurements, the PA change is on the red side of the emission line, with a roughly 20$^\circ$ change. Our measurements now show a double-peaked change, still on the red side of the emission line with a similar wavelength range, but now with a two-fold change: a 15$^\circ$ increase in PA followed by a 15$^\circ$ decrease. The significance of this is best seen in the qu-loops, shown in figure \ref{fig:cfht-loop}. Our plot shows the u changes as both positive and negative with extreme values near -0.4\% and 0.1\% whereas the V05 qu-loops show only a significant decrease in u with a barely noticeable increase.

\begin{figure} [!h]
\centering
\includegraphics[width=0.35\textwidth, angle=90]{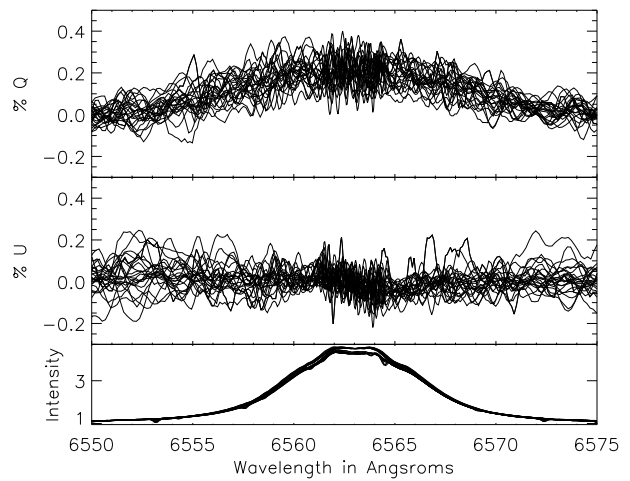}
\caption[$\gamma$ Cas Polarization]{Twenty-five q, u, and intensity spectra of $\gamma$ Cassiopeia taken on many nights from September 2006 to September 2007.  The plane of polarization has been rotated for each spectra to maximize q in the core of the emission line to provide a relative reference frame.  The average of all the polarization spectra is illustrated as the thick solid line showing a broad 0.2\% polarization change across the line.}
\label{fig:gmcas-comp}
\end{figure}

\subsection{WWS and HPOL Comparisons}

	Another comparison between HiVIS and published low to medium resolution spectropolarimetry is possible. The William-Wehlau spectropolarimter (WWS, Eversberg et al. 1998) and the HPOL spectropolarimeter have data on several Be stars with relatively stable spectropolarimetric signatures.  
	
	The WWS has a design very similar to ESPaDOnS, but using two achromatic quarter-wave plates instead of two half-wave and one quarter-wave Fresnel rhombs.  Eversberg et al. 1998 report commissioning observations of $\gamma$ Cas on the 1.6m Observatoire du mont M\'{e}gantic (OMM) telescope that reproduce the 0.2\% polarization change originally seen in a Pockels-cell polarimeter measurement with the University of Western Ontario 1.2m telescope (Poeckert \& Marlborough 1977).  We see a similar smooth and broad signature in $\gamma$ Cas as well, shown in figure \ref{fig:gmcas}.  The observations have all been rotated by an arbitrary angle to maximize the +q spectrum in the core of the emission line.  With the simple, broad form of the polarization signature, this has the effect of producing an effective de-rotation to a common polarimetric frame.  The resulting polarization spectra are averaged to produce our best estimate of the average polarization signature, shown as the thick line in figure \ref{fig:gmcas-comp}.  This average signature shows an amplitude of nearly 0.2\% with a very smooth profile.
	
\begin{figure} [!h]
\centering
\includegraphics[width=0.35\textwidth, angle=90]{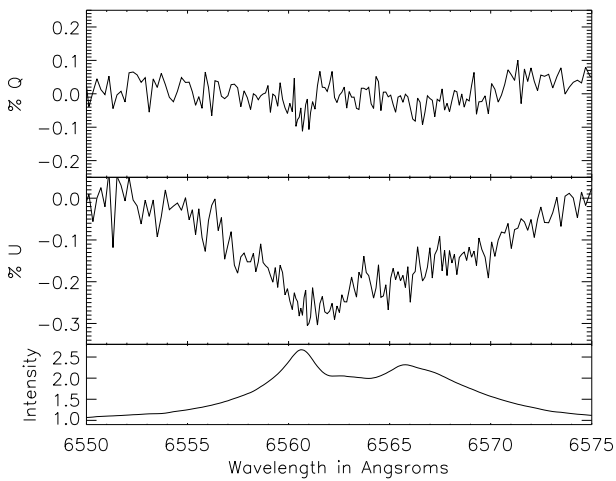}
\caption[$\zeta$ Tau Polarization]{Spectropolarimetry of $\zeta$ Tau. When compared to the low-resolution spectropolaimetry of Quirrenbach et al. 1997, the agreement is quite good. Even the asymmetry across the line is reproduced.}
\label{fig:zettau-comp}
\end{figure}

	In Quirrenbach et al. 1997, low-resolution spectropolarimetric data on $\gamma$ Cas, $\psi$ Per and $\zeta$ Tau is presented. The polarization across the H$_\alpha$ line is plotted with 3-4 resolution elements across the line. $\gamma$ Cas shows a depolarization similar to that seen with WWS, HiVIS, and as described in Poeckert \& Marlborough 1977. $\zeta$ Tau also shows surprising agreement. In Quirrenbach et al. 1997, the H$_\alpha$ line shows an asymmetric depolarization of roughly 0.2\% with the blue continuum side of the line being somewhat higher than the red. In an example of our observations, shown in figure \ref{fig:zettau-comp}, there is a broad change in polarization across the line but the blue side shows the strongest effect. The magnitude of the effect is 0.3\% when our data is binned to lower resolution, the magnitude agrees perfectly with Quirrenbach et al. 1997.

	In summary, HiVIS is capable of reproducing spectropolarimetric measurements made with many other instruments. Nearly simultaneous observations between ESPaDOnS and HiVIS show very good agreement. Comparison with archival ESPaDOnS data show good agreement in overall form and magnitude. However, there are some complications that arise from the polarization properties of the coud\'{e} path that cause some spectropolarimetric measurements to be significantly reduced in magnitude and/or rotated. HiVIS observations are quite good at showing the wavelengths, morphologies, and minimum magnitudes of a spectropolarimetric effect. The reduction in magnitude and rotation caused by the telescope makes our detections lower-limits and the rotation makes the interpretation of the position-angle difficult. Even with these complications, good detection statistics and morphological characterizations are possible, especially when observations are compiled over many different telescope pointings. Comparison between epochs and variability studies are also possible, as long as the observations are taken at the same pointing.

\subsection{Instrument Summary and Conclusions}

	In Chapters 2-5, we have presented the construction and characterization of the spectropolarimeter, our dedicated IDL-based reduction package, the polarization properties of the AEOS telescope and HiVIS spectropolarimeter and a detailed comparison between HiVIS and other spectropolarimeters. The instrument is performing as expected. The reduction package is robust and produces accurate, repeatable results.  The Liber-Esprit package on ESPaDOnS and our scripts give very similar results when both are applied to the same ESPaDOnS raw data. We have found that the telescope induces 0.5\% to 3.0\% continuum polarization in its current arrangement in an unpolarized source with a strong dependence on secondary focus. The telescope can also strongly depolarize a source or rotate the plane of polarization as was shown by highly-polarized twilight observations. All of these telescope polarization effects are functions of altitude, azimuth, and wavelength. Optical ray trace models with Zeemax were presented to illustrate the pointing dependence of the telescope polarization properties at the coud\'{e} focus, but were ultimately shown to be quantitatively inaccurate due to the extra optics as well as the very common problem with modeling the mirror oxide coatings.  
	
	While all these telescope effects complicate the analysis of data taken with this instrument, useful information about the amplitude, morphology and wavelength dependence of the polarization across a spectral line can still be obtained. The performance of the instrument was compared to other spectropolarimeters, particularly ESPaDOnS on CFHT, ISIS on WHT, HPOL on PBO and WWS on OMM. HiVIS found to reproduce the spectropolarimetric signatures well, sometimes strikingly well given the possible complications. This instrument will be able to make sensitive spectropolarimetric observations, despite some complications in the interpretation of the polarization after so many oblique reflections.

\section{The HAe/Be Observing Campaign}

	Since spectropolarimetry is a photon-hungry technique and AEOS is a telescope with pointing-dependent polarization properties, this technique was applied over many pointings and many nights to a number of bright, well-studied Herbig Ae/Be stars. This large data set allows for a clear detection of stellar signatures and robust detection statistics as well as variability studies. In this chapter, the observing campaign, spectroscopy and information about individual Herbig Ae/Be stars will be presented. A presentation of the spectropolarimetry will wait until some theoretical background is developed in chapter 7. 
	
	The spectropolarimetric survey was conducted over three years while simultaneously doing the instrument development and engineering. In early 2004 the spectropolarimeter was being assembled. Some test observations were done during the engineering of the HiVIS spectropolarimeter in late 2004. During this period, the original science camera failed and was pulled off for repairs. In the summer of 2005 the reassembled camera was mounted. There was a large polarization calibration effort in this season associated with the spectropolarimetry of comet 9P/Tempel 1 for the Deep Impact event (Meech et al. 2005, Harrington et al. 2007). In mid-2006 there were a number of instrument fixes to the visible as well as infrared spectrographs. There were then two major Herbig Ae/Be observing seasons once the instrument configuration stabilized: September 2006 to February 2007 and July 2007 to January 2008. Table \ref{obslog} lists the observing and engineering dates. Each line gives a description of what was done on each night.

\onecolumn
\begin{table}
\begin{center}
\caption[Observing Log]{Observing Log \label{obslog}}
\begin{small}
\begin{tabular}{ll|ll}
\hline
{\bf Date} & {\bf Description} & {\bf Date}& {\bf Description}           \\
\hline
\hline
2004 02 27 & Testing Specpol				        & 2006 11 23 & Bad Weather \\									      	
2004 04 29 & Testing Specpol					& 2006 11 28 & Specpol	\\									      
2004 05 26 & Testing Specpol, Waveplate				& 2006 11 29 & Specpol	\\									      
2004 06 23 & Testing Specpol, Waveplate				& 2006 11 30 & Specpol	\\									      
2004 08 18 & Specpol						& 2006 12 09 & Specpol	\\									      
2004 08 19 & Specpol						& 2006 12 20 & Pixis Shutter Died \\							      
2004 08 22 & Specpol						& 2006 12 23 & IR JHK Observing	\\							      
2004 10 04 & HiVIS Shakedown					& 2006 12 27 & Install Spare Pixis, Specpol	\\						      
2004 10 06 & Specpol						& 2006 12 28 & Specpol					\\					      
2004 10 07 & Specpol						& 2006 12 29 & Specpol                                    \\        				      
2004 10 13 & Bad Weather						      & 2007 01 02 & Specpol				         \\			      
2004 10 20 & Specpol							      & 2007 01 03 & Specpol, IR Observing			      	 \\
2004 10 22 & Specpol						      & 2007 01 04 & Bad Weather				      	 \\			      
2004 10 28 & Bad Weather,  IterGain Testing			      & 2007 01 05 & Bad Weather				      	 \\
2004 10 29 & Bad Weather,  IterGain Testing				      & 2007 01 09 & Bad Weather				      	 \\
2004 11 26 & Specpol, Unpol, IterGain Test				      & 2007 01 10 & Bad Weather				      	 \\
2004 11 27 & Specpol, Unpol, IR Dichroic	              	      & 2007 01 17 & Specpol				      	 \\			      
2004 12 02 & Specpol, IR Testing					      & 2007 01 18 & Specpol, IR Observing, IR MilUtil    	 \\	
2004 12 03 & Bad Weather						      & 2007 01 19 & Specpol				      	 \\			      
2004 12 09 & Specpol							      & 2007 01 29 & IR MilUtil 			      	 \\			      
2004 12 10 & Specpol							      & 2007 01 30 & IR MilUtil			      	 \\				      
2004 12 15 & Specpol							      & 2007 02 04 & IR MilUtil			      	 \\				      
2004 12 16 & Bad Weather						      & 2007 02 05 & IR MilUtil			      	 \\				      
&									      & 2007 02 06 & IR MilUtil			      	 \\				      
&  									      & 2007 02 08 & Bad Weather				      	 \\		      
2005 06 27  & Reassemble VisCam, Shakedown.				      & 2007 06 18 & Specpol		      	 \\			      
2005 06 28  & Realign Polarimeter,  Efficiency		                   &2007 06 19 & Specpol				      	 \\			      
2005 06 30  & Unpol Stds, Gain Cal, Sky pol, Tempel			      & 2007 06 20 & Specpol				      	 \\			      
2005 07 01  & Sky pol, Unpol Stds, Tempel 				      & 2007 06 21 & Specpol				      	 \\			      
2005 07 02  & Sky pol, Unpol Stds					      & 2007 07 27 & Sky pol, Bad Weather			      	 \\		      
2005 07 03  & Sky pol, Unpol Stds, Tempel				      & 2007 07 28 & Sky pol, Specpol			      	 \\			      
2005 07 04  & Sky pol, Unpol Stds, Tempel	 	      & 2007 07 31 & Specpol 				      	 \\			      
2005 07 05  & Sky pol, Tempel						      & 2007 08 01 & Specpol				      	 \\			      
2005 07 06  & Tempel						 	        & 2007 08 28 & Specpol				      	 \\			      
2005 07 07  & Tempel, Specphot, Sky glow			                   & 2007 08 29 & Specpol			      	 \\		      
&									      & 2007 08 30 & Specpol				      	 \\			      
&  								                      & 2007 08 31 & Bad Weather				      	 \\			      
2006 05 08 & IR Shakedown - SlitViewer, VidFeed		       	      & 2007 09 01 & Bad Weather   \\		      
2006 05 09 & IR chip testing - Chip noise, 		        	      & 2007 08 17 & Bad Weather				      	 \\		      
2006 05 15 & IR chip testing - Chip noise, J Sensitiv        	    & 2007 09 17 & AEOS Mount broken	      	 \\			
2006 05 16 & IR Observing, Bkgnd, H\&K sensitivity 	       	      & 2007 09 18 & Specpol		      	 \\		      
2006 08 09 & Specpol, Unpol 						      & 2007 09 19 & Bad Weather				      	 \\		      
2006 08 16 & Bad Chip RFI - No Observing				      & 2007 09 20 & Specpol			      	 \\			      
2006 08 21 & Specpol							      & 2007 09 21 & Specpol				      	 \\			      
2006 08 26 & Bad Weather						      & 2007 10 30 & Specpol				      	 \\			      
2006 09 06 & Bad Weather					              & 2007 10 31 & Specpol				      	 \\			      
2006 09 12 & Fix Vis Leach, Specpol, Unpol				      & 2007 11 01 & Bad Weather	    	 \\		      
2006 09 13 & Bad Chip RFI - No Observing			& 2007 11 02 & Bad Weather				       \\				      
2006 09 14 & Bad Chip RFI - No Observing			              & 2007 11 03 & Bad Weather      \\		      
2006 09 18 & Pixis Mounted, Specpol, Unpol			          & 2007 11 21 & Specpol				      	 \\			      
2006 09 19 & Specpol, Unpol                                     & 2007 11 22 & Bad Weather				      	 \\				      
2006 09 21 & Specpol, Unpol						      & 2007 11 23 & Specpol				      	 \\			      
2006 09 22 & Specpol, Unpol						      & 2007 11 24 & Specpol				      	 \\			      
2006 09 28 & Pixis Remounted, Specpol					      & 2007 11 27 & Bad Weather			      	 \\		      
2006 10 27 & Specpol, VisMotor, PolMount, Grating	        	      & 2007 11 28 & Bad Weather				      	 \\		      
2006 11 03 & Bad Weather						      & 2008 01 14 &	Specpol			      	 \\			      
2006 11 07 & Specpol, Unpol Stds				      & 2008 01 15 & Specpol				      	 \\				      
2006 11 17 & Specpol, CalStage Lenses, New Dekkar	    & 2008 01 16 & Specpol					      	 \\
2006 11 18 & Specpol							      & 2008 01 17 & Specpol			      	 \\			      
2006 11 21 & Specpol						      & 2008 01 19 &  Specpol   \\			      
2006 11 22 & Specpol						         & & 								 \\			      
\hline
\end{tabular}
\end{small}
\end{center}
\end{table}

\begin{figure}
\centering
\subfloat[AB Aurigae]{\label{fig:lprof-abaur}
\includegraphics[ width=0.21\textwidth, angle=90]{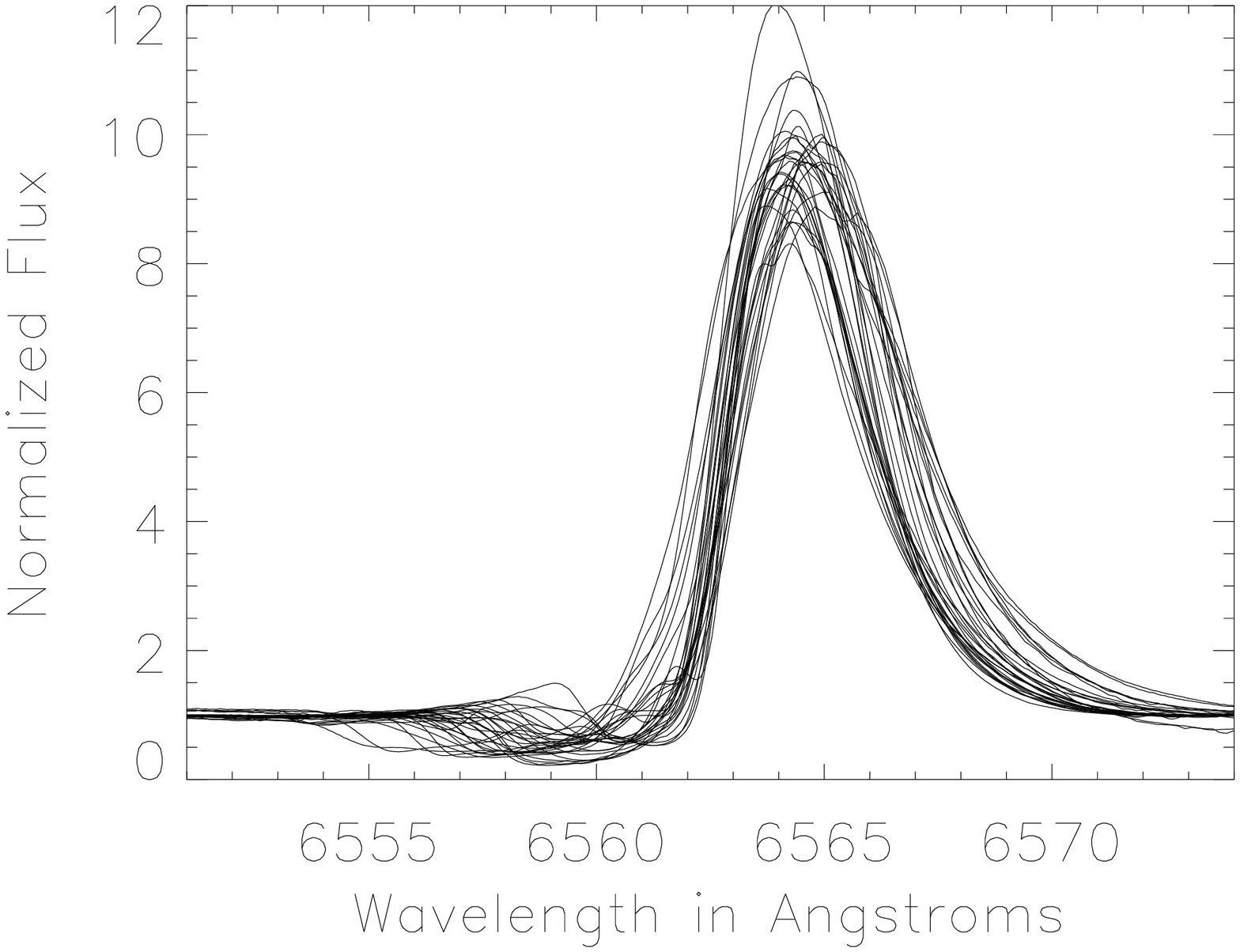}}
\quad
\subfloat[MWC 480]{\label{fig:lprof-mwc480}
\includegraphics[ width=0.21\textwidth, angle=90]{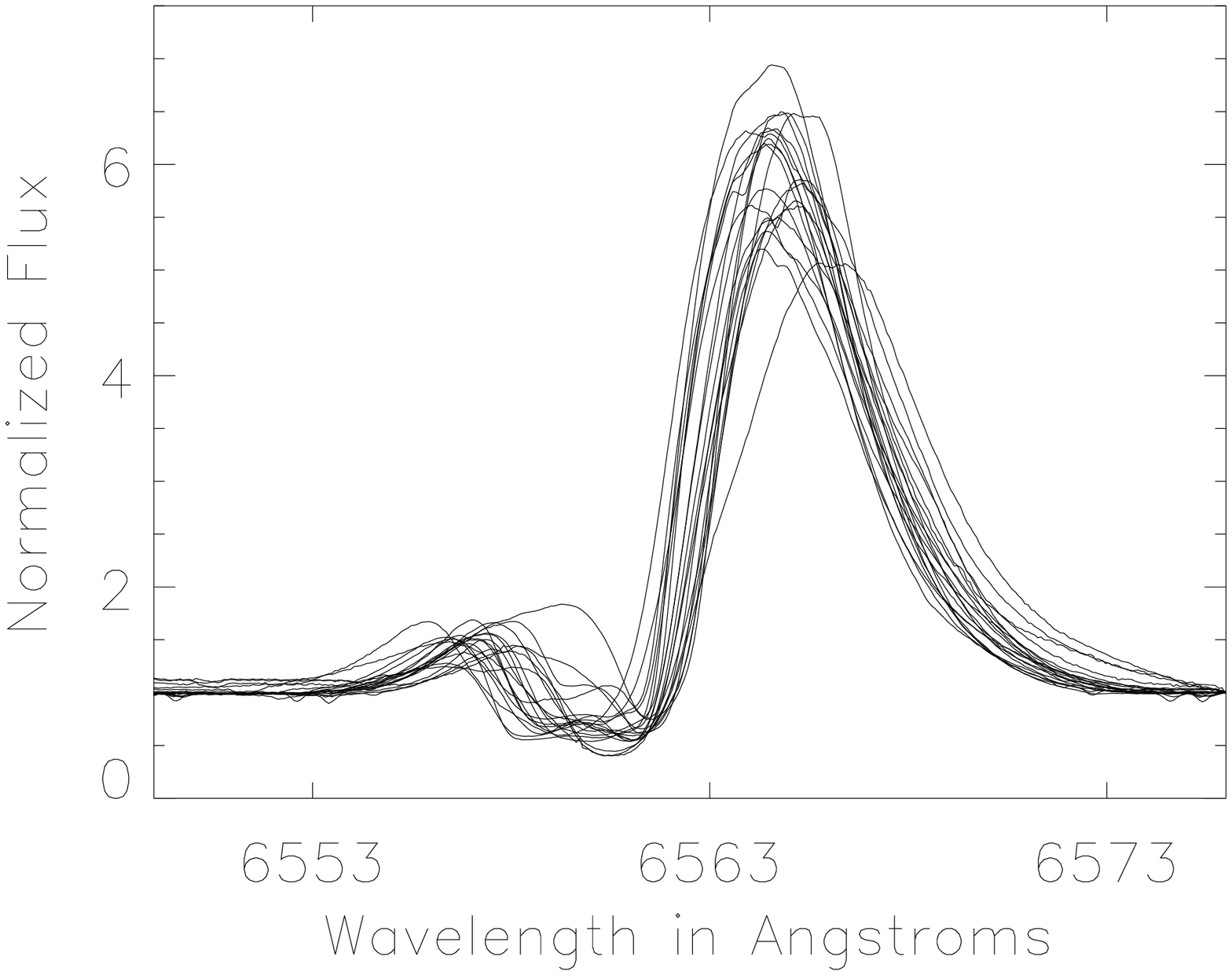}}
\quad
\subfloat[MWC 120]{\label{fig:lprof-mwc120}
\includegraphics[ width=0.21\textwidth, angle=90]{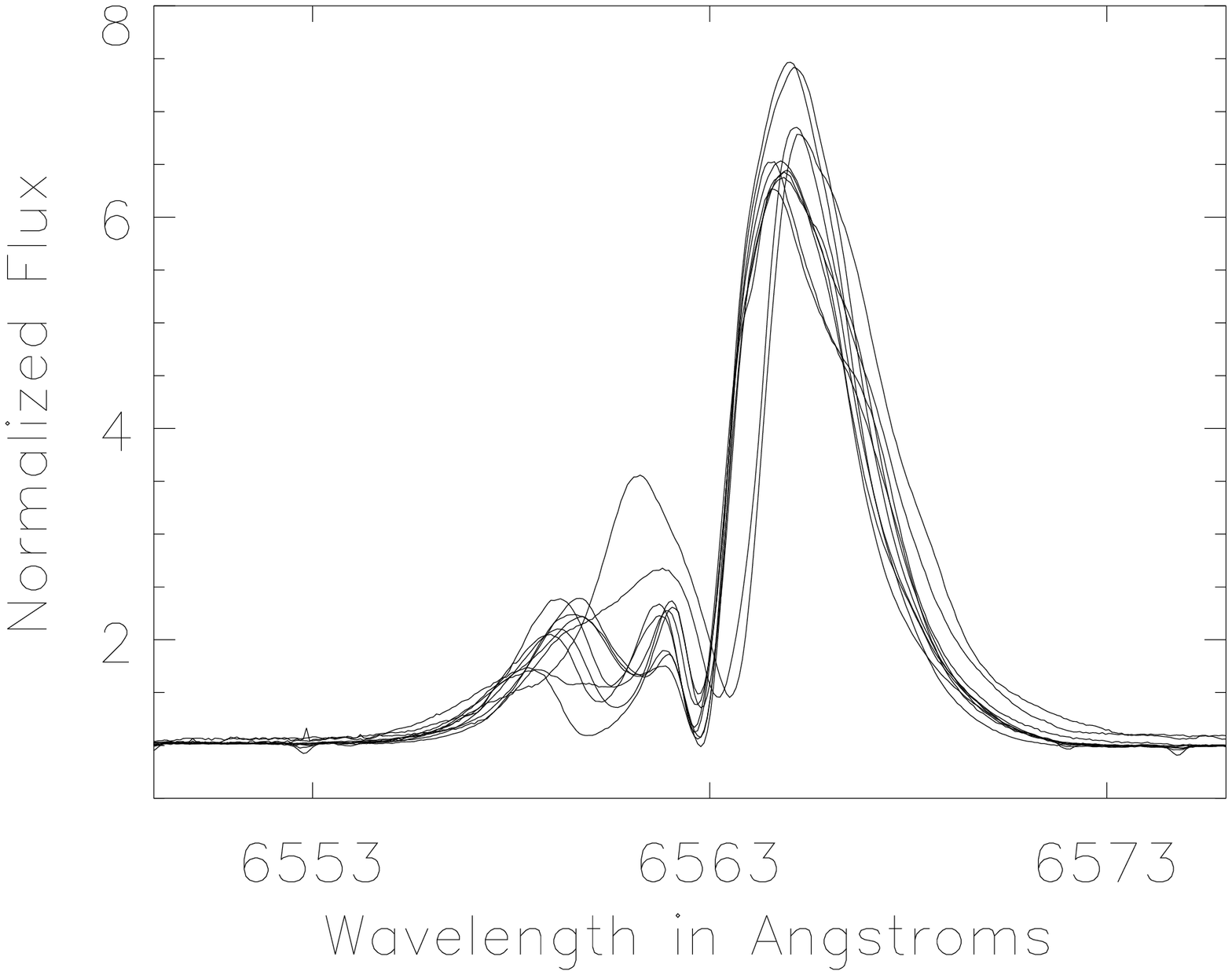}}
\quad
\subfloat[HD 150193]{\label{fig:lprof-hd150}
\includegraphics[ width=0.21\textwidth, angle=90]{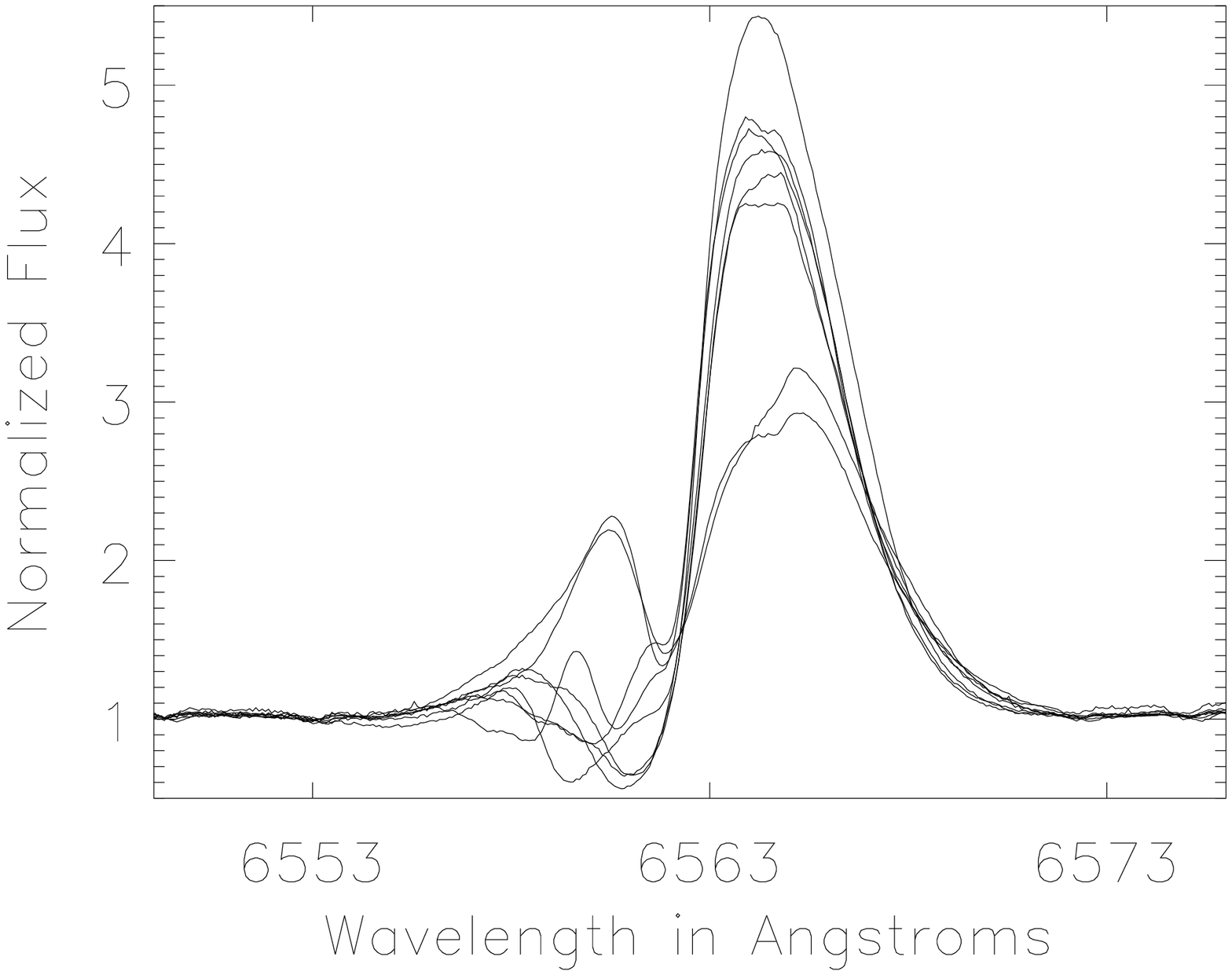}}
\quad
\subfloat[HD 163296]{\label{fig:lprof-hd163}
\includegraphics[ width=0.21\textwidth, angle=90]{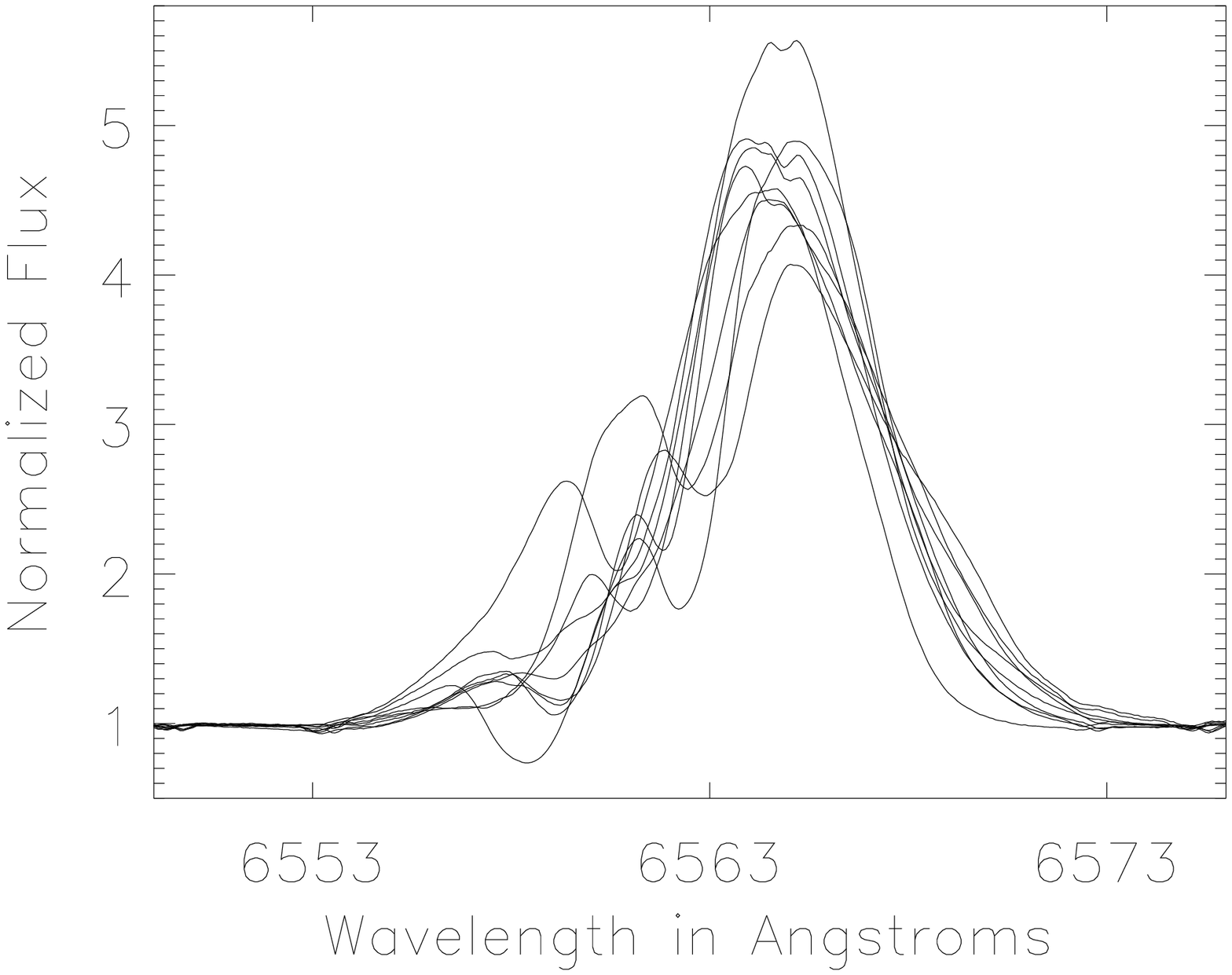}}
\quad
\subfloat[HD 179218]{\label{fig:lprof-hd179}
\includegraphics[ width=0.21\textwidth, angle=90]{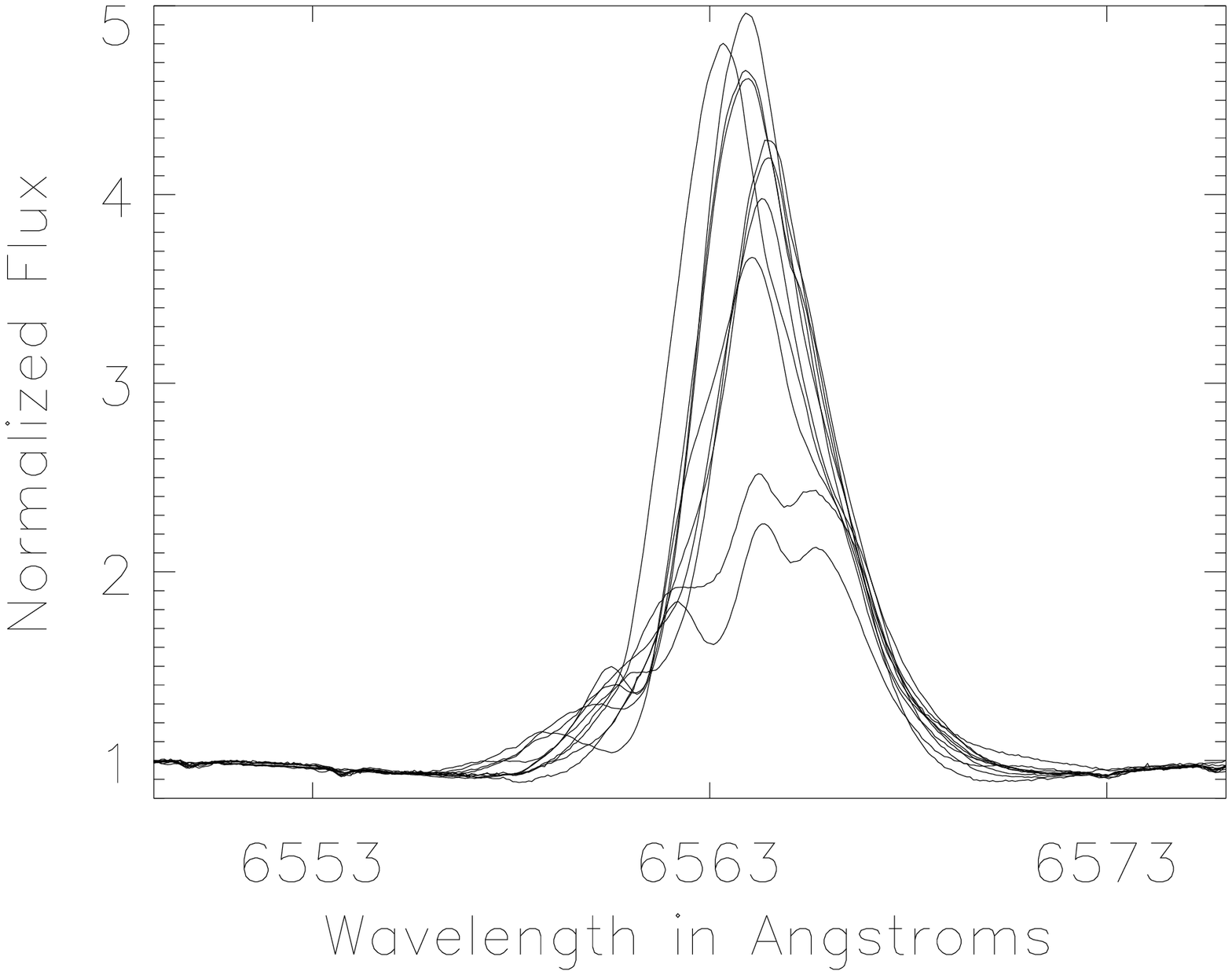}}
\quad
\subfloat[HD 144432]{\label{fig:lprof-hd144}
\includegraphics[ width=0.21\textwidth, angle=90]{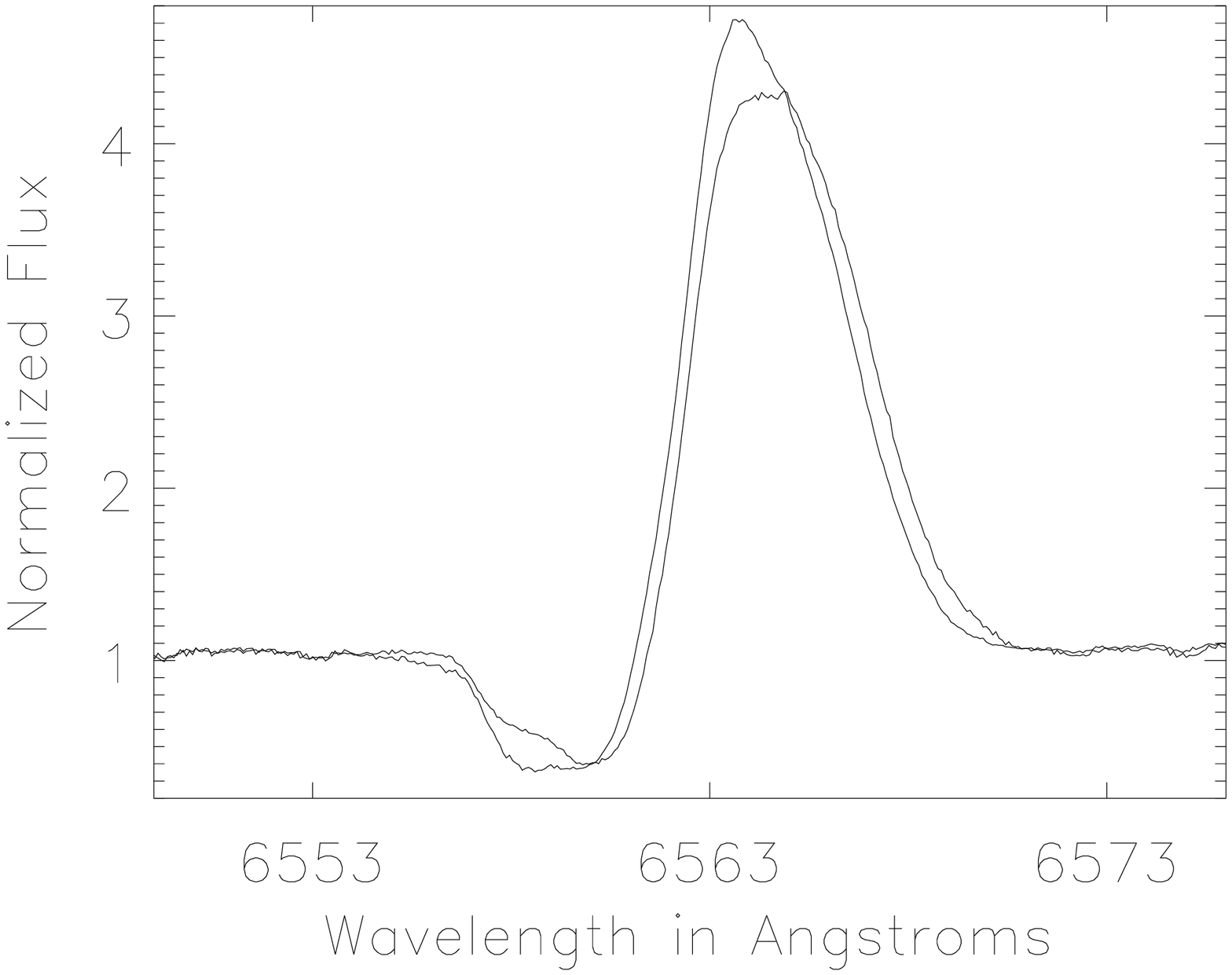}}
\quad
\subfloat[MWC 758]{\label{fig:lprof-mwc758}
\includegraphics[ width=0.21\textwidth, angle=90]{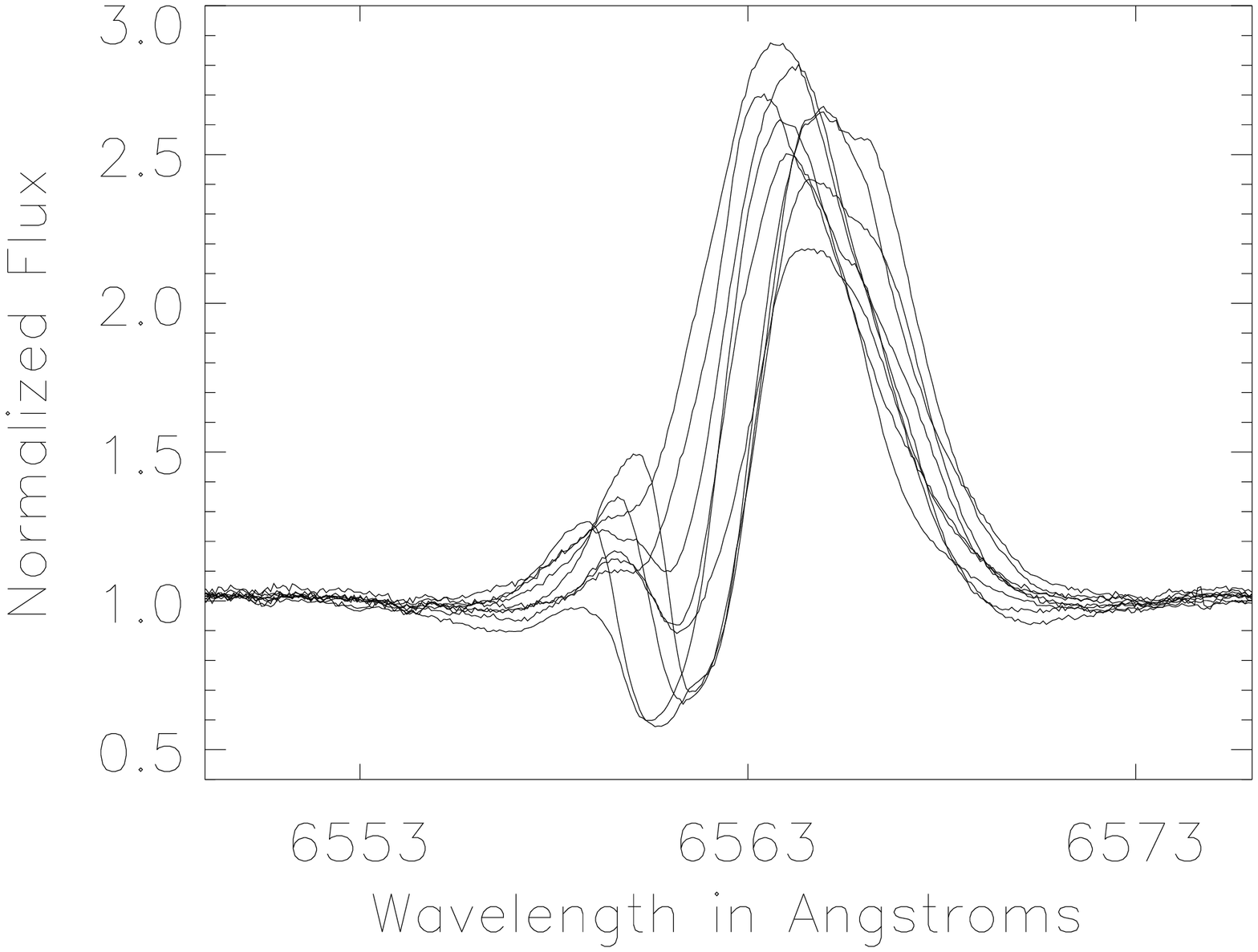}}
\quad
\subfloat[HD 169142]{\label{fig:lprof-hd169}
\includegraphics[ width=0.21\textwidth, angle=90]{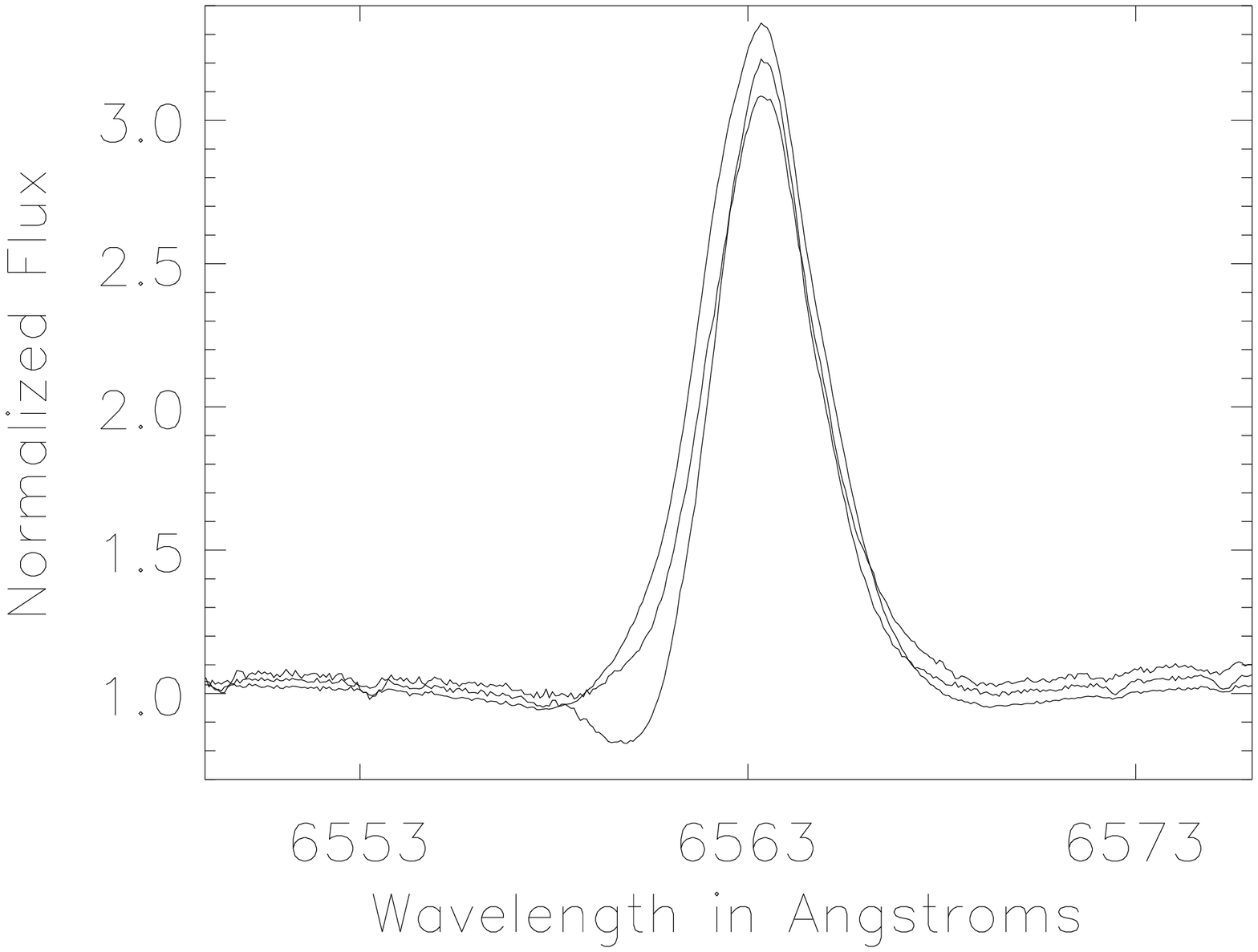}}
\quad
\subfloat[KMS 27]{\label{fig:lprof-kms27}
\includegraphics[ width=0.21\textwidth, angle=90]{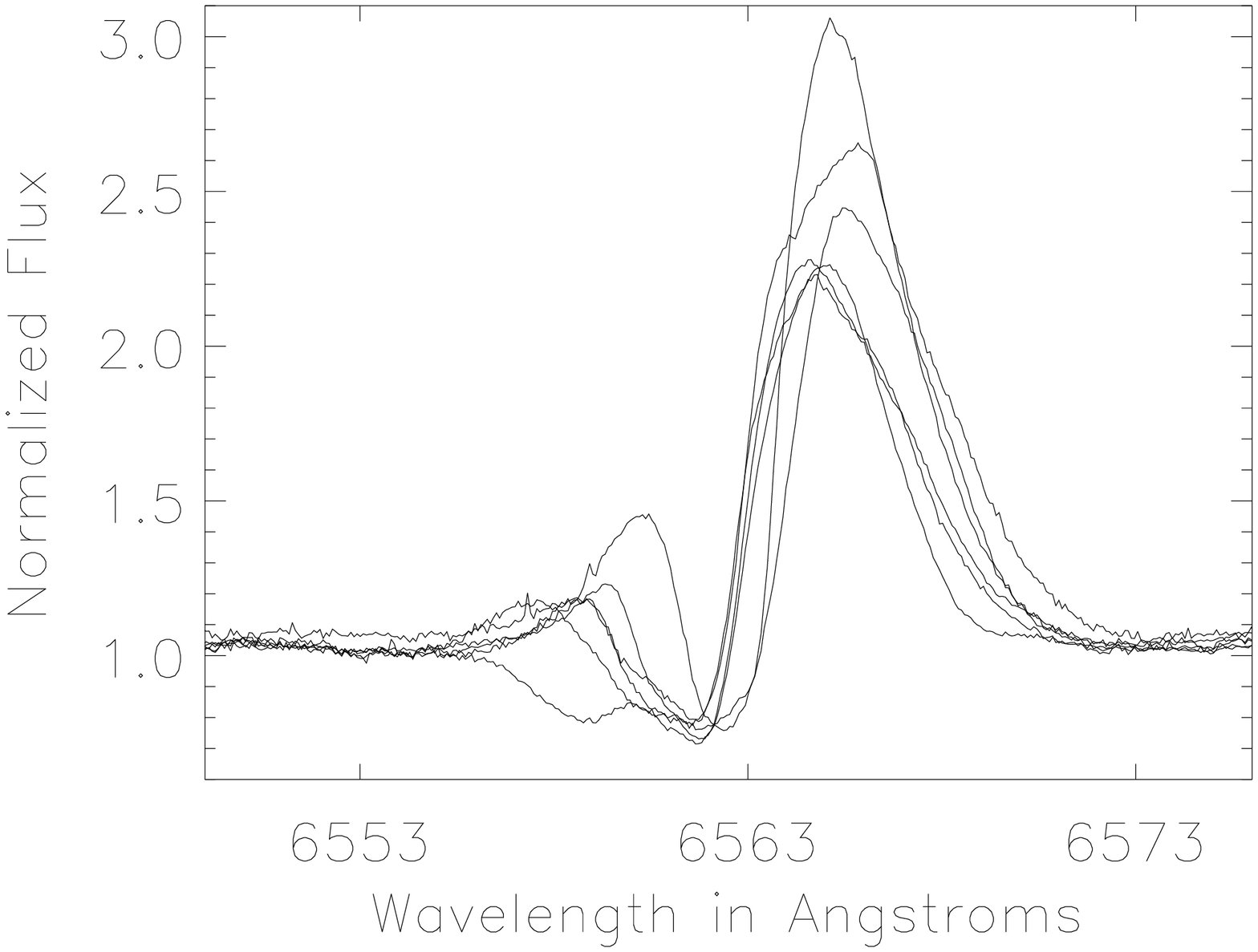}}
\quad
\subfloat[V 1295Aql]{\label{fig:lprof-v1295}
\includegraphics[ width=0.21\textwidth, angle=90]{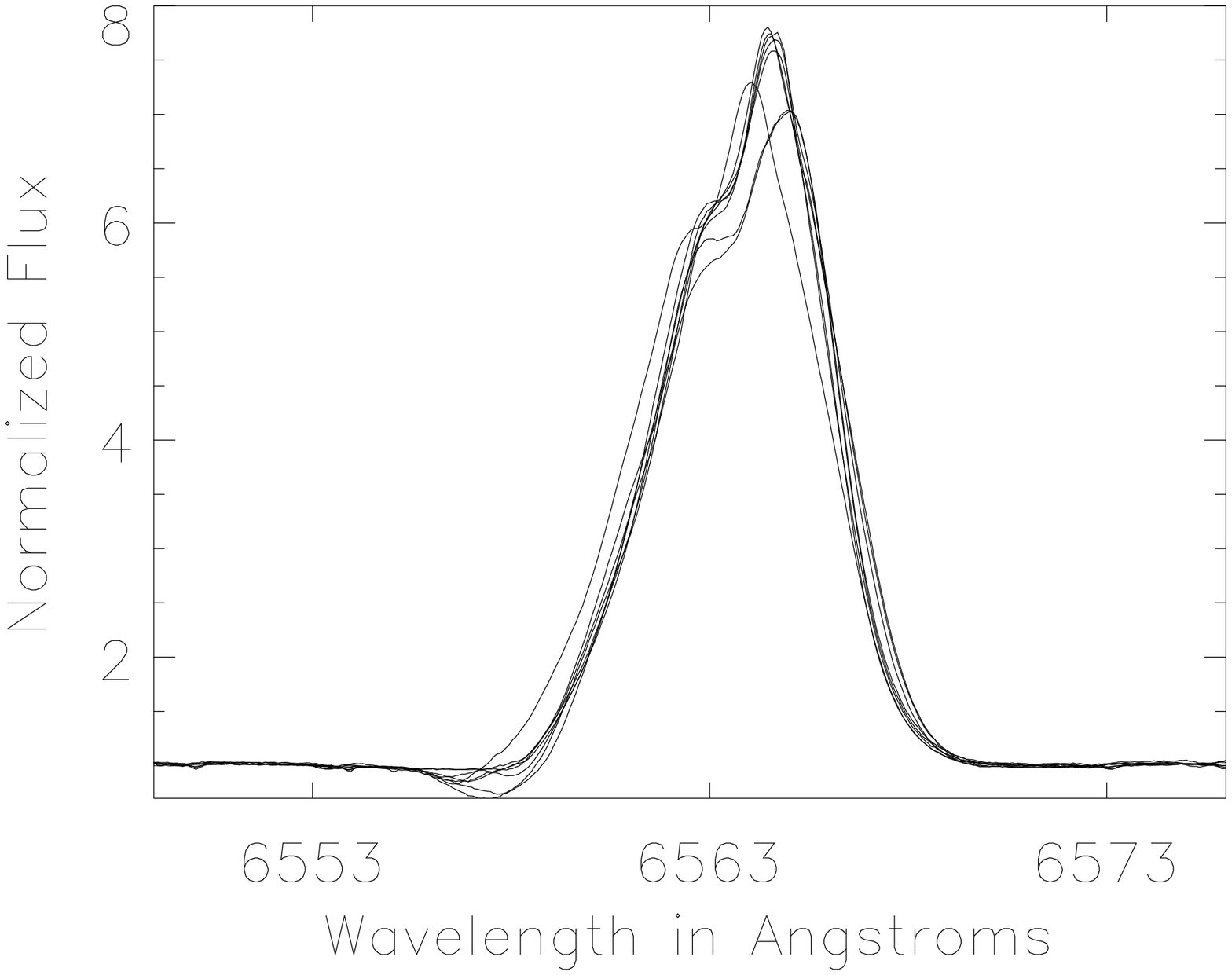}}
\quad
\subfloat[HD 35187]{\label{fig:lprof-hd351}
\includegraphics[ width=0.21\textwidth, angle=90]{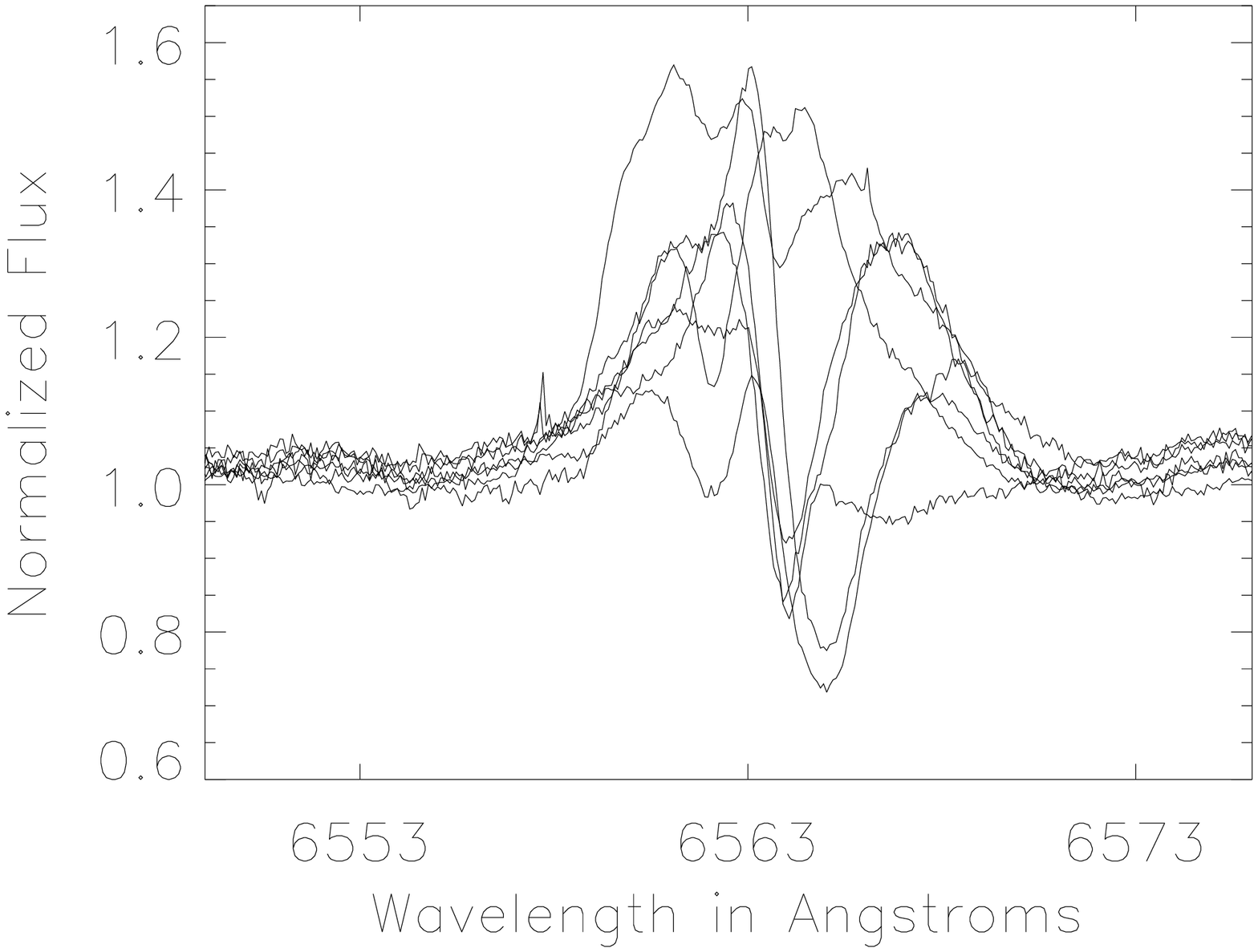}}
\quad
\subfloat[HD 142666]{\label{fig:lprof-hd142}
\includegraphics[ width=0.21\textwidth, angle=90]{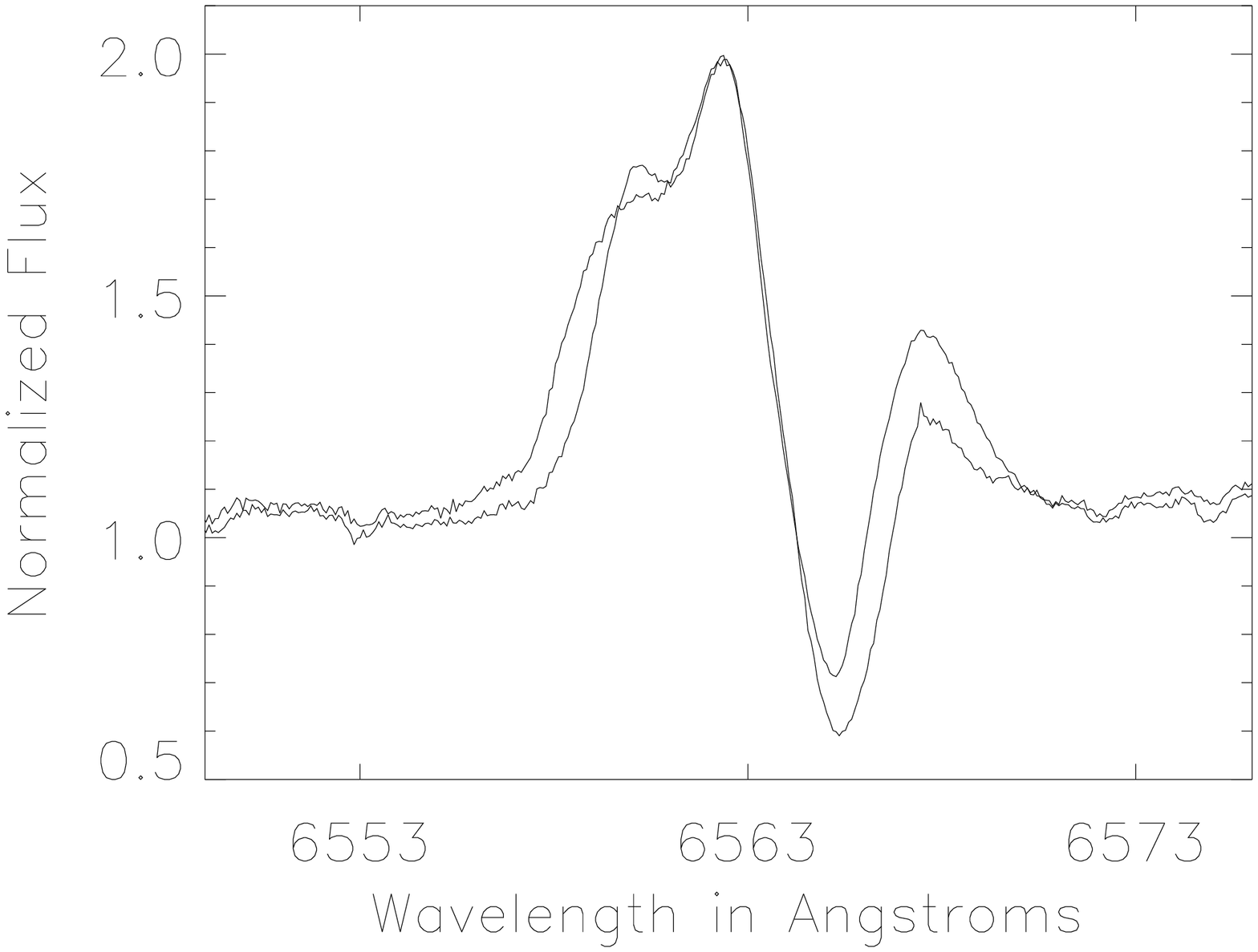}}
\quad
\subfloat[T Ori]{\label{fig:lprof-tori}
\includegraphics[ width=0.21\textwidth, angle=90]{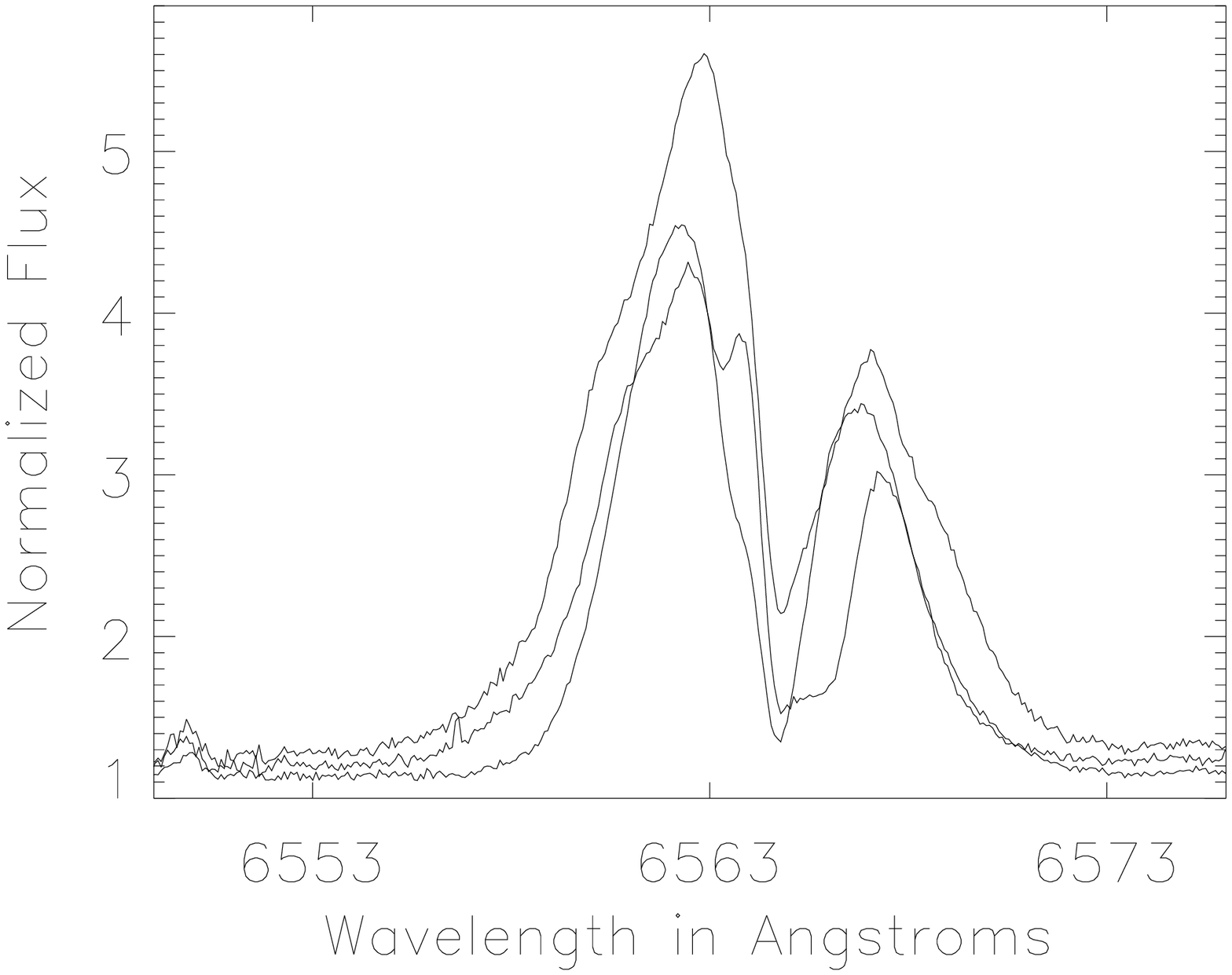}}
\caption[Herbig Ae/Be Line Profiles I]{HAe/Be Line Profiles I}
\label{fig:haebe-lprof1}
\end{figure}

\begin{figure}
\centering
\subfloat[MWC 158]{\label{fig:lprof-mwc158}
\includegraphics[ width=0.21\textwidth, angle=90]{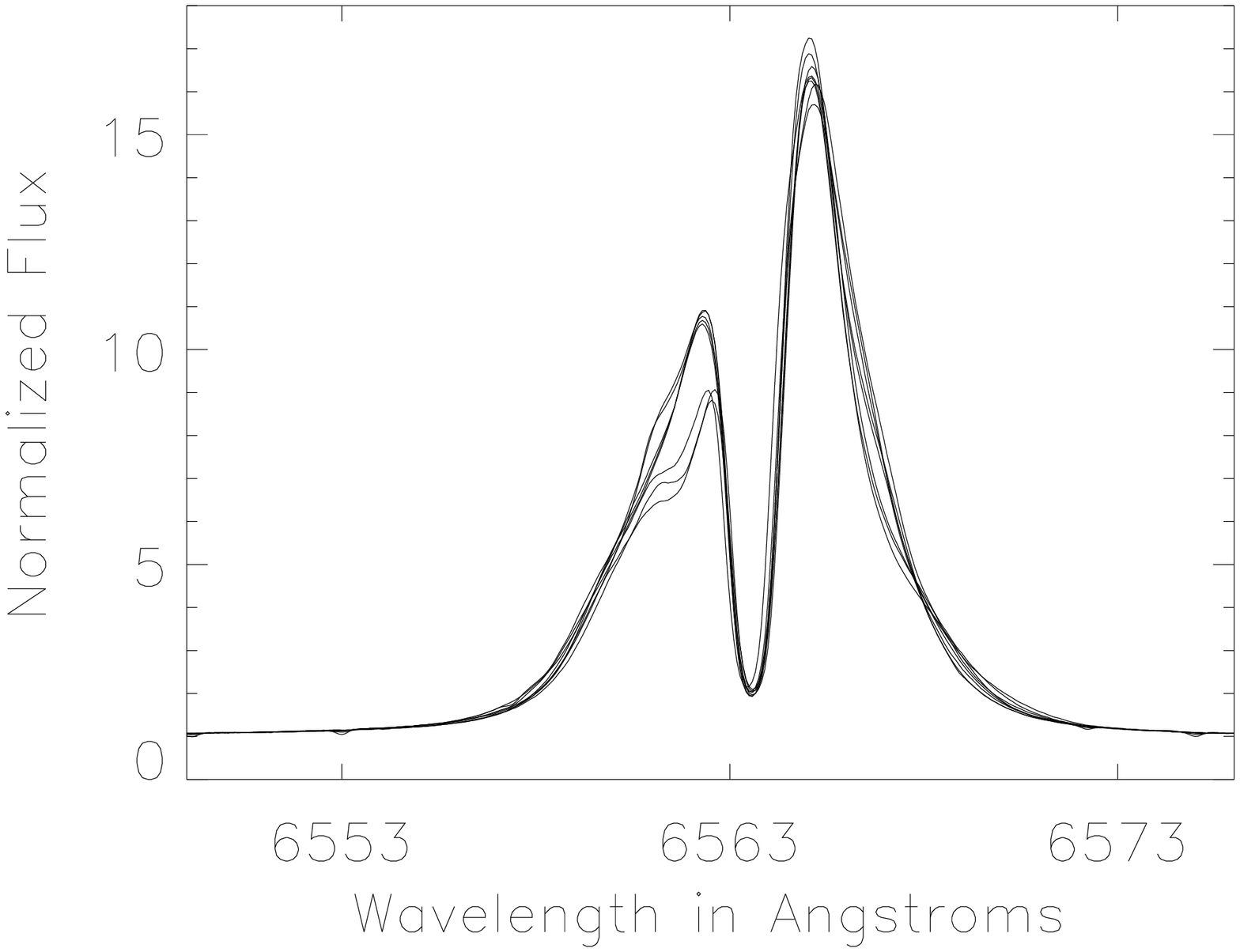}}
\quad
\subfloat[HD 58647]{\label{fig:lprof-hd}
\includegraphics[ width=0.21\textwidth, angle=90]{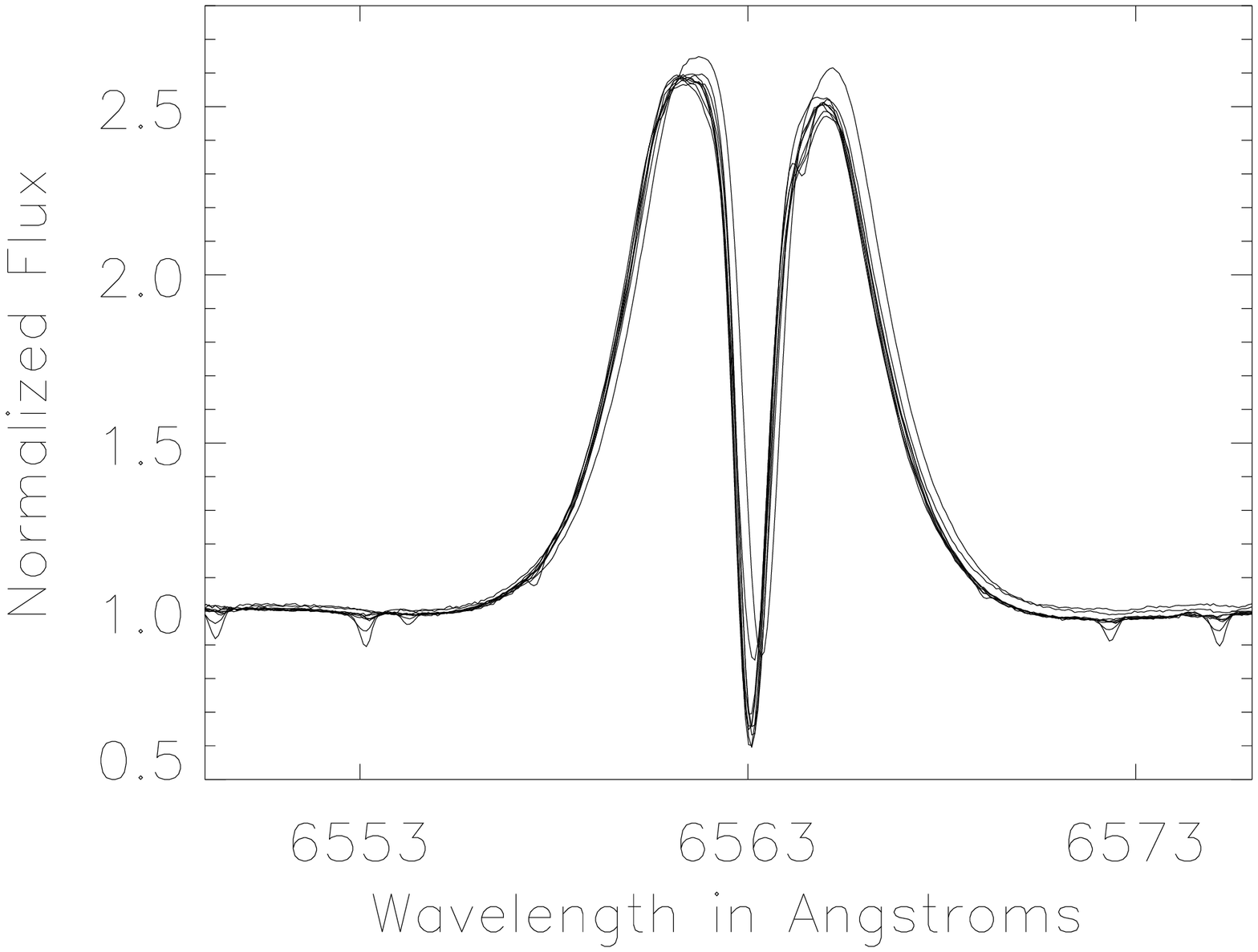}}
\quad
\subfloat[MWC 361]{\label{fig:lprof-mwc361}
\includegraphics[ width=0.21\textwidth, angle=90]{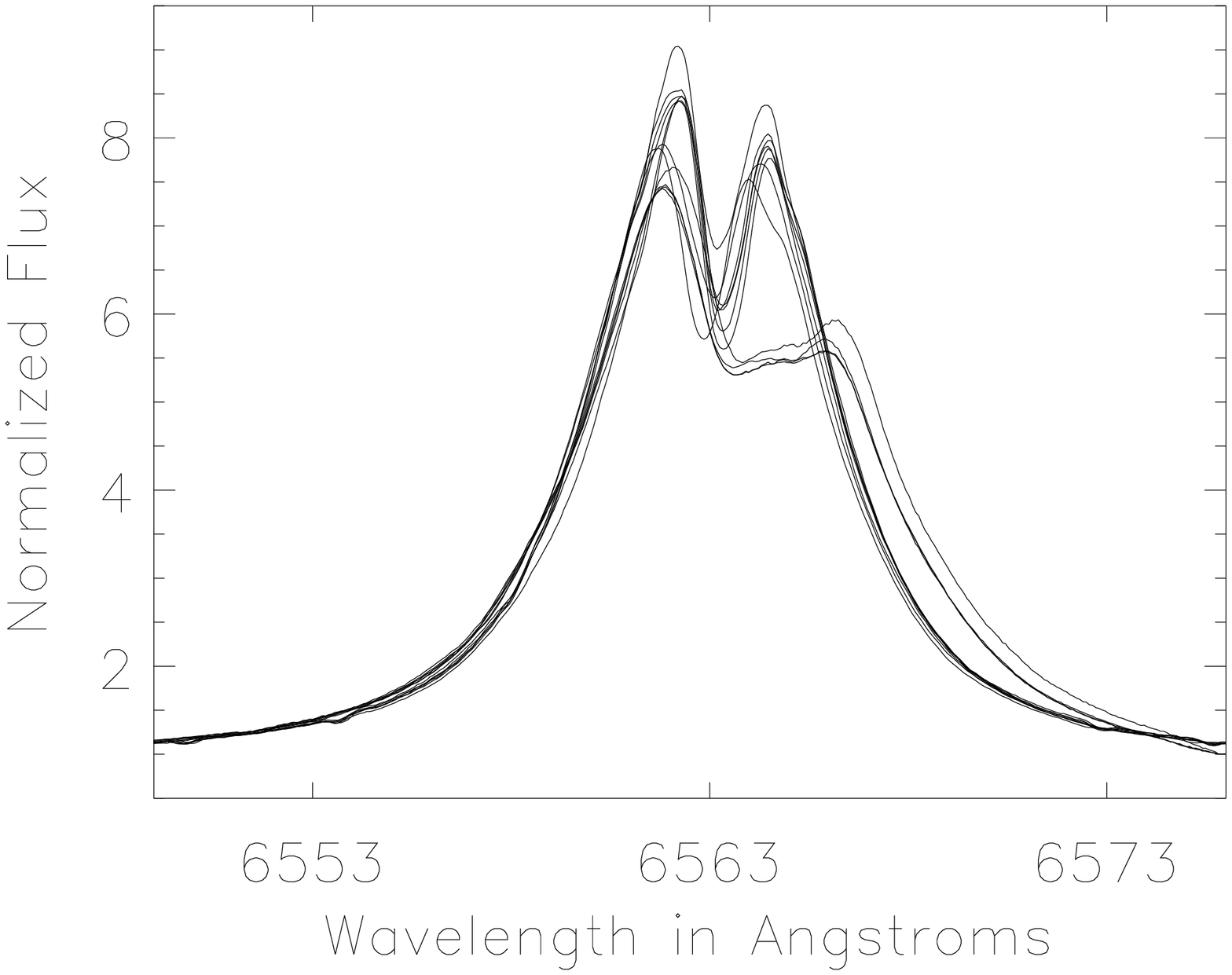}}
\quad
\subfloat[HD 141569]{\label{fig:lprof-hd141}
\includegraphics[ width=0.21\textwidth, angle=90]{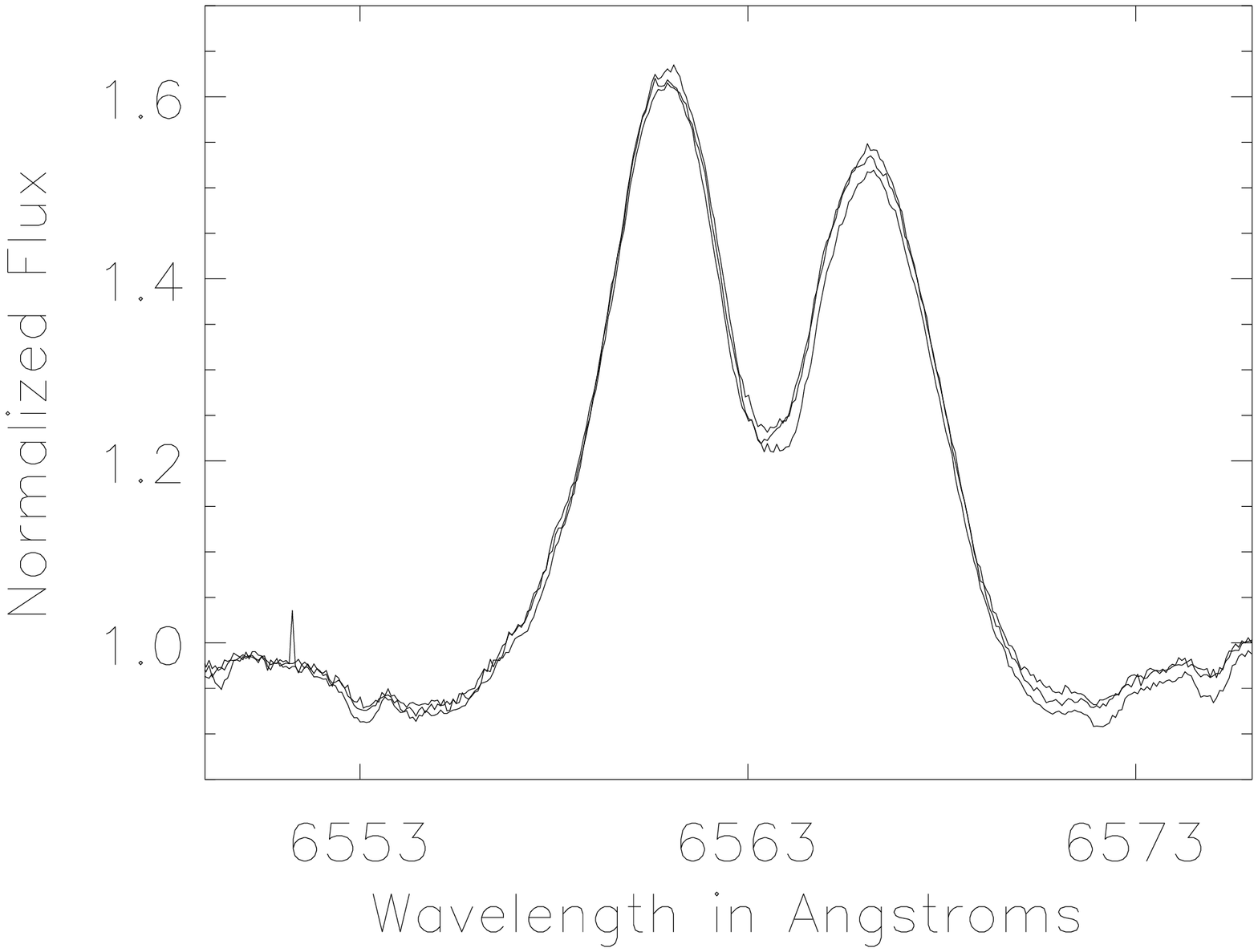}}
\quad
\subfloat[51 Oph]{\label{fig:lprof-51oph}
\includegraphics[ width=0.21\textwidth, angle=90]{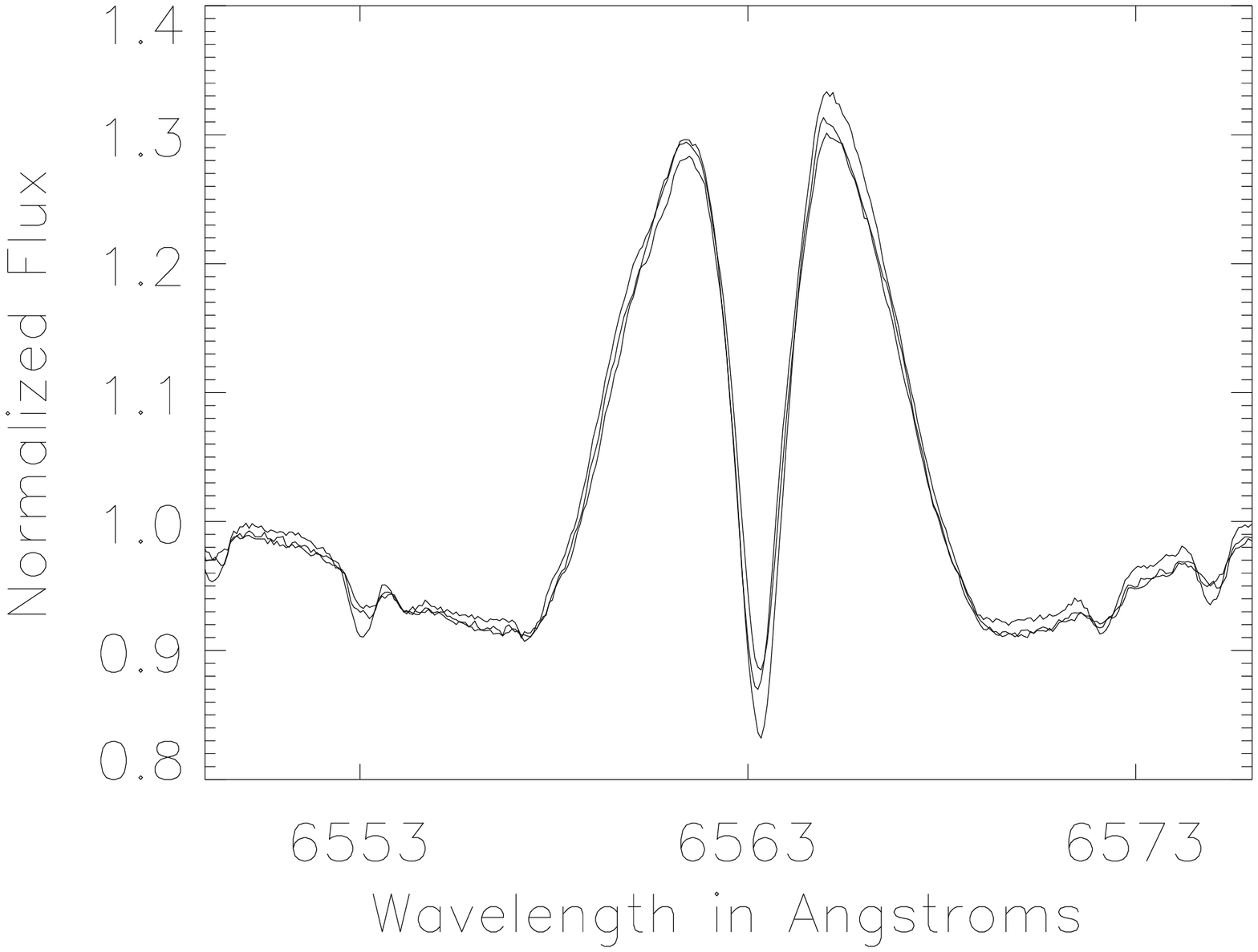}}
\quad
\subfloat[XY Per]{\label{fig:lprof-xyper}
\includegraphics[ width=0.21\textwidth, angle=90]{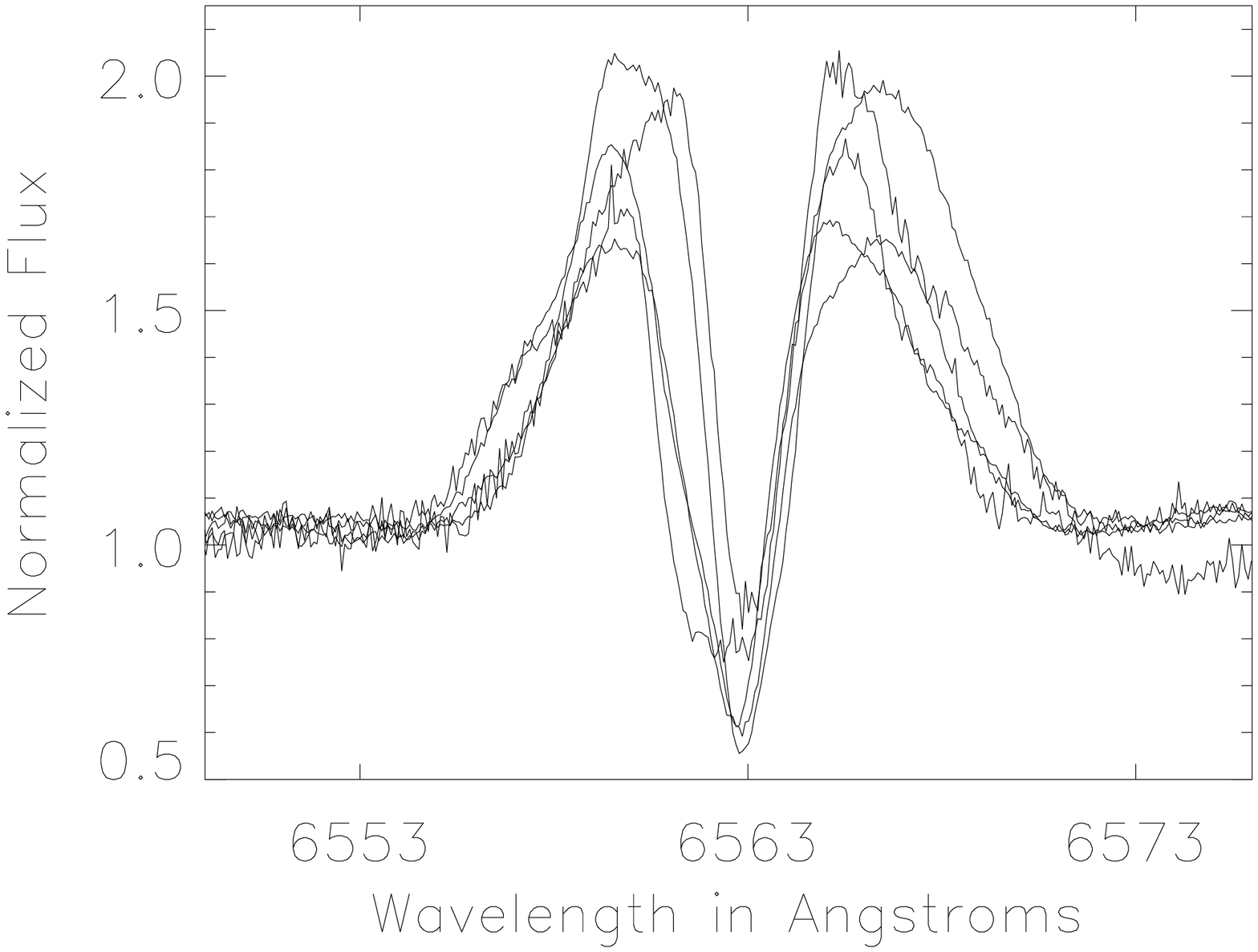}}
\quad
\subfloat[MWC 166]{\label{fig:lprof-mwc166}
\includegraphics[ width=0.21\textwidth, angle=90]{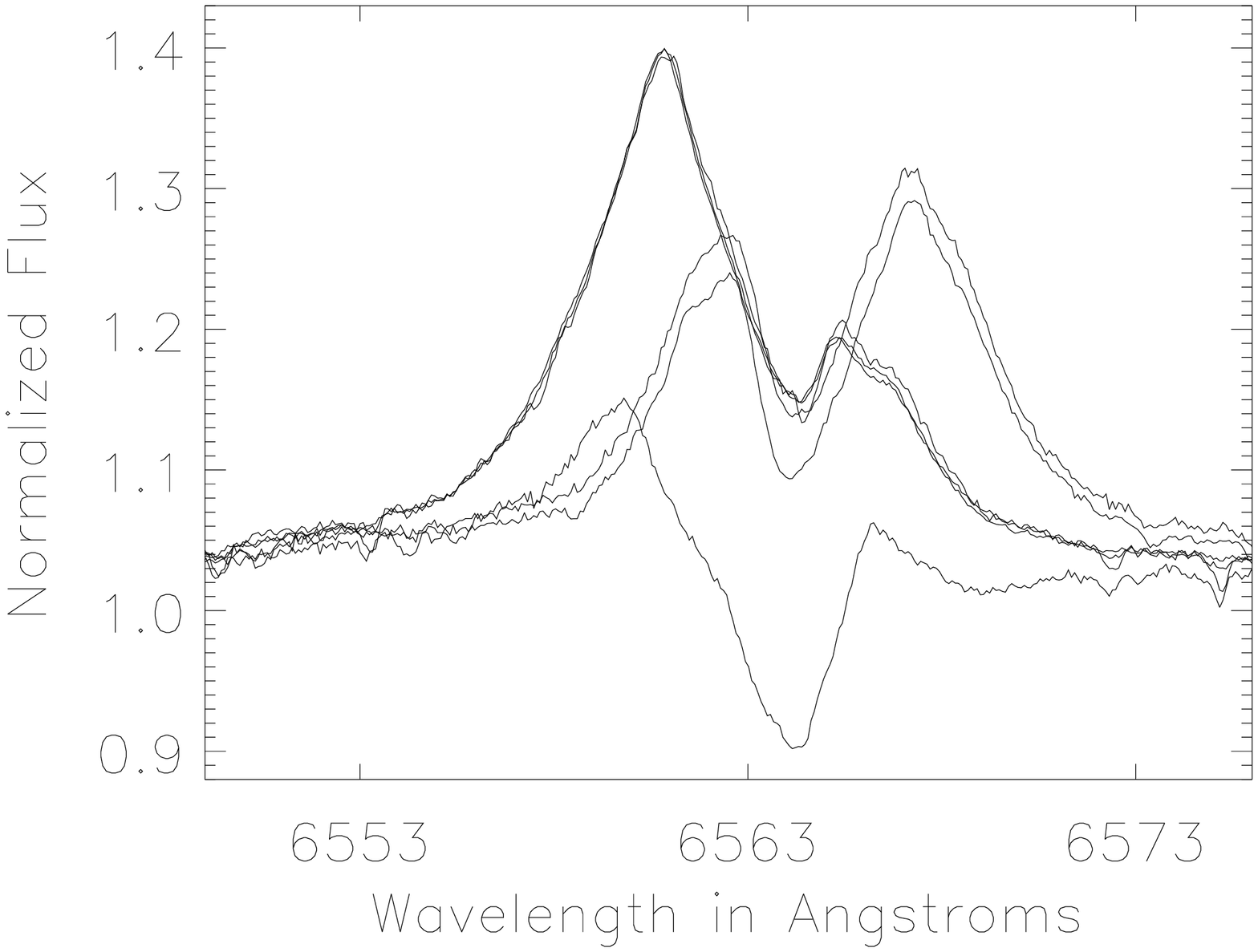}}
\quad
\subfloat[MWC 170]{\label{fig:lprof-mwc170}
\includegraphics[ width=0.21\textwidth, angle=90]{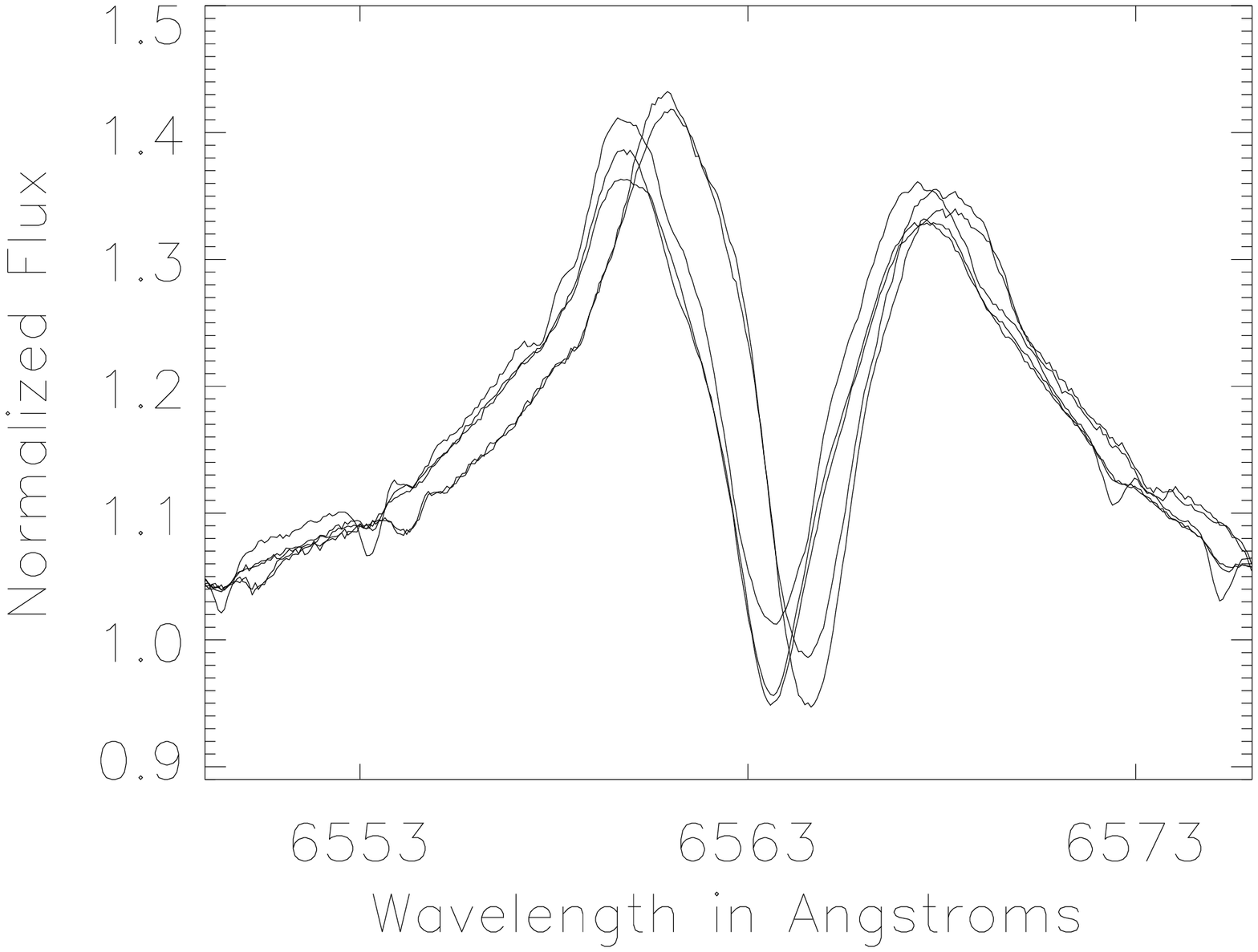}}
\quad
\subfloat[HD 45677]{\label{fig:lprof-hd456}
\includegraphics[ width=0.21\textwidth, angle=90]{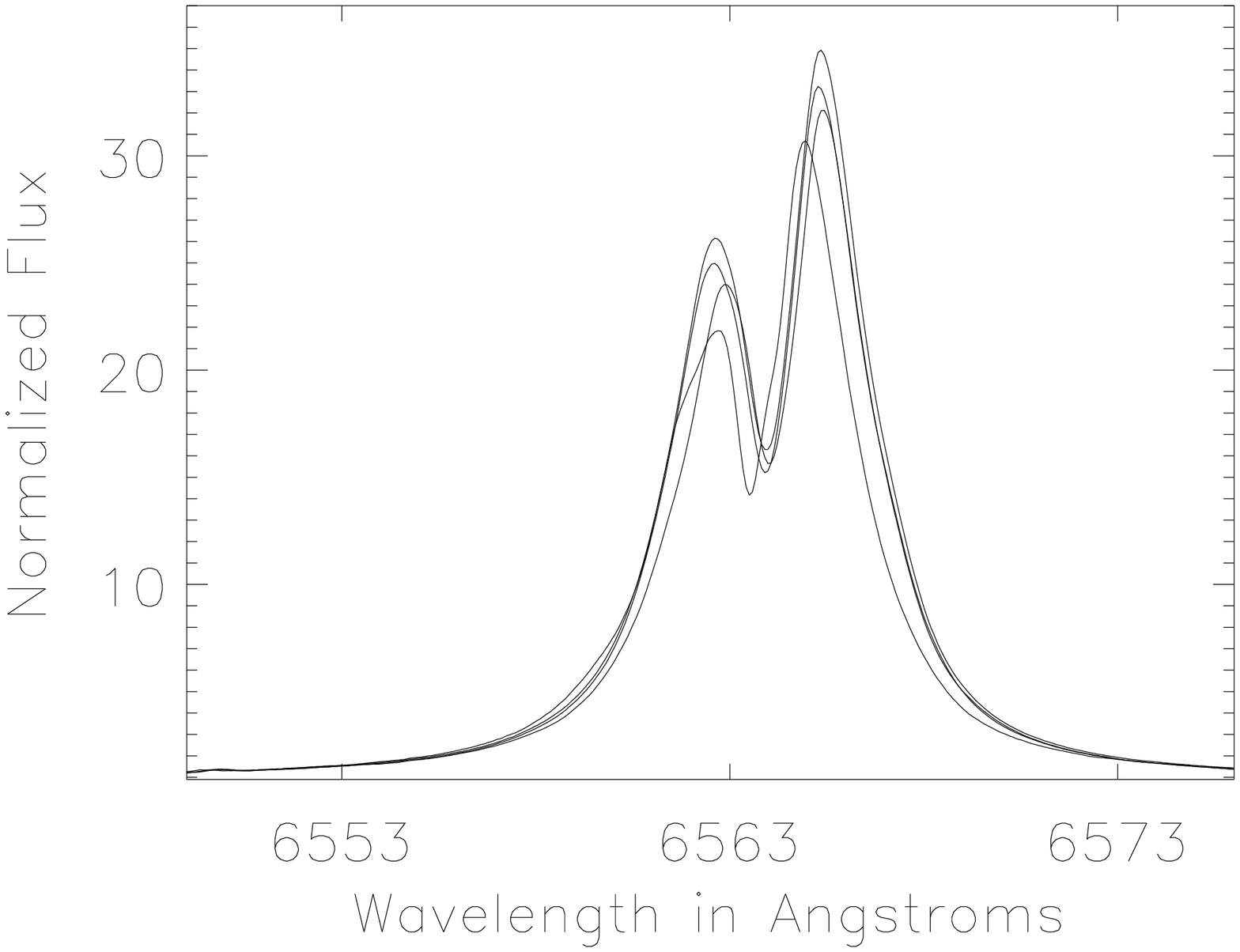}}
\quad
\subfloat[MWC 147]{\label{fig:lprof-mwc147}
\includegraphics[ width=0.21\textwidth, angle=90]{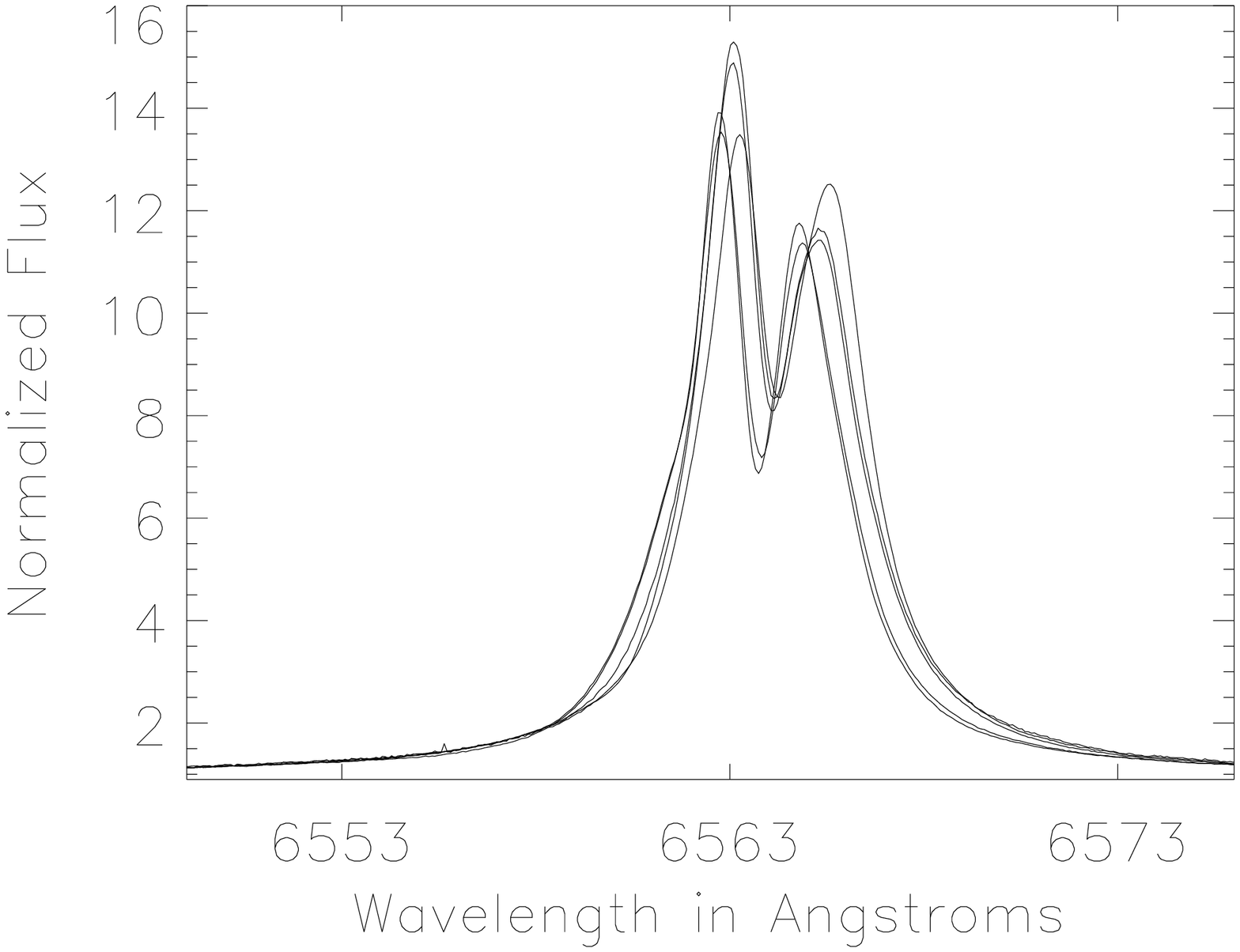}}
\quad
\subfloat[Il Cep]{\label{fig:lprof-ilcep}
\includegraphics[ width=0.21\textwidth, angle=90]{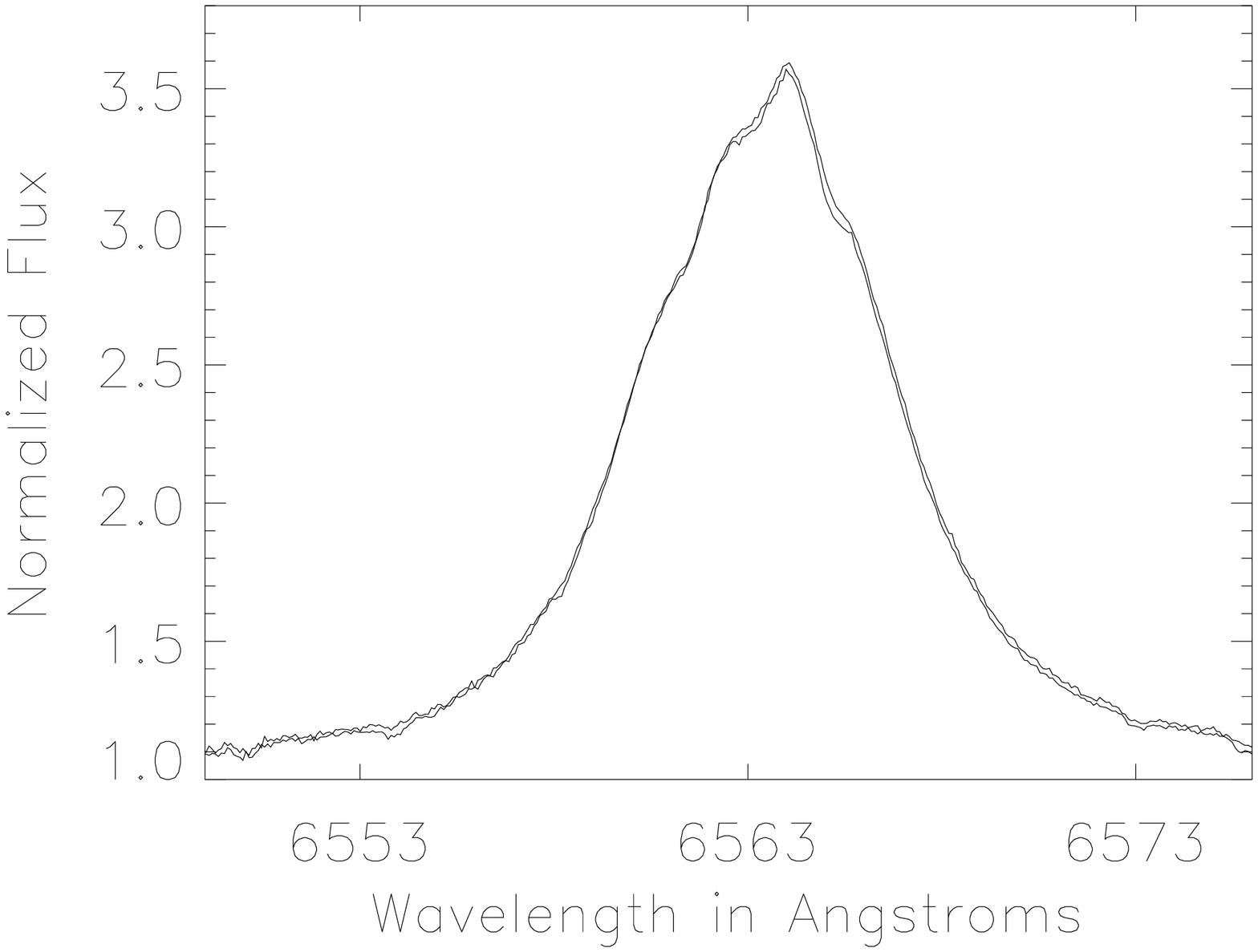}}
\quad
\subfloat[MWC 442]{\label{fig:lprof-mwc442}
\includegraphics[ width=0.21\textwidth, angle=90]{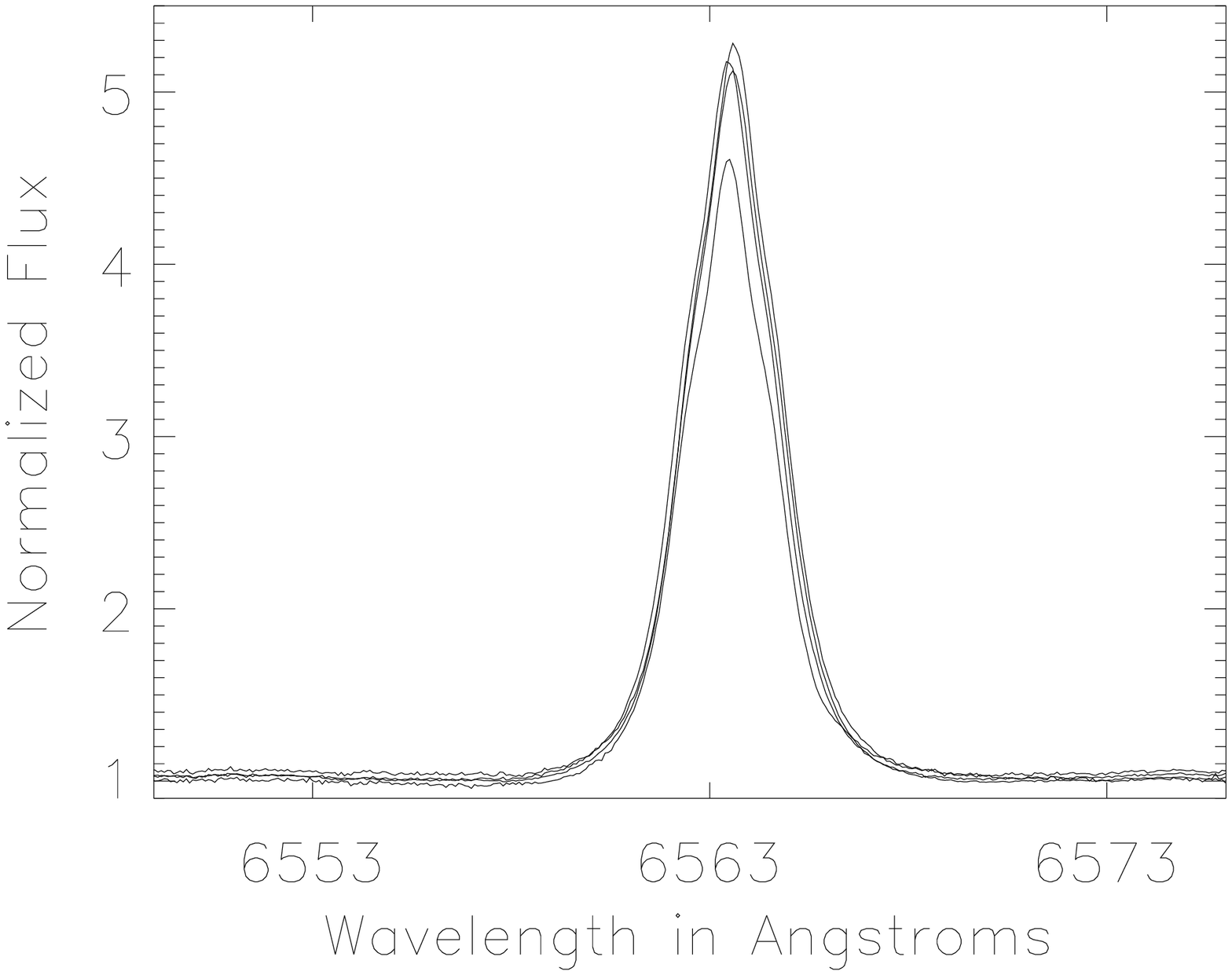}}
\quad
\subfloat[HD 35929]{\label{fig:lprof-hd359}
\includegraphics[ width=0.21\textwidth, angle=90]{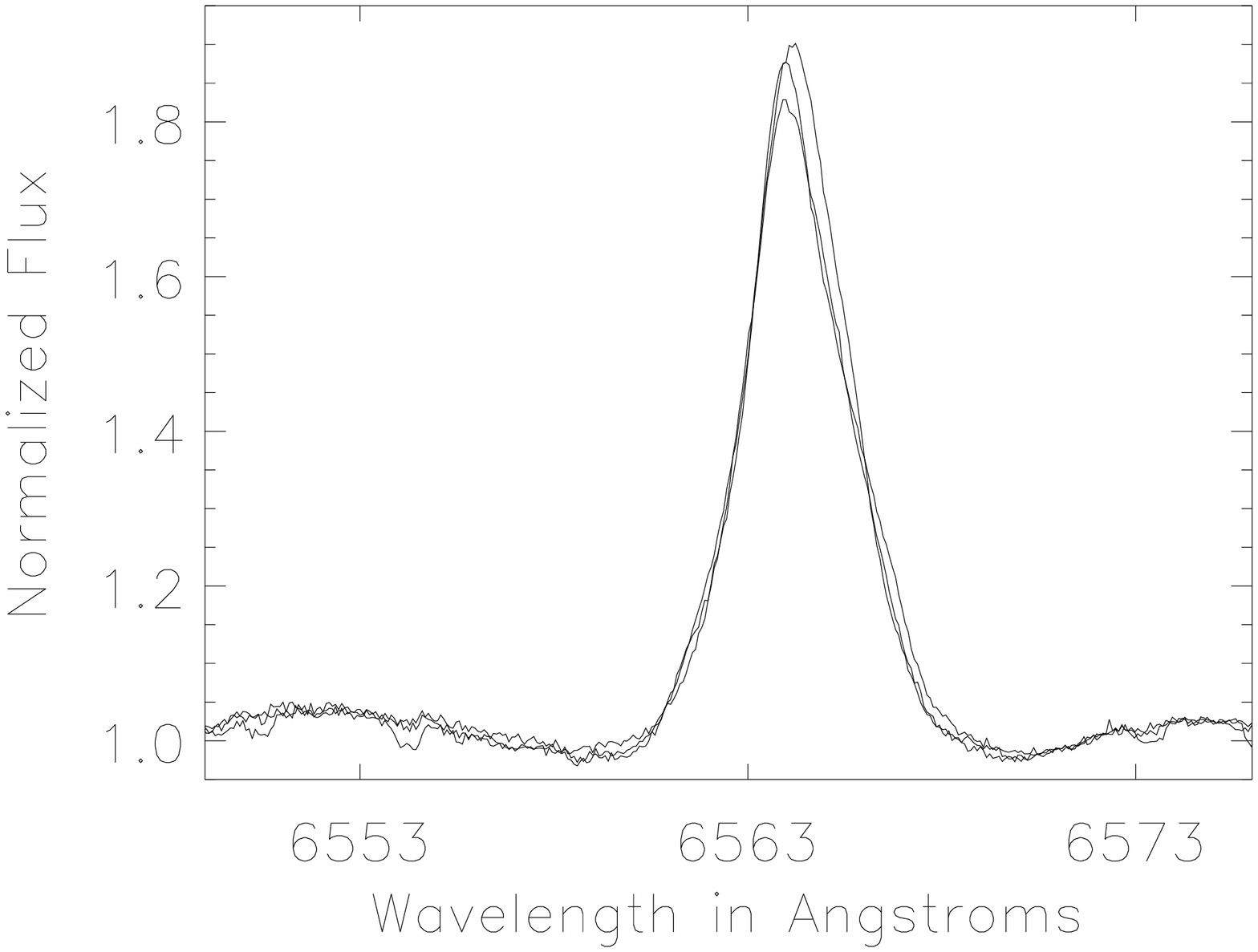}}
\quad
\subfloat[GU CMa]{\label{fig:lprof-gucma}
\includegraphics[ width=0.21\textwidth, angle=90]{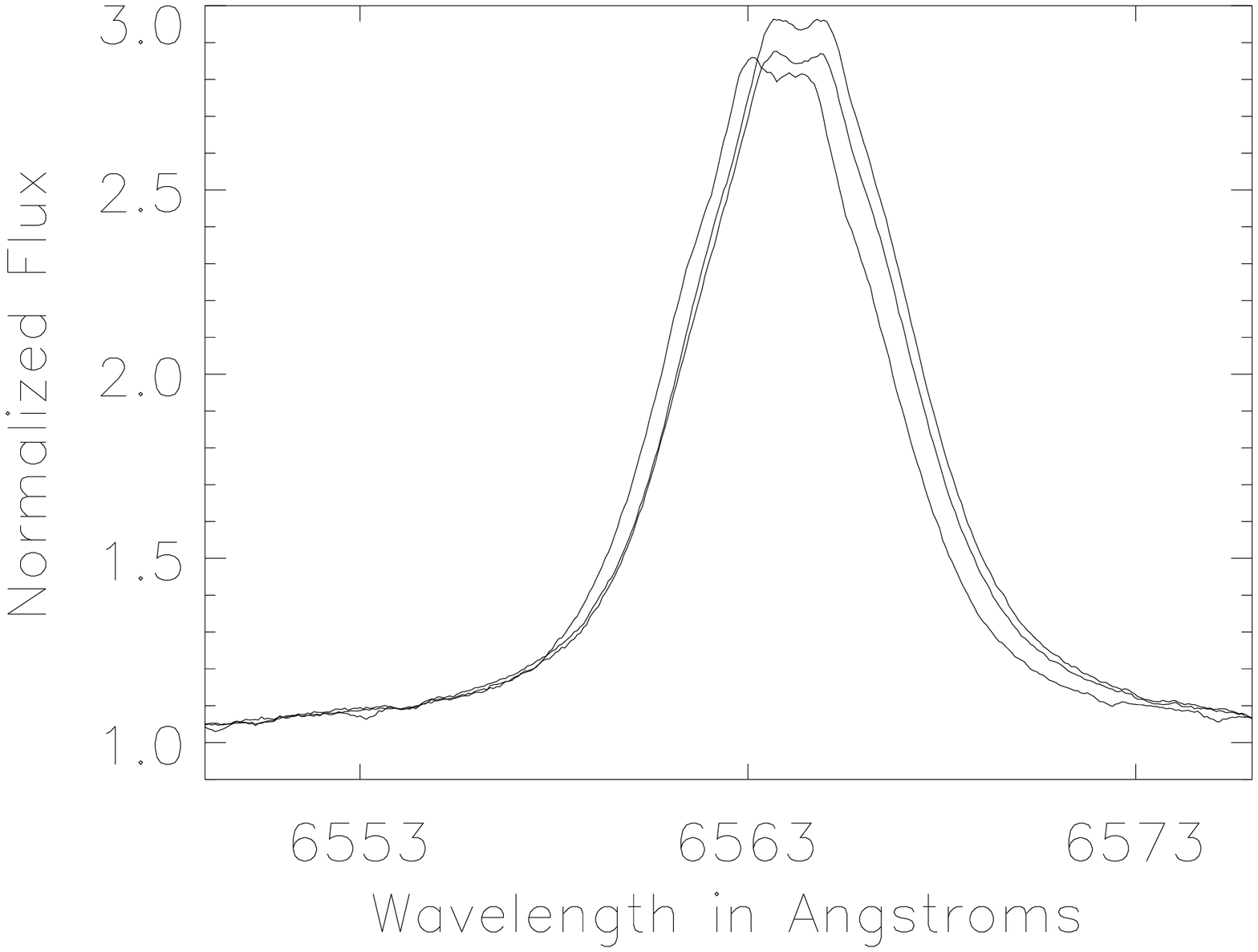}}
\quad
\subfloat[HD 38120]{\label{fig:lprof-hd381}
\includegraphics[ width=0.21\textwidth, angle=90]{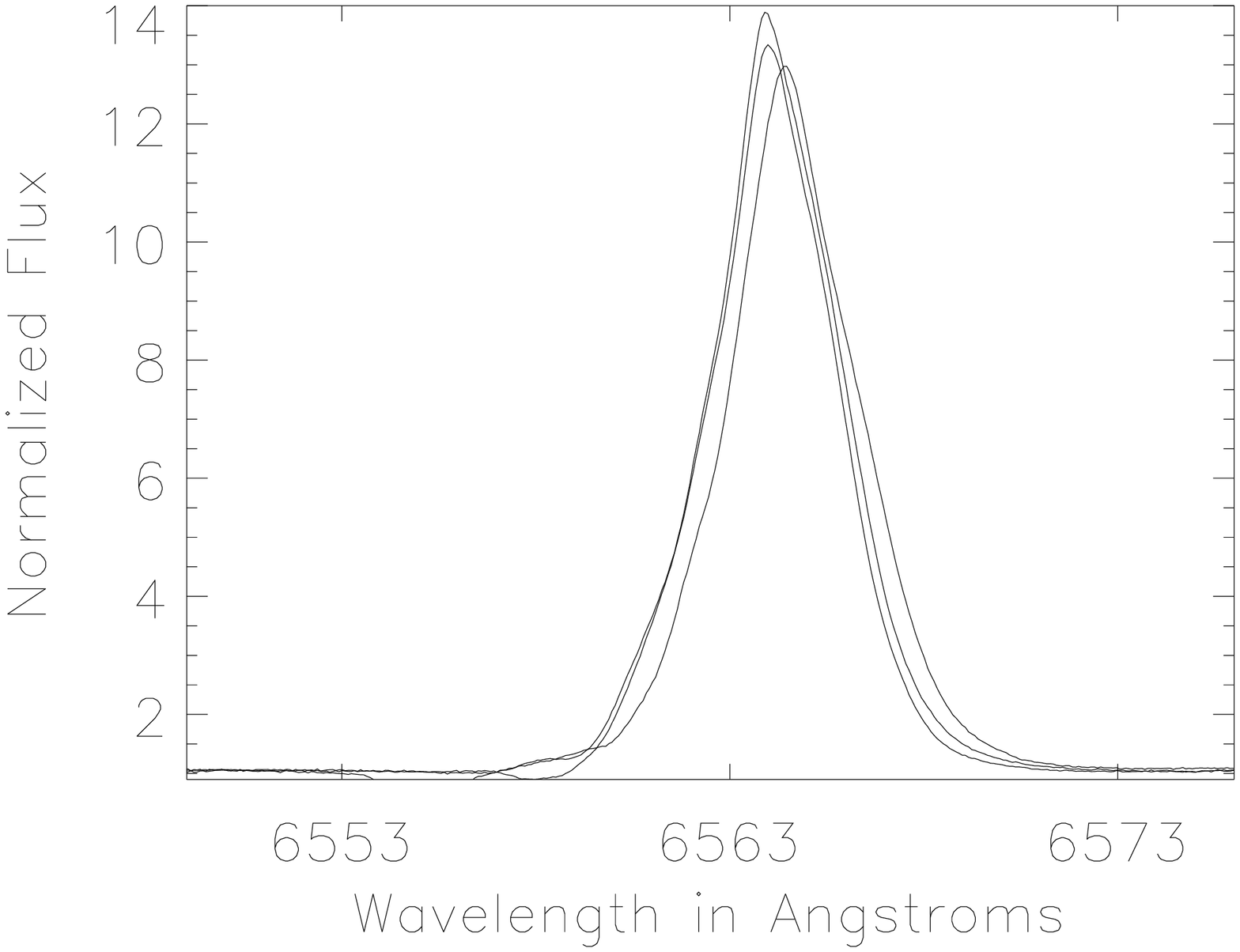}}
\caption[Herbig Ae/Be Line Profiles II]{HAe/Be Line Profiles II}
\label{fig:haebe-lprof2}
\end{figure}

\landscape
\begin{table}[!h,!t,!b]
\begin{center}
\begin{small}
\caption{Herbig Ae/Be Stellar Properties \label{tab-obs}}
\begin{tabular}{lrccccrccccccc}
\hline
\hline
{\bf Name}  & {\bf HD} &  {\bf MWC} & {\bf V} & {\bf ST} & {\bf $T_{eff}$} & {\bf $M_\odot$} & {\bf D} & {\bf B(G)} & {\bf Age} & {\bf R} & {\bf  L} \\
\hline
MWC442    & 13867    &   442    &7.7      &B5Ve           &             &                  &                  &                   &                   &              &             \\
AB Aur        & 31293     &  93       & 7.1     & A0Vpe        &  9.6     & 2.4            & 144         & $<$100    & 2.0            &  2.7v     & 1.68           \\
MWC480    & 31648     &  480    & 7.7     & A3pshe+    & 8.7      &  2.2           & 131         & $<$100    & 2.5             &               & 1.51            \\
HD35187   &  35187    &              &7.8      &A2e             & 9.5       &                  &                  &                  &                    &             & \\
HD35929    & 35989    &               &8.1     &A5               &  7.2      &                   &$>$360   &                  &                    &             & \\
MWC758      & 36112    &   758    &8.3    & A3e            &  8.1      &  2.0            & 200         &                 & 3.1              &             & \\
KMS 27         & 37357    &              &8.9    &A0e             &             &                     &                &                 &                    &             & \\
MWC120     & 37806      &  120    & 7.9    & A0              & 8.9      & 3.0d           & $>$230  & $<$100  & 0.37e 2.6w& 2.4w   & $>$1.51     \\
HD38120     & 38120    &               &9.1     &A0              &             &                    &                 &                 &                    &             & \\
FS CMa       & 45677     &   142    & 8.1     & Bpshe       &  21.4   &                   &$>$300   &                 &                    &             & \\ 
MWC158     & 50138    &  158     & 6.6     & B9               &15.5     &  5.0         & 290        &                  & 0.10          &              &  2.85               \\
GU CMa       & 52721    & 164      &6.6      &B2Vne       & 21.9    &                     &               &                 &                    &            & \\
MWC 166     & 53367    &   166    &7.0     &B0IVe         &            & 31.6            &                &                 &                    &            & \\
27 CMa        & 56014     & 170      &  4.7   &  B3IIIe       &             &                     &                &                 &                    &            & \\
HD58647    & 58647       &            & 6.8     & B9IV          & 10.7   & 4.2              & 280         &                & 0.16           &              & 2.48                \\
HD141569  & 141569    &            & 7.0     & B9.5e         & 6.3      & 2.3            &  99           &                &  $>$10       &            & \\
HD142666  & 142666    &            &8.8      & A8Ve          &             &                  &                  &               &                    &             & \\ 
HD144432  & 144432    &            &8.2       & A9/F0V     &  8.1       &                 & $>$200   &               &                     &            & \\
V2307 Oph & 150193     & 863    &  8.9     & A1Ve         & 9.3       &  2.3          & 150         &               & $>$6.3       &             & 1.47            \\
51Oph          & 158643    &            &4.8       &A0V            &  10.0    &   4.0         & 131         &               &  0.31           &            & \\
MWC275    & 163296     &  275   & 6.9      & A1Ve          & 9.3      & 2.3            & 122        & -60h      & 3.00e          &             &   1.48                  \\
HD169142  & 169142    & 925    &8.2      &B9Ve           &               &                 &                &                &                     &            & \\
MWC614    & 179218     &  614   & 7.2     & B9e             &  10.5     & 4.3          & 240        &                & 1.06e          &             &   2.50                 \\
V1295 Aql   & 190073     & 325   & 7.8     & A2IVpe      & 8.9         & 2.85c       &$>$290  & $<$100 & 1.0e 1.3w & 3.3w    &                           \\
MWC361    & 200775     &  361   & 7.4     &  B2Ve        &  20.4      &  10           & 430        &                & 0.02           &              &    3.89              \\
Il Cep           & 216629    & 1075   &9.3     &B2IV-Vne   &18.6       &                  &$>$160   &              &                      &            & \\
V700 Mon    &259431   & 147      & 8.8     &  B6pe        &              &                   &                &               &                      &            & \\
XY Per         & 275877   &              &  9.4   & A2IIv          & 14.1     &                   &                &                &                      &            & \\
T Ori              &                 &  763     &9.5     & A3V           &               &                   &                &               &                      &            & \\
\hline
\hline
\end{tabular}
\end{small}
\end{center}
The names, HD and MWC catalog numbers are listed for each star. The V magnitudes and spectral types are from the Simbad catalog.  The effective temperature in thousands of Kelvin ($T_{eff}$), mass in solar units ($M_\odot$), distances in parsecs (D), magnetic field strengths in Gauss, B(G), age in millions of years (Age), radius compared to one solar radii (R) and log luminosity in solar units (L), and are from van den Ancker et al. 1998 unless noted.  Magnetic field strengths in Gasus, B(G), are from Wade et al. 2007 unless noted.  c is Catala et al. 2007, d is Bohm \& Catala 1995, e is Manoj et al. 2006, h is Hubrig et al. 2006 and w is Wade et al. 2007
\end{table}
\endlandscape

\twocolumn

	There were a number of nights spent on component testing or experimental programs such as the initial spectrograph, spectropolarimeter and IR spectrograph testing. There were also tests of the IR dichroic, new spectrograph components, and military utility experiments. There are also notes in the table showing when the various improvements, tests, and repairs described in chapter 2 took place.  
	
	In the first part of the survey, few bright targets were chosen for close monitoring.  AB Aurigae and MWC 480 were monitored almost every night over the 2006-2007 winter.  During the same period, MWC 120, MWC 158 and HD 58647 were less well covered, but still monitored heavily. AB Aurigae or MWC 480 were observed continuously for several hours on some nights, and all 5 targets intermittently on others.  There are a total of 148 polarization measurements for AB Aurigae, 58 for MWC 480, 24 for MWC120, 39 for MWC 158, 19 for HD 58647, plus another 33 unpolarized standard star observations taken over the 40 nights in winter 2006-2007.    
	
	The H$_\alpha$ line for these five stars showed significant variability in intensity, width, and profile shape that is entirely consistent with other spectroscopic variability studies (cf. Beskrovnaya et al. 1995, Beskrovnaya \& Pogodin 2004, Catala et al. 1999).  On some nights, AB Aurigae and MWC480 showed dramatic spectroscopic variation on a timescale of minutes to hours, mostly in the blueshifted absorption trough.  Some general line width and line strength variability was also seen on short timescales.  All nights showed much smaller but significant variation.  
	
	The campaign was broadened in July 2007 to include many other Herbig Ae/Be stars. During the summer, HD 163296, HD 179218, HD 150193 and V1295 Aql were chosen for closer study. Another 111 unpolarized standard star observations were also taken since the instrument had changed significantly. Table \ref{tab-obs} shows a list of the Herbig Ae/Be stars and their properties taken from the literature. During the fall of 2007 and winter of 2008, a much larger target list of Herbig Ae/Be stars was used as well as bright Be and emission-line stars for comparison. Throughout the survey MWC 361 was used as a detailed stable comparison between HiVIS and ESPaDOnS. All of the H$_\alpha$ line profiles for the Herbig Ae/Be stars are shown in figures \ref{fig:haebe-lprof1} and \ref{fig:haebe-lprof2}

\subsection{Comments on Individual Stars}
	
	The long-term targets AB Aurigae, MWC 480, MWC 120, MWC 158 and HD 58647 are quite different from one another. Some present stable evidence for strong winds with P-Cygni profiles while others show strong central reversals or variable absorption properties. Each star deserves an in-depth discussion.
			
\subsection{AB Aurigae - HD 31293 - MWC 93}

\begin{figure}
\centering
\includegraphics[ width=0.5\textwidth, angle=90]{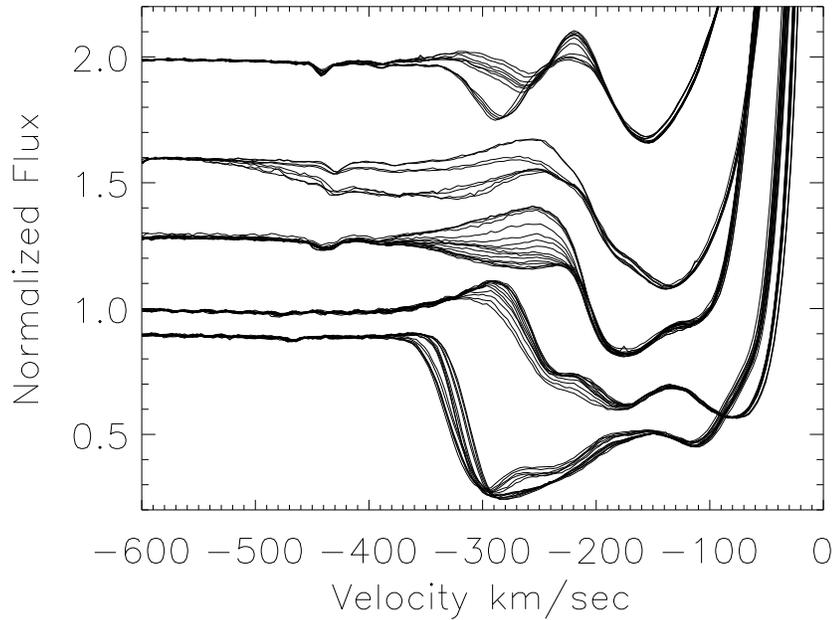}
\caption[Short Timescale Variability of AB Aurigae]{The short-term variability of the H$_\alpha$ P-Cygni absorption trough in AB Aurigae. Each curve represents 8 to 16 minutes. From bottom to top the nights are: 061228, 070117, 061106, 061027, and 061128. These are gaps in the coverage for all but 070117 and 061106.}
\label{fig:var-abaur}
\end{figure}

	The Herbig Ae star, AB Aurigae (HD31293, HIP22910) is the brightest of the Northern hemisphere Herbig Ae stars (V=7.1) and is one of the best studied intermediate-mass young stars. It has a near face-on circumstellar disk resolved in many wavelengths (eg: Grady et al. 2005, Fukagawa et al. 2004). It also has an active stellar wind with it's strong emission lines often showing strong P-Cygni profiles. Spectroscopic measurements put AB Aurigae somewhere between late B and early A spectral types (B9 in Th\'{e} et al. 1994, B9Ve in Beskrovnaya et al. 1995, A0 to A1 Fernandez et al. 1995) with an effective temperature of around 10000K. Bohm \& Catala (1993, 1995) present a complete spectral atlas and followup work on stellar activity. The star has a wind that is not spherically symmetric with a mass loss rate of order $10^{-8} {M_\odot}$ per year, and an extended chromosphere reaching $T_{max}\sim$17000K  at 1.5$R_\ast$ (Catala \& Kunasz 1987, Catala et al. 1999). However, Bouret et al. 1997 also found N V in the wind which require clumps of T$\sim$140,000K material at a filling factor of $\sim10^{-3}$. A short-term variability study done by Catala et al. (1999) showed that an equatorial wind with a variable opening angle, or a disk-wind originating 1.6$R_\ast$ out with a similar opening angle could explain the variability. Bohm et al. 1996 describe strong changes in the P-Cygni absorption with a 32-hour period. Several nights were spent continuously monitoring this star for short-term variability. Figure \ref{fig:var-abaur} shows several of the more variable nights with long monitoring periods. There is evidence presented in Hubrig et al. 2006b showing AB Aurigae has significant inhomogeneities in the distribution of elements in its atmosphere. Baines et al. 2006 present spectroastrometry also showing evidence for binarity. Wade et al. 2007 did not detect a magnetic field and had an upper limit of roughly 50-250G depending on the field type. Rogers 2001 derive an accretion rate of -6.85 Log(M$_\odot$/yr) using Bracket-$\gamma$ emission at 2.166$\mu$m.

\subsection{HD 31648 - MWC 480}

	  MWC 480 showed strong blue-shifted absorption components in the H$_\alpha$ line. Mannings et al. 1997 showed the presence of a circumstellar disk inclined at 30$^\circ$. Kozlova et al. 2003 presented a spectroscopic study of many lines. They found a v-sini of 90km/s. Mannings et al. 1997 found an inclination of roughly 30$^\circ$. They concluded that the stellar wind is the inner layers of the accretion disk. The high velocity component of the wind were interpreted as jets ejected from this region seen in projection. The H$_\alpha$ spectrum and the continuum polarization had also been studied in detail by Beskrovnaya \& Pogodin (2004) who concluded that MWC 480 also had an inhomogenious, azimuthally-structured wind which was variable on short timescales. Kozlova 2006 presented a H$_\alpha$ monitoring program with observations from 1998 to 2005. The H$_\alpha$ line shows continuum-normalized intensities of roughly 4-6 with a complicated blue-shifted absorption profile and variability on the timescale of hours. A low-velocity ($\sim$-100km/s) absorption is almost always present while higher velocity (more blue-shifted) absorptions are more infrequent and more variable. Wade et al. 2007 did not detect a magnetic field and had an upper limit of roughly 50-150G depending on the field type but Hubrig et al. 2006 present a magnetic field measurement of +87G $\pm$22.

\subsection{HD 37806 - MWC 120}

 	MWC 120 showed strong blue-shifted absorption components. H$_\alpha$ line spectropolarimetry from 1995 and 1996 was presented in Oudmaijer et al. 1999. The emission line profiles they observed changed drastically between the years. In January 1995, the line was double-peaked but with a much weaker blue-shifted emission. In December 1996, the emission line was evenly double-peaked. Wade et al. 2007 did not detect a magnetic field and had an upper limit of roughly 50-100G depending on the field type.

\begin{figure}
\centering
\includegraphics[ width=0.5\textwidth, angle=90]{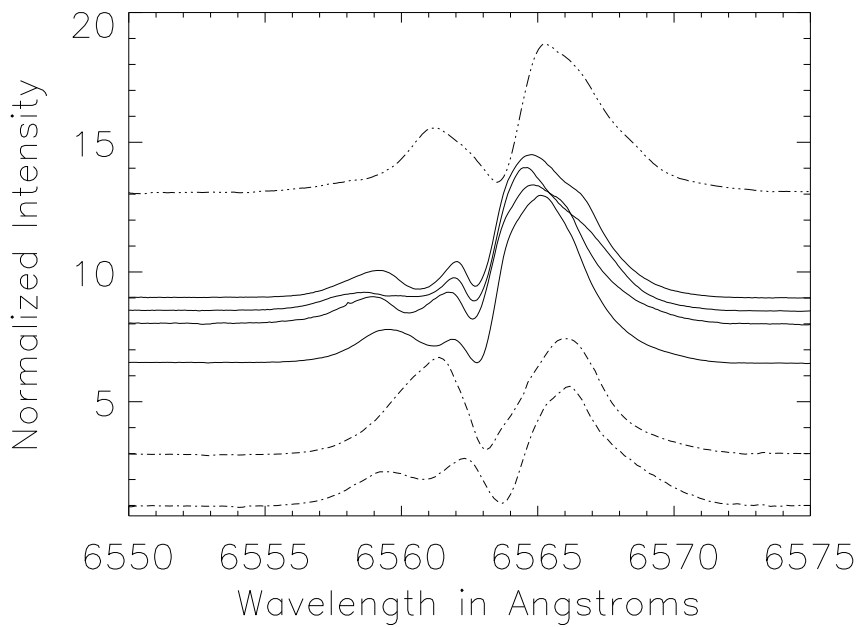}
\caption[Variability of MWC 120]{ The variability of the H$_\alpha$ line for MWC 120 in 2004, 2006 and 2007 vertically offset for clarity. The dashed lines on the bottom show spectra from engineering runs on October 20th (bottom) and December 9th (top) 2004. The solid lines clustered in the middle of the plot from bottom to top are December 27th 2006, January 3rd, 18th, and 19th 2007. The final solid line on top is August 29th 2007.}
\label{fig:var-mwc120}
\end{figure}

	We observed strong morphological changes over the course of three years. Figure \ref{fig:var-mwc120} shows that, while the line maintains a normalized intensity of roughly 5-7 times continuuum, the blue-shifted absorption varies quite strongly.

 \subsection{HD 50138 - MWC 158} 
 
	  MWC 158 is a mid-B type star which had a very strong emission line which showed a very strong central absorption. The peak intensities were roughly 15-17 times continuum while the central absorption was typically 2 times continuum. MWC 158 was previously studied for spectroscopic variability as well as low-resolution spectropolarimetry (Bjorkman et al. 1998, Pogodin 1997, Jaschek \& Andrillat 1998). Pogodin 1997 concluded that there is evidence for winds as well as in-falling matter from the envelope lines with the H$_\alpha$ absorption component existing from -200km/s to +70km/s. Bjorkman et al. 1998 found a nearly wavelength-independent continuum polarization and concluded that electron scattering (rather than dust scattering) had to be the polarizing mechanism. H$_\alpha$ line spectropolarimetry from 1995 and 1996 was presented by Oudmaijer et al. (1999).  Baines et al. 2006 showed spectroastrometry as evidence for binarity. They found a change in centroid and equivalent-width of the point-spread function across H$_\alpha$.  	  

 \subsection{HD58647} 
 
	  HD 58647 is a late B type star (B9 in Th\'{e} et al. 1994).  Baines et al. 2006 presented spectroastrometry claiming this star is a binary. They found a change in centroid and equivalent-width of the point-spread function across H$_\alpha$. Alhough this star was monitored for over 2 years, the H$_\alpha$ line was essentially invariant.

\subsection{HD 163296 - MWC 275}

\begin{figure}
\centering
\includegraphics[ width=0.5\textwidth, angle=90]{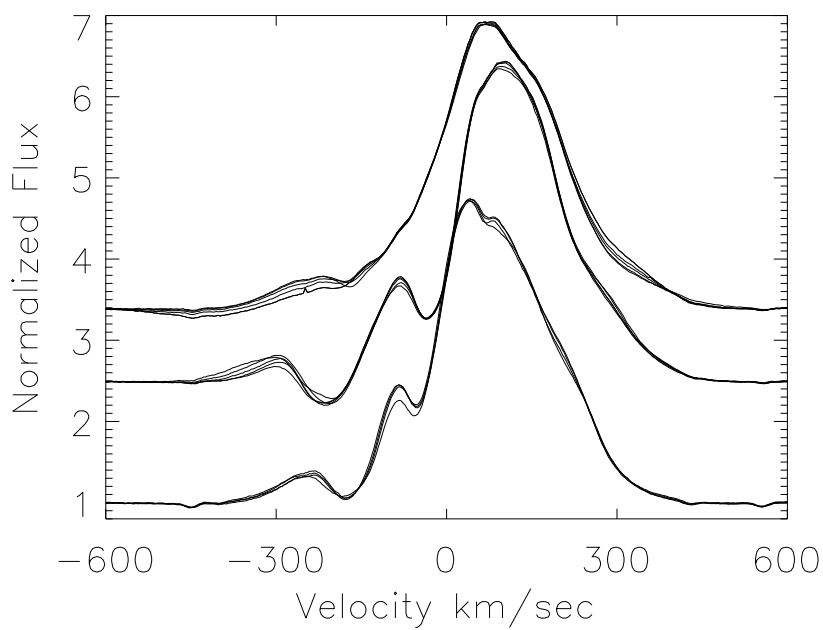}
\caption[Short Timescale Variability of HD 163296]{The H$_\alpha$ for HD 163296 on three of the most variable nights where there was good coverage.}
\label{fig:var-hd163}
\end{figure}

	This star has shown very strong H$_\alpha$ variability across the entire line. Figure \ref{fig:var-hd163} shows three nights during June and July of 2007 with good coverage. On each occasion, there is significant variability in relatively narrow wavelength ranges on both blue-shifted and red-shifted sides of the line.
	
	Periodic variations variations in UV lines showed 35 and 50 hour periodicities (Catala et al. 1989). Pogodin 1994 studied the H$_\alpha$ and H$_\beta$ lines and found an active region of the stellar wind near the star with a stable shell surrounding it. Longer lived rotating jets as well as short-timescale clumps were concluded to cause the variability. The H$_\alpha$ line profiles had roughly the same shape and variability as were observed in figure \ref{fig:lprof-hd163}. Garcia Lopez et al. 2006 derive an accretion rate of -7.12 Log(M$_\odot$/yr) using Bracket-$\gamma$ emission at 2.166$\mu$m. Grady et al. 2000 showed an azimuthally symmetric disk around this star inclined at 60$^\circ$ and chain of Herbig-Haro objects oriented perpendicular to the disk. Catala et al. 1989 reported UV spectroscopy showing a variable wind. Hillenbrand et al. 1992 derived an accretion rate from NIR emission of 1.3 10$^{-6}M_\odot$/yr but Skinner et al. 1993 get $<9.1 10^{-9}M_\odot$/yr using radio continuum emission. Devine et al. 2000 reported images of a bipolar outflow that was traced to within 60mas (7.3AU) of the star. The jets from this star are seen in Ly$_\alpha$ to extend from at least as close as  0.06$''$ (7.3AU) out to 6$''$ (725AU) forming a collimated, bipolar outflow designated HH409. Thus, this star is one of the more complex with claimed detections of a wind, disk, accretion, and a bipolar outflow.

\subsection{HD 179218 - MWC 614}

	This star is an isolated Herbig Ae star with a vsini 60km/s. Miroshnichenko et al. 1999 presented spectroscopy and low-resolution spectropolarimetry from 1995-1997. The H$_\alpha$ line was typically only 2 or 3 times continuum with a complicated absorption structure overlying the emission. They detected an R-band polarization of 0.45\% at an angle of 102$^\circ$ in August and September of 1997. They concluded that the star was 0.9 magnitudes above the zero-age main-sequence and had a luminosity greater than pre-main-sequence stars of similar temperature such as AB Aurigae. They argued for the presence of an inhomogeneous flattened circumstellar envelope. Kozlova 2004 presented spectroscopy of the H$_\alpha$ and Na I D lines from 1999 to 2003. They showed that the typical profile for this period was a 3 or 4 times continuum line with typically small but significant blue-shifted absorption. One extremely absorbed, nearly flat-topped emission line was reported in August 1999, similar to the two low profiles presented in figure \ref{fig:hd179} or those of Miroshnichenko et al. 1999. Our line profiles were typically 4 or 5 times continuum but a consistent small amount of blue-shifted absorption was also observed.
	
	Liu et al. 2007 resolved warm circumstellar material around this star on the 10's of AU scale using 10$\mu$m nulling interferrometry. Garcia Lopez et al. 2006 derive an accretion rate of -6.59 Log(M$_\odot$/yr) using Bracket-$\gamma$ emission at 2.166$\mu$m which is quite high.

\onecolumn
	
\begin{figure}
\centering
\includegraphics[ width=0.5\textwidth, angle=90]{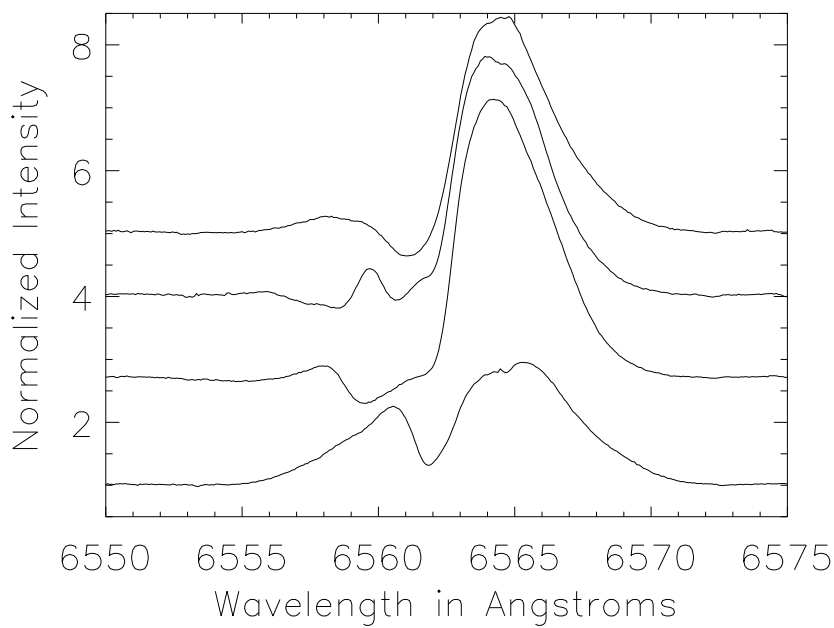}
\caption[Variability of HD 150193]{ The variability of the H$_\alpha$ line for HD 150193 in 2007. The lines from bottom to top are June 20th, July 28th, August 1st and August 29th. The star changed from an emission line with a small, nearly central absorption to a mostly classical P-Cygni structure with a much greater intensity.}
\label{fig:var-hd150}
\end{figure}

\begin{figure}
\centering
\includegraphics[ width=0.5\textwidth, angle=90]{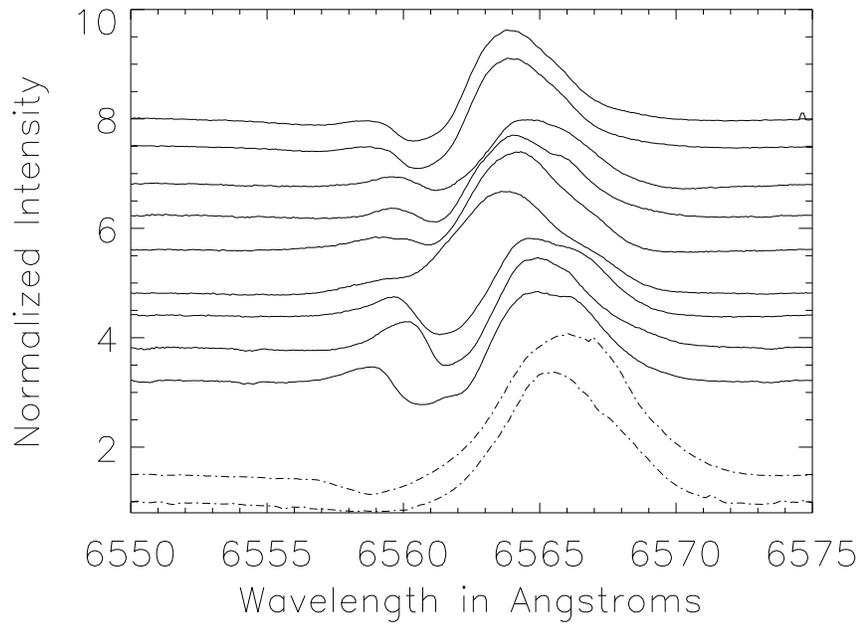}
\caption[Variability of MWC 758]{ The variability of the H$_\alpha$ line for MWC 758 in 2004 and 2007. The dashed lines on the bottom show spectra from engineering runs on October 20th (bottom) and December 15th (top) 2004. The solid lines from bottom to top are August 28 and 29, September 20, October 30 \& 30, November 21 \& 24.}
\label{fig:var-mwc758}
\end{figure}

\begin{figure}
\centering
\includegraphics[ width=0.5\textwidth, angle=90]{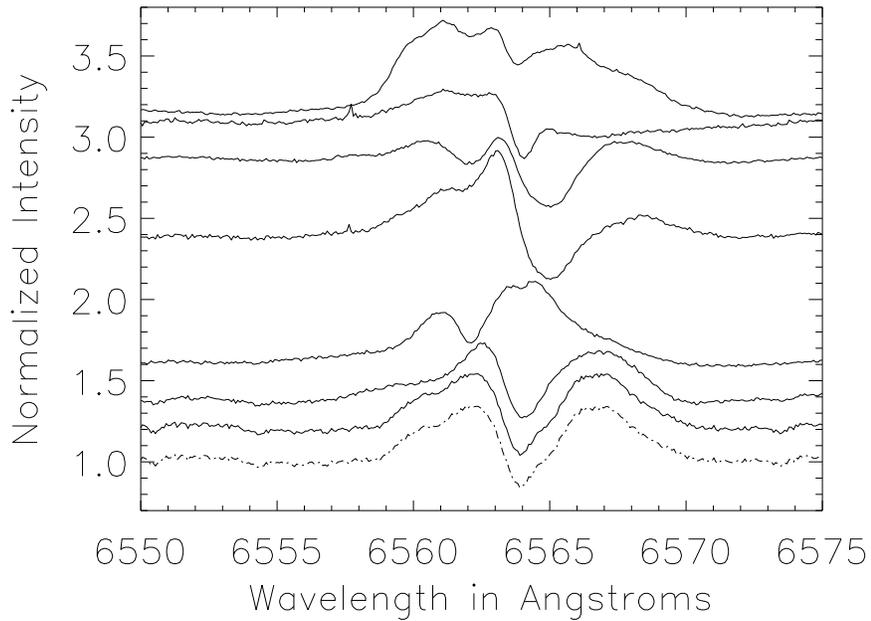}
\caption[Variability of HD 35187]{ The variability of the H$_\alpha$ line for HD 35187 in 2007. From bottom to top the dates are August 28 and 29, September 20, October 30 \& 30, and finally November 21.}
\label{fig:var-hd351}
\end{figure}

\twocolumn

\subsection{HD 150193 - MWC 863}

	This star underwent a quite strong change in line profile. Figure \ref{fig:var-hd150} shows the change from a low-intensity line with a relatively narrow, mildly blue-shifted absorption to a full P-Cygni profile. Carmona et al. 2007 reported FORS2 spectroscopy of the TTauri-component: HD 150193B at a distance of 1.1$''$ from the primary. The companion had a spectral type of F9Ve and had H$_\alpha$ emission of roughly 2.5 times continuum while the primary had a ration of 2.2:1. Garcia Lopez et al. 2006 derive an accretion rate of -7.29 Log(M$_\odot$/yr) using Bracket-$\gamma$ emission at 2.166$\mu$m.

\subsection{V1295 Aql - HD 190073 - MWC 325}

	This star is far away from the well known star formation regions, but shows a large far-IR excess from cool dust and is more luminous than other Herbig Ae stars (Sitko 1981). Pogodin et al. 2005 presented a spectroscopic study of this star from 1994-2002 showing a very pronounced unstable wind with an optically thick equatorial disk. The star displays a wealth of lines in emission. The Pogodin et al. 2005 H$_\alpha$ profiles looked fairly similar to our observations presented in figure \ref{fig:lprof-v1295} though with slightly stronger and more complicated absorption as well as a slightly lower intensity. They did not find any short-term variability, and derived a vsini of 12km/s. Wade et al. 2007 did not detect a magnetic field with an upper limit of roughly 50G but Catala et al. 2007 did claim a detection of $\sim$70G. Both of these detections are too small to produce any measurable linear polarization signature across any individual line. Catala et al. 2007 fit the H$_\alpha$ line with a 1.4 10$^{-8}M_\odot$/yr wind having a terminal velocity of 290km/s and a base temperature of roughly 18000K. Baines et al. 2006 suggested this is a possible binary star based on a change in width of the psf across the H$_\alpha$ line without a corresponding change in centroid location.

\subsection{HD 200775 - MWC 361}

	This Herbig Be star has a fairly stable, double-peaked emission line except when it undergoes a period of strong variability every 3.68-years (Beskrovnaya et al. 1994, Miroshnichenko et al. 1998, Pogodin et al. 2000, 2004) and is a known binary at 2.25$''$ and 164$^\circ$ (Pirzkal et al. 1997). Ismailov \& Aliyeva 2005 reported H$_\alpha$ lines of 8.5-10 times continuum with no significant short-timescale variability but a slow drift over 20 days of individual line components. Pogodin et al. 2004 used radial velocity measurements to derive an orbital solution for the binary as an eccentric binary at e$\sim$0.3 with the mean separation at 1000R$_\odot$ with the secondary most likely to be a $\sim$3.5M$_\odot$ pre-main-sequence star.

\subsection{HD 36112 - MWC 758}

	This star had a significantly variable H$_\alpha$ line. Figure \ref{fig:var-mwc758} shows a progression from 2004 to 2008 as both the intensity and absorption change significantly. Beskrovnaya et al. 1999 presented spectroscopy as well as multi-color photometry and polarimetry. They conclude that the star is an unreddened A8V star of 5Myr age, close to the ZAMS. They present evidence for a gaseous dusty envelope and a variable stellar wind with an extended acceleration zone. The wind dominates the inner envelope and there is a high temperature zone, likely chromospheric in origin. They find small but irregular photometric variability (0.2m). A P-Cygni type H$_\alpha$ line was observed in the 1994-1996 period with rapid variability, thought to be caused by jet-like inhomogeneities. They also concluded that this envelope is the inner part of a disk that is close to edge on. A vsini was calculated as 60km/s $\pm$6 and the H$_\alpha$ line had an intensity of roughly 2-3. They report the SED is 'flat' in the IR region which is typical of younger HAe/Be stars like AB Aurigae with hot dust. This star also has barium and silicon abundance anomalies common in chemically-peculiar stars.

\subsection{HD 35187}

	This star is quite variable on monthly timescales. Figure \ref{fig:var-hd351} shows the line profiles in timeseries. The star is an obvious binary. Wade et al. 2007 did not detect a magnetic field in either binary component and had an upper limit of roughly 50-100G depending on the field type. Rodgers 2001 derive an accretion rate of -7.29 Log(M$_\odot$/yr) using Bracket-$\gamma$ emission at 2.166$\mu$m.
 
\subsection{HD 35929}

	HD 35929 is a pre-main-sequence Herbig Ae star candidate because it exhibits H$_\alpha$ emission and a weak IR excess. Miroshnichenko et al. 2004 presented spectroscopic analysis that concluded that this star is F2 III spectral type with a T$_{eff}$ of around 6900, vsini of 70km/s with a mass of 2.3M$_\odot$ at roughly 350pc. They supported an argument by Marconi et al. 2000 that this star is not a young object, but post main-sequence giant in the instability strip with significant mass loss.

\subsection{HD 144432}

  	HD 144432 is a binary at 1.4$''$ with a K2 to K7 type star (Carmona et al. 2007). They reported that the H$_\alpha$ emission of the primary had a peak of 3.4:1 and the secondary also had H$_\alpha$ emission at roughly 2 times continuum. 

\subsection{HD 45677}

	In the line profiles presented in Oudmaijer et al. 1999, the H$_\alpha$ line in this star is single peaked and asymmetric. This is quite different than the double-peaked line profile of figure \ref{fig:lprof-hd456}. Grady et al. 1993 reported the presence of redshifted absorption lines caused by infalling bodies. Baines et al. 2006 claimed this star to be a binary based on their spectroastrometry.

\begin{figure} [!h]
\centering
\includegraphics[ width=0.3\textwidth, angle=90]{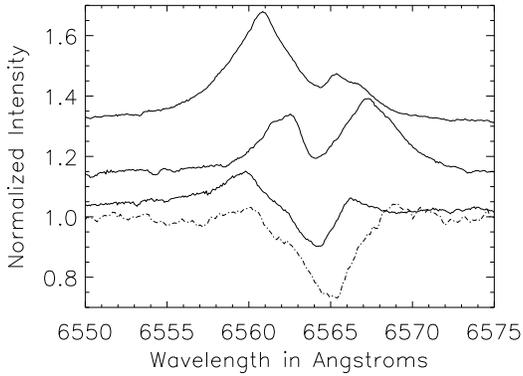}
\caption[Variability of MWC 166]{ The variability of the H$_\alpha$ line for MWC 166. The dashed line on the bottom shows a single spectrum from an engineering run on October 20th 2004. The solid line in the middle is December 27th 2006 and the top line is September 20th 2007.}
\label{fig:var-mwc166}
\end{figure}

\subsection{ HD 53367 - MWC 166 }

	This star is a known binary at a separation of 0.65$''$ (Fu et al. 1997). Baines et al. 2006 used this star as a test of their spectroastrometry and showed a strong change in both point-spread function width and centroid across the H$_\alpha$ line. This star showed a strong change from 2004 to 2008 as shown in figure \ref{fig:var-mwc166}. The bottom dashed line from 2004 is very similar to the spectrum observed in December 2003 by Vink et al. 2005. However, the emission increased in 2007. In January 2008, the emission on the blue-shifted side of the line increased dramatically, while the red-shifted component was significantly reduced compared to 2007.

\subsection{Summary}

	In this chapter a summary of the observing campaign was presented with the H$_\alpha$ spectroscopy and detailed descriptions of individual stars. The collection of stars in this sample is quite diverse. There is a wide range of morphologies and H$_\alpha$ line types. Some showed remarkable stability in line-profile shape over more than a year. HD 58647 showed an essentially invariant line profile from December of 2006 to January of 2008. Others were quite variable and some even showed a wide range of morphologies. Many of the targets are known or suspected binaries. Some are significantly variable on very short time-scales. In general, the blue-shifted absorption is much more variable. This is entirely expected because the absorption comes from a comparatively small volume. However, some stars showed significant variability across the entire line, particularly HD 35187. This large collection allows a solid characterization of each star and will be useful in discussing the spectropolarimetric effects in later chapters.

\section{Current Theory of Spectropolarimetric Line Profiles - Thomson Scattering}

	Before discussing any specific results, it is useful to review several current theories and methods for calculating spectropolarimetric line profiles. This will give us a framework for comparing and contrasting the observations which do fit these theories from those that don't.

	The main framework currently used to describe spectropolarimetric effects in hot stars is scattering of stellar light off electrons in asymmetric circumstellar environments, though many of these models can be adapted to any other type of particles. Scattered light is linearly polarized with a dependence on the scattering angle. Asymmetries in the circumstellar environment lead to an incomplete cancellation of the polarized scattered light from different regions around the star, leaving a net polarization. A wavelength dependence is imparted by the motion of the scattering particles.

	Initially, some of these theories were developed to explain Be-star observations (cf. McLean \& Brown 1978, McLean \& Clarke 1979, McLean 1979, Poeckert \& Marlborough 1976, 1977, 1978, 1979). The theories included the asymmetric-envelope theory for continuum polarization and the ``depolarization'' effect, especially the theories developed in McLean 1979. Analytical calculations were then done in Wood et al. 1993 and Wood \& Brown 1994 to add a thin-disk and equatorial-wind scattering theory. This theory was then updated to a numerical Monte-carlo model in Vink et al. 2005a that was subsequently used to explain observations of Herbig Ae/Be stars in Vink et al. 2002, 2005b and Mottram et al. 2007 as well as hot-massive stars in Oudmaijer 2005.

	This chapter explores a simple analytical theory for calculating spectropolarimetric signatures from a thin equatorial disk. A few related effects that have been used to explain spectropolarimetric signatures in Be and Herbig Ae/Be stars will also be reviewed. Special attention will be paid to the rotating-disk scattering model as well as the depolarization model as applied to line profiles with significant absorptive effects as these morphologies are the majority of the Herbig Ae/Be H$_\alpha$ line profiles.

\subsection{Scattering Theory}

  The theory to calculate spectropolarimetric line profiles can be broken down into a few basic parts. The key pieces to calculating the polarization from light scattered by circumstellar material are:  the distribution and motion of the scattering particles, the geometry of the scattering, and the physics of scattering off those particles.  An initial stellar photospheric source is assumed to be an emission or absorption line. Then, given the motion/distribution of the particles, the polarization of the scattered light is calculated at each wavelength from each point around the star.  This relies heavily on the theory developed in Wood et al. 1993, but will be more pedagogical and will have more examples and more steps explicitly shown. Vectors will be bold ($\hat{{\bf k_\ast}}$) and tensors/matricies will have a bar over them ($\overline{{\bf R}}$).

\subsection{Scattering Geometry}

Calculating the polarization caused by the particles around a star is done knowing the geometry of the scattering events and how particles polarize the scattered light.  The geometry is calculated by tracing a ray of light from every point on the photosphere of the star to every point of the circumstellar material and calculating the scattering angle and the orientation on the sky of the scattering event.  Once the geometry of each event is known, you then must do the polarized radiative transfer.  Given a certain type of scattering particle (electrons, dust, atoms, etc), one must model how the Stokes vectors transform (${\bf I_{in}} \rightarrow {\bf I_{out}}$).  The last thing to do then is integrate over all possible ray paths from the telescope to all the scattering particles to add up all the photons in each polarization state at each wavelength to calculate what the average polarization would be for the entire system, since it will be unresolved on the sky.  \\

The three components are thus:

\noindent 1) Geometry - Star-Particle-Telescope paths.  \\
2) Scattering Physics - ${\bf I_{in}} \rightarrow {\bf I_{out}}$  \\
3) Integrate Over Space - Sum all photons  \\

There are three main geometrical integrations. 1) all photospheric source locations ($d\Omega_\ast$)  2) all lines of sight from the telescope through the circumstellar material (dl), and 3) all pointings of the telescope to the disk ($d\Omega$).  

The Stokes vector of light scattered at some angle can be expressed in terms of the density of the scatterers ($\rho$) and the geometrical scattering coefficients (${\bf j}$), called the Stokes emission coefficients: 

\begin{eqnarray}
  {\bf I_{out}} = \left( \begin{array} {c}  I \\ Q \\ U \\ V \end{array} \right) = \rho {\bf j}
\end{eqnarray}

The emission coefficients (${\bf j}$) hold all the geometrical and physical details.  The coefficients are an integral over all incident rays (photospheric sources, $d\Omega_\ast$) to the particle.  The amount of scattered light is set by the initial intensity at the photosphere, the opacity of the material (scattering probability), and the geometry of the scattering event (scattering efficiency).  The polarization of the scattered light is mainly controlled by the scattering geometry since continuum and line emission from the photosphere is generally unpolarized ($<0.01\%$ in the sun).  The source-particle vector, $\hat{{\bf k}}$, and the particle-telescope vector, $\hat{{\bf k_\ast}}$, specify the geometry of the event.

\subsection{Scattering Physics}

At a single point and for a single incident ray, ignoring doppler shifts and wavelength dependent materials properties, the Stokes emission coefficients, which describe the output stokes vector, can be written as:

\begin{equation} 
{\bf j}=opacity \times scattering\times input=\kappa\overline{{\bf R}} \bullet {\bf I_{in}}
\end{equation}

The scattering efficiency and polarized radiative transfer is contained in the phase matrix ($\overline{{\bf R}}$). This is a 4$\times$4 matrix that describes how each of the 4 incident Stokes numbers gets transformed into the outgoing Stokes vectors.  Each term specifies how one Stokes coefficient transfers to another ($Q_{in} \rightarrow U_{out}$ or simply QU). The phase matrix is different for each type of particle and is typically measured in the lab or calculated theoretically from light-scattering models. Given a specific geometry (scattering angle) and type of particle, you can write the phase matrix as a 4$\times$4 array:    

\begin{small}
\begin{eqnarray}
  \overline{{\bf R}} =     \left ( \begin{array}{cccc}
           II   &   QI    &  UI    &  VI  \\
           IQ  &  QQ  &  UQ   &  VQ  \\
           IU  &  QU  &  UU   &  VU  \\
           IV  &  QV  &  UV   &  VV  
           \end{array} \right )  
\end{eqnarray}
\end{small}

For instance, Thomson scattering (electron scattering) phase matrix is independent of wavelength and is particularly simple.  It only depends on the scattering angle (star-scatterer-telescope angle, $\chi$).  You can define an angle, $\xi=90^\circ-\chi$, as the dot product between the star-scatter vector ($ \hat{{\bf k_\ast}}$) and scatter-observer vector ($\hat{{\bf k}}$).

\begin{equation}
  \cos\xi= \hat{{\bf k_\ast}}\bullet \hat{{\bf k}}
\end{equation}

\begin{figure}
\centering
\includegraphics[width=0.3\textwidth]{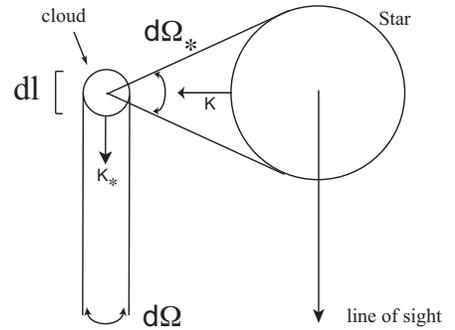}
\caption[Scattering Geometry and Variables]{ The three main geometrical parameters.  The stellar disk subtends a solid angle $d\Omega_\ast$. The scattering depth along a single line of sight is dl.  The solid angle of the cloud as seen by the detector is $d\Omega$.  The three integrations ($d\Omega_{\ast}, dl, d\Omega$) make up the scattered light detected for a given parcel of circumstellar material.}
\label{fig:scatgeo}
\end{figure}

\begin{table}[!h,!t,!b]
\begin{center}
\begin{small}
\caption{Variables \label{tab-var}}
\begin{tabular}{ll}
\hline
\hline
Name                     &             \\
$\chi$                     &   Scattering angle: star-cloud-telescope           \\         
$\phi$                     &   Orbital phase: Line-of-sight=0$^\circ$            \\         
$i$                           &    Inclination angle \& +Q reference frame           \\         
$\psi$                       &   Angle between scattering plane +Q and telescope +Q       \\         
$\beta_r$                &   Radial doppler shift in stellar frame       \\         
$\beta_\phi$           &   Orbital doppler shift in stellar frame       \\         
$\hat{{\bf k}}$          &  Cloud-telescope unit vector       \\         
$\hat{{\bf k_\ast}}$  &  Star-cloud unit vector       \\         
\hline
\hline
\end{tabular}
\end{small}
\end{center}
\end{table}

With this definition of the angle $\chi$, the Thomson scattering phase matrix becomes:

\begin{eqnarray}
  \overline{{\bf R}}= \frac{3}{8\pi}
           \left ( \begin{array}{cccc}
              \frac{1}{2}(1+\cos^2\chi) & \frac{1}{2}\sin^2\chi      & 0       & 0 \\
              \frac{1}{2}\sin^2\chi     & \frac{1}{2}(1+\cos^2\chi)  & 0       & 0  \\
              0                        & 0                         & \cos\chi & 0 \\
              0                        & 0                         & 0       & \cos\chi 
           \end{array} \right )
\end{eqnarray}

The phase matrix always defines the +Q direction as perpendicular to the scattering plane, so this phase matrix is inherently dependent on the scattering geometry of each individual scattering event. As an example, if the scattering angle is $\chi=0^\circ$, as is the case for forward scattering, then $\cos\chi$=1, $\sin\chi=0$, and the phase matrix becomes the identity matrix (neglecting the normalization constant) and there is no polarization induced in the scattered light. If the scattering angle is 90$^\circ$ then $\cos\chi$=0, $\sin\chi=1$ only non-zero terms are the Q terms. This matrix shows that incident light of any polarization state scattered at 90$^\circ$ becomes totally polarized in the scattering plane: Q/I = (I+Q)/(I+Q) = 1.  

\begin{eqnarray*}
{\bf I_{out}} = \overline{{\bf R}} \bullet {\bf I_{in}} & = &        
       \left( \begin{array} {cccc} 
                                   1 & 1 & 0 & 0 \\ 
                                   1 & 1 & 0 & 0 \\
                                   0 & 0 & 0 & 0 \\
                                   0 & 0 & 0 & 0  \end{array} \right)
       \left( \begin{array} {c}  I \\ Q \\ U \\ V \end{array} \right) \\ & = &
       \left( \begin{array} {c}  I+Q \\ I+Q \\ 0 \\ 0 \end{array} \right)
\end{eqnarray*}

For a single ray coming from the photosphere to an electron in the circumstellar disk scattered at 90$^\circ$ to the telescope, you can expect 100\% linear polarization perpendicular to the scattering plane: +Q in the scattering-frame. This individual scattering plane is in general not the defined +Q axis. Typically, +Q is taken in theoretical calculations as the position angle of the stellar rotation axis. The phase matrix for each individual scattering event must be rotated to the stellar frame. Since the phase matrix is computed in the scattering plane, the plane containing $\hat{{\bf k_\ast}}$ and $\hat{{\bf k}}$, this computed polarization must be rotated by the angle $\psi$, defined as the angle between the the projected scattering plane normal and stellar rotation axis, to get the phase matrix into the reference frame for the entire system. By applying a rotation matrix, $\overline{{\bf L}}$, to the phase matrix the general form of the 90$^\circ$ electron-scattering phase matrix is:

\begin{small}
\begin{eqnarray}
  \overline{{\bf R_{rot}}}=\overline{{\bf L}} \bullet \overline{{\bf R}} 
            \left ( \begin{array}{cccc}
              1 & 0      & 0       & 0 \\
              0     & cos2\psi  & sin2\psi       & 0  \\
              0                        & -sin2\psi                        & \cos2\psi & 0 \\
              0                        & 0                         & 0       & 1
           \end{array} \right )
    \bullet                                   
 \left( \begin{array} {cccc} 
                                   1 & 1 & 0 & 0 \\ 
                                   1 & 1 & 0 & 0 \\
                                   0 & 0 & 0 & 0 \\
                                   0 & 0 & 0 & 0  \end{array} \right)
\end{eqnarray}
\end{small}

\begin{eqnarray}
  \overline{{\bf R_{rot}}}=
            \left ( \begin{array}{cccc}
              1 &          0      & 0       & 0 \\
              0 &  cos2\psi  & 0     & 0  \\
              0 & -sin2\psi   & 0     & 0 \\
              0 &      0           & 0      & 0
           \end{array} \right )
\end{eqnarray}

	Now, instead of both the II and QQ terms being 1, there is rotational effect that says $Q_{out}=Q_{in}cos2\psi$ and $U_{out}=Q_{in}sin2\psi$. The rotation introduces a U term. Since an unpolarized stellar source is typically assumed, the product $\overline{{\bf L}} \bullet \overline{{\bf R}} \bullet {\bf I}$ can be computed as the Stokes emission coefficients for a single wavelength given an unpolarized source as:

\begin{eqnarray}
{\bf P}=\overline{{\bf L}} \bullet \overline{{\bf R}} \bullet {\bf I}
\end{eqnarray}

\begin{eqnarray*}
{\bf P}        = \frac{3}{8\pi} 
        \left ( \begin{array}{cccc}
           1  & 0                     & 0                      & 0 \\
           0  & cos2\psi  & sin2\psi    & 0  \\
           0  & -sin2\psi  & \cos2\psi  & 0 \\
           0  & 0                      & 0                      & 1
        \end{array} \right ) 
        \bullet \\
        \left ( \begin{array}{cccc}
         \frac{1}{2}(1+\cos^2\chi) & \frac{1}{2}\sin^2\chi      & 0       & 0 \\
          \frac{1}{2}\sin^2\chi     & \frac{1}{2}(1+\cos^2\chi)  & 0       & 0  \\
          0                        & 0                         & \cos\chi & 0 \\
          0                        & 0                         & 0       & \cos\chi 
           \end{array} \right )
       \bullet
        \left( \begin{array} {c}  I_0 \\ 0 \\ 0 \\ 0   \end{array} \right)
\end{eqnarray*}

\begin{eqnarray}
{\bf P}=\frac{3I_0}{16\pi}\left ( \begin{array}{c}
           1 + \cos^2\chi  \\ \sin^2\chi\cos2\psi \\ -\sin^2\chi\sin2\psi \\ 0 
        \end{array} \right )        
\end{eqnarray}

This gives us the output Stokes vector at a given scattering angle accounting for the arbitrary alignment of the individual scattering plane on the sky ($\phi$).

\subsection{Circumstellar Matter - Doppler Shifts \& Distributions}

To calculate a spectropolarimetric line profile, the distribution and motion of scattering particles around a star must be known. This circumstellar matter is typically described as a disk and/or a wind. The matter will be assumed to take either two forms: a disk in Keplarian orbit or a wind flowing radially outward. The surface-density of the circumstellar material is usually assumed to follow an inverse square law and the orbital velocity is Keplarian:

\begin{equation}
 \Sigma(r)=\Sigma_0(\frac{R_0}{r})^2
\end{equation}

\begin{equation}
 v_\phi(r)=v_0(\frac{R_0}{r})^\frac{1}{2}
\end{equation}

Here, $R_0$ is the stellar radius and $r$ is the actual distance of the cloud. With a wind, the density of the outflowing material is assumed to follow a 1/r distribution with the surface density being related to the stellar mass loss rate ($\dot{M}$) and the wind velocity, $v_r(r)$. The functional form of the velocity can be taken from radiation-driven stellar wind theory using two parameters: the velocity at the stellar surface ($v_\ast$) and at infinity ($v_\infty$). Using the mass of the proton as $m_p$, the surface density and motion becomes:

\begin{equation}
\Sigma(r)=\frac{\dot{M}}{m_p2\pi v_r r}
\end{equation}

\begin{equation}
v_r(r)=v_\ast+v_\infty(1-\frac{R_0}{r})^\frac{1}{2}
\end{equation}

Once the density, distribution, and velocity of the scattering particles is set up, the amount of light scattered from these particles can be calculated. The wavelength changes induced in the scattered light will be determined by the relative doppler shifts between the star and particle as well as between the particle and the observer. How the doppler-shifts redistribute the polarized light in wavelength can now be calculated. At any observed wavelength, $\lambda$, radiation can be scattered into your line of sight from other wavelengths $\lambda_\ast$ by particles moving at the correct velocity. To properly calculate the scattered light at one wavelength an integration across the scattered light of all wavelengths must now be done to count the radiation dopler-shifted into the wavelength of interest. The doppler-shift from the photospheric rest wavelength ($\lambda$) to the doppler-shifted wavelength ($\lambda'$) can be written as:

\begin{equation}
\lambda=\lambda' \gamma(1-\beta\hat{\bf v} \bullet \hat{{\bf k_\ast}})
\end{equation} 

Where $\gamma$ is the relativistic gamma-factor, and $\beta$ is the particles velocity in gaussian units (v/c).  Once the doppler-shift from the scattering event is known, one must then doppler-shift the scattered radiation into the observers frame through the motion of the particle with respect to the observer. The wavelength observed by the telescope ($\lambda_{obs}$) can be related to the wavelength at the cloud ($\lambda_\ast$) as:

\begin{equation}
\lambda_{obs}=\lambda_\ast \frac{1-\beta\hat{\bf v} \bullet \hat{{\bf k}}}{1-\beta\hat{\bf v} \bullet \hat{{\bf k_\ast}}}
\end{equation} 

Applying the dot products with the known vectors, the relation in the stellar frame can be written. Using $i$ as the inclination between the stellar rotation axis and the line of sight and $\phi$ as the orbital phase of the cloud, the relation can be written as:

\begin{equation}
\lambda_\ast=\lambda \left( \frac{1-\beta_r}{1-\beta_r\sin i\cos\phi+\beta_\phi\sin i\sin\phi} \right)
\end{equation}

Note that each of the velocities ($\beta_r, \beta_\phi$) are dependent on the star-particle distance (r). In order to put this in to the stellar source function, which can have a range of intensities, one must define a Doppler-redistribution function. Using the line center wavelength ($\lambda_0$) and the emission/absorption line with as $\lambda_{wid}$, the redistribution function is written as:

\begin{equation}
A(r,\phi,\lambda)^2=A^2=\left( \frac{\lambda_\ast-\lambda_0}{\lambda_{wid}} \right)
\end{equation}

This will let us write the wavelength-dependent stellar source and scattered intensities in integrable form. Using the doppler-shifts computed for $\lambda_\ast$ the doppler-redistribution function is:

\begin{equation}
A=\frac{ (1-\beta_r)\lambda - (1-\beta_r\sin i\cos\phi + \beta_\phi\sin i\sin\phi)\lambda_0}{(1-\beta_r\sin i\cos\phi+\beta_\phi\sin i\sin\phi)\lambda_{wid}}
\end{equation}

Now there is a way of describing ${\bf I(\lambda)}$ and one can compute the polarized flux. As an example, consider an edge-on system. In this thin equatorial disk case, all scattering events will be in the the stellar frame since all scattering plane normal vectors will align with the stellar rotation axis. Using sin(i)=1 one can compute the redistribution function for forward scattering at an orbital phase of $\phi$=0. With forward scattering, all sin terms are zero leading to a cancellation of all $\beta$ terms.

\begin{equation}
A=\frac{ \lambda - \lambda_0}{\lambda_{wid}}
\end{equation}

This states that there is no wavelength shift of the photons in the forward scattering case, as is expected. It also nicely shows how the redistribution function is really a normalized shift parameter - it gives the doppler shift in units of the emission line width, $\lambda_{wid}$. Using the 90$^\circ$ scattering case at an orbital phase of $\phi$=90 and sin(i)=1, the redistribution function simplifies to:

\begin{equation}
A=\frac{ (1-\beta_r)\lambda - (1+ \beta_\phi)\lambda_0}{(1+\beta_\phi)\lambda_{wid}}
\end{equation}

In this situation, radial movement of material with respect to the star, through $\beta_r$, will contribute a shift to the stellar spectrum incident on the cloud ($\lambda$). Orbital motion, $\beta_\phi$, will produce a shift in the light scattered to the line of sight through the clouds motion with respect to the telescope.

For a face-on disk, sin(i)=0 and only the radial motion will induce a doppler shift since all orbital motion will be tangential to both the star and the line-of-sight. The redistribution function is:

\begin{equation}
A=\frac{ (1-\beta_r)\lambda - \lambda_0}{\lambda_{wid}}
\end{equation}

	The phase matrix is now wavelength dependent because of the relative velocities of the source, particle, and observer. To account for this effect, you must calculate how much light is scattered into the wavelength of interest ($\lambda$) at the particle from the original wavelength at the photosphere ($\lambda_\ast$) given the geometry ($\hat{{\bf k_\ast}}$ and  $\hat{{\bf k}}$) and the motion of the scattering particle ($v\hat{{\bf v}}$). For simplicity, ignore the rotation of the star because the doppler shift caused by the rotation of the star at each point on the photosphere is small compared to the orbital/wind velocities. This is not entirely justified as the rotational velocity can be of order 200km/sec for an A star, but this will be ignored for now to keep the calculation simple. The wavelength-dependent version of the equation for the Stokes emission coefficients is:

\begin{equation}
{\bf j_\lambda}=\int_{\Omega_\ast} \kappa_\lambda \overline{{\bf R}}(\lambda_\ast,\hat{{\bf k_\ast}},\lambda,\hat{{\bf k}}) \bullet {\bf I_{\lambda_\ast}}  d\Omega_\ast
\end{equation}

\subsection{Integrating Over The Circumstellar Material}

	Once one knows how to do a basic polarized scattering problem at a single scattering angle, an integration of the scattered light over all the photospheric sources to the particle must be done. If one introduces the opacity of the material, $\kappa$, the Stokes emission coefficients become an integral over all the incoming radiation sources, $d\Omega_\ast$, i.e. all points on the photosphere: 

\begin{equation}
{\bf j}=\int_{\Omega_\ast} \kappa \overline{{\bf R}} \bullet {\bf I} d\Omega_\ast
\end{equation}

It has been implicitly assumed in the integration that the scattering phase matricies are all rotated into a single frame. Integrating through the disk along a line of sight (l) to the the polarization at one telescope pointing gives:

\begin{equation}
{\bf I}=\int_{l}\rho{\bf j}dl = \int_{l}\int_{\Omega_\ast} \rho \kappa \overline{{\bf R}} \bullet {\bf I} d\Omega_\ast dl
\end{equation}

Once one knows how to get the intensity integrated through the disk, the polarized flux, {\bf F}, that the telescope will recieve by integrating over all telescope pointings can be calculated. This will be the total flux at a single wavelength for the unresolved disk:

\begin{equation}
{\bf F}=\int_{Disk}{\bf I} d\Omega = \int_\Omega\int_{l}\rho{\bf j} dl d\Omega 
\end{equation}

\begin{equation}
{\bf F}=\int_\Omega\int_{l}\int_{\Omega_\ast} \rho \kappa \overline{{\bf R}} \bullet {\bf I} d\Omega_\ast dl d\Omega
\end{equation}
 
Now, $\Omega$ is the solid angle of the disk as seen by the telescope so that the integration across $\Omega$ is an integration across all telescope pointings. Typically disks are unresolved so that this integration calculates the total polarization at each wavelength for the entire disk, as would be seen in an unresolved image. The integration through the disk, l, counts up all photons in the line of sight at a single pointing. By noting that the integral over $dl d\Omega$ is a volume integral across the disk, and noting that astronomical sources are very distant such that the distances between points in the disk is much smaller than the distance to the source (small angle approx) one gets:

\begin{equation}
{\bf F}=\frac{1}{D^2}\int_{Vol}\rho{\bf j} dV
\end{equation}

This is intuitive: the polarized flux you recieve is just the summation of all the individual fluxes over the volume of the disk. The density of the material is contained in $\rho$ and the scattering physics is in {\bf j}.

\subsection{Single Clump Scattering}

At this point it is useful to start by considering a very simple example of scattering by a single clump of material at some location away from a star assuming all it's properties are known. In this example, most of the formalism will be ignored. This may be an overly simplistic case but it serves to illustrate the underlying concepts. Lets assume that there is a cloud with the geometry of figure \ref{fig:scatgeo} so that the scattering angle (and orbital phase) is at 90$^\circ$. This would be the case for an edge-on equatorial disk. Let's further assume that the star is a point source to neglect the integration over d$\Omega_\ast$. The stellar source will be a Gaussian emission line with a line/continuum ratio of 3:1 and a width of 4{\AA}. Assume the cloud is small and thin so that one can also neglect the spatial integrations dl and d$\Omega$. Further assume that the cloud is moving at 200km/s towards the telescope and that this cloud scatters the stellar light towards the line of sight with an amplitude of 0.1\% of the stellar continuum. The scattering plane normal in this geometry will be aligned with the stellar axis so no rotation of the phase-matrix is necessary. 

\begin{figure}
\centering
\subfloat[Stellar and Scattered Intensity]{\label{fig:scat-intens}
\includegraphics[width=0.35\textwidth, angle=90]{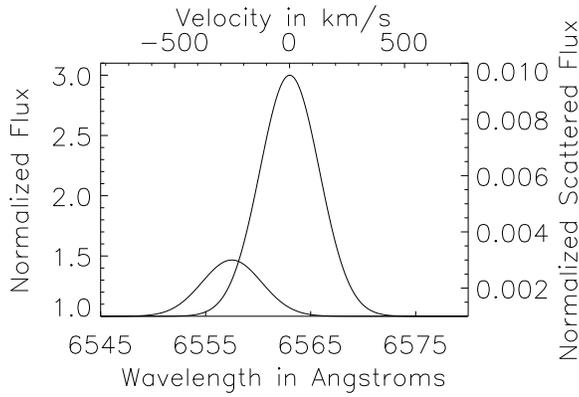}}
\quad
\subfloat[Clump Scattered Polarization]{\label{fig:scat-pol}
\includegraphics[width=0.35\textwidth, angle=90]{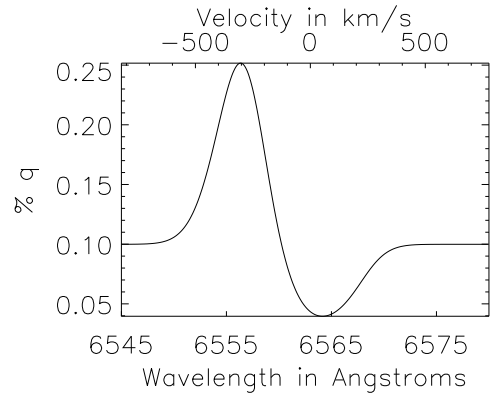}}
\caption[Single Clump Scattering Polarization]{  {\bf a)}  The continuum-normalized intensity from the stellar point source is a Gaussian with a line-continuum ratio of 3:1 and a width of 4{\AA}. The scattered flux is blue-shifted, 100\% polarized (+Q) and has an amplitude of 0.1\%. {\bf b)} The computed spectropolarimetric profile is simply the scattered light divided by the summed stellar and scattered light.}
\label{fig:clumpscat}
\end{figure}

One can easily calculate the scattered light properties. The cloud will scatter the stellar spectrum blueshifted by 200km/s with an amplitude of 0.001. This light will be 100\% polarized since it is scattered at a 90$^\circ$ angle and it will be +Q because of the choice of coordinate system (edge-on). Figure \ref{fig:scat-intens} shows the original stellar spectrum and the scattered light on a separate scale. One can compute the polarization spectrum simply as Q/I which is the scattered spectrum divided by the sum of the scattered and stellar spectra. The polarization spectrum is shown in figure \ref{fig:scat-pol}. The continuum polarization is 0.1\% because the blob scatters 0.1\% of the stellar light at all wavelengths, continuum included. Since there is a significant blue-shift of the scattered light, the maximum polarization achieved in the blue wing of the stellar line where the scattered line flux increases. In the core of the stellar emission line, there is a depolarization with respect to the 0.1\% continuum because the scattered emission had been shifted outside the line center and the original unpolarized stellar emission is able to dilute the blue-shifted scattered light.

\onecolumn

\begin{figure}
\centering
\subfloat[Disk Geometry]{\label{fig:scat-geo}
\includegraphics[width=0.38\textwidth]{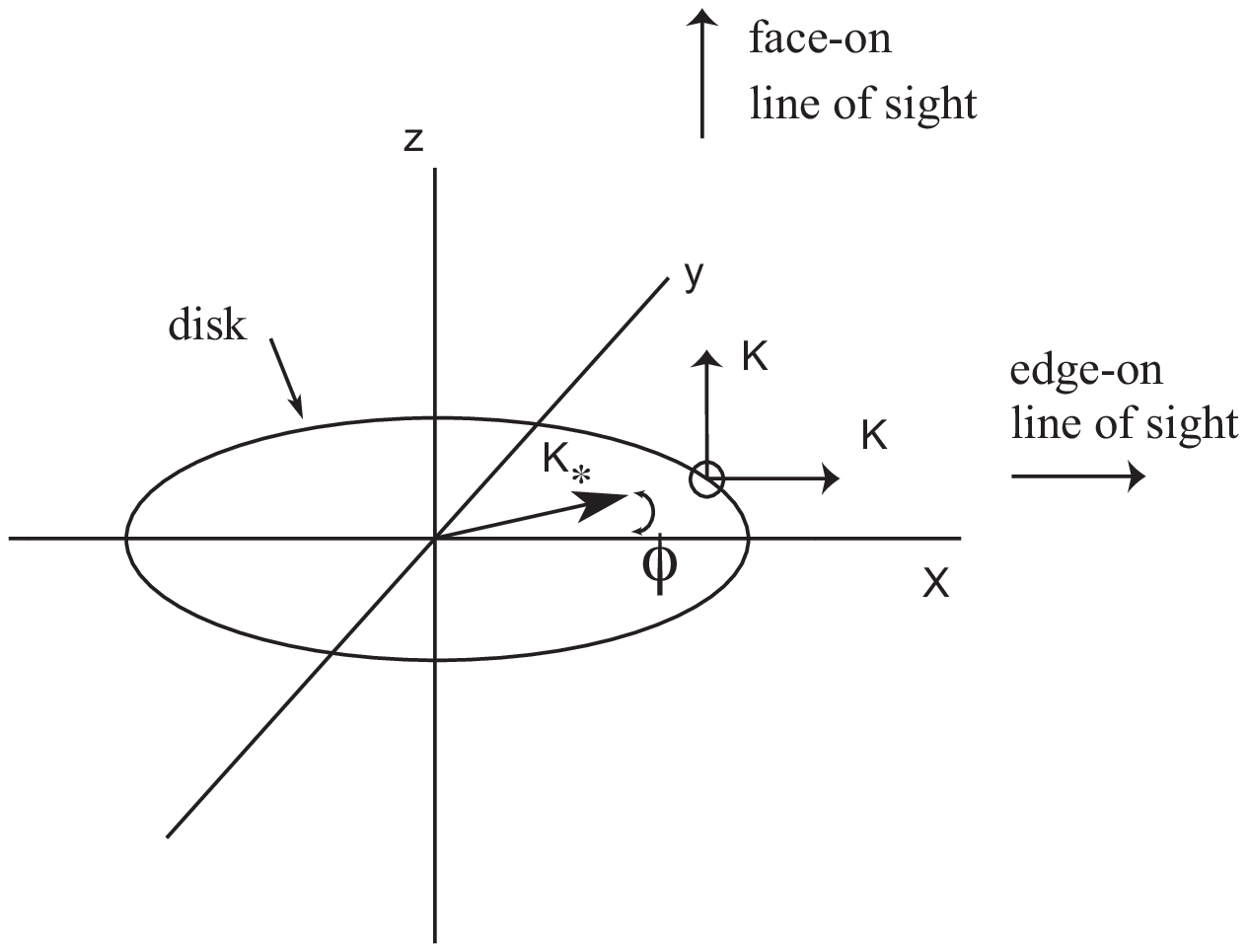}}
\quad
\subfloat[Face-on Disk Geometry]{\label{fig:scat-face-geo}
\includegraphics[width=0.38\textwidth]{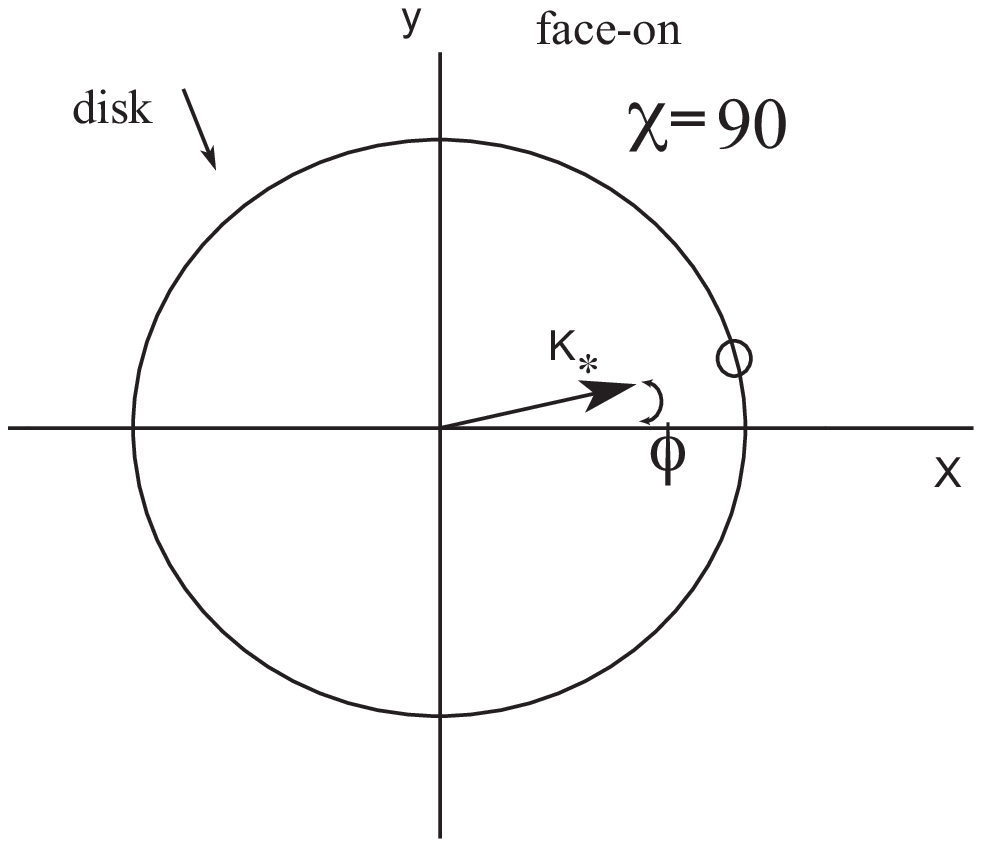}}
\caption[Disk Geometry]{  {\bf a)}  The coordinate system used in this example.  {\bf b)}  Face-on geometry.}
\label{fig:diskgeo}
\end{figure}
\twocolumn

This example shows the main effects of circumstellar scattering for a single cloud of material. These clouds contribute to the continuum polarization as well as polarization effects across a line. The task now is to integrate the effects of these individual parcels over various geometries and distributions to calculate polarized spectra for more realistic situations. It should be pointed out that this clump scattering idea has been explored for a variety of clumpy-wind scenarios in hot stars for line effects and continuum polarization (cf. Harries 2000, Davies et al. 2007).

\subsection{Thin-ring Electron Scattering Example}

An extension of this simple case is to consider a thin ring of material around a star, but this time using more of the formalism. The stellar photosphere is again an unpolarized source with the Stokes vector:

\begin{eqnarray}
{\bf I_{in}}=\left( \begin{array}{c} I_0 \\ 0 \\ 0 \\ 0 \end{array} \right)
\end{eqnarray}

We'll also assume that an emission line is the wavelength dependent stellar source (${\bf I_{in}}(\lambda_\ast)$) with a line/continuum ratio of $\alpha$, a width of $\lambda_{wid}$ and a rest wavelength $\lambda_0$: 

\begin{equation}
I(\lambda_\ast)=I_0 \left( 1+\alpha e^{-(\frac{\lambda_\ast-\lambda_0}{\lambda_{wid}})^2} \right)
\end{equation}

Since we've assumed that the scattering will be dominated by electron scattering $\rho\kappa=n\sigma_T$. Starting with the Stokes emission coefficients as an integration over stellar photosphere sources, one can write the coefficients as:

\begin{equation}
\rho{\bf j}=\int_{\Omega_\ast}\rho\kappa\overline{{\bf R}} \bullet {\bf I}d\Omega_\ast
\end{equation}

This example uses coordinates where the +Q axis coincides with the stellar rotational axis. The disk will be thin and equatorial so that the angle between the stellar pole and the scattering particles is always $\Theta=\frac{\pi}{2}$. The scattering particles location can then be described by a radius from the center of the star, r, and an orbital phase, $\phi$, with $\phi=0$ corresponding to the direction of the telescope. You can write the star-particle and particle-telescope vectors in a cartesian coordinate system as:

\begin{equation}
\hat{{\bf k_\ast}}=(\cos\phi, \sin\phi, 0)
\end{equation}

\begin{equation}
\hat{{\bf k}}=(\sin i, 0, \cos i)
\end{equation}

The telescope reference frame, shown in figure \ref{fig:scat-geo}, lies in the xz-plane and the line-of-sight, $\hat{{\bf k}}$ is set by the inclination. The equatorial disk is in the xy-plane and the coordinate of the scattering cloud as seen by the star, $\hat{{\bf k_\ast}}$, only depends on the orbital phase, $\phi$. Given the disk's inclination angle (i) to the line of sight ($\hat{{\bf k}}$), you can write a relation between the scattering angle ($\chi$) and the angle between +Q and the normal to the scattering plane ($\psi$):

\begin{equation}
\cos\chi=\sin i \cos\phi
\end{equation}

\begin{equation}
\sin\chi\cos\psi=\sin\phi
\end{equation}

\begin{equation}
\sin\chi\sin\psi=-\cos i \cos\phi
\end{equation}

The relation will allow you to relate the bulk disk and particle properties (inclination, particle orbit, doppler-shift) to scattering properties (scattering angle, polarization orientation, wavelength-dependence).  For example, in a face-on disk, the scattering angle is always $\chi=90^\circ$ since $\sin i=0$ and the scattered light is always 100\% polarized. The orbital phase, $\phi$, determines the position angle of polarization. Similarly, in an edge-on disk, the scattering angle is the orbital phase, $\chi=\phi$ because $sini=1$. Figure \ref{fig:diskgeo} shows a face-on disk example.

The doppler shifts can also be broken into the radial/wind ($\beta_r$) and azimuthal/disk ($\beta_\phi$) components. Since the disk is thin and vertical motion is being neglected, the doppler-shift components will only lie in the xy-plane of figure \ref{fig:diskgeo}. The doppler-shift vector in this coordinate system is:

\begin{equation}
{\bf \beta}=(\beta_r\cos\phi - \beta_\phi sin\phi,   \beta_r\sin\phi+\beta_\phi\cos\phi,   0)
\end{equation}

\subsection{60$^\circ$ Inclined Ring Example}

To illustrate actual solutions, we'll pick a ring at an inclination of 60$^\circ$. Using the relations between the scattering angle, $\chi$, orbital phase, $\phi$, and scattering-plane normal rotation angle, $\psi$, one can solve for the scattering and rotation angles as a function of a cloud's orbital phase:   

\begin{equation}
\cos\chi=\frac{\sqrt{3}}{2}\cos\phi, \sin\chi\cos\psi=\sin\phi, \sin\chi\sin\psi=\frac{1}{2}\cos\phi
\end{equation}

\begin{equation}
\chi=\cos^{-1}(\frac{\sqrt{3}}{2}\cos\phi)
\end{equation}

\begin{equation}
\psi=\tan^{-1}\frac{1}{2tan\phi}
\end{equation}

\begin{figure}
\centering
\subfloat[$\psi$ vs $\phi$]{\label{fig:60ring-rang}
\includegraphics[width=0.35\textwidth, angle=90]{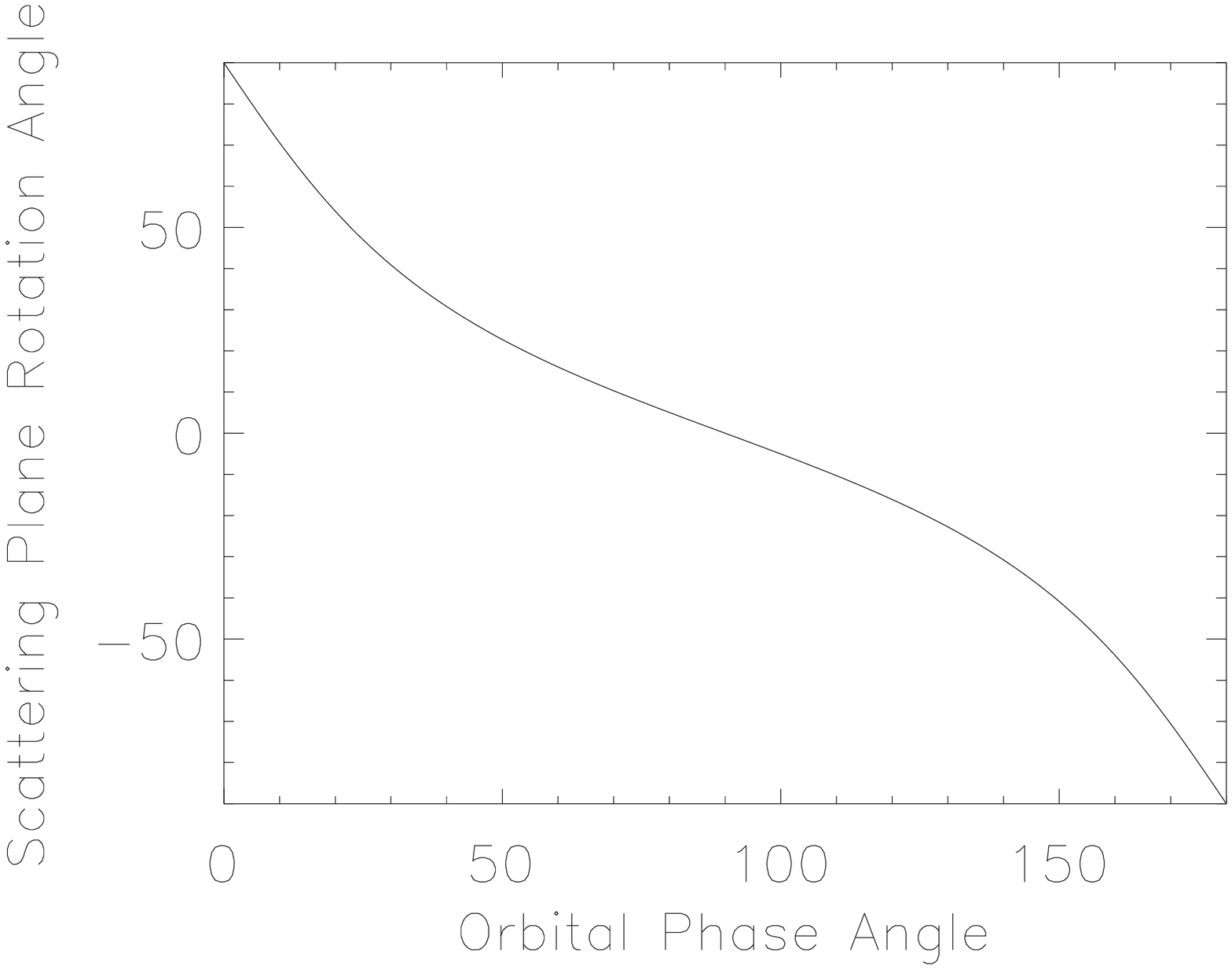}}
\quad
\subfloat[$\chi$ vs $\phi$]{\label{fig:60ring-ang}
\includegraphics[width=0.35\textwidth, angle=90]{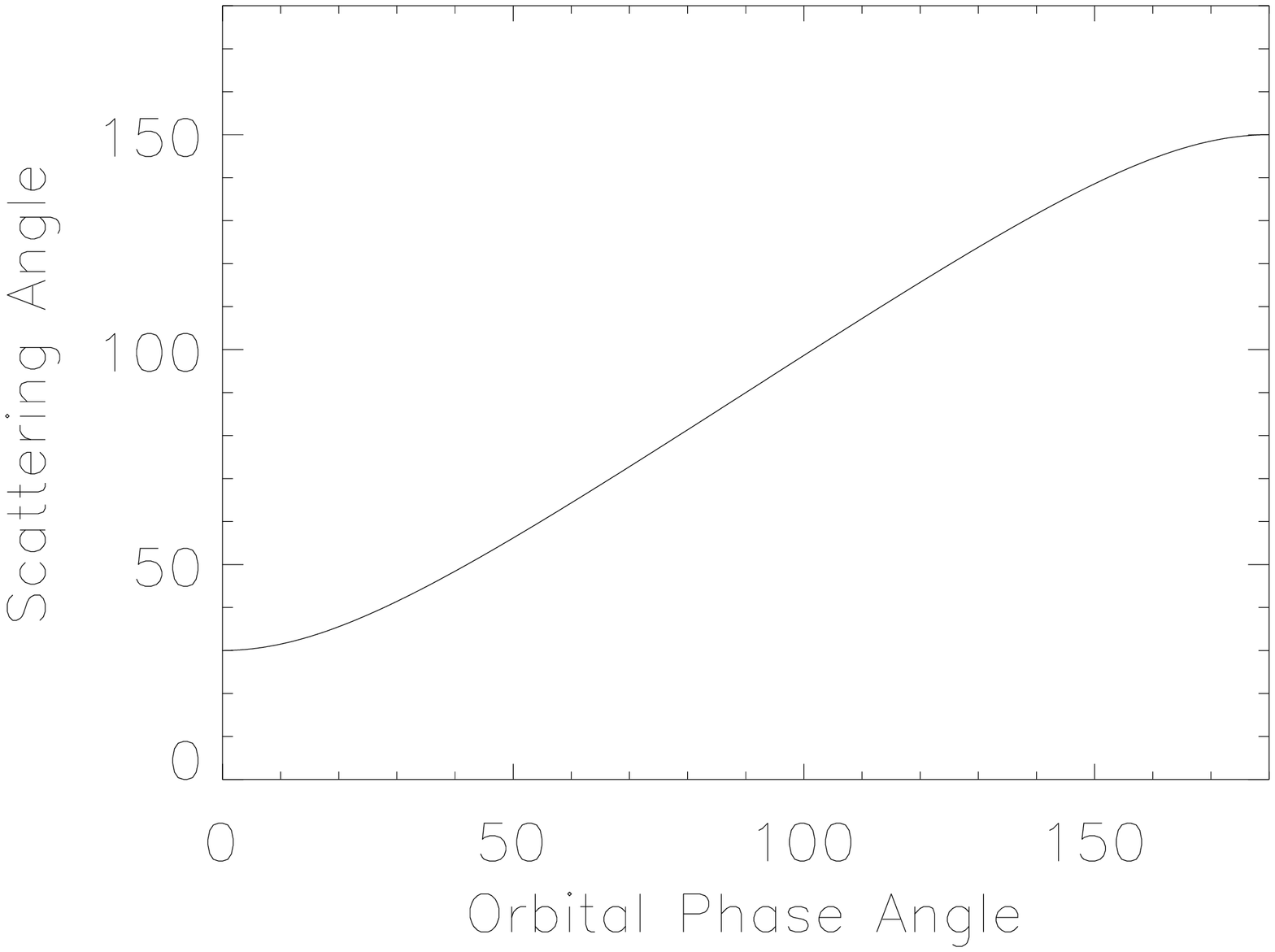}}
\quad
\subfloat[\%q vs $\phi$]{\label{fig:60ring-pq}
\includegraphics[width=0.35\textwidth, angle=90]{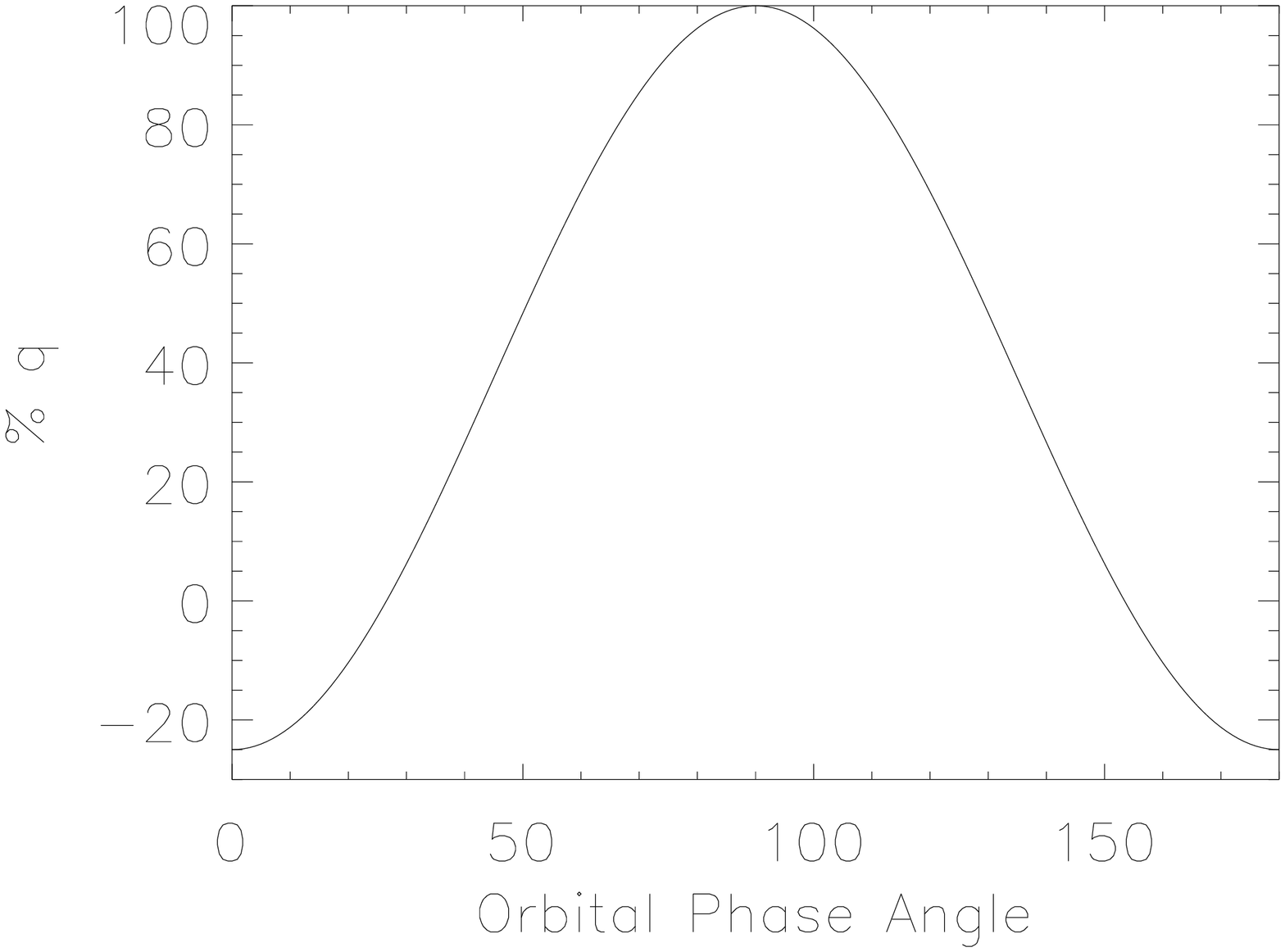}}
\quad
\subfloat[\%u vs $\phi$]{\label{fig:60ring-pu}
\includegraphics[width=0.35\textwidth, angle=90]{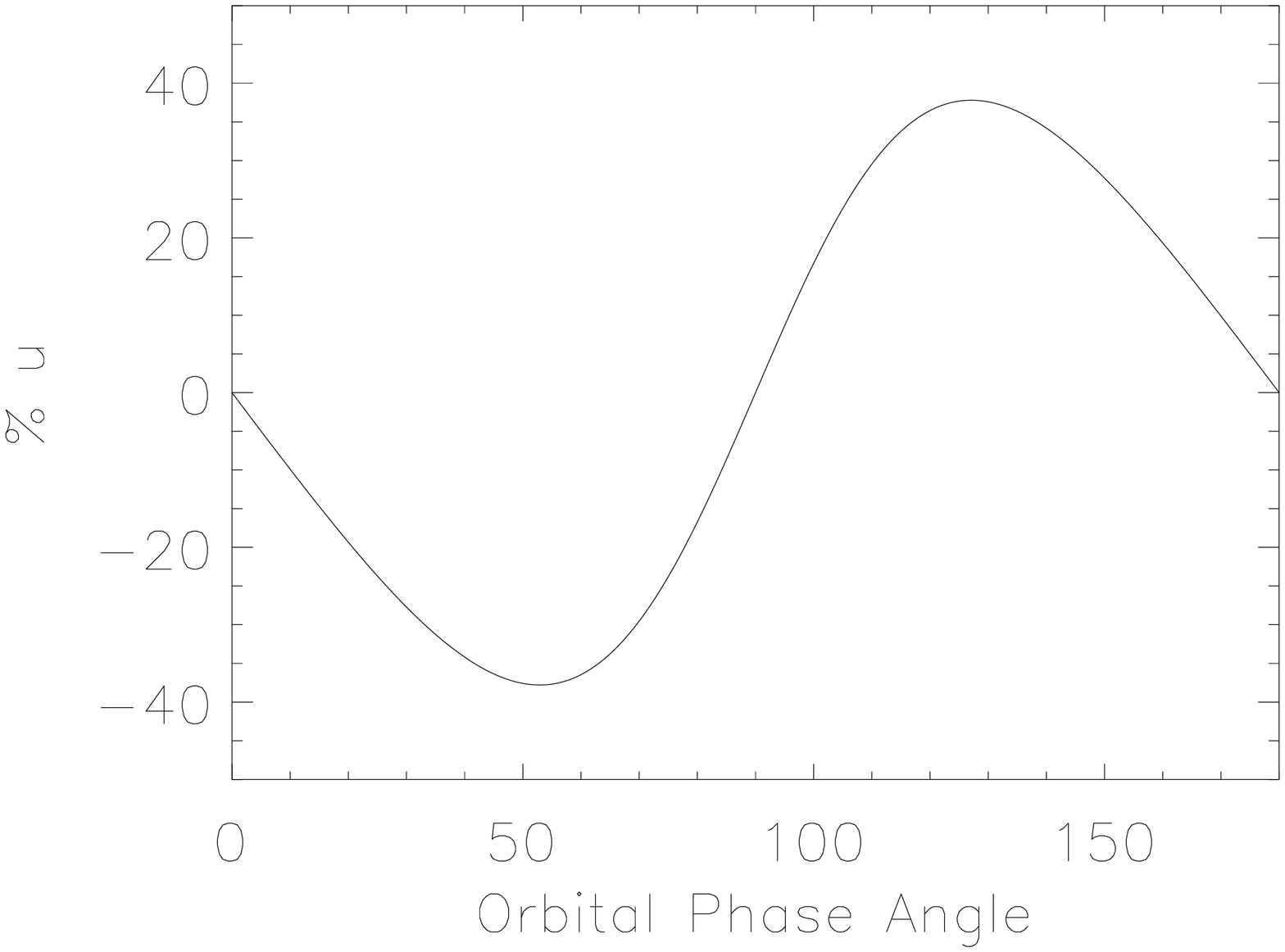}}
\quad
\subfloat[\% Polarization vs $\phi$]{\label{fig:60ring-pp}
\includegraphics[width=0.35\textwidth, angle=90]{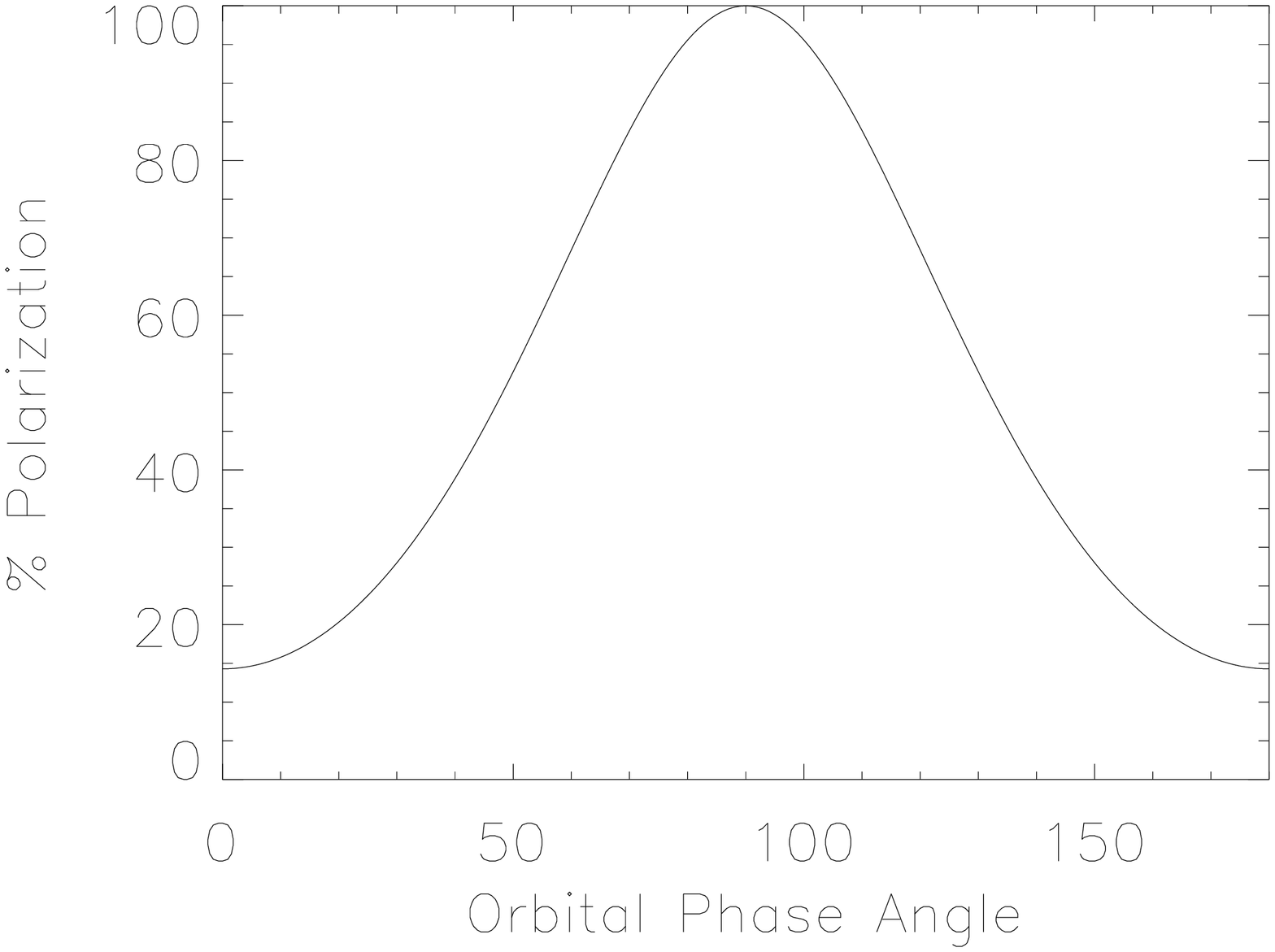}}
\quad
\subfloat[$\theta$ vs $\phi$]{\label{fig:60ring-pa}
\includegraphics[width=0.35\textwidth, angle=90]{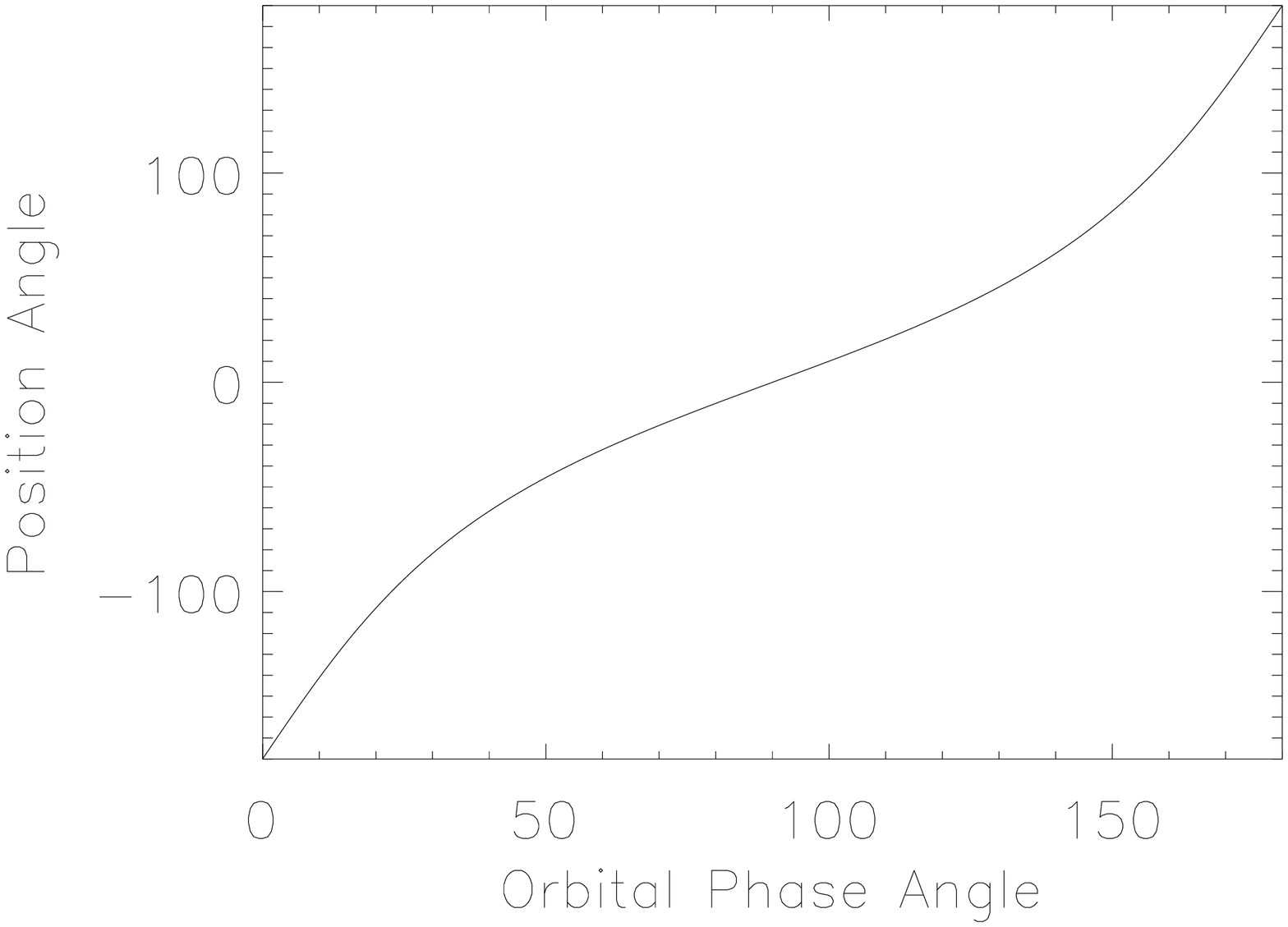}}
\caption[60$^\circ$ Inclined Ring Properties]{The properties of the coordinate system and scattered light for a 60$^\circ$ inclined ring. {\bf a)} The rotation angle between the scattering plane and the stellar rotation axis, $\psi$ as a function of orbital phase. {\bf b)} The scattering angle as a function of orbital phase. {\bf c)} \%q as a function of orbital phase. {\bf d)} \%u as a function of orbital phase. {\bf e)} The degree of polarization as a function of orbital phase.{\bf c)} The position angle of polarization as a function of orbital phase.}
\label{fig:ring60}
\end{figure}

Figures \ref{fig:60ring-rang} and \ref{fig:60ring-ang} show that the scattering angle goes from 30$^\circ$ 150$^\circ$ and the scattering plane rotates by 180 degrees as you go from the front of the disk ($\phi=0$) to the back ($\phi=180$). Since there are explicit solutions for the scattering geometries, one can solve for the polarization properties of the scattered light - the rotated phase matrix:

\onecolumn

\begin{figure}
\centering
\subfloat[A(6550{\AA}) vs $\phi$]{\label{fig:60ring-redis}
\includegraphics[width=0.35\textwidth, angle=90]{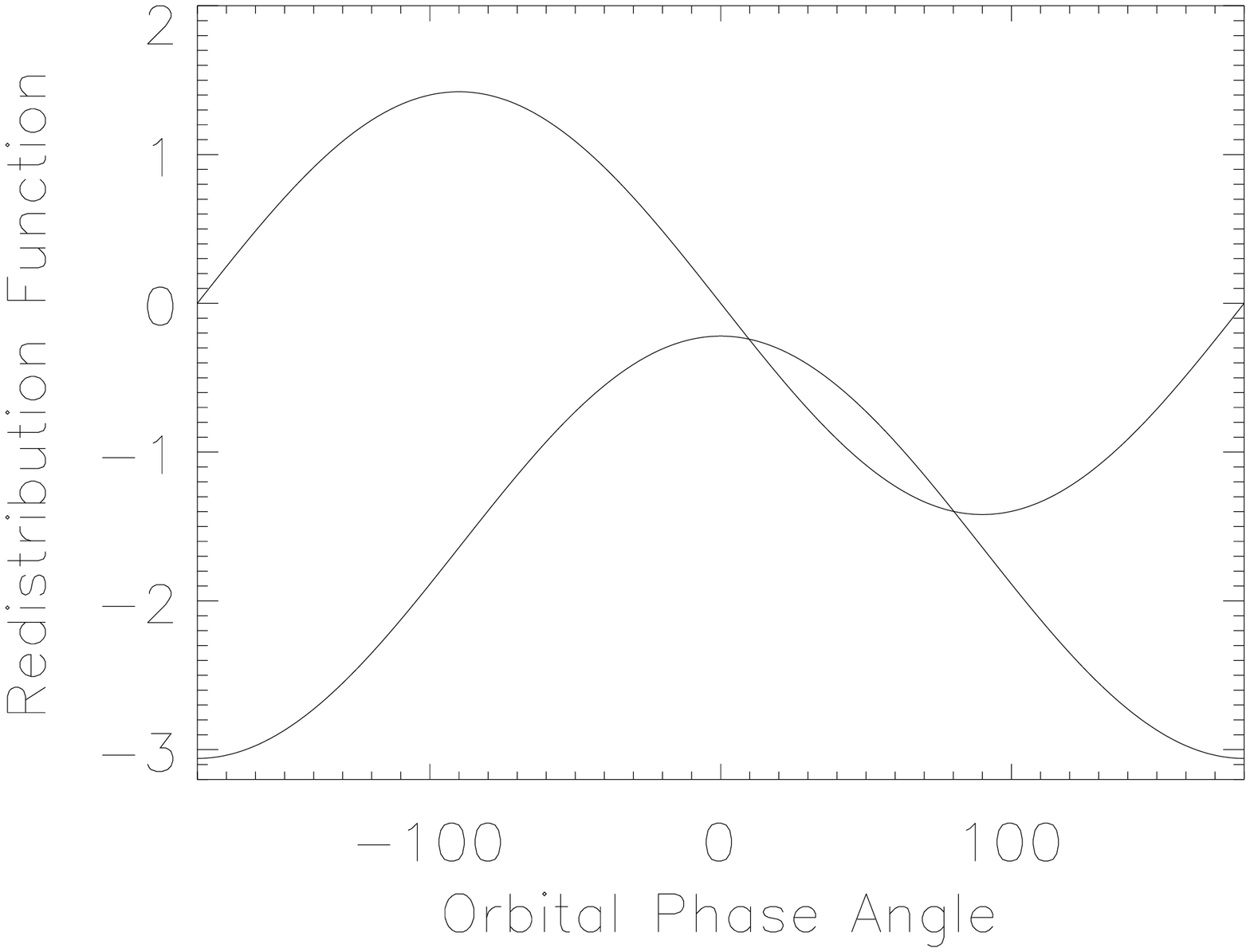}}
\quad
\subfloat[{\bf P$_{QU}$} vs $\phi$]{\label{fig:60ring-pmat}
\includegraphics[width=0.35\textwidth, angle=90]{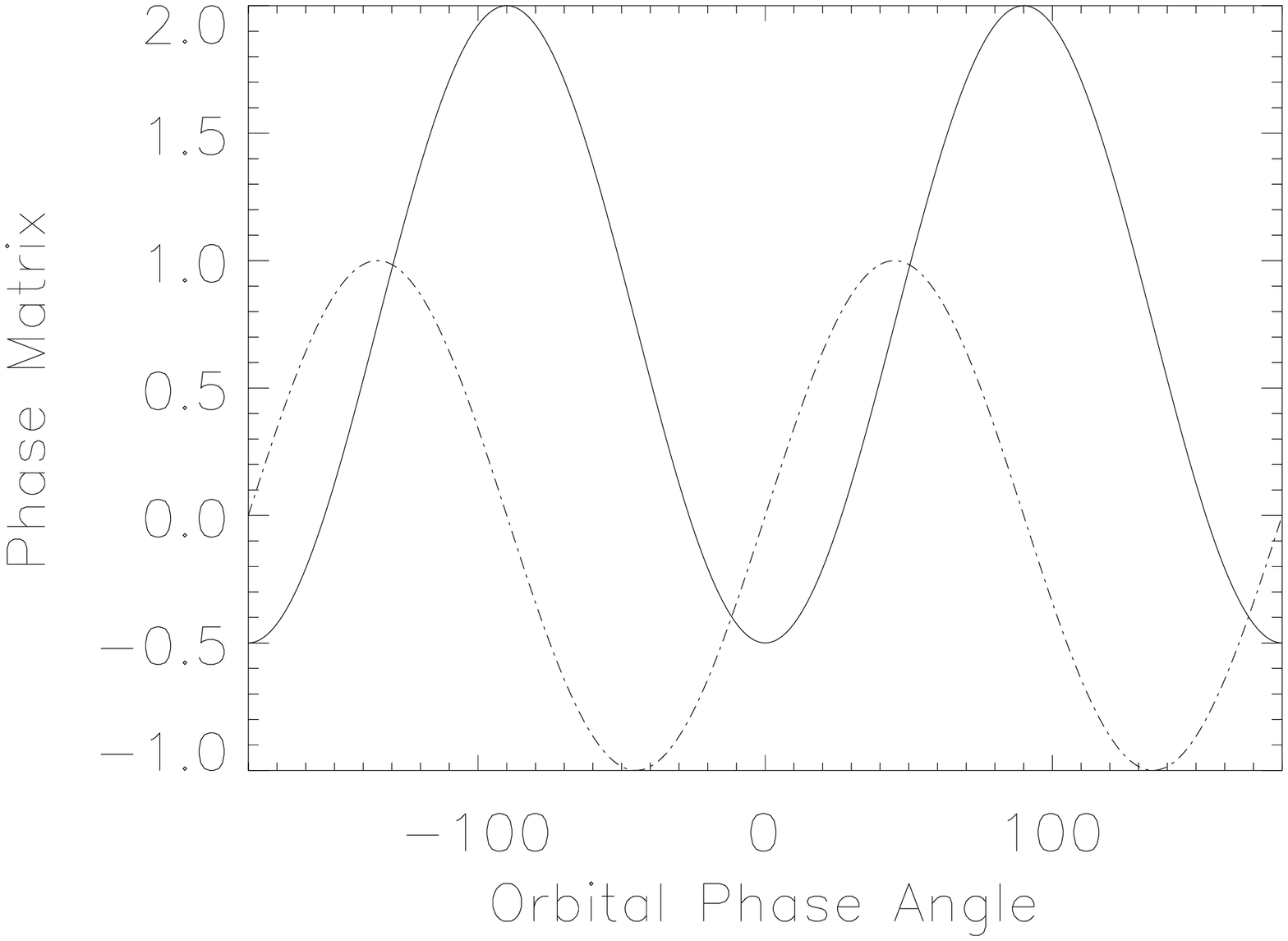}}
\quad
\subfloat[Orbital Q and U vs $\phi$ for 6550{\AA}]{\label{fig:60ring-pfxo}
\includegraphics[width=0.35\textwidth, angle=90]{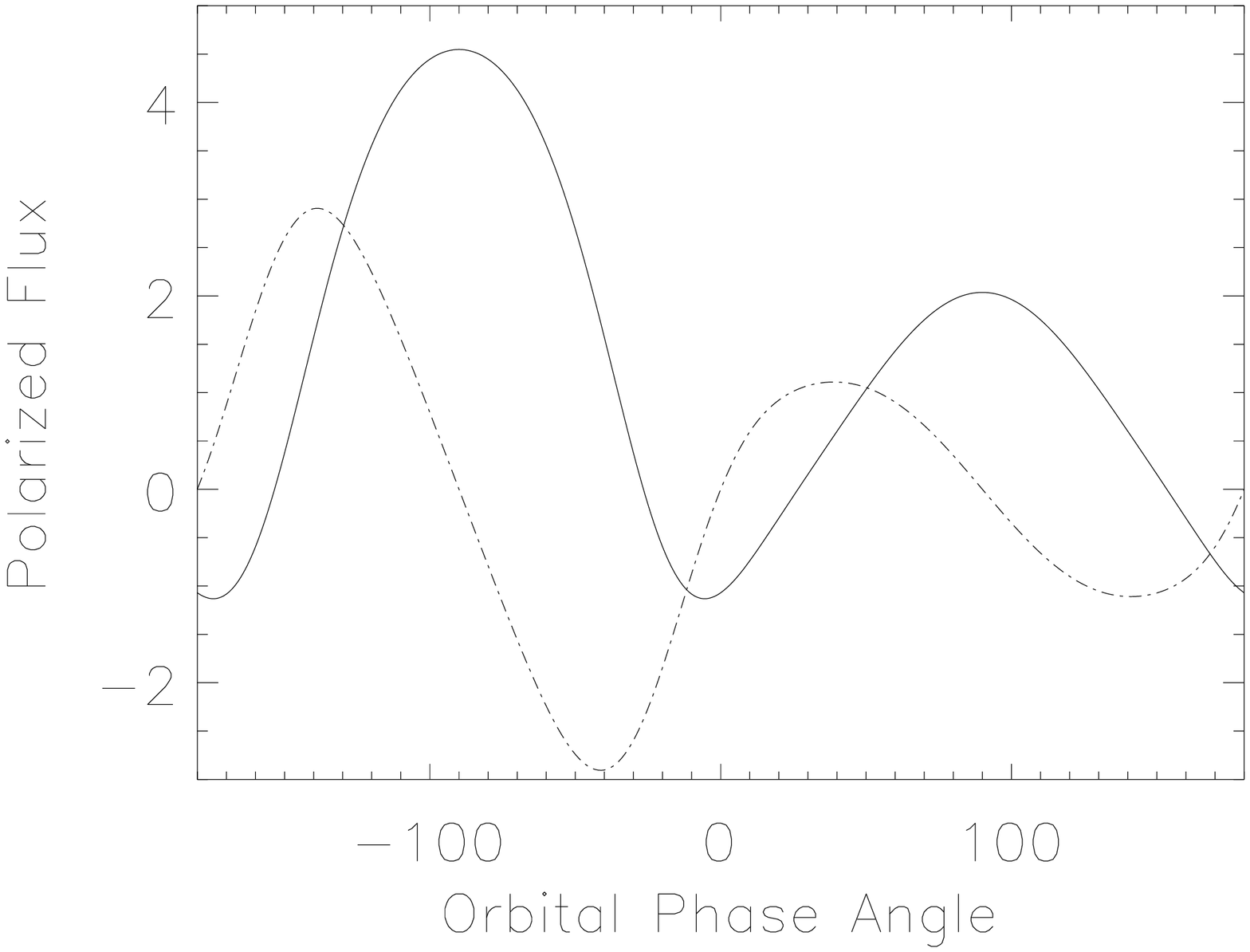}}
\quad
\subfloat[Radial Q and U vs $\phi$ for 6550{\AA}]{\label{fig:60ring-pfxr}
\includegraphics[width=0.35\textwidth, angle=90]{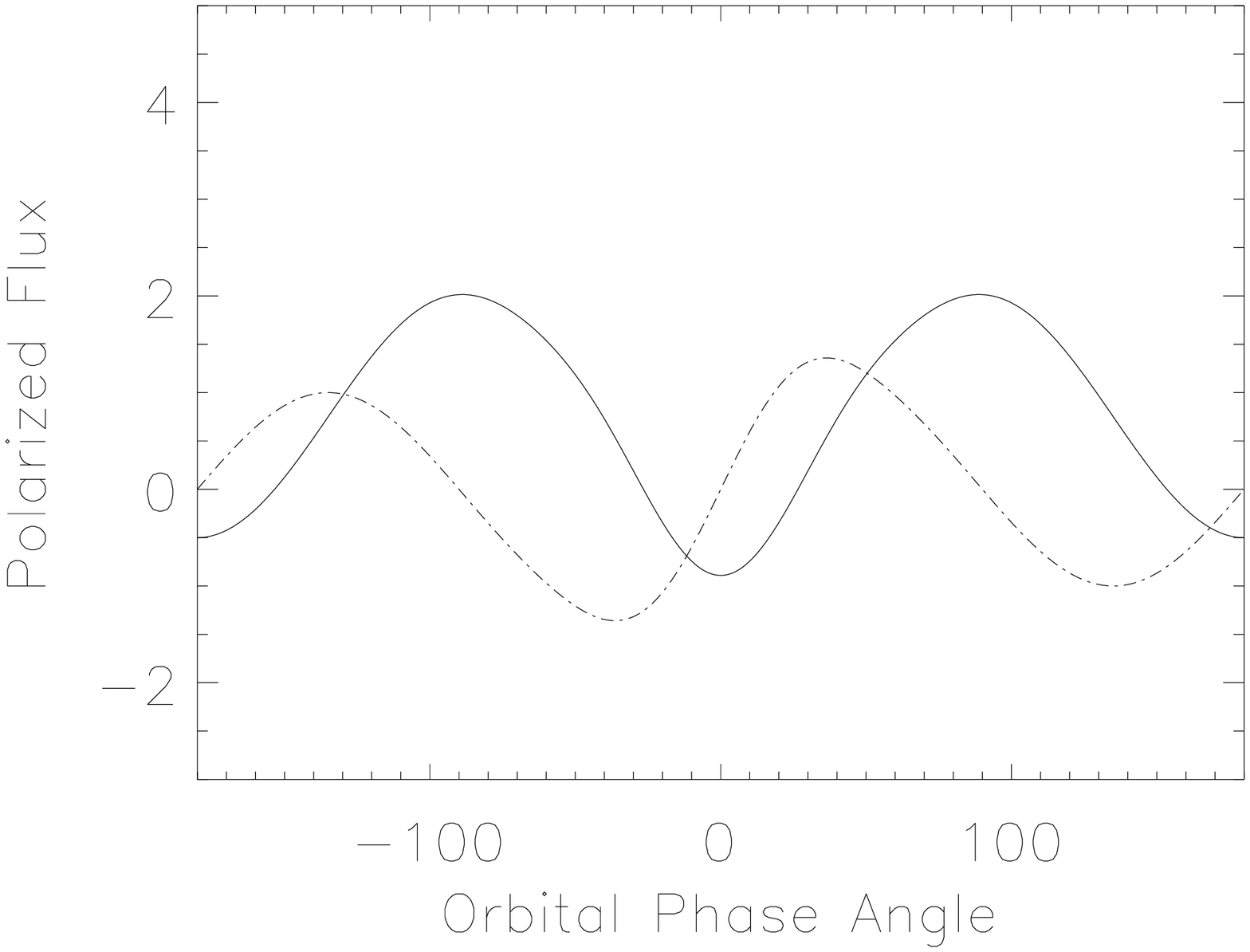}}
\quad
\subfloat[Orbital \% Polarization]{\label{fig:60ring-iquo}
\includegraphics[width=0.35\textwidth, angle=90]{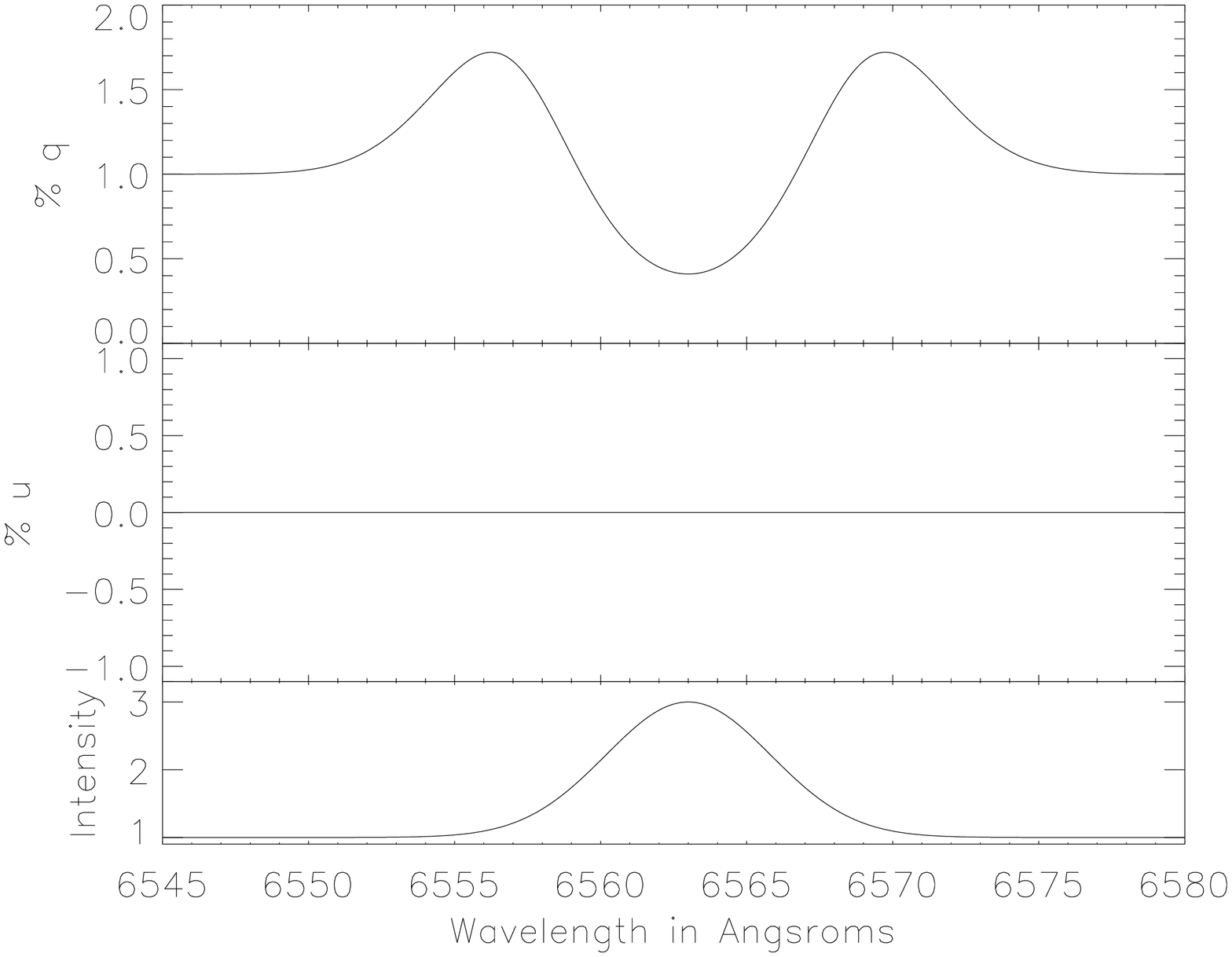}}
\quad
\subfloat[Radial \% Polarization]{\label{fig:60ring-iqur}
\includegraphics[width=0.35\textwidth, angle=90]{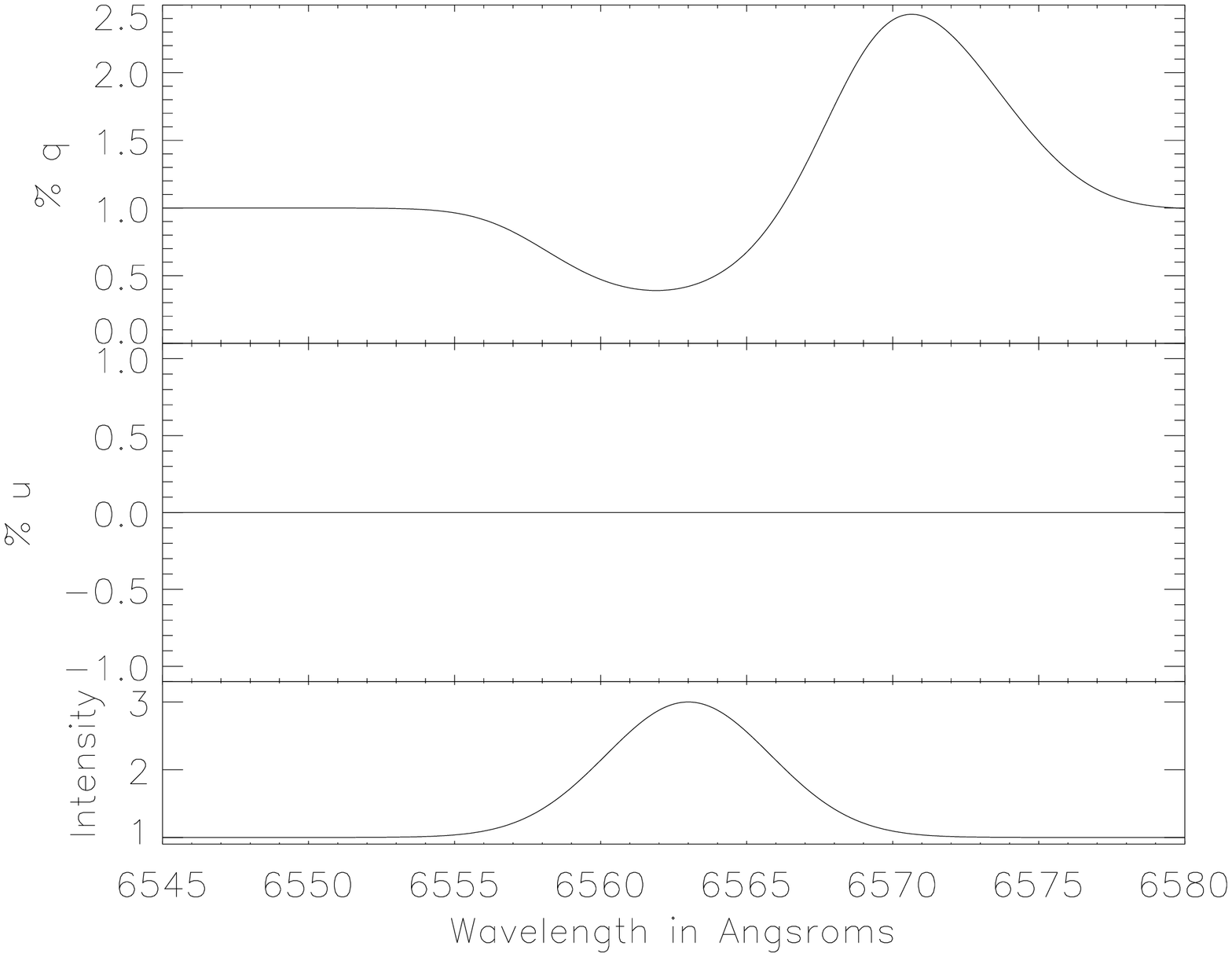}}
\caption[60$^\circ$ Inclined Ring Polarization]{The spectropolarimetric model for a 60$^\circ$ inclined ring. {\bf a)} The doppler redistribution function A at 6560{\AA} as a function of orbital phase. {\bf b)} The q and u phase matrix components as a function of orbital phase. {\bf c) \& d)} Q \& U as a function of orbital phase. {\bf e) \& f)} The resulting spectropolarimetry.}
\label{fig:ring60pol}
\end{figure}

\twocolumn

\begin{eqnarray}
{\bf P}=\frac{3I_0}{16\pi}\left ( \begin{array}{c}
           1 + \cos^2\chi  \\ \sin^2\chi\cos2\psi \\ -\sin^2\chi\sin2\psi \\ 0 
        \end{array} \right )        
\end{eqnarray}

The scattered light is $\sim$20\% polarized in the -q direction at orbital phase of zero and increases to 100\% +q polarization at 90$^\circ$ orbital phase (and scattering angle) as shown in figure \ref{fig:60ring-pq}. Figure \ref{fig:60ring-pu} shows the more complex structure of Stokes u. This parameter is maximized when the scattering plane is rotated by $\pm$22.5$^\circ$ which occurs at intermediate orbital phases for this inclined system. The corresponding degree of polarization and position-angle are shown in figures \ref{fig:60ring-pp} and \ref{fig:60ring-pa}. They show maximum polarization at 90$^\circ$ orbital phase angles (and 90$^\circ$ scattering angle) with the position angle of polarization at 0$^\circ$ (+q).

If one assumes this ring is only moving radially, in stellar winds or accretion, the redistribution function is:

\begin{equation}
A=\frac{ (1-\beta_r)\lambda - (1-\beta_r\sin i\cos\phi)\lambda_0}{(1-\beta_r\sin i\cos\phi)\lambda_{wid}}
\end{equation}

And conversely, if the ring is only orbiting, the redistribution function is:

\begin{equation}
A=\frac{\lambda - (1-\beta_\phi\sin i\sin\phi)\lambda_0}{(1+\beta_\phi\sin i\sin\phi)\lambda_{wid}}
\end{equation}

In the stellar point-source approximation, the total flux (direct flux from the star) can be estimated as the distance-modulated stellar source. The spatial integration across the disk is simply the stellar emission line divided by the distance to the telescope:

\begin{equation}
F_\lambda=\int_\Omega I_\lambda d\Omega=\frac{\pi R^2}{D^2}I_\lambda 
\end{equation}

This is the just total intensity at each wavelength ($F_I$). This approximation also ignores the scattered light contributed to the total flux, but this is typically very small and will be neglected for simplicity. This coefficient, $F_I$,  will allow us to normalize the computed Stokes vector to get fractional polarizations. Since the Thompson-scattering phase-matrix produces no circular polarization (Stokes V) and it has been assumed that there is no circular polarization coming from the star, $F_V$ is set to zero. The normalized Stokes vector is:

\begin{footnotesize}
\begin{eqnarray}
\frac{{\bf F_\lambda}}{F_I}=\frac{1}{F_I} \left( \begin{array}{c} F_I \\ F_Q \\ F_U \\ F_V \end{array} \right) = 
   \left( \begin{array}{c} 1 \\ F_Q/F_I \\  F_U/F_I \\ 0 \end{array} \right) = 
   \left( \begin{array}{c} 1 \\ Q/I \\ U/I \\ 0 \end{array} \right) = 
   \left( \begin{array}{c} 1 \\ q \\ u \\ 0 \end{array} \right)
\end{eqnarray}
\end{footnotesize}

Normalizing the computed Stokes emission from before and noting that the $R^2$ and $D^2$ terms cancel. In the point-source approximation where you treat the photosphere as a single point, the integration over $d\Omega_\ast$ reduces to a $\frac{1}{r^2}$ function:

\begin{equation}
{\bf F}= \frac{3\sigma_T}{16\pi D^2}\int_V n I_0{\bf P} \frac{\pi R^2}{r^2} dV
\end{equation}

This leaves the resulting in the polarized spectra:

\begin{equation}
\frac{{\bf F_\lambda}}{F_{I_\lambda}}= \frac{1}{F_{I_\lambda}}\frac{3\sigma_T}{16\pi D^2}\int_V n I_{\lambda_\ast} \pi R^2 {\bf P}\frac{dV}{r^2}=\frac{3\sigma_T}{16\pi I_\lambda} \int_V n I_{\lambda_\ast} {\bf P} \frac{dV}{r^2}
\end{equation}

We know that the stellar source is just an emission line ($I_\nu$). The redistributed scattered light ($I_{\nu_\ast}$) is also an emission line but with a modified exponent ($A(r,\phi,\lambda)^2$) and a polarization controlled by the phase matrix {\bf P}. If you define $G=\frac{\lambda_\ast-\lambda_0}{\lambda_{wid}}$ then you can rewrite this equation in terms of the emission line parameters and the redistribution function:

\begin{equation}
{\bf F_\lambda}=\frac{3\sigma_T}{16\pi} (1+\alpha e^{-G^2})^{-1} \int_V (1+\alpha e^{-A^2}) {\bf P} \frac{dV}{r^2}
\end{equation}

The flux normalization is done with the original unshifted emission line $(1+\alpha e^{-G^2})$ but the scattered flux is redistributed by A, $(1+\alpha e^{-A^2})$, and polarized {\bf P}. Through the geometric relations between ($\chi,\psi$) and ($i,\phi$) one can rewrite the Q and U coefficients of the scattering matrix {\bf P}:

\begin{eqnarray*}
{\bf P_{QU}} & = & \frac{3I_0}{16\pi}
        \left ( 
        \begin{array}{c}
        \sin^2\chi\cos2\psi \\ 
       -\sin^2\chi\sin2\psi
        \end{array} 
         \right )     \\ & = & 
        \left ( 
        \begin{array}{c}
       \sin^2i-(1+\cos^2i)\cos2\phi \\ 
       2\cos i\sin2\phi
        \end{array} 
         \right )                
\end{eqnarray*}

We can then write the general expression for the normalized Stokes vectors of the disk at a single wavelength as the integrals across the disk:

\begin{footnotesize}
\begin{equation}
q_\lambda=\frac{P_0}{1+\alpha e^{-G^2}}\int_V (1+\alpha e^{-A^2})f(r)\left[\sin^2i-(1+\cos^2i)\cos2\phi \right] \frac{dV}{r^2}
\end{equation}
\end{footnotesize}

\begin{equation}
u_\lambda=\frac{P_0}{1+\alpha e^{-G^2}}\int_V (1+\alpha e^{-A^2})f(r)\left[2\cos i\sin2\phi \right] \frac{dV}{r^2}
\end{equation}

The f(r) and 1/$r^2$ terms are left in place to reflect the radial dependence of the redistribution function and source intensity respectively. Since the current example is a ring at a fixed distance, this volume integral is simplified to a line integral along the ring at all orbital phases. Absorbing coefficients into P$_0$ gives the polarization integrals:

\begin{equation}
q_\lambda=\frac{P_0}{1+\alpha e^{-G^2}}\int_\phi (1+\alpha e^{-A^2})\left[\sin^2i-(1+\cos^2i)\cos2\phi \right] d\phi
\end{equation}

\begin{equation}
u_\lambda=\frac{P_0}{1+\alpha e^{-G^2}}\int_\phi (1+\alpha e^{-A^2})\left[2\cos i\sin2\phi \right] d\phi
\end{equation}

The results of a numerical solution are shown in figure \ref{fig:ring60pol}. The parameters $\alpha=2$ and either $\beta_r=0.001$ or $\beta_\phi=0.001$ were used as inputs. The redistribution function at a wavelength of 6550{\AA} is shown for the radial and orbital cases in figure \ref{fig:60ring-redis}. The orbital motion produces a sinusoidal curve in orbital phase since one side of the disk is blue-shifted while the other is red-shifted. In the wind case, the redistribution function is always negative. It is near zero at zero orbital phase, but the scattering angle is 30$^\circ$ as was shown in figure \ref{fig:60ring-ang}. This means that the red-shift of the star as seen by the cloud is not completely canceled by the blue-shift of the cloud with respect to the line-of-sight. At the rear of the disk, $\phi$=180, there are two red-shifts leading to a very negative A value. The phase matrix components, shown in figure \ref{fig:60ring-pmat}, show the asymmetry and 22.5$^\circ$ phase shift. The polarized flux at 6550{\AA}, normalized arbitrarily, is shown for orbital motion in figure \ref{fig:60ring-pfxo} and for radial motion in figure \ref{fig:60ring-pfxr}. In the orbital case, the approaching side of the disk dominates the contribution to the polarized flux as the blue-shifted stellar emission line will contribute substantially more polarized flux. In the radial-motion case, there are basically equal contributions to the polarized flux from both sides of the disk because of the symmetry. The overall amplitude is lower because none of the stellar emission will be blue-shifted when scattered. The corresponding spectropolarimetry is shown for orbital motion in figure \ref{fig:60ring-iquo} and for radial motion in figure \ref{fig:60ring-iqur}. Both cases show no net u polarization because of the symmetry - every +U component is canceled by a -U component. The \%q spectra for orbital motion is symmetric about line center.  It shows that both the blue and red sides of the disk remove light from line center and produce the characteristic symmetric profile. The wind case shows only a red-shifted polarization increase with a central decrease in polarization. This arises because every cloud in a wind sees a redshifted source. Though there is a blue-shift for material moving toward the telescope, these blue and red shifts cancel in the wind case along the line of sight, leaving no doppler shift, and blue side of the spectrum remains unaffected. It should also be noted that the overall amplitude of the effect simply scales with the amount of scattered light. 

This example shows what a spectropolarimetric signature looks like for a ring of material undergoing either orbital motion or outflow. The geometry controls the polarization through the phase matrix. The orientation and type of motion controlls how the doppler shifts redistribute the stellar light in the cloud frame and how the scattered light is shifted into the telescope frame.

\subsection{ Edge-on and Face-on Rings} 

In the phase matrix and the geometric relations between ($\chi,\psi$) and ($i,\phi$) one can rewrite the Q and U coefficients of the scattering matrix {\bf P} in very simple forms. In the face-on case, the scattering phase matrix is: 

\begin{eqnarray*}
{\bf P_{QU}} & = & \left( \begin{array} {c} P_Q \\ P_U \end{array} \right) = \left( \begin{array}{c} \sin^2i-(1+\cos^2i)\cos2\phi \\ 2\cos i\sin2\phi \end{array} \right) \\ & = & \left( \begin{array}{c} 2cos2\phi \\ 2sin2\phi \end{array} \right)
\end{eqnarray*}

The Stokes emission coefficients now only have an orbital phase dependence. The polarization at any wavelength from a single ring of material will become an integral of sin or cos around a ring, which is always zero. In the face-on case with azimuthal symmetry, there is no spectropolarimetric effect.

\begin{equation}
q_ \lambda =\frac{P_0}{1+\alpha e^{-G^2}} \int_V (1+\alpha e^{-A^2})f(r)\left[ 2\cos2\phi \right] \frac{dV}{r^2}=0
\end{equation}

\begin{equation}
u_\lambda=\frac{P_0}{1+\alpha e^{-G^2}}\int_V (1+\alpha e^{-A^2})f(r)\left[2\sin2\phi \right] \frac{dV}{r^2}=0
\end{equation}

In the edge-on case, the scattering phase matrix also has a simple form: 

\begin{eqnarray*}
{\bf P_{QU}} & = & \left( \begin{array} {c} P_Q \\ P_U \end{array} \right) = \left( \begin{array}{c} \sin^2i-(1+\cos^2i)\cos2\phi \\ 2\cos i\sin2\phi \end{array} \right) \\ & = & \left( \begin{array}{c} 1-cos2\phi \\ 0 \end{array} \right)
\end{eqnarray*}

This shows that there is no Stokes u polarization from an edge on disk or wind. The phase matrix shows that for all orbital phases the scattering plane normal is aligned with the stellar rotation axis and scattered light is always $\pm$Q. The Stokes emission coefficients now become:

\begin{equation}
q_ \lambda =\frac{P_0}{1+\alpha e^{-G^2}} \int_V (1+\alpha e^{-A^2})f(r)\left[1- \cos2\phi \right] \frac{dV}{r^2}=0
\end{equation}

\begin{equation}
u_\lambda=0
\end{equation}

The orbital phase controls the degree of polarization: forward and backward scattered light is unpolarized while orbital phases of $\pm$90$^\circ$ give maximum q$_\lambda$.

\onecolumn
\begin{figure}
\centering
\subfloat[A(6550{\AA}) vs $\phi$]{\label{fig:90ring-redis}
\includegraphics[width=0.35\textwidth, angle=90]{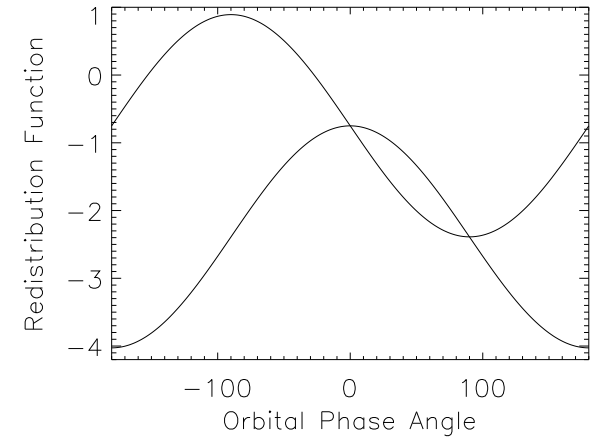}}
\quad
\subfloat[{\bf P$_{Q}$} vs $\phi$]{\label{fig:90ring-pmat}
\includegraphics[width=0.35\textwidth, angle=90]{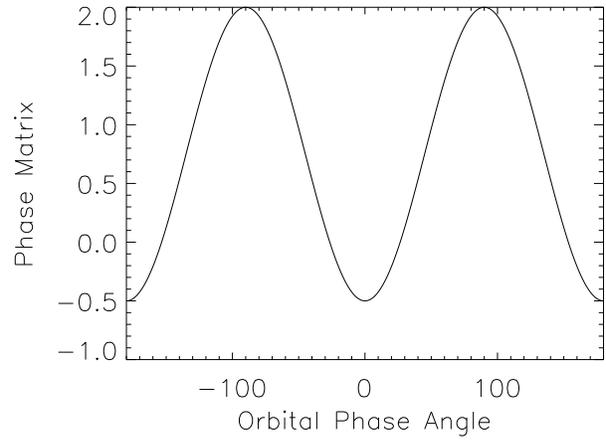}}
\quad
\subfloat[Orbital \% Polarization]{\label{fig:90ring-iquo}
\includegraphics[width=0.35\textwidth, angle=90]{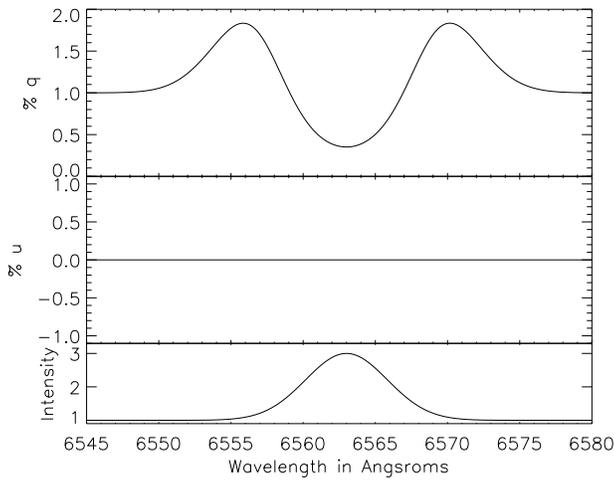}}
\quad
\subfloat[Radial \% Polarization]{\label{fig:90ring-iqur}
\includegraphics[width=0.35\textwidth, angle=90]{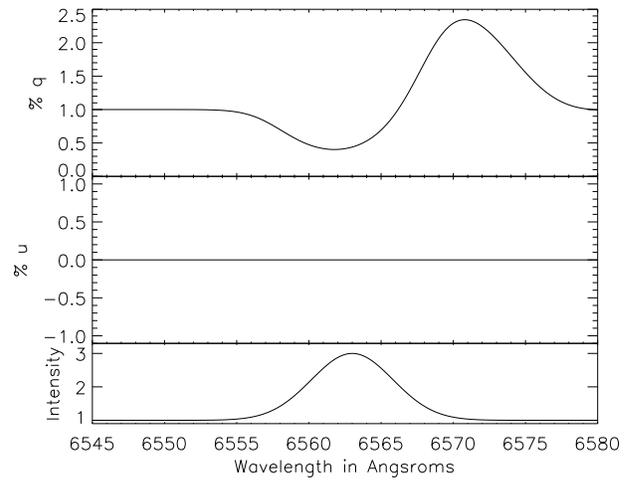}}
\caption[60$^\circ$ Inclined Ring Polarization]{The spectropolarimetric model for an edge-on ring. {\bf a)} The doppler redistribution function A at 6560{\AA} as a function of orbital phase. {\bf b)} The q and u phase matrix components as a function of orbital phase. {\bf c) \& d)} The resulting spectropolarimetry.}
\label{fig:ring60pol}
\end{figure}

\begin{figure}
\centering
\subfloat[$\beta_r$ and $\beta_\phi$ for an Edge-on System]{\label{fig:edge-beta}
\includegraphics[width=0.3\textwidth, angle=90]{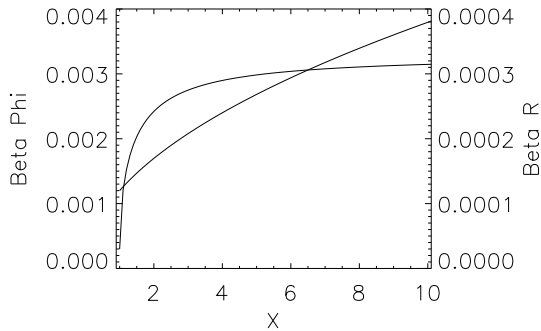}}
\quad
\subfloat[The Redistribution Function for an Edge-on System]{\label{fig:edge-a}
\includegraphics[width=0.3\textwidth, angle=90]{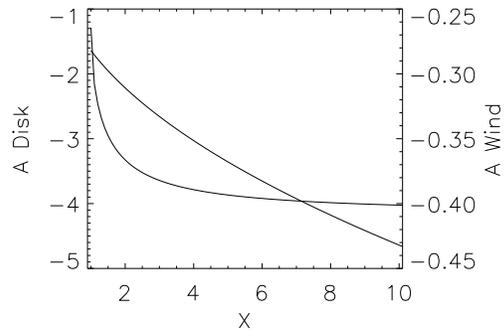}}
\caption[Edge-on System Doppler and Redistribution Functions]{  {\bf a)}  The velocity shifts computed as a function of stellar distance x=r/$R_\odot$ for radial and orbital motion.  {\bf b)}  The doppler redistribution function, A, for radial and orbital motion of a cloud at an orbital phase of 45$^\circ$ and an observed wavelength of 6562{\AA}.}
\label{fig:edge-pars}
\end{figure}
\twocolumn

\subsection{Thin-Disk Electron Scattering Example}

Now that the integral over orbital phase for a single radius has been done, the distributions of material and the spatial integrations can be discussed. The phase matrix gets inserted into the integral over all photospheric source points so one can calculate the Stokes emission from a single particle at a single wavelength at a single scattering angle as:

\begin{equation}
\rho{\bf j}=\frac{3\sigma_T}{16\pi}\int_{\Omega_\ast}nI_0\overline{{\bf P}}d\Omega_\ast
\end{equation}

To compute the total Stokes emission over the entire disk at a single wavelength, add an integration over all particles in the the disk (V) to get:

\begin{eqnarray}
{\bf F}=\left( \begin{array}{c} F_I \\ F_Q \\ F_U \\ F_V \end{array} \right) = \frac{3\sigma_T}{16\pi D^2}\int_V\int_{\Omega_\ast} n I_0{\bf P}d\Omega_\ast dV
\end{eqnarray}

At any single wavelength, the observed polarized flux is an integral over every scattering particle (V) and over every photospheric source ($d\Omega_\ast$). To integrate these equations across the disk, one must introduce a convenient integral operator. If a dimensionless distance parameter is defined as $x=r/R$, the distance of the scattering particle in units of stellar radii, one can write:

\begin{equation}
\phi(x)=\sin^{-1} \left[ \frac{(1-x^{-2})^{\frac{1}{2}}}{\sin i} \right]
\end{equation}

Since an infinitessimally thin equatorial disk has been assumed while still including the effects of occultation, the volume integral can be rewritten as:

\begin{small}
\begin{equation}
\int_{Area}=\int_{x=1}^{\alpha_D} \int_{\phi=0}^{2\pi} - \int_{x=0}^\delta \int_{\phi=\frac{\pi}{2}+\phi(x)}^{\frac{3\pi}{2}-\phi(x)}=
     \int(Disk) - \int (Occulted)
\end{equation}
\end{small}

Where $\alpha_D$ is the disk radius in units of stellar radii, and we've defined the parameter $\delta=\sec i$ for $\alpha_D > \sec i$ and $\delta=\alpha_D$ for $\alpha_D < \sec i$. The integral runs from the stellar photosphere ($x=1$) to the edge of the disk ($x=\alpha_D$) over all orbital phases ($\phi=0,2\pi$). The occulted part of the disk is behind the star ($\frac{\pi}{2} < \phi < \frac{3\pi}{2}$) but the occulted area depends on the inclination. A pole-on disk will not have any occultation whereas an edge-on disk will have a wedge occulted (all orbital phases at x=1, only $\phi=\pi$ at $x=\infty$). 

Our photosphere is wavelength-dependent and scattered intensities have two doppler shifts so the spatial integrations at any one wavelength must have the incident and scattered fluxed shifted properly. Using the doppler-shifts computed for $\lambda_\ast$ one can use the doppler-redistribution function and some assumptions about disks and winds to make integrable scattered spectra. In the simple disk or wind cases, one can get the density (n) from the equations for the surface density ($\Sigma$).  One can define equations for the density in terms of a constant ($P_0$) and a radial function f(r):

\begin{equation}
A=\frac{ (1-\beta_r)\lambda - (1-\beta_r\sin i\cos\phi + \beta_\phi\sin i\sin\phi)\lambda_0}{(1-\beta_r\sin i\cos\phi+\beta_\phi\sin i\sin\phi)\lambda_{wid}}
\end{equation}

\begin{equation}
\frac{3\sigma_T}{16\pi}n=\frac{3\sigma_T}{32\pi}\Sigma_0 \left( \frac{R}{r} \right)^2=P_0 f(r)
\end{equation}

\begin{equation}
\frac{3\sigma_T}{16\pi}n=\frac{3\sigma_T}{32\pi} \frac{\dot{M}}{m_p2\pi Rv_\infty} \left[ \frac{v_1}{v_\infty}+\left(1-\frac{R}{r}\right)^{\frac{1}{2}} \right]^{-1}=P_0 f(r)
\end{equation}

For a Be star, as used in Wood et al (1993) $\alpha$=2, $\lambda_{wid}$=4{\AA} (about 180$\frac{km}{s}$ broadening), $\lambda_0$=6563{\AA} for the H$_\alpha$ line. The orbital velocity at the stellar surface is $v_0$=360$\frac{km}{s}$. The wind can be fit with $v_\infty$=90$\frac{km}{s}$ and $v_1$=0.1$v_\infty$. The mass-loss rate can be estimated at $\dot{M}=10^{-7}M_\odot$/yr and R=5R$_\odot$. These assumed parameters let us estimate the normalization coefficient ($P_0$) for the wind:

\begin{equation}
P_0=\frac{3\sigma_T}{32\pi} \frac{\dot{M}}{m_p2\pi Rv_\infty}=\frac{1}{250}=0.4\%
\end{equation}

Note that when you're mixing wind \& disk parameters, as will be done later to illustrate radial motion in a disk, the normalization constants ($P_0$) must match between disk and wind, since they are the same particles. This implies that $\Sigma_0 = \dot{M}/2\pi m_pRv_\infty$. With these parameters, one can calculate the doppler shifts at every point on the disk:

\begin{small}
\begin{equation}
\beta_\phi(r)=\frac{v_\phi}{c}=\frac{v_0}{c} \left( \frac{R}{r} \right)^{\frac{1}{2}}=1.2\times10^{-3}x^{-\frac{1}{2}}
\end{equation}

and at every point for a wind:

\begin{equation}
\beta_r(r)=\frac{v_1}{c}+\frac{v_\infty}{c} \left( 1-\frac{R}{r} \right)^{\frac{1}{2}}=
              3\times10^{-5}+3\times10^{-4} \left( 1-\frac{1}{x} \right)^{\frac{1}{2}}
\end{equation}
\end{small}

With those doppler shifts, one can calculate the redistibuted Stokes emission ($A(r,\phi,\lambda)$). Computing the f(r) dependencies of the velocities:

\begin{equation}
f_{disk}(r)=\left( \frac{R}{r} \right)^{-2}=x^{-2}
\end{equation}

\begin{equation}
f_{wind}(r)=\left[ \frac{v_1}{v_\infty} + \left( 1-\frac{R}{r} \right)^{\frac{1}{2}} \right]^{-1}=\left[0.1+\left(1-\frac{1}{x} \right)^{\frac{1}{2}} \right]^{-1}
\end{equation}

Now there is a way to compute the integral over all the disk at any wavelength: the scattering matrix (${\bf P_{QU}}$), the radial dependencies, f(r), the stellar source intensities ($1+\alpha e^{-G^2}$), the doppler-shifted scattered intensities ($1+\alpha e^{-A^2}$), normalization constants ($P_0$) and the integral operator that integrates across the disk ($\int x=1,\alpha_D \phi=0,2\pi$) and includes occulatation ($\int x=1,\delta \phi=\frac{\pi}{2}+\phi(x),\frac{3\pi}{2}-\phi(x)$) are all known.

\subsection{Pole-on Disk or Wind}

If one takes i=0$^\circ$, the redistributed Stokes emisssion coefficients ($A(r,\phi,\lambda)$) with $\sin(i)=0$ and all orbital motion perpindicular to the line-of-sight (no $\phi$ doppler shift) is calculated:

\begin{equation}
A=\frac{ (1-\beta_r)\lambda - \lambda_0}{\lambda_{wid}}
\end{equation}

This only has a dependence on $\beta_r$, which will only occur for a wind, where there is a radial doppler-shift between the star and the scattering particle. Since in a pole-on disk there is no motion of the particles along your line-of-sight, there would be no spectropolarimetric signature for a disk. There will also be no occultation so that the integration across the disk just becomes a simple double integral.  

\begin{equation}
\int_{Area}=\int_{x=1}^{\alpha_D} \int_{\phi=0}^{2\pi}
\end{equation}

The scattering phase matrix and Stokes parameters take on a simple form: 

\begin{eqnarray*}
{\bf P_{QU}} & = & \left( \begin{array} {c} P_Q \\ P_U \end{array} \right) = \left( \begin{array}{c} \sin^2i-(1+\cos^2i)\cos2\phi \\ 2\cos i\sin2\phi \end{array} \right) \\ & = & \left( \begin{array}{c} 2cos2\phi \\ 2sin2\phi \end{array} \right)
\end{eqnarray*}

\begin{equation}
q_\lambda=\frac{P_0}{1+\alpha e^{-G^2}} \int_{x=1}^{\alpha_D} \int_{\phi=0}^{2\pi} (1+\alpha e^{-A^2})f(r)\left[ 2\cos2\phi \right] dxd\phi
\end{equation}

\begin{equation}
u_\lambda=\frac{P_0}{1+\alpha e^{-G^2}} \int_{x=1}^{\alpha_D} \int_{\phi=0}^{2\pi}  (1+\alpha e^{-A^2})f(r)\left[2sin2\phi \right] dxd\phi
\end{equation}

At each wavelength, you can integrate across x and $\phi$ knowing the x dependence of the redistribution function (A) and the x dependence of the wind velocity, f(r) \& $\beta_r(r)$ to compute the Stokes emission.  

\begin{equation}
q_\lambda=\frac{0.8}{1+2e^{-(\frac{\lambda-6563}{4})^2}} \int_{x=1}^{10} \int_{\phi=0}^{2\pi}  \frac{(1+2e^{-A^2})}
           {\left[0.1+\left(1-\frac{1}{x} \right)^{\frac{1}{2}}\right]}    \cos2\phi dxd\phi
\end{equation}

\begin{equation}
u_\lambda=\frac{0.8}{1+2e^{-(\frac{\lambda-6563}{4})^2}} \int_{x=1}^{10} \int_{\phi=0}^{2\pi} \frac{(1+2e^{-A^2})}
            {\left[0.1+\left(1-\frac{1}{x} \right)^{\frac{1}{2}}\right]}    \sin2\phi dxd\phi
\end{equation}

With the explicit function A:

\begin{small}
\begin{equation}
A=\frac{ \left( 1-3\times 10^{-5}+3\times 10^{-4}(1-\frac{1}{x})^{\frac{1}{2}} \right) \lambda -6563}{4}
\end{equation}
\end{small}

In this case, since both integrals contain a sin2$\phi$ or cos2$\phi$ term, the integrals are strictly zero and there is no spectropolarimetric effect. This comes about because, in the uniform wind case, there is no asymmetry or occultation. Every part of the circumstellar disk that produces some +Q scattered light at some wavelength will have a corresponding part that produces -Q scattered light, nullifying the polarization. Regardless of the radial velocity dependence, in a face-on system without azimuthal asymmmetry there will be no spectropolarimetric signature.

\subsection{Edge-on Disk or Wind}

If one lets i=90$^\circ$, one can calculate the redistributed Stokes emisssion coefficients ($A(r,\phi,\lambda)$) with $\sin(i)=1$ and all orbital motion perpindicular to the line-of-sight (no $\phi$ doppler shift):

\begin{equation}
A=\frac{ (1-\beta_r)\lambda - (1-\beta_r \cos\phi + \beta_\phi \sin\phi)\lambda_0}{(1-\beta_r \cos\phi+\beta_\phi \sin\phi)\lambda_{wid}}
\end{equation}

This now has a dependence on $\beta_r$ and $\beta_\phi$ as well as $\phi$ itself. For simplicity, ignore the occultation so that the integration across the disk is still a simple double integral. The scattering phase matrix still has a simple form and as in the ring case there is no Stokes U polarization from an edge on system. The phase matrix and Stokes emission coefficients are: 

\begin{eqnarray*}
{\bf P_{QU}} & = & \left( \begin{array} {c} P_Q \\ P_U \end{array} \right) = \left( \begin{array}{c} \sin^2i-(1+\cos^2i)\cos2\phi \\ 2\cos i\sin2\phi \end{array} \right) \\ & = & \left( \begin{array}{c} 1-cos2\phi \\ 0 \end{array} \right)
\end{eqnarray*}

\begin{equation}
q_\lambda=\frac{P_0}{1+\alpha e^{-G^2}} \int_{x=1}^{\alpha_D} \int_{\phi=0}^{2\pi} (1+\alpha e^{-A^2})f(r)\left[ 1-\cos2\phi \right] dxd\phi
\end{equation}

\begin{equation}
u_\lambda=0
\end{equation}

At each wavelength, you can integrate across x and $\phi$ knowing the x dependence of the redistribution function (A) and the x dependence of the wind velocity, f(r) \& $\beta_r(r)$ to compute the Stokes emission.  

\begin{footnotesize}
\begin{equation}
q_\lambda=\frac{0.4}{1+2e^{-(\frac{\lambda-6563}{4})^2}} \int_{x=1}^{10} \int_{\phi=0}^{2\pi}  \frac{(1+2e^{-A^2})}
           {\left[0.1+\left(1-\frac{1}{x} \right)^{\frac{1}{2}}\right]}    (1-\cos2\phi) dxd\phi
\end{equation}
\end{footnotesize}

\onecolumn
\begin{figure}
\centering
\subfloat[Depolarization Intensities]{\label{fig:depol-intens}
\includegraphics[width=0.35\textwidth, angle=90]{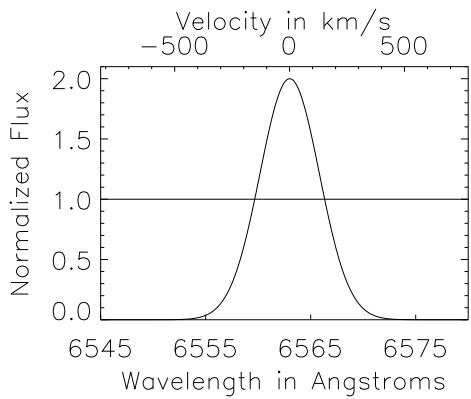}}
\quad
\subfloat[Depolarization Effect]{\label{fig:depol-pol}
\includegraphics[width=0.35\textwidth, angle=90]{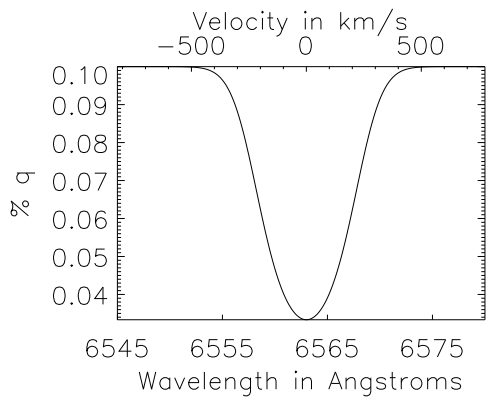}}
\caption[The Depolarization Effect]{  {\bf a)}  The intensities from the stellar continuum and the broad line formation region. This simple example assumes a simple continuum, line-free spectrum that is 0.1\% polarized. An emission line with the line:continuum ratio of 2:1 forms outside the polarizing region and is entirely unpolarized.  {\bf b)}  The polarization spectrum assuming the star is 0.1\% polarized in +Q. The unpolarized emission dilutes the continuum polarization.}
\label{fig:depol}
\end{figure}

\begin{figure}
\centering
\subfloat[Absorption + Depolarization Intensities]{\label{fig:abspol-intens}
\includegraphics[width=0.35\textwidth, angle=90]{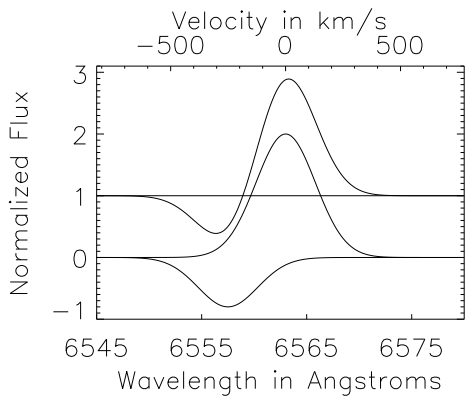}}
\quad
\subfloat[Absorption + Depolarization Polarization]{\label{fig:abspol-pol}
\includegraphics[width=0.35\textwidth, angle=90]{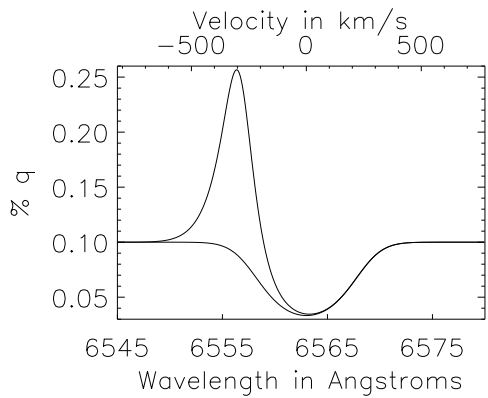}}
\caption[The Effect of Disk or Wind Absorption on Depolarization]{  {\bf a)}  A P-Cygni type profile as the sum of intensities from the stellar continuum and the broad line formation region as well as blue-shifted absorption. This simple example assumes a simple continuum, line-free spectrum that is 0.1\% polarized. An emission line with the line:continuum ratio of 2:1 forms outside the polarizing region and is entirely unpolarized. The blue-shifted absorption removes unpolarized light from the system.  {\bf b)}  The polarization spectrum assuming the stellar continuum is 0.1\% polarized entirely in +Q. The unpolarized emission dilutes the continuum polarization but the absorptive component removes unpolarized light, increasing the polarization across the absorptive component. The original depolarization signature is overplotted to illustrate the effect of the absorption in the depolarized trough.}
\label{fig:abspol}
\end{figure}

\twocolumn

At this point, the solution can be simply described by noting that the solution is simply a sum of many edge-on ring solutions with varying velocity parameters. The full solutions look very similar to the single-ring case, as can be seen in Wood et al. 1993. 

	These examples serve to show the morphology for these disk scattering effects. For orbital motion, the signatures are symmetric about line center and are broader than the intensity profile. In the wind/outflow case, the spectropolarimetric effect is predominantly in the center and red-shifted components of the line.

\subsection{The Depolarization Effect}

	In the case of emission lines that form over broad regions, a depolarization effect is possible. If somehow the stellar continuum polarization forms interior to the line-formation region, the less-polarized emission will depolarize the stellar light across the emission line producing a broad decrease in polarization. This was initially discussed in Poeckert \& Marlborough 1977 and McLean 1979. There is a very significant body of literature that uses the depolarization effect in Be stars to separate intrinsic and interstellar polarization. Electron scattering models that assume asymmetric envelopes fit low-resolution spectropolarimetric observations quite well. The depolarization effect was conceived in this context - the H$_\alpha$ emission region is thought to be quite large in Be stars, and flattened, large emission regions have been resolved interferometrically (cf. Quirrenbach et al. 1994). The electron scattering optical depth estimates showed that the bulk of the intrinsic continuum polarization came from the inner region of the flattened envelope. Since the H$_\alpha$ emission formed outside this region, it would be unpolarized.
	
	In another simplistic example, if you assume that a 0.1\% continuum polarization is formed near the star by scattering in an asymmetric envelope and assume that an emission line is formed entirely outside this polarizing region, a simple depolarization effect is calculated as the continuum polarized flux added to the unpolarized emission line flux. For simplicity, assume the asymmetric interior region produces only +Q polarization. The corresponding intensities and polarization spectra are shown in figure \ref{fig:depol}. 

	There is a method for producing a polarization change in the absorptive components of an emission line as well that involves absorption of the stellar continuum by occulting material. The H$_\alpha$ photons come from outside this asymmetric region with net polarization, and they will be less polarized than the stellar continuum since atomic emission is inherently unpolarized. This depolarization is centered on the emission and is directly proportional to it. However, the presence of material around the star typically leads to absorption by cooler material further out in the envelope. This is what gives rise to the typical ``disky" central absorptions or the ``windy" P-Cygni type profiles. To produce a change in the absorptive component of the line, this depolarizing atomic emission as well as the stellar continuum must be absorbed preferentially by material directly along the line of sight. This preferential absorption of unpolarized emission has the effect of allowing more scattered light to reach the detector in the absorptive component of the line. 
	
	Another simple example of this effect can be constructed for a P-Cygni profile while using the 0.1\% continuum polarization and geometrical assumptions from the depolarization effect, following McLean 1979. We'll start by assuming that the blue-shifted absorption only removes unpolarized light. In this model, the absorption occurs only along the lines of sight to the photosphere and emission region and that the continuum polarization must come from a broad asymmetric envelope inside the emission region. The observed line profile is built as a supposition of stellar continuum, emission and a smaller blue-shifted absorption, as in figure \ref{fig:abspol-intens}. An assumed absorption blue-shifted by 250km/s with the same width as the stellar spectrum is imposed on the continuum and emission. Since both the absorption and emission are only effecting unpolarized light, and the continuum is assumed to be polarized at a constant 0.1\%, the observed spectropolarimetric effect is essentially 1/I scaled by 0.1\%. The resulting spectropolarimetric effect is shown in figure \ref{fig:abspol-pol} with the corresponding simple depolarization effect for reference. Essentially, the removal of unpolarized light increases the polarization from continuum values by a factor of 1/I. This model is quite simple, but a more advanced model incorporating more realistic scenario's has not been created. An important thing to note is the morphology of the depolarization effect for a P-Cygni profile. The effect follows a 1/I type curve and there must be a significant spectropolarimetric effect across the entire line for the depolarization effect to be at work.

\begin{figure} [!h]
\centering
\subfloat[Circular Polarization via Zeeman Splitting]{\label{fig:zeem-v}
\includegraphics[width=0.3\textwidth, angle=90]{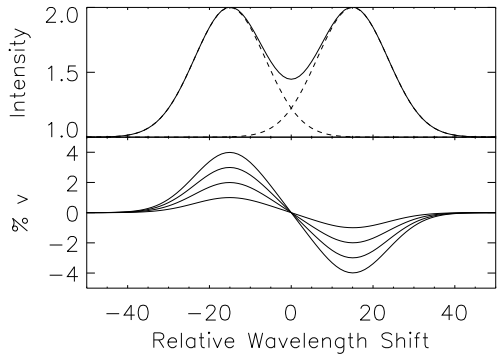}}
\quad
\subfloat[Linear Polarization via Zeeman Splitting]{\label{fig:zeem-q}
\includegraphics[width=0.3\textwidth, angle=90]{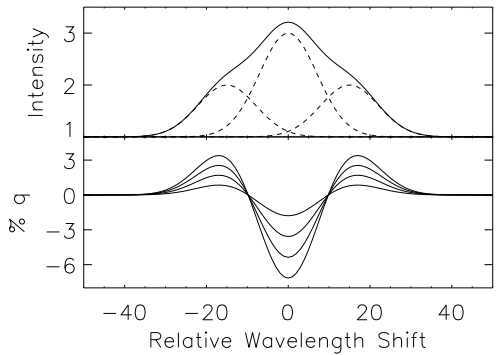}}
\caption[The Effect Magnetic Fields - Zeeman Splitting]{Magnetic fields can cause circular and linear polarization. {\bf a)} A radial magnetic field can cause a wavelength shift between $\pm$V states. The field induces a wavelength between two lines of equal intensity. {\bf b)} A tangential (azimuthal) magnetic field can cause a wavelength shift between three line components: Two +Q states are shifted and a central -Q line of double intensity is unshifted giving rise to the classic symmetric q signature. }
\label{fig:zeeman}
\end{figure}

\subsection{Magnetic Fields and Zeeman Splitting}

	Magnetic fields are another mechanism that can cause polarization across atomic and molecular lines. Most discussion about spectropolarimetry in the night-time astronomical community revolves around circular polarization from magnetic fields. Magnetic fields cause magnetically sensitive transitions to show wavelength shifts between the different polarized components of the transition. Figure \ref{fig:zeeman} shows both linear and circular polarization effects for a simple emission line with significant splitting. A radial magnetic field causes a line to split into $\pm$V states with the shift proportional to the field strength. The two states have equal intensity and give rise to the antisymmetric V profiles seen in figure \ref{fig:zeem-v}. A tangential (azimuthal) magnetic field causes a line to split into three components: two +Q states are shifted and there is an unshited -Q line of double intensity. This gives rise to the classic symmetric q signature of figure \ref{fig:zeem-q}. 	

	Though this effect is routinely observed in many hot stars, the amplitudes of the v signatures are almost always 0.1\% or lower across magnetically sensitive lines. In most cases, sophisticated least-squares deconvolution outines are needed to sum spectropolarimetric signatures across hundreds of lines to achieve the required signal-to-noise ratio's. The linear polarization signatures are even smaller. This is because even for stars with simple dipolar fields there is significant smearing when the polarized light from all photospheric regions is summed.

\subsection{Theory Summary}

	This chapter has presented a summary of scattering theory for a simple thin-disk case for orbital and radial motion. The depolarization effect has been discussed with an extension to absorptive effects and P-Cygni profiles. The morphologies are very distinct. Disk scattering causes symmetric polarization effects that span the entire width of the spectral line. Radial wind motion causes the spectropolarimetric signatures to shift toward the red side of the line, since the scattering particles all scatter a red-shifted sourc. The depolarization effect causes a broad decrease in polarization inversely related to the emission/absorption amplitude, 1/I. This effect can cause significant effects in absorptive components of a line profile if the absorption selectively removes unpolarized light. A very significant point is that this selective removal must be present in conjunction with a broad depolarization effect which also spans the entire width of the line, although at a possibly smaller amplitude. There is also a relation between the amount of scattered light and the expected amplitude of these polarimetric effects. The amount of scattered light controls the amplitude of these effects.
	
	Magnetic field effects were also discussed, but given the large amount of literature on circular polarization effects, the linear polarization signatures are much smaller than any of the signatures detected in this survey.

\section{Herbig Ae/Be Spectropolarimetry}

	In this chapter,  the spectropolarimetry of the Herbig Ae/Be stars will be presented. The spectropolarimetric measurements of most targets are incompatible with current scattering theory framework presented in the previous chapter. This will require very detailed and thorough presentation of the observations. The compiled spectropolarimetry for each target will be presented as an overview followed by a discussion of individual targets and individual observations. A recurring theme in the spectropolarimetric morphology of Herbig Ae/Be stars is the presence of spectropolarimetric signatures in and around absorptive components of the emission line and the detected amplitudes are 0.2\% to 2\%. Scattering theory shows no natural amplitude for a spectropolarimetric signature. The detailed morphology of individual lines will highlight this ``polarization-in-absorption" theme. 
		
	There is at least one good measurement for each star with 3-10 observations per star being typical. Since the polarization properties of the telescope conspire to rotate the plane of polarization in a pointing-dependent way, no attempt to de-rotate the results in any major systematic way will be made. There are a large number of observing conditions and source-magnitudes. In figures \ref{fig:haebe-specpol1} and \ref{fig:haebe-specpol2} all good spectropolarimetric data sets are shown binned-by-flux to a continuum-threshold of 5. Since the individual emission lines have continuum-normalized intensities of 0.2 to nearly 40, this binning will give a much more uniform signal-to-noise ratio for each measurement while preserving good wavelength coverage. Table \ref{aebe-res} shows the stars observed, the type and strength of the H$_\alpha$ line and the detected polarization effect. The actual signal-to-noise ratios in a polarization spectrum are dependent on four individual exposures, and measurements were performed through a wide range of observing conditions. The ratio's range from 100 to over 1000 depending on the seeing, exposure time, and binning. Individual examples and a discussion of the stars with significant detections will follow.

\begin{table}[!h,!t,!b]
\begin{center}
\caption{Herbig Ae/Be H$_\alpha$ Results \label{aebe-res}}
\begin{tabular}{lcccc}
\hline    
\hline    
{\bf Name}    &{\bf H$_\alpha$}    &  {\bf Effect?}          & {\bf Mag}            & ${\bf Type}$                  \\
AB Aur           & 8-11 &   Y                                           &    1.5\%               &   Wind                             \\
MWC480        & 4-7  &    Y                                        &    1.5\%                &   Wind                              \\
MWC120        & 5-8  &    Y                                        &   1.0\%                 &   Wind                            \\
HD163296    & 6-8  &  Y                                           &     1.0\%                &   Wind                          \\
HD179218    & 2-5   &   Y                                          &   0.5\%                 &   Wind*                         \\
HD150193     &3-5   & Y                                             &  0.5\%                 &  Wind*                            \\
MWC758        &2.5-3 & Y                                            &  0.5\%                   & Wind                             \\
HD144432     & 4       & Y                                            &  2.0\%                  & Wind                              \\
MWC158        & 15-17 &  Y                                          &    1.0\%                &   Disk                             \\
HD58647       & 2.6      &  Y                                           &    0.5\%                &   Disk                            \\
MWC361        & 8-10   &   Y                                          &   0.3\%                 &  Disk                            \\
51 Oph           & 1.3    & Y                                             &  0.3\%                  & Disk                             \\
HD45677      & 36    & Y                                              & 1.0\%                  &  Disk                            \\   
MWC147       & 15    & Y                                           &  0.3\%                   & Disk                            \\
MWC170       & 1.4  & Y                                              & 0.3\%                  & Disk                            \\
\hline
V1295Aql      &7-8   & N                                            &   $<$0.2\%                        &  Wind                           \\
KMS 27          &2.5   & N                                            &   $<$0.3\%                         & Wind                          \\
HD169142    &3.1-3.4 & N                                        &  $<$0.3\%                       &  Wind                            \\
HD35989      &1.9    & N                                            &  $<$0.2\%                     &  Wind                            \\
HD141569    &1.6   & N                                             &  $<$0.2\%                    &  Disk                            \\
XY Per           &2.0    & N                                            &   $<$0.5\%                    &  Disk                            \\
MWC166      &1.3     & N                                            &   $<$0.1\%                    &  Disk                            \\
Il Cep            &3.5     & N                                            &    $<$0.2\%                     &  Emis                            \\
MWC442      &6.2    & N                                             &  $<$0.2\%                   &  Emis                            \\
HD38120     &13    & N                                              &   $<$0.4\%                   &  Emis                            \\
GU CMa       &2.9    & N                                             &  $<$0.1\%                   &  Emis                            \\
HD35187      &1.1-1.6  & N                                       &  $<$0.5\%                   &  Var                            \\
HD142666   &2.0    & N                                            &   $<$0.5\%                  &  Acc                            \\
T Ori              &4-6    & N                                            &   $<$7\%                    & Disk**                           \\
\hline
\hline
\end{tabular}
\end{center}
The name and typical amplitude of the H$_\alpha$ line are presented along with detection statistics. The H$_\alpha$ column is the normalized line intensity, Effect? is the presence (Y) or absence (N) of a spectropolarimetric effect, Mag is the amplitude of the spectropolarimetric effect and Type is the H$_\alpha$ line type. 
\end{table}

\subsection{Comments on Individual Targets}

	In this section, detailed descriptions of the HiVIS, ESPaDOnS and literature spectropolarimetric results will be discussed for each individual star. Examples of spectropolarimetric effects will be shown for each and a description of the morphology will be given.

\subsection{AB Aurigae - HD 31293 - MWC 93}

	There were only two previous spectropolarimetric observations showing 0.4\% to 0.7\% polarization across the H$_\alpha$ line. One observation performed in 1999 was presented in two different papers showing two somewhat different results (Pontefract et al. 2000 and Vink et al. 2002). It is unclear what caused this difference in results when using the same data, but both publications show a line effect. Another set of observations taken in 2004 was presented in Mottram et al. 2007 showing a change from the 1999 observations presented in both the Pontefract et al. 2000 and Vink et al. 2002 publications. The optical continuum polarization was 0.8\% $\pm$ 0.1 January 9th 1999 (Ashok et al. 1999).  Beskrovnaya et al. 1995 report R-band 0.3\% in 1993 and 0.15\% in 1988 with substantial variability on a daily basis. 
	
\begin{figure}
\centering
\subfloat[AB Aurigae]{\label{fig:abaur}
\includegraphics[ width=0.21\textwidth, angle=90]{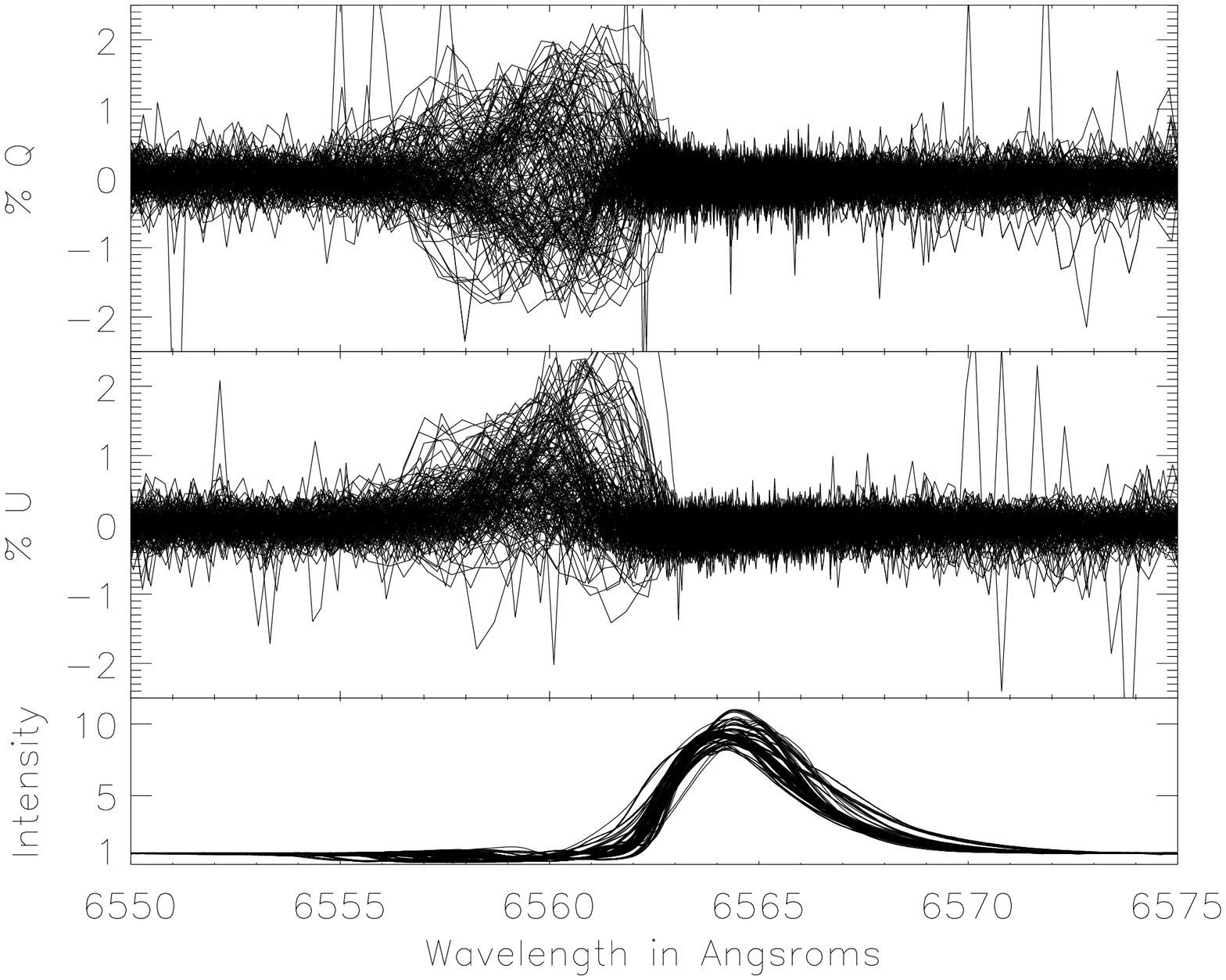}}
\quad
\subfloat[MWC 480]{\label{fig:mwc480}
\includegraphics[ width=0.21\textwidth, angle=90]{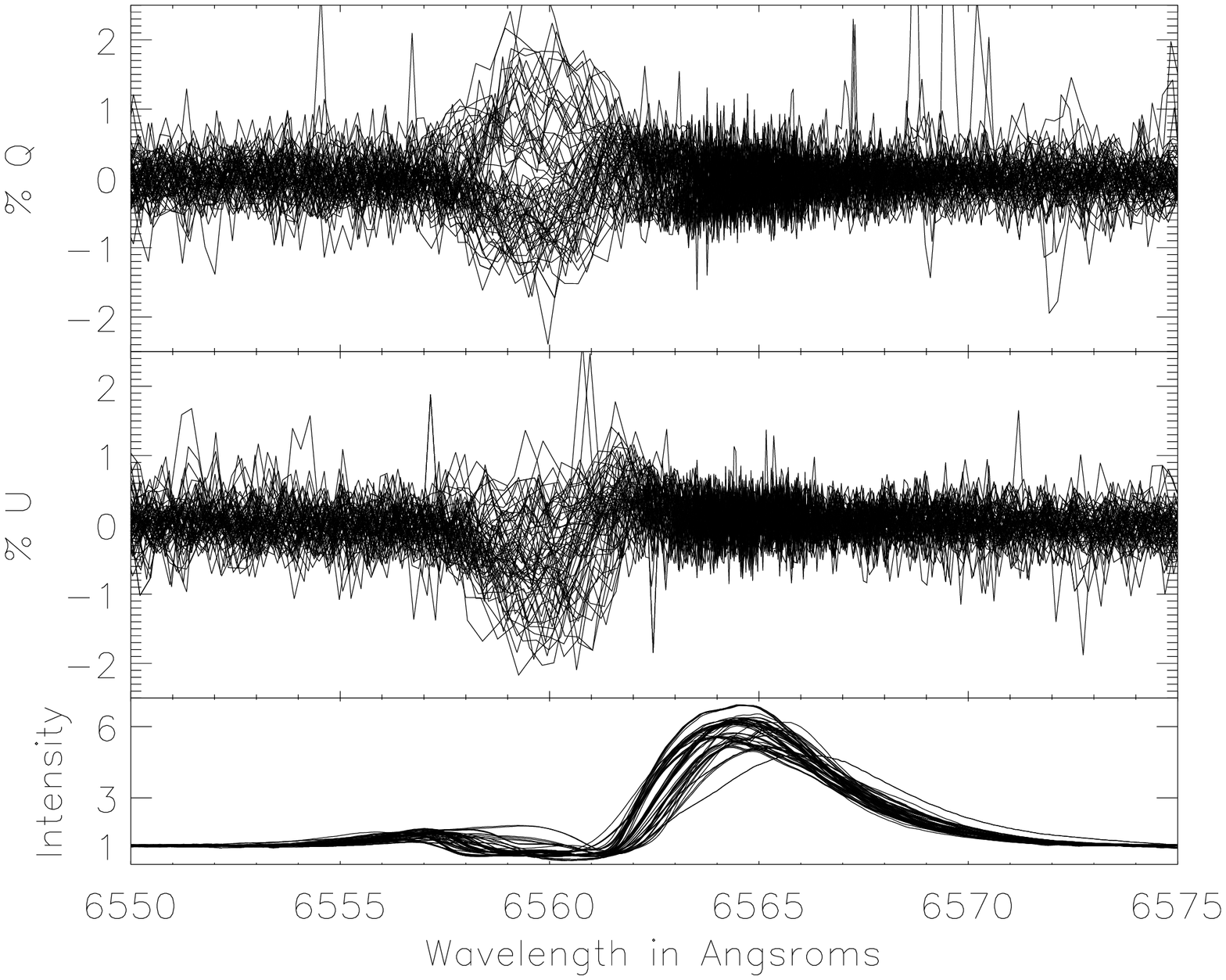}}
\quad
\subfloat[MWC 120]{\label{fig:mwc120}
\includegraphics[ width=0.21\textwidth, angle=90]{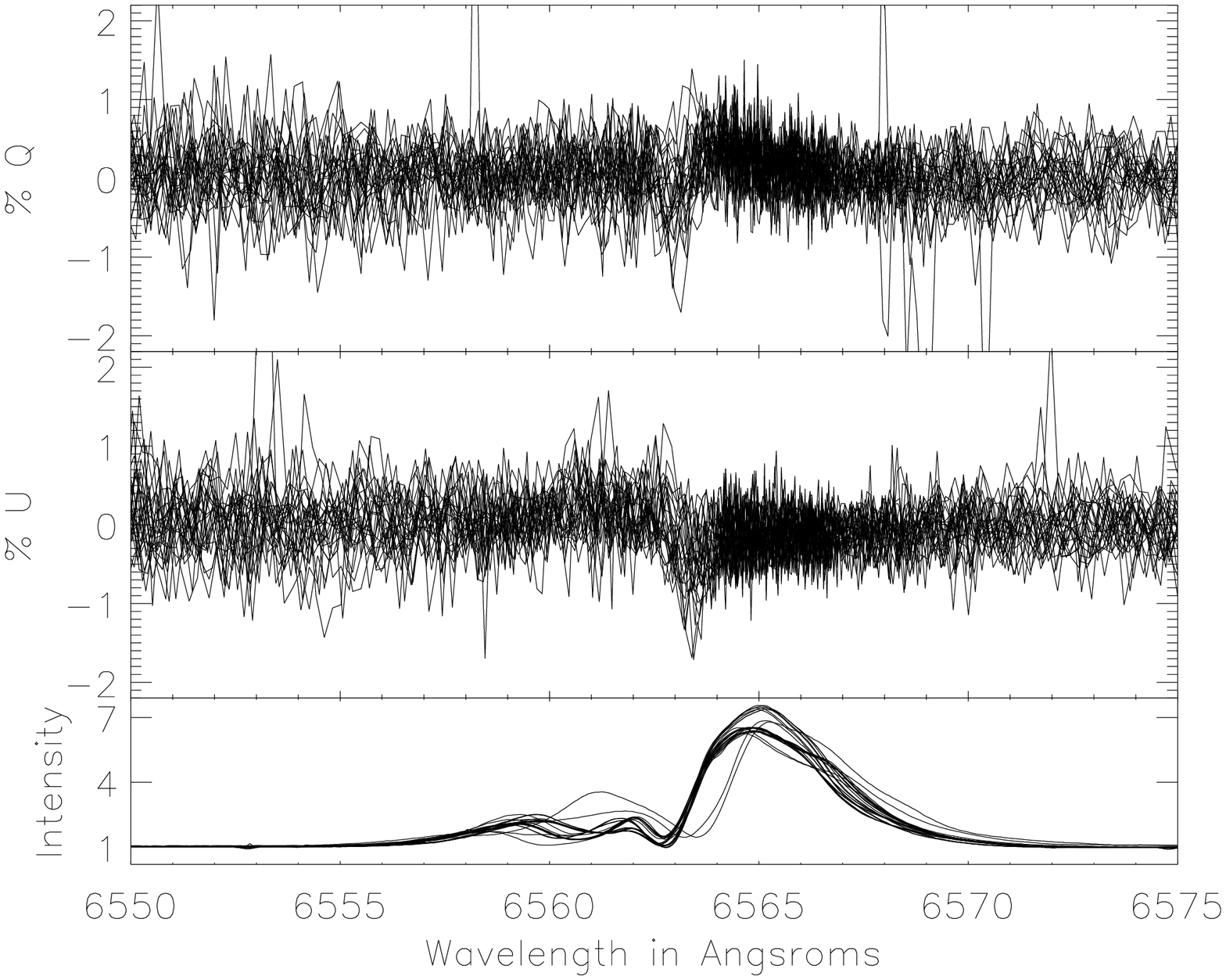}}
\quad
\subfloat[HD 150193]{\label{fig:hd150}
\includegraphics[ width=0.21\textwidth, angle=90]{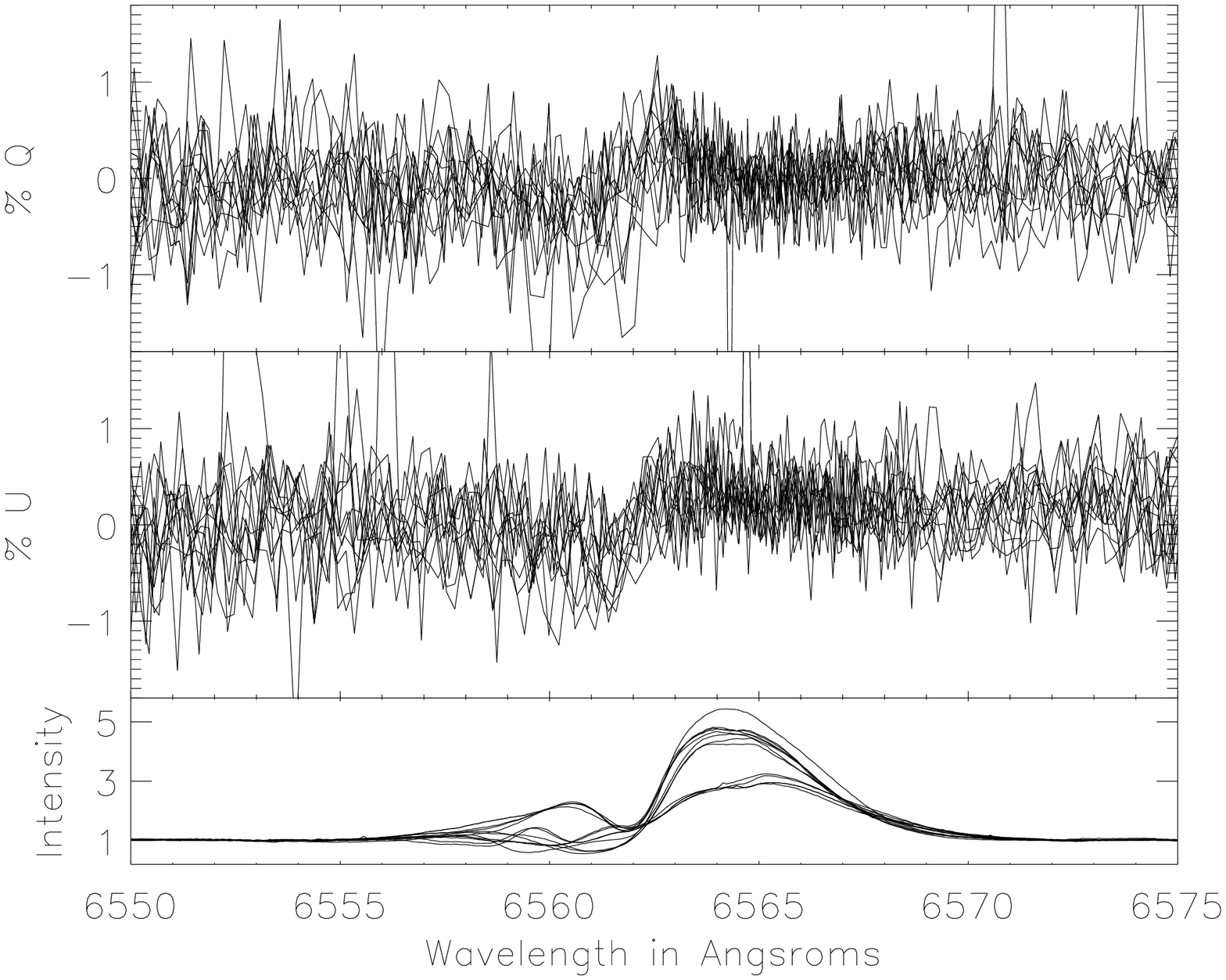}}
\quad
\subfloat[HD 163296]{\label{fig:hd163}
\includegraphics[ width=0.21\textwidth, angle=90]{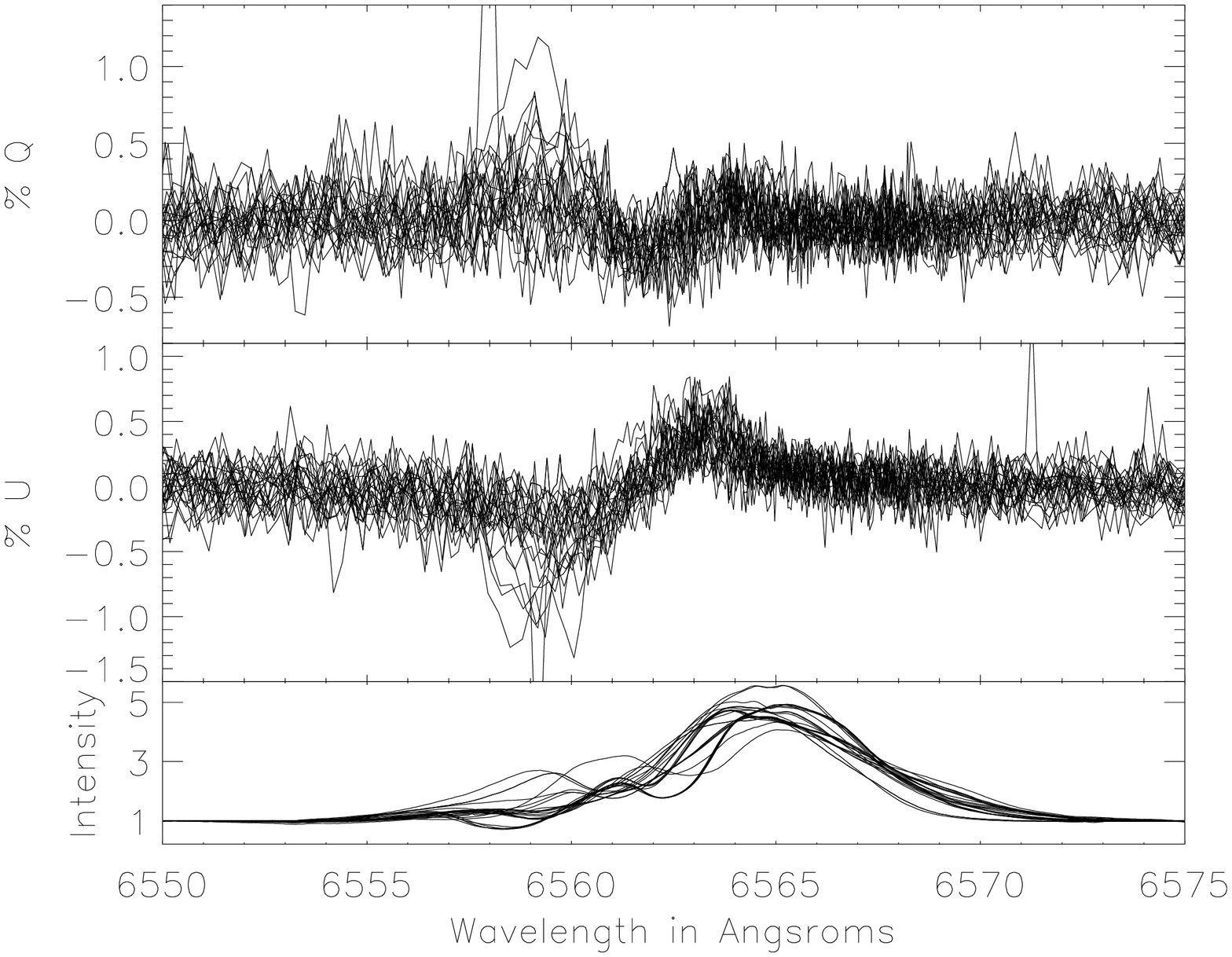}}
\quad
\subfloat[HD 179218]{\label{fig:hd179}
\includegraphics[ width=0.21\textwidth, angle=90]{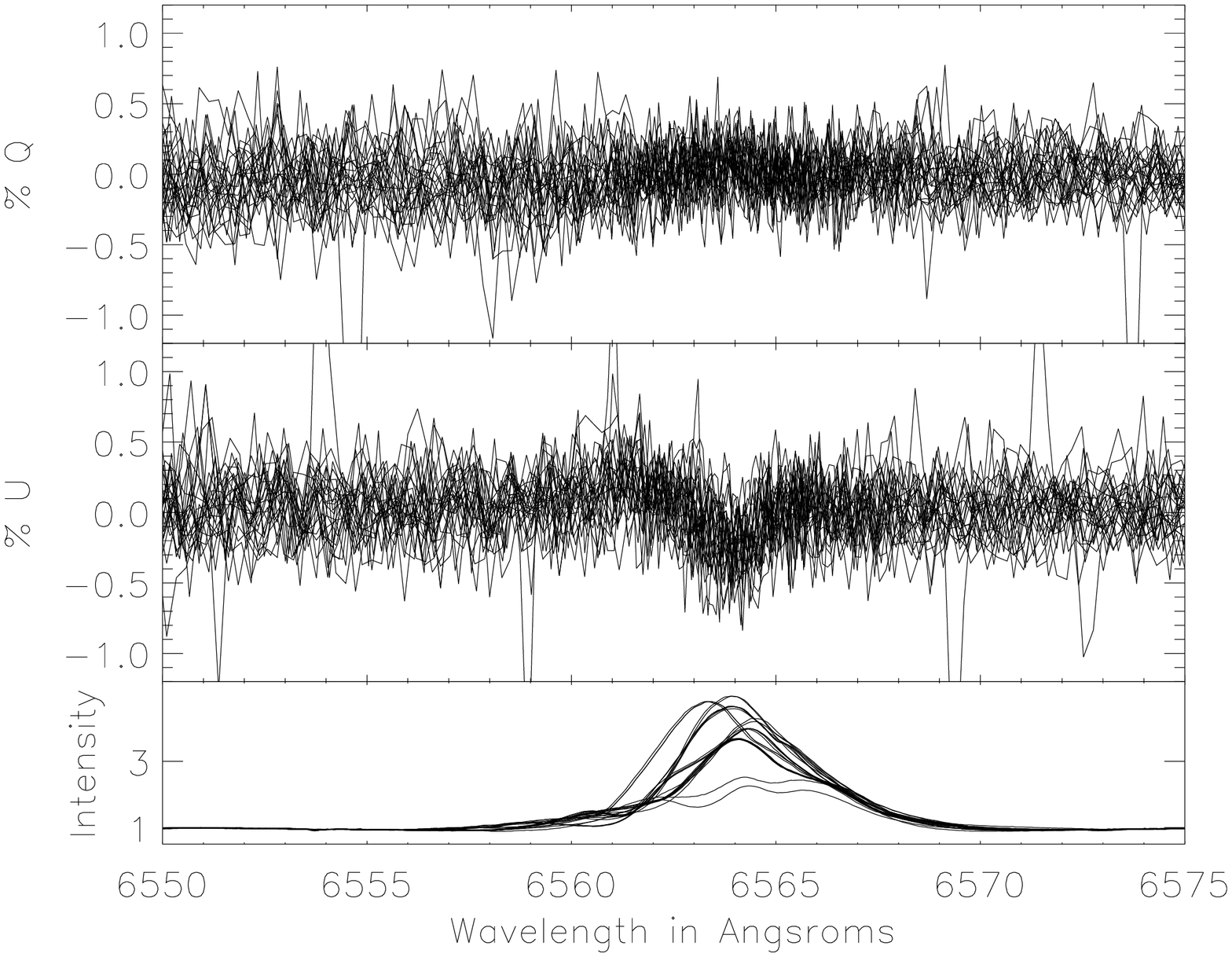}}
\quad
\subfloat[HD 144432]{\label{fig:hd144}
\includegraphics[ width=0.21\textwidth, angle=90]{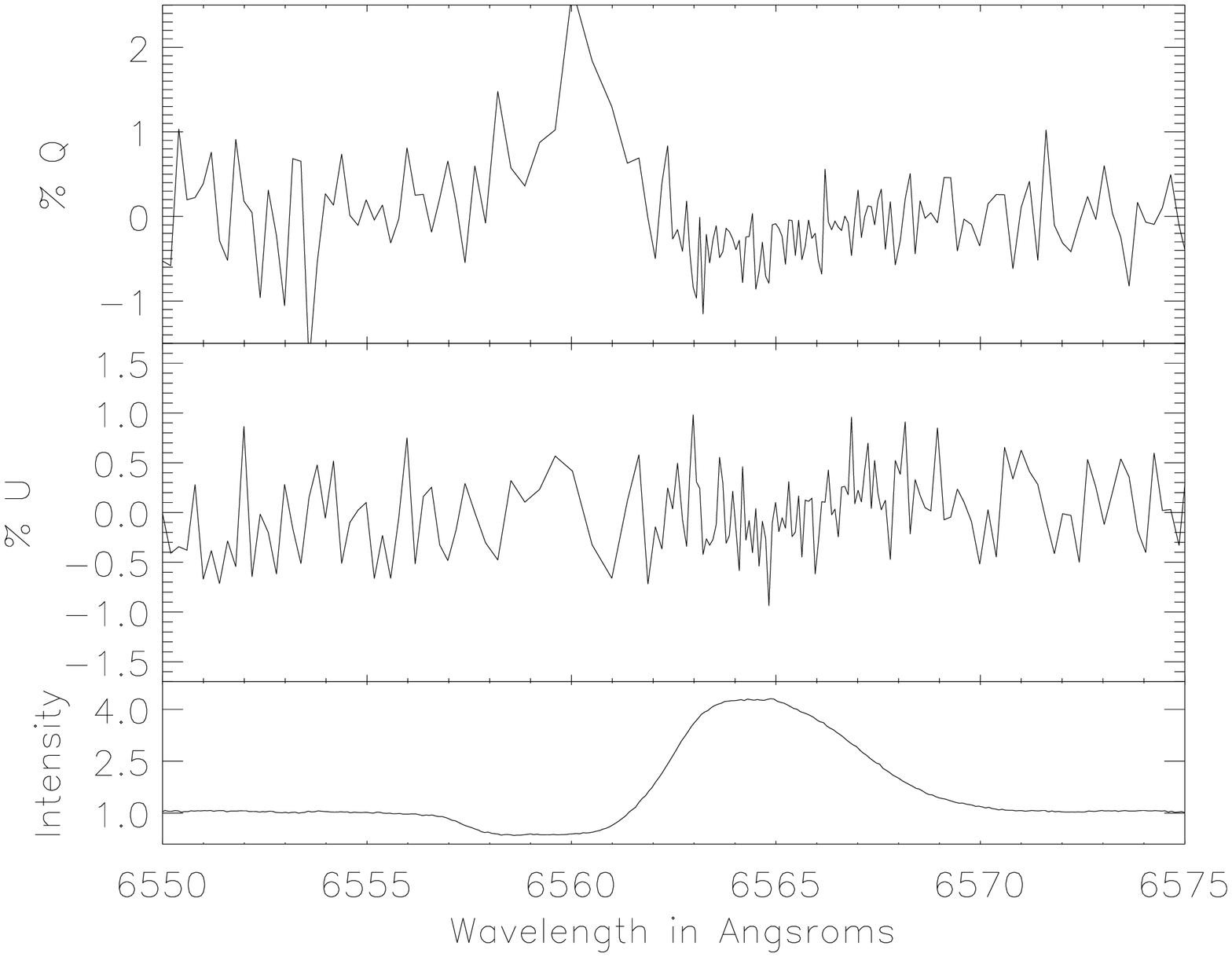}}
\quad
\subfloat[MWC 758]{\label{fig:mwc758}
\includegraphics[ width=0.21\textwidth, angle=90]{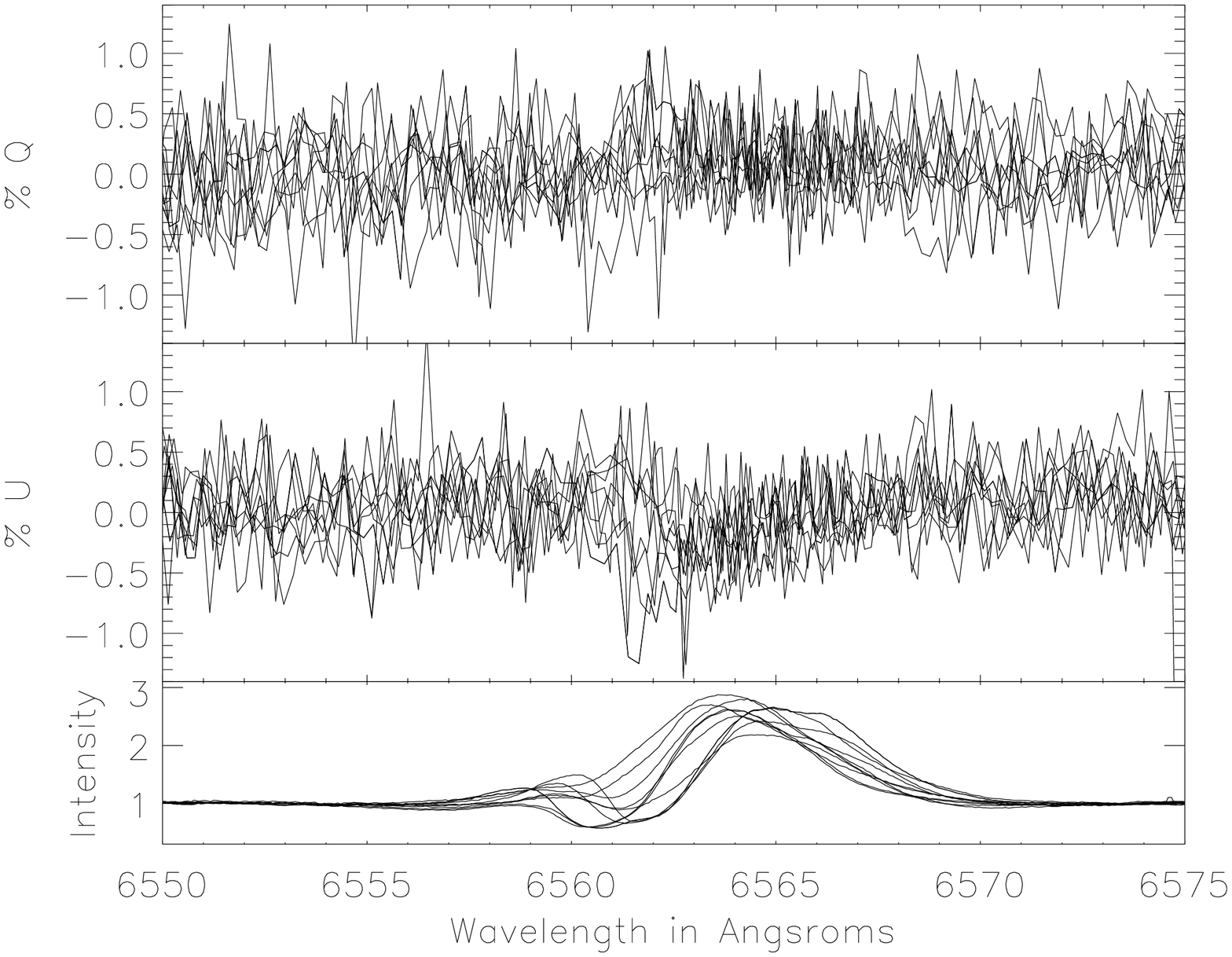}}
\quad
\subfloat[HD 169142]{\label{fig:hd169}
\includegraphics[ width=0.21\textwidth, angle=90]{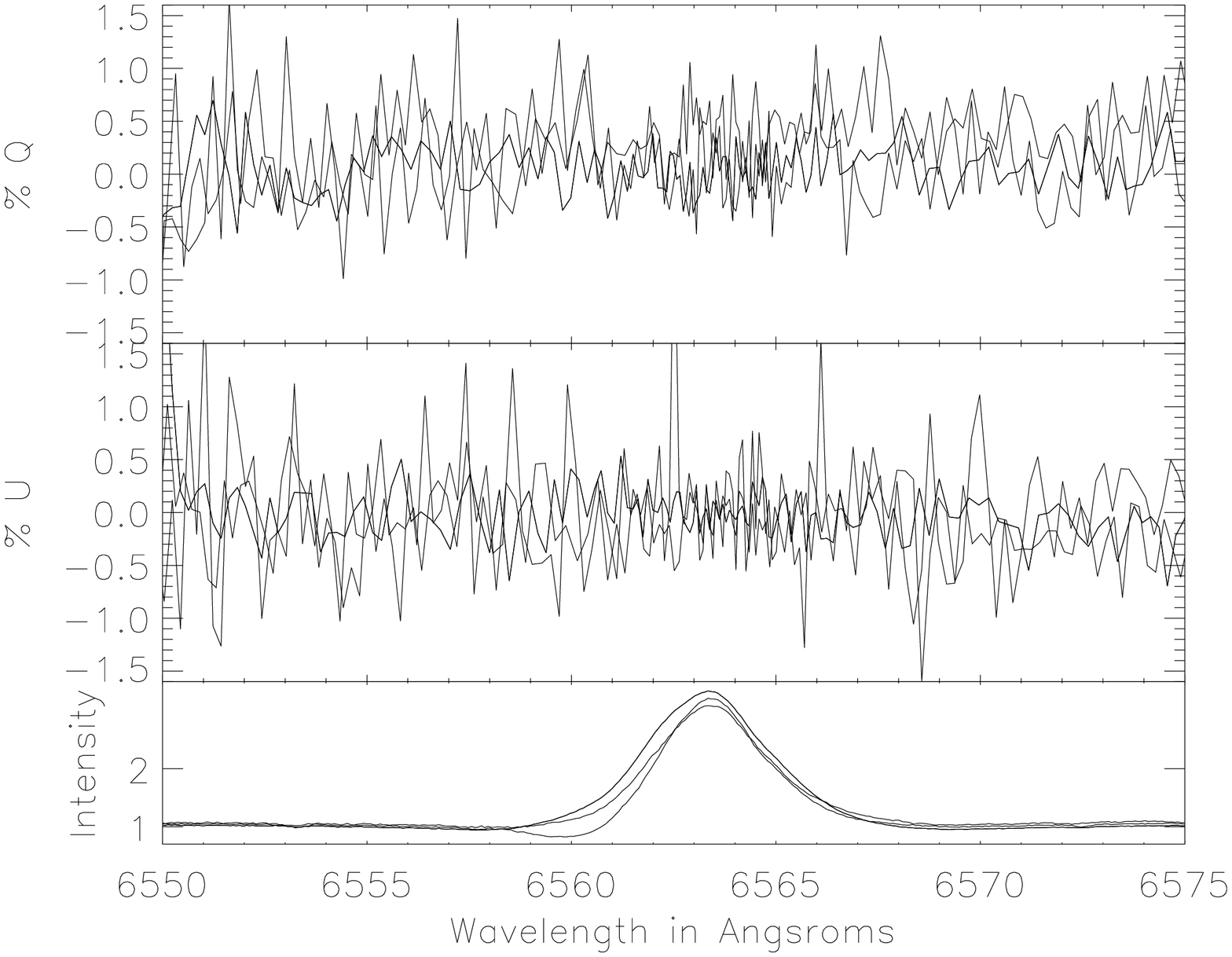}}
\quad
\subfloat[KMS 27]{\label{fig:kms27}
\includegraphics[ width=0.21\textwidth, angle=90]{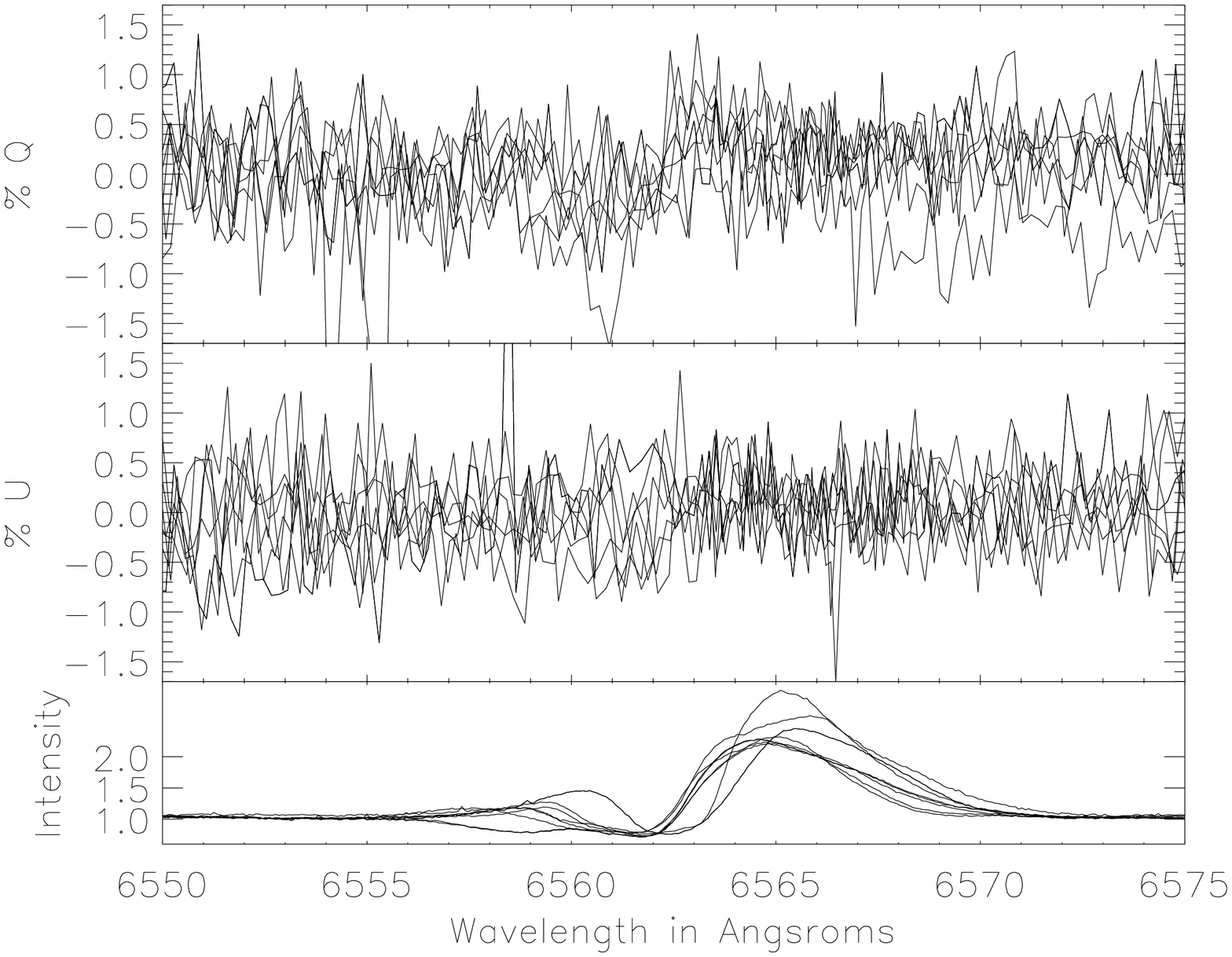}}
\quad
\subfloat[V 1295Aql]{\label{fig:v1295}
\includegraphics[ width=0.21\textwidth, angle=90]{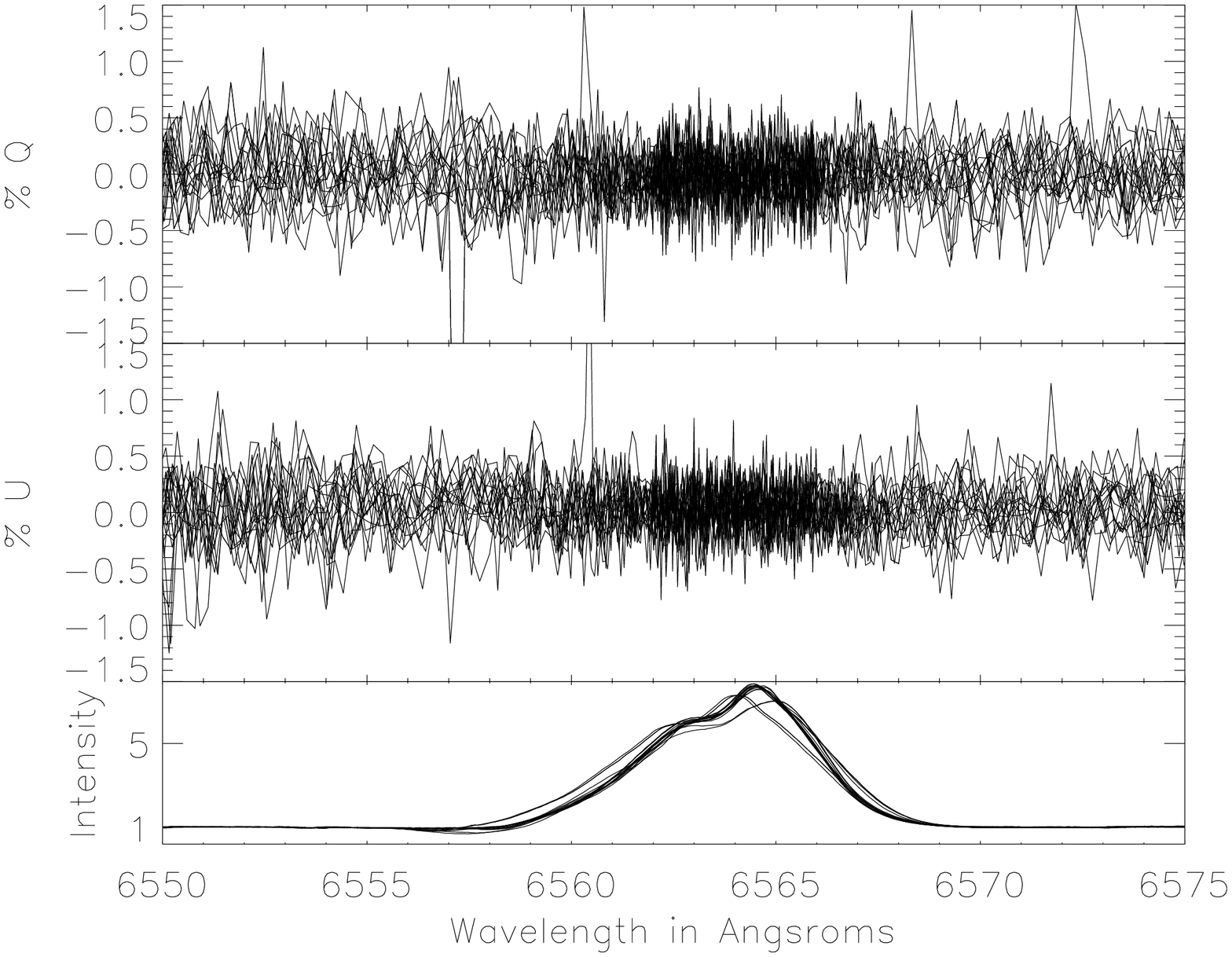}}
\quad
\subfloat[HD 35187]{\label{fig:hd351}
\includegraphics[ width=0.21\textwidth, angle=90]{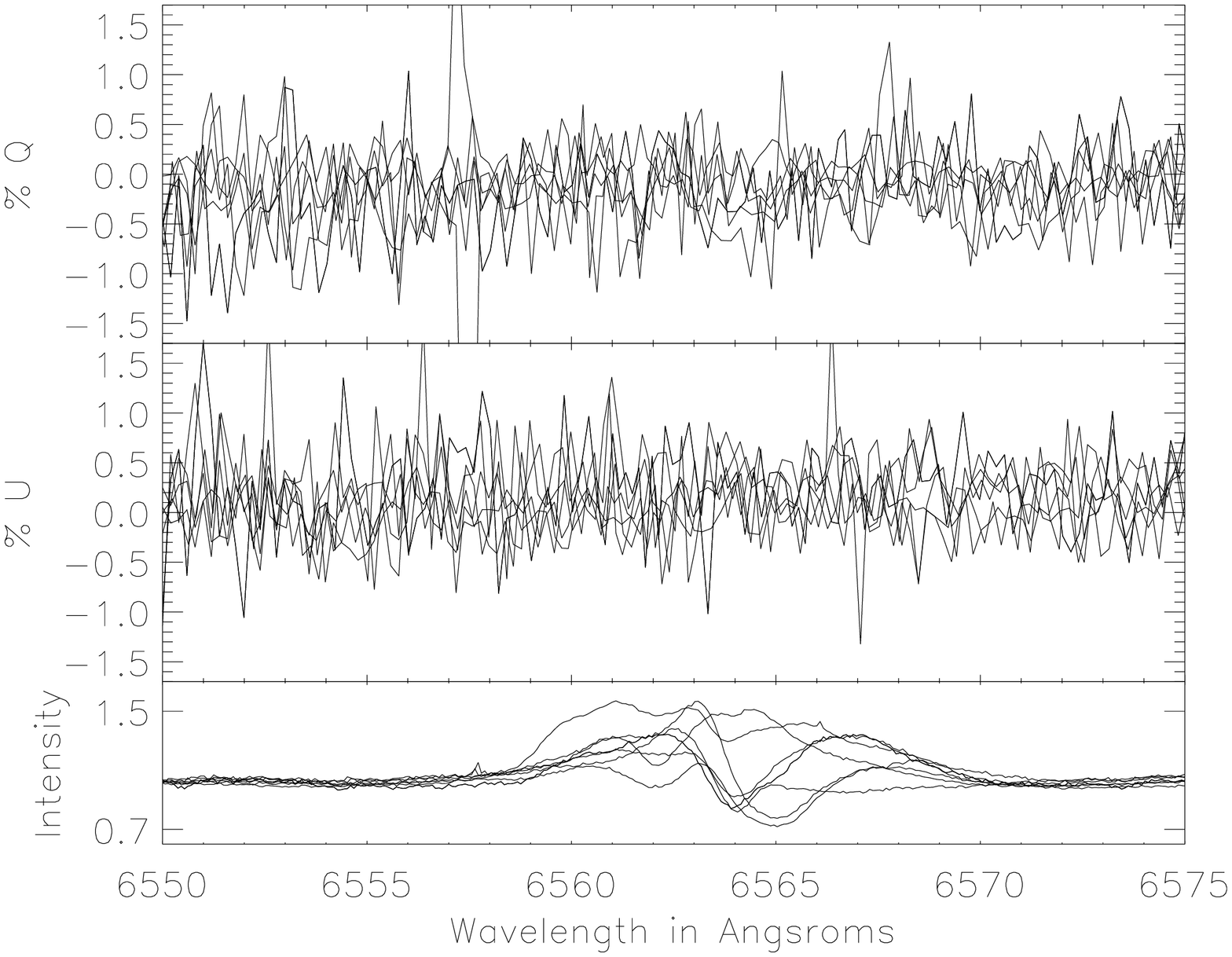}}
\quad
\subfloat[HD 142666]{\label{fig:hd142}
\includegraphics[ width=0.21\textwidth, angle=90]{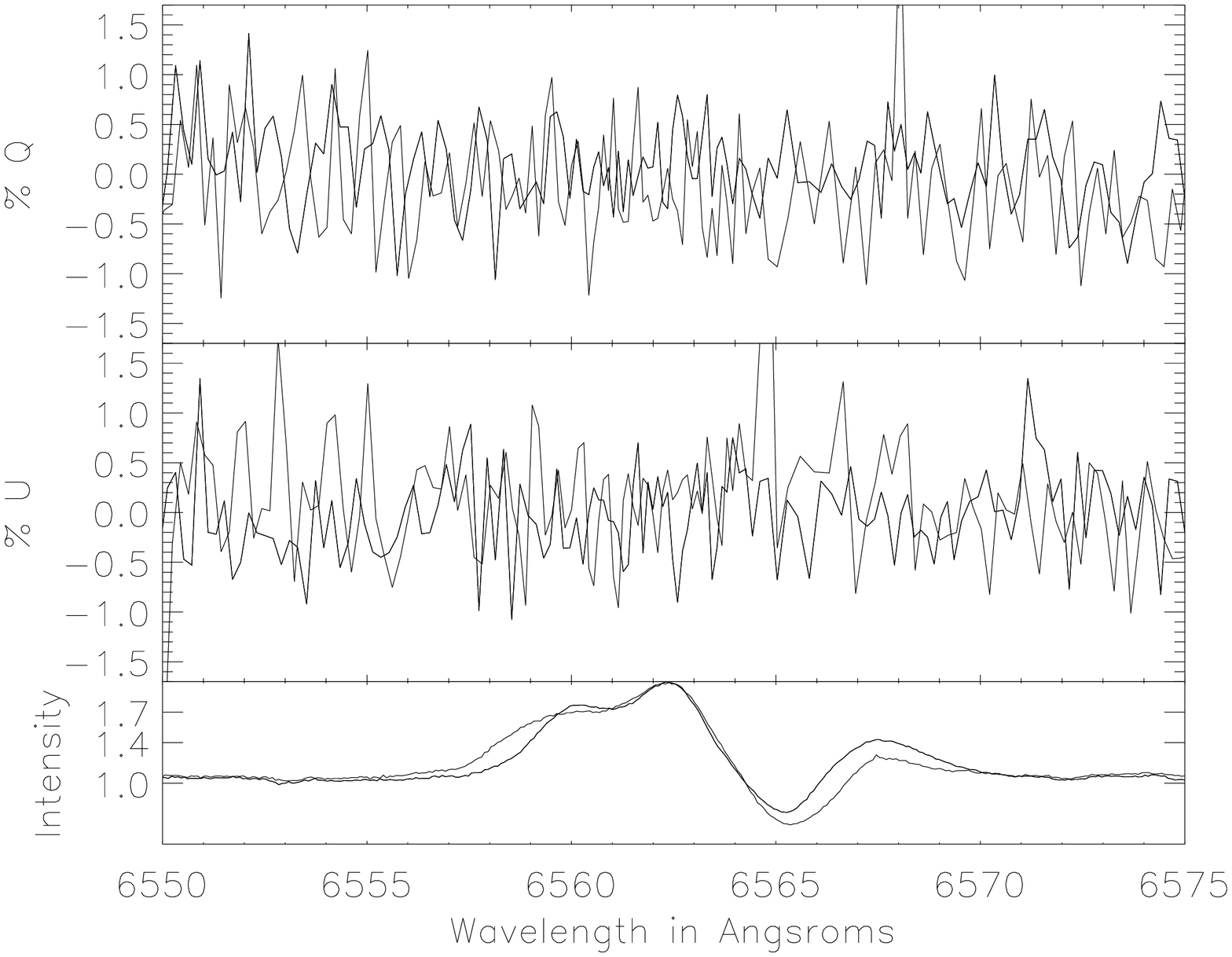}}
\quad
\subfloat[T Ori]{\label{fig:tori}
\includegraphics[ width=0.21\textwidth, angle=90]{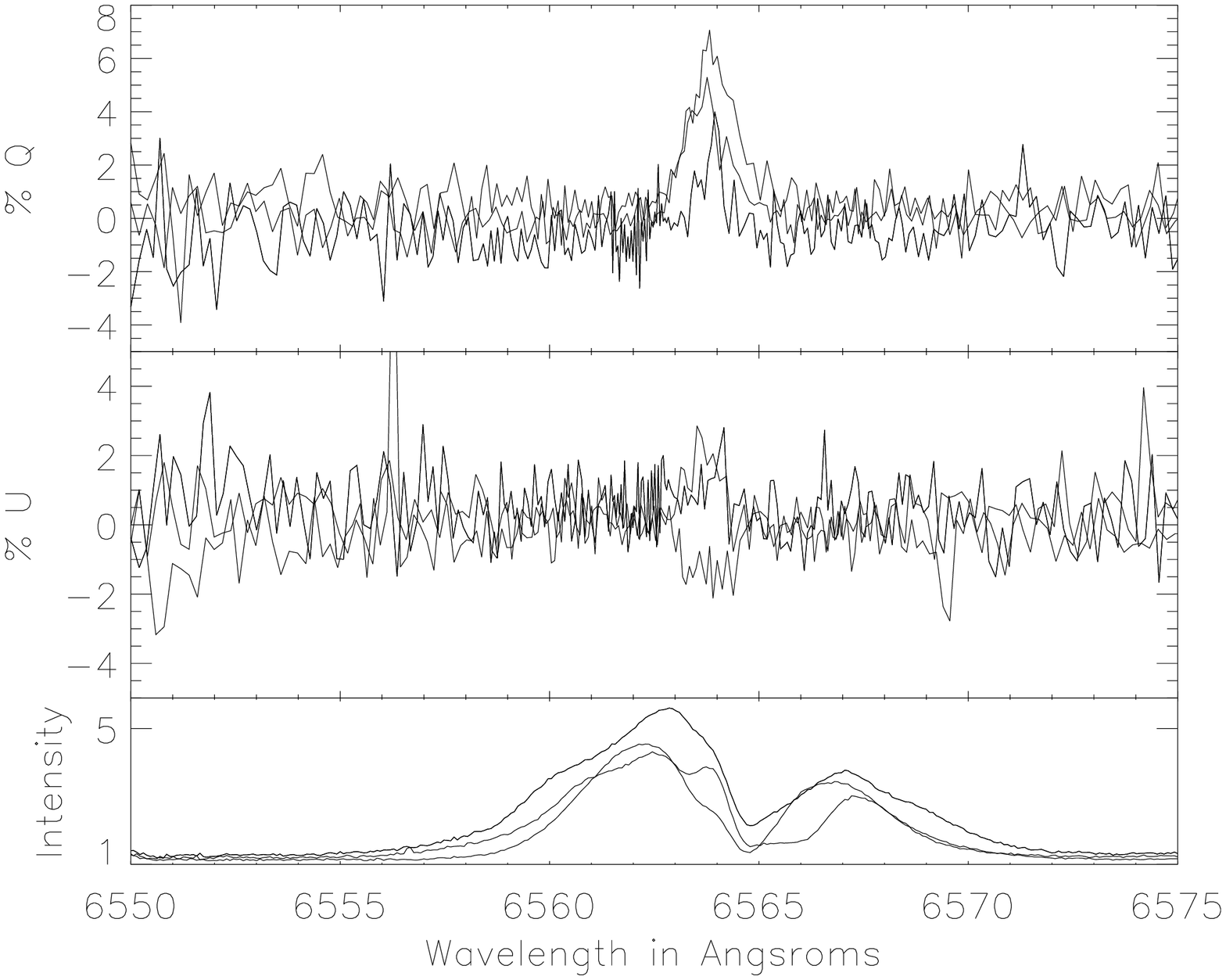}}
\caption[Herbig Ae/Be Spectropolarimetry I]{HAe/Be Spectropolarimetry I}
\label{fig:haebe-specpol1}
\end{figure}

	There are 166 polarization measurements presented in figure \ref{fig:abaur}. Figure \ref{fig:swp-abaur-indiv} shows an individual polarized spectrum that shows a very clear detection of polarization at the 1\% level in both q and u. The polarization change is largest in the absorptive component of the P-Cygni profile and the polarization at maximum emission is indistinguishable from the continuum polarization. Figure \ref{fig:swp-abaur-qu} shows the corresponding qu-plot for the line, 6547.9 to 6564.1{\AA}. There is a knot of points near (0,0) that represents the continuum. As wavelength increases, q and u both increase in amplitude but not entirely at the same wavelengths.  This gives rise to the loop - if both q and u increased simultaneously, the qu-plot would show a line. 

	With a strong P-Cygni profile having absorption/emission ratio around 20, the possibility of a 1/I systematic error must be explored. Figure \ref{fig:swp-abaur-pfx} shows the normalized flux polarized change, basically the difference between the detected Stokes Q and Stokes Q for a line with a constant degree of polarization. The polarized flux for a purely 1/I error would be completely flat. That the polarized flux has both positive and negative components with different structure from Q to U shows that this is not a systematic effect. No shift in continuum polarization before computing q*I could cause this Q spectrum.

	Polarization in absorption was the only type of signature detected for this system. Figure \ref{fig:swp-abaur-polabs} shows the polarization for three different epochs with three very different absorption spectra. The polarization shows a clear detection correlated with the absorption. The detection was always around 1\% but the exact value of the position angle of the change is dependent on the rotation of the plane of polarization by the telescope as well as the source. Though the width of the polarization effect in all HiVIS observations is strongly correlated with the width of the absorptive component, there are a number of observations which show stronger polarization closer to line center, such as in the third (bottom) panel of figure \ref{fig:swp-abaur-polabs}.

	Curiously, the ESPaDOnS archive observations shown in figure \ref{fig:swp-abaur-esp}, only a small polarization is detected at the 0.2\% level with a more complicated morphology. Given the overwhelming number of detections with HiVIS over a wide range of conditions, and the variability of other sources detected with ESPaDOnS, it must be concluded that this star is strongly variable in it's spectropolarimetric signatures.

\onecolumn

\begin{figure}
\centering
\subfloat[MWC 158]{\label{fig:mwc158}
\includegraphics[ width=0.21\textwidth, angle=90]{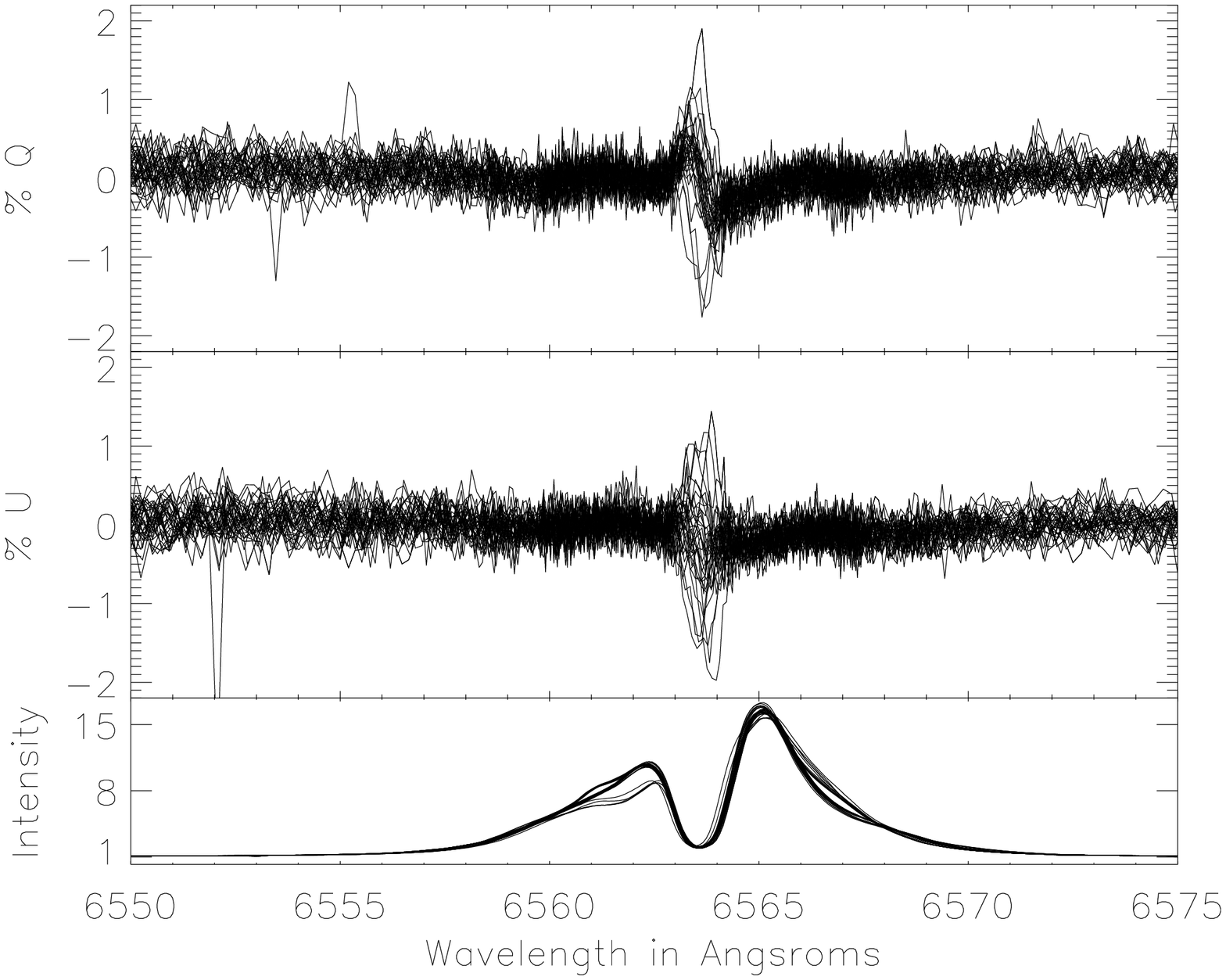}}
\quad
\subfloat[HD 58647]{\label{fig:hd}
\includegraphics[ width=0.21\textwidth, angle=90]{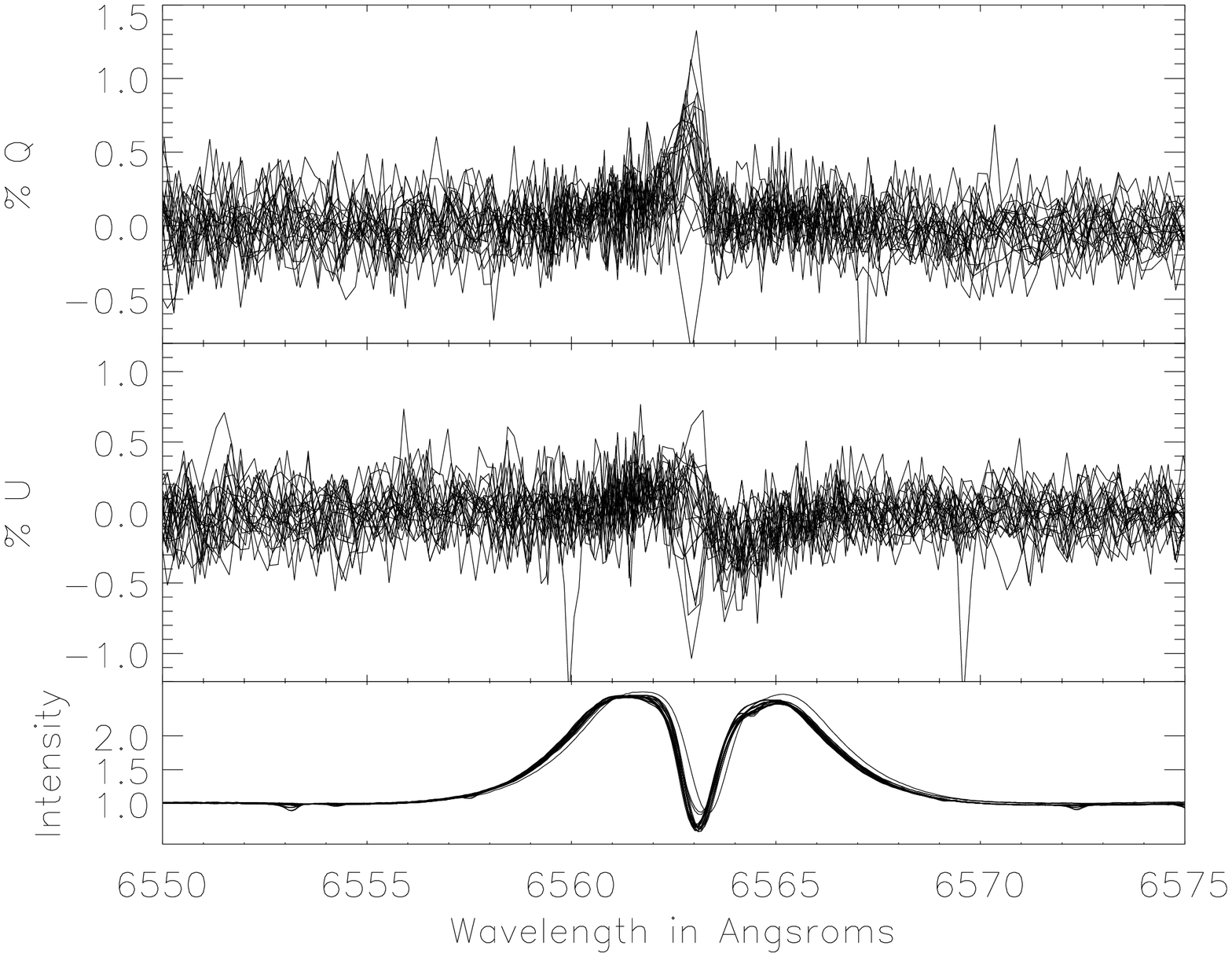}}
\quad
\subfloat[MWC 361]{\label{fig:mwc361}
\includegraphics[ width=0.21\textwidth, angle=90]{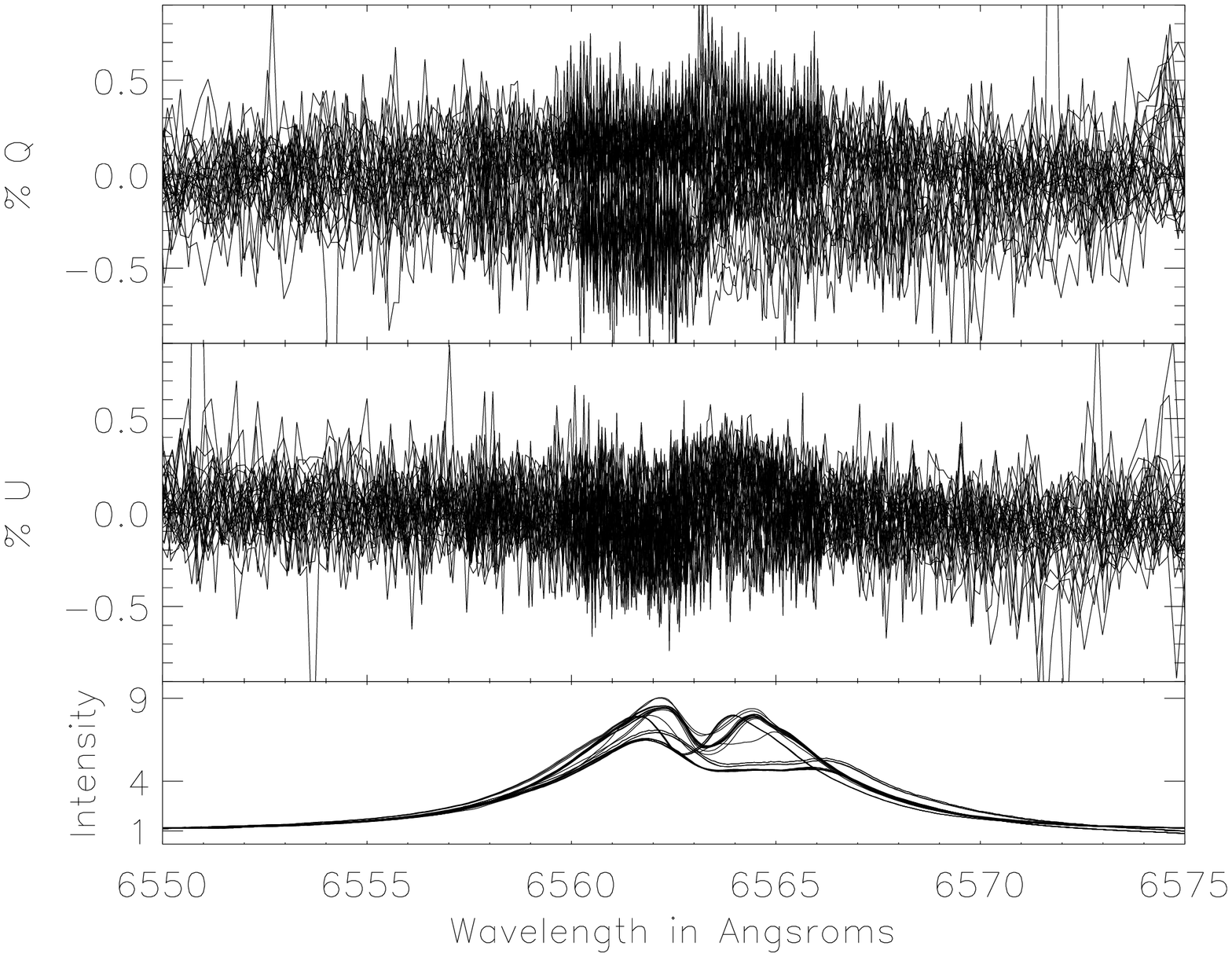}}
\quad
\subfloat[HD 141569]{\label{fig:hd141}
\includegraphics[ width=0.21\textwidth, angle=90]{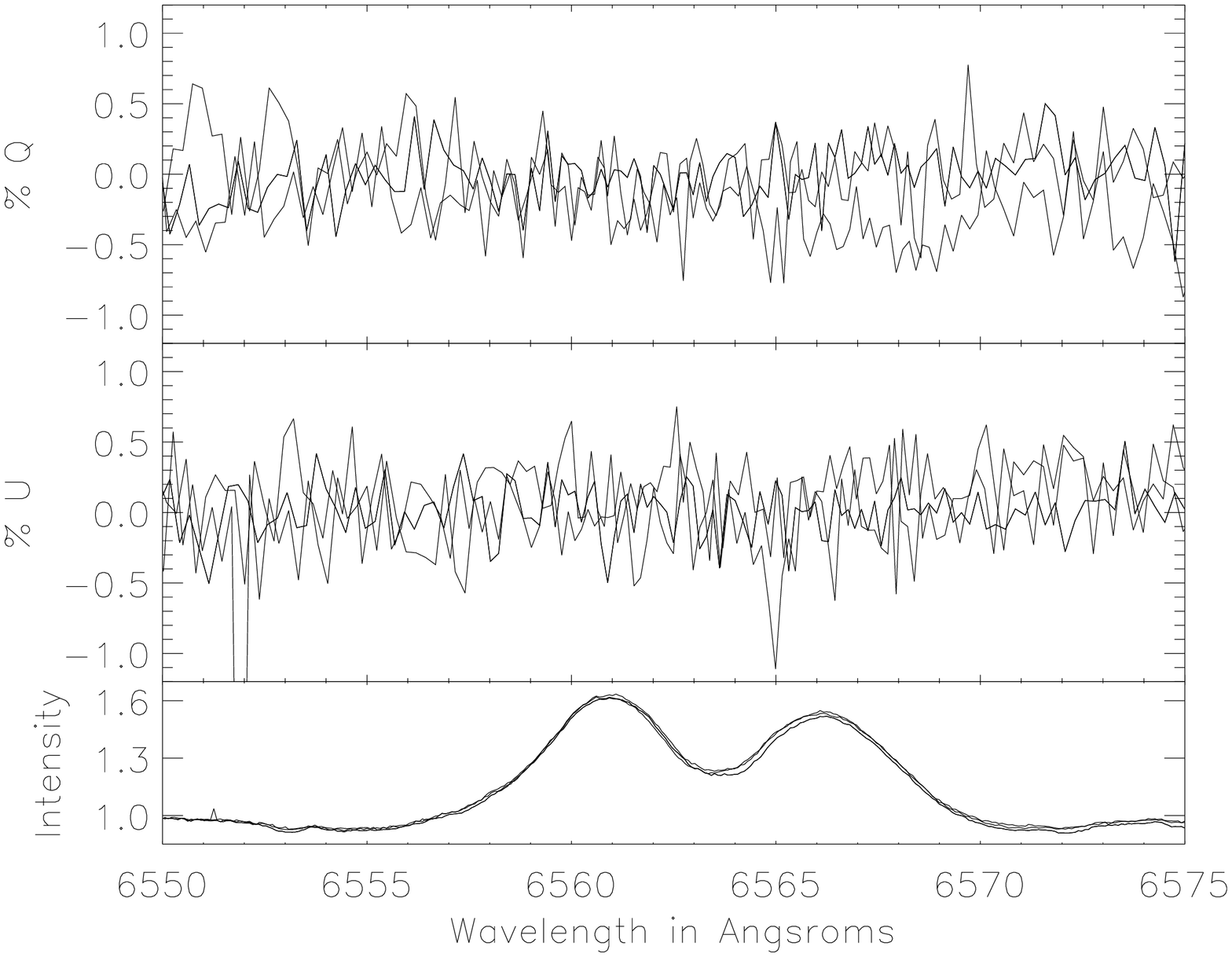}}
\quad
\subfloat[51 Oph]{\label{fig:51oph}
\includegraphics[ width=0.21\textwidth, angle=90]{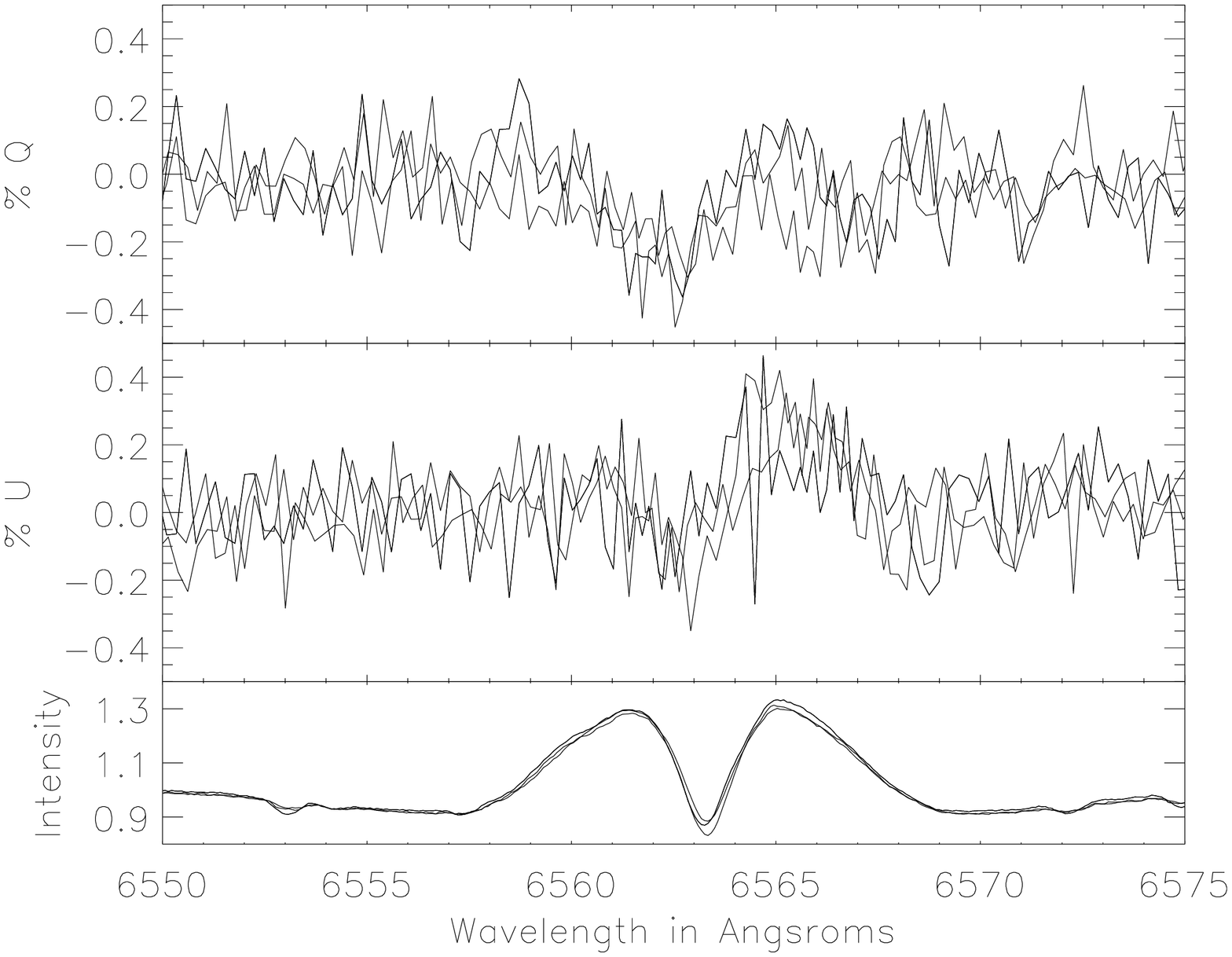}}
\quad
\subfloat[XY Per]{\label{fig:xyper}
\includegraphics[ width=0.21\textwidth, angle=90]{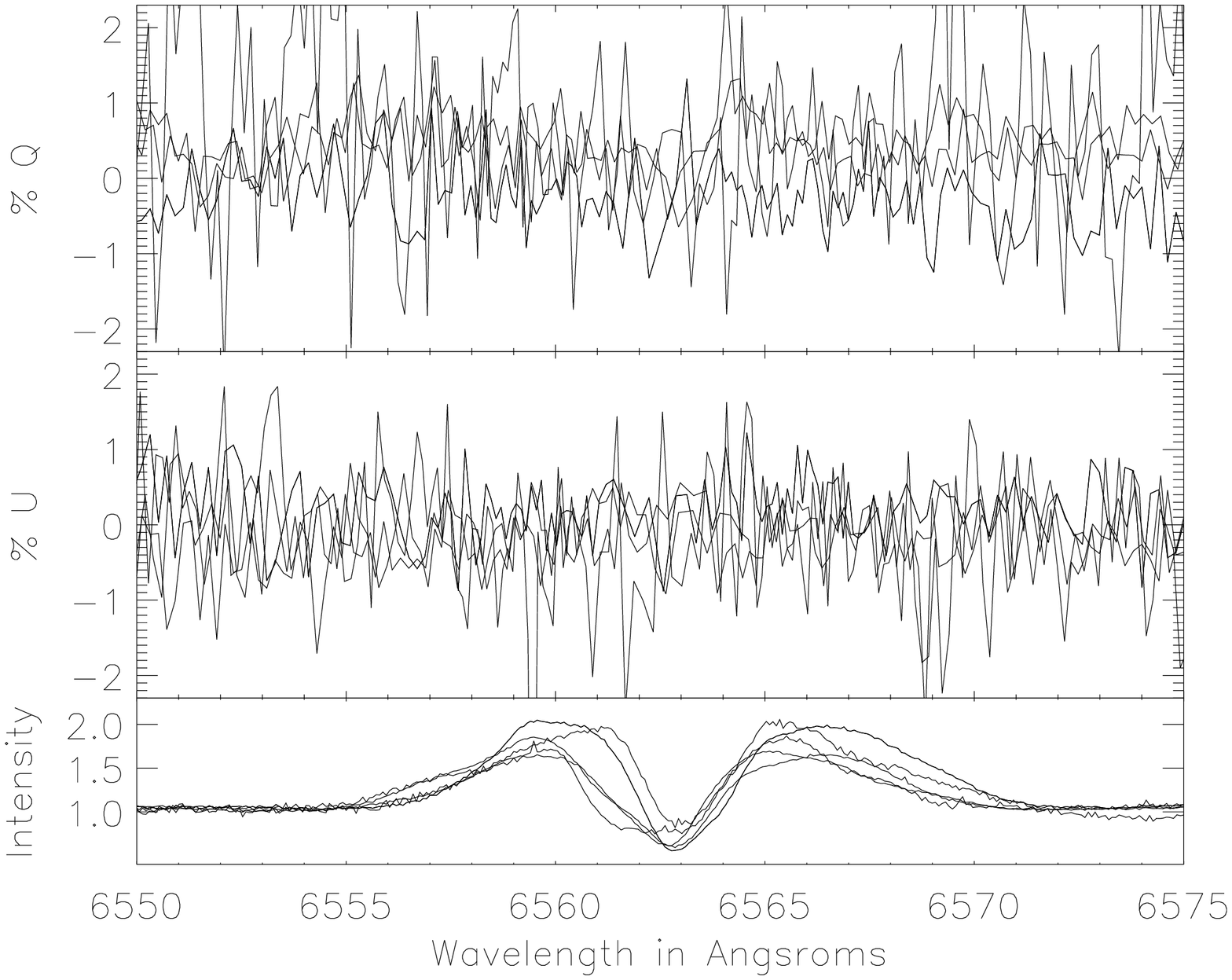}}
\quad
\subfloat[MWC 166]{\label{fig:mwc166}
\includegraphics[ width=0.21\textwidth, angle=90]{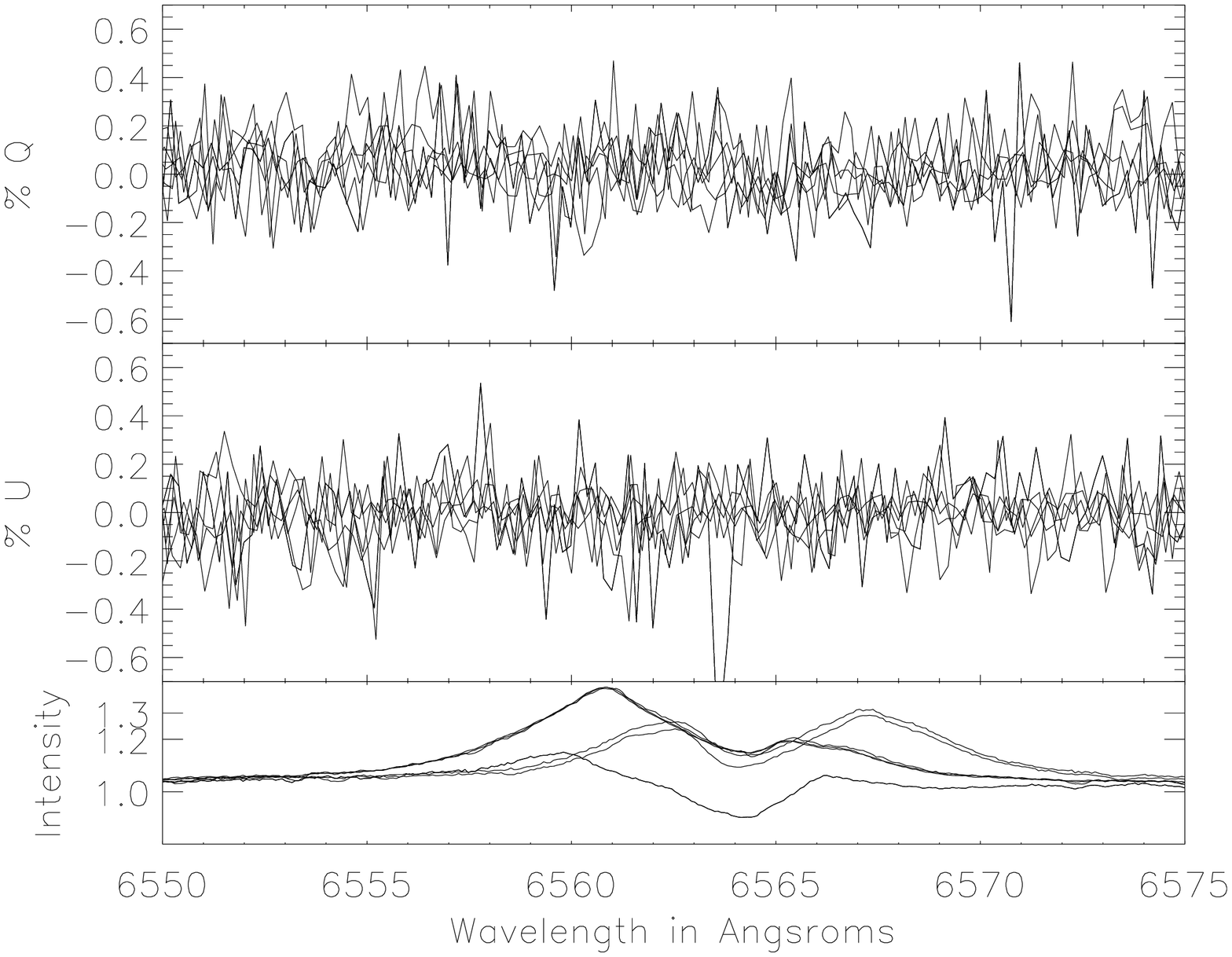}}
\quad
\subfloat[MWC 170]{\label{fig:mwc170}
\includegraphics[ width=0.21\textwidth, angle=90]{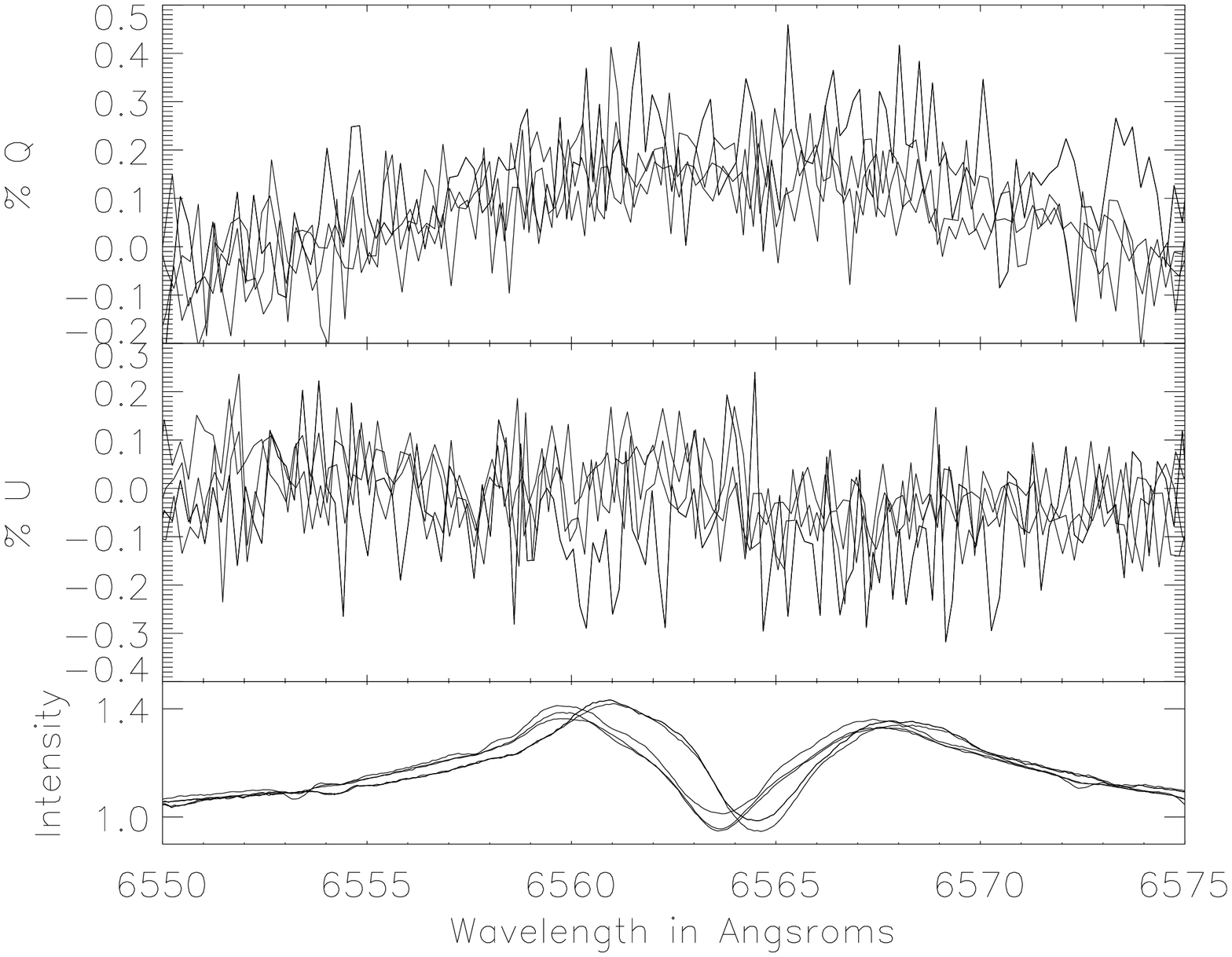}}
\quad
\subfloat[HD 45677]{\label{fig:hd456}
\includegraphics[ width=0.21\textwidth, angle=90]{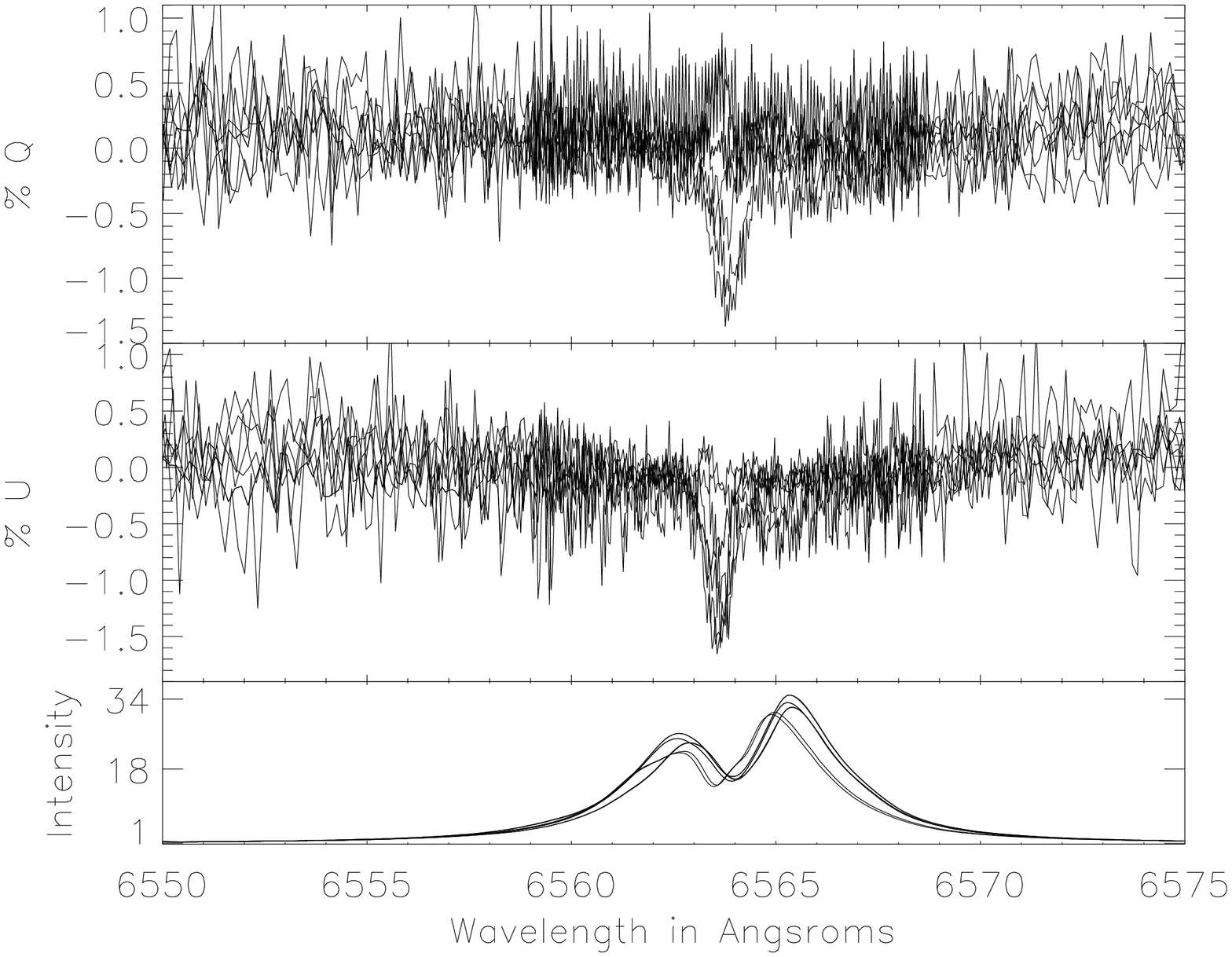}}
\quad
\subfloat[MWC 147]{\label{fig:mwc147}
\includegraphics[ width=0.21\textwidth, angle=90]{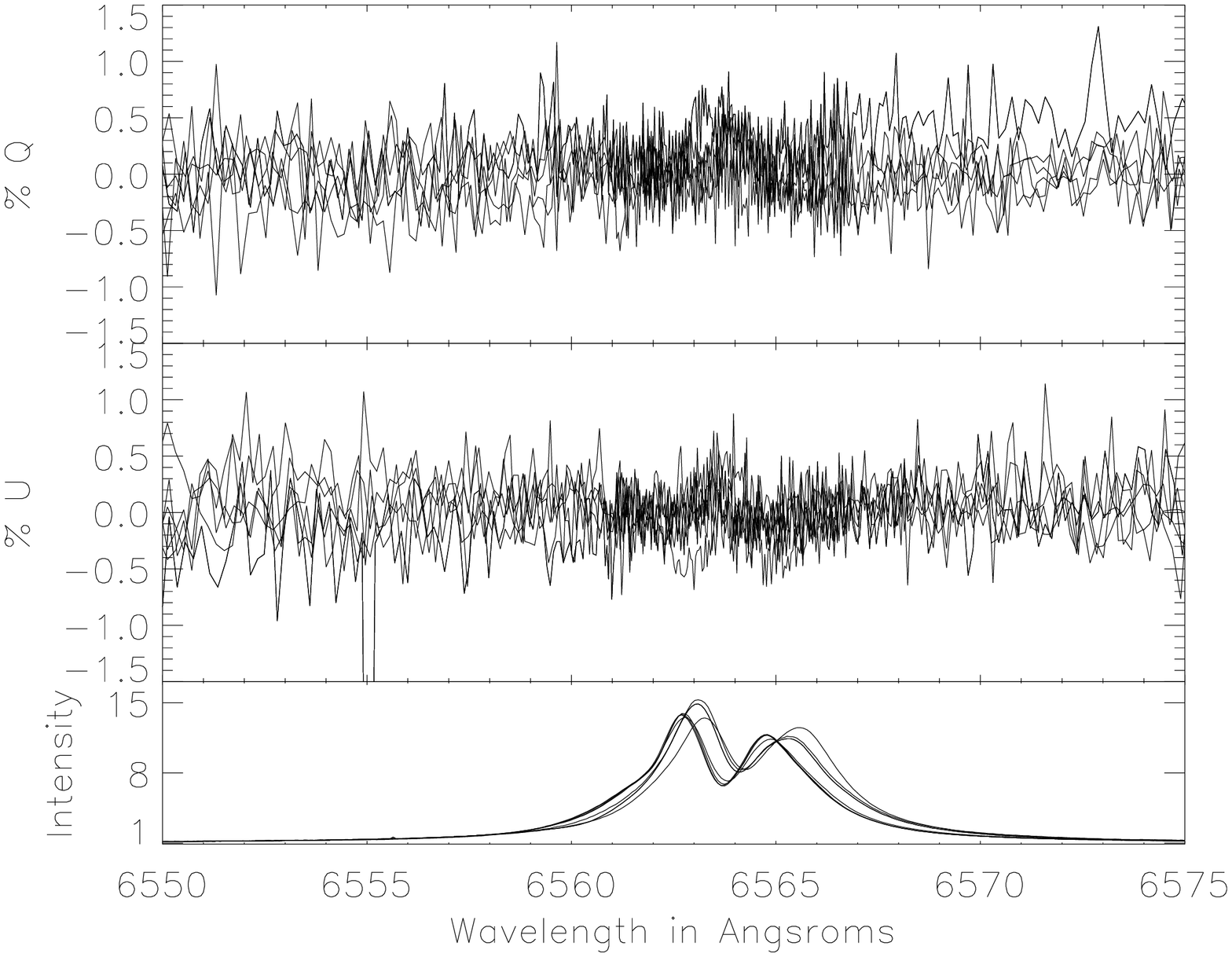}}
\quad
\subfloat[Il Cep]{\label{fig:ilcep}
\includegraphics[ width=0.21\textwidth, angle=90]{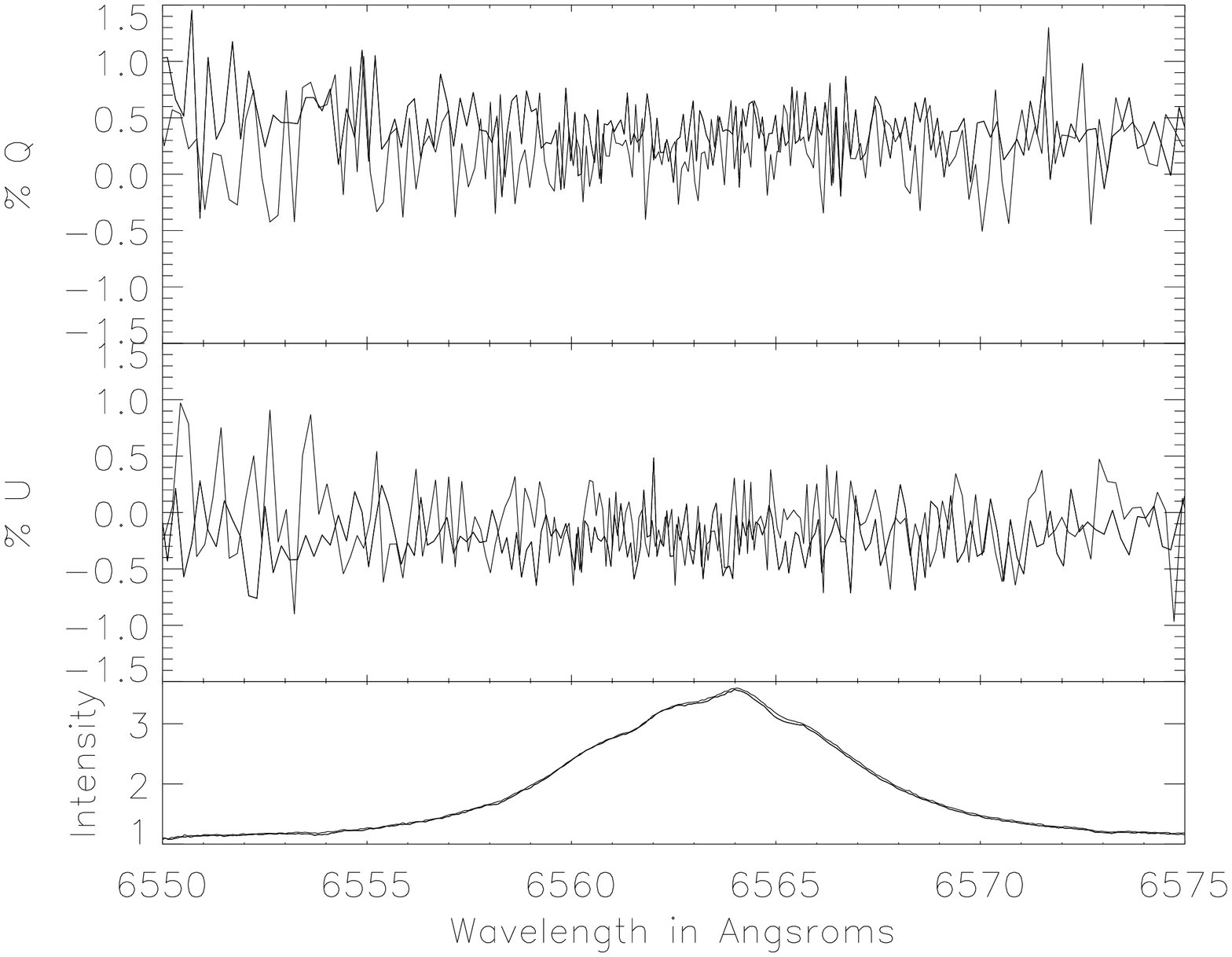}}
\quad
\subfloat[MWC 442]{\label{fig:mwc442}
\includegraphics[ width=0.21\textwidth, angle=90]{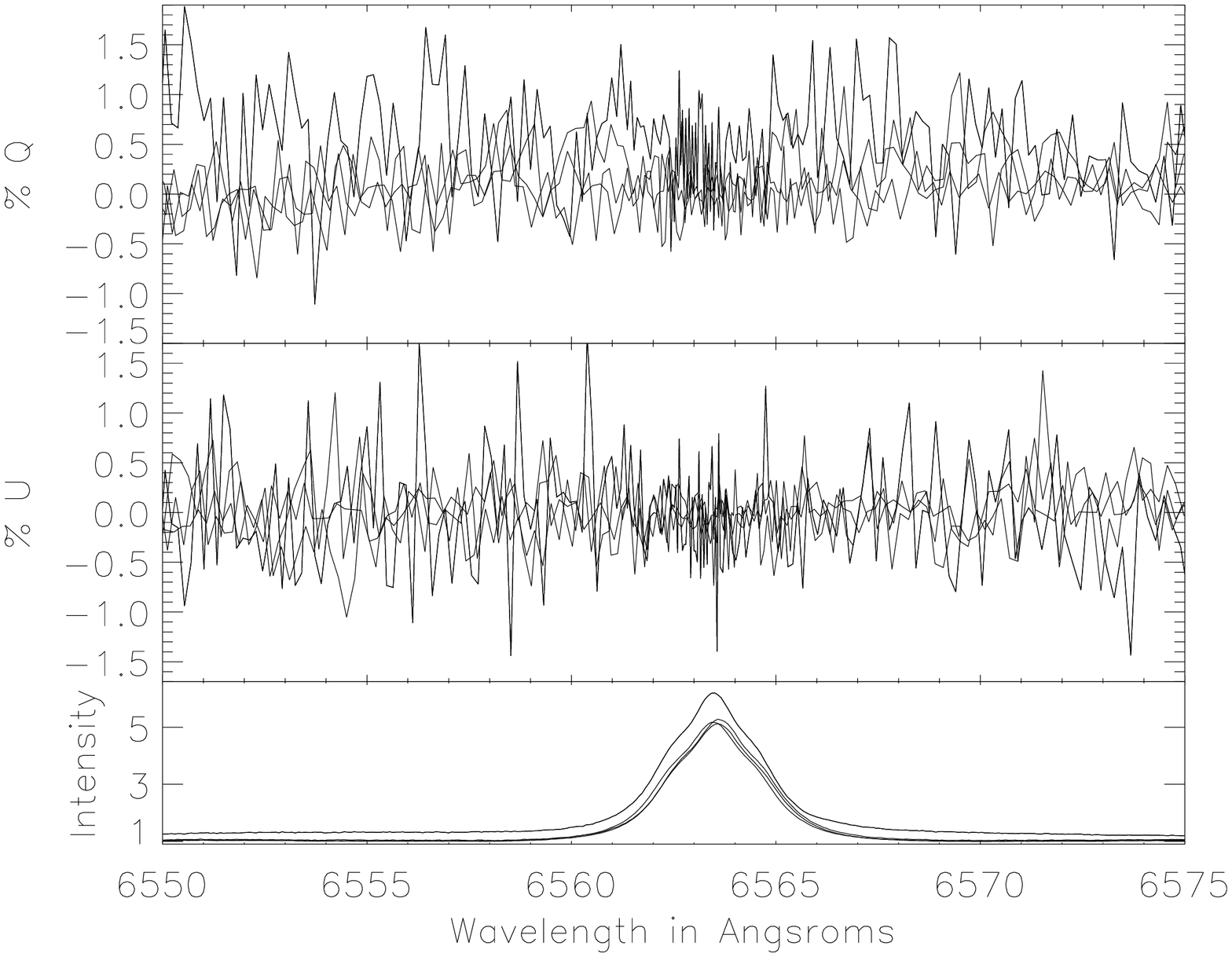}}
\quad
\subfloat[HD 35929]{\label{fig:hd359}
\includegraphics[ width=0.21\textwidth, angle=90]{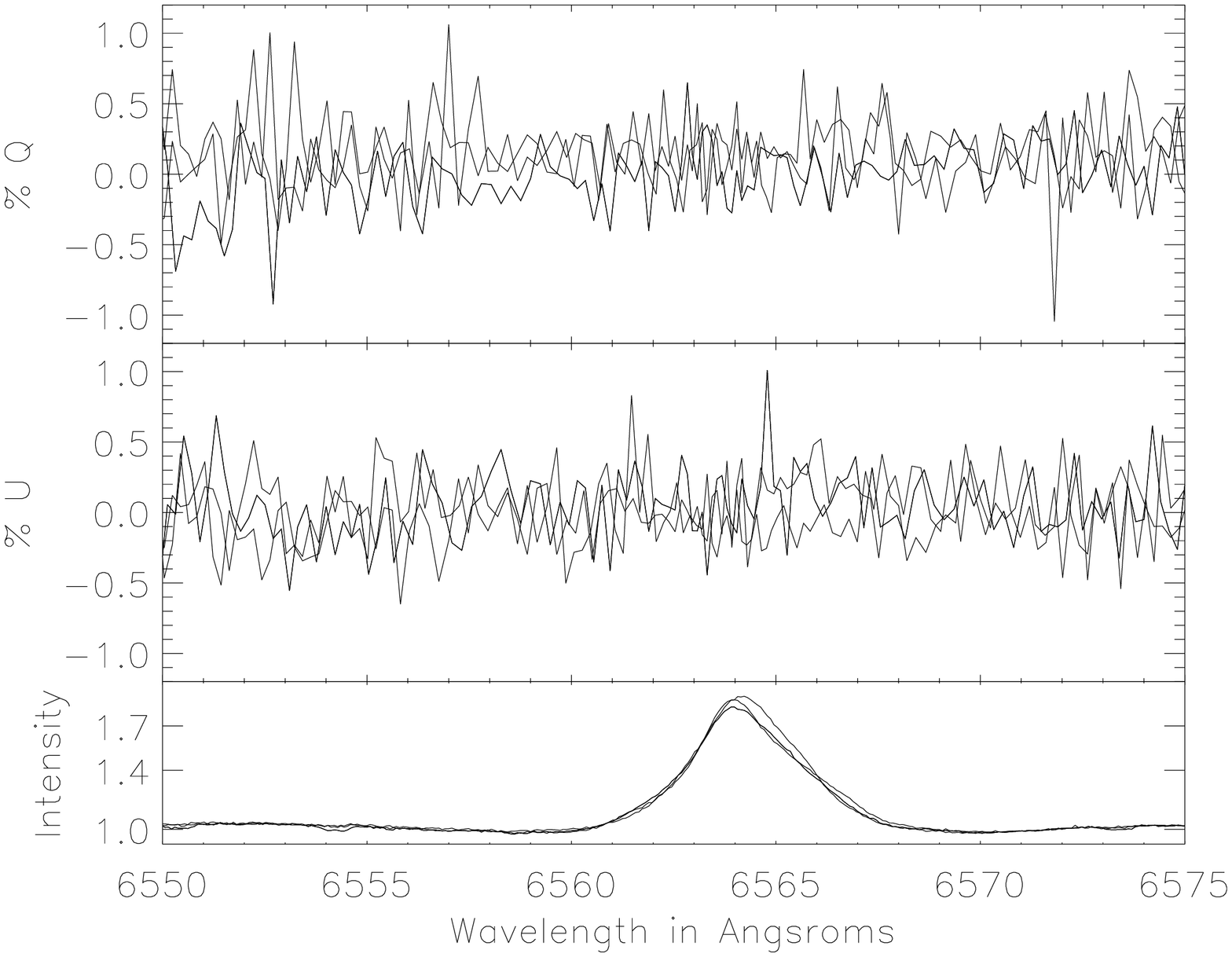}}
\quad
\subfloat[GU CMa]{\label{fig:gucma}
\includegraphics[ width=0.21\textwidth, angle=90]{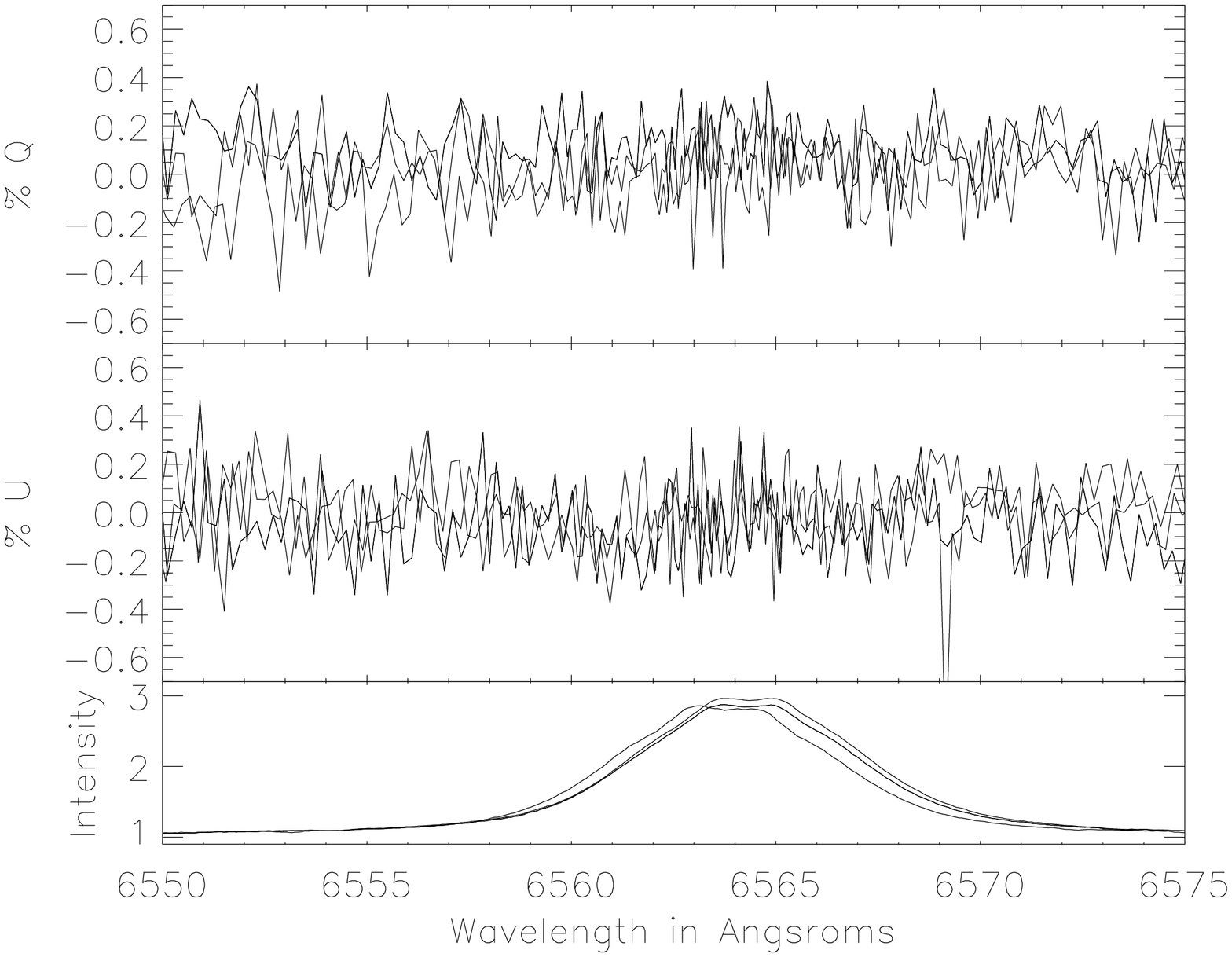}}
\quad
\subfloat[HD 38120]{\label{fig:hd381}
\includegraphics[ width=0.21\textwidth, angle=90]{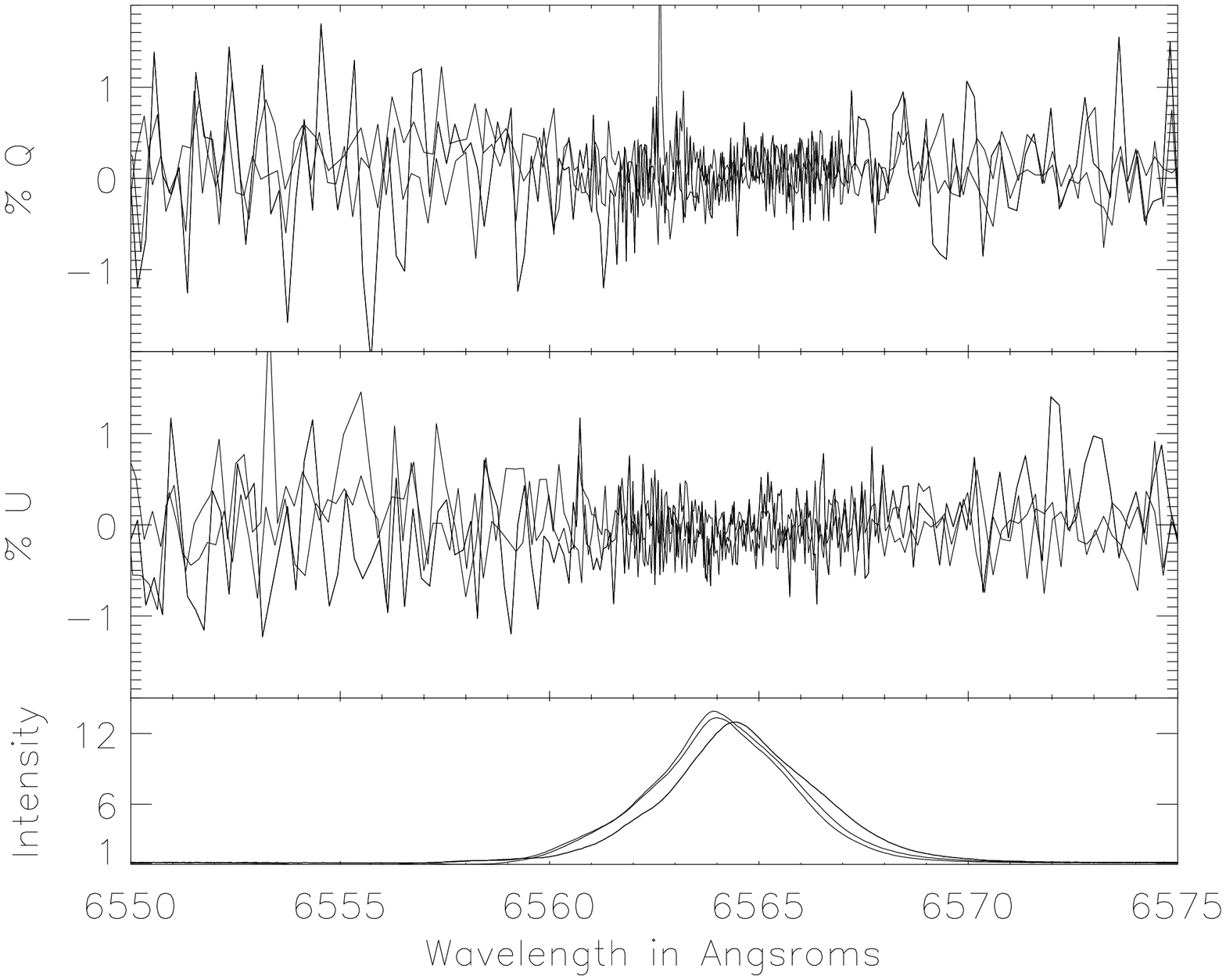}}
\caption[Herbig Ae/Be Spectropolarimetry II]{HAe/Be Spectropolarimetry II}
\label{fig:haebe-specpol2}
\end{figure}

\begin{figure}
\centering
\subfloat[AB Aurigae Polarization Example]{\label{fig:swp-abaur-indiv}
\includegraphics[ width=0.35\textwidth, angle=90]{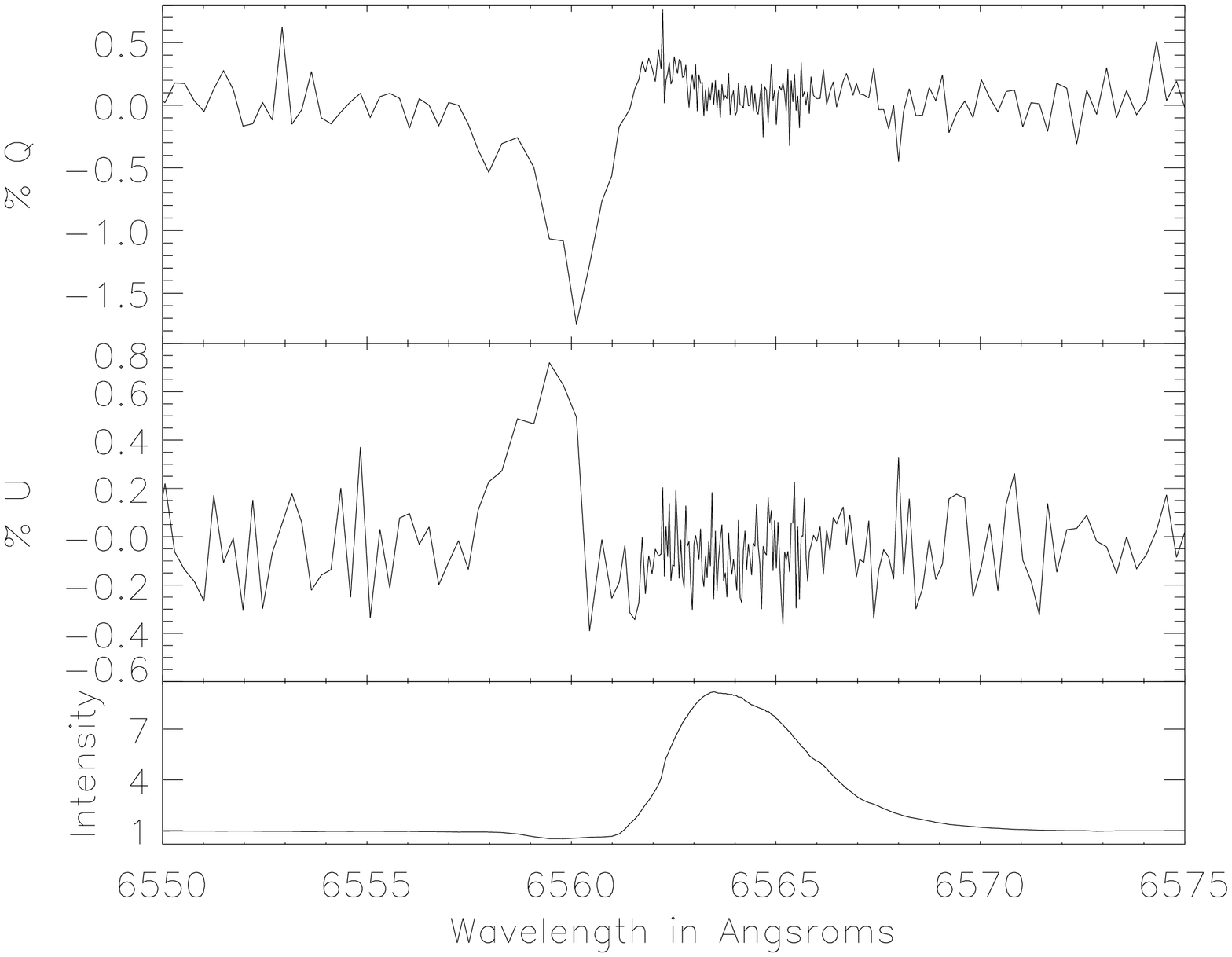}}
\quad
\subfloat[AB Aurigae QU Plot]{\label{fig:swp-abaur-qu}
\includegraphics[ width=0.35\textwidth, angle=90]{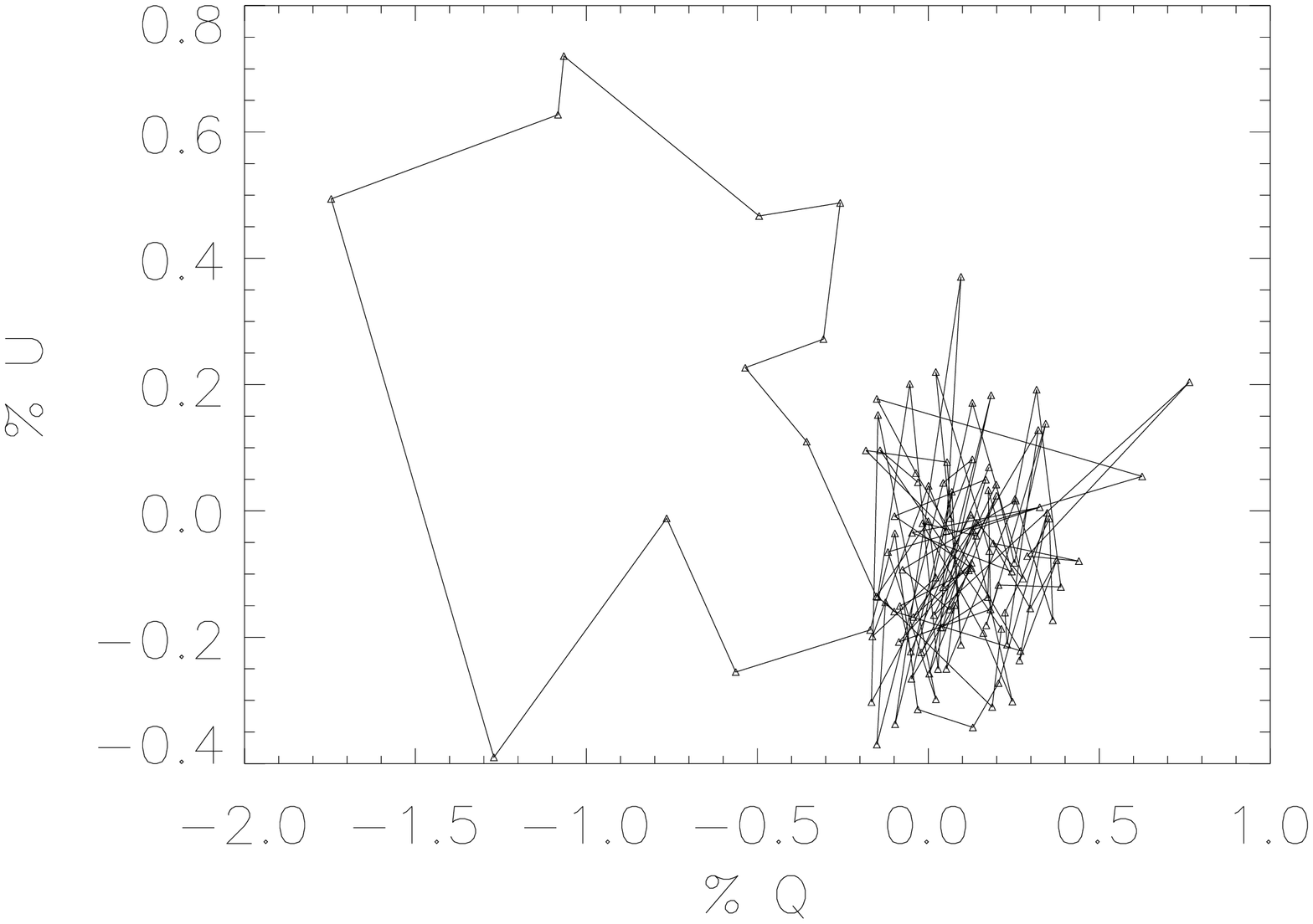}}
\caption[AB Aurigae Polarization]{{\bf a)} An example polarized spectrum for the AB Aurigae H$_\alpha$ line. The spectra have been binned to 5-times continuum. The top panel shows Stokes q, the middle panel shows Stokes u and the bottom panel shows the associated normalized H$_\alpha$ line. There is clearly a detection in the blue-shifted absorption of -1.5\% in q and 0.7\% in u. {\bf b)} This shows q vs u from 6547.9{\AA} to 6564.1{\AA}.  The knot of points at (0.0,0.0) represents the continuum.}
\label{fig:swp-abaur}
\end{figure}

\begin{figure}
\centering
\includegraphics[ width=0.5\textwidth, angle=90]{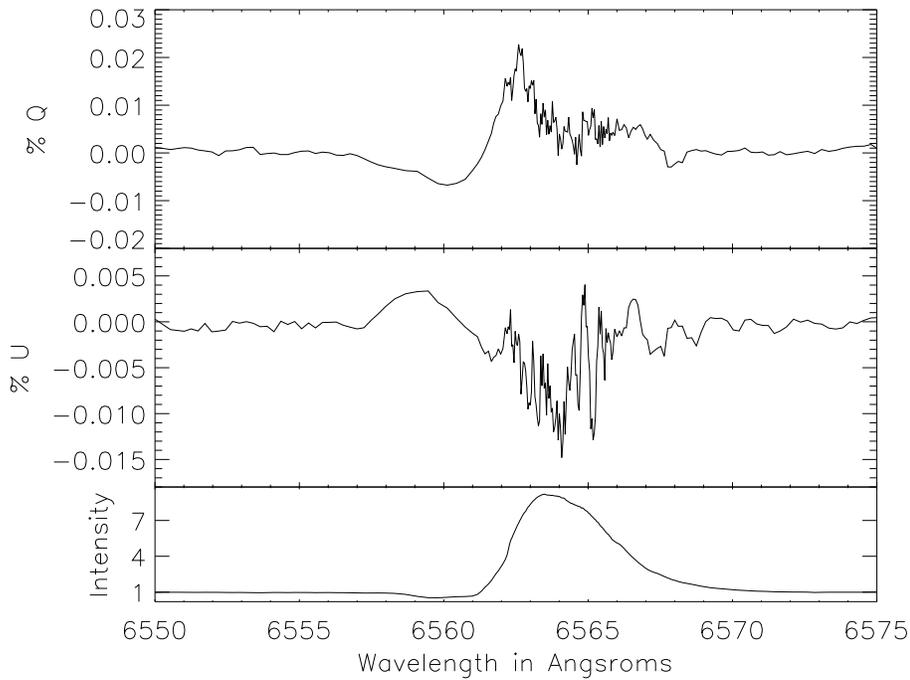}
\caption[AB Aurigae Polarized Flux]{The polarized flux (q*I) for the AB Aurigae spectropolarimetry in the previous plot.}
\label{fig:swp-abaur-pfx}
\end{figure}

\begin{figure}
\centering
\includegraphics[ width=0.45\textwidth, angle=90]{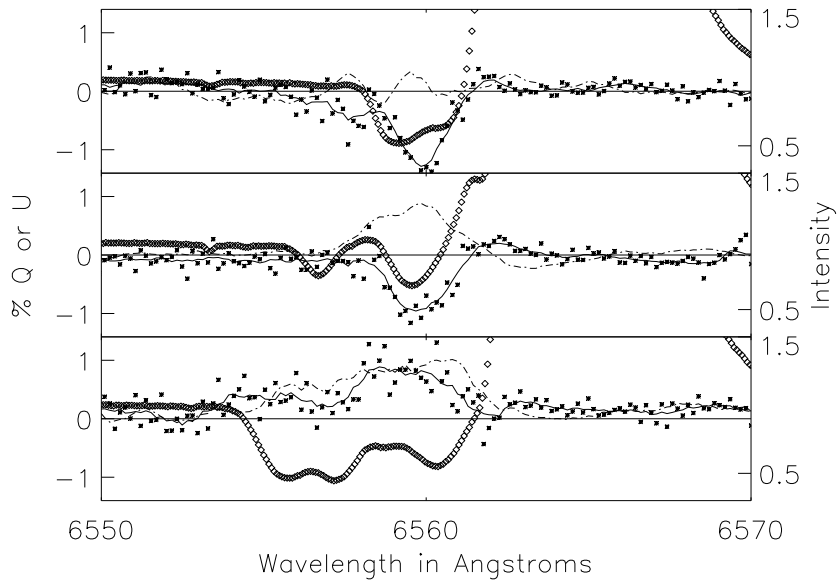}
\caption[AB Aurigae Absorptive Polarization]{Examples of spectropolarimetry for the AB Aurigae H$_\alpha$ line. The polarization is shown before any frame-rotation or flux-dependent binning to illustrate the data quality. The star symbols show the raw Stokes q spectrum binned 4:1 for clarity and the raw Stokes u is not shown. The solid and dashed curves are the smoothed Stokes q and u spectra. The diamonds show the normalized intensity of the line.  Each individual measurement has a raw continuum S/N of 300-800.  The polarization signature in the absorptive part of the line is clearly visible, and traces the absorptive component of the H$_\alpha$ line.}
\label{fig:swp-abaur-polabs}
\end{figure}

\begin{figure}
\centering
\includegraphics[width=0.45\textwidth, angle=90]{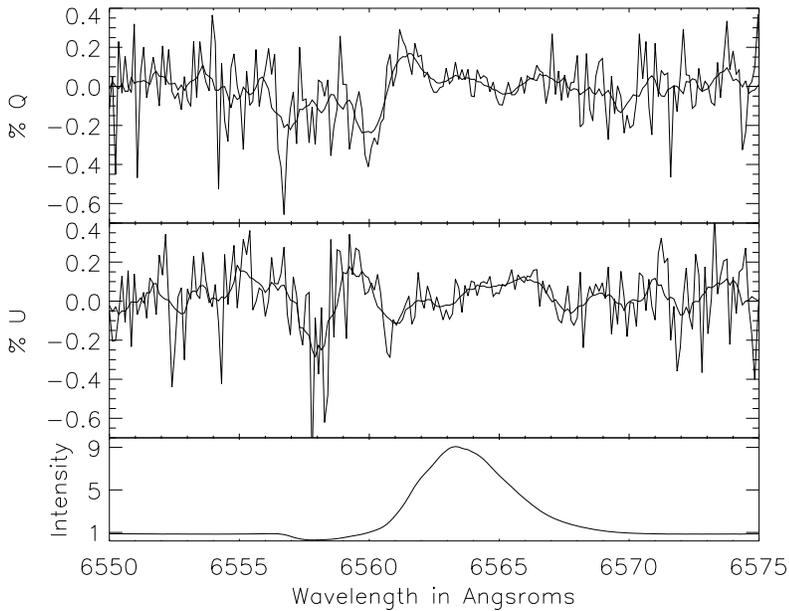}
\caption[AB Aurigae Archive ESPaDOnS Spectropolarimetry.]{The ESPaDOnS archive data for AB Aurigae on February 7th 2006. The solid lines in the spectropolarimetry show a lower resolution smoothed data set. There is a clear, low-amplitude detection across the blue-shifted absorption.}
\label{fig:swp-abaur-esp}
\end{figure}

\twocolumn

Another thing that must be pointed out is the very unique morphology of the spectropolarimetric effects. Most of the HiVIS observations, when showing a clear spectropolarimetric signature, are of the same morphology as figures \ref{fig:swp-abaur-polabs} or \ref{fig:swp-abaur-indiv}. The polarization in the center of the line and on the red-shifted side of the line is identical to continuum.  It is very difficult to match these morphologies with either the depolarization effect or the disk-scattering effects from scattering theory.

\subsection{HD 31648 - MWC 480}

	  This star had a large amplitude ($\sim$0.9\%) polarization increase in the blue-shifted absorption as well as a 0.3\% decrease across the emission line from a continuum of 0.4\% in Vink et al. 2002. In Vink et al. 2005b, the polarization increase in the absorption was 0.8\% with the continuum varying from 0.18\% to 0.30\%. The star showed a 0.4\% increase in the blue-shifted absorption with no signature in the emission line on a continuum of 0.2\% in Mottram et al. 2007, pointing to variability. Beskrovnaya \& Pogodin 2004 also found R-band polarization of roughly 0.3\% in most of their measurements.
	  
\begin{figure}
\centering
\subfloat[MWC 480 Polarization Example]{\label{fig:swp-mwc480-indiv}
\includegraphics[ width=0.35\textwidth, angle=90]{figs-swap-indiv-indivswap-rebin-mwc480.eps}}
\quad
\subfloat[MWC 480 QU Plot]{\label{fig:swp-mwc480-qu}
\includegraphics[ width=0.35\textwidth, angle=90]{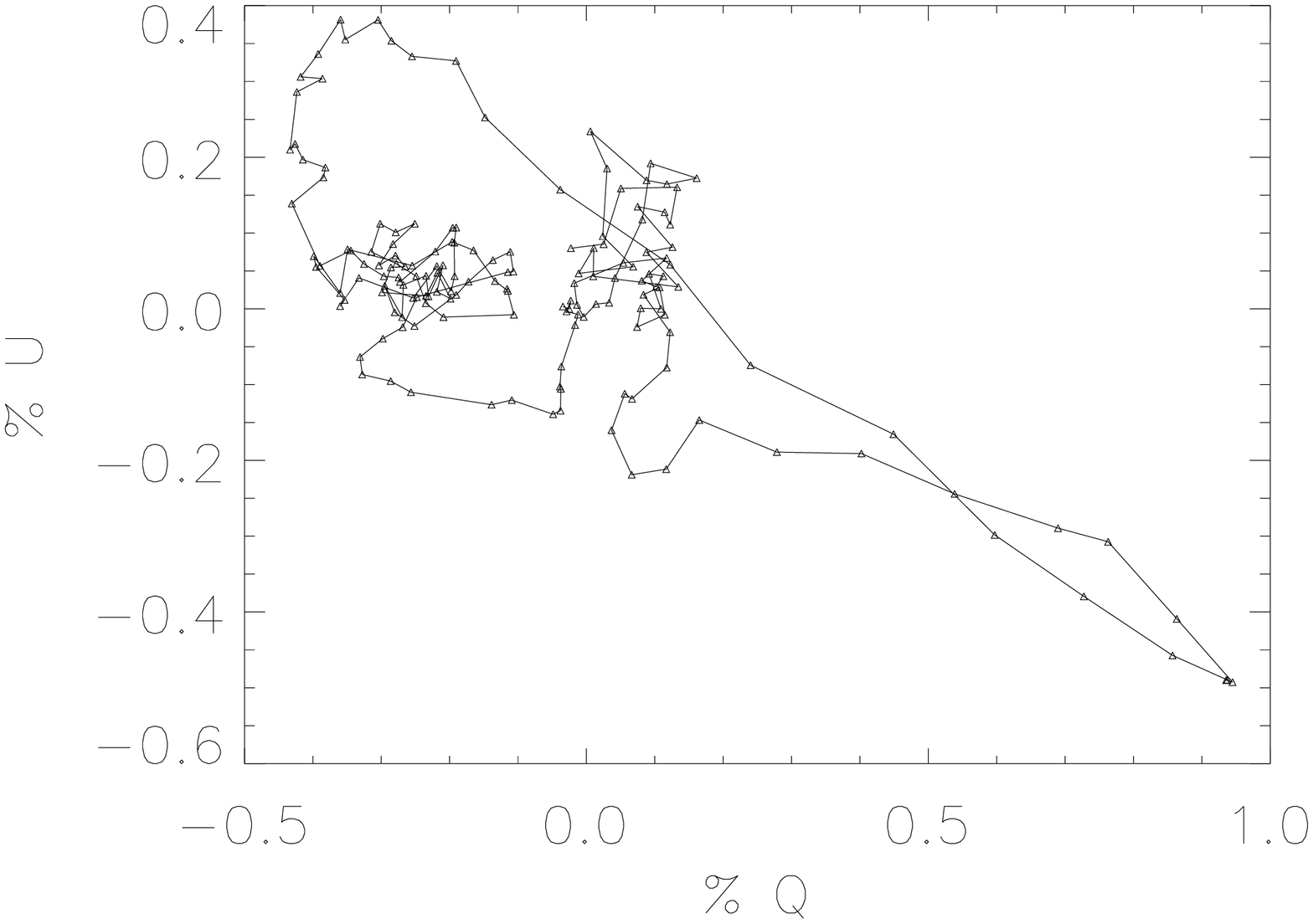}}
\caption[MWC 480 Polarization]{An example polarized spectrum for the MWC 480 H$_\alpha$ line. The spectra have been binned to 5-times continuum. The top panel shows Stokes q, the middle panel shows Stokes u and the bottom panel shows the associated normalized H$_\alpha$ line. There is clearly a detection in the blue-shifted absorption of 1.0\% in q and -0.6\% in u. {\bf b)} This shows q vs u from 6545.7{\AA} to 6566.6{\AA}. The knot of points at (0.0,0.0) represents the continuum. There is another knot of points near (-0.2,0.0) which represents the non-zero q value at the emissive peak. The non-linear wavelength coverage created in the bin-by-flux procedure highlights this effect and shows distinctly that the emissive peak does have a small non-zero polarization.}
\label{fig:swp-mwc480}
\end{figure}

	  This star also showed polarization in absorption but the change was significantly wider and did extend toward the emissive peak. There are 67 measurements shown in figure \ref{fig:mwc480}, most of which show polarization in the absorption. Figure \ref{fig:swp-mwc480-indiv} shows a polarization spectrum where there is a large 1\% change in the absorptive component but with the change in Stokes u extending much wider to the blue and both q and u showng a change on the blue side of the emission peak. Figure \ref{fig:swp-mwc480-qu} shows the qu-plot for this polarization spectrum from 6545.7 to 6566.6{\AA}. There is a continuum knot near (0,0) but there is another knot near (-0.2,0) which corresponds to the change in q across the emission peak. The qu-plot is non-linear in wavelength because of the flux-dependent binning. The qu-plot shows a linear extension of (+q, -u) from the absorption trough but it also shows a smaller (-q, +u) loop from the blue side of the emission peak. The qu-plot does not return to continuum until the red side of the emission line near 6566.6{\AA}

\begin{figure}
\centering
\includegraphics[ width=0.4\textwidth, angle=90]{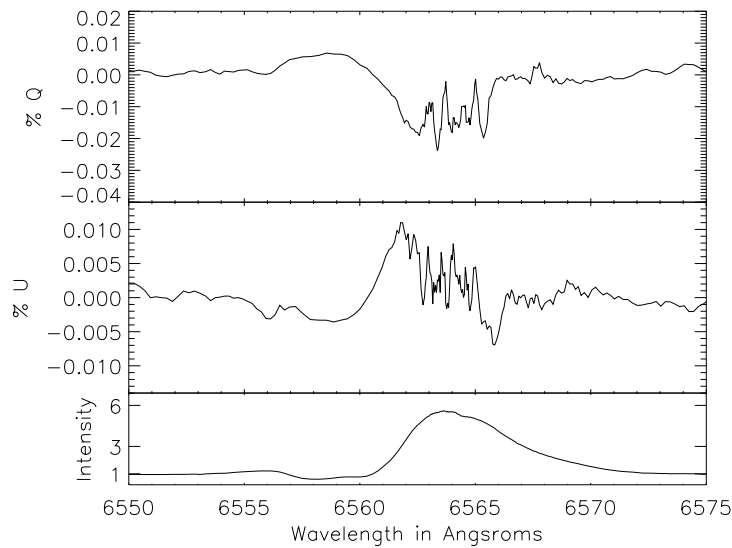}
\caption[MWC 480 Polarized Flux]{The polarized flux for the MWC 480 spectropolarimetry in the previous figure.}
\label{fig:swp-mwc480-pfx}
\end{figure}

	This star also has a strong absorption/emission ratio and a strong polarization-in-absorption so an exploration of the polarized flux is in order. Figure \ref{fig:swp-mwc480-pfx} shows the normalized QU change (q*I). There is significant deviation across the entire line profile. Across the emission peak the change in U goes from positive to negative. This also cannot be caused by a systematic error.

\begin{figure}
\centering
\includegraphics[width=0.5\textwidth, angle=90]{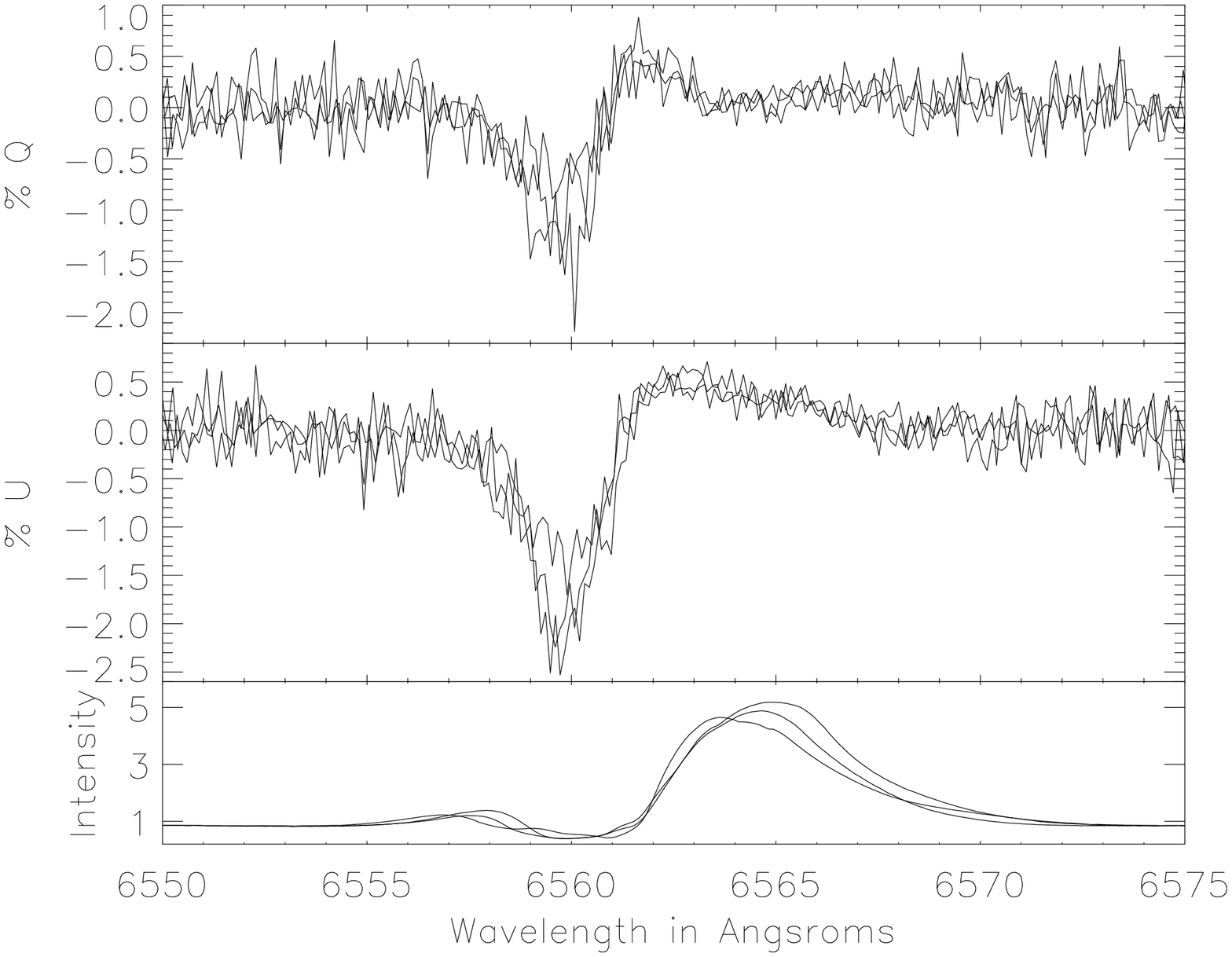}
\caption[MWC 480 Archive ESPaDOnS Spectropolarimetry.]{The ESPaDOnS archive data for MWC 480 on February 7th, 8th, and August 13th 2006. The two polarized spectra from February are nearly identical and show a larger magnitude signature in the absorption trough. The August observations show a similar shape but the magnitude of the signature is smaller.}
\label{fig:swp-mwc480-esp}
\end{figure}

	There is a very clear detection in ESPaDOnS archival observations as well. Three separate observations on February 7th, 8th, and August 13th 2006 all showed very strong spectropolarimetric signatures. The line profiles were similar in shape to those observed with HiVIS, but with a slightly lower amplitude. The spectropolarimetry, shown in figure \ref{fig:swp-mwc480-esp}, shows a very strong variability of the polarization even though the line profile itself is not. The two February observations show a roughly 1.5\% decrease in q and a 2\% decrease in u, both centered on the blue-shifted side of the absorption trough. There is a sign-change to a 0.5\% increase in both q and u, with the zero-point occurring on the red-shifted side of the absorption trough, though still on the blue side of the emission peak. By the center of the emission peak, both polarizations are quite close to continuum, although Stokes u does not reach continuum until 6567{\AA}, on the red-shifted side of the emission peak.

\begin{figure}
\centering
\subfloat[MWC 480 ESPaDOnS Polarization Change]{\label{fig:swp-mwc480-espqu}
\includegraphics[ width=0.35\textwidth, angle=90]{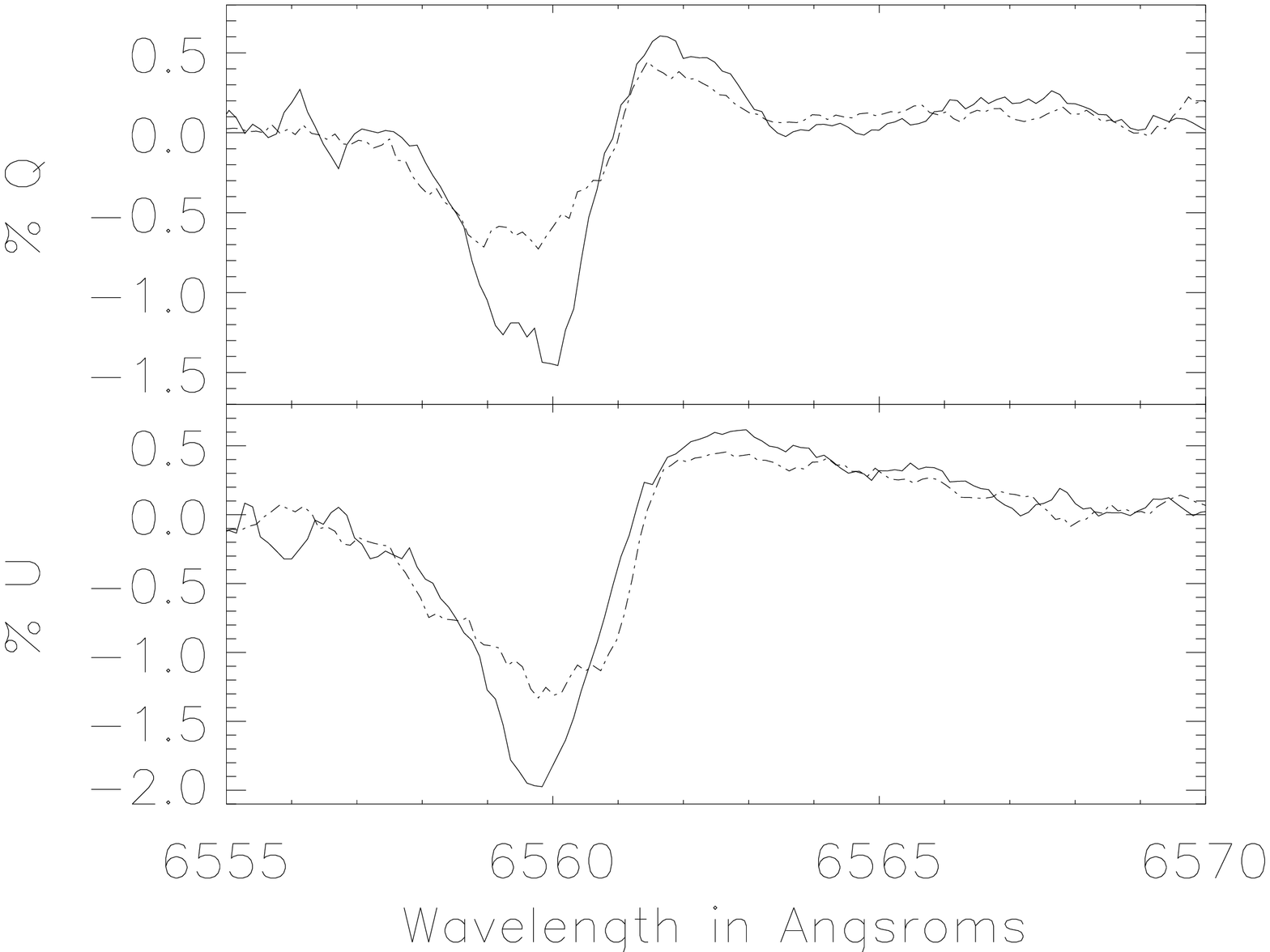}}
\quad
\subfloat[MWC 480 ESPaDOnS Intensity Change]{\label{fig:swp-mwc480-espi}
\includegraphics[ width=0.35\textwidth, angle=90]{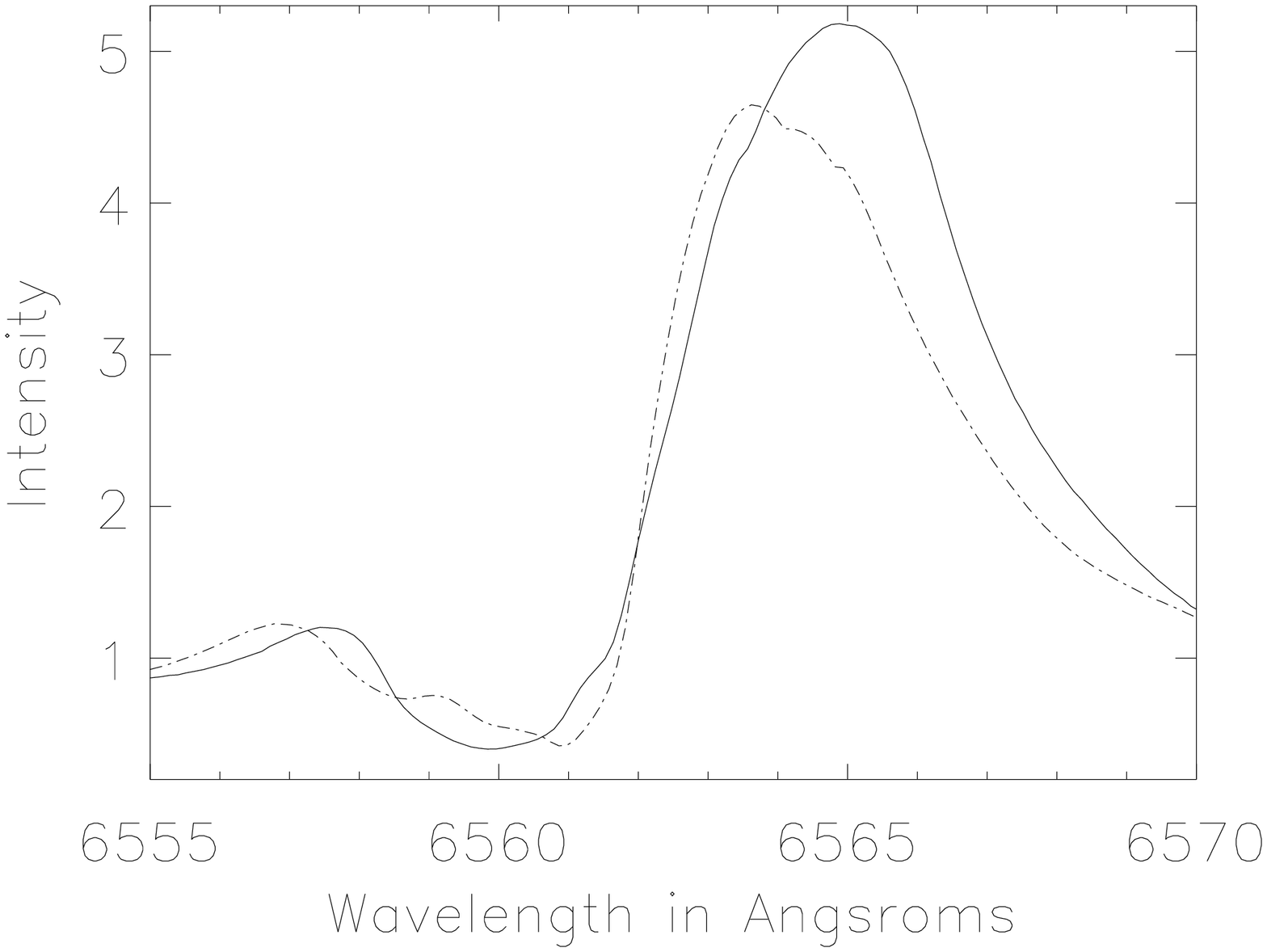}}
\caption[MWC 480 6 Month Change]{Archival data for  MWC 480 with ESPaDOnS showing the change over 7 months - from February 7th (solid) to August 13th 2006 (dashed). {\bf a)} Shows the polarization change across the line, smoothed with a 5-pixel boxcar. The difference is strongest in the P-Cygni absorption where the magnitude of polarization change decreases by a factor of two, from (-1.5,-2.0) to (-0.7,-1.0). {\bf b)} Shows the corresponding line profiles. In the August observations, the absorption was not quite as deep but was more broad.}
\label{fig:swp-mwc480espdif}
\end{figure}

\begin{figure}
\centering
\includegraphics[width=0.3\textwidth, angle=90]{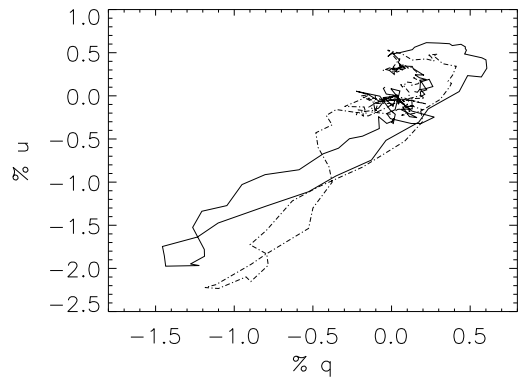}
\caption[MWC 480 Archive ESPaDOnS Spectropolarimetry QU Loop.]{The QU-loops for the ESPaDOnS archive data for MWC 480 on February 7th, and August 13th 2006. The two polarized spectra showed similar shapes when plotted as spectra but in qu-space they show some significant differences.}
\label{fig:swp-mwc480-esp-quloop}
\end{figure}

	These differences are highlighted in figures \ref{fig:swp-mwc480-espqu} and \ref{fig:swp-mwc480-espi}. The H$_\alpha$ line profiles in \ref{fig:swp-mwc480-espi} show a much more significant change in the emissive component of the line whereas the blue-shifted absorption only mildly varies shape with the depth being roughly 0.5 in both. In contrast, the polarization values in the blue-shifted absorption vary by a factor of two over the 6-month baseline. Though this profile at first glance might be compatible with the depolarization effect, there are a number of complications. A depolarization effect is a linear feature in qu-space. In figure \ref{fig:swp-mwc480-esp-quloop} the qu-loops show a significant width on the emissive peak (the increase in q and u). The polarization in the absorption, though mostly linear in the wavelengths of greatest polarization, does show some significant width in one of the observations. If the depolarization was acting, the unpolarized flux from the emission would simply create a linear extension in qu-space away from the intrinsic+interstellar polarization (the continuum value which has been set to zero) toward the interstellar polarization value. Since the qu-loop across the emission has a very significant width, and one of the loops two from the absorptive excursion also show significant width, the depolarization effect has difficulty explaining these observations as well. This star is a very clear example of an object with polarization only where there is significant absorptive effects.

\subsection{HD 37806 - MWC 120}

 	H$_\alpha$ line spectropolarimetry from 1995 and 1996 was presented in Oudmaijer et al. 1999. The emission line profiles they observed changed drastically between the years. In January 1995, the line was double-peaked but with a much weaker blue-shifted emission. In December 1996, the emission line was evenly double-peaked. They observed a change in polarization angle only in both cases. In their December 1996 observations the change in PA is clearly seen in the central absorption. In the December 1995 observations the angle change is wider, but still centered on the absorption. The continuum polarization is 0.29\% and 0.36\% for the two data sets. Vink et al. 2002 present observations that show a double-peaked emission line with a 0.4\% increase in polarization on the red red-shifted emission peak off a continuum of 0.4\%. Mottram et al. 2007 report 2004 observations that show a 0.2\% decrease in polarization from a continuum of 0.4\% on the red-shifted emission peak. The line was much more symmetric, having a narrow mildly blue-shifted absorption, but nothing resembling the strong windy profiles observed in 2006.

\onecolumn

\begin{figure}
\centering
\subfloat[MWC 120 Polarization Example]{\label{fig:swp-mwc120-indiv}
\includegraphics[ width=0.35\textwidth, angle=90]{figs-swap-indiv-indivswap-rebin-mwc120.eps}}
\quad
\subfloat[MWC 120 QU Plot]{\label{fig:swp-mwc120-qu}
\includegraphics[ width=0.35\textwidth, angle=90]{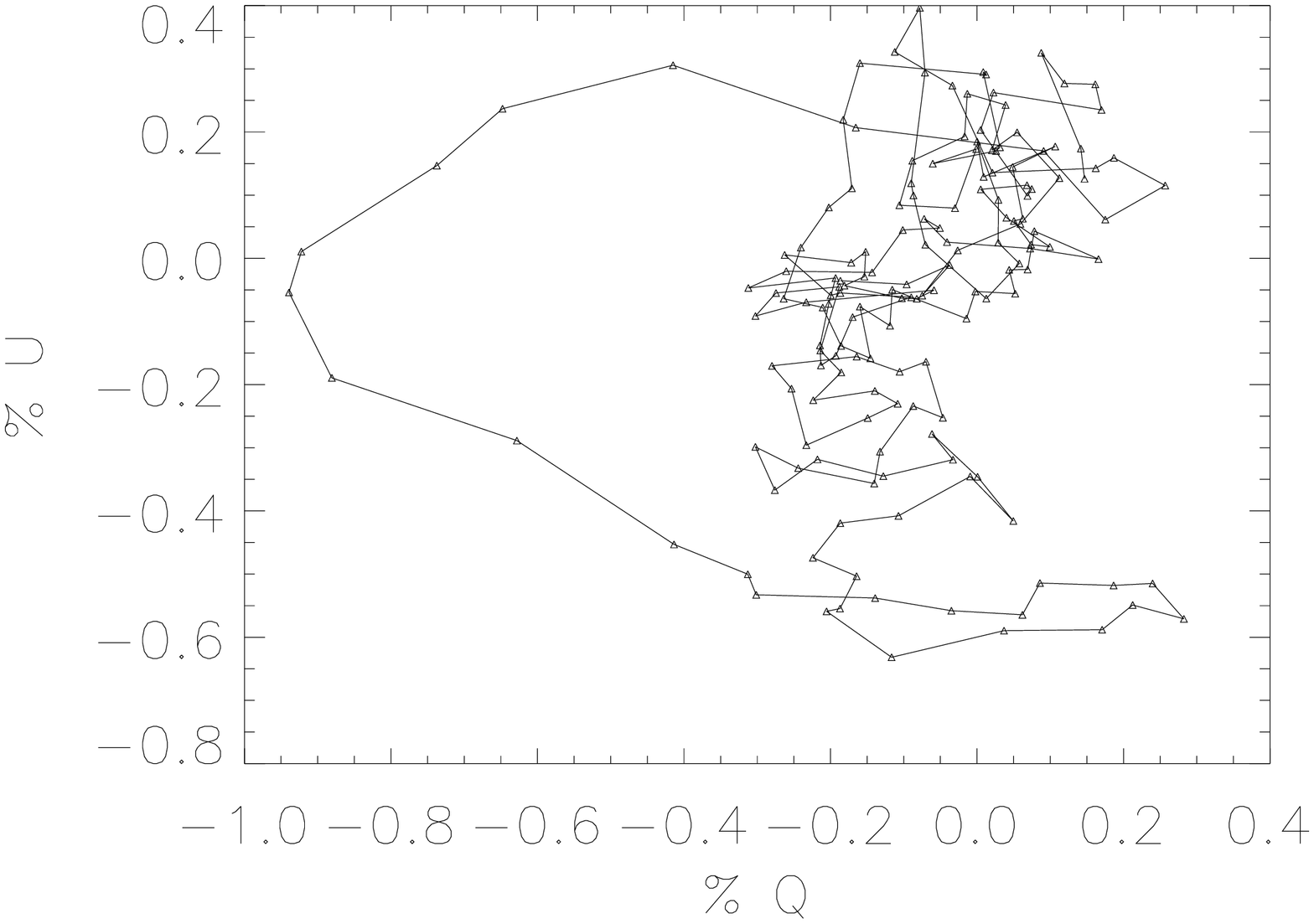}}
\quad
\subfloat[MWC 120 ESPaDOnS Archive Data]{\label{fig:swp-mwc120-indiv-esp}
\includegraphics[width=0.35\textwidth, angle=90]{figs-esp-espadonsswap-mwc120.eps}}
\quad
\subfloat[MWC 120 ESPaDOnS QU Plot]{\label{fig:swp-mwc120-qu-esp}
\includegraphics[width=0.35\textwidth, angle=90]{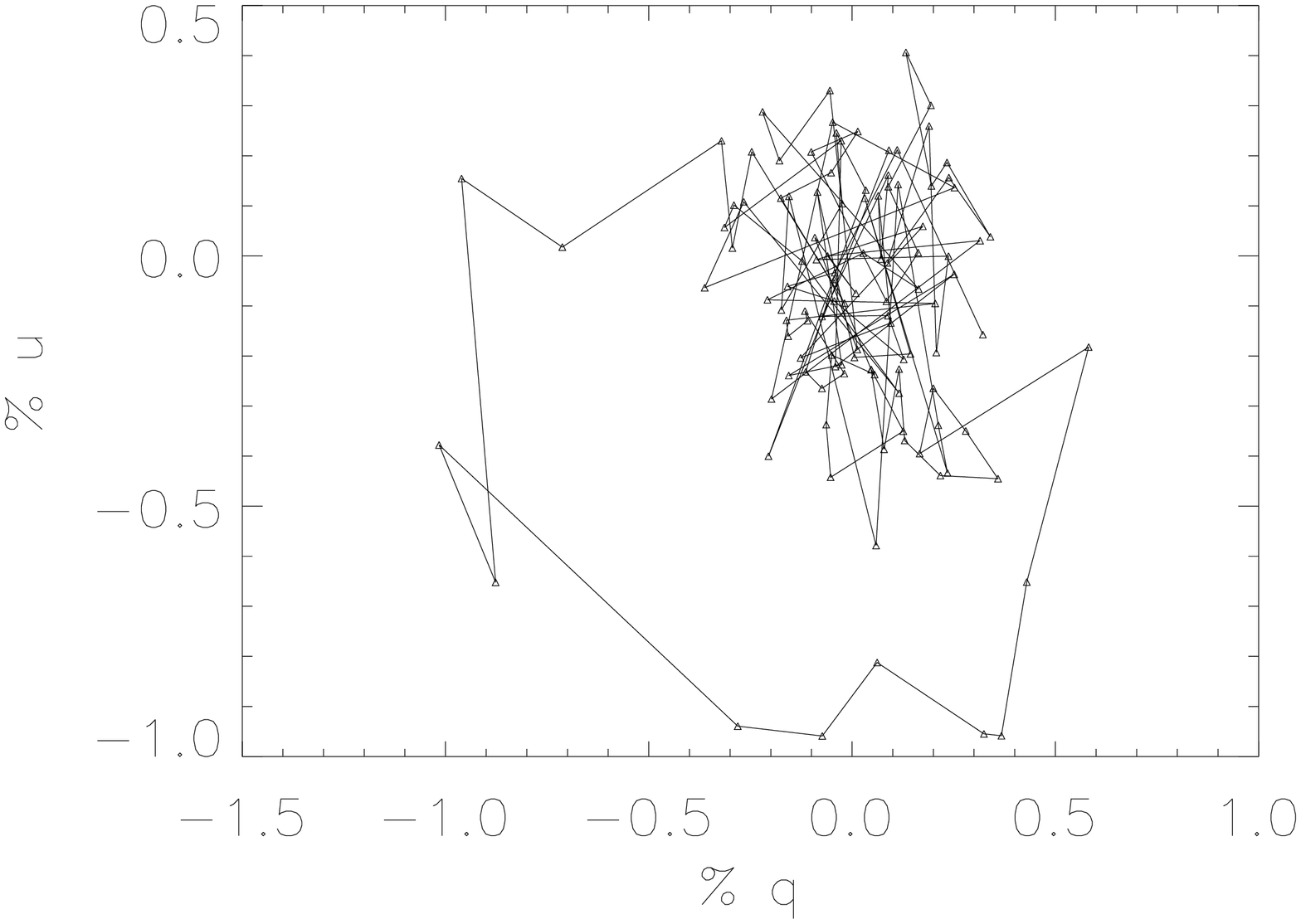}}
\caption[MWC 120 Individual Spectropolarimetry]{An example polarized spectrum for the MWC 120 H$_\alpha$ line. The spectra have been binned to 5-times continuum. The top panel shows Stokes q, the middle panel shows Stokes u and the bottom panel shows the associated normalized H$_\alpha$ line. There is clearly a detection in the central absorption of -1.5\% in q. {\bf b)} This shows q vs u from 6557.2{\AA} to 6573.1{\AA}.  The knot of points at (0.0,0.0) represents the continuum. {\bf c)} The ESPaDOnS archive data for MWC 120 on February 9th, 2006. There is a very strong, narrow polarization change near the line center with a more broad change on the red wing of the line. {\bf d)} Shows the corresponding qu-loop, demonstrating that the changes in q and u are out of phase.}
\label{fig:swp-mwc120}
\end{figure}

\begin{figure}
\centering
\includegraphics[ width=0.3\textwidth, angle=90]{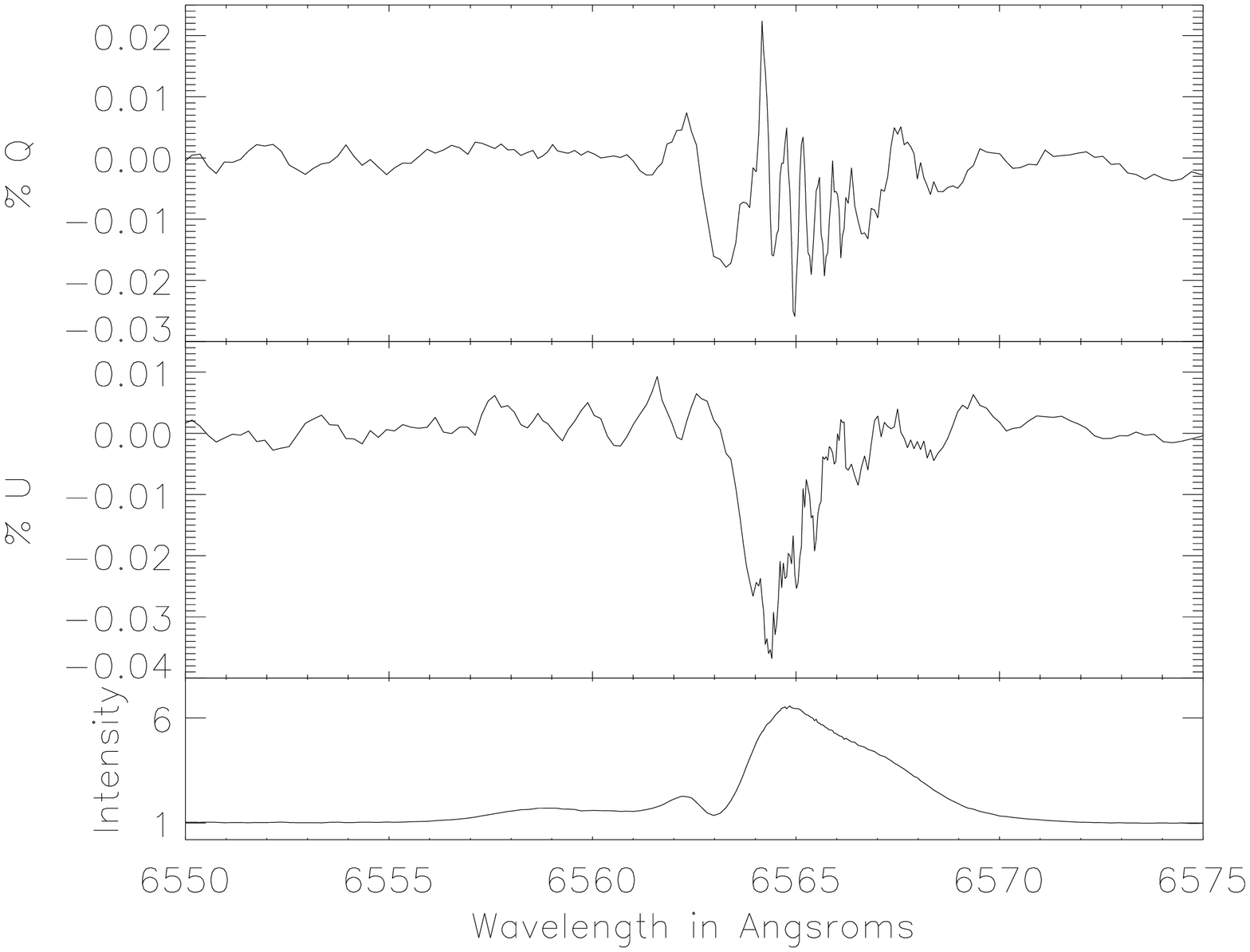}
\caption[MWC 120 Polarized Flux]{The polarized flux (q*I) for the MWC 120 spectropolarimetry in the previous figure.}
\label{fig:swp-mwc120-pfx}
\end{figure}

\begin{figure}
\centering
\subfloat[MWC 120 Polarization Example]{\label{fig:swp-mwc120-indiv2}
\includegraphics[ width=0.35\textwidth, angle=90]{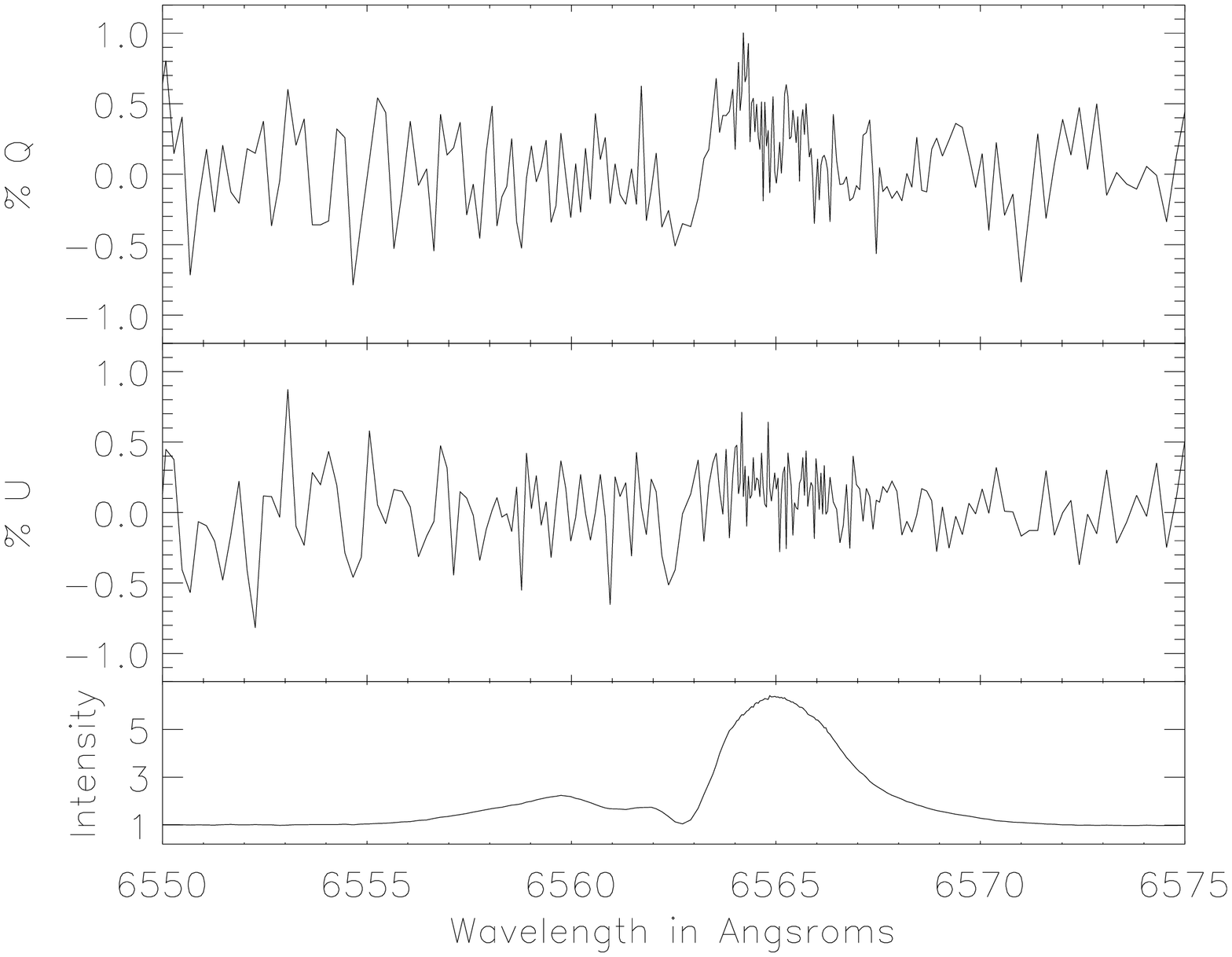}}
\quad
\subfloat[MWC 120 QU Plot]{\label{fig:swp-mwc120-qu2}
\includegraphics[ width=0.35\textwidth, angle=90]{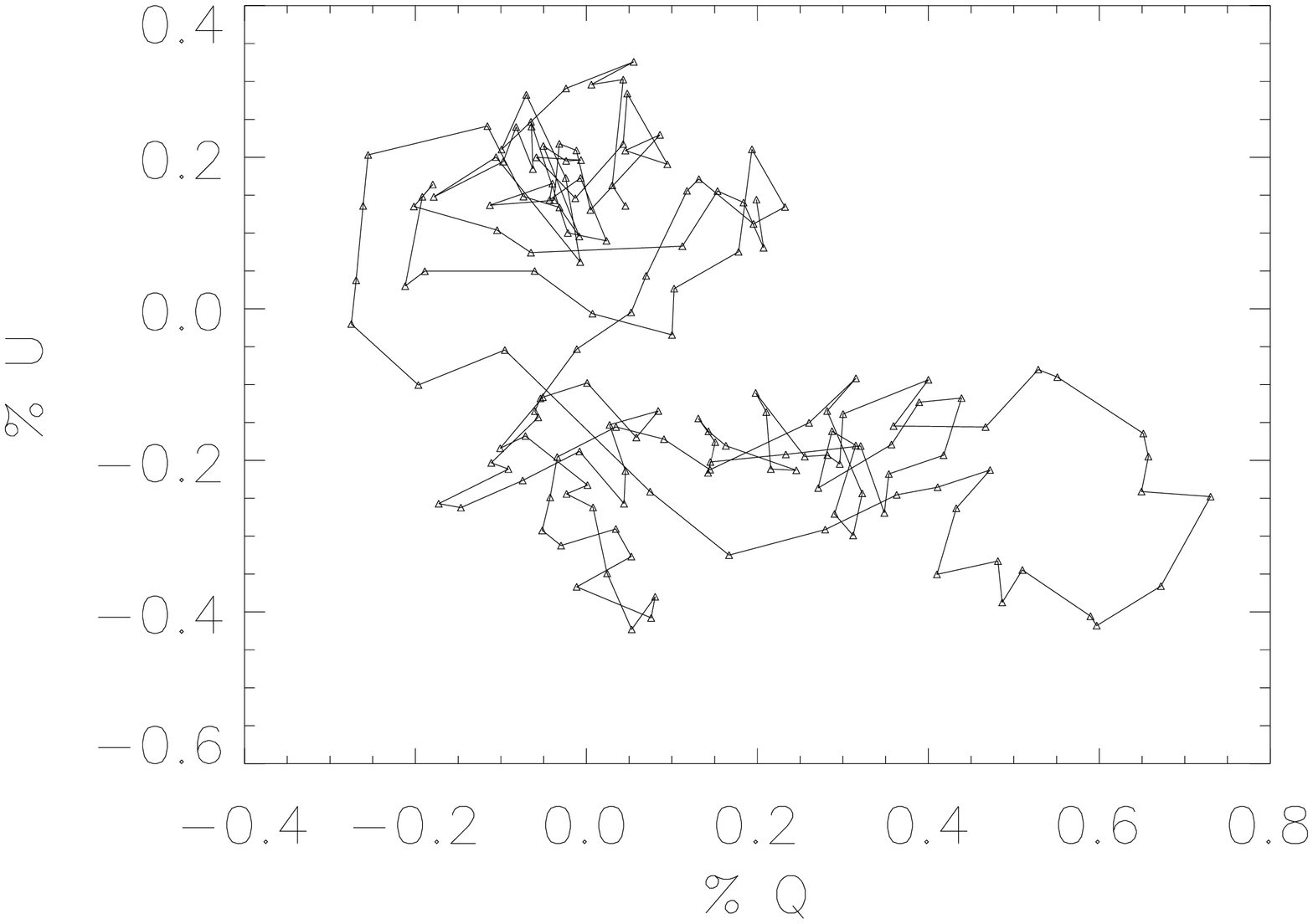}}
\caption[MWC 120 Polarization]{An example polarized spectrum for the MWC 120 H$_\alpha$ line. The spectra have been binned to 5-times continuum. The top panel shows Stokes q, the middle panel shows Stokes u and the bottom panel shows the associated normalized H$_\alpha$ line. There is clearly a detection of an antisymmetric 0.5\% signature in q. {\bf b)} This shows q vs u from 6557.2{\AA} to 6573.1{\AA}.  The knot of points at (0.0,0.0) represents the continuum.}
\label{fig:swp-mwc120-2}
\end{figure}

\begin{figure}
\centering
\includegraphics[ width=0.3\textwidth, angle=90]{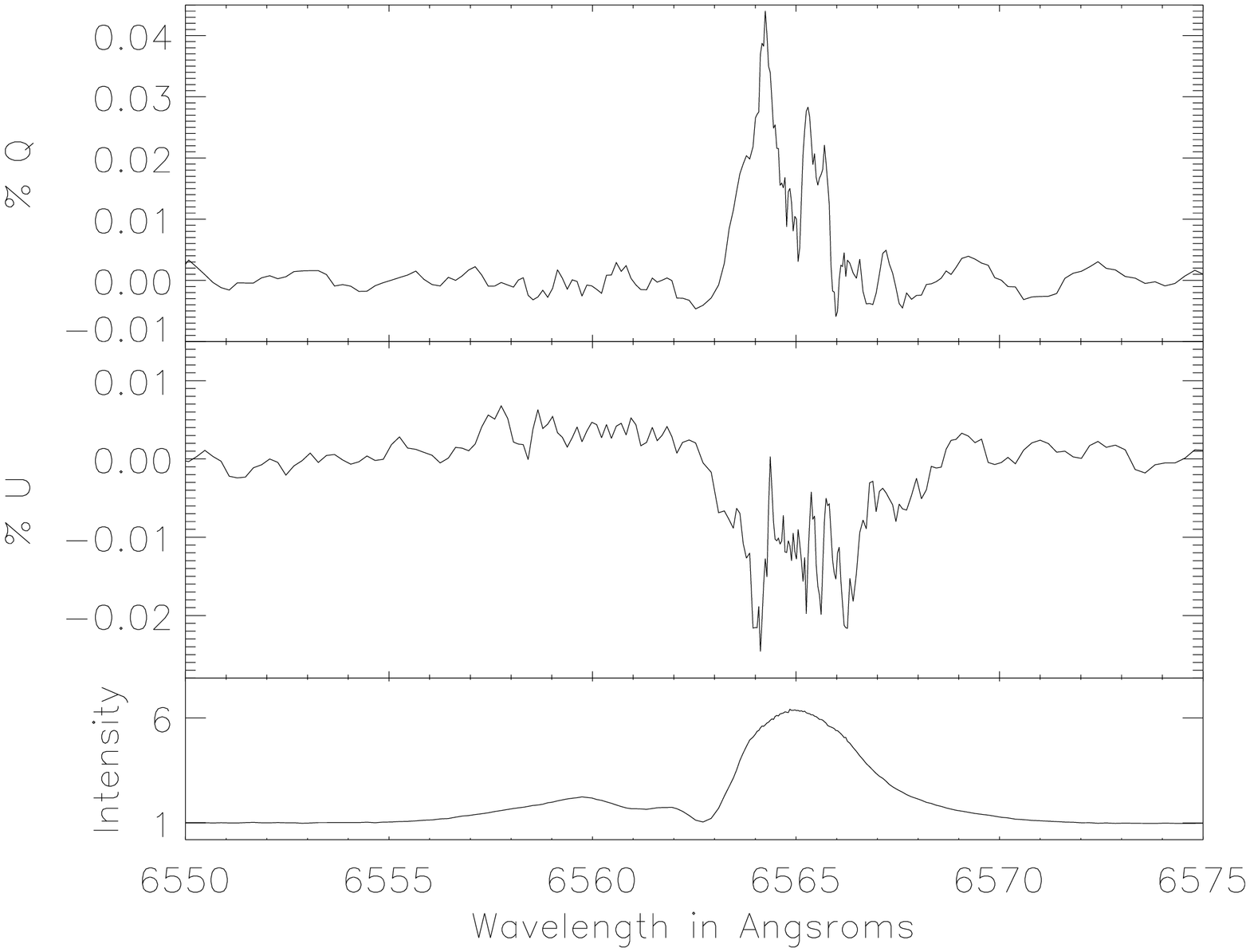}
\caption[MWC 120 Polarized Flux]{The polarized flux (q*I) for the MWC 120 spectropolarimetry in the previous figure.}
\label{fig:swp-mwc120-2}
\end{figure}

\begin{figure}
\centering
\subfloat[MWC 158 Polarization Example]{\label{fig:swp-mwc158-indiv}
\includegraphics[width=0.33\textwidth, angle=90]{figs-swap-indiv-indivswap-rebin-mwc158.eps}}
\quad
\subfloat[MWC 158 QU Plot]{\label{fig:swp-mwc158-qu}
\includegraphics[ width=0.33\textwidth, angle=90]{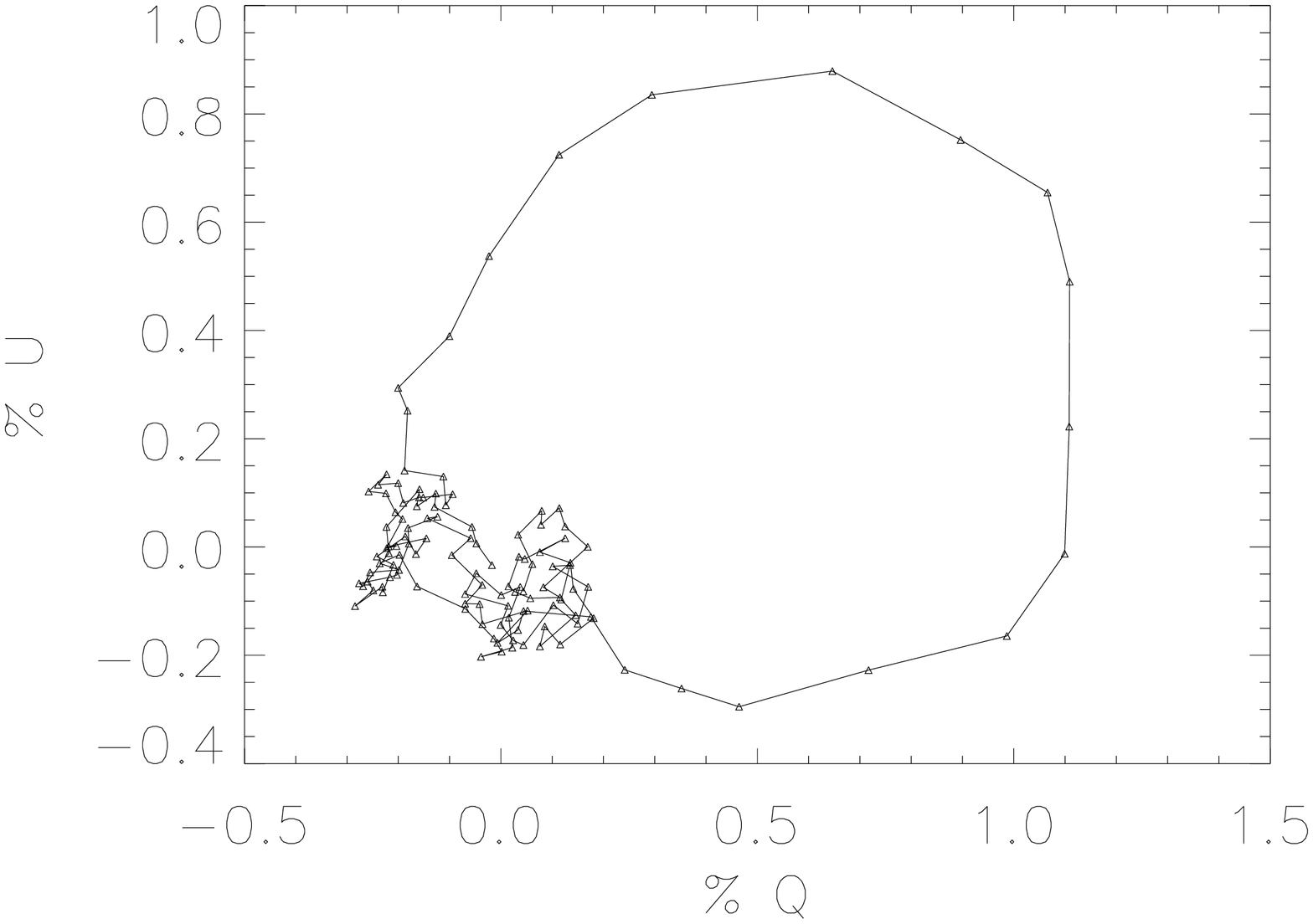}}
\quad
\subfloat[MWC 158 Archive ESPaDOnS Spectropolarimetry]{\label{fig:swp-mwc158-indiv-esp}
\includegraphics[width=0.33\textwidth, angle=90]{figs-esp-espadonsswap-mwc158.eps}}
\quad
\subfloat[MWC 158 Archive ESPaDOnS QU-Plot]{\label{fig:swp-mwc158-qu-esp}
\includegraphics[width=0.33\textwidth, angle=90]{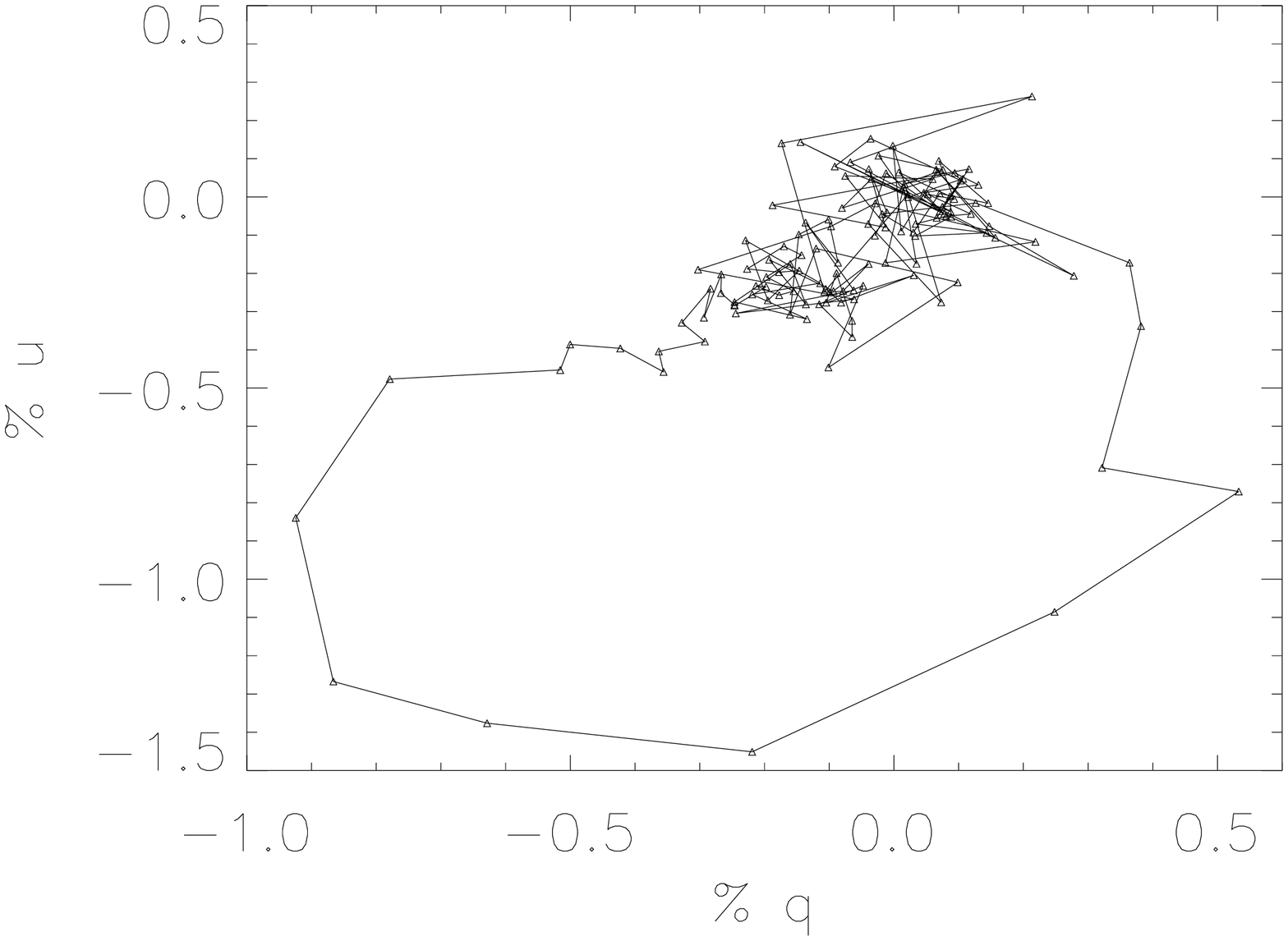}}
\caption[MWC 158 Individual Spectropolarimetry]{An example polarized spectrum for the MWC 158 H$_\alpha$ line. The spectra have been binned to 5-times continuum. The top panel shows Stokes q, the middle panel shows Stokes u and the bottom panel shows the associated normalized H$_\alpha$ line. There is clearly a detection in the central absorption of 2.0\% in q and an more complex 1.5\% detection in u. {\bf b)} This shows q vs u from 6560.1{\AA} to 6565.6{\AA}. The knot of points at (0.0,0.0) represents the continuum. {\bf c)} The ESPaDOnS archive data for MWC 158 on February 9th, 2006.  {\bf d)} This shows q vs u. The knot of points at (0.0,0.0) represents the continuum. Since both the magnitudes and widths of q and u are not similar, the qu-loop is wide and flat. The small, broad change seen on the red side of the emission in both q and u is seen as a separate cluster of points adjacent to the continuum knot in towards the lower left (-q, -u).}
\label{fig:swp-mwc158-esp}
\end{figure}

\begin{figure}
\centering
\includegraphics[ width=0.3\textwidth, angle=90]{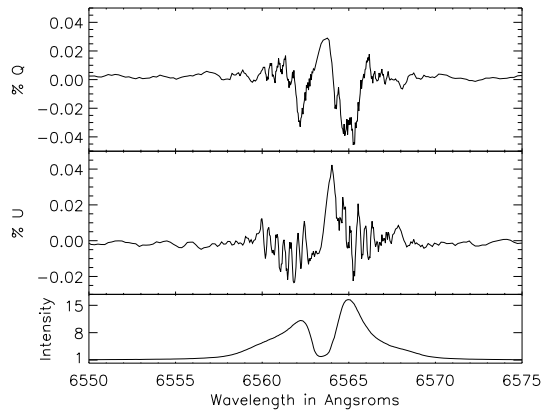}
\caption[MWC 158 Polarized Flux]{The polarized flux (q*I) for the MWC 158 spectropolarimetry of the previous figure.}
\label{fig:swp-mwc158-pfx}
\end{figure}

\twocolumn

	This star had a more complex polarization spectrum, with strong polarization changes in the blue-shifted absorption, but also showed some significant change across the emission line. This can been seen in the overall shape of q and u in figure \ref{fig:mwc120} which shows 26 epochs. Figure \ref{fig:swp-mwc120-indiv} shows one polarized spectrum where there is a strong change in q over a very narrow wavelength range. The change is on the blue side of the emission line but is not coincident with the strong notch near 6563{\AA}. There is a subtle change in u across the emission line that is better seen in the qu-plot of figure \ref{fig:swp-mwc120-qu}. There is a strong change in q of -1\% that forms the wide loop but there is a significant elongation in u across a much wider range of wavelengths. The change in q starts near u of -0.5\% and ends near 0.2\%. The u values don't return to continuum until the 6573{\AA}. This morphology is also seen very clearly in the ESPaDOnS archive observations of figure \ref{fig:swp-mwc120-indiv-esp} from February 9th 2006. The signal-to-noise ratio and spectral resolution are higher and show the complicated morphology of spectropolarimetric effect. The qu-loop for the archive data is dominated by the two large 1\% excursions that form the loop in figure \ref{fig:swp-mwc120-qu-esp}. There is significant polarization seen at low levels in a wider wavelength range that are lost among the continuum-knot in the qu-plot but are seen in the original polarized spectra. The polarized flux for the HiVIS observations, shown in figure \ref{fig:swp-mwc120-pfx}, shows a Q that is significantly lower than continuum across most of the emission line. The U change however is dominated by the change near line-center.

	An entirely different example from HiViS can be seen in another example spectrum shown in figure \ref{fig:swp-mwc120-indiv2}. In this measurement, the change in q is much more broad with a decrease on the blue side of the line and an increase at the emissive peak. There is a small but significant change in u across the emission peak as well. In the qu-plot of figure \ref{fig:swp-mwc120-qu2} there is a complicated structure. Simplistically, it shows that both q and u decrease then increase together, shown by the flattened elongation in qu-space, but there is no single dominating change at any one wavelength. The change in polarized flux also shows the more complex structure. Q is basically flat up until a small dip on the blue side of the emission line but U shows a broad gentle rise up to the emission peak. In the emission line there is a big change in Q and U flux as both q and u showed significant deviation across the line in figure \ref{fig:swp-mwc120-indiv2}. However, Q is more narrow and U is flat-topped. 

	In the first HiVIS example, as well as the ESPaDOnS archive data, the strong, narrow change in polarization in the absorption poses a problem for the disk-scattering and depolarization effects. Even though there is polarization across a broad range of wavelengths at a low amplitude, the red-shifted side of the line profile shows significant deviation from continuum while the blue does not. If one assumes that the absorption removes only unpolarized light, as in the depolarization effect, the central absorption of the ESPaDOnS archive data would be linear in qu-space, not a loop as it is. Also, the difference between the absorptive changes in q and u are significantly different. If only unpolarized light was removed, the sign of the spectropolarimetric change would be opposite in absorption and emission - Stokes u would not decrease in emission and then decrease further in absorption. Also, the broad changes in q and u are also out of phase, u being more on the red side and q being more blue. As far as the thin-disk scattering effect is concerned, the broad monotonic changes are not created by the disk effect. Scattered light is redistributed in wavelength providing decreases at some parts of the line profile while creating increases elsewhere. Both theories have difficulties explaining these observations.

 \subsection{HD 50138 - MWC 158} 
 
	  H$_\alpha$ line spectropolarimetry from 1995 and 1996 was presented by Oudmaijer et al. (1999). The line profiles they observed were similar to the HiVIS observations and a line effect was detected in both observations. Their Jan. 1995 observtions showed a 0.4\% decrease in polarization across the red emission peak while the Dec. 1996 observations showed a depolarization of the same magnitude but with more complex form. They report continuum polarizations of 0.71\% and 0.65\% respectively. Bjorkman et al. 1998 found a continuum polarization of 0.7\% at 160$^\circ$ with an ``almost flat" wavelength dependence from 4000 to 9000{\AA}. From the flatness they concluded that electron scattering, not dust scattering is the polarizing mechanism. They found significant polarization in nearby stars of 0.1\% to 0.8\% and concluded that the nearby environment was complex and variable. Using a number of methods, including an estimate based on the 'depolarization' mechanism, they conclude that the inter-stellar polarization towards this star is 0.2\%. Vink et al. 2002 report a double-peaked line profile with complex signature with a 0.3\% amplitude on a continuum polarization of 0.65\%.

	This star had a very strong change in the central absorptive component of the line a much smaller change in double-peaked emission. Figure \ref{fig:mwc158} shows 35 HiVIS measurements and there is a very clear change in the line center detected in nearly every data set. Figure \ref{fig:swp-mwc158-indiv} shows one such example. There is a 1.5\% change in both q and u in the central absorptive component, even thought the absorption does not go below continuum. The qu-plot of figure \ref{fig:swp-mwc158-qu} shows a very strong loop indicating that the change in q and u are not in phase (not at the same wavelengths) and that there is a significant asymmetric component to the Stokes u measurement. First q increases with a small decrease in u, then u increases strongly. There are ESPaDOnS archival observations from February 9th 2006 that show this effect very clearly. In this observation set, q is strongly asymmetric with a 0.5\% rise followed by a 0.8\% drop in the central absorption. Stokes u shows a strong decrease of 1.3\% that is centered on the red side of the central absorption. In both ESPaDOnS observations, there is also a small, broad decrease in both q and u of roughly 0.3\%, but only on the red side of the line. The polarization on the blue-shifted emission is identical to that of continuum. In terms of polarized flux, shown in figure \ref{fig:swp-mwc158-pfx} for the HiVIS observations, the change in Q shows a decrease in the double-peaked emission, but an increase in the absorptive component. The change in U is much more confined to the absorptive component but is off center towards the red.
	
	In the HiVIS data, and very clearly in the ESPaDOnS archive data, the spectropolarimetric effect is again inconsistent with the depolarization effect. Since the removal of unpolarized light in absorption will act opposite to to the broad depolarization signature the HiVIS u and archival q observations of figures \ref{fig:swp-mwc158-indiv} and \ref{fig:swp-mwc158-indiv-esp} cannot be simple depolarization. In addition, if somehow the depolarization effect was acting only on the red-shifted emission peak, the removal of unpolarized light in the central absorption would have to change the sign of the spectropolarimetric signature in absorption. This is clearly not the case in the archival u observations of \ref{fig:swp-mwc158-indiv-esp}. As far as the disk-scattering effect is concerned, since the broad polarization effect is present in both q and u for the red-shifted emission, a case can be made that this star, ignoring the absorptive component could be consistent with the ``expanding" disk case. From chapter 7, in that case there is a decrease in polarization at line center with an increase in only the red-shifted emission because the scattered light is all red-shifted by the out-flowing scatterers. One can claim that the absorptive effects complicate and mask the ``decrease" in central polarization and that the broad effect on the red-shifted side is from the out-flowing disk case. However, one would still expect that the preferential removal of unpolarized light at line-center would act in a sense opposite to that in the wings.

 \subsection{HD 58647} 
 	
\begin{figure}
\centering
\subfloat[HD 58647 Polarization Example]{\label{fig:swp-hd-indiv}
\includegraphics[ width=0.3\textwidth, angle=90]{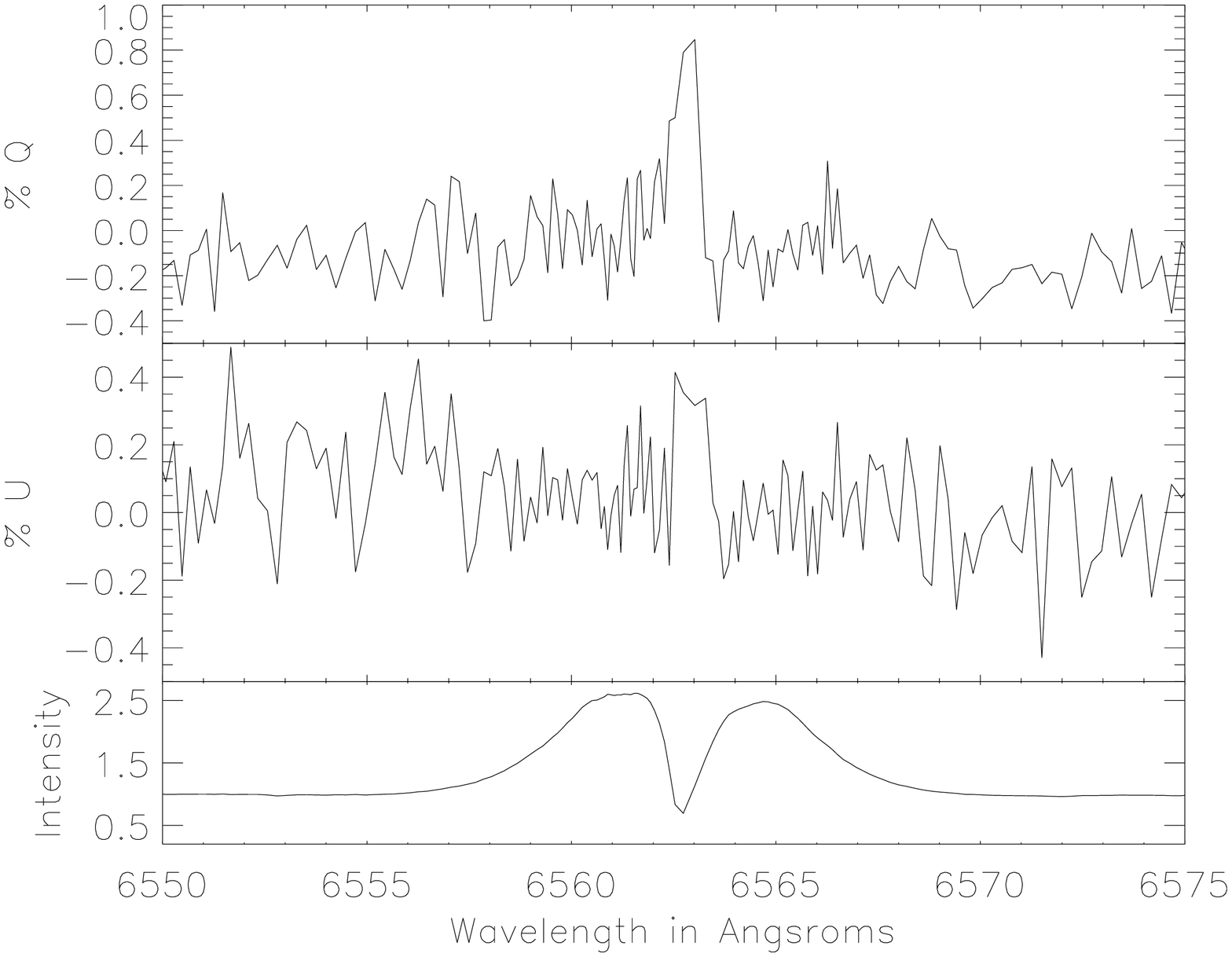}}
\quad
\subfloat[HD 58647 QU Plot]{\label{fig:swp-hd-qu}
\includegraphics[ width=0.3\textwidth, angle=90]{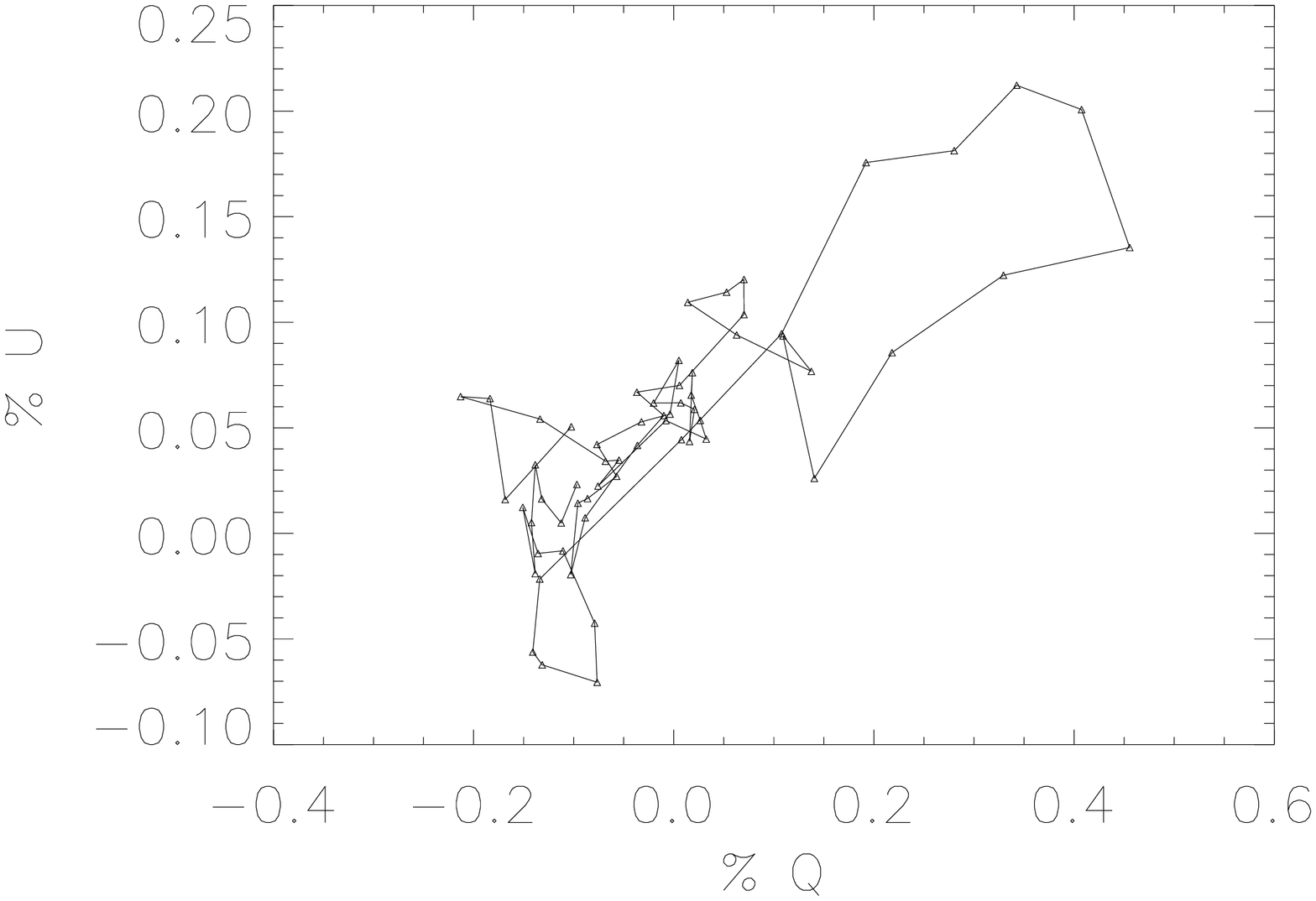}}
\caption[HD 58647 Polarization]{An example polarized spectrum for the HD 58647 H$_\alpha$ line. The spectra have been binned to 5-times continuum. The top panel shows Stokes q, the middle panel shows Stokes u and the bottom panel shows the associated normalized H$_\alpha$ line. There is clearly a detection in the central absorption of 0.8\% in q and 0.4\% in u. {\bf b)} This shows q vs u from 6557.9{\AA} to 6565.3{\AA}.  The knot of points at (0.0,0.0) represents the continuum.}
\label{fig:swp-hd}
\end{figure}

	Vink et al. 2002 report a 0.6\% increase in polarization in a very narrow range  (single resolution element) in the central absorption on a continuum of 0.1\%. This star is very similar to MWC 158 in that there is a large clear polarization in the central absorption trough, and a smaller change seen in the double-peaked emission. Figure \ref{fig:hd} shows 22 measurements where this can be seen. The magnitude of the signature is smaller, at best 1\% with 0.5\% being more typical. However, the intensity of the line is more than 4 times smaller than MWC 158. Figure \ref{fig:swp-hd-indiv} shows a typical spectrum - a 0.8\% change in q and a 0.4\% change in u. However, in this star the change in q and u are in phase producing a linear extension in the qu-plot of figure \ref{fig:swp-hd-qu}.

	Oudmaijer et al. 2001 presented multi-wavelength polarimetry as evidence for a disk-like structure in the ionized gas surrounding the object. The polarization was nearly flat with wavelength near 0.15\% but had U and I-band excesses of 0.05-0.1\%. Since dust alone would not produce both U and I polarization excesses, they conclude that the Serkowski-misfitting would not imply a dusty disk.  
	
	In the context of the scattering models, this set of observations is somewhat inconclusive. The polarization-in-absorption is the individual example of figure \ref{fig:swp-hd-indiv} seen very clearly, but there is no detected polarization outside this central absorption at the 0.2\% level. A broader signature may or may not be present but it must be a small signature. In qu-space, the changes are in-phase, producing a linear deviation. This can be consistent with the depolarization effect. The possibility of a disk-effect is left open as well, however the disk theories presented so far have not included absorptive effects. Polarization-in-absorption is the only detected effect in this example. 
	
	However, in the compilation plot of figure \ref{fig:hd}, the Stokes u plot shows a clear $\sim$ 0.2\% asymmetric signature in many of the observations. An asymmetric signature at line-center is not predicted in disk-theory for either wind or orbital motion. It is also not consistent with the depolarization effect, which must act in one direction and be somewhat proportional to the emission. This asymmetric u signature is also confined to the central absorption and the polarization at the red and blue emission peaks is identical to continuum within the noise.

\subsection{HD 163296 - MWC 275}
	  
	  There are several continuum polarization measurements of this star that show variability. Beskrovnaya et al. 1998 report R-band polarizations of roughly 0.3\% with a 7.5 day cyclic variation of 0.1\% in the July 17-30 1995 period. However, Oudmaijer et al. 2001 found a non-variable V-band polarization of only 0.02\% on 7 observations over the 1998-1999 period.
	  
\begin{figure}
\centering
\subfloat[HD 163296 Polarization Example]{\label{fig:swp-hd163-indiv}
\includegraphics[ width=0.3\textwidth, angle=90]{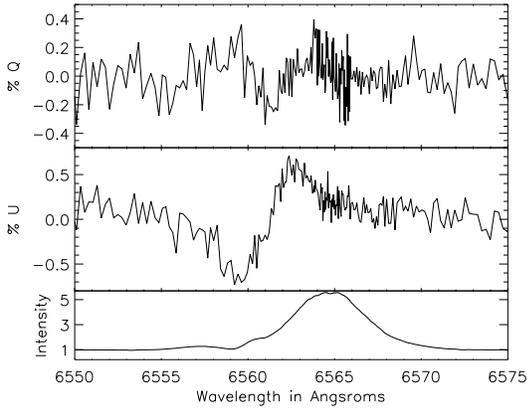}}
\quad
\subfloat[HD 163296 QU Plot]{\label{fig:swp-hd163-qu}
\includegraphics[ width=0.3\textwidth, angle=90]{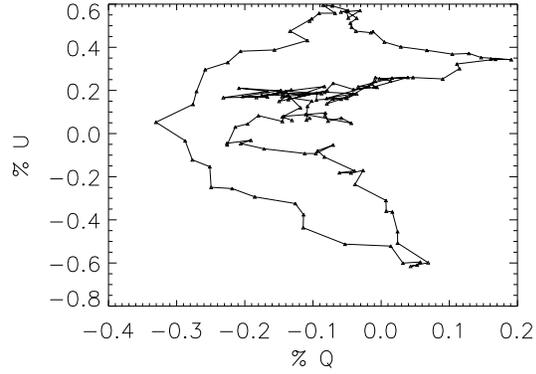}}
\quad
\subfloat[HD 163296 ESPaDOnS Archive Data]{\label{fig:swp-hd163-esp-archive}
\includegraphics[width=0.3\textwidth, angle=90]{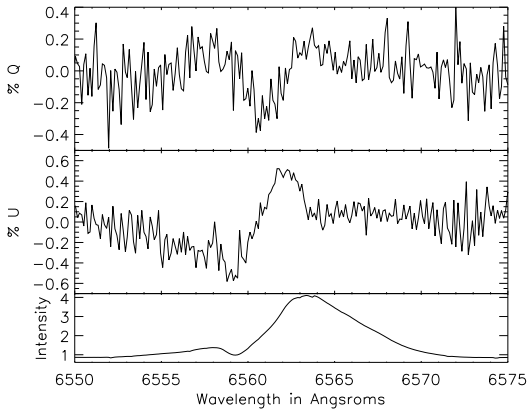}}
\quad
\subfloat[HD 163296 ESPaDOnS Observations]{\label{fig:swp-hd163-esp-swap}
\includegraphics[ width=0.3\textwidth, angle=90]{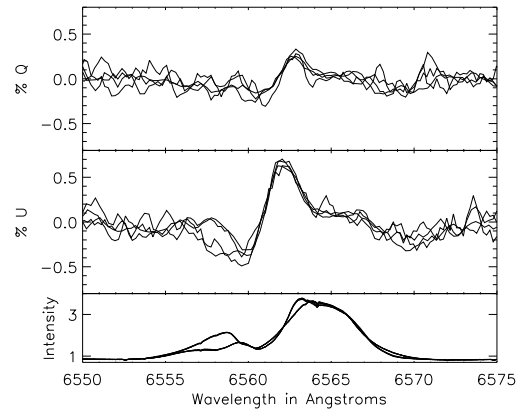}}
\caption[HD 163296 ESPaDOnS Observations]{An example polarized spectrum for the HD 163296 H$_\alpha$ line. The spectra have been binned to 5-times continuum. The top panel shows Stokes q, the middle panel shows Stokes u and the bottom panel shows the associated normalized H$_\alpha$ line. There is clearly a detection in the blue-shifted absorption of a complex 0.2\% signature in q and a 0.7\% antisymmetric signature in u. {\bf b)} This shows q vs u from 6549.9{\AA} to 6565.8{\AA}.  The knot of points at (0.0,0.0) represents the continuum. {\bf c)} Shows archive ESPaDOnS data from August 13th 2006 again with a complex 0.2\% q and antisymmetric 0.5\% u. {\bf d)} shows the HiVIS observations on June 23 and 24th 2007.}
\label{fig:swp-hd163-esp}
\end{figure}

	  This star had a very clear change in polarization in the blue side of the line profile, but the change extended all the way to the center of the emission. Figure \ref{fig:hd163} shows 24 measurements all with similar form in q and u. A typical example is shown in figure \ref{fig:swp-hd163-indiv}. The q spectrum shows a small increase in the blue shifted absorption as well as one centered on the emission. The u spectrum shows a very clear decrease then increase with the polarization returning to continuum by the emissive peak. The plot in qu-space, shown in figure \ref{fig:swp-hd163-qu} is very complex. The two increases in q manifest themselves as the two right-most excursions while the u spectrum gives the vertical extent. The polarized flux also shows a very strong clear change. Both Q and U change retain their overall shape but with amplification at the emissive peak.

	This star also presents problems with both the depolarization and disk theories. The asymmetry of the line shows the strong absorption on the blue-shifted side of the line profile. The polarization on the red side of the emission line is essentially the continuum value. The polarization on the blue-shifted side of the line profile shows a very strange morphology in both the qu-space and the polarized spectra. For disk-scattering theory to work, the scattering particles must be inflowing to produce a significant blue-shift in the scattered light. However, this star has ample evidence for stellar winds (out-flow) from other studies outlined in chapter 6. The depolarization effect also has difficulties because of the very complex morphology. In both HiVIS and ESPaDOnS archive observations (figures \ref{fig:swp-hd163-indiv} and \ref{fig:swp-hd163-esp-archive}) on the far blue component, q increases while u decreases. Moving towards the red, the Stokes u signature reverses sign while q returns to zero then increases without changing sign. If one argues that removal of unpolarized light is all that that is occurring, this morphology is very difficult to explain.

\subsection{HD 179218 - MWC 614}
	
	This star shows a small change in polarization extending from the blue-shifted absorption across the entire emission line. The change small, typically 0.3\% to 0.5\%. Figure \ref{fig:hd179} shows 25 observations with the change clearest in the Stokes u spectra. An example spectrum is shown in figure \ref{fig:swp-hd179-indiv}. There is a 0.3\% increase in u with a 0.2\% decrease in q in the blue-shifted absorption. This then reverses to a 0.3\% decrease in u and a 0.2\% increase in q centered on the emissive peak. In the qu-plot of figure \ref{fig:swp-hd179-qu} this shows up as two linear extensions away from the continuum-knot at (0,0).

	Though this star does show evidence for significant absorption, this signature is consistent, at least in overall form, with the depolarization effect. The q and u signatures are in-phase, showing up as linear extensions in qu-space. In the blue-shifted absorption, the q and u effects are also phased and in the opposite direction as the changes across the emission peak. Since the disk-theory does not include absorptive effects, it is difficult to compare the results. However, one of the overall results of disk-theory is that the polarization effects will be significantly wider than the emission line, and this does not seem to be the case for this star.

\subsection{HD 150193 - MWC 863}

	This star shows a small but significant detection extending from the blue shifted absorption to the emission profile in the 12 measurements of figure \ref{fig:swp-hd150}. This can be seen as a small deviation in Stokes u in figure \ref{fig:hd150}. This star is a bit noisier than many, but the signature is seen clearly in both q and u as the strong change on the blue side of the emission peak. An example of this signature is shown in figure \ref{fig:swp-hd150-indiv}. There is a clear change in q of 0.5\% first decreasing in the blue-shifted absorption and then increasing on the blue side of the emission peak. The u spectrum shows a much more broad change across the emission peak. Chavero et al. 2006 measured a R-band polarization of 5.0$\pm$0.5\% and show that the polarization deviates significantly from the traditional Serkowski law. This star had previously been identified as a polarized standard in Whittet et al. 1992 with an R-band polarization of 5.19$\pm$0.05\%.

	This signature also poses problems for both disk and depolarization effects. There is a marginal broad effect seen over the emission in Stokes u, but the change in q is asymmetric and blue-shifted. If one argues for depolarization being the underlying effect, the wavelength offset between q and u across the emission peak is difficult to explain. Also, in absorption, the q profile is asymmetric. It is negative on the far blue side with a strong, rapid change to positive on the blue edge of the transition between absorption and emission. Though disk-scattering theory has not yet included absorption, the central position of the broad increase in u and the lack of any q or u signature on the red side of the line are difficult to explain in this context.

\subsection{HD 200775 - MWC 361}
	
	This star has been very stable in HiVIS polarization signature over the past few years. A broad spectropolarimetric signature was detected in all 32 measurements in figure \ref{fig:mwc361}. In Harrington \& Kuhn 2008, HiVIS and ESPaDOnS observations that showed the signature didn't change in over a year. The HiVIS observations did not show significant morphological changes from telescope polarization effects. When all HiVIS observations were rotated to a common frame, the morphology of the spectropolarimetric effects match the ESPaDOnS observations very well. Figure \ref{fig:mwc361} shows all the HiVIS data unrotated. In figure \ref{fig:swp-mwc361-indiv}, a least-squares rotation was applied to all measurements and then the average was taken. A very clear, very high signal-to-noise polarized spectrum shows a very broad 0.25\% change in q with a much more narrow asymmetric flip in u reaching -0.1\% and +0.15\%. The u measurements are slightly smaller in amplitude than that detected with ESPaDOnS, but the overall form matches perfectly.

	The qu-plot of figure \ref{fig:swp-mwc361-pfx} illustrates the change very clearly - q increases, u goes low then high, and finally q returns to zero. The polarized flux of figure \ref{fig:swp-mwc361-pfx} shows that the Q flux is essentially a widened and flattened copy of the H$_\alpha$ line. The U spectrum has basically the same shape as the u spectrum but with an amplification.

	In the context of scattering theory, the broad q signature centered on line-center suggests the depolarization mechanism. The emission line is quite strong while the polarization is relatively small, less-strongly peaked and significantly broader than the emission line. The asymmetric change seen in u complicates the picture. If depolarization is the only effect acting, it only produces linear extensions in qu-space. In this star, q does provide a strong linear extension, but while q is nearly constant in the emission peak, the u profile complicates this explanation. In terms of disk-theory, the broadness of the detection fits while the actual profiles do not. At line center, the projected scattering angles are orthogonal to those on the red and blue wings. A symmetric q profile can not be produced. Though this star does show evidence for absorption as the center of the line is notched, in many observations asymmetrically so, the u polarization is much broader than the central notch. This star, though complicated, may be showing a polarization-in-absorption effect.

\onecolumn
	  
\begin{figure}
\centering
\includegraphics[ width=0.45\textwidth, angle=90]{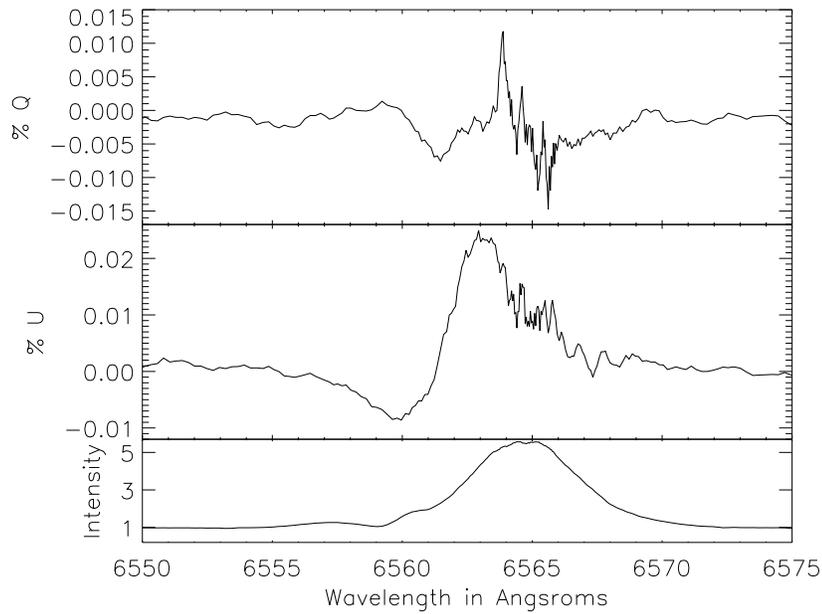}
\caption[HD 163296 Polarized Flux]{The polarized flux (q*I) for the HD 163296 spectropolarimetry shown in panel a) of the previous figure.}
\label{fig:swp-hd163-pfx}
\end{figure}

\begin{figure}
\centering
\subfloat[HD 179218 Polarization Example]{\label{fig:swp-hd179-indiv}
\includegraphics[ width=0.35\textwidth, angle=90]{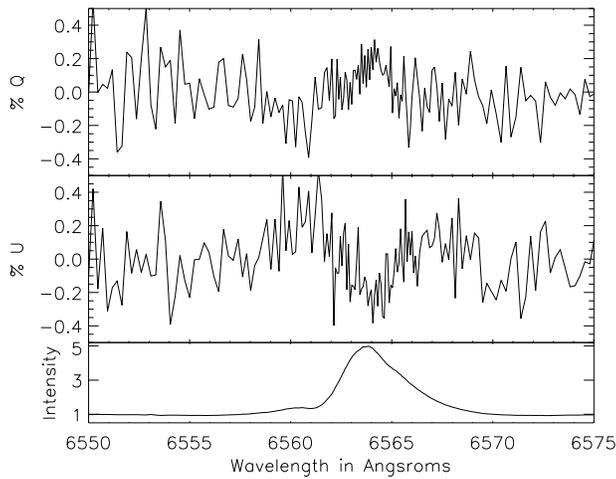}}
\quad
\subfloat[HD 179218 QU Plot]{\label{fig:swp-hd179-qu}
\includegraphics[ width=0.35\textwidth, angle=90]{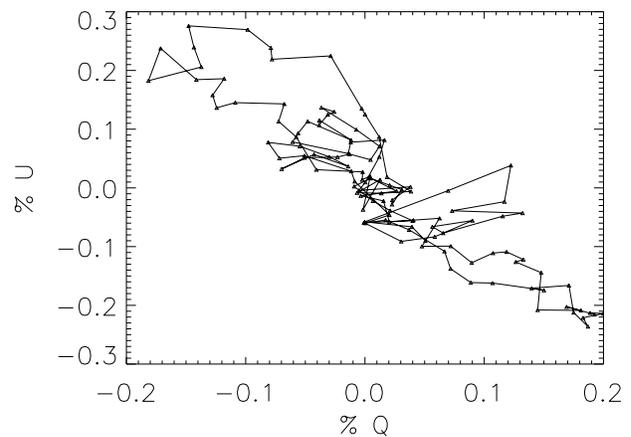}}
\caption[HD 179218 Polarization]{An example polarized spectrum for the HD 179218 H$_\alpha$ line. The spectra have been binned to 5-times continuum. The top panel shows Stokes q, the middle panel shows Stokes u and the bottom panel shows the associated normalized H$_\alpha$ line. There is clearly a detection across the entire line of 0.2\% in q and u. {\bf b)} This shows q vs u from 6551.9{\AA} to 6568.9{\AA}.  The knot of points at (0.0,0.0) represents the continuum.}
\label{fig:swp-hd179}
\end{figure}

\begin{figure}
\centering
\subfloat[HD 150193 Polarization Example]{\label{fig:swp-hd150-indiv}
\includegraphics[ width=0.3\textwidth, angle=90]{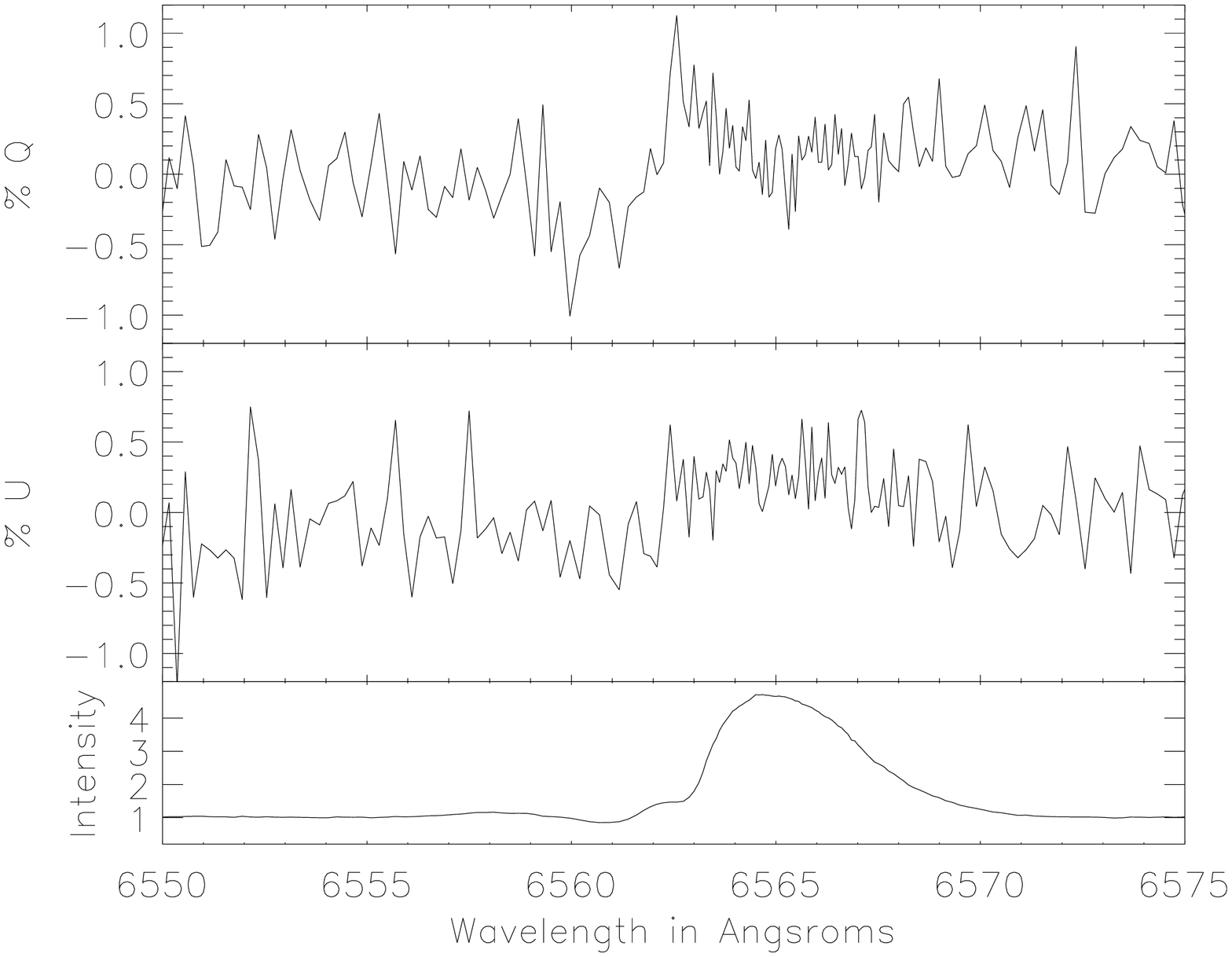}}
\quad
\subfloat[HD 150193 QU Plot]{\label{fig:swp-hd150-qu}
\includegraphics[ width=0.3\textwidth, angle=90]{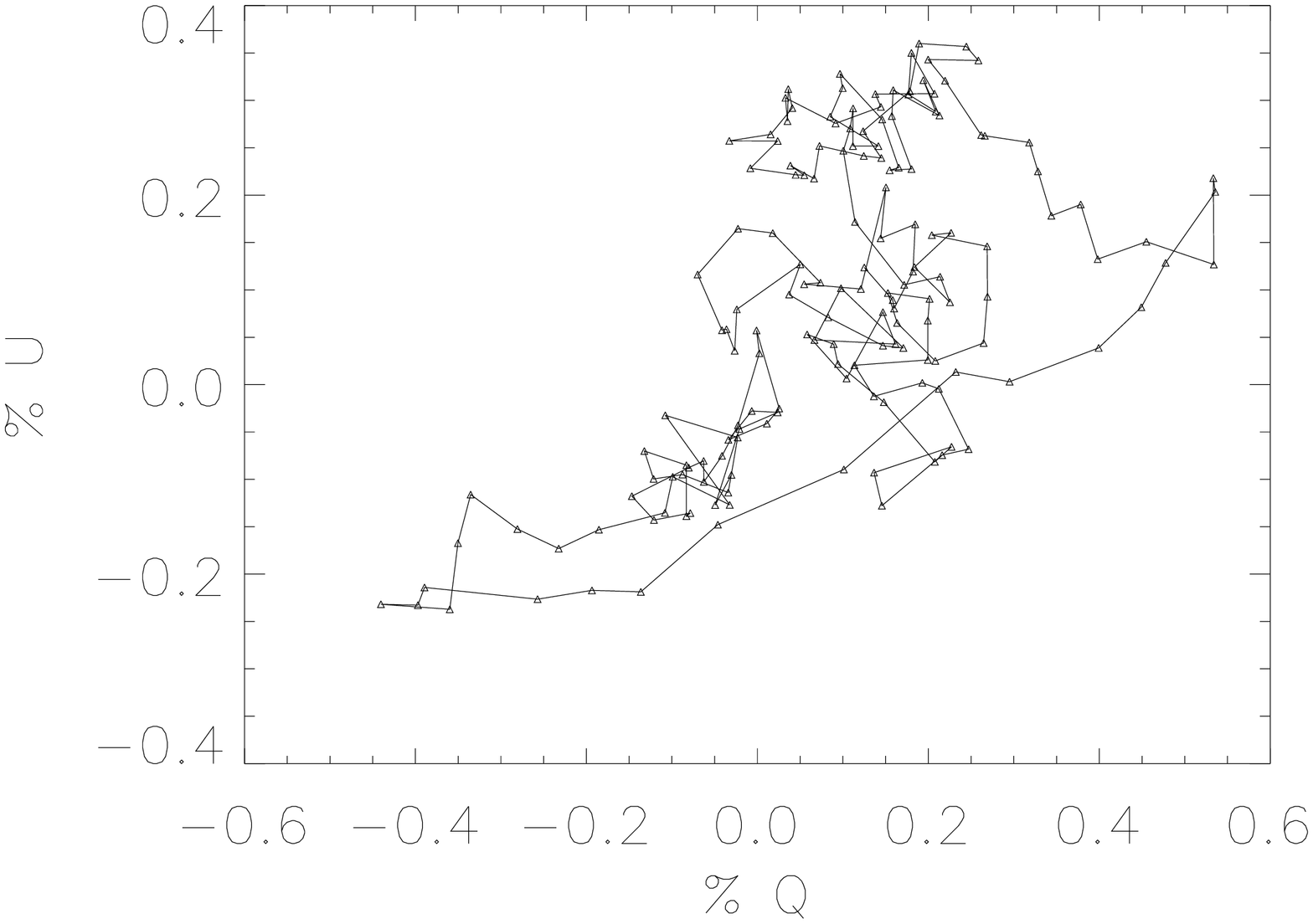}}
\caption[HD 150193 Polarization]{An example polarized spectrum for the HD 150193 H$_\alpha$ line. The spectra have been binned to 5-times continuum. The top panel shows Stokes q, the middle panel shows Stokes u and the bottom panel shows the associated normalized H$_\alpha$ line. There is clearly a detection in the blue-shifted absorption of $\pm$0.5\% in q. {\bf b)} This shows q vs u from 6553.1{\AA} to 6578.3{\AA}. The knot of points at (0.0,0.0) represents the continuum.}
\label{fig:swp-hd150}
\end{figure}

\begin{figure}
\centering
\subfloat[MWC 361 Polarization Example]{\label{fig:swp-mwc361-indiv}
\includegraphics[ width=0.3\textwidth, angle=90]{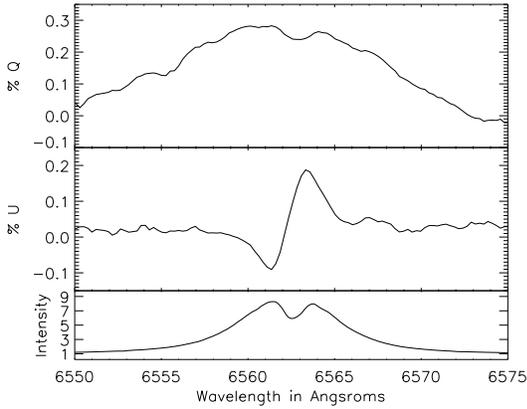}}
\quad
\subfloat[MWC 361 QU Plot]{\label{fig:swp-mwc361qu}
\includegraphics[ width=0.3\textwidth, angle=90]{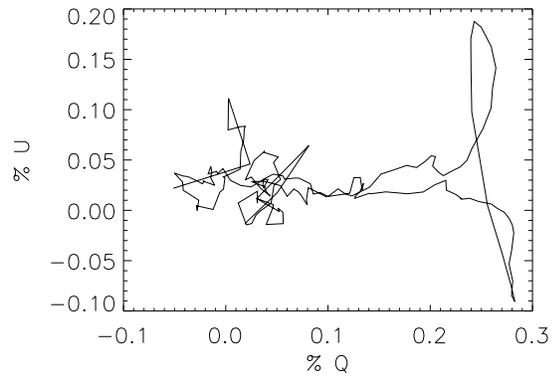}}
\quad
\subfloat[MWC 361 Polarized Flux]{\label{fig:swp-mwc361-pfx}
\includegraphics[ width=0.3\textwidth, angle=90]{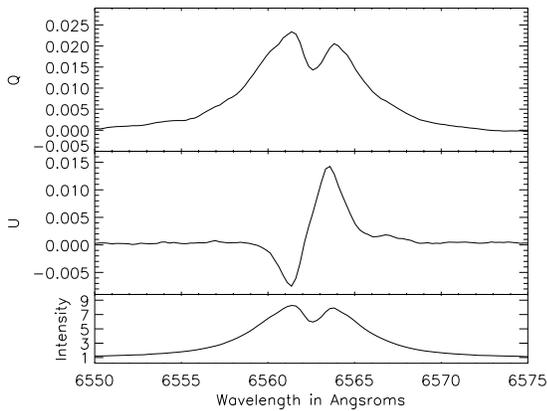}}
\quad
\subfloat[MWC 361 ESPaDOnS Spectropolarimetry]{\label{fig:swp-mwc361-esp}
\includegraphics[width=0.3\textwidth, angle=90]{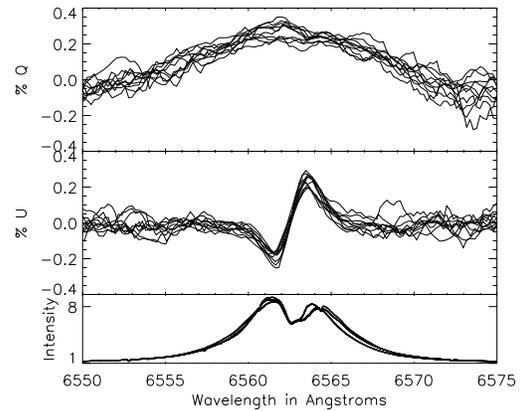}}
\caption[MWC 361 Polarization]{All polarized spectra averaged for the MWC 361 H$_\alpha$ line. The spectra have been individually rotated to a common frame and then averaged. The top panel shows Stokes q, the middle panel shows Stokes u and the bottom panel shows the associated averaged and normalized H$_\alpha$ line. {\bf b)} This shows q vs u from 6550{\AA} to 6575{\AA}. The knot of points at (0.0,0.0) represents the continuum. {\bf c)} The polarized flux q*I for the HiVIS observations. {\bf d)} ESPaDOnS observations for MWC 361 on August 1st and 3rd, 2006 and June 23rd and 24th 2007.}
\label{fig:swp-mwc361}
\end{figure}

\begin{figure}
\centering
\subfloat[MWC 758 Polarization Example]{\label{fig:swp-mwc758-indiv}
\includegraphics[ width=0.35\textwidth, angle=90]{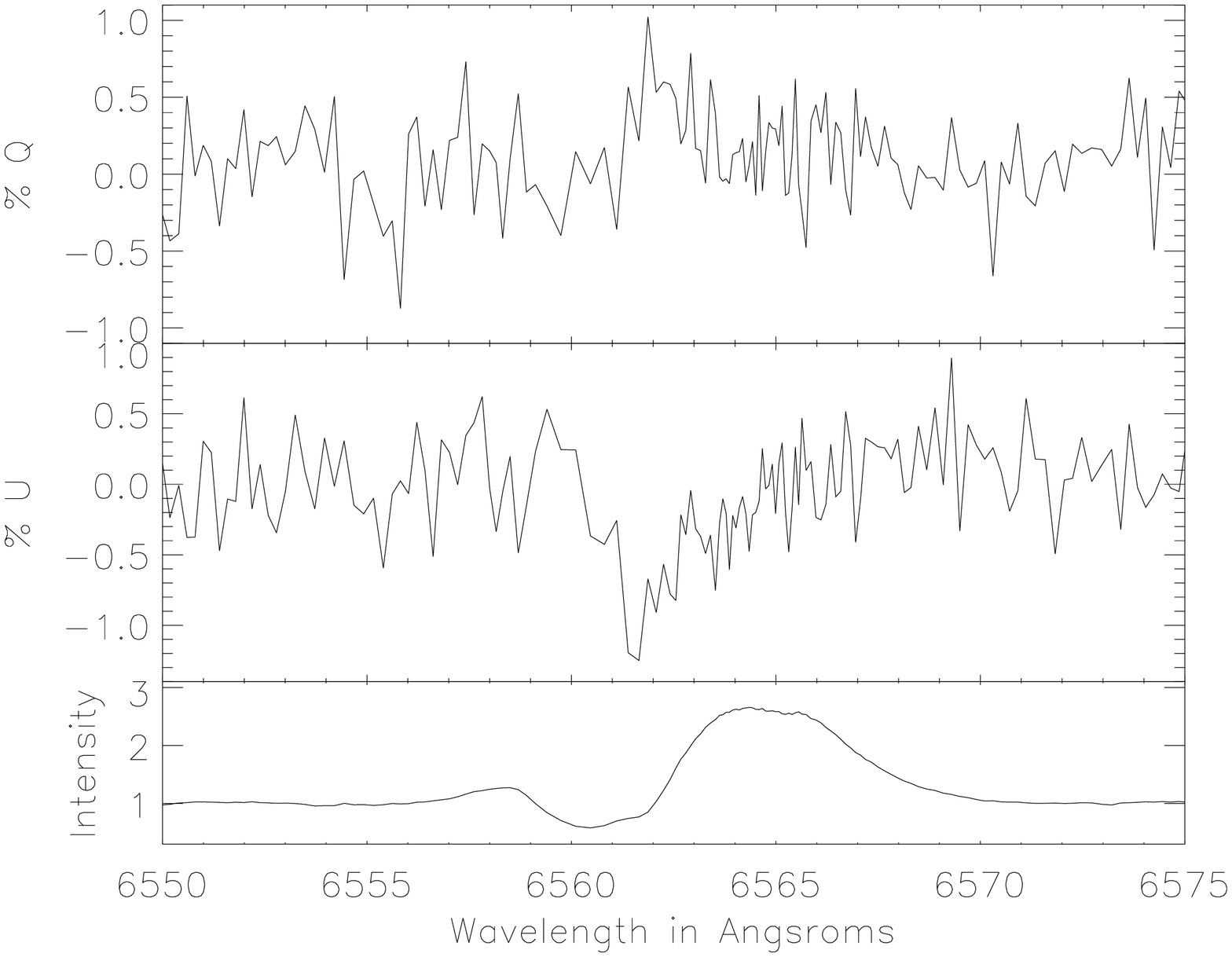}}
\quad
\subfloat[MWC 758 QU Plot]{\label{fig:swp-mwc758-qu}
\includegraphics[ width=0.35\textwidth, angle=90]{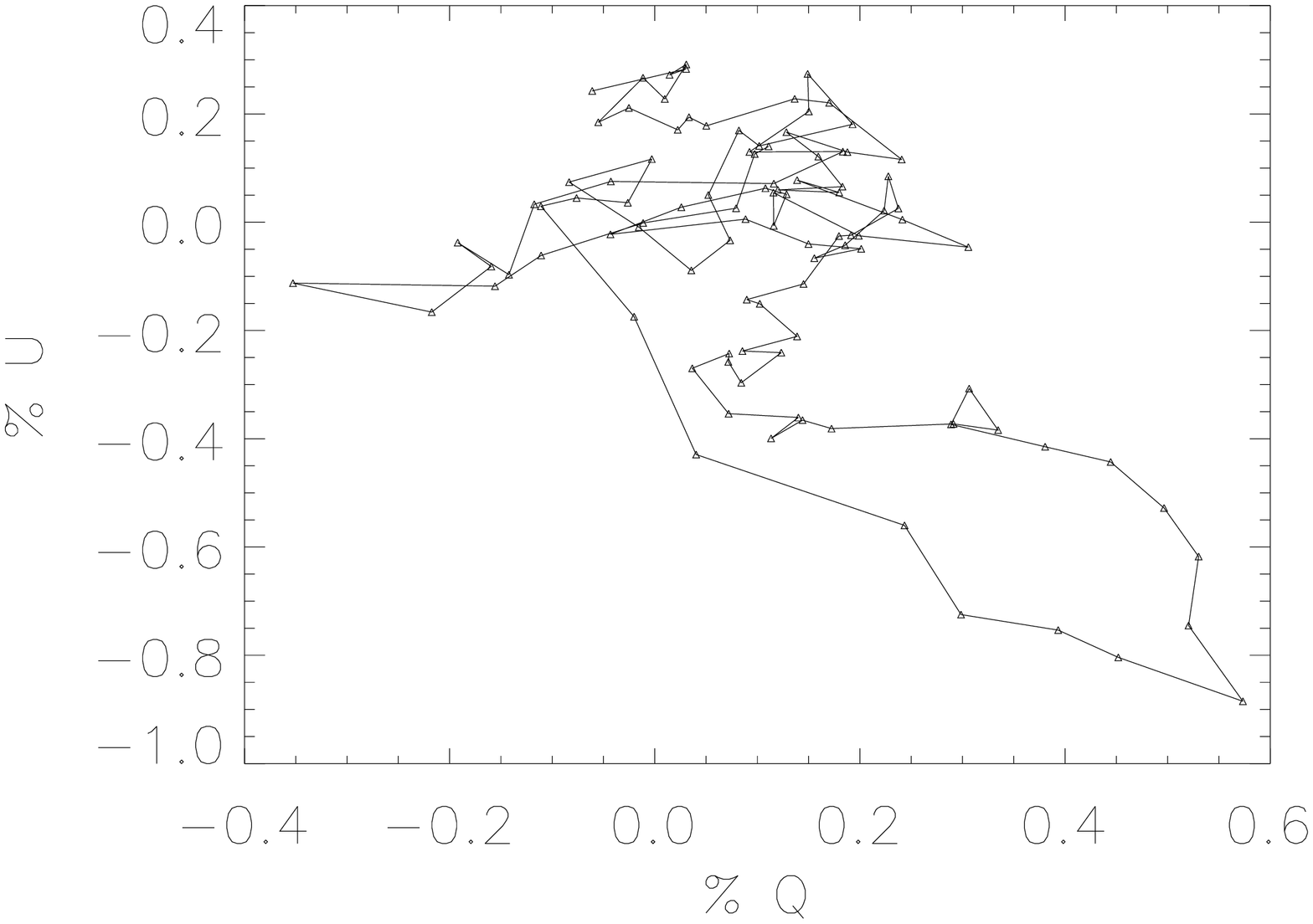}}
\quad
\subfloat[MWC 758 ESPaDOnS Archive Data]{\label{fig:swp-mwc758-esp}
\includegraphics[width=0.5\textwidth, angle=90]{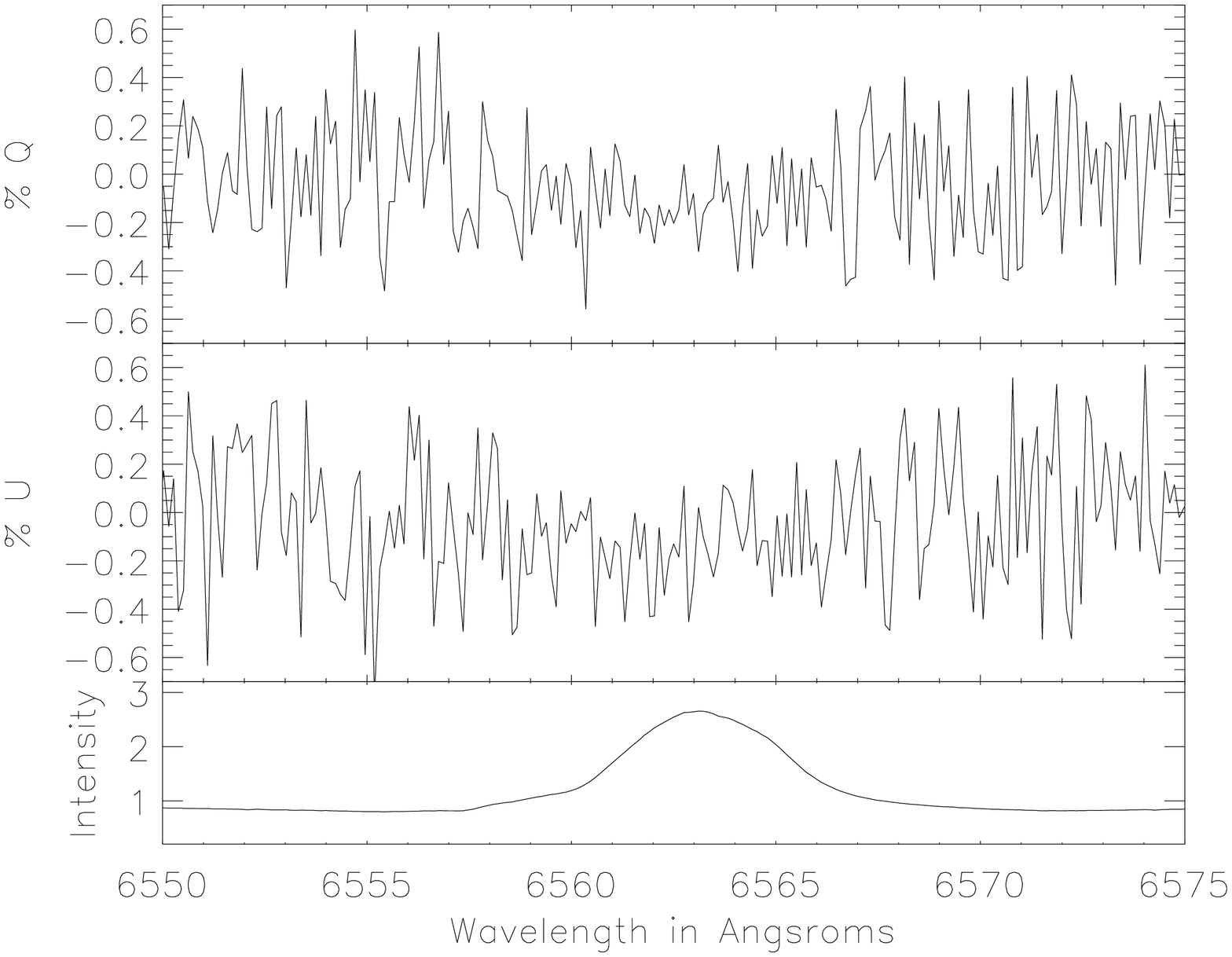}}
\caption[MWC 758 Spectropolarimetry]{{\bf a)} An example polarized spectrum for the MWC 758 H$_\alpha$ line. The spectra have been binned to 5-times continuum. The top panel shows Stokes q, the middle panel shows Stokes u and the bottom panel shows the associated normalized H$_\alpha$ line. There is clearly a detection in the blue-shifted absorption through line center of -1.0\% in u. {\bf b)} This shows q vs u from 6553.7{\AA} to 6569.7{\AA}.  The knot of points at (0.0,0.0) represents the continuum. {\bf c)} Archived ESPaDOnS observations for MWC 758 on February 9th, 2006.}
\label{fig:swp-mwc758}
\end{figure}

\begin{figure}
\centering
\subfloat[HD 45677 Polarization Example]{\label{fig:swp-hd456-indiv}
\includegraphics[ width=0.35\textwidth, angle=90]{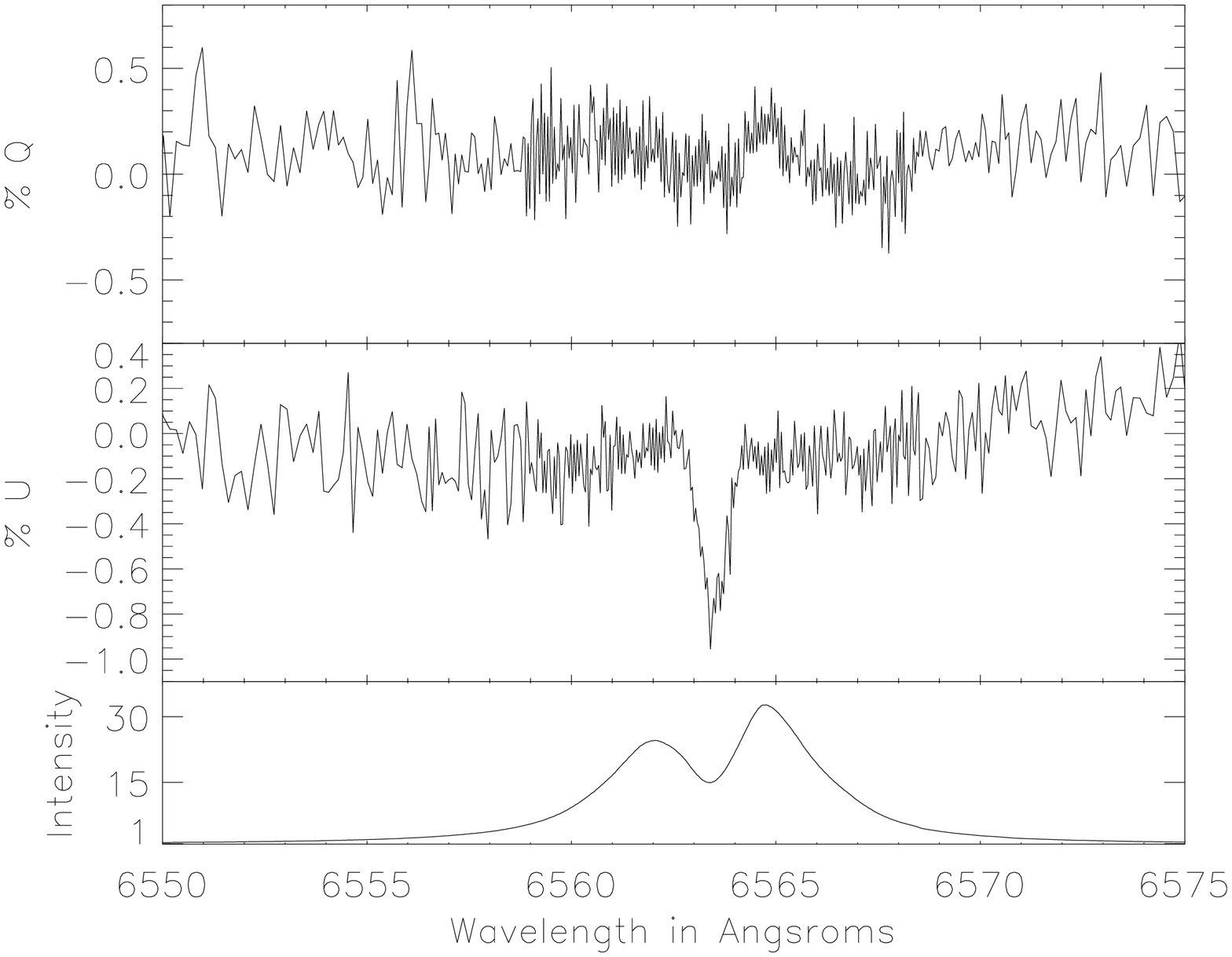}}
\quad
\subfloat[HD 45677 QU Plot]{\label{fig:swp-hd456-qu}
\includegraphics[width=0.35\textwidth, angle=90]{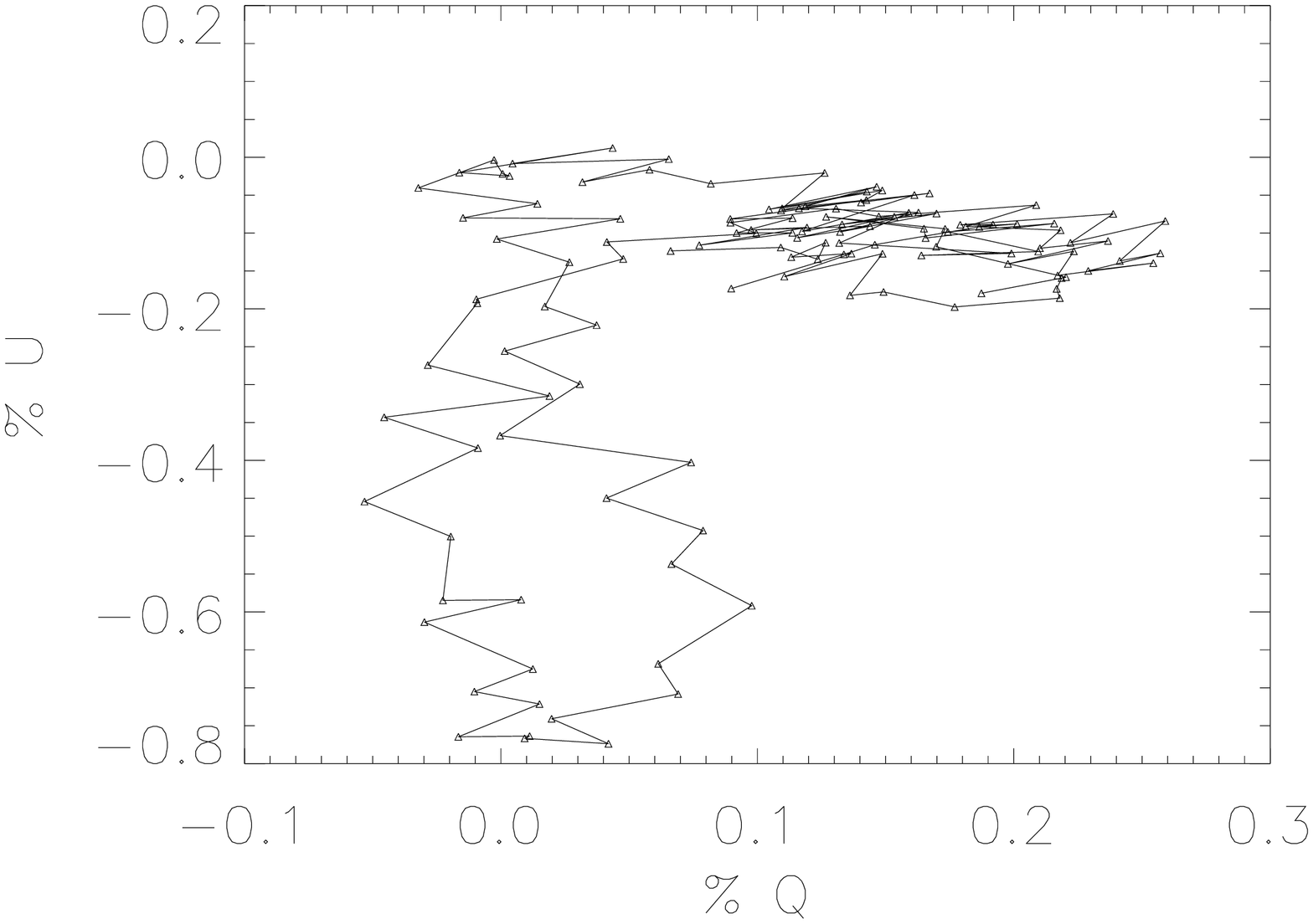}}
\caption[HD 45677]{An example polarized spectrum for the HD 45677 H$_\alpha$ line. The spectra have been binned to 5-times continuum. The top panel shows Stokes q, the middle panel shows Stokes u and the bottom panel shows the associated normalized H$_\alpha$ line. There is clearly a detection of -0.2\% in q and -0.8\% in u. {\bf b)} This shows q vs u from 6560.0{\AA} to 6565.2{\AA}.  The knot of points at (0.0,0.0) represents the continuum.}
\label{fig:swp-hd456}
\end{figure}

\begin{figure}
\centering
\includegraphics[width=0.45\textwidth, angle=90]{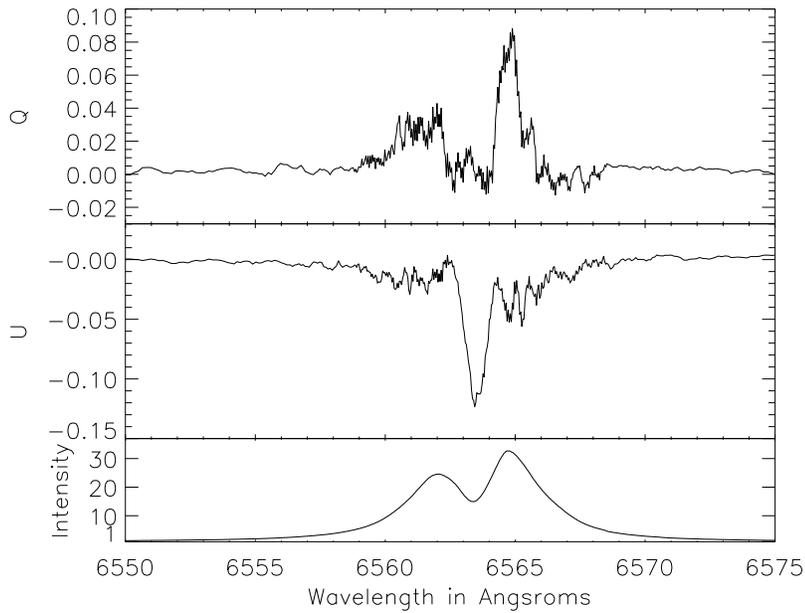}
\caption[HD 45677 Polarized Flux]{The polarized flux (q*I) for the HD 45677 of the previous figure.}
\label{fig:swp-hd456-pfx}
\end{figure}

\twocolumn

\subsection{HD 36112 - MWC 758}

	This star did not have the highest signal-to-noise but it has a detectable signature. Figure \ref{fig:mwc758} shows all 11 measurements and there is a subtle effect that can be seen in the u spectrum. However, in an individual, higher signal-to-noise spectrum of figure \ref{fig:swp-mwc758} you can see a 1\% change in u. There is also a smaller effect across the q spectrum. The qu-plot of figure \ref{fig:swp-mwc758-qu} shows the qu change clearly. The continuum knot is clear and there is a mostly linear extension out to (+0.6,-1.0). In the ESPaDOnS archival data from February 9th 2006, this star was a non-detection but the H$_\alpha$ line showed significantly less absorption than in almost all of the HiVIS observations. Beskrovnaya et al. 1999 report a continuum polarization of near 0.15\% but with very significant nightly variability of  0.4\%. This star did not show a spectropolarimetric effect in Vink et al. 2002.
		
	In the context of both depolarization and disk theories, this star is somewhat difficult to explain. The polarization change occurs mainly on the blue-shifted transition from absorption to emission, and is not centered in the absorptive trough. One could attempt to attribute the broad change in polarization, and even the strong decrease in u to the depolarization effect with the blue side of the absorption precisely ``filling in" the depolarization to the continuum value. However, the maximum decrease in u occurs not where there is maximum emission, but just at the edge of the blue-shifted absorption. In qu-space, the changes are close to being in-phase but are not quite matched. There is significant width to the qu-loop arising from the small increase in q. In disk theory, a spectropolarimetric effect must be present on both sides for orbital motion, and this is not seen.

\subsection{HD 45677}

	In the line profiles presented in Oudmaijer et al. 1999, the H$_\alpha$ line in this star is single peaked and asymmetric, quite different than the double-peaked line profile observed. They report a continuum polarization of 0.33\% in January 1995 and 0.14\% in December 1996 with line/continuum ratios of 35 and 34 respectively. They also report aline-effect as an increase in the degree of polarization of roughly 0.4\% in both observations. Grady et al. 1993 presented evidence of accreting gas in this system based on the UV spectra.

	This star has a large and clear signature in the central absorption of the emission line. It is detected at various amplitudes in all 7 measurements of figure \ref{fig:hd456}. Figure \ref{fig:swp-hd456-indiv} shows a single example of this. The q spectrum shows a small amplitude change on the red emission peak and the u spectrum shows a nearly 1\% decrease in the central absorption. In the qu-plot of figure \ref{fig:swp-hd456-qu} the u decrease causes the 1\% vertical extension with a complete return to continuum before the small increase in q causes the horizontal extension. The polarized flux shows that there is a significant amount of Q flux in both emission peaks, but none in the central absorption. The U flux is much larger and dominated by the flux in the central absorption.

	In the context of scattering theory, this star is also difficult to explain. The detection is quite strong and is mostly confined to the central absorptive notch in the emission line. As can be clearly seen in the qu-plot as well as the polarized flux, there is significant polarization across the entire line, but the functional form is quite strange. A strong change in q at line center, and a small, independent change in u only on the red-shifted emission peak. Depolarization seems unlikely as an explanation because there is essentially no detectable broad signature, and the absorption in line-center is not that strong. Similarly, disk theory produces blue and red shifted effects, which are not seen.

\subsection{HD 158643 - 51 Oph}

\begin{figure}
\centering
\subfloat[51 Oph Polarization Example]{\label{fig:swp-51oph-indiv}
\includegraphics[ width=0.28\textwidth, angle=90]{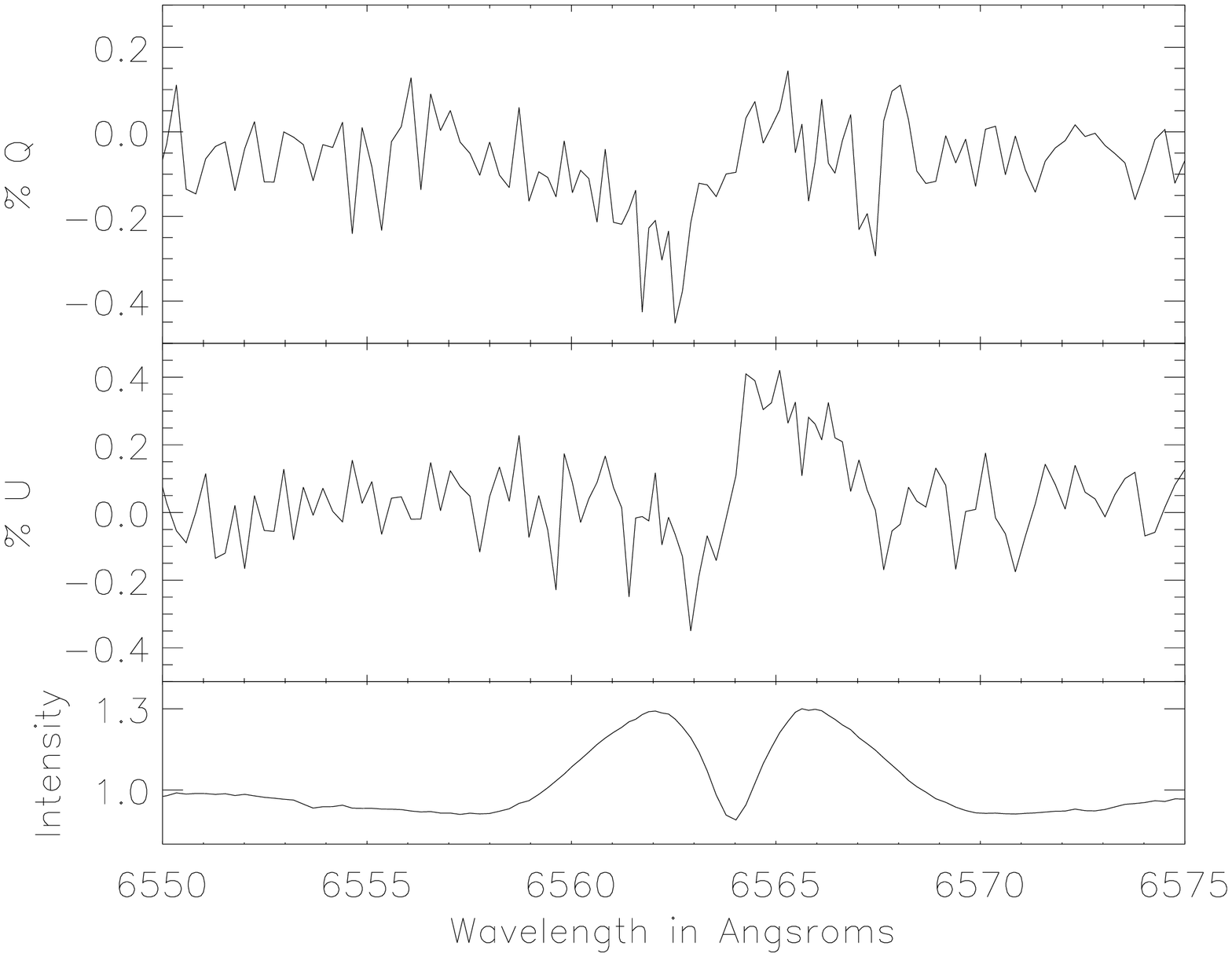}}
\quad
\subfloat[51 Oph QU Plot]{\label{fig:swp-51oph-qu}
\includegraphics[ width=0.28\textwidth, angle=90]{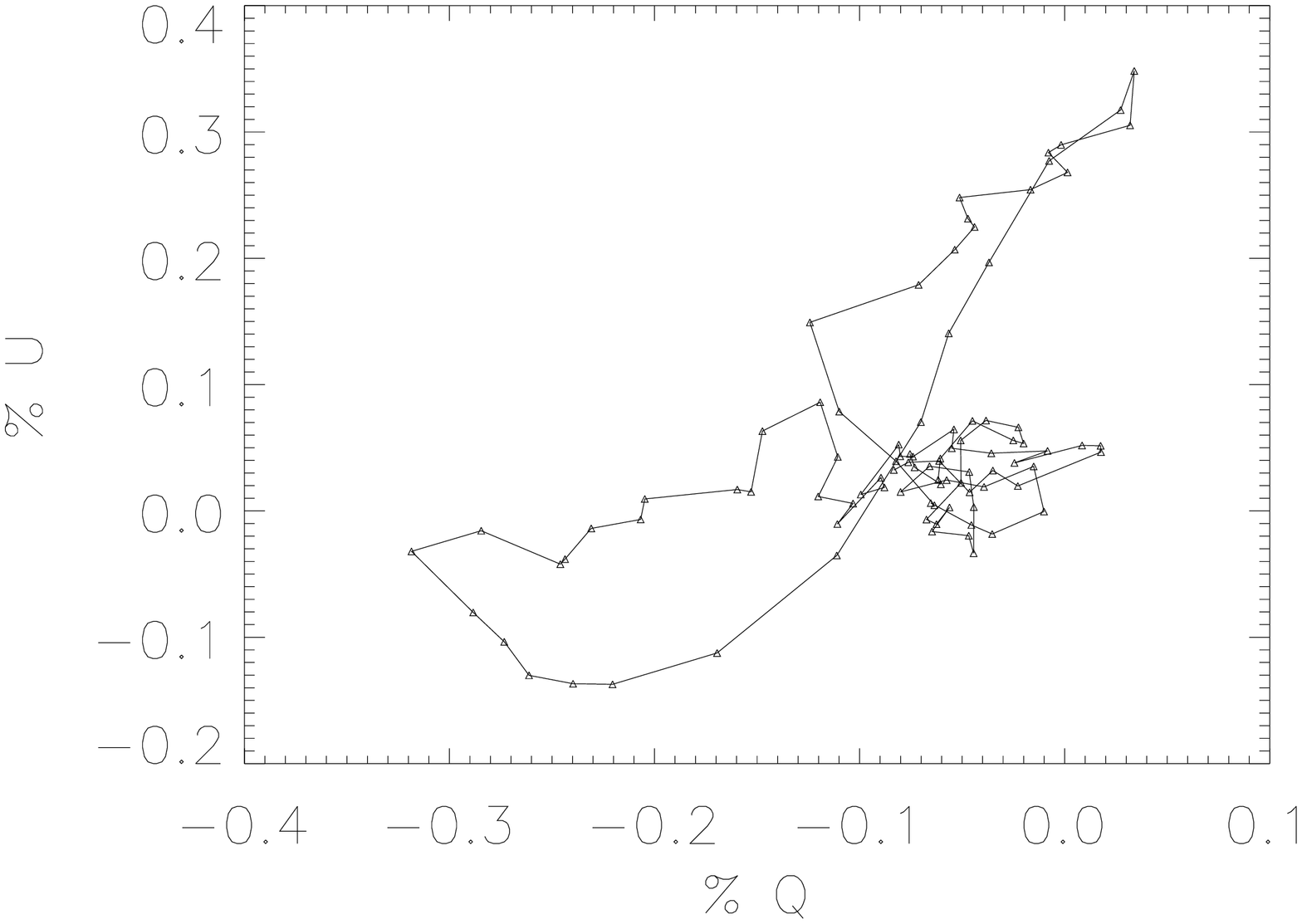}}
\quad
\subfloat[51 Oph Archive ESPaDOnS Spectropolarimetry]{\label{fig:swp-51oph-esp}
\includegraphics[width=0.28\textwidth, angle=90]{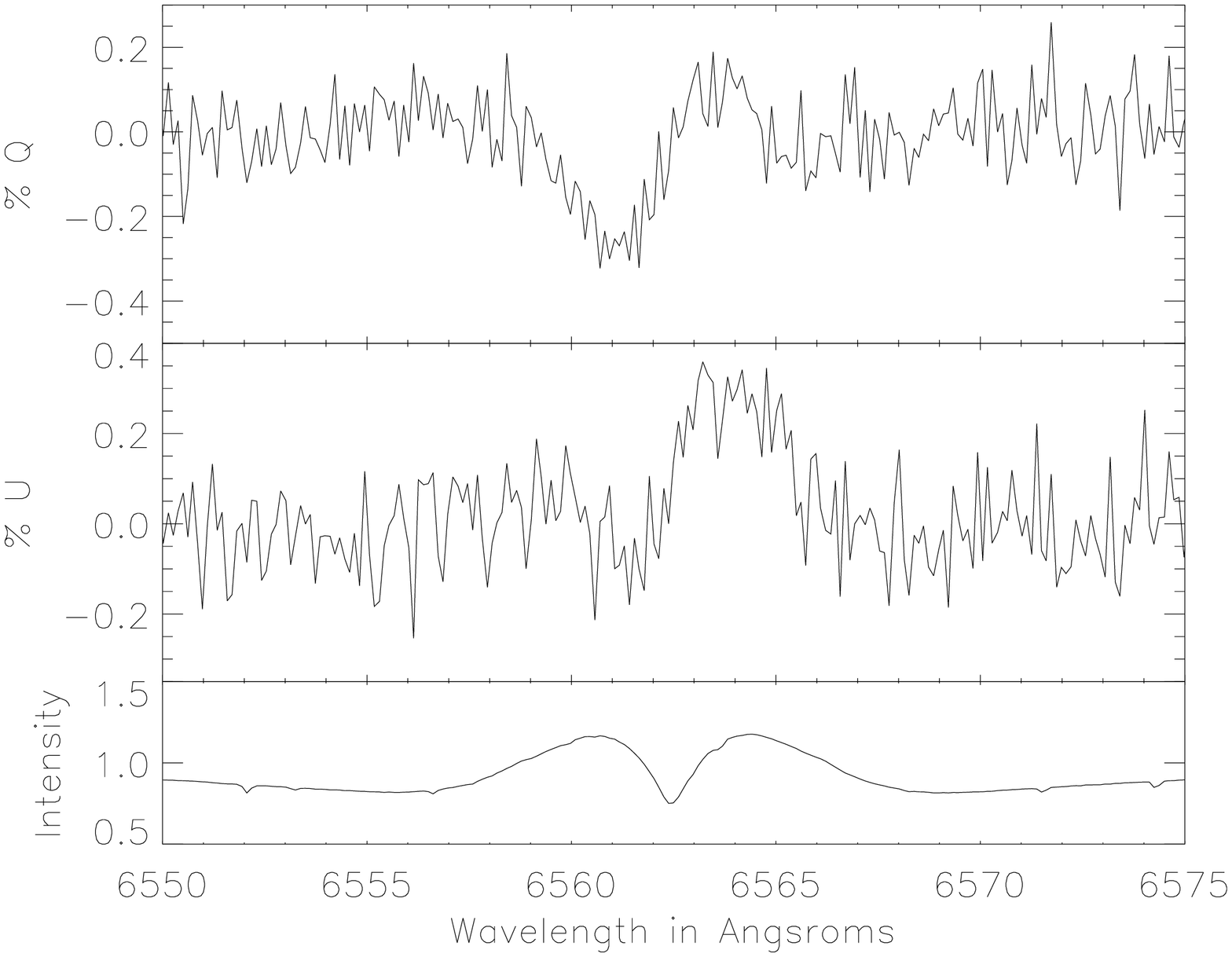}}
\quad
\subfloat[51 Oph Archive ESPaDOnS QU Pllot]{\label{fig:swp-51oph-esp-qu}
\includegraphics[width=0.28\textwidth, angle=90]{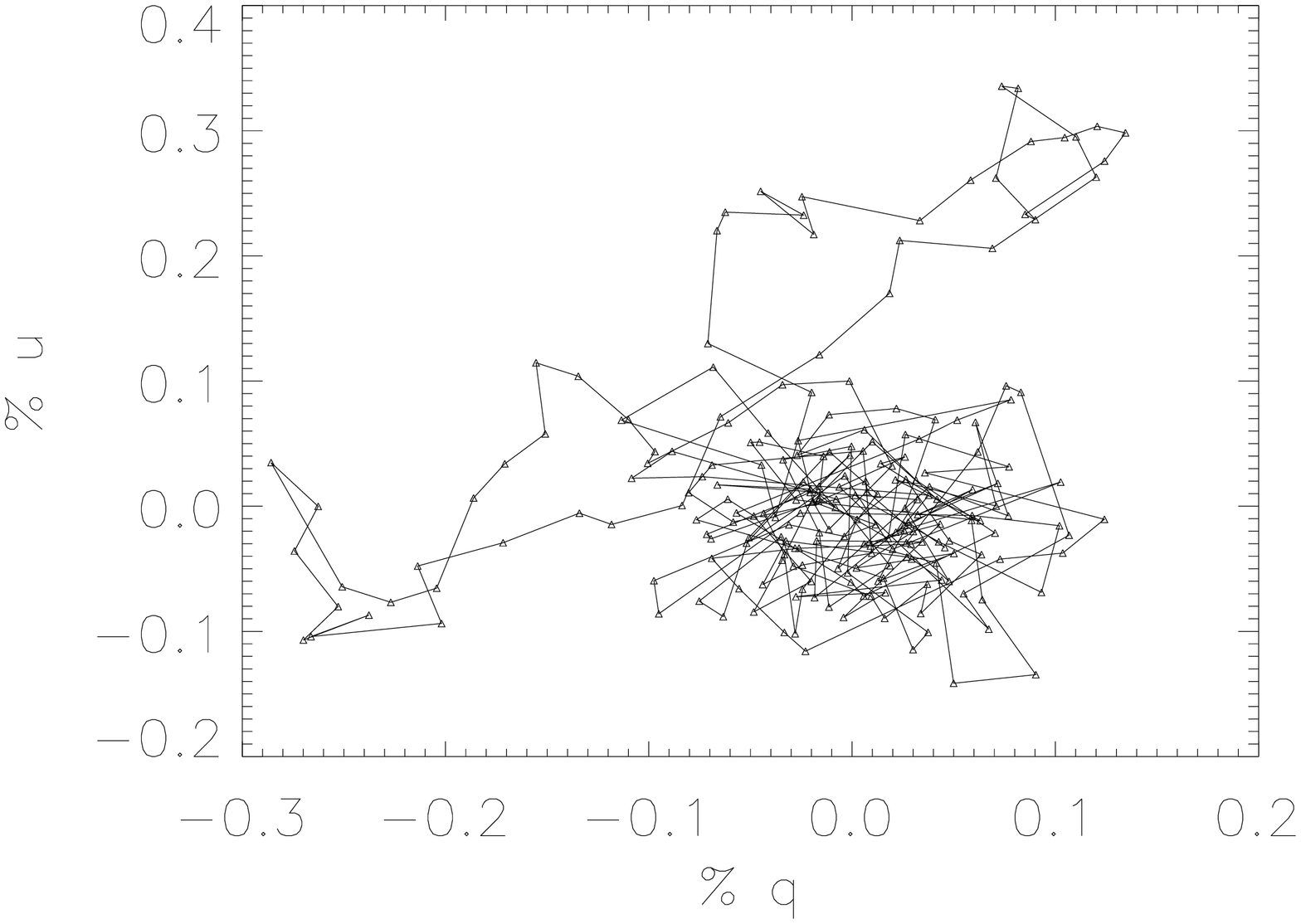}}
\caption[51 Oph Polarization]{An example polarized spectrum for the 51 Oph H$_\alpha$ line. The spectra have been binned to 5-times continuum. {\bf a)} The top panel shows Stokes q, the middle panel shows Stokes u and the bottom panel shows the associated normalized H$_\alpha$ line. Both q and u show an antisymmetric signature of roughly 0.3\% {\bf b)} This shows q vs u from 6554.9{\AA} to 6574.3{\AA}. The knot of points at (0.0,0.0) represents the continuum. {\bf c)} The ESPaDOnS archive data for 51 Oph on August 13th, 2006. Both q and u again show antisymmetric signatures of 0.3\% {\bf d)} The corresponding qu plot, which looks fairly similar to that of HiVIS.}
\label{fig:swp-51oph}
\end{figure}

	This star has a significant but small change across the entire line in all 3 measurements of figure \ref{fig:51oph}. The amplitude is roughly 0.1\% in all measurements of figure \ref{fig:51oph}. Figure \ref{fig:swp-51oph-indiv} shows an individual example where the change can be seen as a decrease in both q and u on the blue-shifted emission and then an increase in both q and u on the red-shifted emission. The qu-plot illustrates this as nearly linear extensions in qu-space.  Chavero et al. 2006 report an R-band polarization of 0.35$\pm$0.05\% and show that a Serkowski law does not fit the stars polarization. Oudmaijer et al. 2001 found 0.47\% V-band polarization and a similar Serkowski mis-fit. They note that this star is sometimes listed as a Vega-type star. The polarization increases towards the blue, going from 0.3\% at 7000-8500{\AA} to 0.55\% at 3500{\AA} and 4500{\AA}. From this they conclude that circumstellar material (likely disks) cause the intrinsic polarization. They argue that the dust responsible for the polarization has small grains and is the same dust responsible for the IR emission.
	
	In the context of the scattering theories, again this star is difficult to interpret. The red-shifted emission peak shows an increase in both q and u while the blue-shifted emission peak shows a decrease in both q and u. The transition between positive and negative changes occurs right at line center. While the changes are both in phase and show linear extensions in qu-space, the opposite directions with wavelength is difficult to explain in both disk and depolarization contexts. Depolarization is unidirectional with absorption acting in the opposite direction. In disk scattering with orbital motion, the profiles are symmetric. Neither morphology fits this star as well. However, the polarization-in-absorption is still a possible fit.

\subsection{HD 259431 - MWC 147 - V700 Mon}

	The observations of Oudmaijer et al. 1999 show a marginal decrease in polarization of 0.2\% across the line with a continuum of 1.06\% and a line/continuum ratio of 11. In Mottram et al. 2007, the observations also showed a similar signature but with only six resolution elements across the bulk of the emission.

\onecolumn

\begin{figure}
\centering
\subfloat[MWC 147 Polarization Example]{\label{fig:swp-mwc147-indiv}
\includegraphics[ width=0.28\textwidth, angle=90]{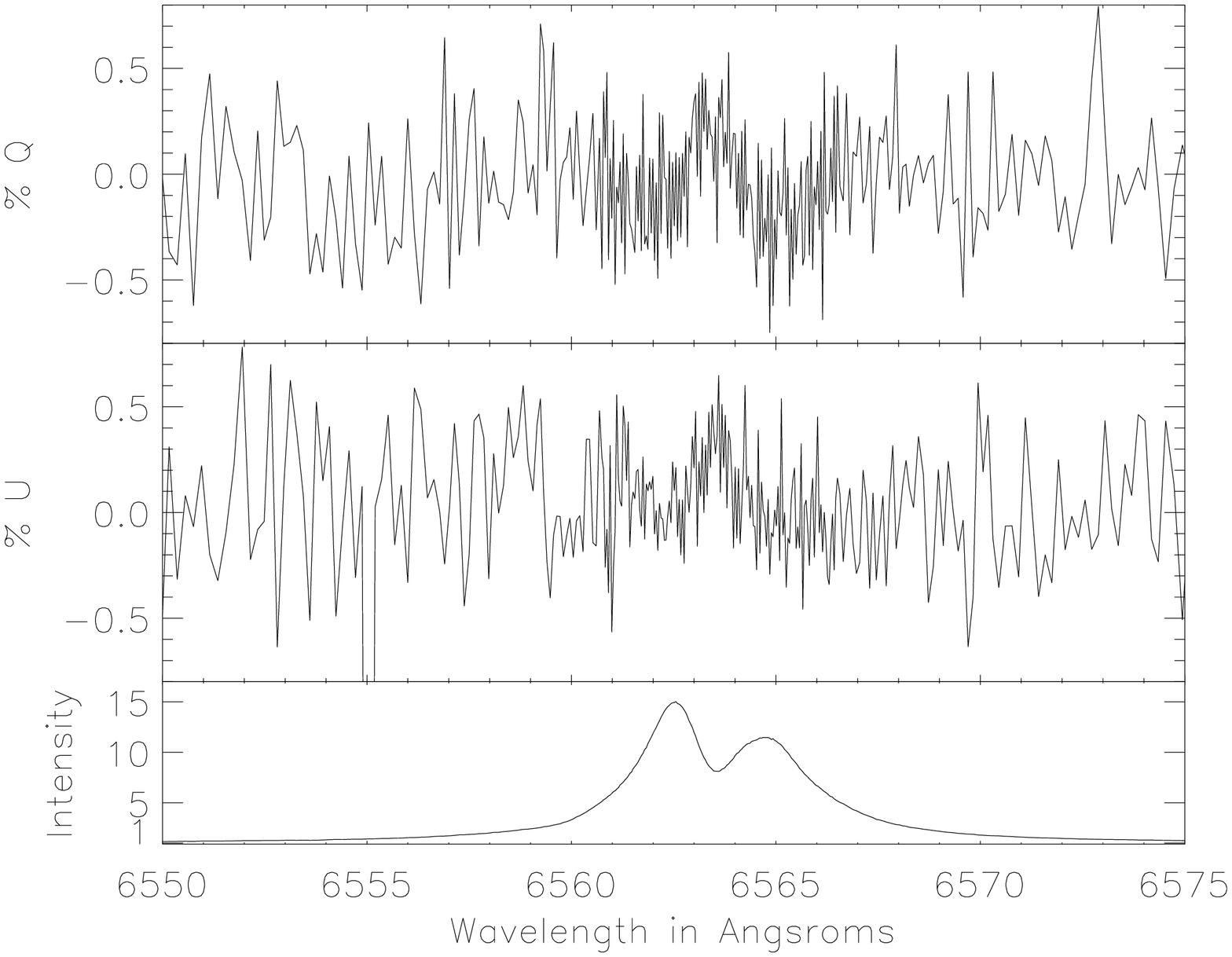}}
\quad
\subfloat[MWC 147 QU Plot]{\label{fig:swp-mwc147-qu}
\includegraphics[ width=0.28\textwidth, angle=90]{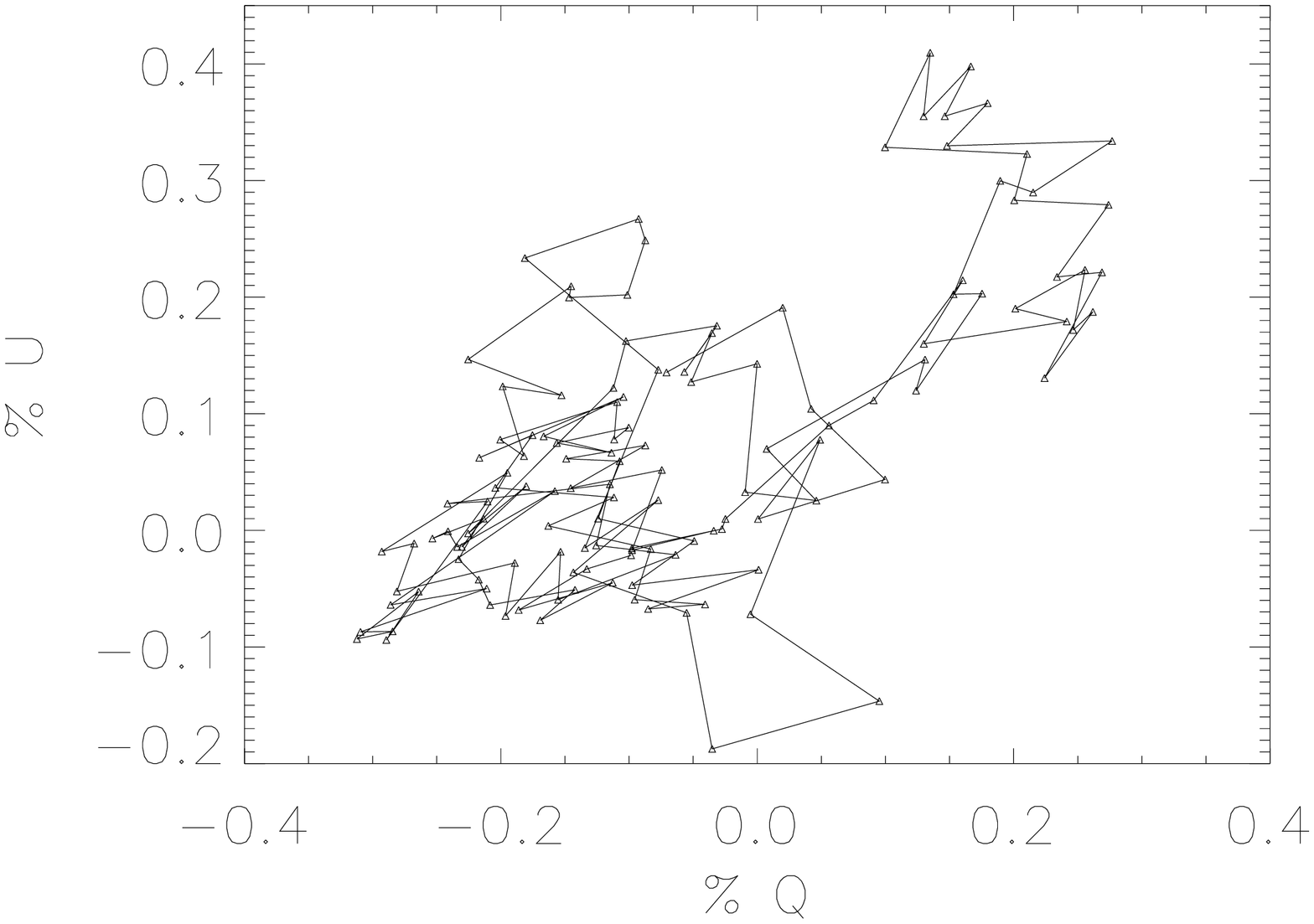}}
\caption[MWC 147 Polarization]{{\bf a)} An example polarized spectrum for the MWC 147 H$_\alpha$ line. The spectra have been binned to 5-times continuum. The top panel shows Stokes q, the middle panel shows Stokes u and the bottom panel shows the associated normalized H$_\alpha$ line. There is a fairly clear detection across the center of the absorption line. {\bf b)} This shows q vs u from 6561.4{\AA} to 6565.9{\AA}. The knot of points at (-0.2,0.0) represents the polarization on both emission peaks, which is significantly less than the average continuum. Since such a narrow wavelength range was chosen to avoid the noisy data outside the line core, the continuum knot is non-existant and is represented by the larger swarm of points generally around zero.}
\label{fig:swp-mwc147}
\end{figure}

\begin{figure}
\centering
\subfloat[HiVIS Spectropolarimetry]{\label{fig:swp-hd144-indiv}
\includegraphics[ width=0.35\textwidth, angle=90]{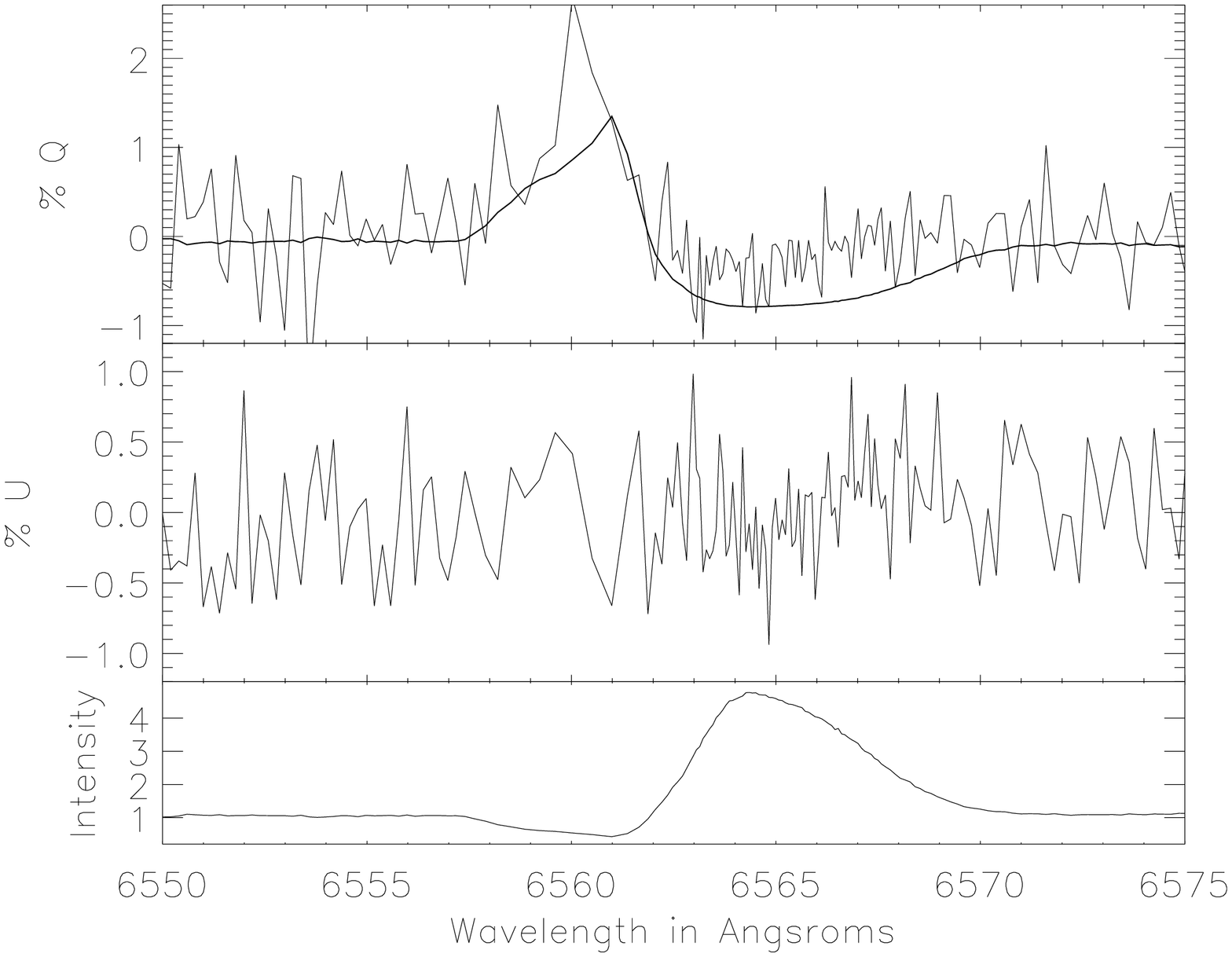}}
\quad
\subfloat[ESPaDOnS Archive Spectropolarimetry]{\label{fig:swp-hd144-esp}
\includegraphics[width=0.35\textwidth, angle=90]{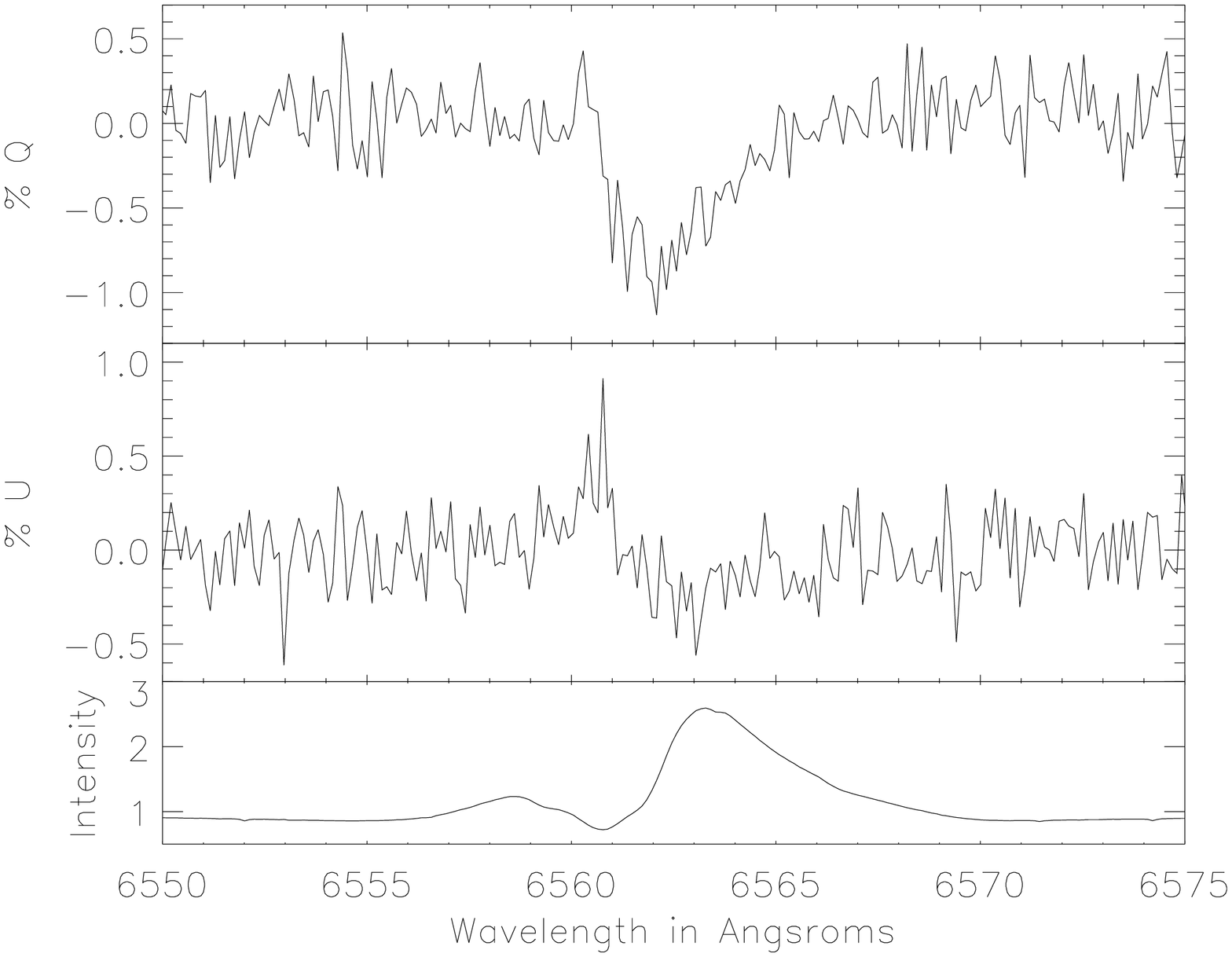}}
\quad
\subfloat[HiVIS Q Adjusting]{\label{fig:swp-hd144-pfix}
\includegraphics[ width=0.35\textwidth, angle=90]{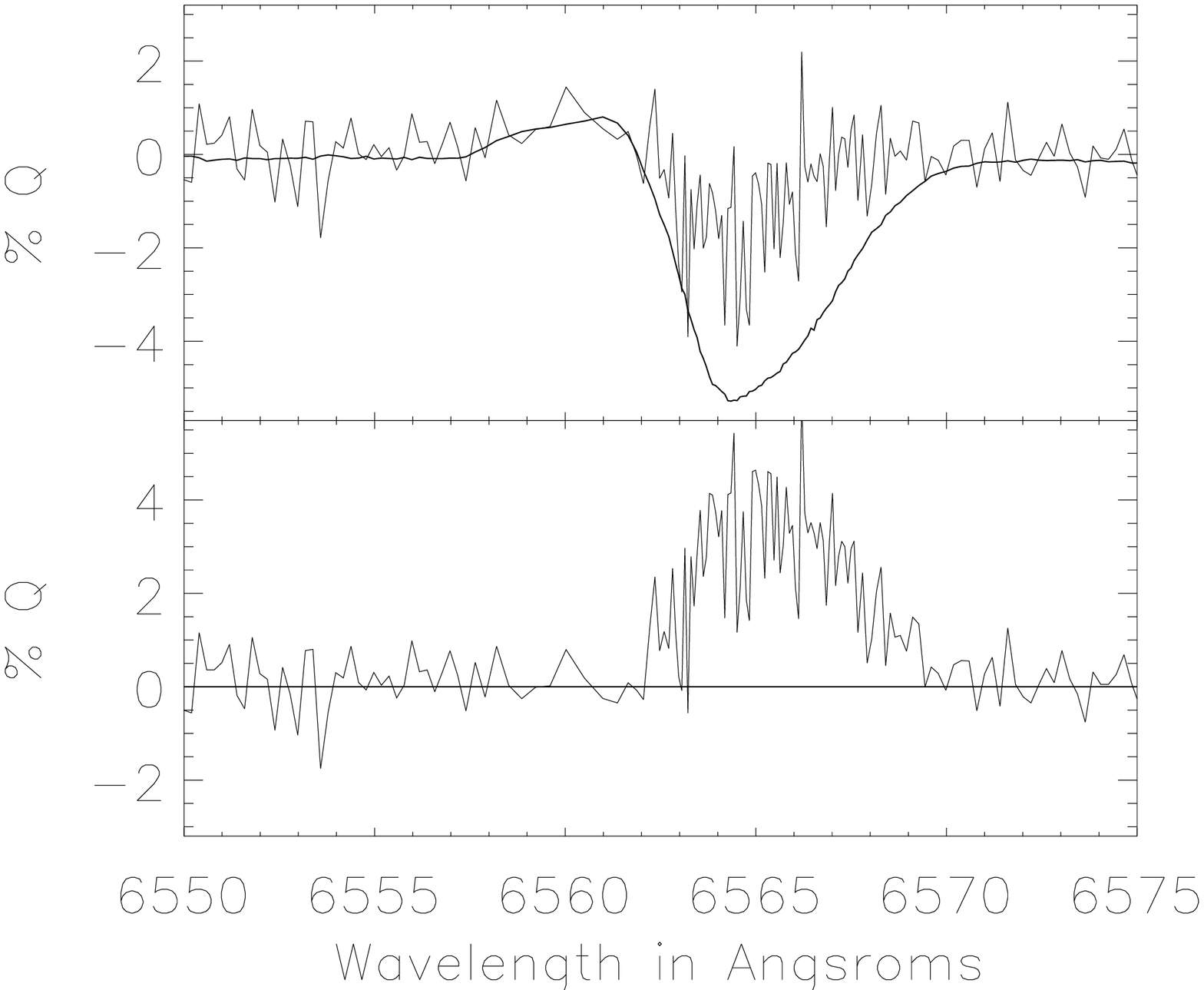}}
\quad
\subfloat[HiVIS q Adjusting]{\label{fig:swp-hd144-qfix}
\includegraphics[ width=0.35\textwidth, angle=90]{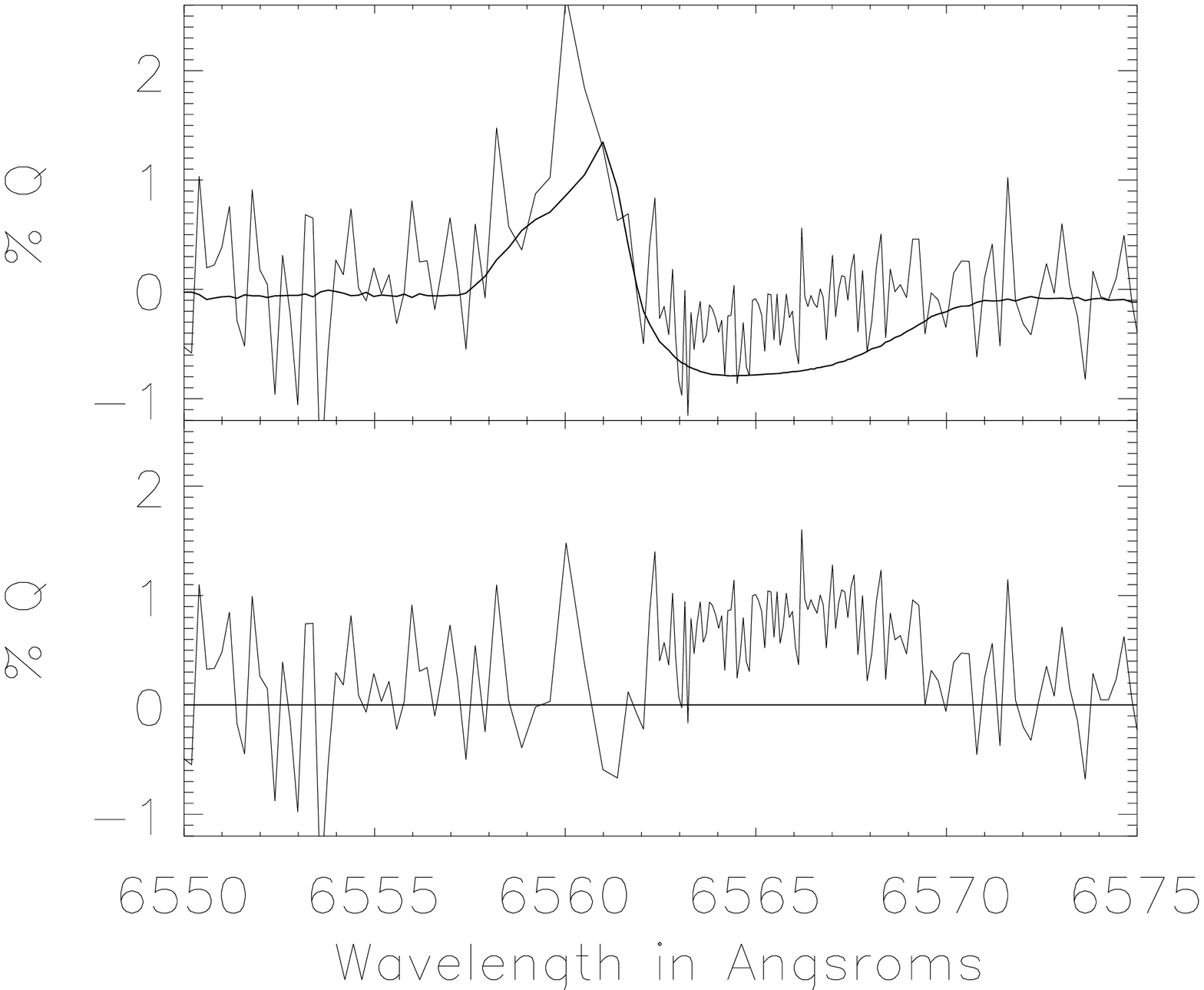}}
\caption[HD 144432 Spectropolarimetry]{Polarized spectra for the HD 144432 H$_\alpha$ line from  {\bf a)} ESPaDOnS on August 14th 2006 and  {\bf b)} HiVIS on August 29th, 2007. The HiVIS polarized spectra have been binned to 5-times continuum. The top panel iin a) and b) shows Stokes q, the middle panel shows Stokes u and the bottom panel shows the associated normalized H$_\alpha$ line. There is clearly a detection in the blue-shifted absorption of -1.0\% in q and 0.5\% in u for the ESPaDOnS data and 2\% in q for the HiVIS data. {\bf c)} A 1/I subtraction of Stokes Q - the polarized flux shown in the top box was adjusted to remove the 2\% polarization in the blue-shifted absorption. The dark line shows the fitted Stokes Q. After removal, the resulting Stokes Q is shown in the bottom box with a significant polarized flux on the side of the emission line. {\bf d)} Stokes q after the adjustment. The top box shows the original Stokes q with the removed 1/I signature. The bottom box shows the resulting Stokes q with the red-shifted residual.}
\label{fig:swp-hd144-all}
\end{figure}

\begin{figure}
\centering
\includegraphics[width=0.45\textwidth, angle=90]{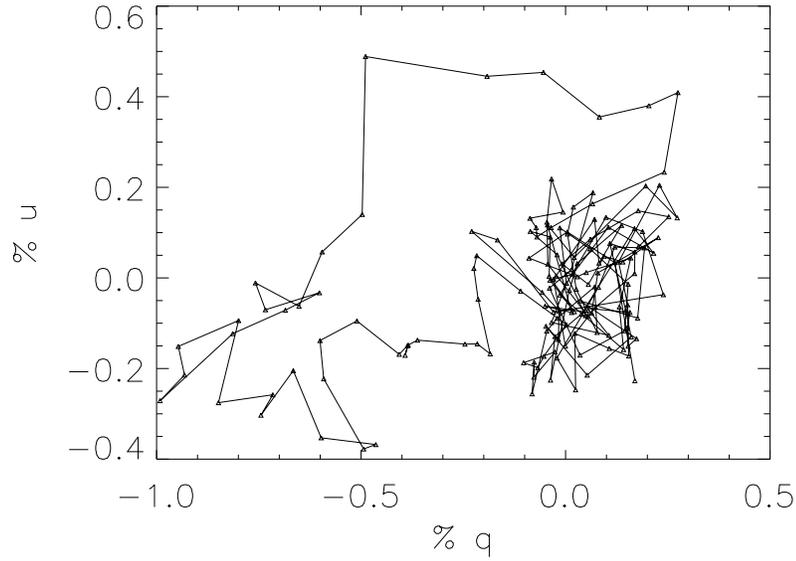}
\caption[HD 144432 ESPaDOnS Archive QU Plot]{The qu loop for the HD144432 ESPaDOnS archive observations.}
\label{fig:swp-esp-hd144-qu}
\end{figure}

\begin{figure}
\centering
\subfloat[V1295 Aql Polarization Example]{\label{fig:swp-v1295-indiv}
\includegraphics[ width=0.35\textwidth, angle=90]{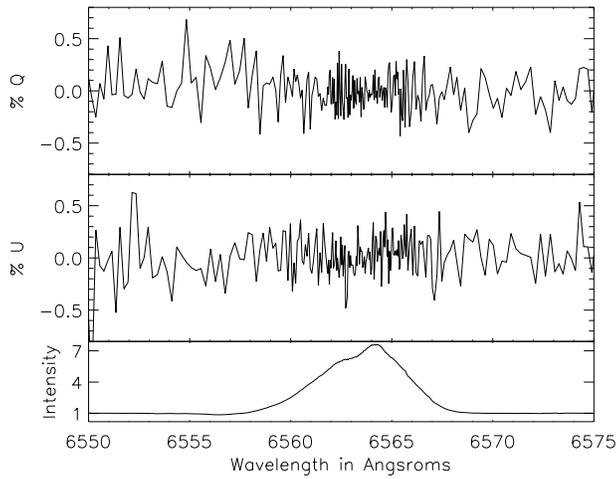}}
\quad
\subfloat[V1295 Aql QU Plot]{\label{fig:swp-v1295-qu}
\includegraphics[ width=0.35\textwidth, angle=90]{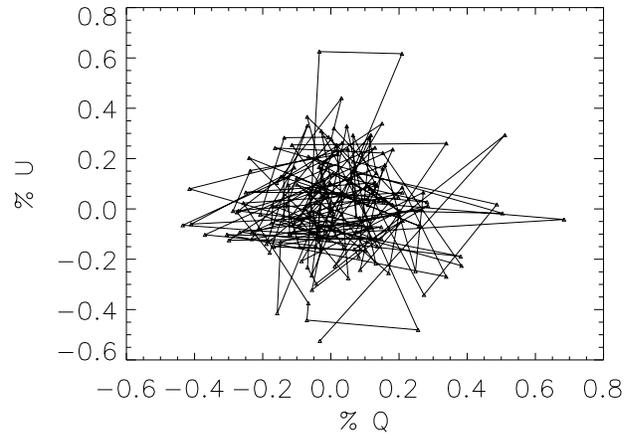}}
\caption[V1295 Aql Polarization]{{\bf a)} An example polarized spectrum for the V1295 Aql H$_\alpha$ line. The spectra have been binned to 5-times continuum. The top panel shows Stokes q, the middle panel shows Stokes u and the bottom panel shows the associated normalized H$_\alpha$ line. {\bf b)} This shows q vs u from 6551.3{\AA} to 6565.6{\AA}.  The knot of points at (0.0,0.0) represents the continuum.}
\label{fig:swp-v1295}
\end{figure}

\begin{figure}
\centering
\subfloat[V1295 Aql ESPaDOnS Archive Data]{\label{fig:swp-v1295-esp-archive}
\includegraphics[width=0.35\textwidth, angle=90]{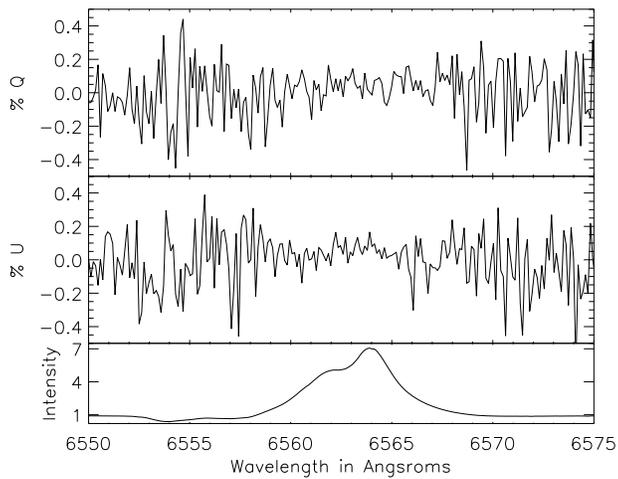}}
\quad
\subfloat[GU CMa ESPaDOnS Archive Data]{\label{fig:swp-gucma-esp-swap}
\includegraphics[width=0.35\textwidth, angle=90]{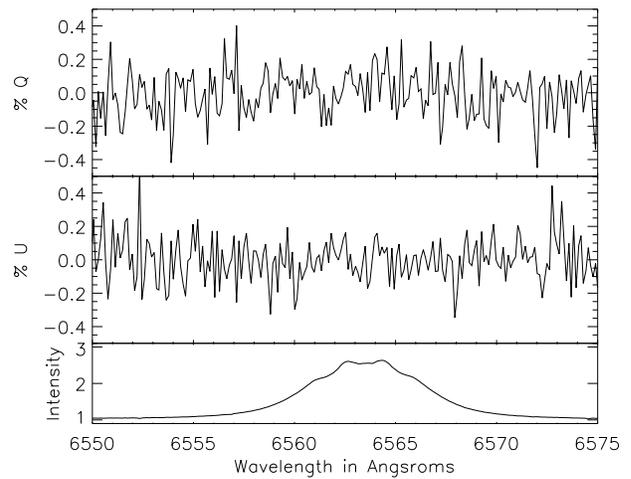}}
\quad
\subfloat[HD 141569 ESPaDOnS Archive Data]{\label{fig:swp-hd141-esp-swap}
\includegraphics[width=0.35\textwidth, angle=90]{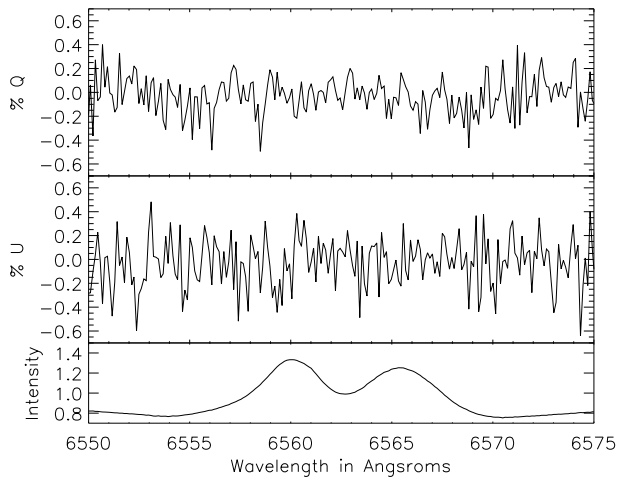}}
\quad
\subfloat[HD 35929 ESPaDOnS Archive Data]{\label{fig:swp-hd359-esp-swap}
\includegraphics[width=0.35\textwidth, angle=90]{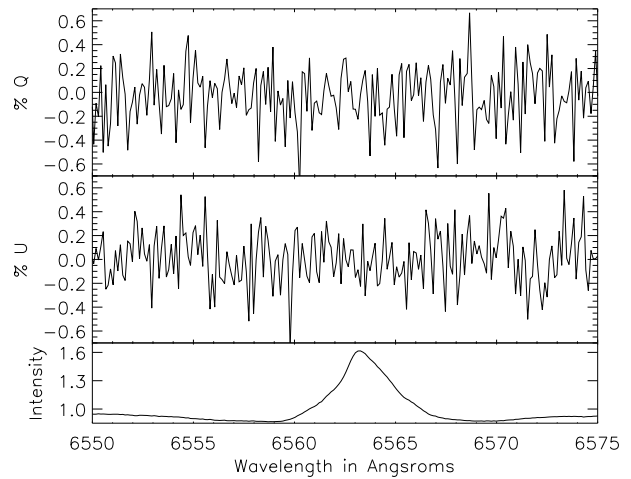}}
\caption[Archive ESPaDOnS Non-Detections: V1295 Aql, GU Cma, HD 141569, HD 35929]{Non-detections in the polarized spectra from the ESPaDOnS archive. Each panel shows Stokes q, Stokes u and the normalized intensity profile from top to bottom. {\bf a)} V1295 Aql from August 13th 2006. {\bf b)} GU Cma from February 9th 2006. {\bf c)} HD 141569 and {\bf d)} HD 35939 both from February 8th 2006.}
\label{fig:swp-nd-esp}
\end{figure}

\twocolumn

	This star had a marginal detection in the center of the emission line in some but not all of the data sets. It's hard to see in the 8 measurements of figure \ref{fig:mwc147}, but in the single example of figure \ref{fig:swp-mwc147-indiv} the change can be seen as small amplitude changes in both q and u across the core of the emission line. The q spectrum shows a small decrease in q in both emission peaks of roughly 0.2\% with a return to  The qu-plot of figure \ref{fig:swp-mwc147-qu} shows a small wavelength range, 6561.4 to 6565.9{\AA}, where the clear change across the central-absorption is seen. The aparent continuum-knot is not centered at zero because the wavelength range selected only includes the line core and both the blue-shifted and red-shifted emission peaks cluster at polarizations of (-0.2,0.0). Adding a greater wavelength range would, with these noise statistics, make the small change from (0,0) across the emission and absorption invisible.

	The morphology of this detection is difficult to discuss because of its smallness and narrowness. In qu-space the extension in the central absorption is nearly linear but with a significant deviation away from linear at the very tip. The polarization in both blue and red emission peaks is only marginally different from zero. With the weakness and narrowness of this detection, it is difficult to interpret this signature for or against any one theory. The change in q and u does seem to be phased, but there is width to the qu-loop. All that can be concluded at this point that there definitely is a small narrow signature.

\subsection{HD 144432}

	This star had only one good observation that is shown in figure \ref{fig:swp-hd144-indiv} from August 29th, 2007. There is a clear change in Stokes q of roughly 2\% in the absorptive component and 0.3\% in the blue side of the emission line. There is also archival ESPaDnS data from August 14th 2006 shown in figure \ref{fig:swp-hd144-esp}. In the year between observations, the absorption in H$_\alpha$ line increased and widened greatly to become a full P-Cygni profile. There is clearly a detection in the blue-shifted absorption of -1\% in q and 0.5\% in u for the ESPaDOnS data and 2\% in q for the HiVIS data. Initially, a 1/I error seemed like a likely candidate, but the calculations shown in figures \ref{fig:swp-hd144-pfix} and \ref{fig:swp-hd144-qfix} show that a 1/I signature does not fit the polarization spectrum. To show this, a fit to the polarized flux was performed to remove the 2\% signature. The 2\% signature in absorption was removed completely but only after introducing a significant red-shifted residual. The archival ESPaDnS data also provided an independent measurement showing that the HiVIS results were robust since the magnitude and wavelengths of the polarization change were the same.

	In the context of scattering theory, there is a resemblance between this star fits the morphology of the depolarization effect, however there are some problems. In the ESPaDOnS archive data, there is a very significant decrease in q and a smaller decrease in u across the center of the emission line. The blue-shifted absorption does shows a change in the opposite sense of the change across the emission line. But, inspecting the peak decrease in q shows that this also occurs not in the emission peak but in the transition from emission to absorption. Also, there is some width to the qu-loop as seen in figure \ref{fig:swp-esp-hd144-qu}. The increases in q and u are not phased together, suggesting that the removal of unpolarized light by absorption is more complex than this simple assumption. These two factors suggest that the depolarization effect may not be the correct theory for this star. As with almost all of the other stars, the single change in polarization in the center and blue-shifted components argues against the simple disk scattering theory.

\subsection{HD 190073 - MWC 325 - V1295 Aql}

	This star had no detectable polarization signature in all 17 observations. The line profiles were consistently asymmetric with the blue-shifted absorption as ample evidence for outflowing material. However, even in heavily binned profiles with 0.2\% detection thresholds there was no significant polarization change. Figure \ref{fig:v1295} shows all the data and examples of an individual spectrum is shown in figure \ref{fig:swp-v1295-indiv}. The qu-plot shows a simple continuum-knot centered at (0,0) with no significant asymmetry or deviation. It was also a non-detection in the archival ESPaDOnS data of Aug 13th 2006 shown in figure \ref{fig:swp-v1295-esp-archive} even though there is evidence for stronger absorption from the decreased line strength, deeper blue-shifted absorption, and a strongly blue-shifted absorption at 6554{\AA}.

\subsection{Non-Detections and Other Notes}

	GU CMa was a non-detection in the archival ESPaDOnS data of Feb 9th 2006 shown in figure \ref{fig:swp-gucma-esp-swap}. It was also a non-detection in Oudmaijer et al. 1999 with a continuum of 1.15\% and a line/continuum ratio of 3.  HD 141569 was a non-detection in the archival ESPaDOnS data of Feb 8th 2006 shown in figure \ref{fig:swp-hd141-esp-swap}. The emission from this star was always weak, being 1.3 in the archival data and 1.6 in the HiVIS observations. HD 35929 was a non-detection in the archival ESPaDOnS data of Feb 8th 2006 shown in figure \ref{fig:swp-hd359-esp-swap}. The star shows blue-shifted absorption though the emission from this star was weak. In the HiVIS observations, the line was somewhat stronger, 1.9 vs 1.6, with less clear absorption. It was also a non-detection in Vink et al. 2002. This star, as noted in chapter 6, has been studied in-depth spectroscopically and was, in 2004, determined to be an F2 III post-MS giant on the instability strip.

\subsection{Comments on Other Systems}
 	
	MWC 166 was a clear detection in Oudmaijer et al. 1999 of a 0.2\% decrease in polarization with a continuum of 1.15\% and a line/continuum ratio of 2.6. Mottram et al. 2007 show a decrease in polarization of 0.2\% across a continuum of 0.5\%. but with a singly-peaked asymmetric line profile. In two observations shown in figure \ref{fig:mwc166}, no effect at the 0.1\% magnitude is detected, though telescope polarization effects could be the cause as there are only a few observations at a few pointings. It should be noted that Oudmaijer et al. 2001 classify HD142666, HD 141569, and HD 163296 as HAeBe but also zero-age main-sequence. Il Cep was a non-detection in Mottram et al. 2007 as well as in this survey. XY Per showed a 0.5\% decrease in polarization in the central absorption from a continuum of 1.5\% in Vink et al. 2002 though HiVIS did not detect any signature. T Ori was reported to have a spectropolarimetric effect of 0.8\% on a continuum of 0.4\% in Vink et al. 2002. The polarization increase is detected at the wavelengths where there is spatially resolved, diffuse H$_\alpha$ emission. HiVIS detected a much larger amplitude signature but at higher resolution. Given the difference in relative flux contribution between the star and diffuse H$_\alpha$, these observations are compatible. However, since it cannot be certain what ontribution is intrinsic to the star, this star has not been included in the detections statistics. In Vink et al. 2002, KMS 27 showed a broad 0.2\% increase in polarization across the entire line on a continuum of 0.1\% with a P-Cygni type profile. This star is a non-detection in the survey, but with a lower signal-to-noise.

\subsection{Summary of Herbig Ae/Be Spectropolarimetry}

	This chapter presented the Herbig Ae/Be spectropolarimetry as well as detailed individual examples. The HiVIS survey has been complemented by ESPaDOnS observations as well as archival ESPaDOnS data. Most of the windy sources and more than half the disky sources showed spectropolarimetric signatures. The signatures show very different morphologies even among stars of similar H$_\alpha$ line type.
	
	In the windy sources, especially AB Aurigae and MWC 480, polarization changes were only seen in the P-Cygni absorption trough. The emission peak and red wing had the same polarization as the continuum. In other windy sources, such as HD 163296, MWC 120, MWC 758, HD 150193, and HD 144432 showed large polarization signatures, 1\% to 2\%, that were greatest in blue-shifted absorptive components even if those components did not go below continuum. There were several stars that were non-detections and one, HD 179218, that showed polarization across the entire line. Although there are difficulties classifying stars in this loose way and making rigorous statistical conclusions from such a small number of stars, more than half the windy stars showed significant polarization. 
	
	In the disky systems, such as MWC 158, HD 58647 and HD 45677, similar spectropolarimetric effects are seen in absorption. The polarization in the absorptive component, near line center, show polarization signatures of 1\% to 2\% while the the emissive components have a polarization at or very near continuum. In other systems, like MWC 147 and 51 Oph, the polarization signature is much smaller but it spans the entire width of the line. 
	
	There are several main morphological considerations when trying to establish the presence of a certain effect. The depolarization effect should show a broad change in polarization whenever there is emission. Absorption is said to preferentially remove unpolarized flux so the depolarization effect in absorption must act with an opposite sign of that in emission. The disk scattering effect also makes concrete morphological predictions that were compared to the observations. In most cases, there were several morphological arguments against these scattering theories. In all stars, there was much greater chance polarization effects in absorption, even if there was no polarization at any other part of the line. There are also amplitude arguments that must be considered. The disk-scattering effect does not have any natural amplitude and simply scales with the amount of scattered light. The depolarization effect is always acting on the underlying continuum polarization, which is measurable for these stars. Before advancing a new theory, observations of Be stars that do fit scattering theory will be presented to clarify the morphological differences between stars that do show polarization-in-absorption (Herbig Ae/Be's) and those that don't (Be's, B-Type Emission line stars).

\section{Be \& Emission-Line Star Comparison - Spectroscopy \& Spectropolarimetry}

	After observing so many stars that had very significant polarization effects that did not match current scattering theory, a shorter observing campaign of Be and other emission line stars was done for comparison and cross-checking. There is a much larger set of observations and theoretical work in the literature on Be stars. These stars are quite bright and have been the target of spectropolarimetric work since at least the 1970's. Since these stars are bright and have readily available data and theories, a set of comparative observations was done on these stars in 2007-2008. The extension of the program has provided a very useful comparison between star types. The spectropolarimetry is sigificantly different even between H$_\alpha$ lines that are quite similar.

	Classical Be stars are simply stars of B-type showing emission lines with a wide variety of  sub-types and polarizing mechanisms that must be considered (cf. Porter \& Rivinius 2003). Classical Be stars are rapidly rotating near-main-sequence stars with a gaseous circumstellar disk or envelope. There is no dust present in the disk so the assumed polarizing mechanism is electron scattering. However, the sub-type B[e] stars do have a strong dust signatures. Many of these stars have been resolved by interferometry and show flattened envelopes as well as intrinsic polarization perpendicular to the long-axis of the envelope (Quirrenbach et al. 1997). Even the stars themselves have been rotationally flattened as was observed for Achernar (Domiciano de Souza et al. 2003). 

	In the context of optically thin envelopes around hot stars, a few models to compute the continuum polarization were developed for asymmetric envelopes (Brown \& McLean 1977, McLean \& Brown 1978, Brown et al. 1978). These models were then applied to bright emission-line stars and the depolarization model was created (McLean \& Clark 1979, McLean 1979). These models have been extended and expanded over the years to include such things as finite-sized sources with limb darkening or winds (Cassinelli et al. 1987,  Brown et al. 1989, Hiller 1990, 1991, 1994, 1996). The models have been applied to specific stars to derive geometrical properties of circumstellar material (cf. Wood et al. 1997).

\subsection{Be Star H$_\alpha$ Line Profiles}

\begin{figure}
\centering
\subfloat[$\gamma$ Cas]{\label{fig:gmcas}
\includegraphics[ width=0.21\textwidth, angle=90]{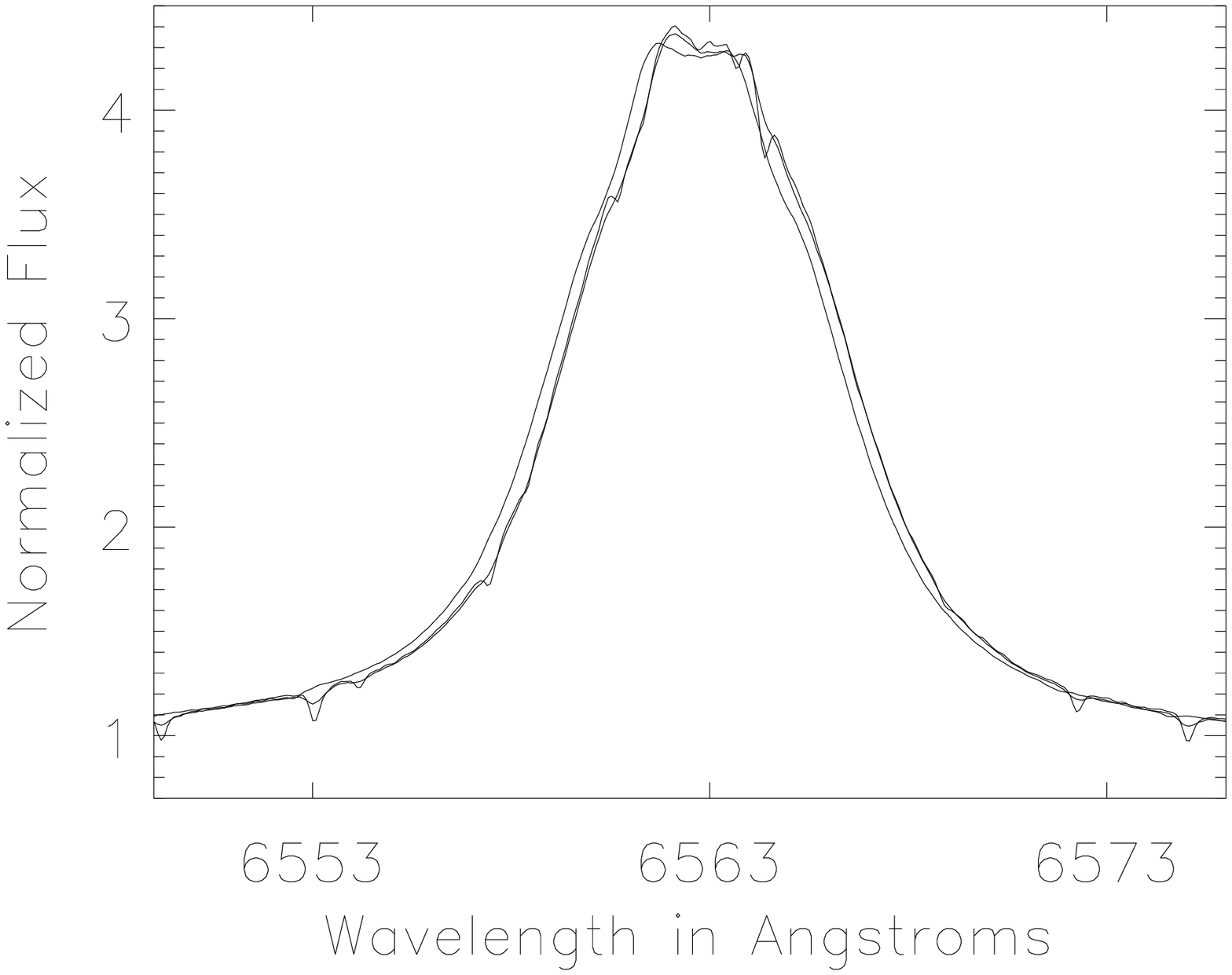}}
\quad
\subfloat[25 Ori]{\label{fig:lprof-25ori}
\includegraphics[ width=0.21\textwidth, angle=90]{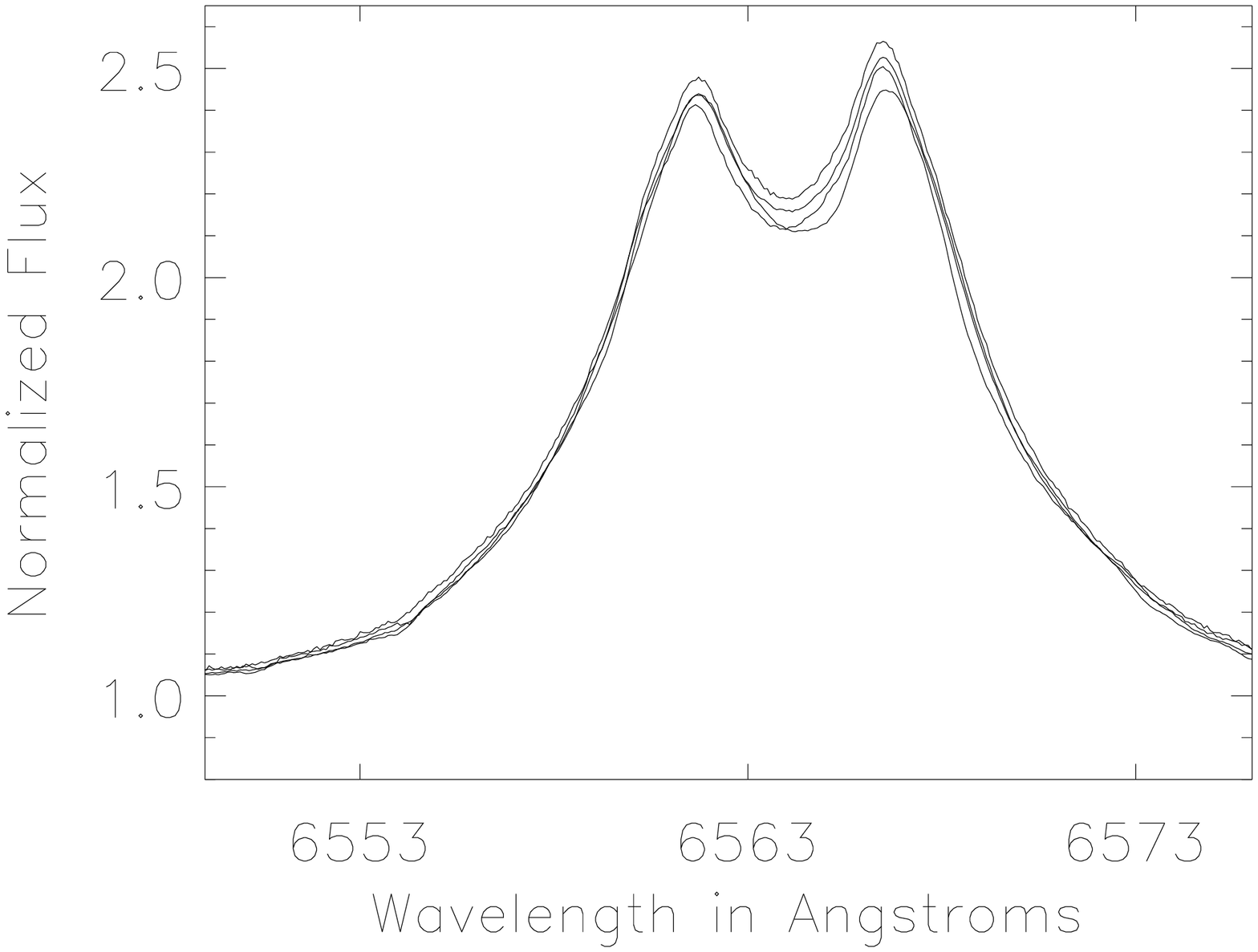}}
\quad
\subfloat[$\psi$ Per]{\label{fig:lprof-psiper}
\includegraphics[ width=0.21\textwidth, angle=90]{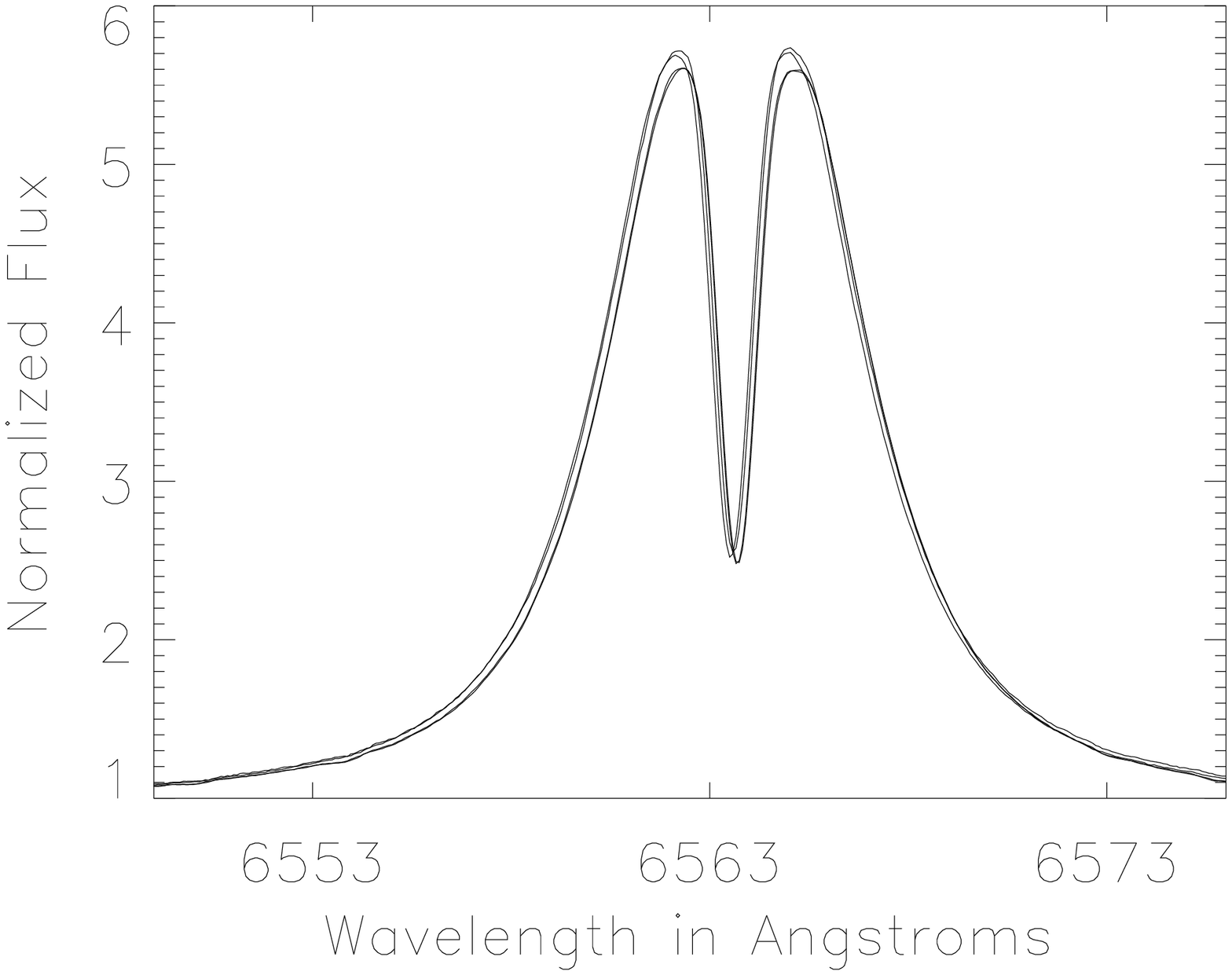}}
\quad
\subfloat[$\eta$ Tau]{\label{fig:lprof-etatau}
\includegraphics[ width=0.21\textwidth, angle=90]{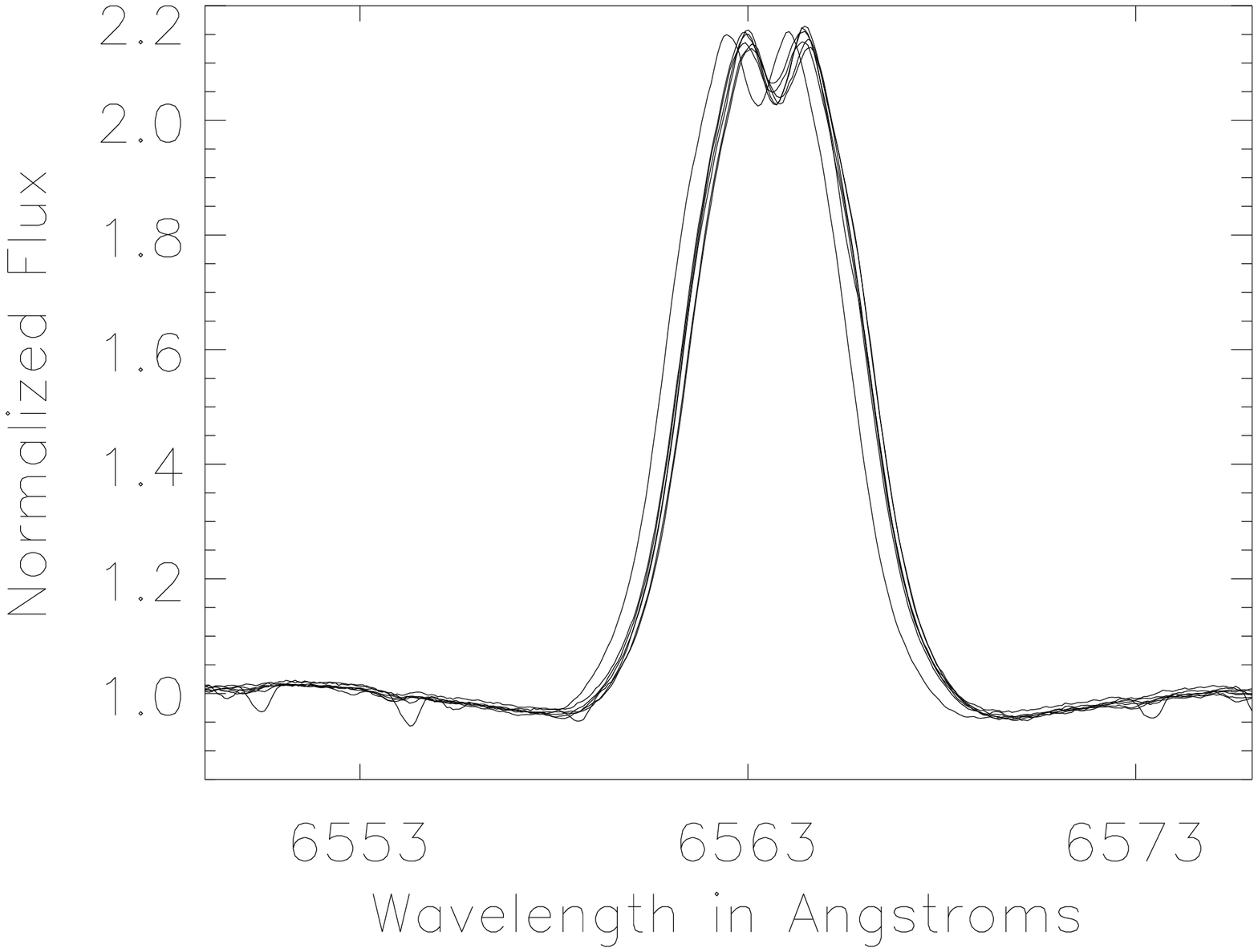}}
\quad
\subfloat[$\zeta$ Tau]{\label{fig:zetatau}
\includegraphics[ width=0.21\textwidth, angle=90]{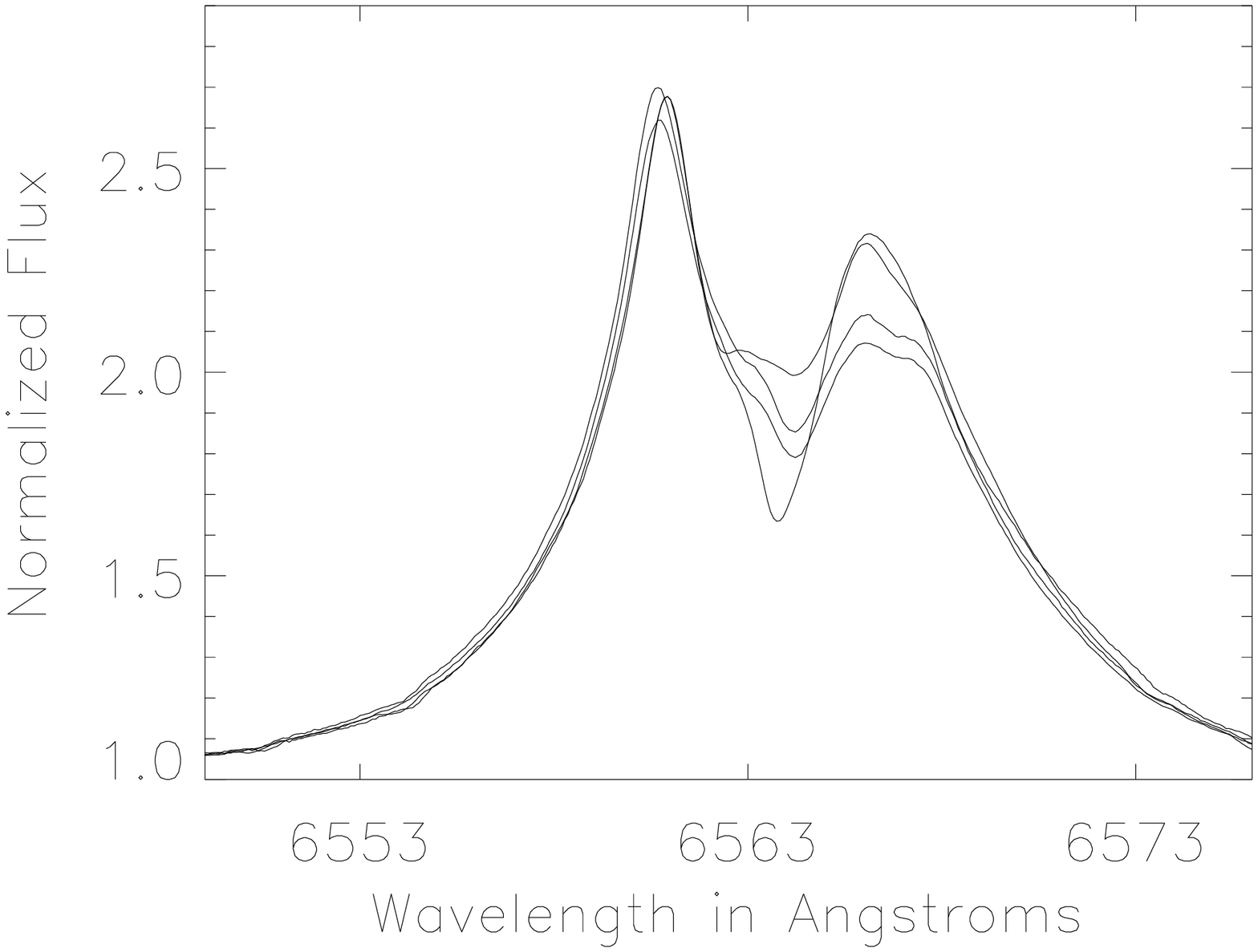}}
\quad
\subfloat[MWC 143]{\label{fig:mwc143}
\includegraphics[ width=0.21\textwidth, angle=90]{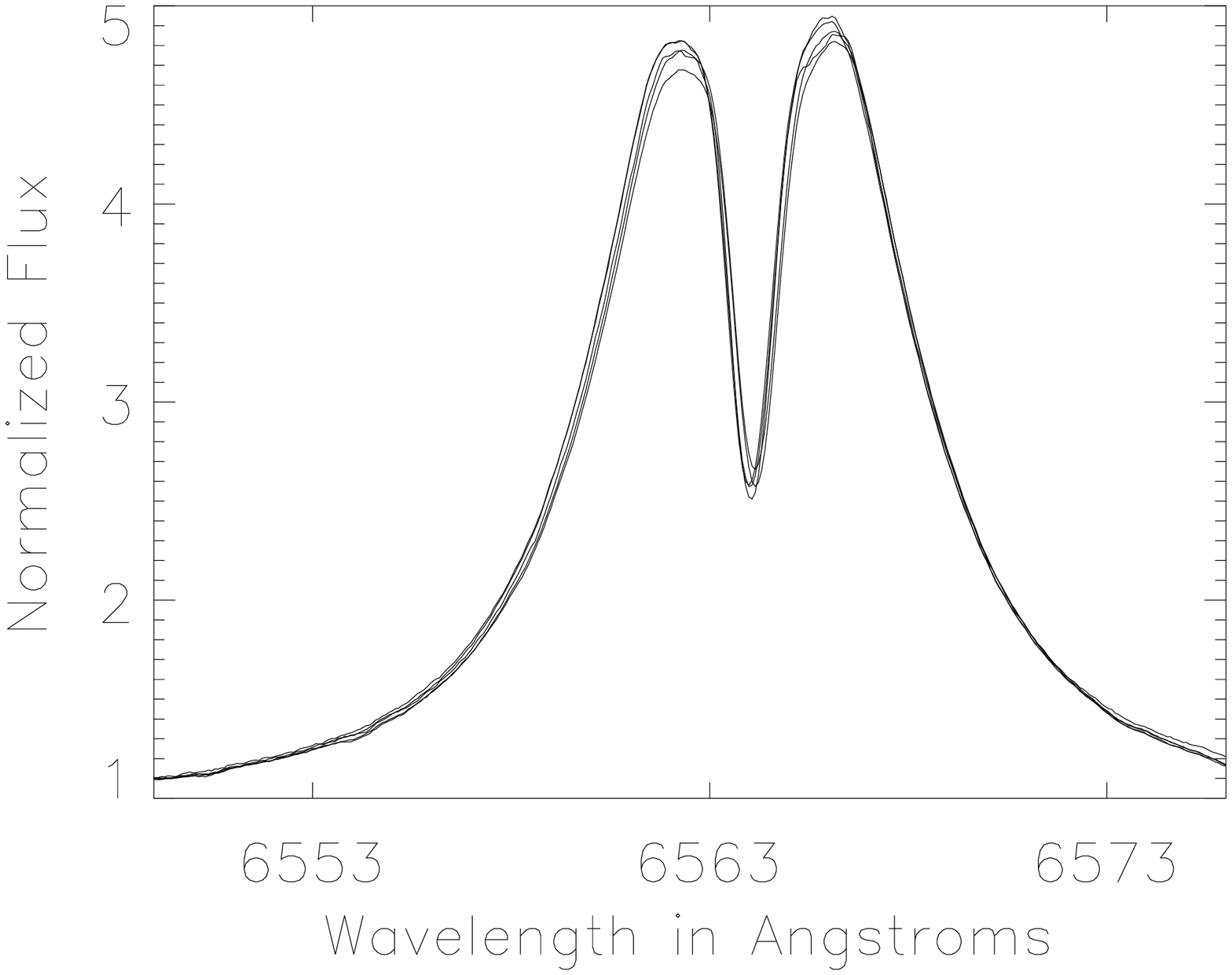}}
\quad
\subfloat[$\omega$ Ori]{\label{fig:omori}
\includegraphics[ width=0.21\textwidth, angle=90]{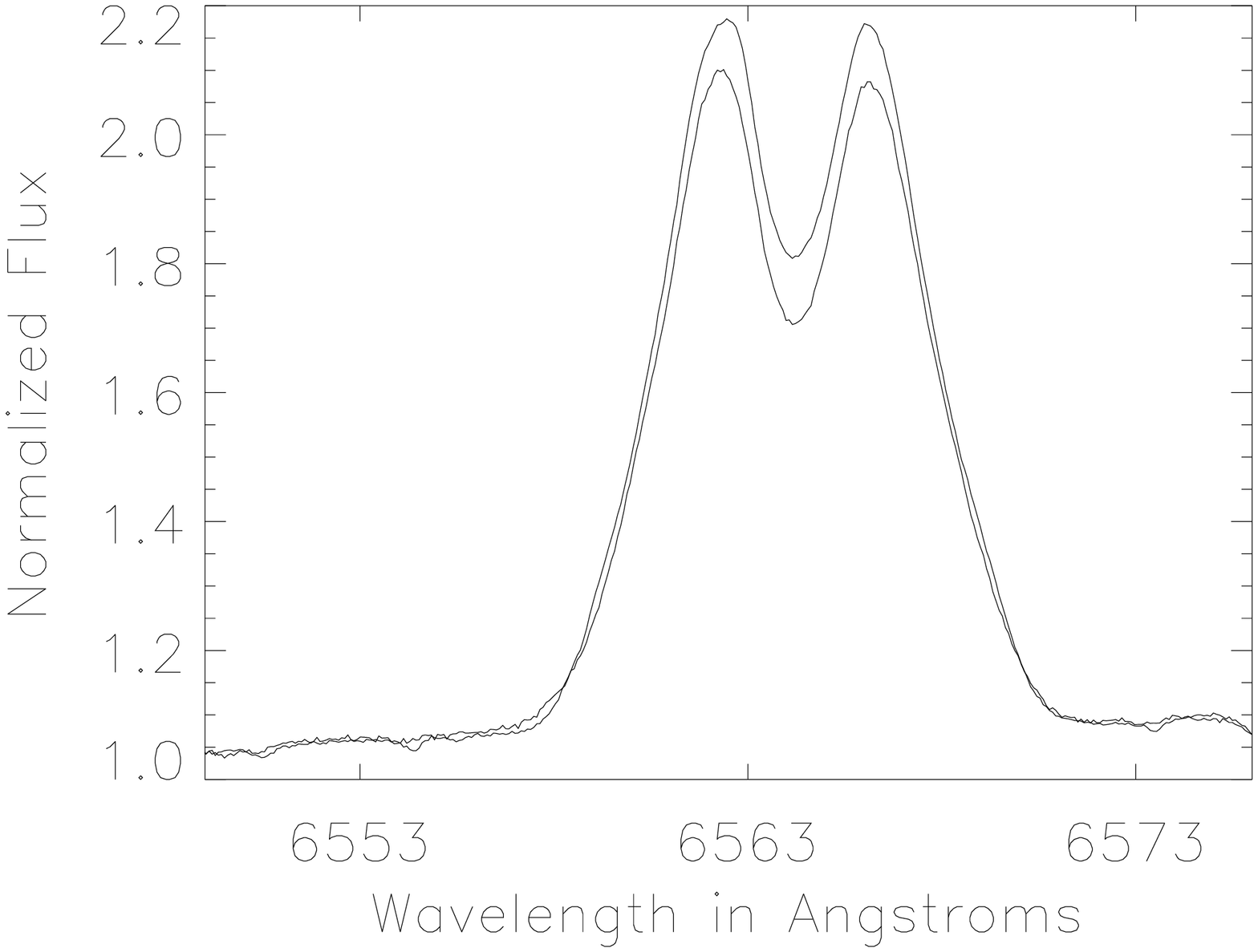}}
\quad
\subfloat[Omi Pup]{\label{fig:ompup}
\includegraphics[ width=0.21\textwidth, angle=90]{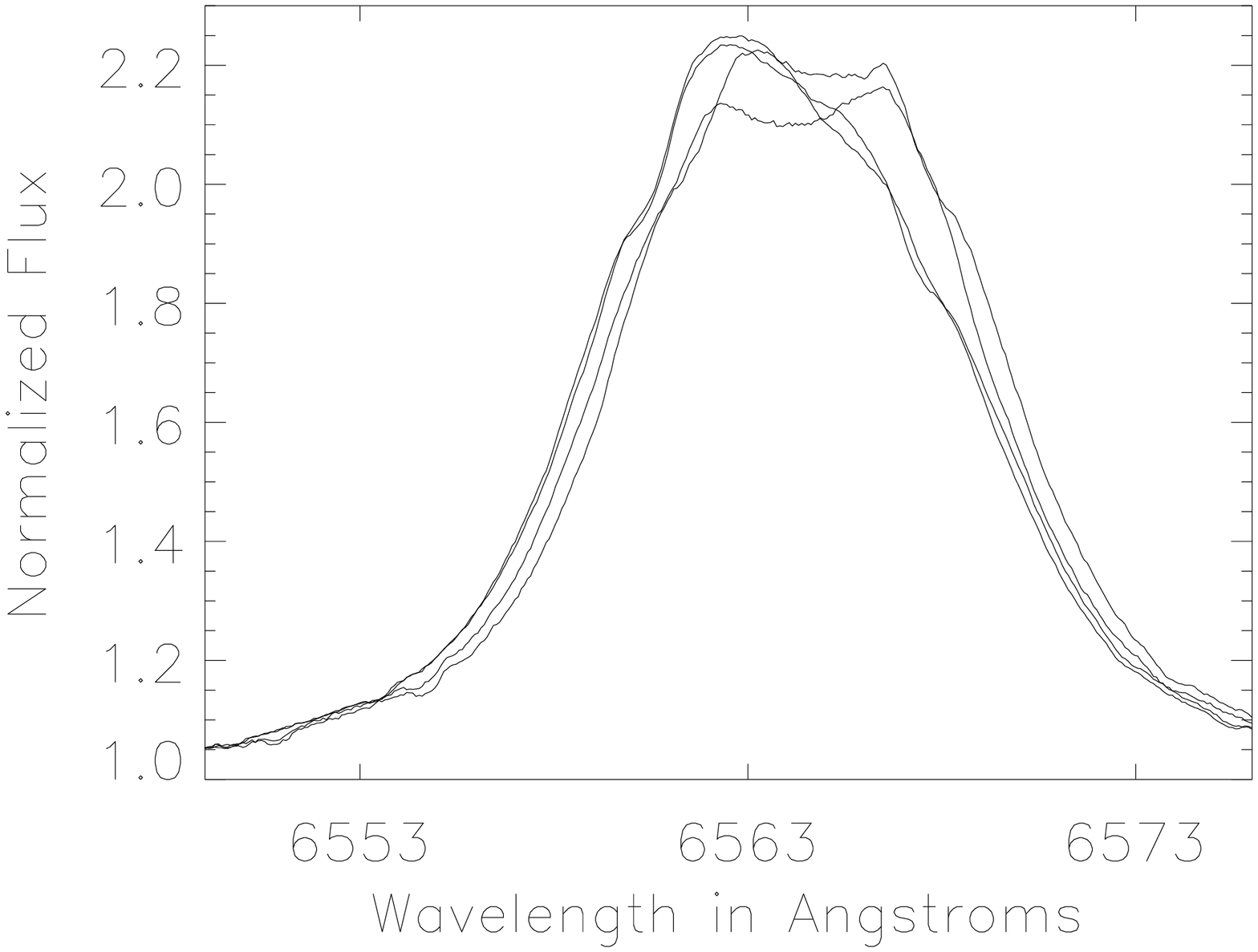}}
\quad
\subfloat[10 CMa]{\label{fig:10cma}
\includegraphics[ width=0.21\textwidth, angle=90]{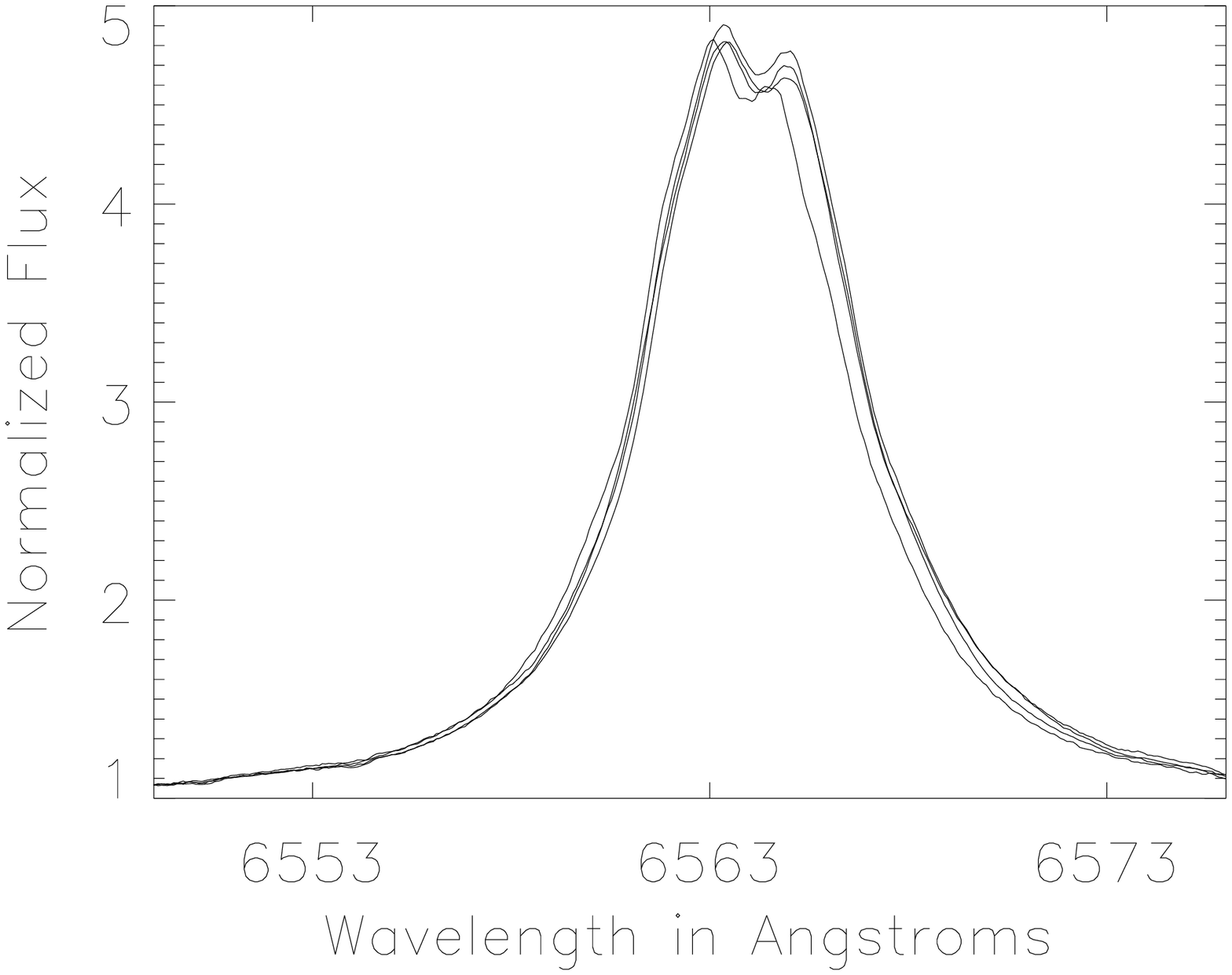}}
\quad
\subfloat[Omi Cas]{\label{fig:omicas}
\includegraphics[ width=0.21\textwidth, angle=90]{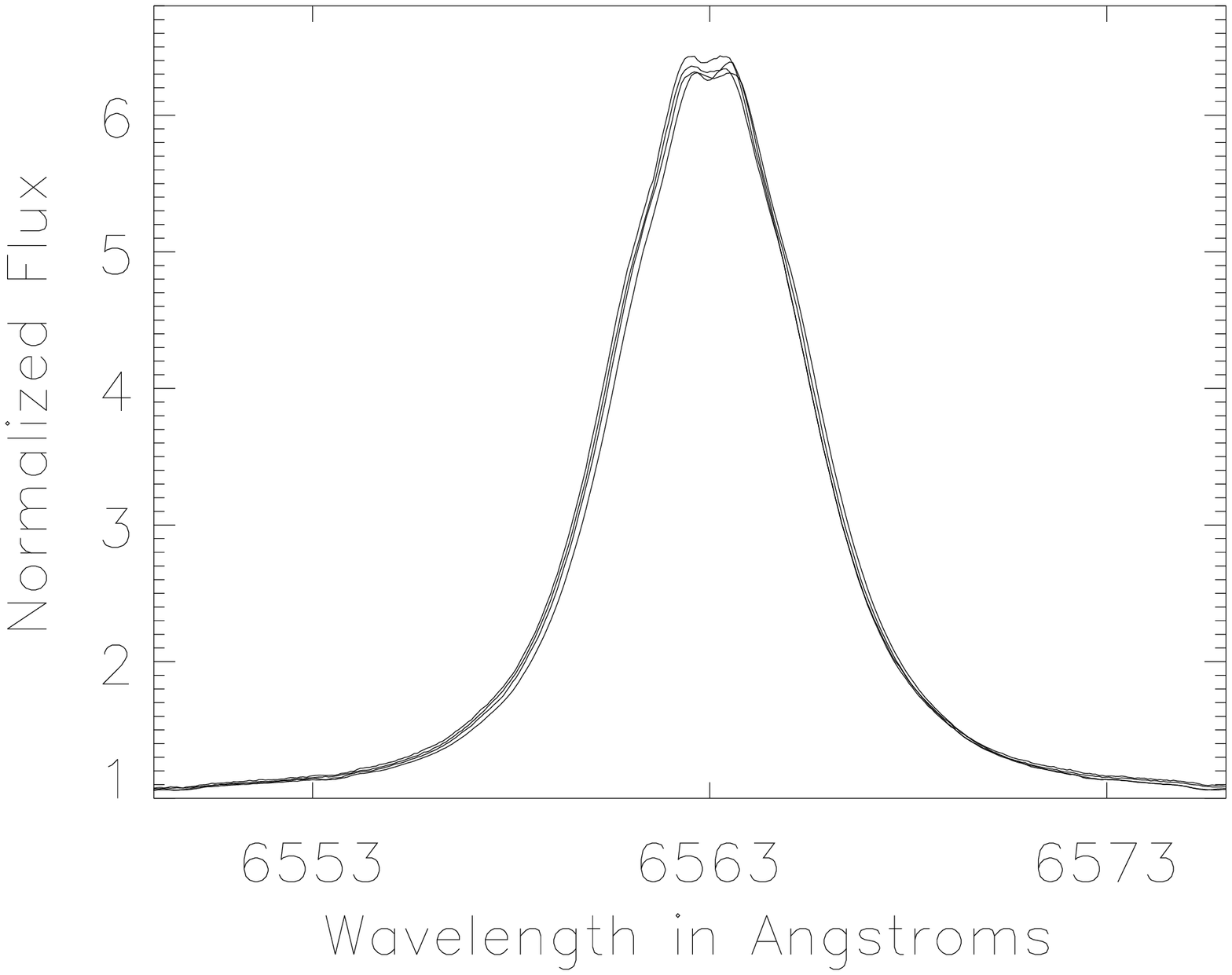}}
\quad
\subfloat[$\kappa$ CMa]{\label{fig:lprof-kapcma}
\includegraphics[ width=0.21\textwidth, angle=90]{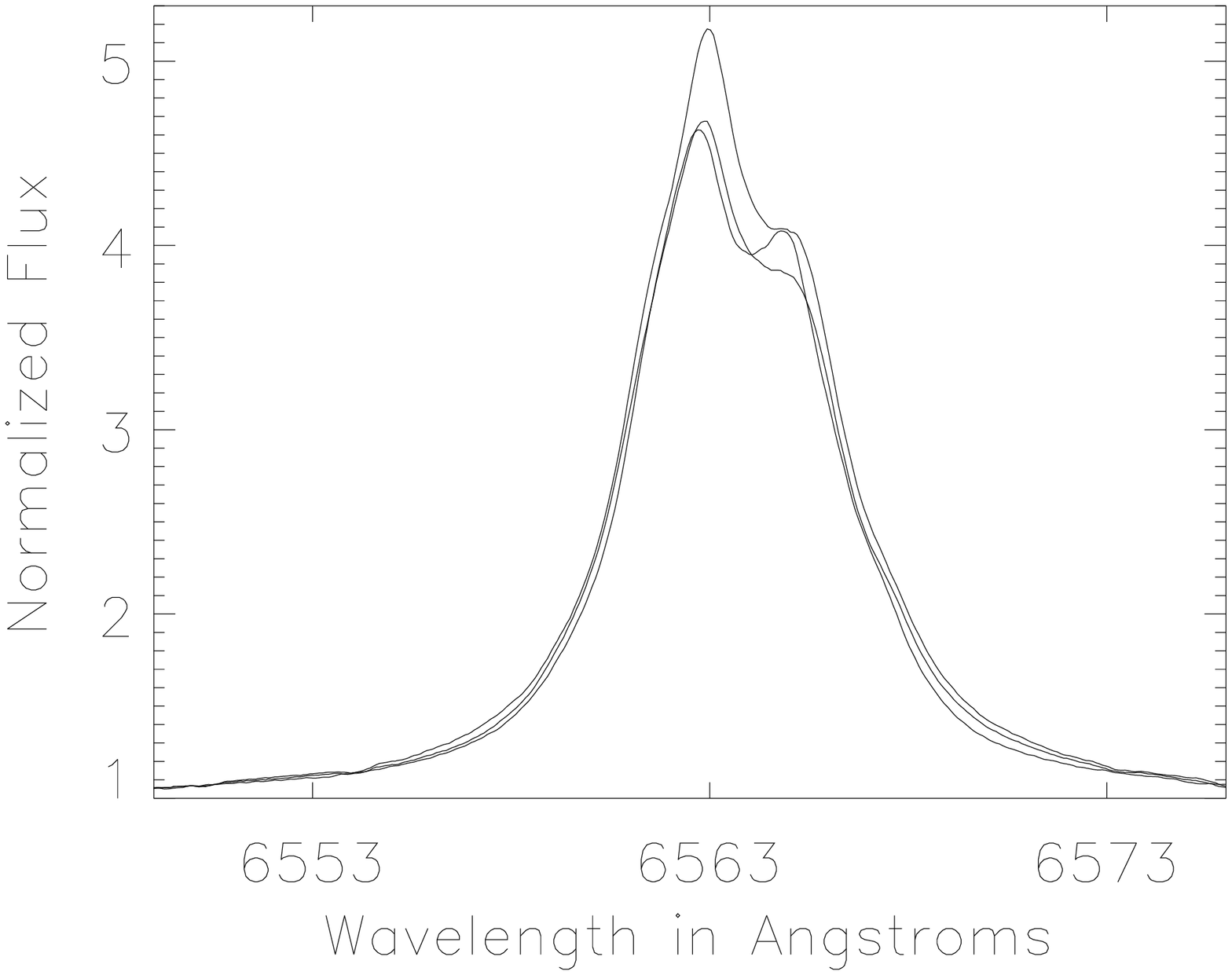}}
\quad
\subfloat[18 Gem]{\label{fig:18gem}
\includegraphics[ width=0.21\textwidth, angle=90]{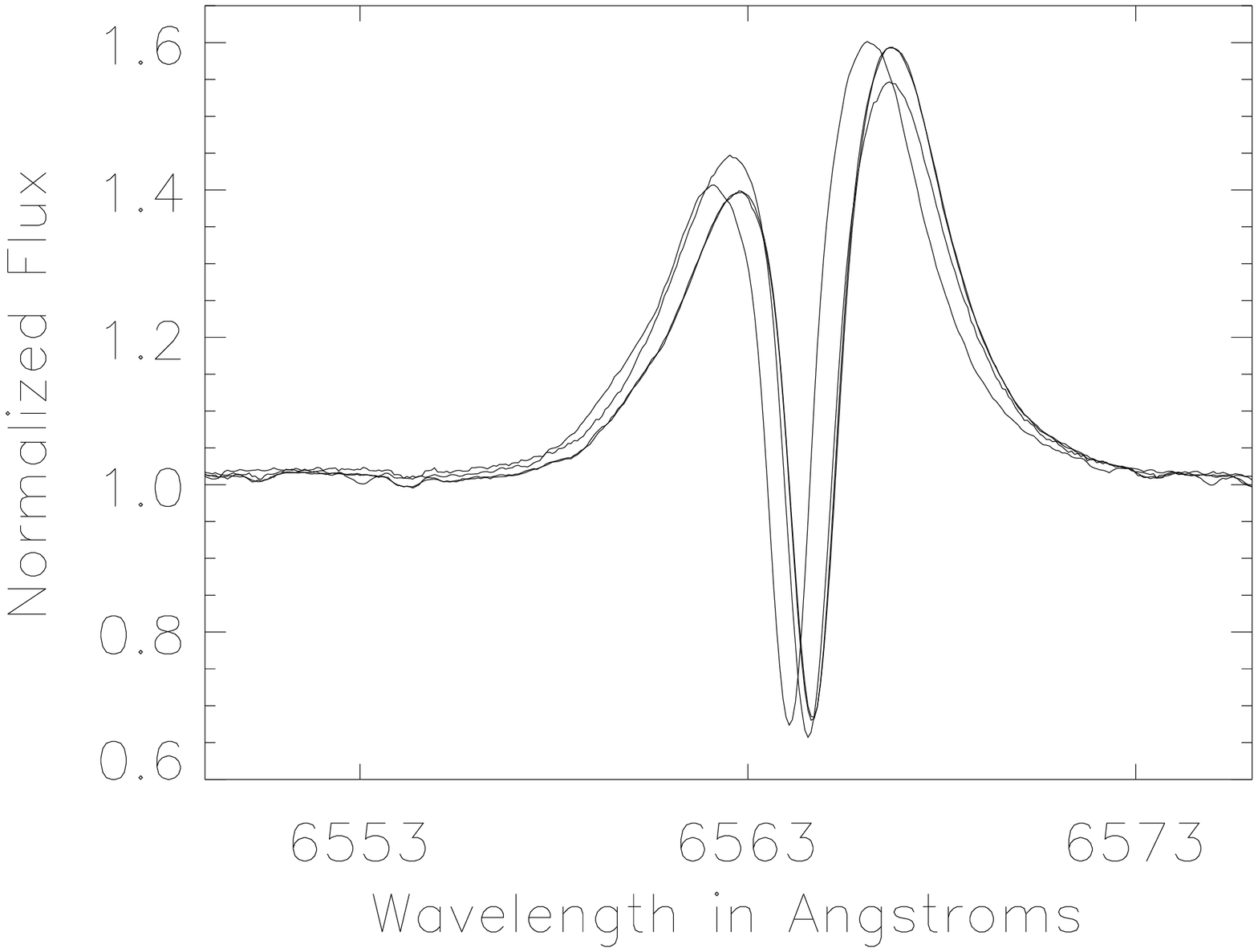}}
\quad
\subfloat[Alf Col]{\label{fig:alfcol}
\includegraphics[ width=0.21\textwidth, angle=90]{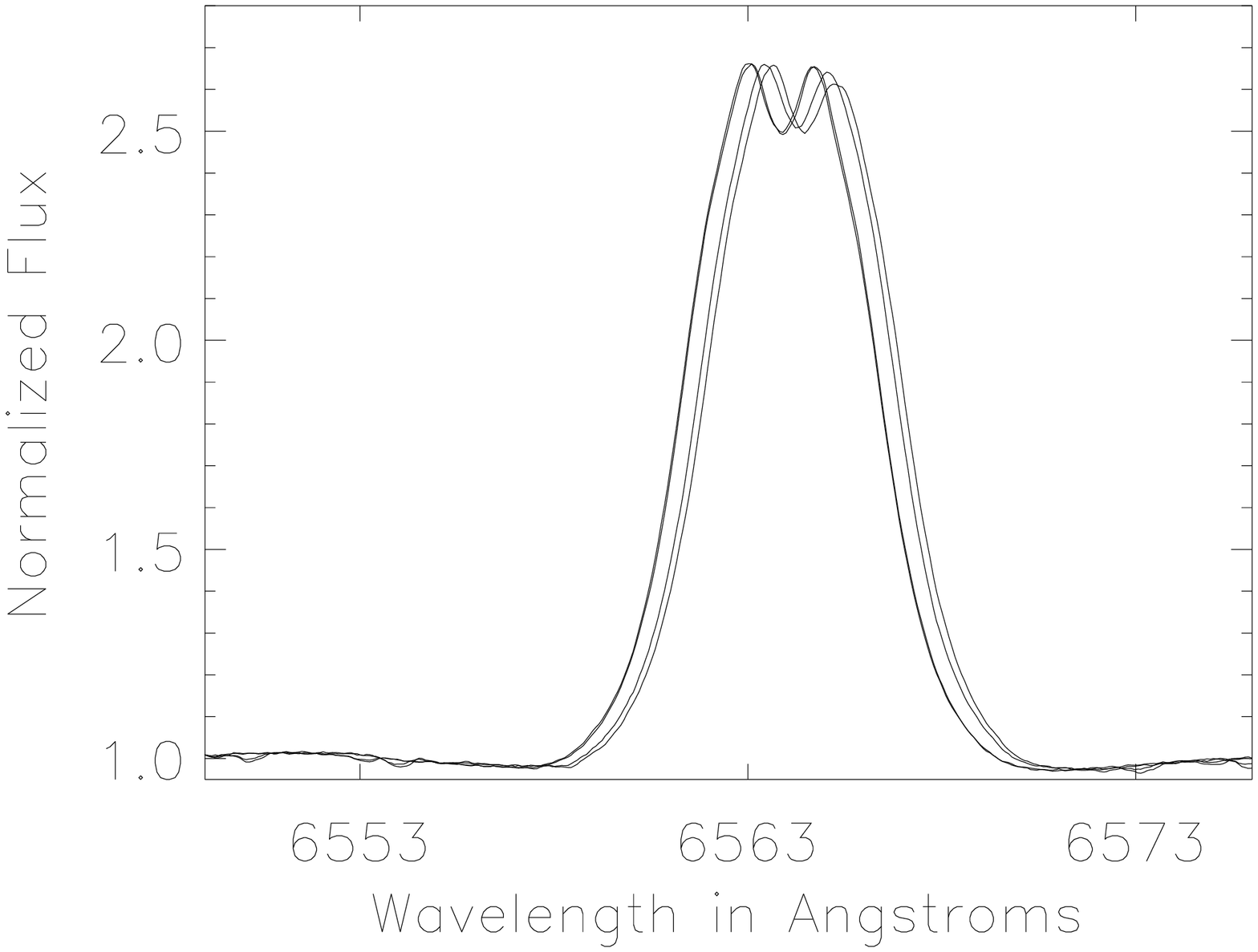}}
\quad
\subfloat[66 Oph]{\label{fig:66oph}
\includegraphics[ width=0.21\textwidth, angle=90]{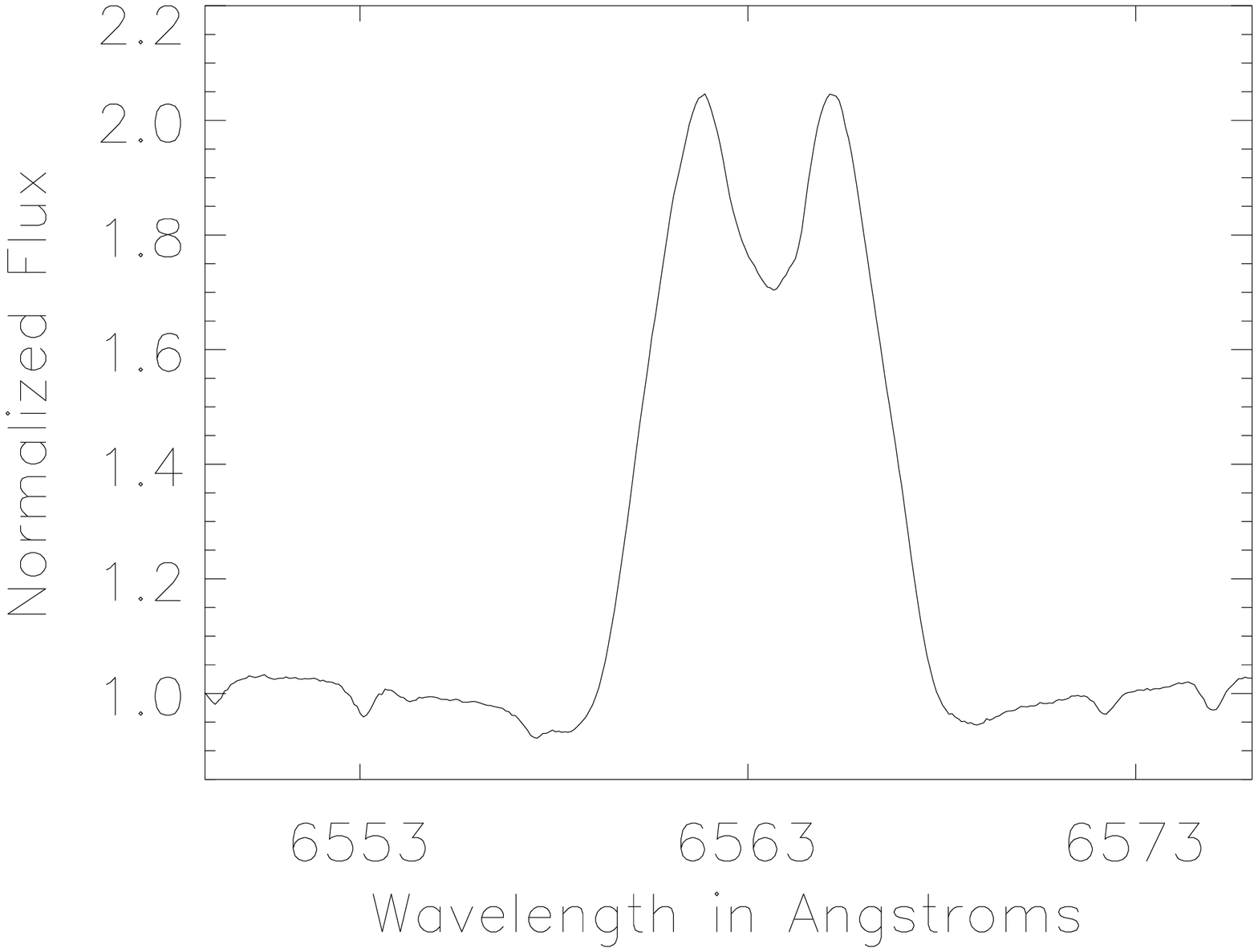}}
\quad
\subfloat[$\beta$ CMi]{\label{fig:bcm}
\includegraphics[ width=0.21\textwidth, angle=90]{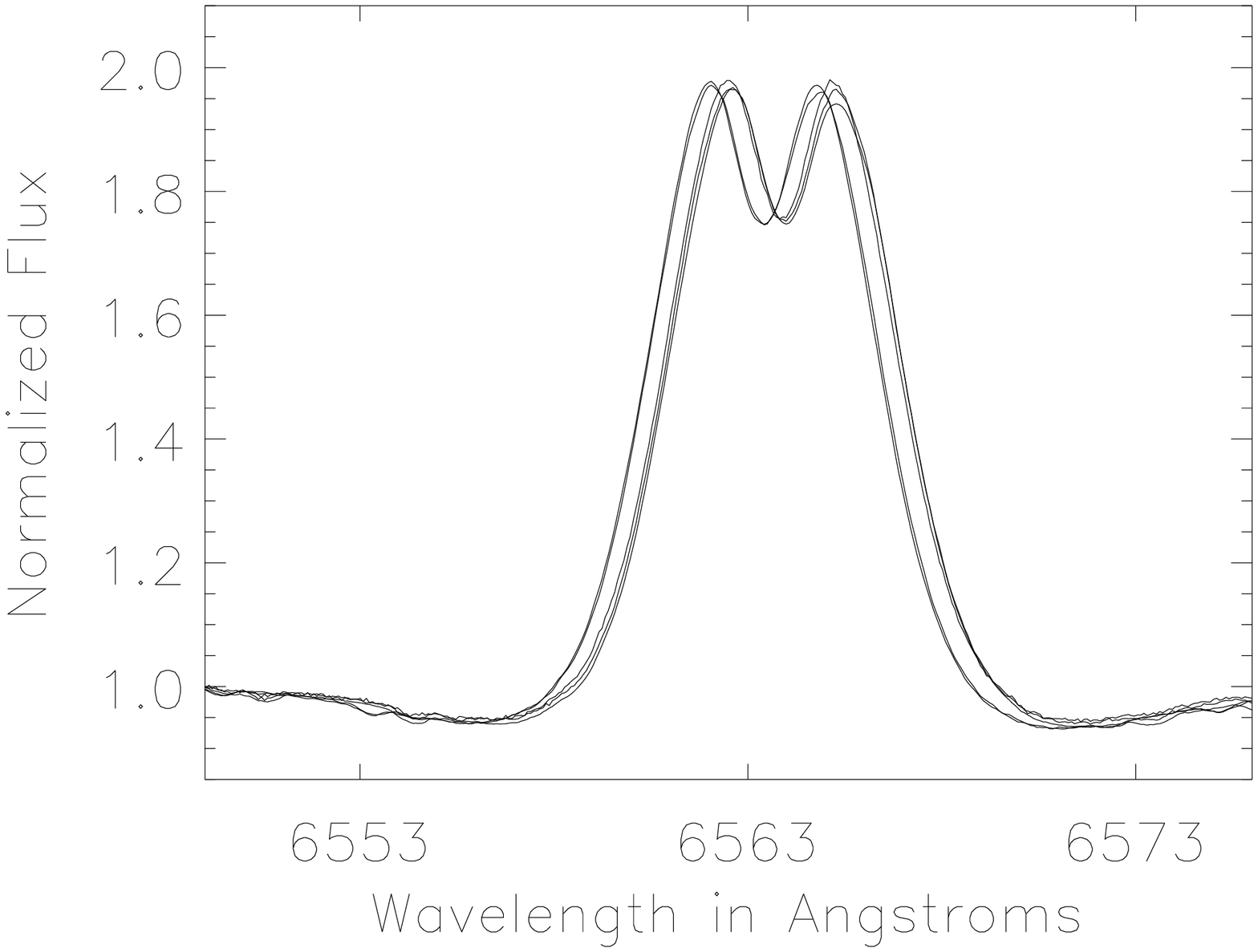}}
\caption[Be Line Profiles I]{Be Line Profiles I}
\label{fig:be-lprof1}
\end{figure}

\begin{figure}
\centering
\subfloat[31 Peg]{\label{fig:31peg}
\includegraphics[ width=0.21\textwidth, angle=90]{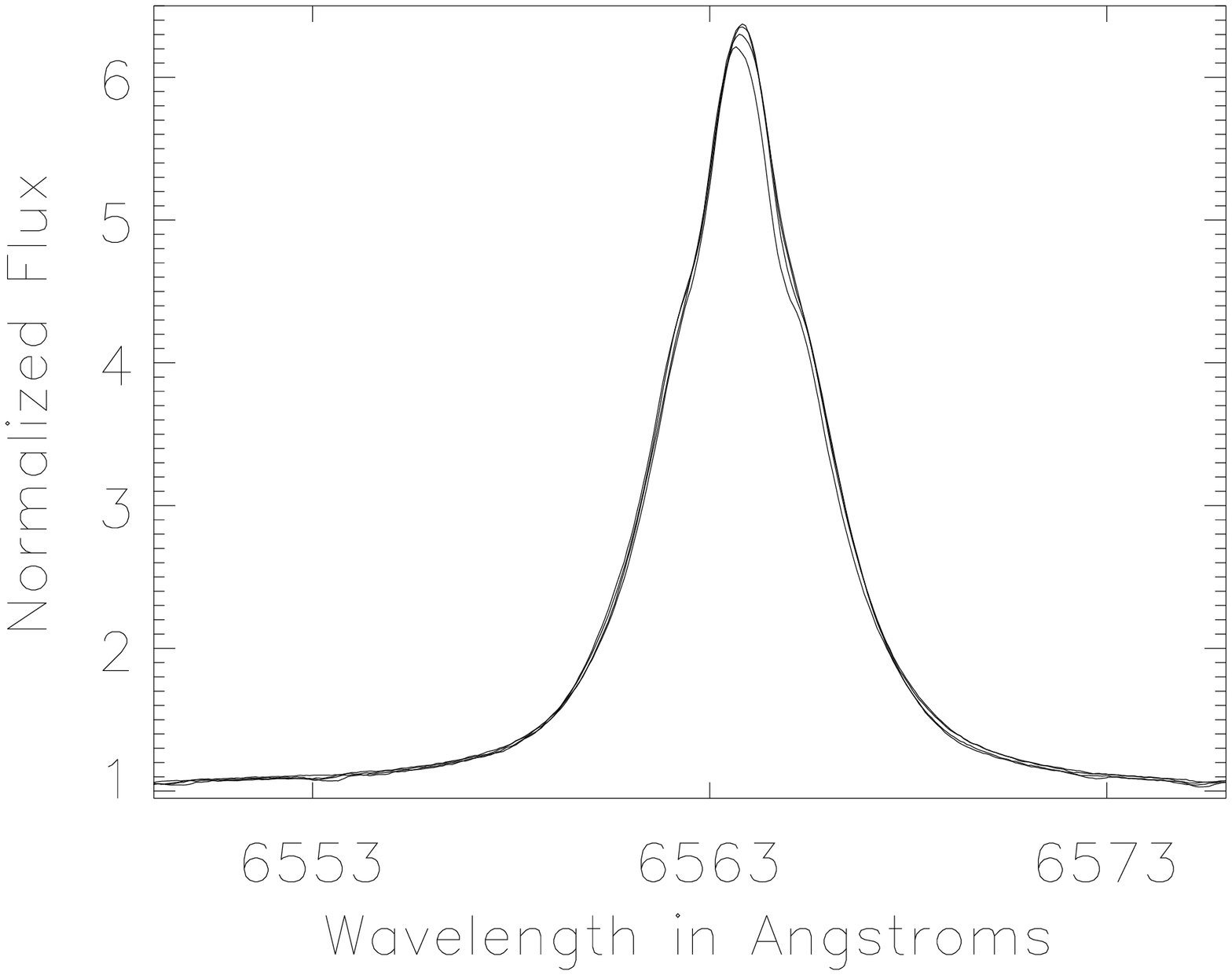}}
\quad
\subfloat[11 Cam]{\label{fig:lprof-11cam}
\includegraphics[ width=0.21\textwidth, angle=90]{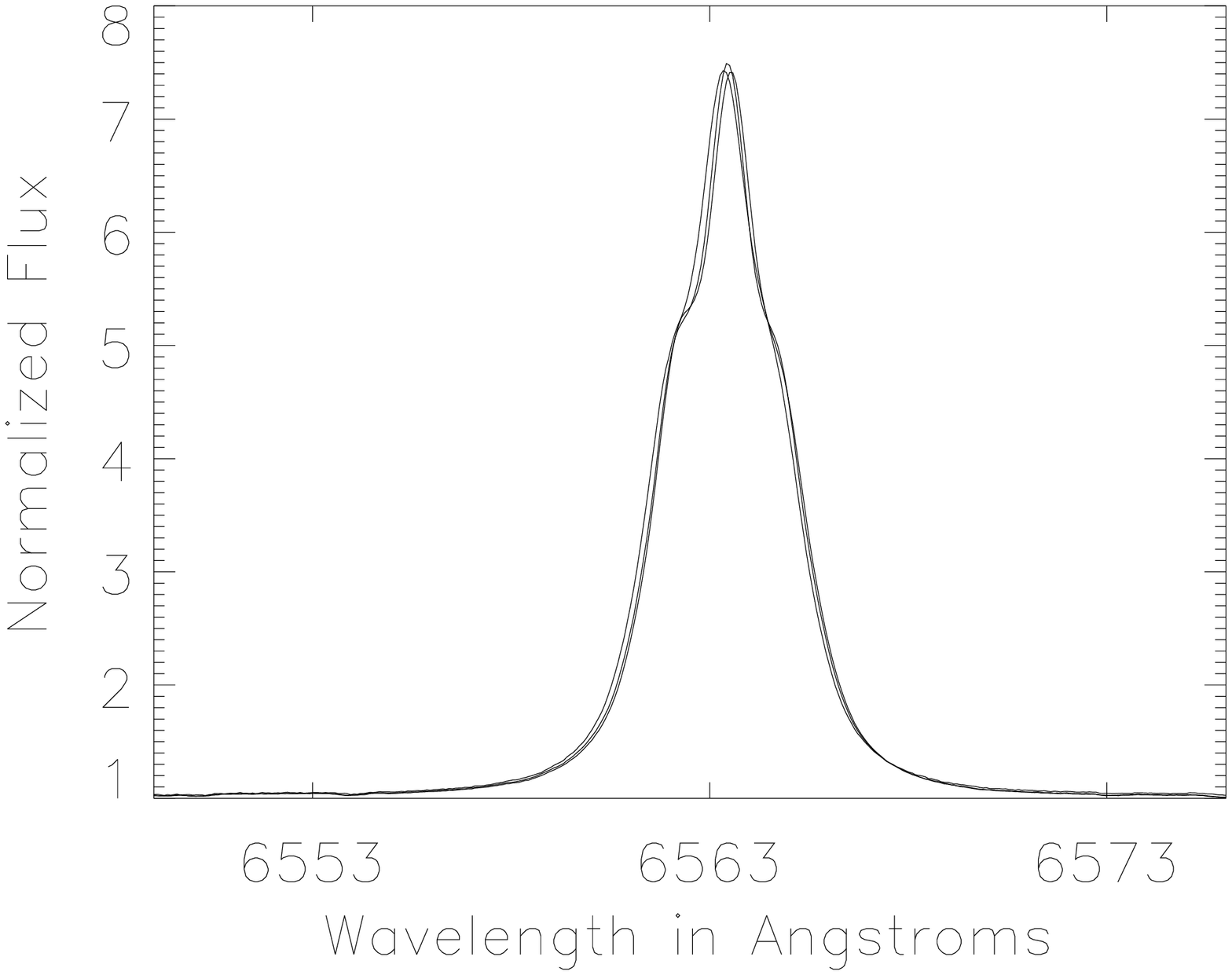}}
\quad
\subfloat[C Per]{\label{fig:lprof-cper}
\includegraphics[ width=0.21\textwidth, angle=90]{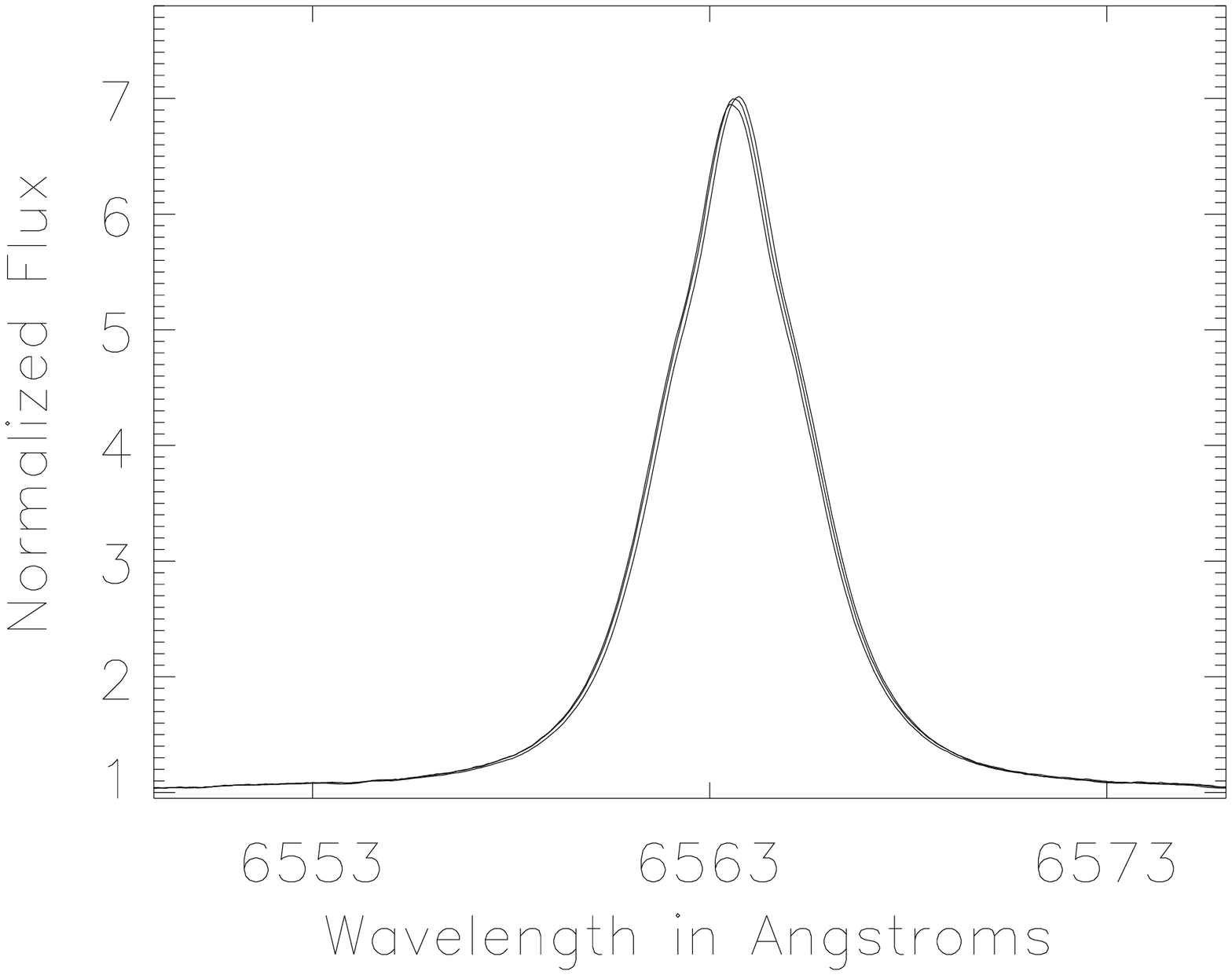}}
\quad
\subfloat[MWC 192]{\label{fig:mwc192}
\includegraphics[ width=0.21\textwidth, angle=90]{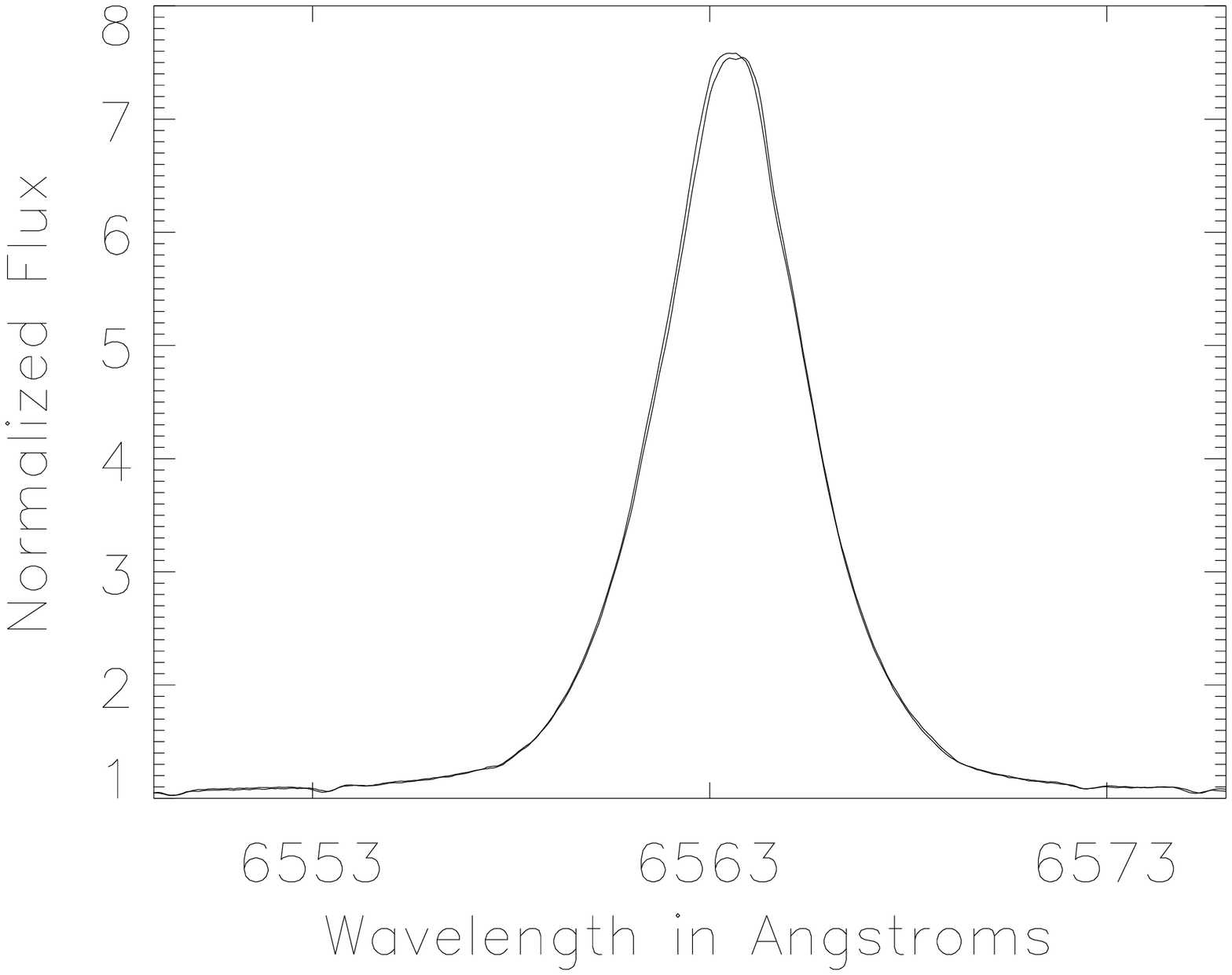}}
\quad
\subfloat[$\kappa$ Dra]{\label{fig:kapdr}
\includegraphics[ width=0.21\textwidth, angle=90]{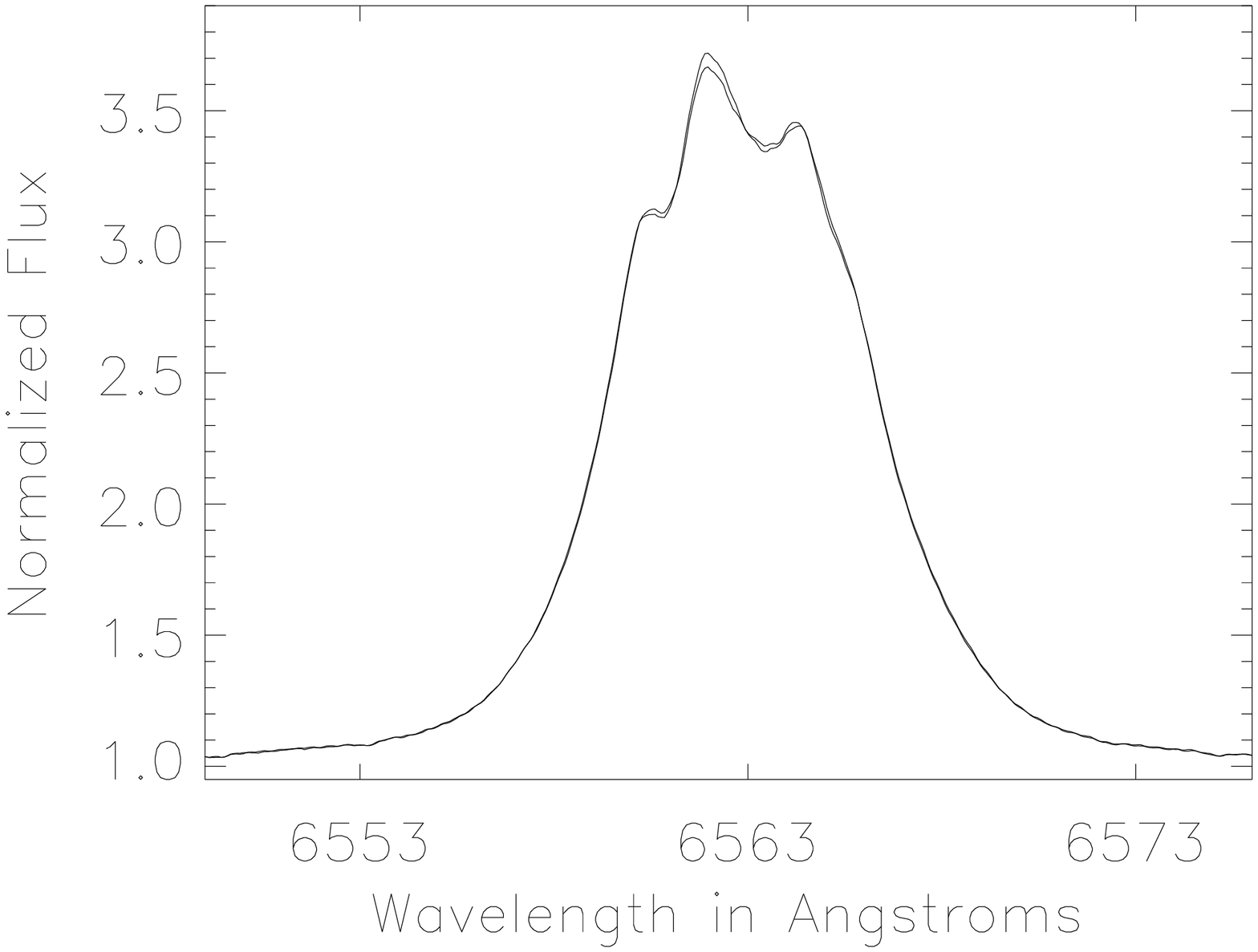}}
\quad
\subfloat[$\kappa$ Cas]{\label{fig:lprof-kapcas}
\includegraphics[ width=0.21\textwidth, angle=90]{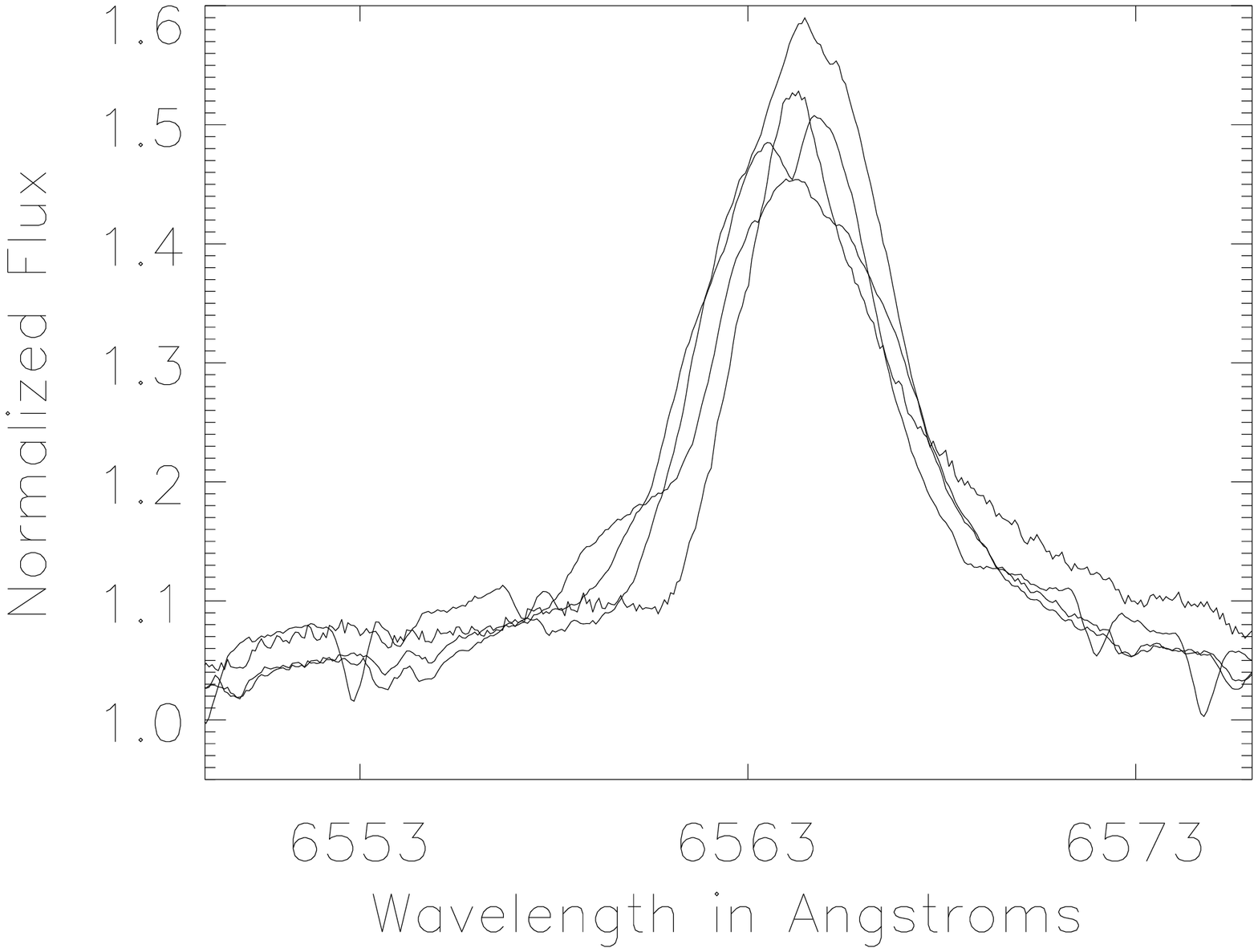}}
\quad
\subfloat[MWC 92]{\label{fig:mwc92}
\includegraphics[ width=0.21\textwidth, angle=90]{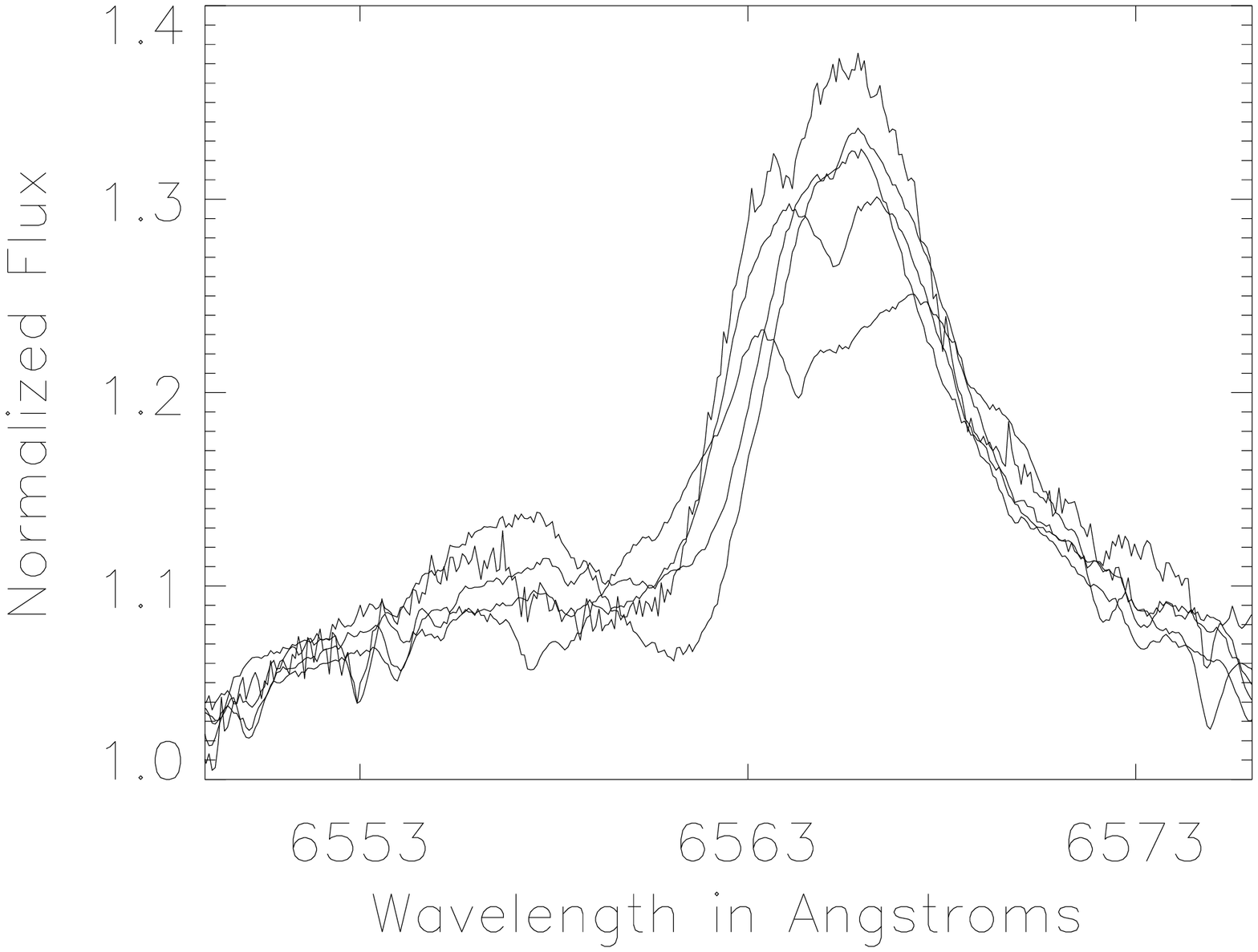}}
\quad
\subfloat[12 Vul]{\label{fig:12vul}
\includegraphics[ width=0.21\textwidth, angle=90]{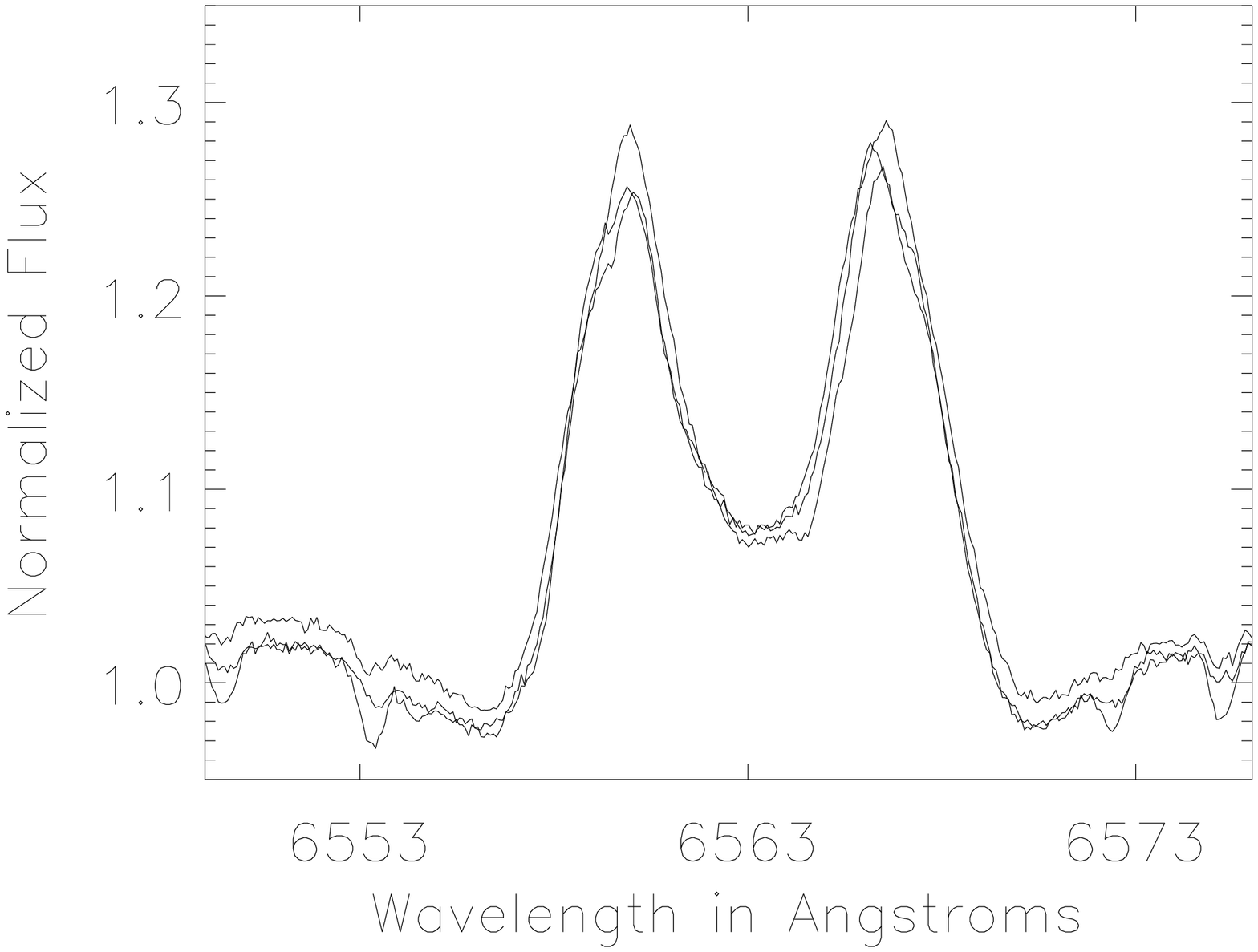}}
\quad
\subfloat[$\phi$ And]{\label{fig:phiand}
\includegraphics[ width=0.21\textwidth, angle=90]{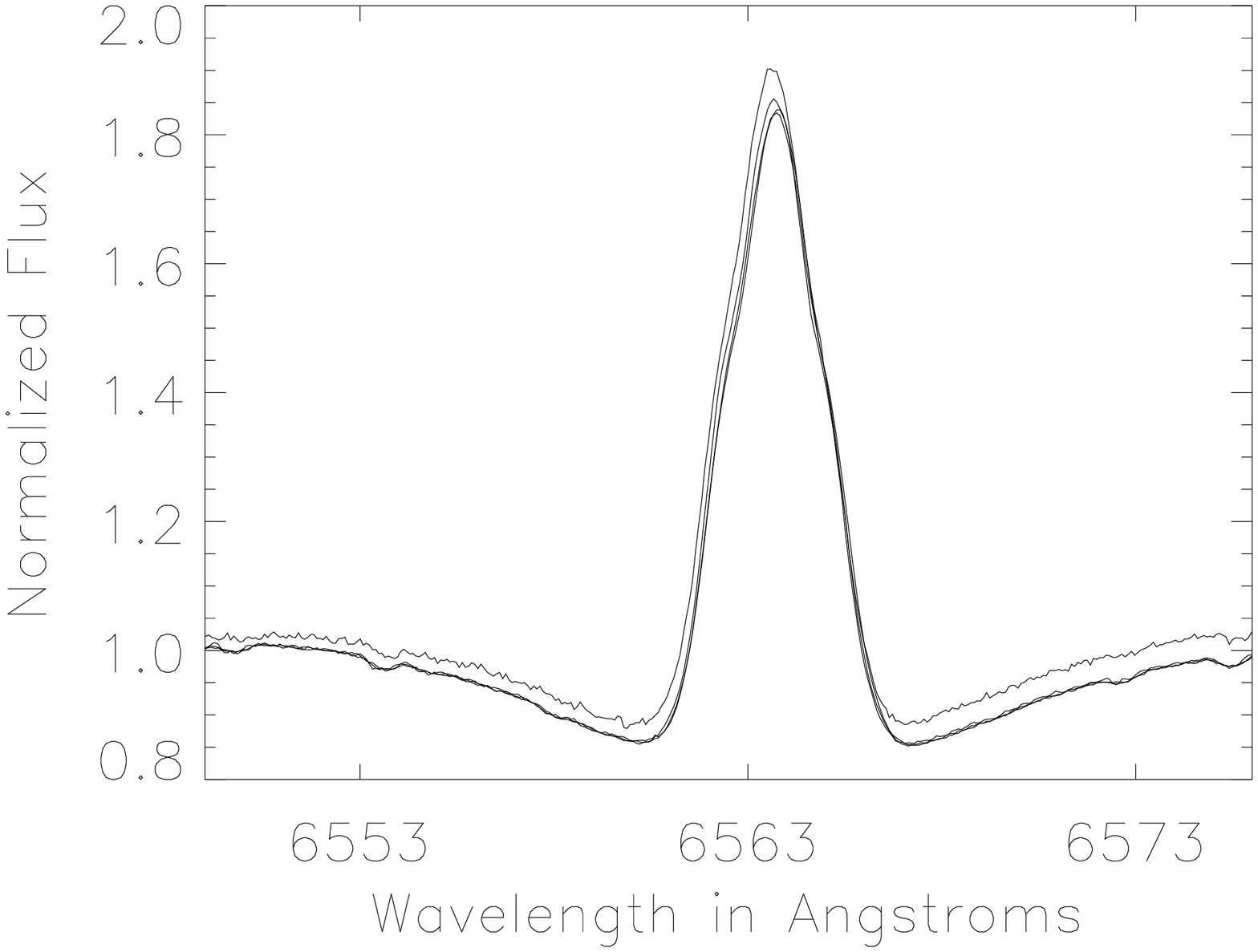}}
\quad
\subfloat[MWC 77]{\label{fig:mwc77}
\includegraphics[ width=0.21\textwidth, angle=90]{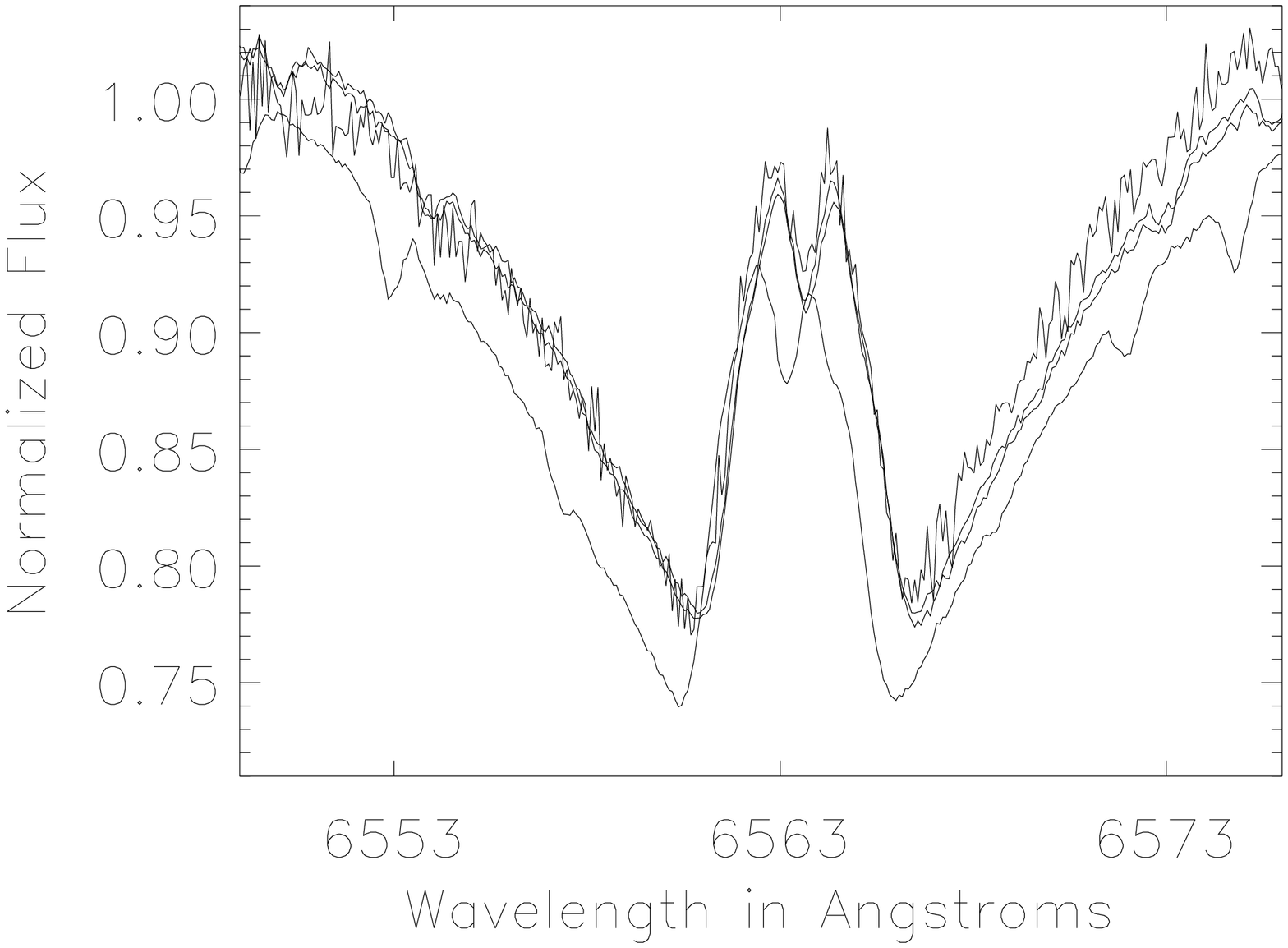}}
\quad
\subfloat[HD 36408]{\label{fig:hd364}
\includegraphics[ width=0.21\textwidth, angle=90]{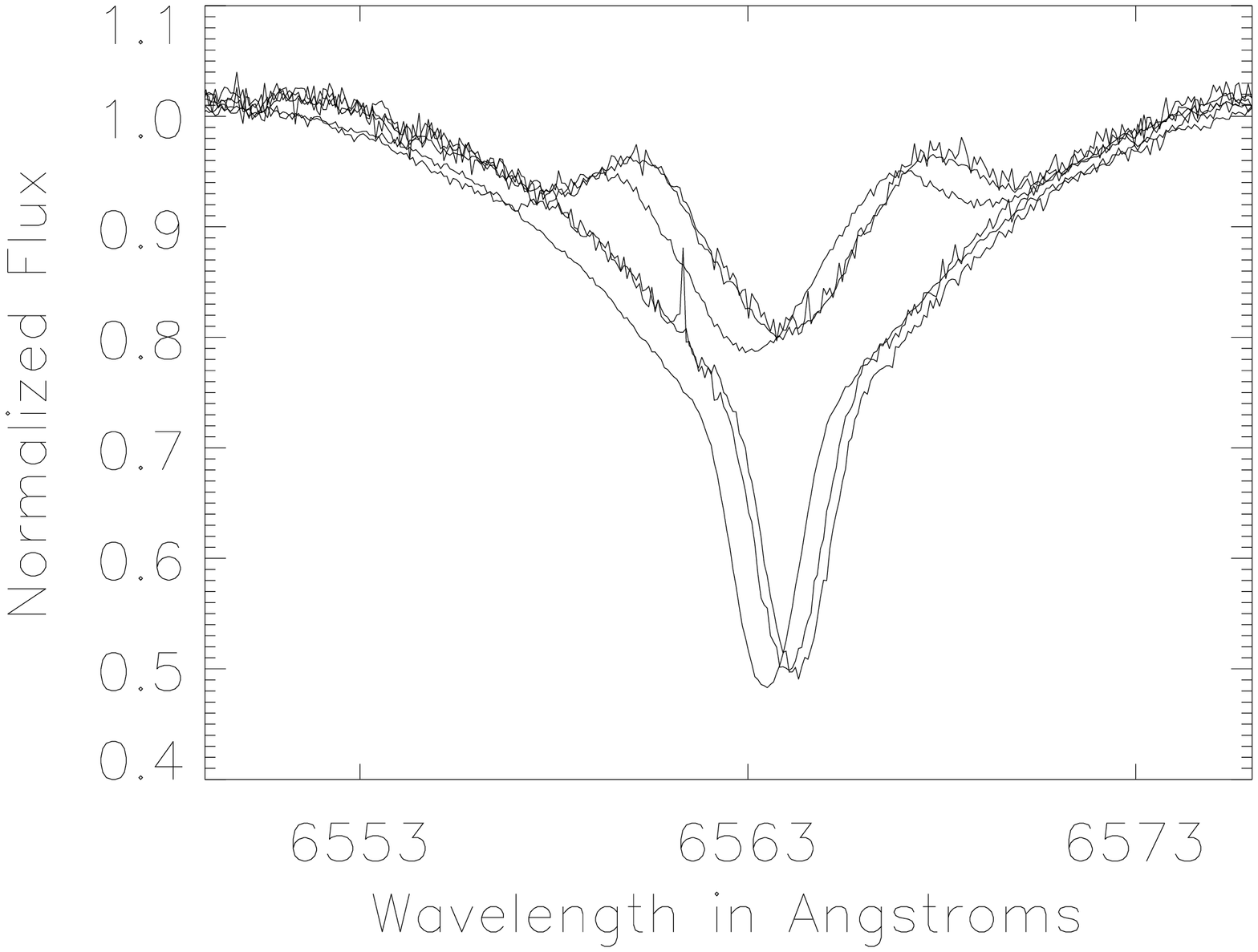}}
\quad
\subfloat[Phecda]{\label{fig:phecd}
\includegraphics[ width=0.21\textwidth, angle=90]{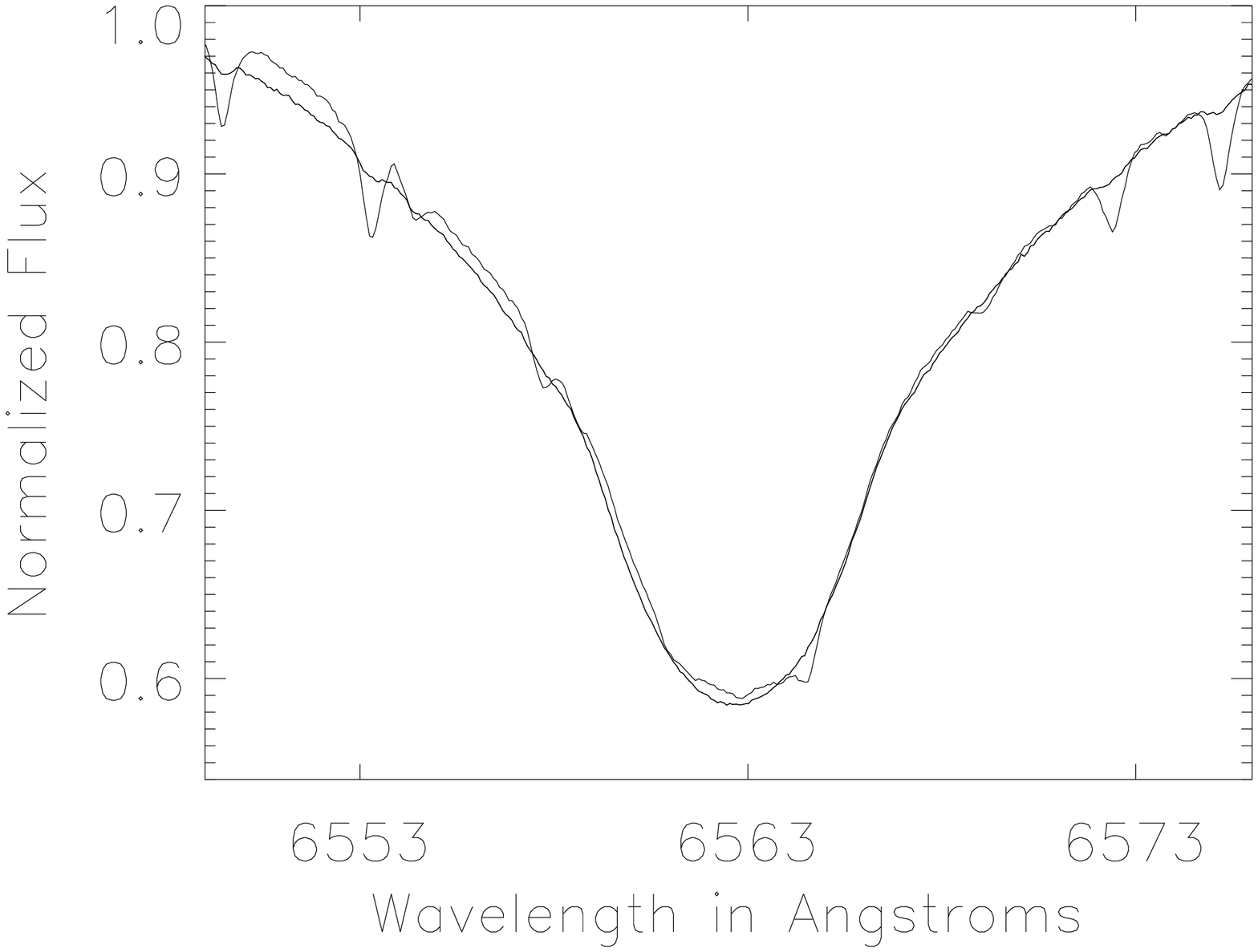}}
\quad
\subfloat[$\lambda$ Cyg]{\label{fig:lamcyg}
\includegraphics[ width=0.21\textwidth, angle=90]{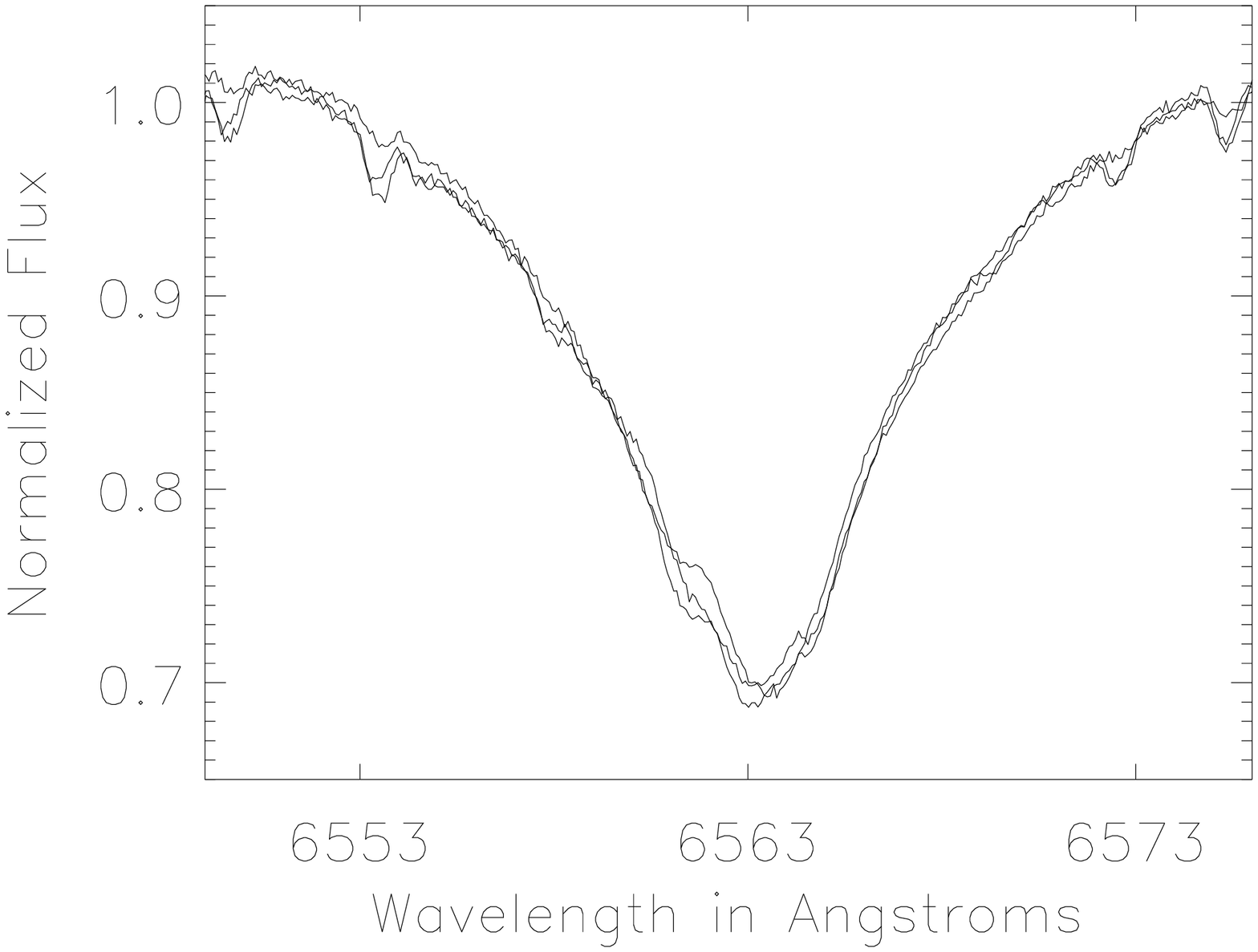}}
\quad
\subfloat[QR Vul]{\label{fig:qrvul}
\includegraphics[ width=0.21\textwidth, angle=90]{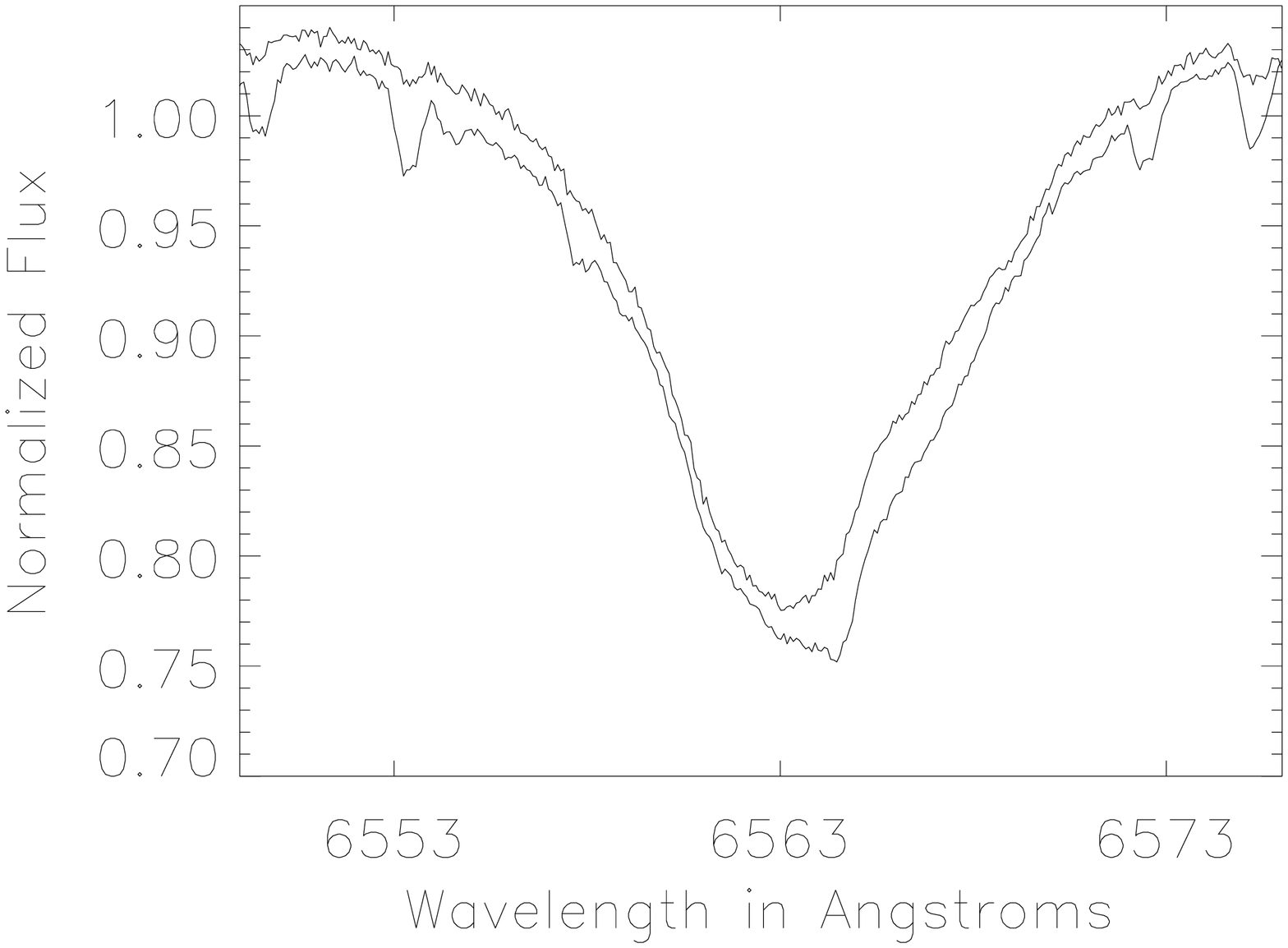}}
\quad
\subfloat[$\xi$ Per]{\label{fig:ksiper}
\includegraphics[ width=0.21\textwidth, angle=90]{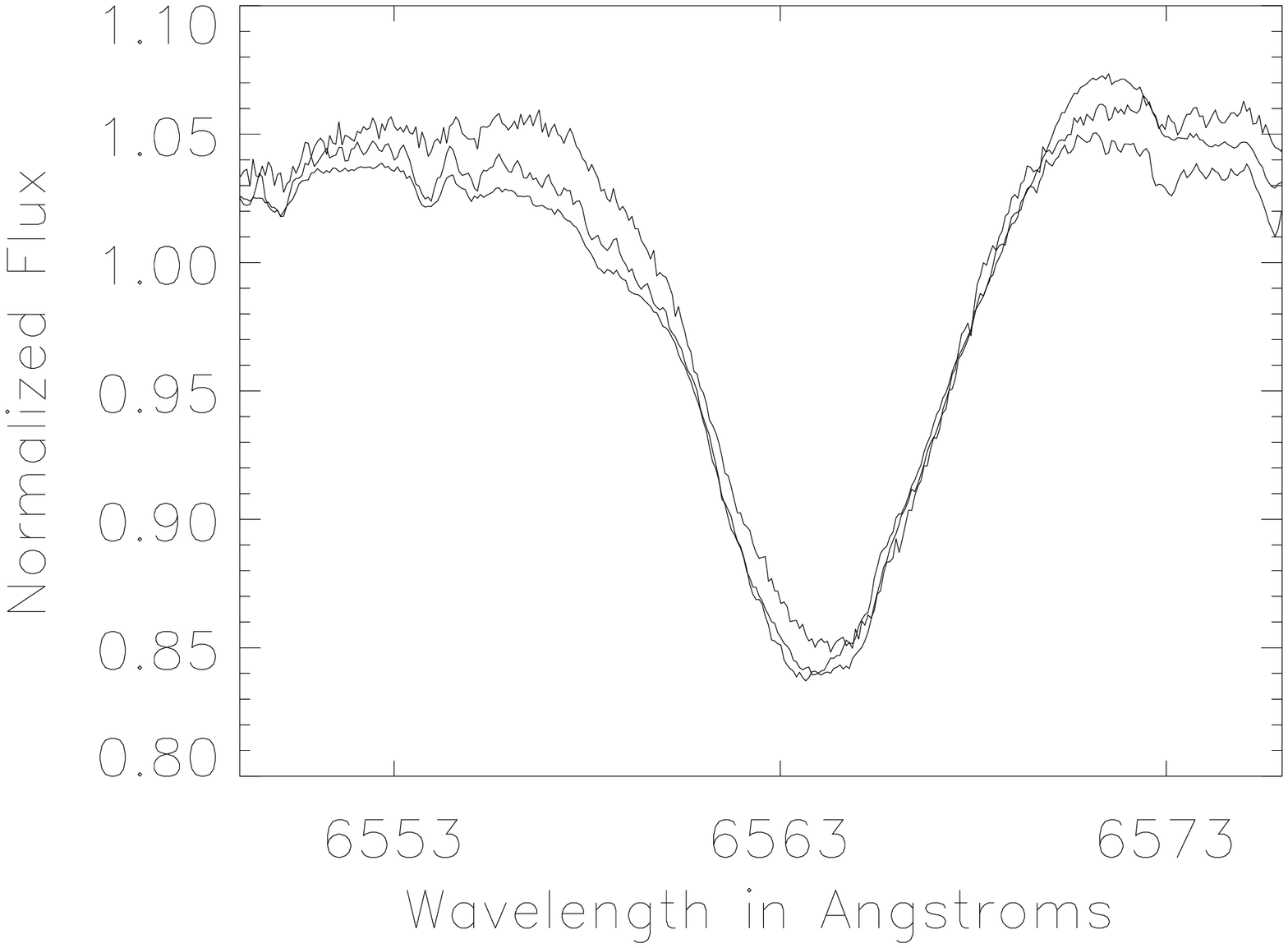}}
\caption[Be Line Profiles II]{Be Line Profiles II}
\label{fig:be-lprof2}
\end{figure}

	Our observing campaign first focused on $\gamma$ Cas in November of 2006 but was then widened to include 27 targets. The H$_\alpha$ line profiles for these stars show a wide variety of profiles. Figures \ref{fig:be-lprof1} and \ref{fig:be-lprof2} show all the profiles compiled for each stars. Some show strong ``disky" signatures like $\psi$ Per and MWC 143. Others show a more simple symmetric emission-line such as 11 Cam or 31 Peg.

\begin{table}[!h,!t,!b]
\begin{center}
\caption{Be/Emission-Line Stellar Properties \label{tab-obsbe}}
\begin{tabular}{lrccc}
\hline
\hline
Name                     & HD           &MWC & V          & ST                \\
$\gamma$ Cas    &   5394     &  9         & 2.4      &B0IVpe            \\         
25 Ori                    & 35439     &  110     & 4.9     & B1Vpe               \\
$\eta$ Tau            &  23630    & 74      & 2.9      &B7IIIe                    \\
11 Mon                & 45725    & 143   & 4.6     & B3Ve                      \\
$\omega$ Ori       & 37490    & 117    & 4.6    & B3IIIe                   \\
Omi Pup                & 63462    & 186    & 4.5     & B1IV:nne             \\
$\kappa$ CMa    & 50013     & 155    & 3.5     & B1.5Ve             \\
$\alpha$ Col       & 37795    & 119      & 2.6      &B7IVe             \\
66 Oph                & 164284    & 278    & 4.8     & B2Ve               \\
31 Peg                & 212076       & 387  & 4.8     & B2IV-Ve               \\
11 Cam              & 32343    &  96  & 5.0       & B2.5Ve             \\
12 Vul                 & 187811    &  323  & 4.9       & B2.5Ve                 \\
HD 36408           & 36408     &      & 5.5     & B7IIIe                  \\
$\lambda$ Cyg  & 198183     & 352   & 4.6     & B5Ve            \\
QR Vul                 & 192685     &      & 4.8     &  B3Ve              \\
\hline
$\psi$ Per             & 22192     &  69      & 4.3       & B5Ve           \\
10 CMa                 & 48917      &  152    & 5.2    & B2IIIe                     \\
Omi Cas               & 4180    &   8  & 4.5     & B5IIIe                     \\
18 Gem                & 45542    &  141     & 4.1     & B6IIIe                    \\
$\alpha$ Cam    &  30614     & 92    &  4.3     & O9.5Iae             \\
$\beta$ CMi        & 58715     & 178      &  2.9   & B8Ve             \\
C Per                  & 25940    &  81  & 4.0      & B3Ve              \\
$\kappa$ Cas   & 2905     &  7  & 4.2       & B1Iae         \\
$\phi$ And         & 6811     &  420   & 4.3      & B7Ve              \\
MWC 77              & 24479    & 77    & 4.9      & B9.5Ve                \\
$\xi$ Per            & 24912    &     & 4.0     & O7.5IIIe         \\
R Pup              &  68980      & 192   & 4.8  & B1.5IIIe   \\
3 Pup                  &62623    &  570     & 4.0     & A3Iabe           \\
\hline
\hline
\end{tabular}
\end{center}
The Be and Emission-line stars. The columns list the name, HD catalog number, MWC catalog number, V magnitude and spectral type (ST) for each star. All information is from the Simbad Online Database.
\end{table}

	Table \ref{tab-obsbe} lists the basic stellar properties. Essentially all the stars are B-type and are 5th magnitude or brighter. Although there is only a month or two baseline between observations for variability studies, most systems showed only small profile changes, if any. The most variable stars were $\zeta$ Tau, with its strong emission, $\kappa$ Cas and MWC 92 with their weaker emission lines. Many stars showed small variability of the emission, such as 25 Ori, $\omega$ Ori, $\kappa$ CMa, or 18 Gem, but the overall structure of the emission line did not change. This shows that during the observing window, the circumstellar region responsible for the emission did not change its structure significantly.  
	
	For the purposes of this study, which is concerned with the circumstellar material, the stars can be broken into subclasses based on their H$_\alpha$ line profiles. There are strong, symmetric emission lines such as 11 Cam in figure \ref{fig:lprof-11cam} that have line:continuum ratios of 5-8 and don't show evidence for intervening absorption. There are the classical ``disky" systems such as MWC 143 that have strong emission with a strong, narrow central absorptive component. Another class will be distinguished in table \ref{tab-obsbe} called Disk* - these show less absorption than the classical ``disky" systems, but do show central, often variable absorption. The last type does not show significant emission, such as QR Vul.

\subsection{Be and Emission-Line Star Spectropolarimetry}

	In contrast to the detected HAe/Be morphologies, most Be and Emission-line stars showed only broad, smooth polarization changes spanning the entire width of the line with the polarization change centered on the line. Although none of these stars show P-Cygni profiles and direct comparison with many HAe/Be detections is not possible, there are several  ``disky" systems provide a robust comparison.
	
	The magnitude of the polarization change in these stars was also smaller on average, with 0.5\% or less being typical. Some stars showed signatures over 1\%, but the polarization signature was much broader than the line and extended out into the line wings. Even though many of the H$_\alpha$ lines in systems with broad spectropolarimetric detections had strong absorption features, there were only a few stars that showed significant deviations from the broad polarization morphology. In 10 very clear detections, the broad polarization change can be seen.

\begin{figure}
\centering
\subfloat[$\gamma$ Cas]{\label{fig:gmcas}
\includegraphics[ width=0.21\textwidth, angle=90]{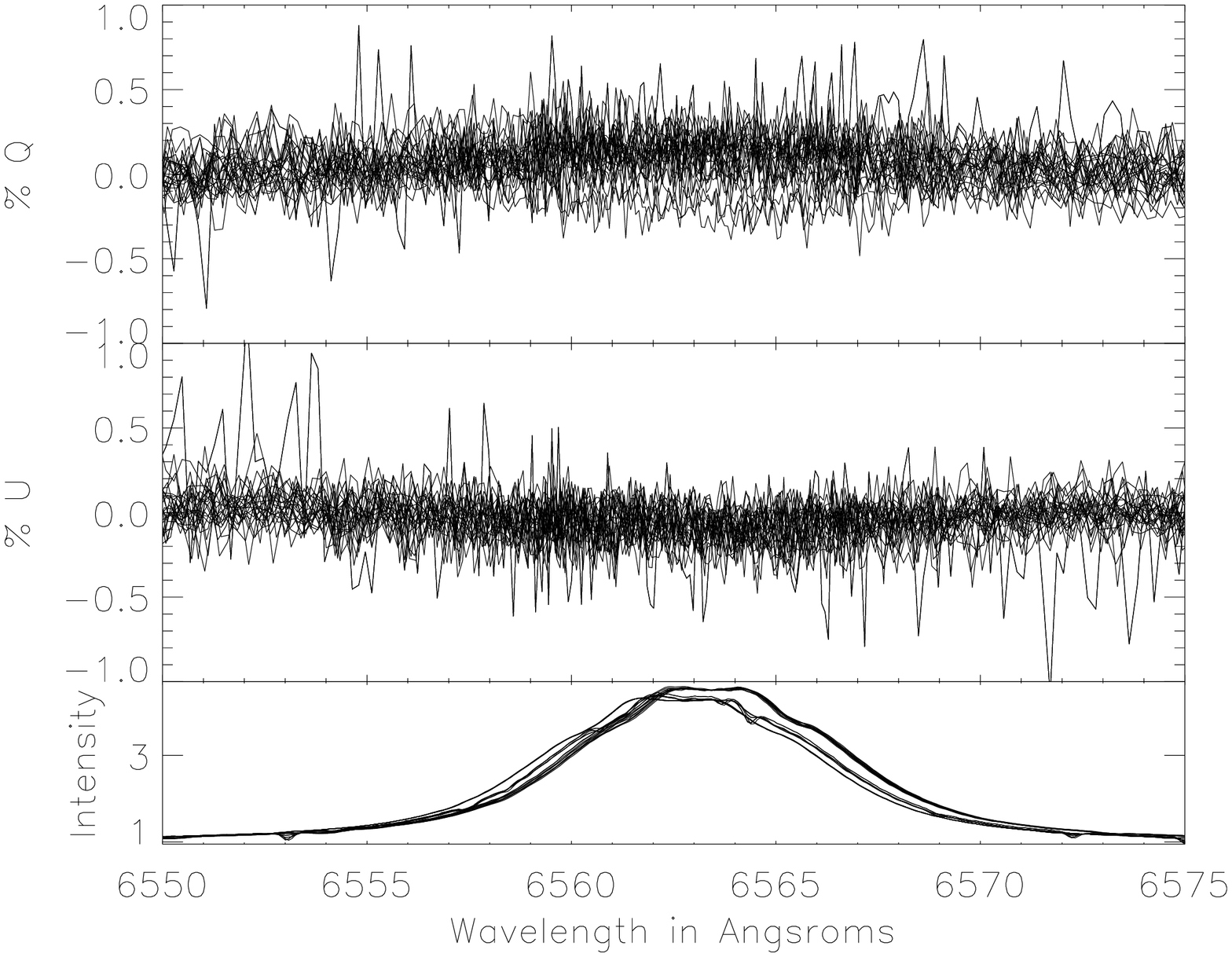}}
\quad
\subfloat[25 Ori]{\label{fig:swap-rebin25ori}
\includegraphics[ width=0.21\textwidth, angle=90]{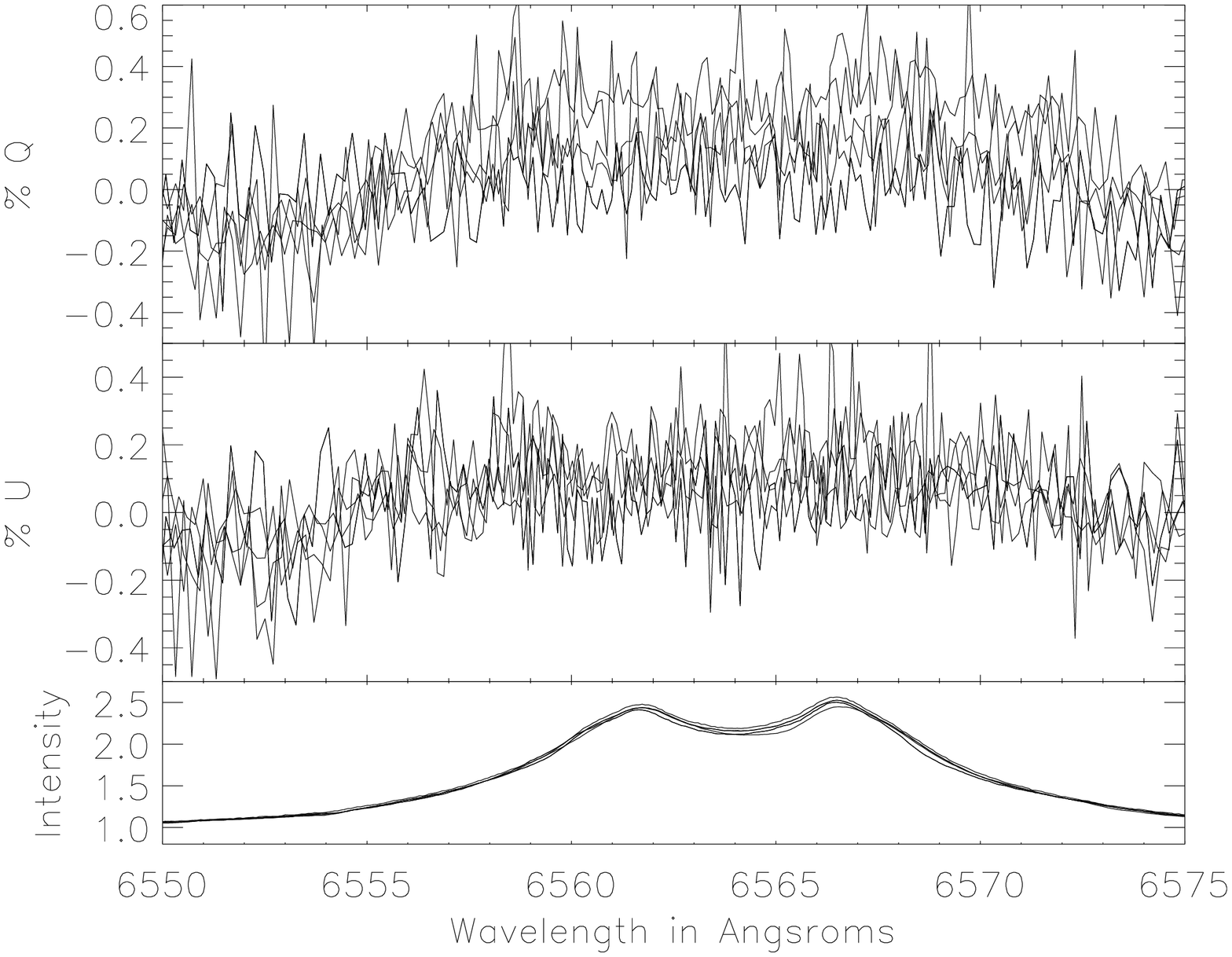}}
\quad
\subfloat[$\psi$ Per]{\label{fig:swap-rebinpsiper}
\includegraphics[ width=0.21\textwidth, angle=90]{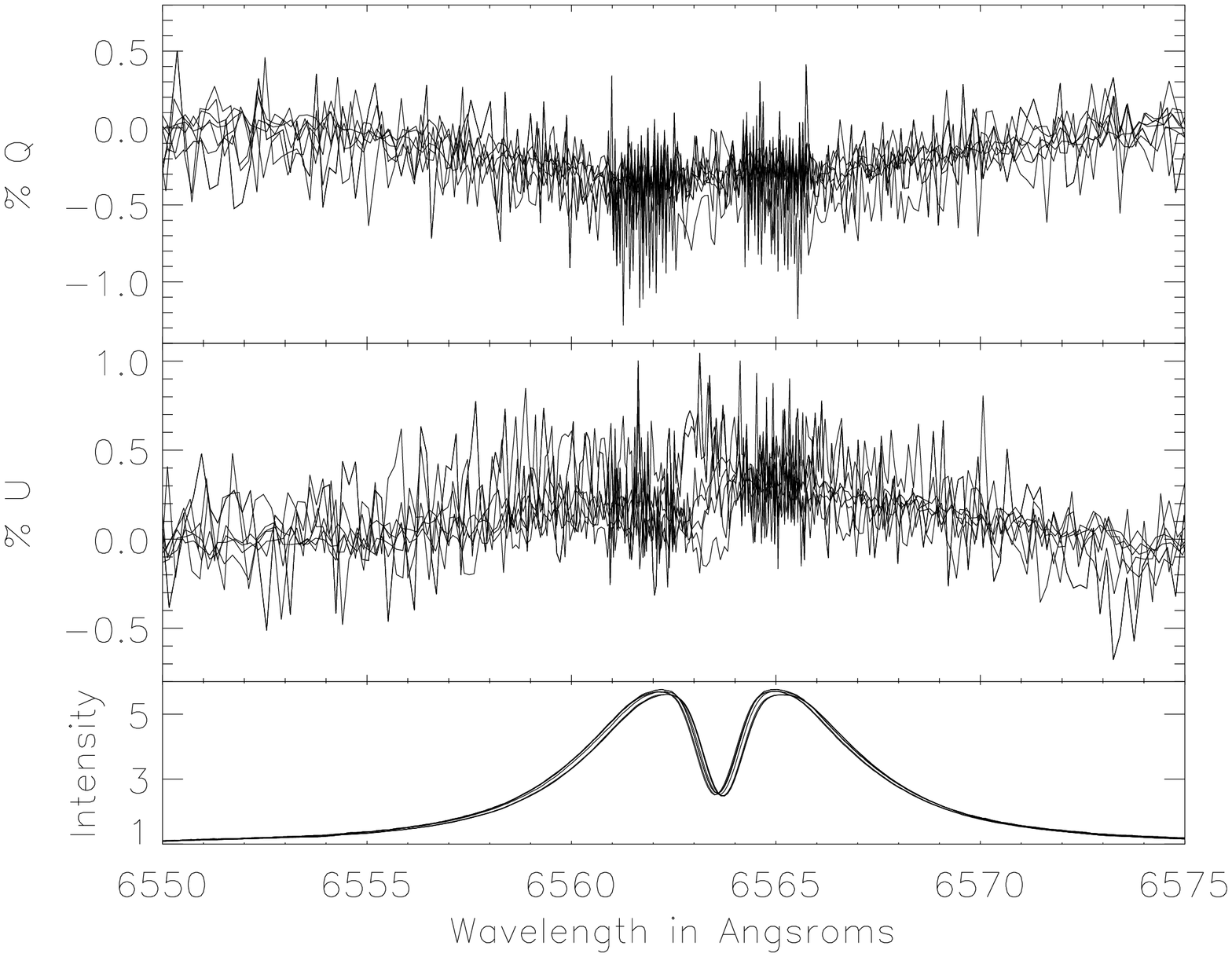}}
\quad
\subfloat[$\eta$ Tau]{\label{fig:swap-rebinetatau}
\includegraphics[ width=0.21\textwidth, angle=90]{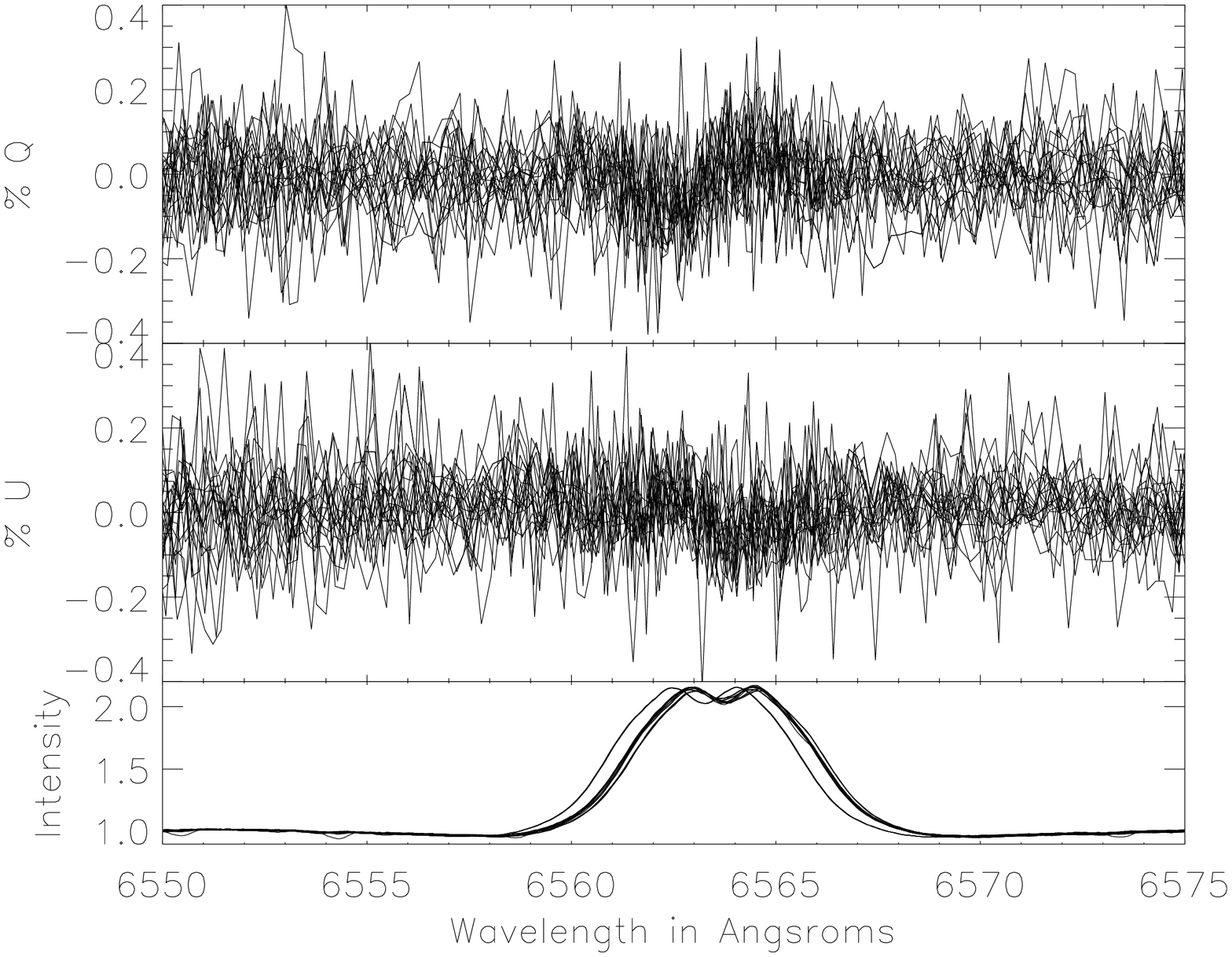}}
\quad
\subfloat[$\zeta$ Tau]{\label{fig:zetatau}
\includegraphics[ width=0.21\textwidth, angle=90]{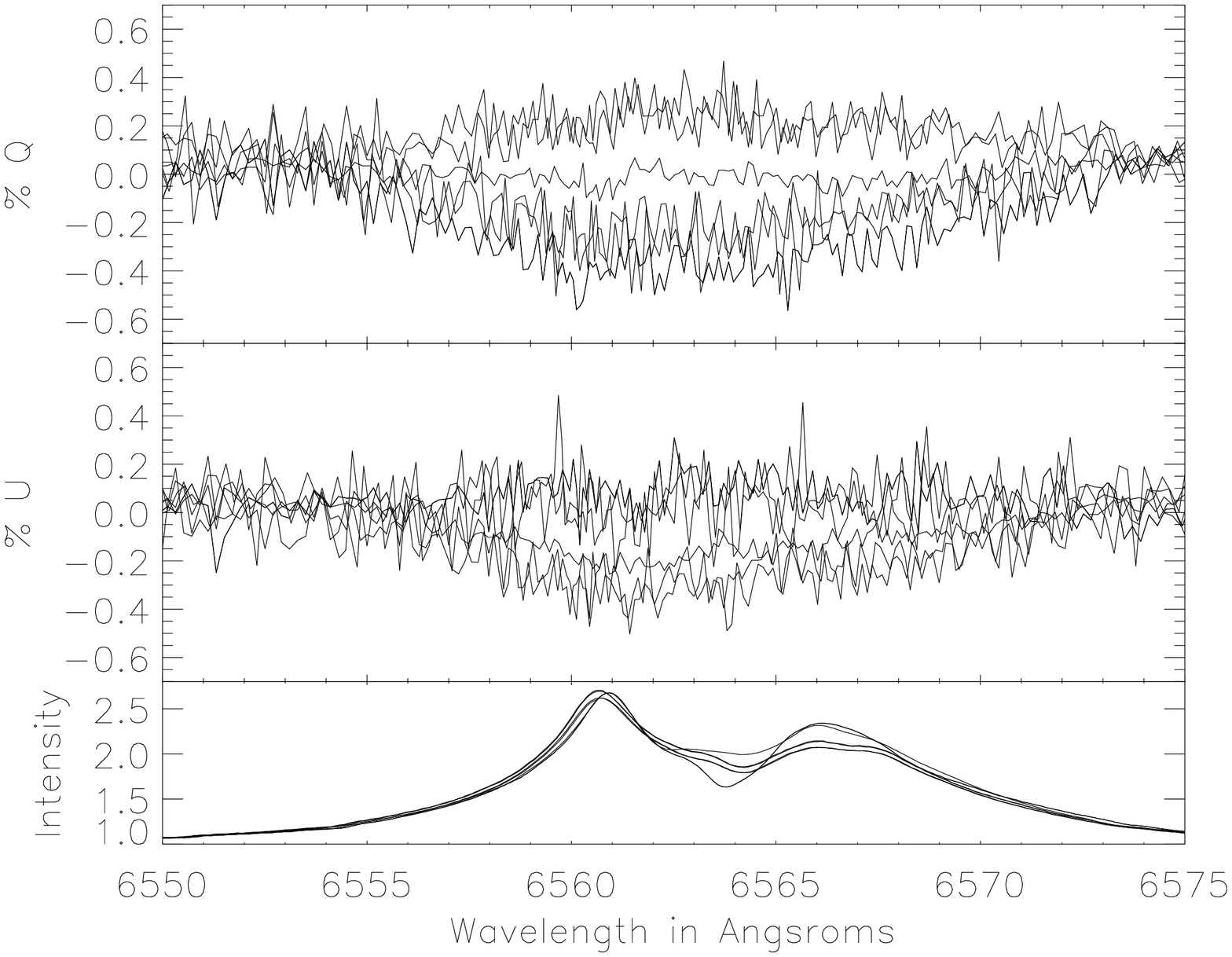}}
\quad
\subfloat[MWC 143]{\label{fig:mwc143}
\includegraphics[ width=0.21\textwidth, angle=90]{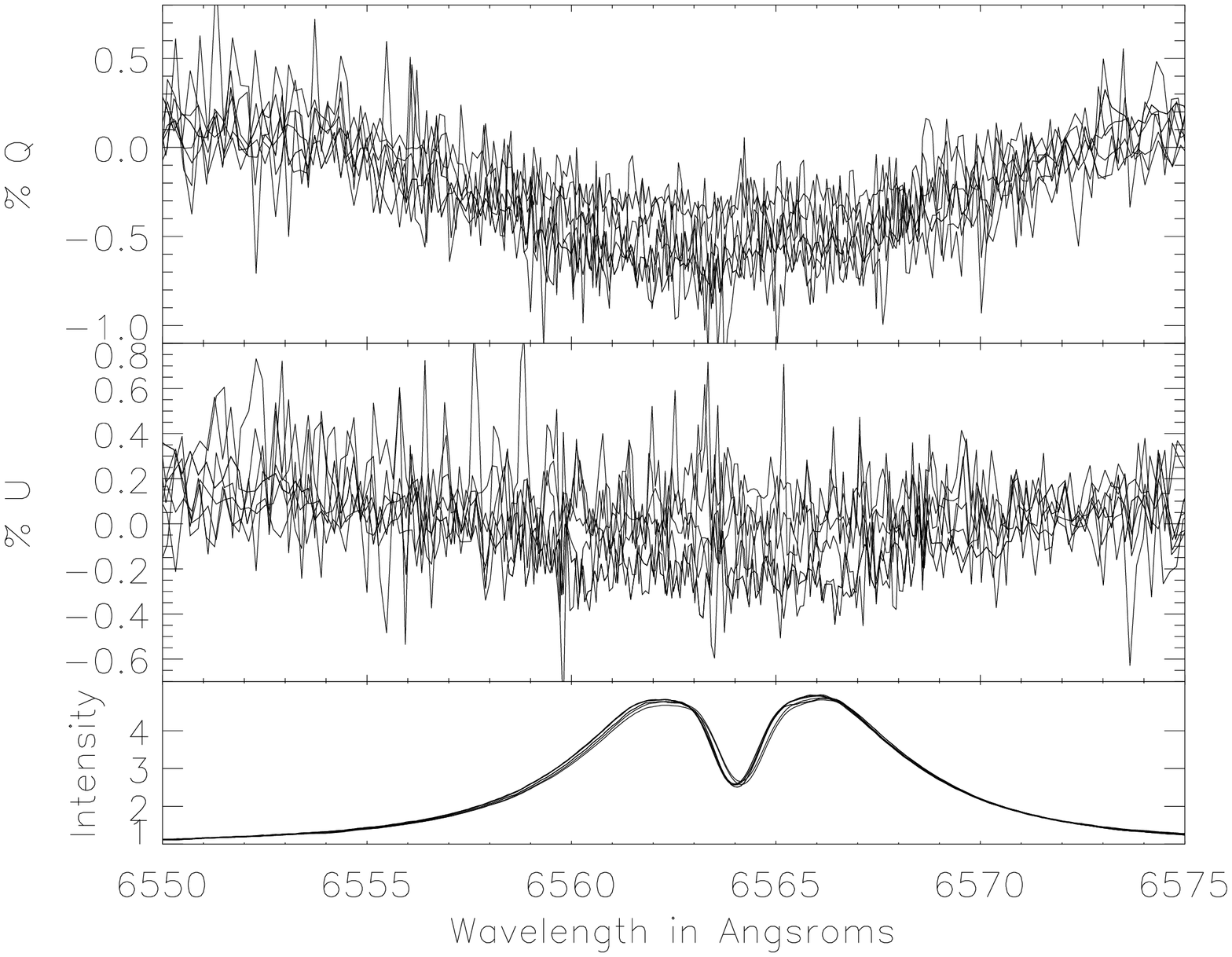}}
\quad
\subfloat[$\omega$ Ori]{\label{fig:omori}
\includegraphics[ width=0.21\textwidth, angle=90]{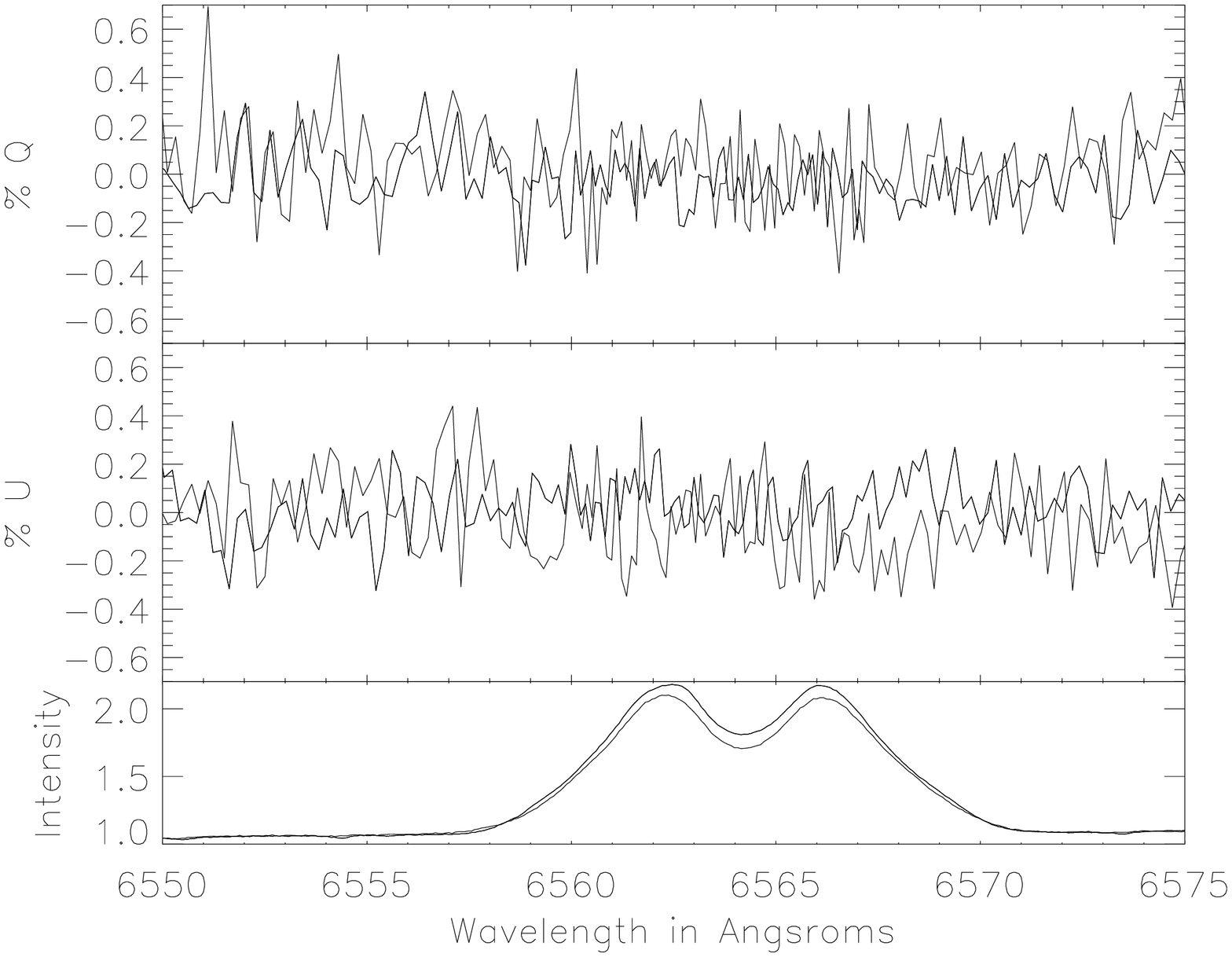}}
\quad
\subfloat[Omi Pup]{\label{fig:ompup}
\includegraphics[ width=0.21\textwidth, angle=90]{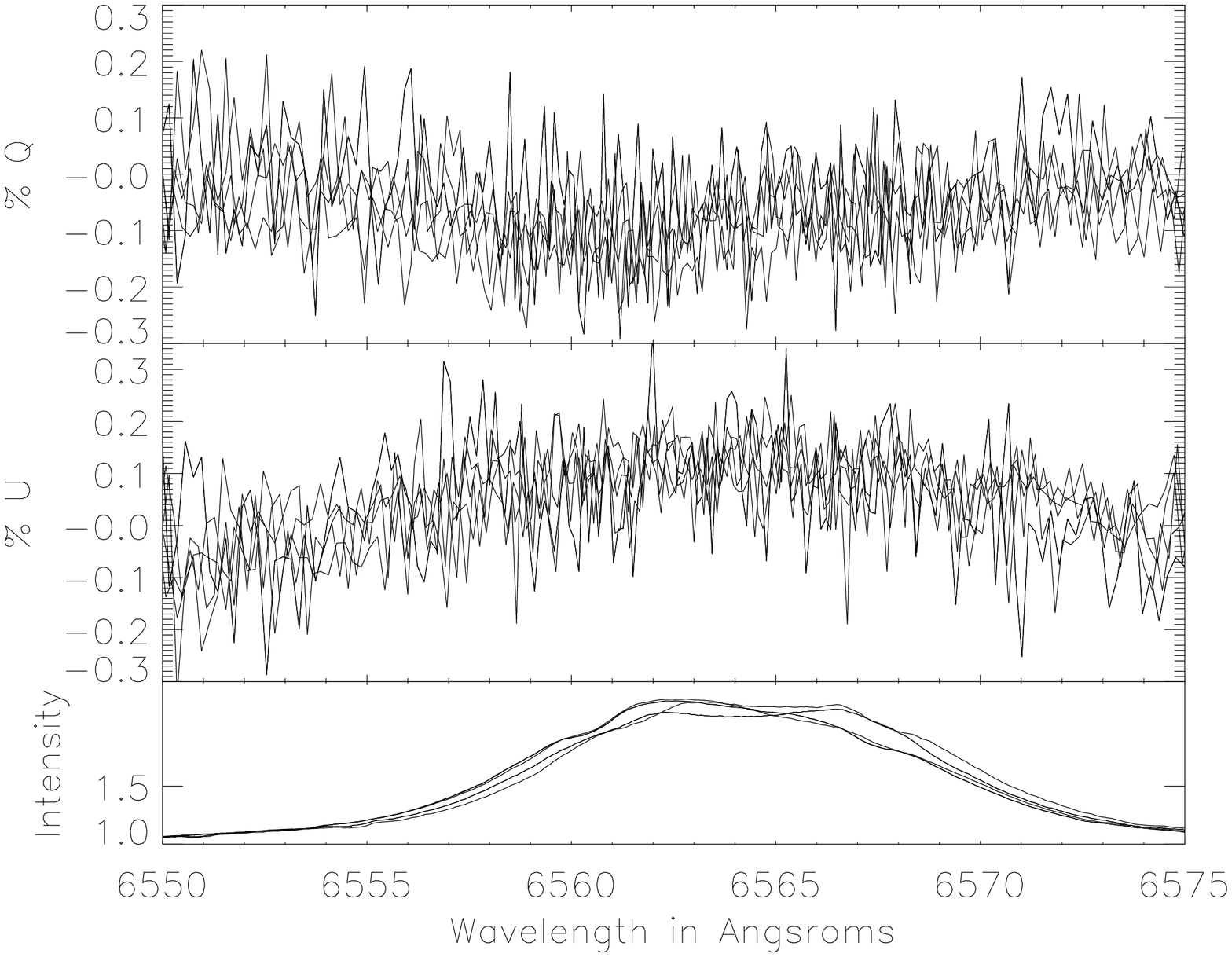}}
\quad
\subfloat[10 CMa]{\label{fig:10cma}
\includegraphics[ width=0.21\textwidth, angle=90]{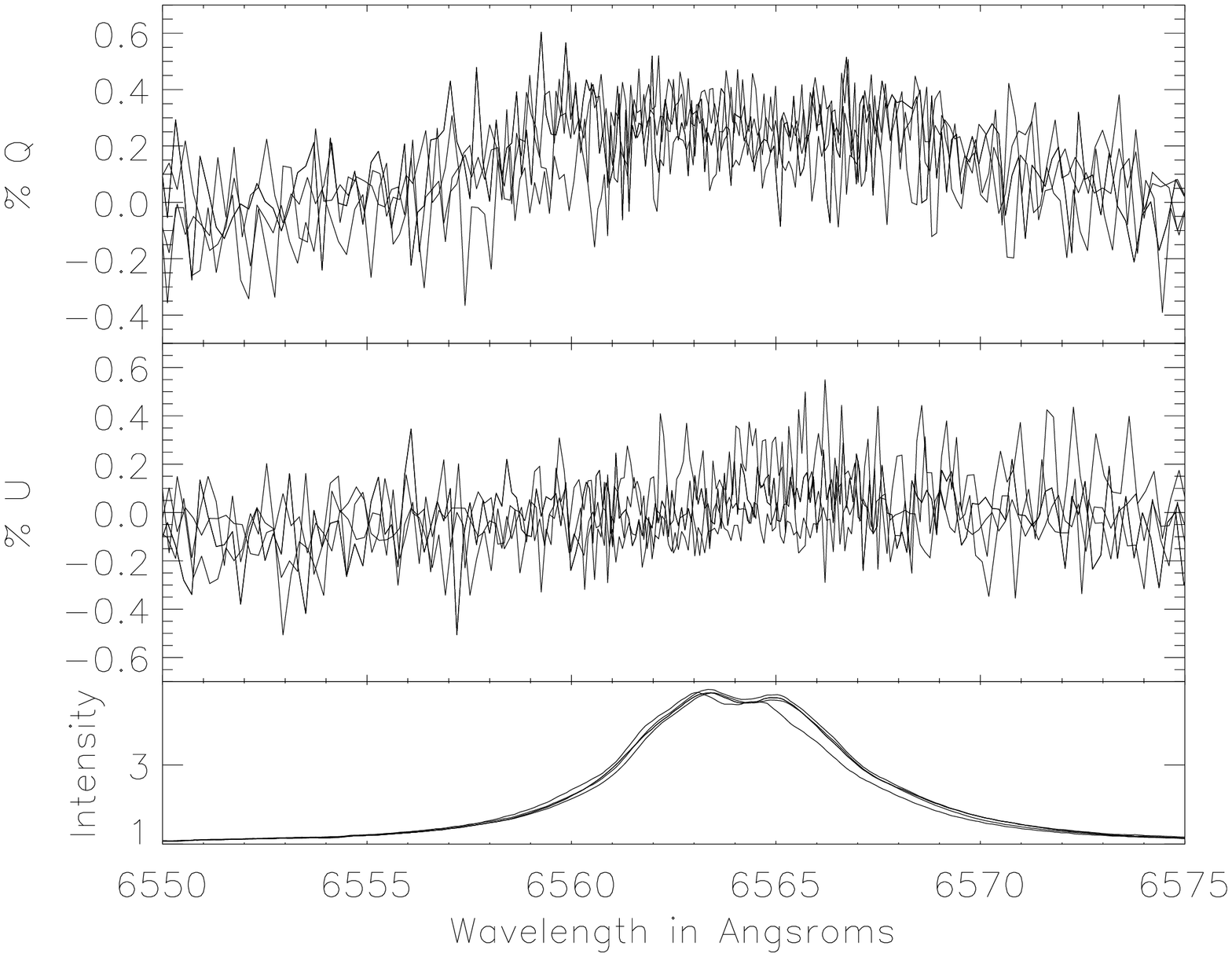}}
\quad
\subfloat[Omi Cas]{\label{fig:omicas}
\includegraphics[ width=0.21\textwidth, angle=90]{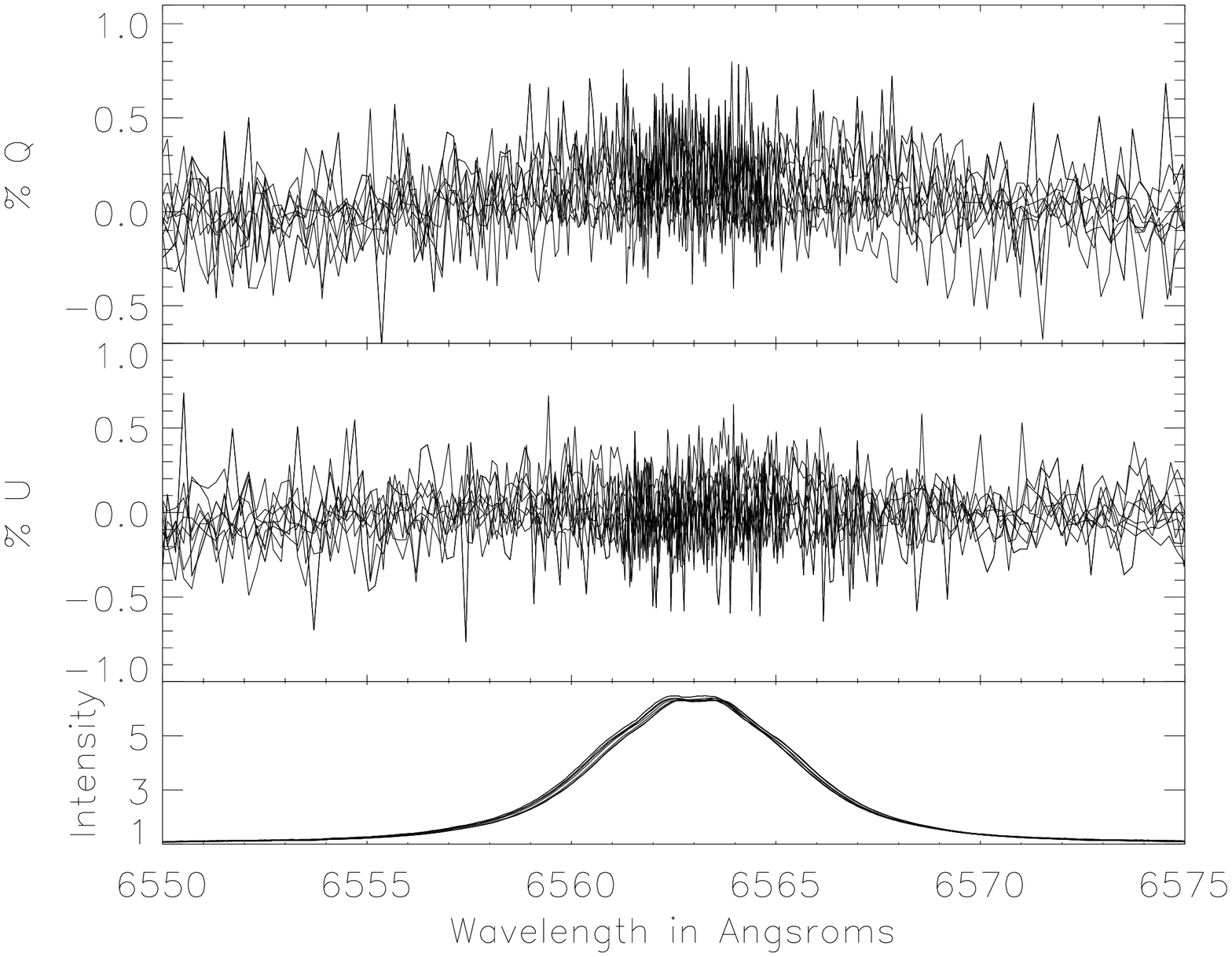}}
\quad
\subfloat[$\kappa$ CMa]{\label{fig:swap-rebinkapcma}
\includegraphics[ width=0.21\textwidth, angle=90]{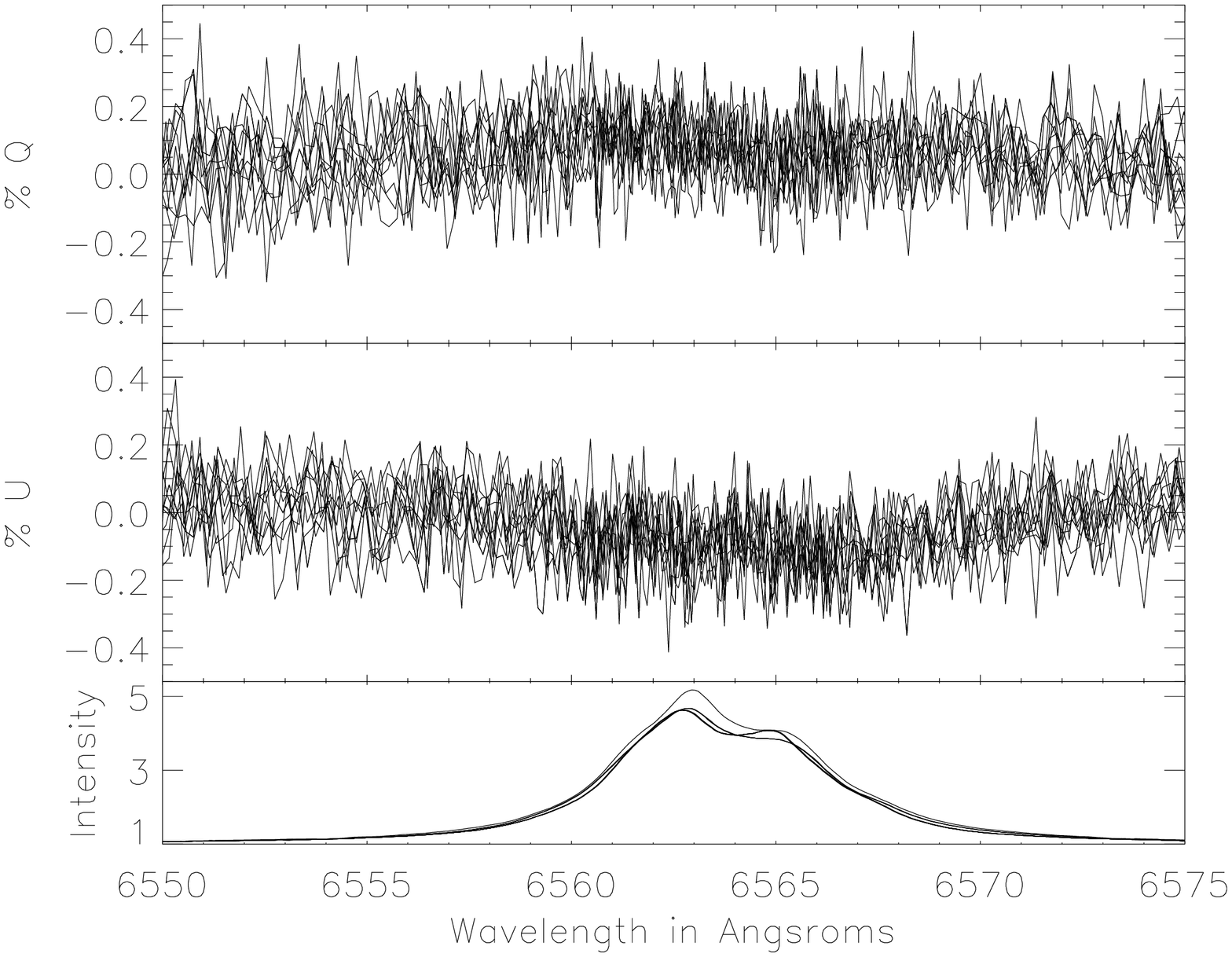}}
\quad
\subfloat[18 Gem]{\label{fig:18gem}
\includegraphics[ width=0.21\textwidth, angle=90]{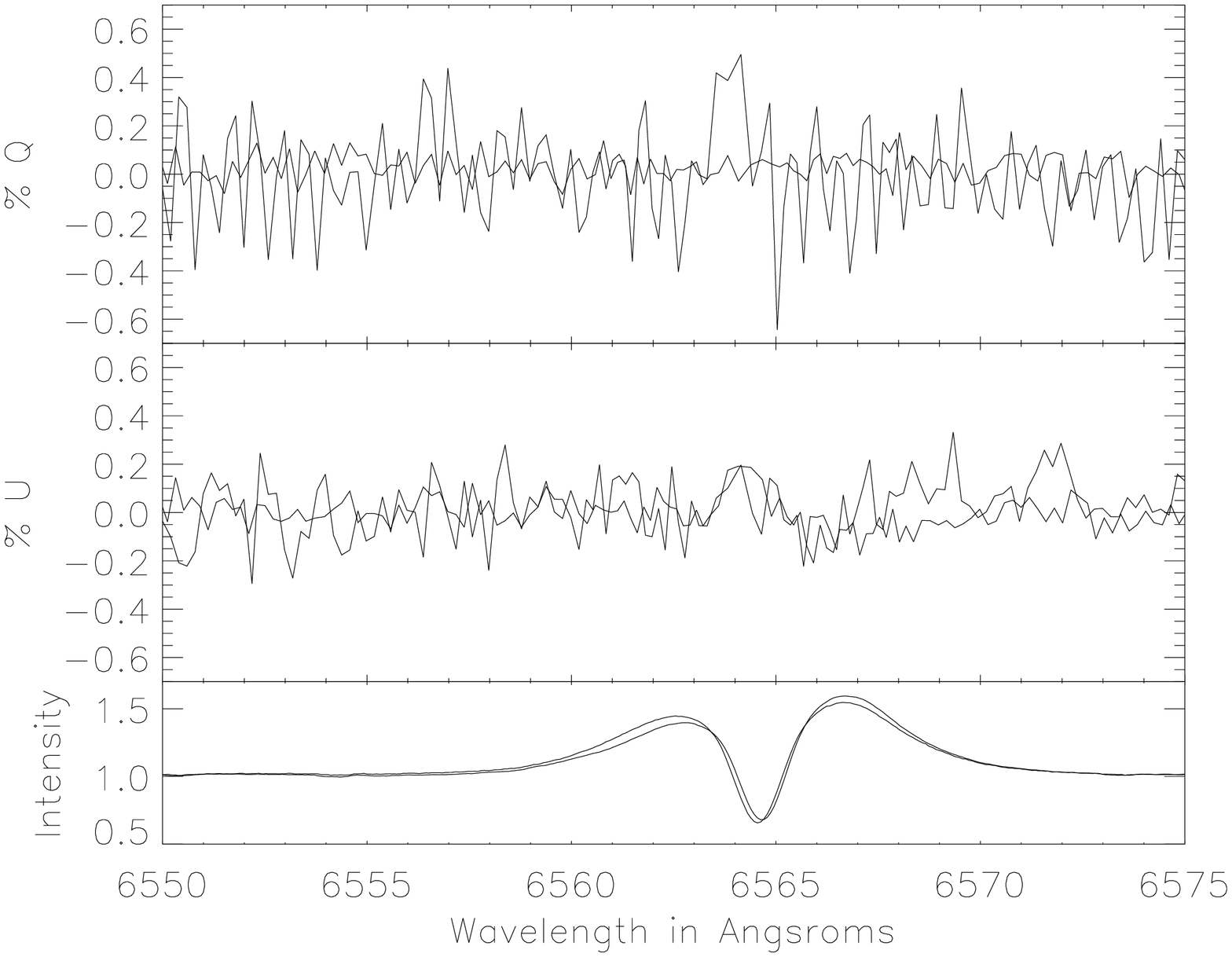}}
\quad
\subfloat[Alf Col]{\label{fig:alfcol}
\includegraphics[ width=0.21\textwidth, angle=90]{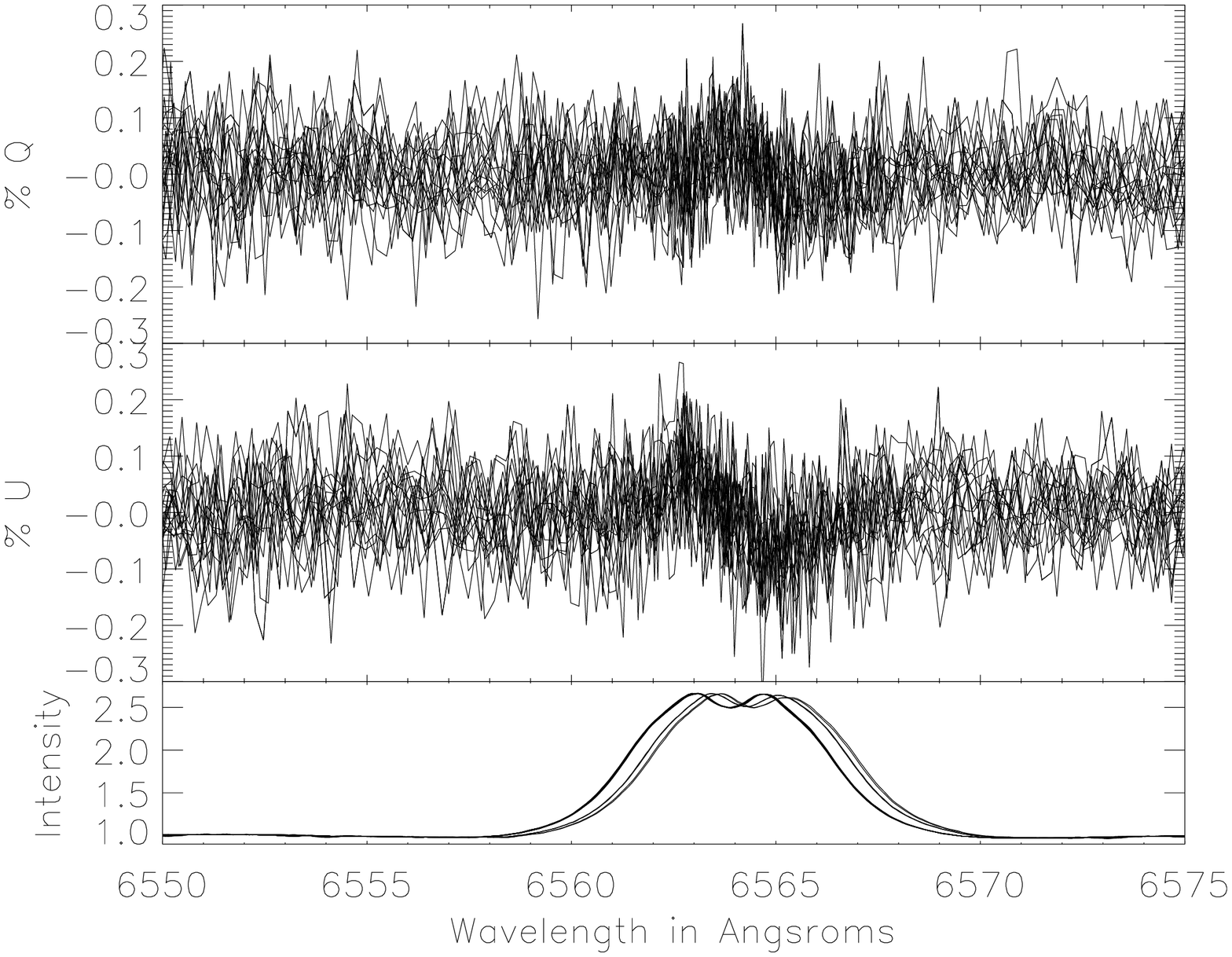}}
\quad
\subfloat[66 Oph]{\label{fig:66oph}
\includegraphics[ width=0.21\textwidth, angle=90]{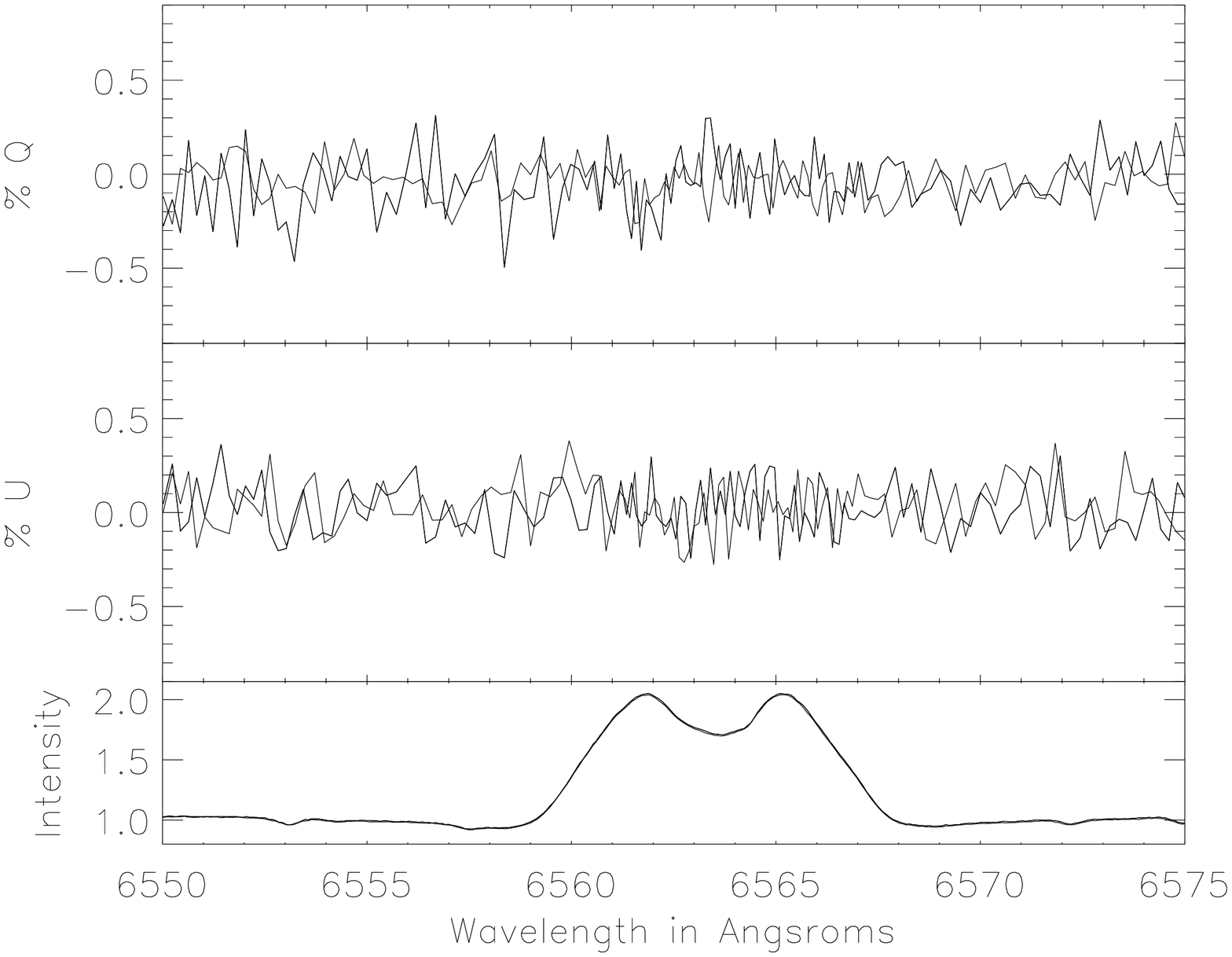}}
\quad
\subfloat[$\beta$ CMi]{\label{fig:bcm}
\includegraphics[ width=0.21\textwidth, angle=90]{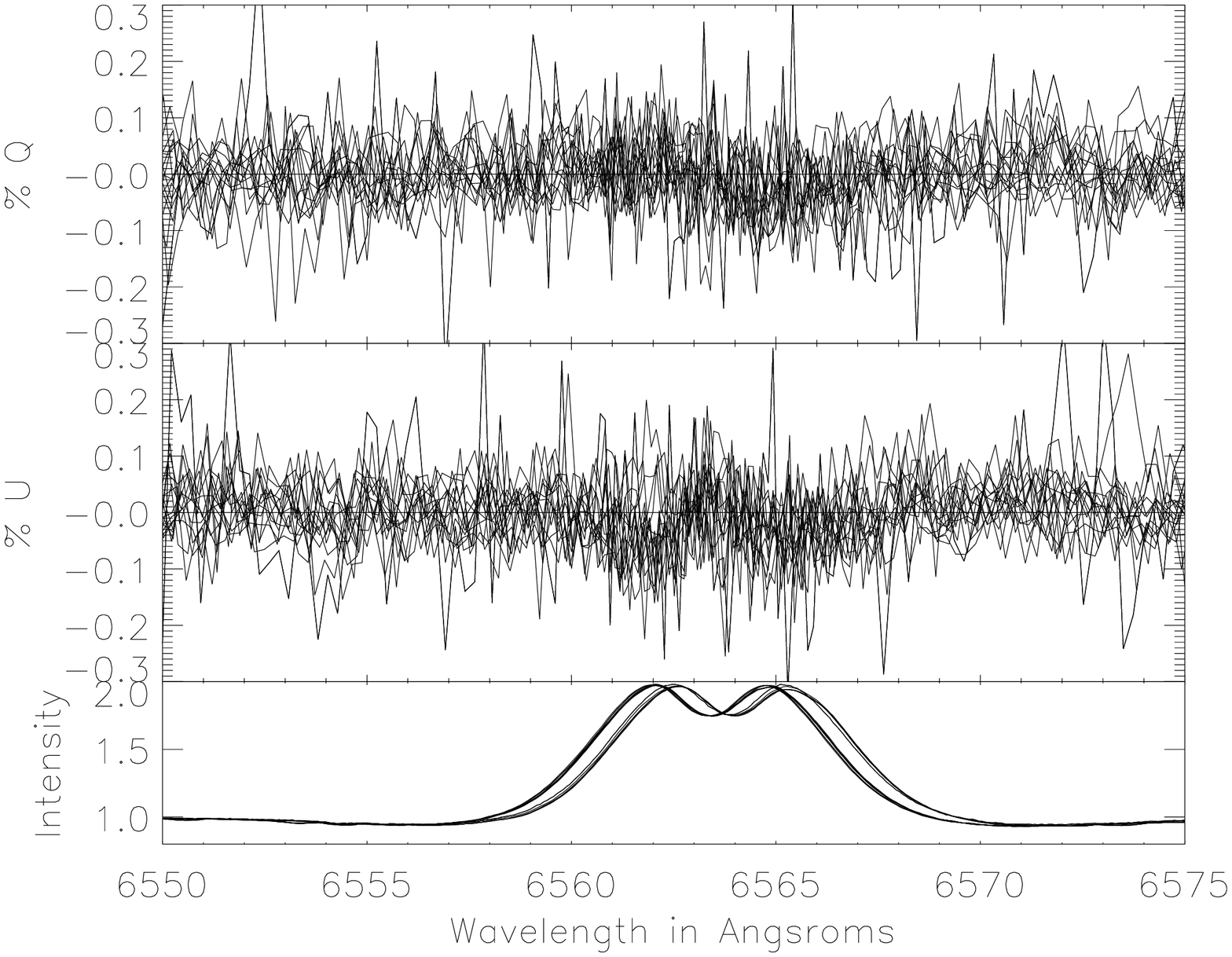}}
\caption[Be Spectropolarimetry I]{Be Spectropolarimetry I}
\label{fig:be-specpol1}
\end{figure}

\begin{figure}
\centering
\subfloat[31 Peg]{\label{fig:31peg}
\includegraphics[ width=0.21\textwidth, angle=90]{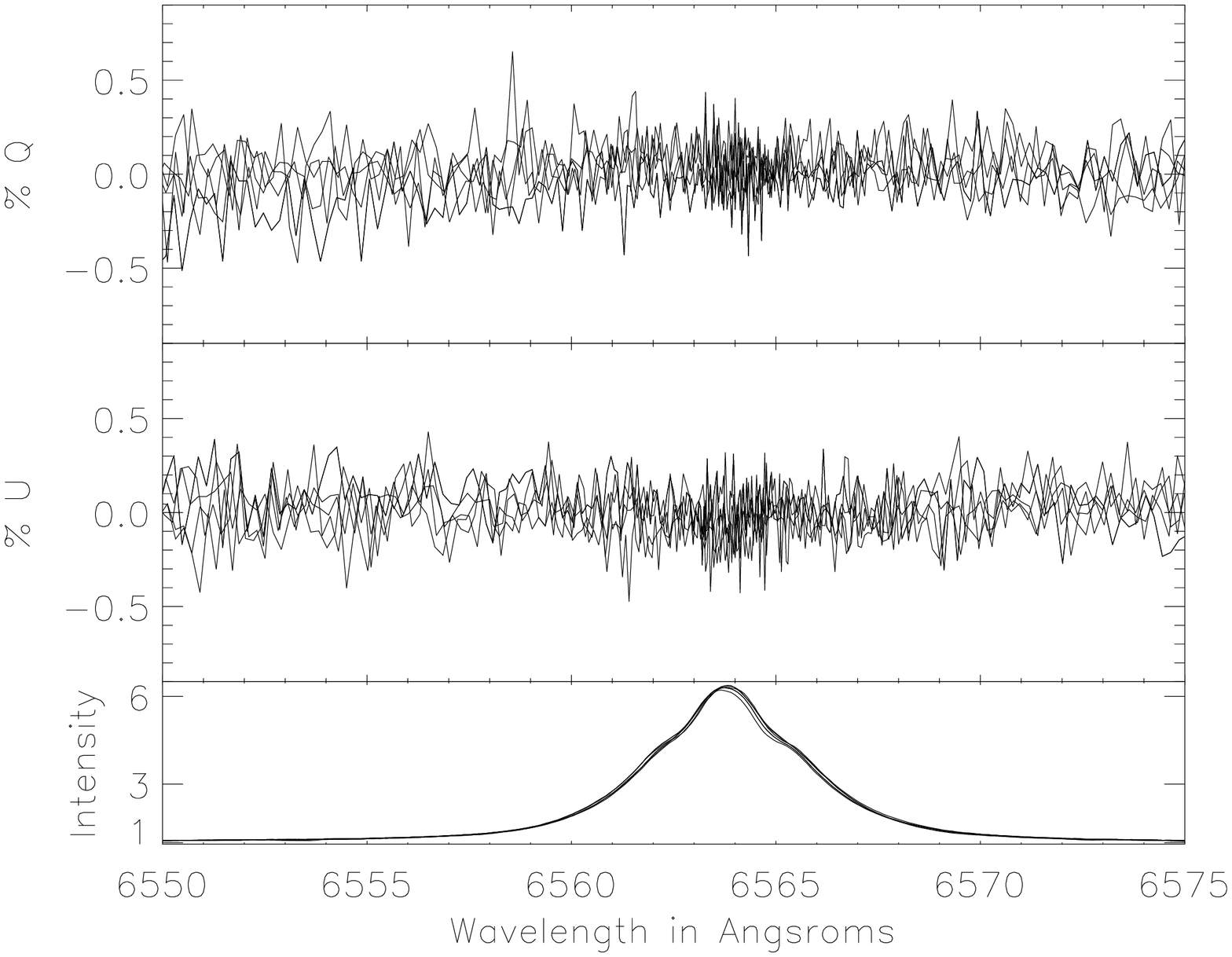}}
\quad
\subfloat[11 Cam]{\label{fig:swap-rebin11cam}
\includegraphics[ width=0.21\textwidth, angle=90]{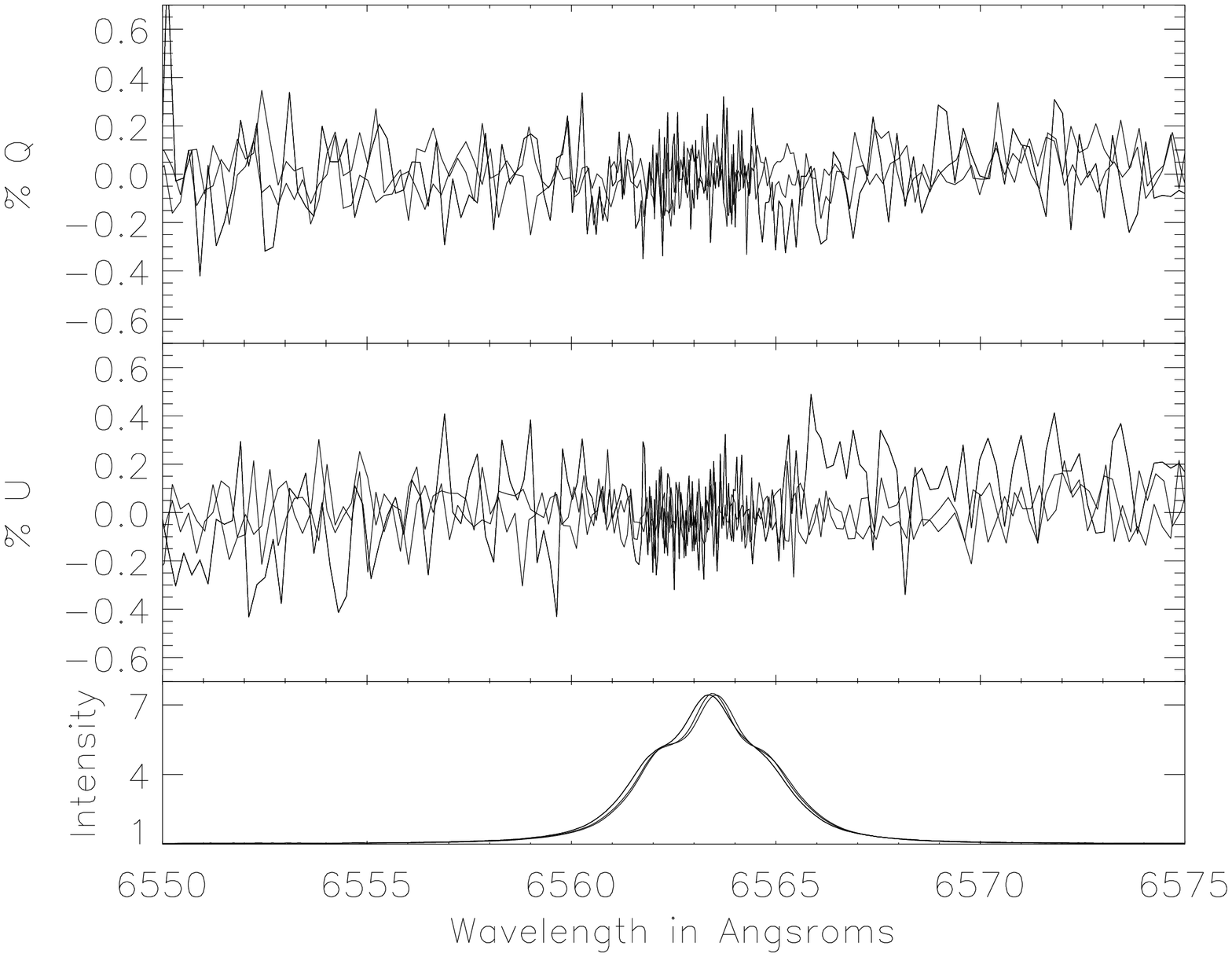}}
\quad
\subfloat[C Per]{\label{fig:swap-rebincper}
\includegraphics[ width=0.21\textwidth, angle=90]{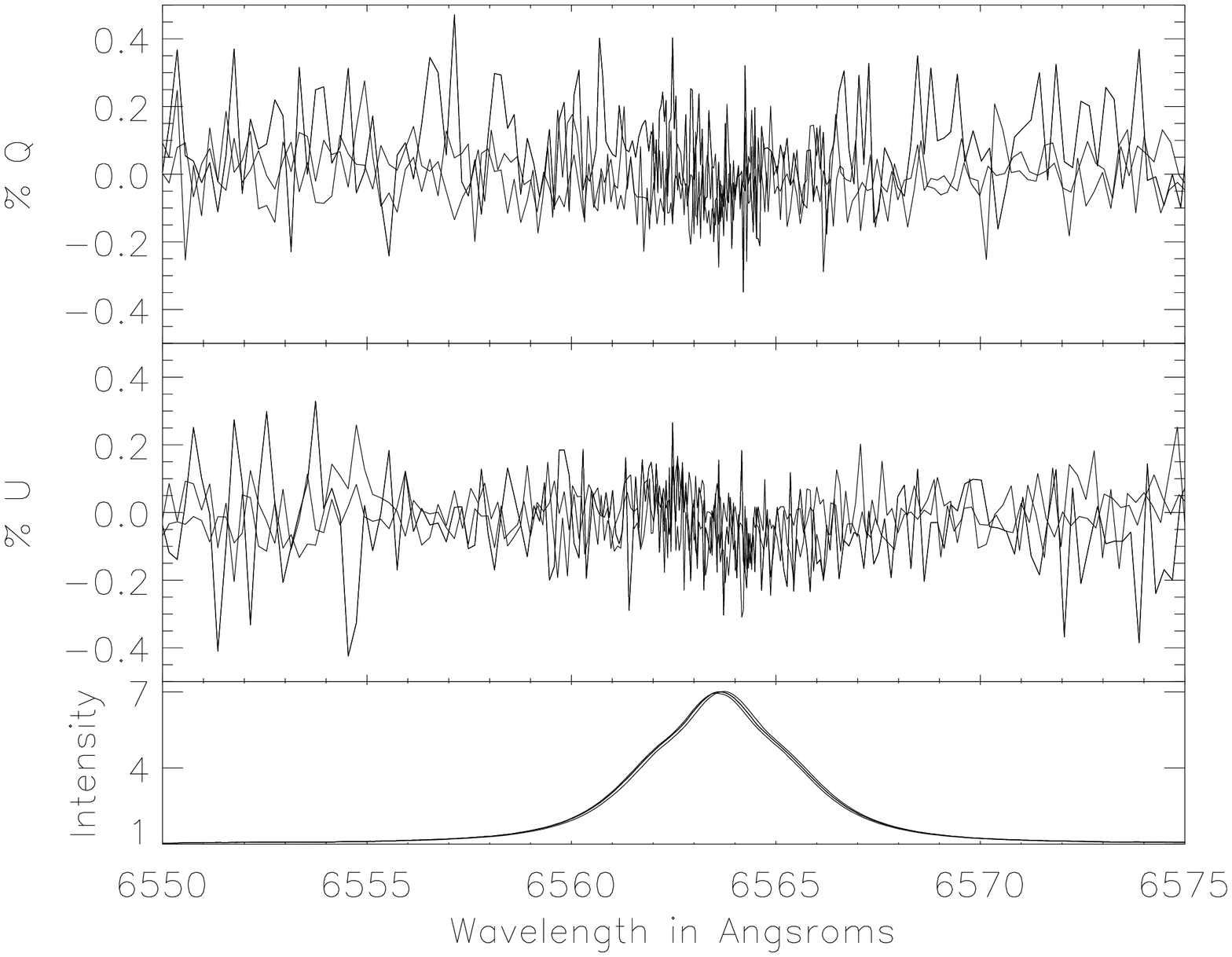}}
\quad
\subfloat[MWC 192]{\label{fig:mwc192}
\includegraphics[ width=0.21\textwidth, angle=90]{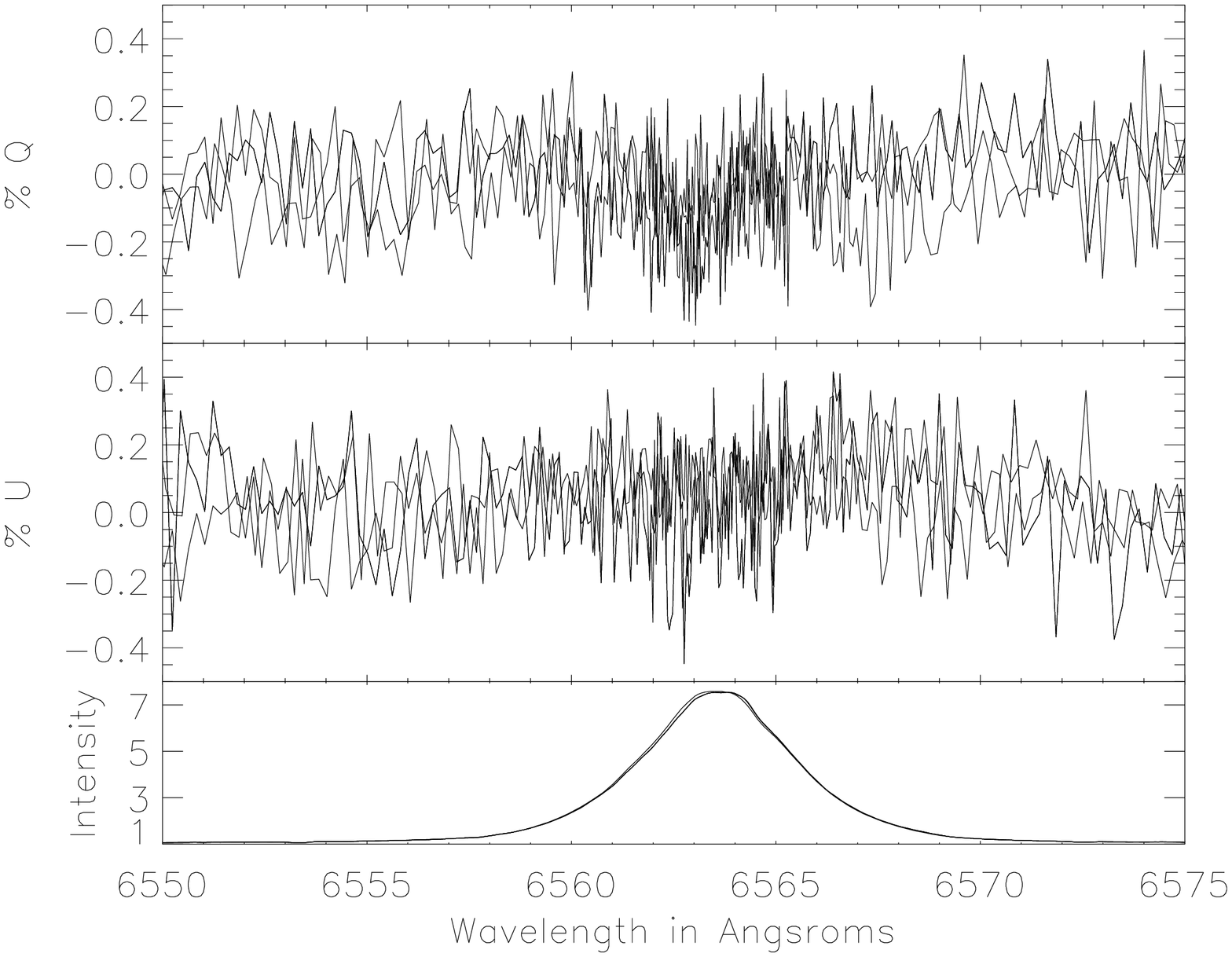}}
\quad
\subfloat[$\kappa$ Cas]{\label{fig:swap-rebinkapcas}
\includegraphics[ width=0.21\textwidth, angle=90]{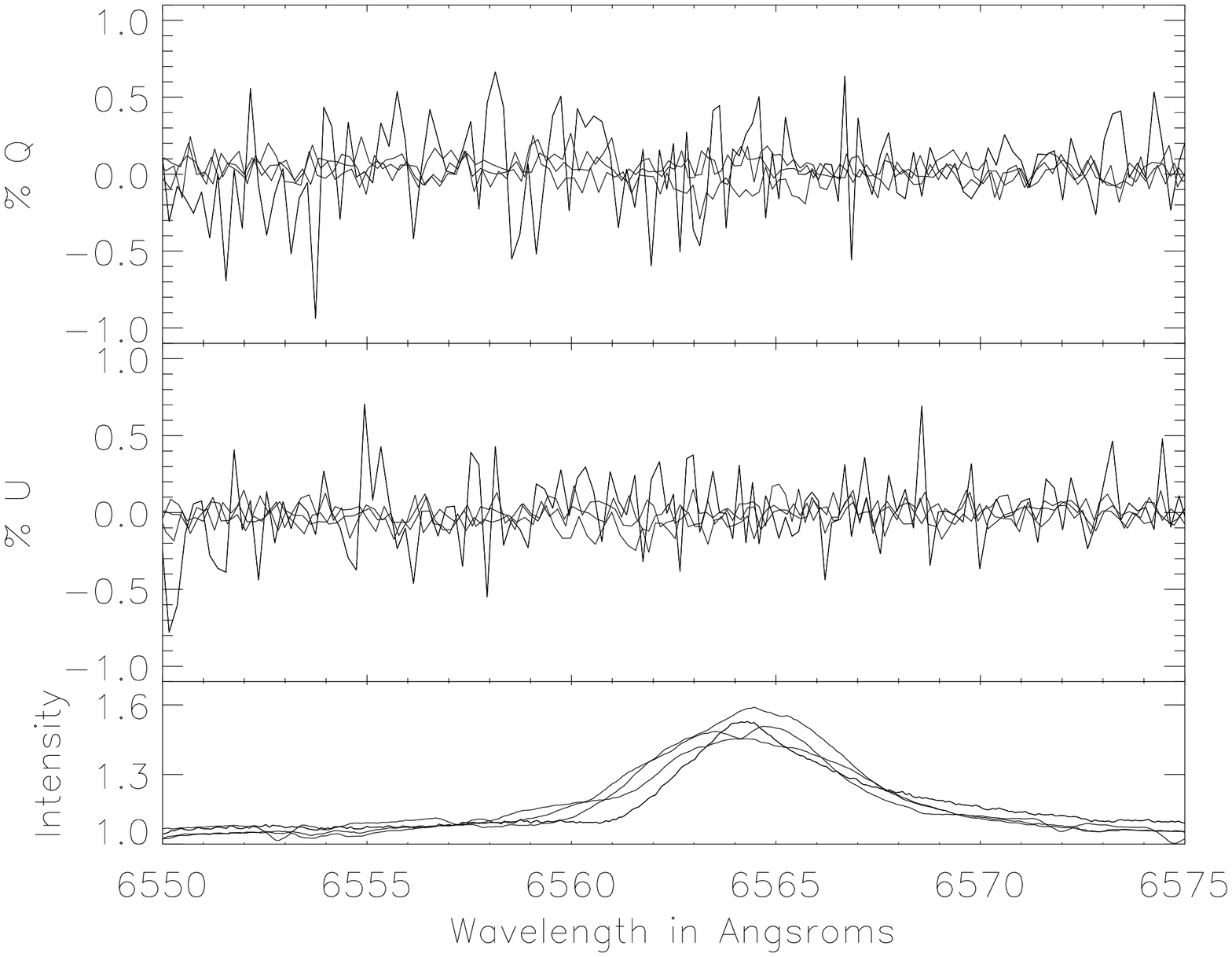}}
\quad
\subfloat[MWC 92]{\label{fig:mwc92}
\includegraphics[ width=0.21\textwidth, angle=90]{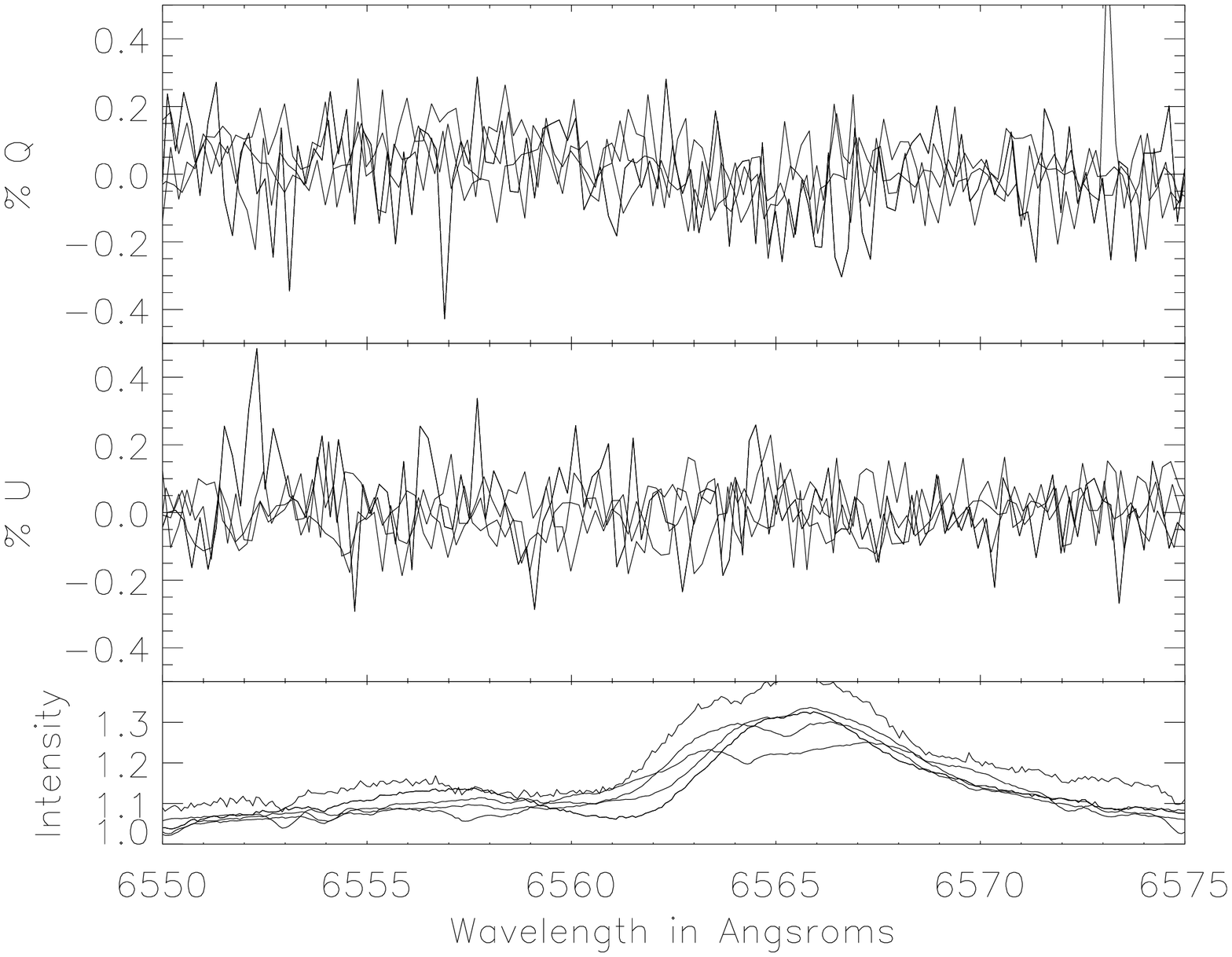}}
\quad
\subfloat[$\kappa$ Dra]{\label{fig:kapdra}
\includegraphics[ width=0.21\textwidth, angle=90]{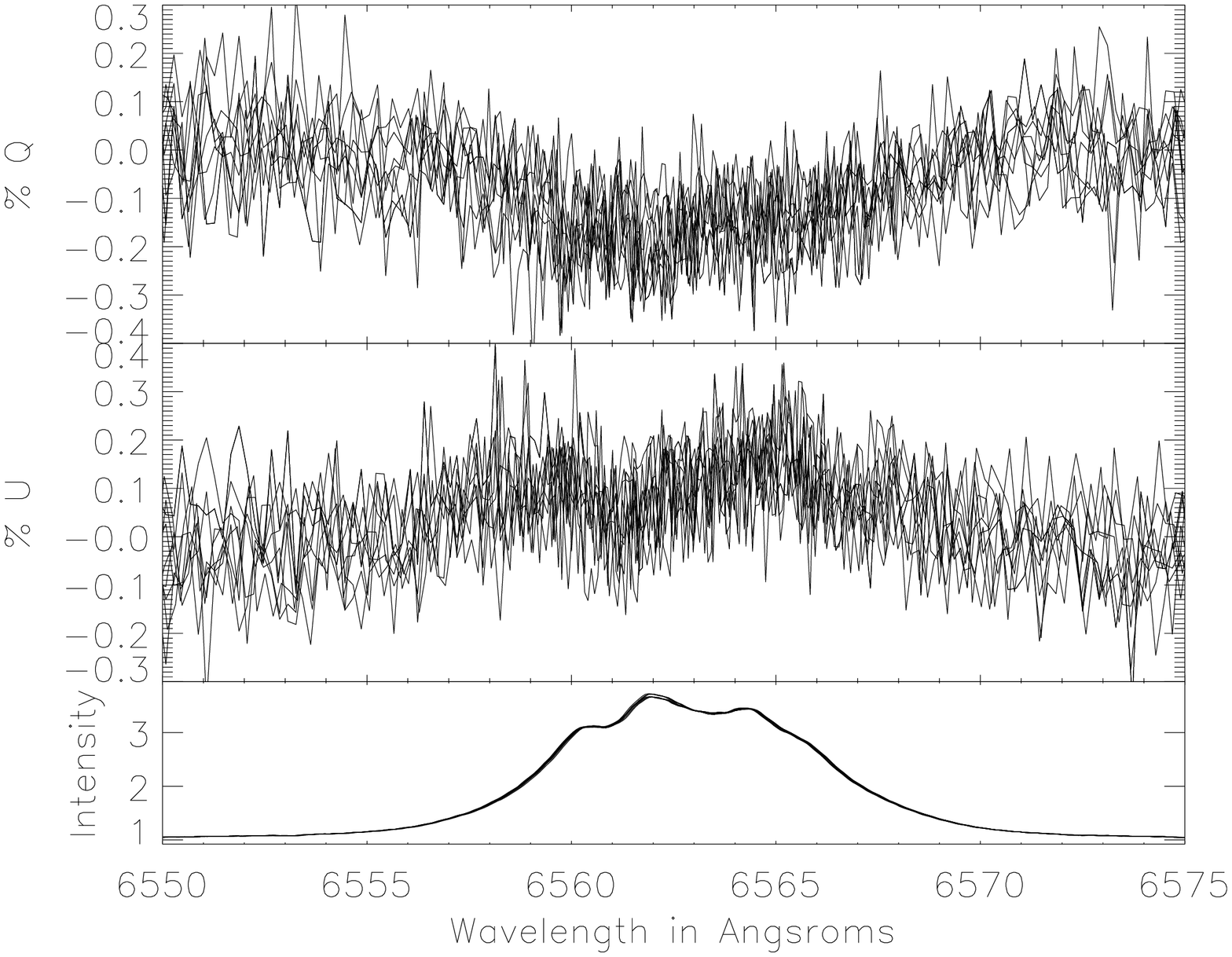}}
\quad
\subfloat[12 Vul]{\label{fig:12vul}
\includegraphics[ width=0.21\textwidth, angle=90]{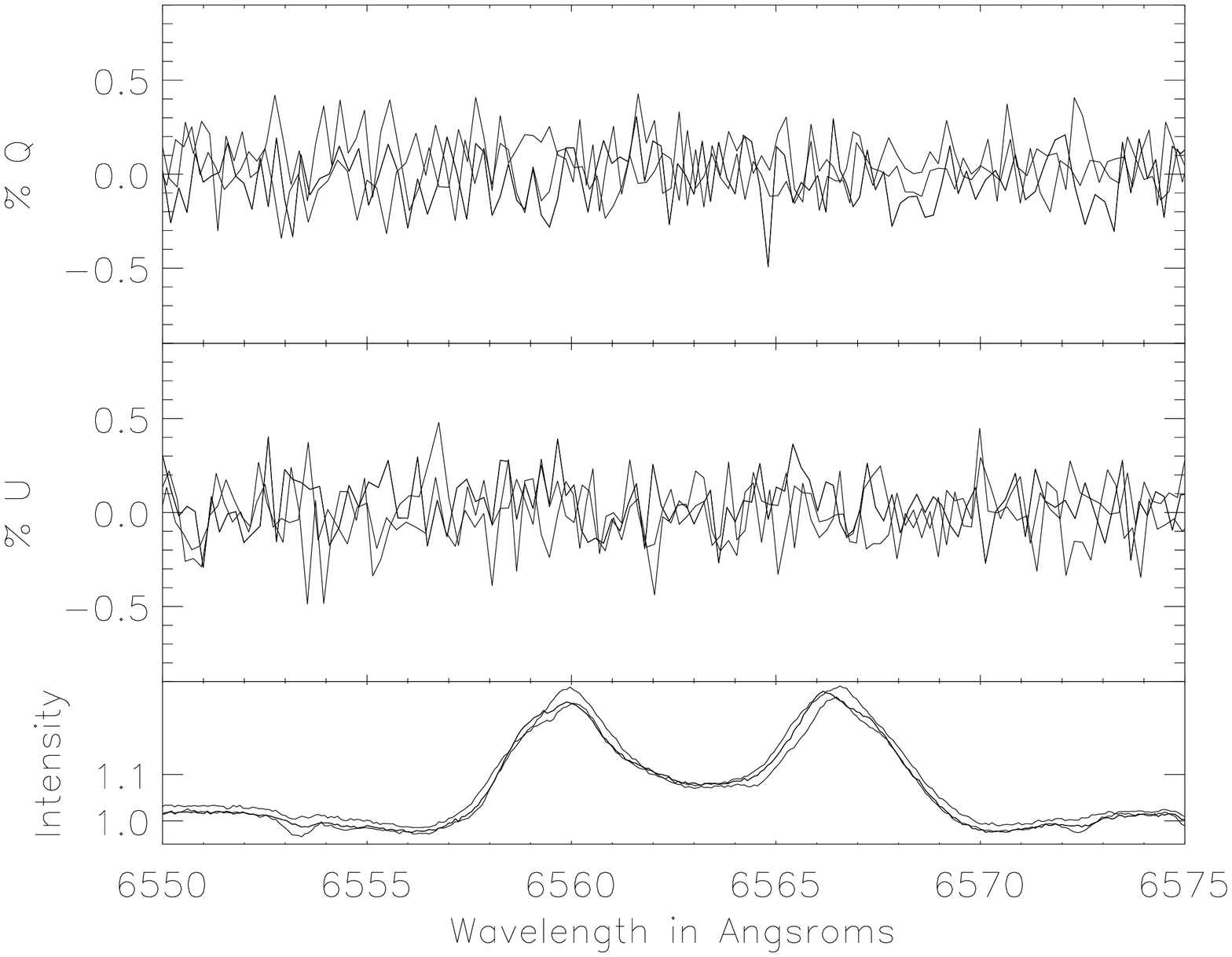}}
\quad
\subfloat[$\phi$ And]{\label{fig:phiand}
\includegraphics[ width=0.21\textwidth, angle=90]{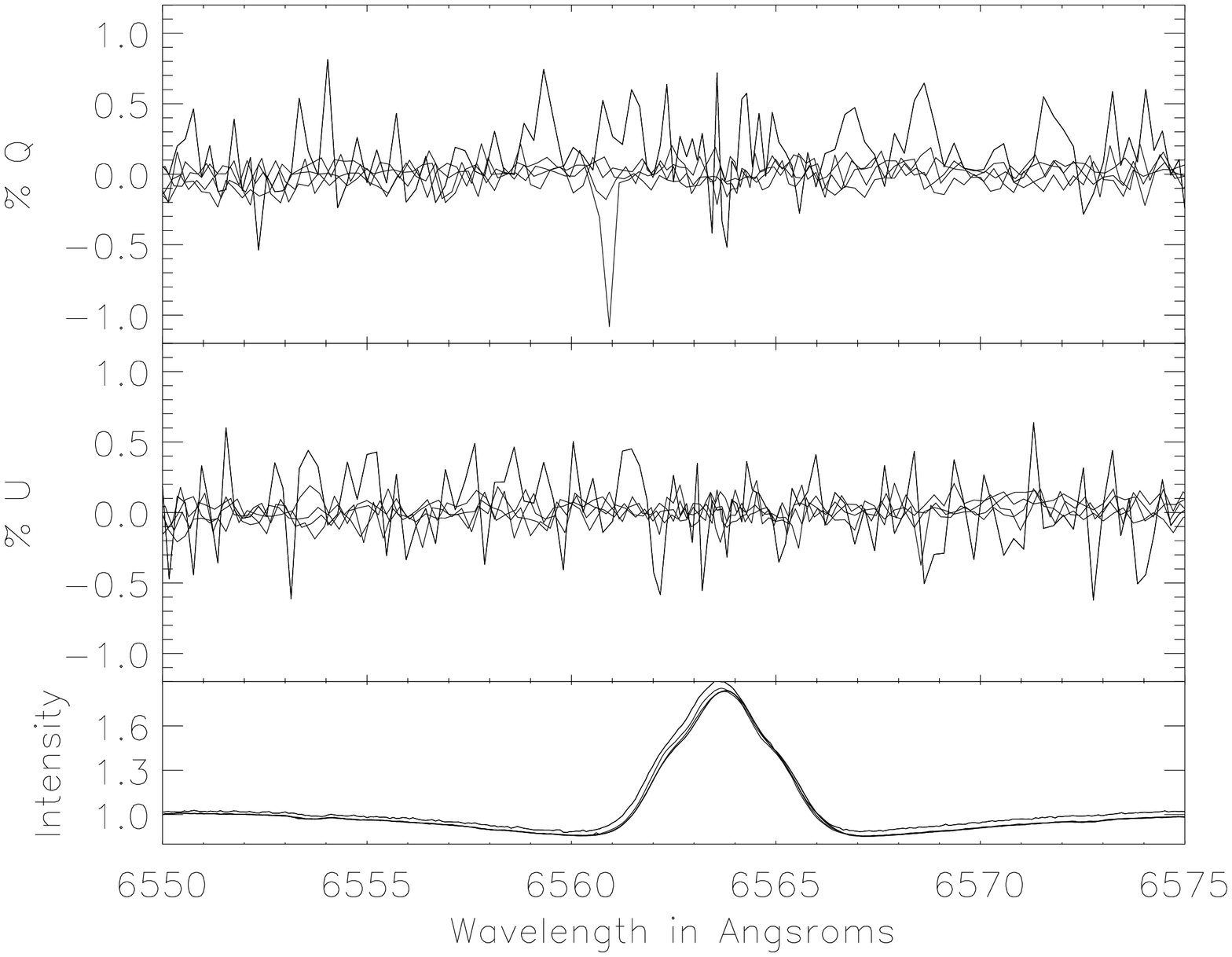}}
\quad
\subfloat[MWC 77]{\label{fig:mwc77}
\includegraphics[ width=0.21\textwidth, angle=90]{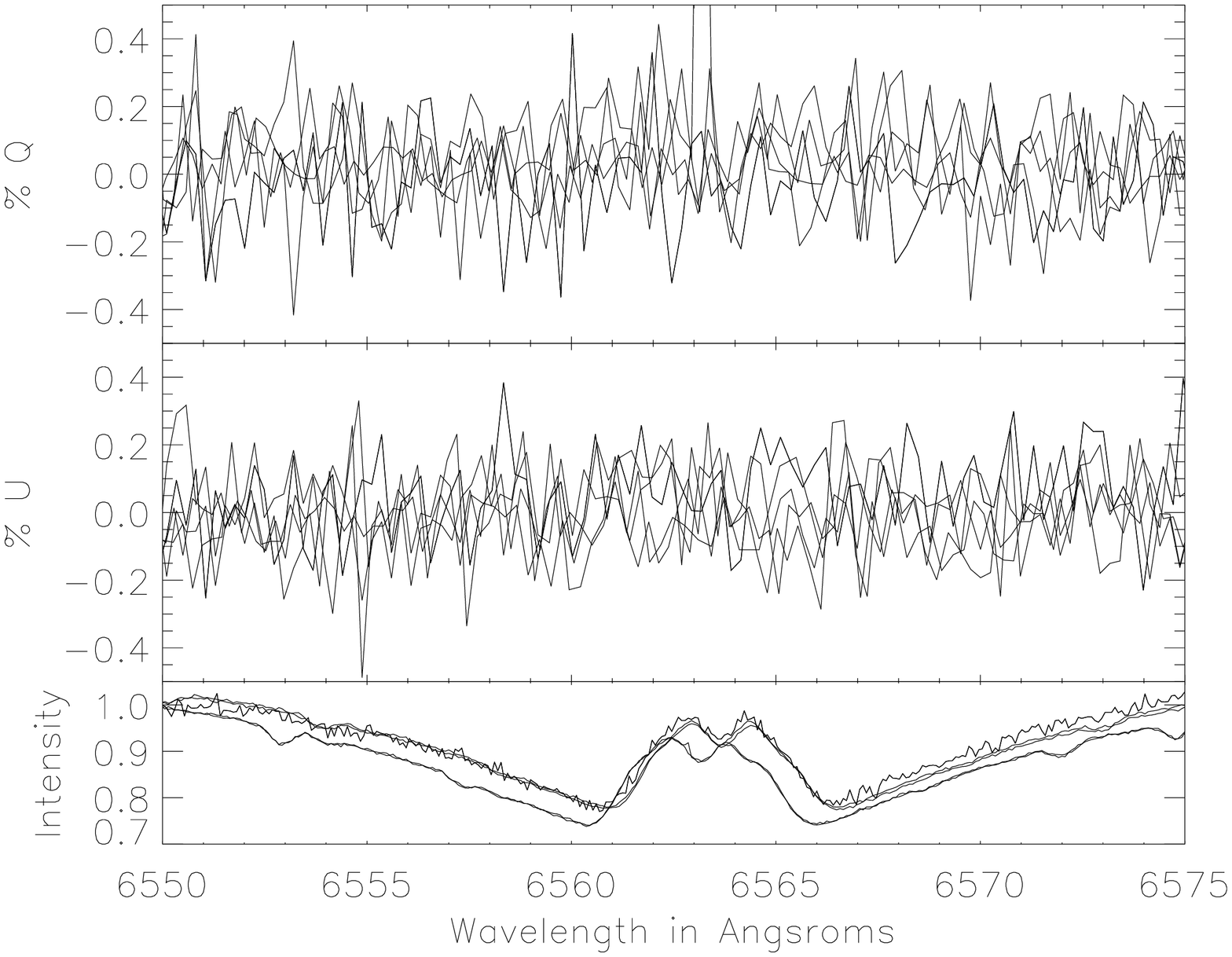}}
\quad
\subfloat[HD 36408]{\label{fig:hd364}
\includegraphics[ width=0.21\textwidth, angle=90]{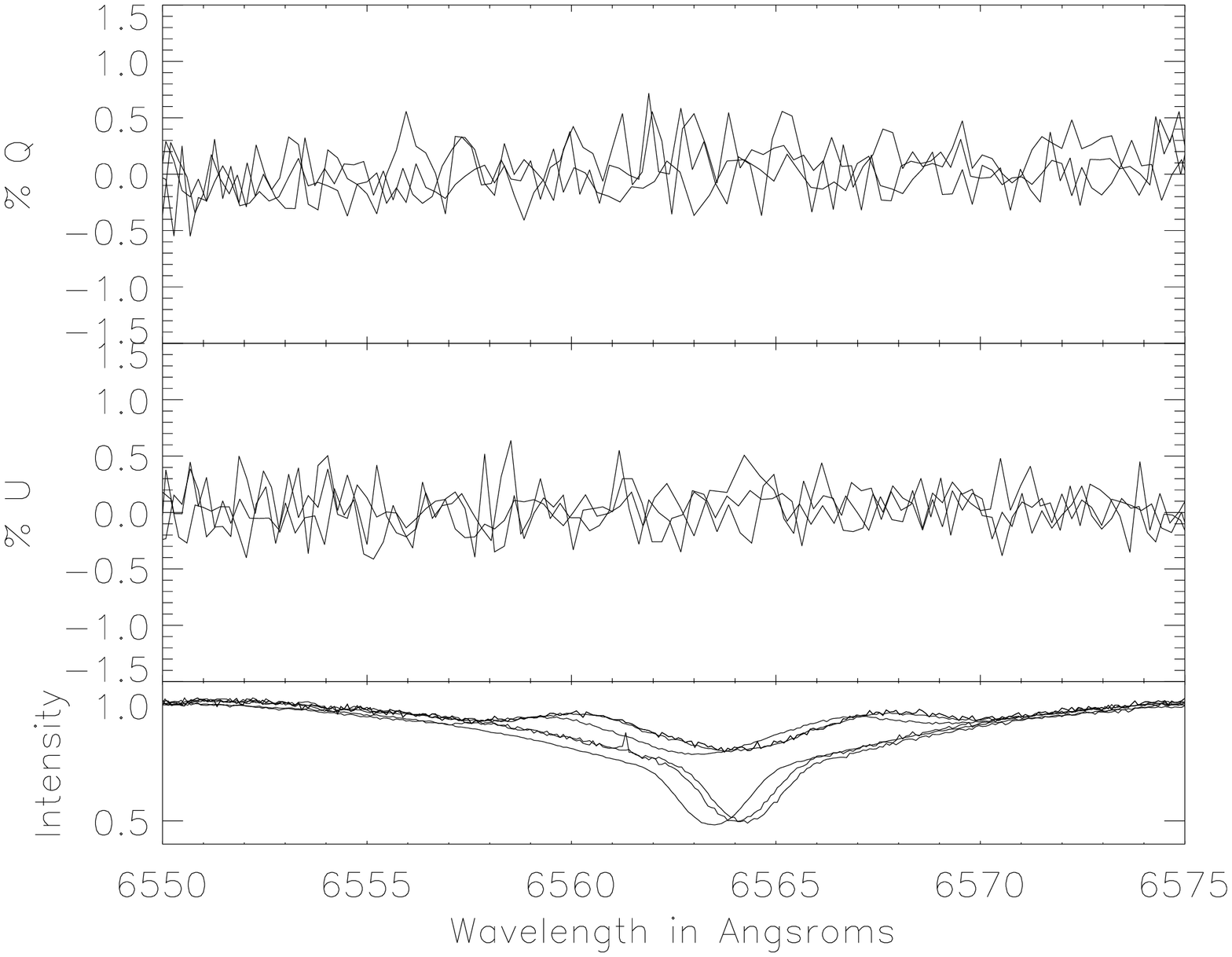}}
\quad
\subfloat[Phecda]{\label{fig:phecda}
\includegraphics[ width=0.21\textwidth, angle=90]{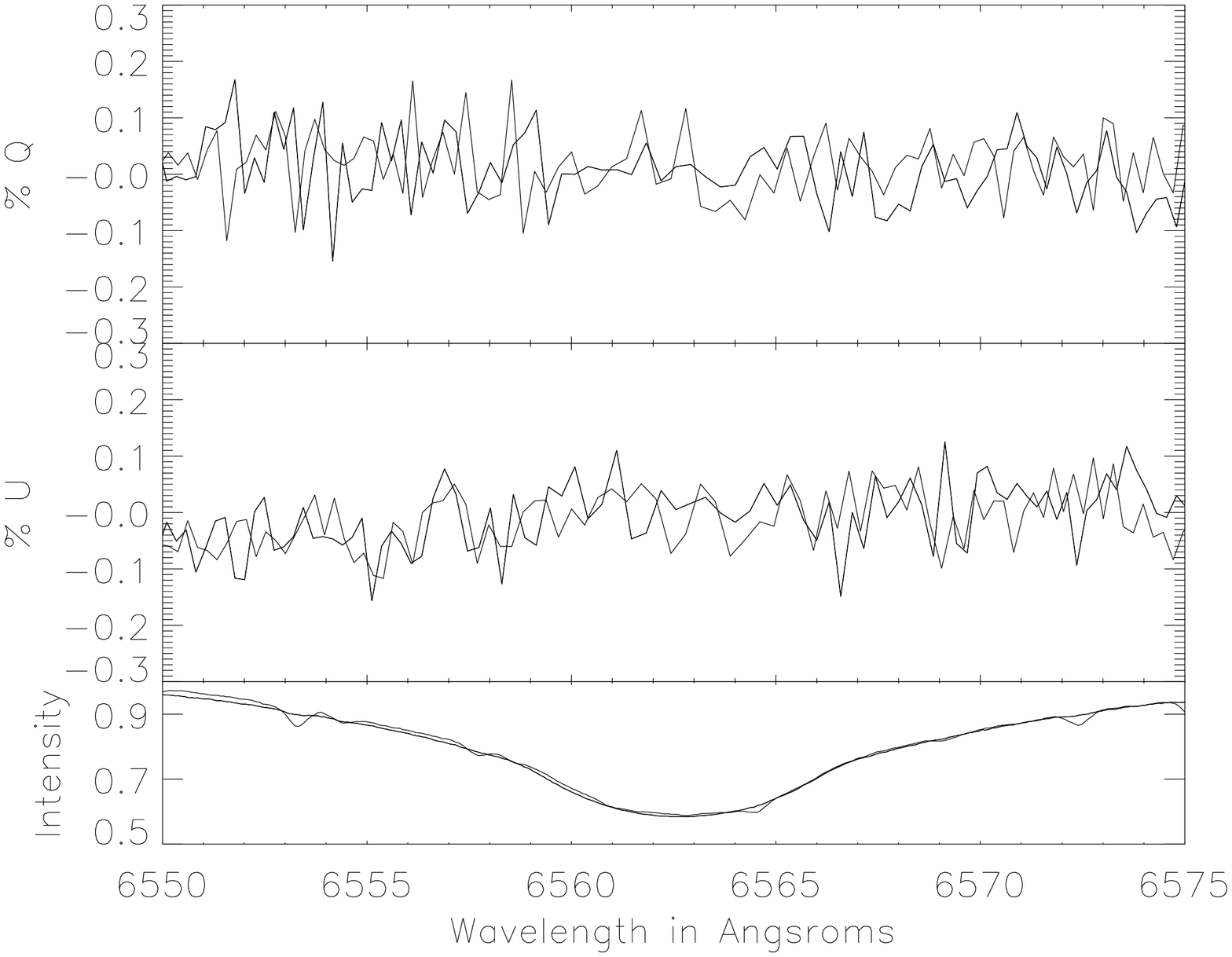}}
\quad
\subfloat[$\lambda$ Cyg]{\label{fig:lamcyg}
\includegraphics[ width=0.21\textwidth, angle=90]{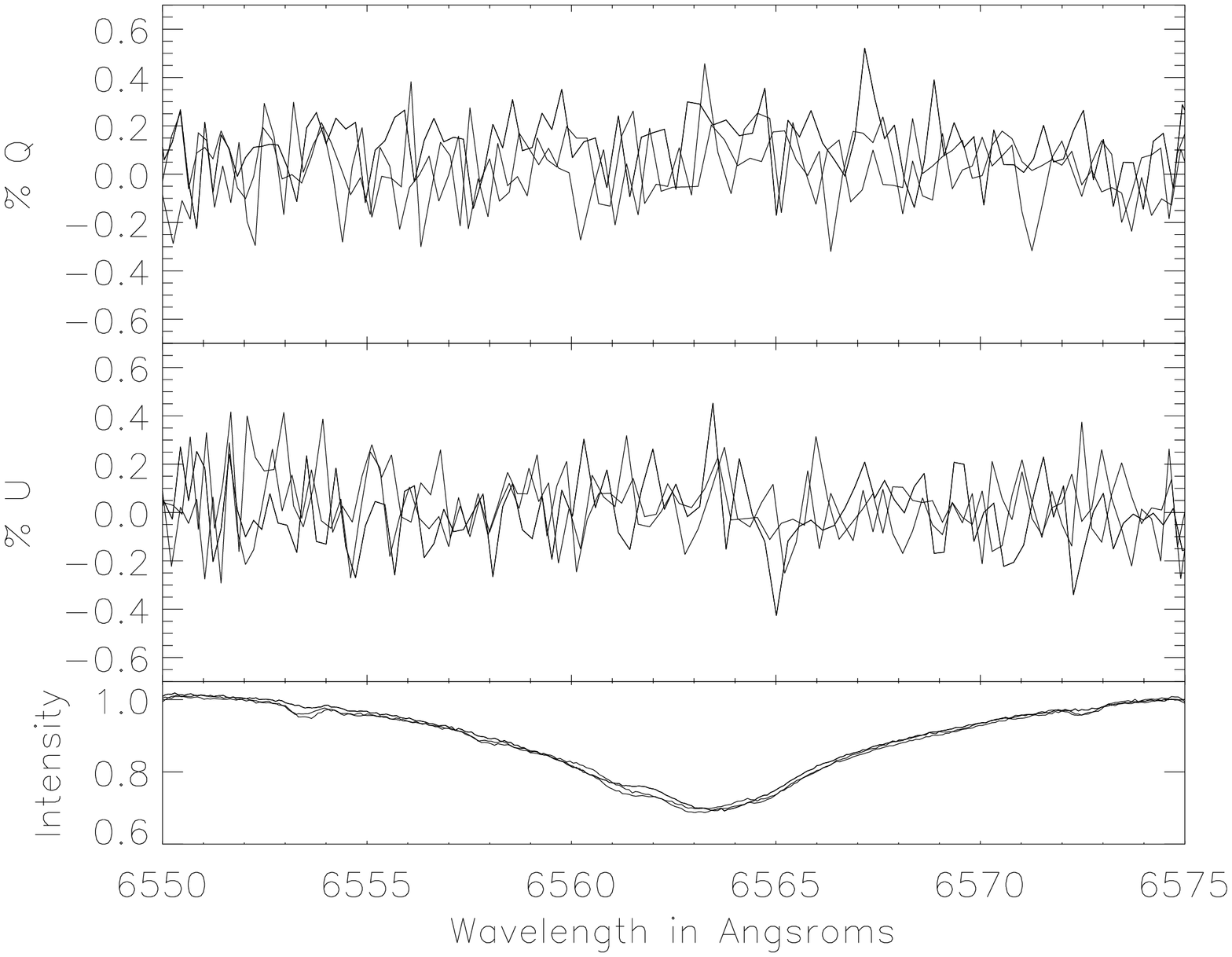}}
\quad
\subfloat[QR Vul]{\label{fig:qrvul}
\includegraphics[ width=0.21\textwidth, angle=90]{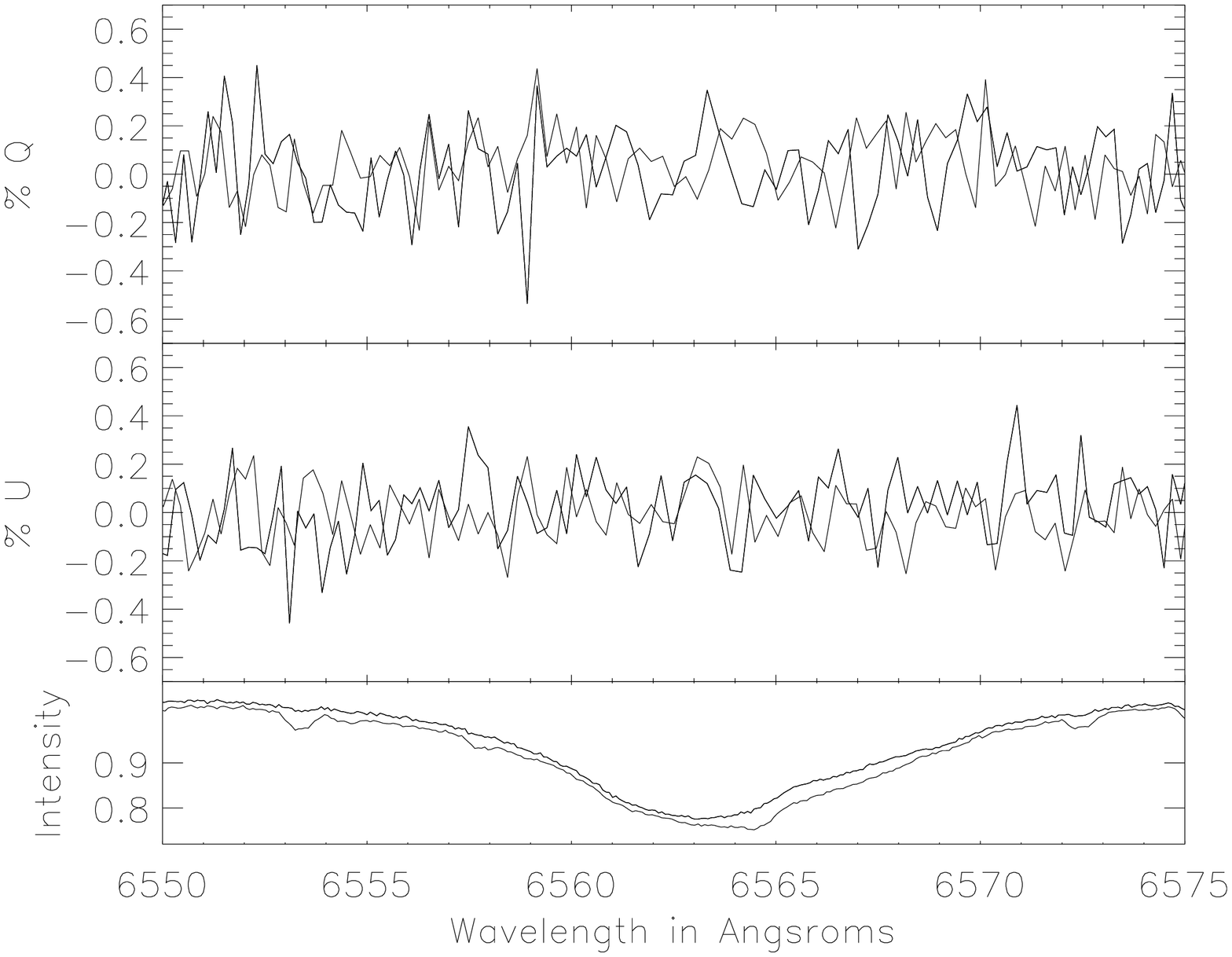}}
\quad
\subfloat[$\xi$ Per]{\label{fig:ksiper}
\includegraphics[ width=0.21\textwidth, angle=90]{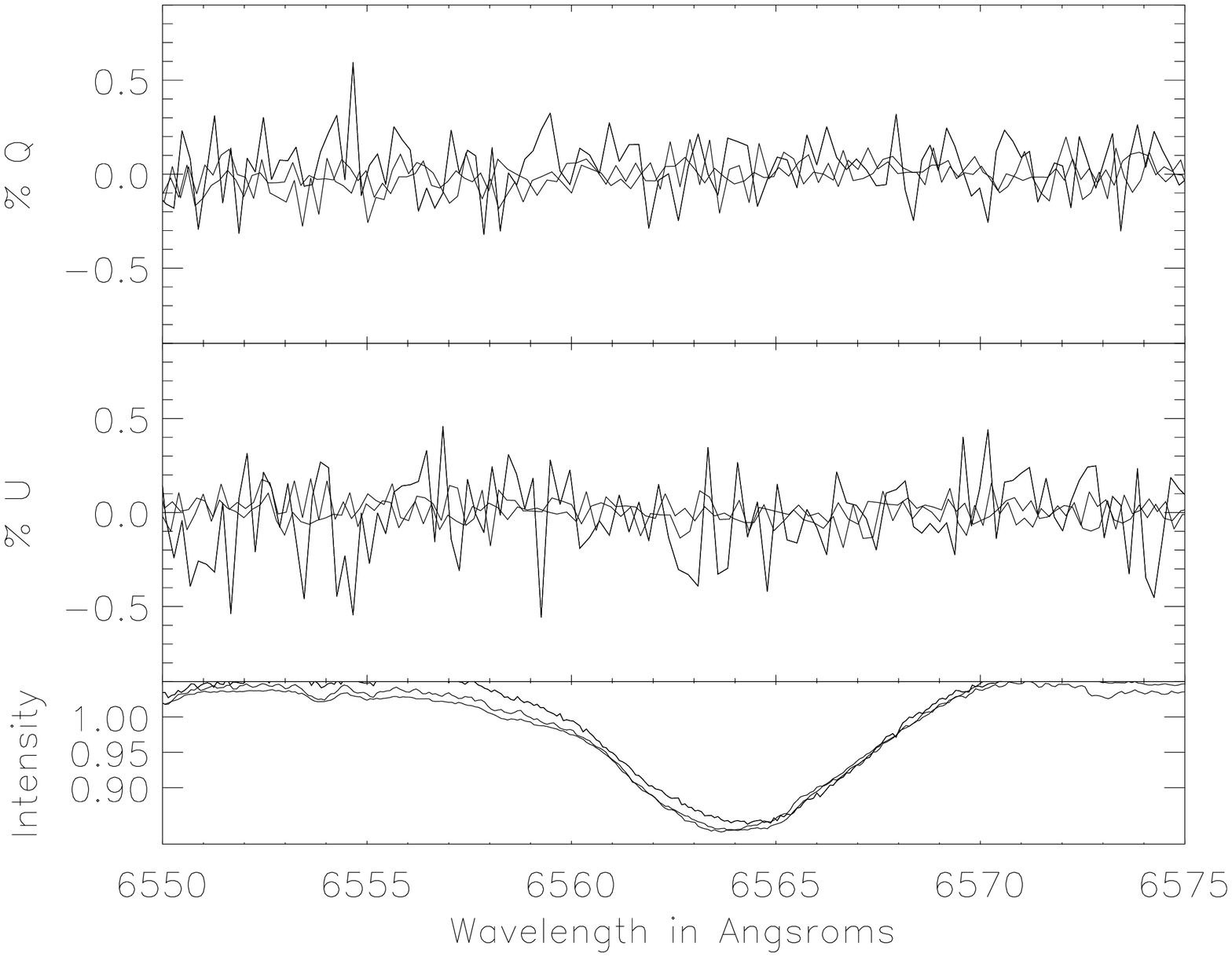}}
\caption[Be Spectropolarimetry II]{Be Spectropolarimetry II}
\label{fig:be-specpol2}
\end{figure}

	The compiled spectropolarimetry is shown in figures \ref{fig:be-specpol1} and \ref{fig:be-specpol2}. The detections are outlined in table \ref{be-res}. There are 30 total targets. Of these, nine stars that show the broad signature, five stars show narrow low-amplitude polarization effects, and the other 16 stars are non-detections at the 0.05\% to 0.1\% level. This detection rate can be broken down into subclasses. The broad signatures range in magnitude from a barely detectable 0.1\% to 1.1\%. All of the stars with detected signatures show evidence of intervening absorption in their line profiles, either by the flattened or notched line centers, or by having a full-blown ``disky" line.

\onecolumn

\twocolumn

	The stars with broad spectropolarimetric effects do show morphological deviations from the typical depolarization profile, but the polarization is quite different from the polarization-in-absorption of the Herbig Ae/Be systems. The five narrow detections do not fit the depolarization description and will be discussed in more depth later.

\begin{table}[!h,!t,!b]
\begin{center}
\caption{Be and Emission-line Star H$_\alpha$ Results \label{be-res}}
\begin{tabular}{lcccc}
\hline        
\hline
{\bf Name}    &{\bf H$_\alpha$}      &  {\bf Line Effect?}      & {\bf Mag}            & ${\bf Type}$                  \\
\hline
\hline
$\zeta$ Tau           & 2.7         & Y                                             &  0.5\%                 &  Disk                            \\
MWC 143               &4.8          & Y                                             &  1.1\%                 &  Disk                            \\
25 Ori                     &2.4         & Y                                            &  0.5\%                  &  Disk                           \\
$\psi$ Per             &5.8          & Y                                             &  0.5\%                 &  Disk                            \\      
10 CMa                  &4.8         & Y                                             &  0.5\%                 &  Disk*                            \\
$\kappa$ CMa      &4.7         & Y                                             &  0.2\%                 &  Disk*                            \\
$\kappa$ Dra      & 4              & Y                                             &  0.2\%                 &  Disk*                            \\
Omi Cas                &6.4          & Y                                             &  0.2\%                 &  Disk*                            \\
Omi Pup                 &2.2          & Y                                             &  0.2\%                 &  Disk*                            \\
$\gamma$ Cas    &4.4          & Y                                             &  0.2\%                 &  Disk*                          \\
\hline
\hline
18 Gem                  &1.5         & Y                                           &   0.2\%                     & Disk                               \\
$\eta$ Tau             &2.2         & Y                                             &  0.15\%                 &  Disk                            \\
$\beta$ CMi           &7.4          & Y                                             &  0.07\%               &  Disk                            \\
$\alpha$ Col          &2.7         & Y                                             & 0.1\%                 &  Disk                             \\
R Pup                      & 7           &  Y                                            & 0.2\%                  &Disk*                            \\
\hline
\hline
31 Peg                   & 6.4         & N                                           &  $<$0.1\%                           &   Emis                            \\
11 Cam                   &7.5         & N                                            &   $<$0.1\%                          &   Emis                         \\
C Per                      &7.0          &  N                                         &     $<$0.1\%                        &  Emis                            \\
12 Vul                    & 1.3          & N                                            &   $<$0.1\%                         &   Disk                              \\
$\omega$ Ori        &2.2          &  N                                           &  $<$0.1\%                         &   Disk                            \\
66 Oph                   &2.1          & N                                           &    $<$0.1\%                           &   Disk                          \\
MWC 77                &0.8          & N                                           &     $<$0.1\%                          & Disk**                       \\
HD 36408             &0.8          & N                                            &    $<$0.1\%                           & Disk**                        \\
QR Vul                  & 0.75         & N                                          &     $<$0.1\%                        &   Abs                             \\
$\lambda$ Cyg     & 0.7         & N                                           &     $<$0.1\%                         &   Abs                             \\
$\xi$ Per                 &0.85       & N                                            &     $<$0.1\%                          &  Abs                           \\
Phecda                  & 0.6          & N                                            &    $<$0.1\%                            & Abs                            \\
MWC 92                  &1.35       &  N                                           &     $<$0.1\%                         &  Other                            \\
$\kappa$ Cas       &1.5           &    N                                          &    $<$0.1\%                          &  Other                            \\ 
\hline
\hline
\end{tabular}
\end{center}
The columns show the name, H$_\alpha$ line strengths, presence of a detectable spectropolarimetric line effect (Line Effect?), magnitude of the polarization effect (Mag) and H$_\alpha$ line morphology (Type). Where there were no detections, upper limits are typically 0.05\% to 0.1\% depending on observing conditions and binning. The top portion shows the 10 broad-signature spectropolarimetric detections. The middle portion shows the 5 smaller, more complex detections. The lower portion shows the non-detections.
\end{table}

\subsection{$\gamma$ Cas - A Clear Example}

	$\gamma$ Cas is a very well-studied target that has an extended envelope and a strong H$_\alpha$ emission line. Quirrenbach et al. 1993 resolved the H$_\alpha$ emission region using interferrometry. An elliptical Gaussian model fits the observations with an axial ratio of 0.74 and an angular FWHM of 3.2mas. This flattened envelope is exactly the kind imagined in the original studies on continuum polarization from circumstellar envelopes.

\begin{figure} [!h]
\centering
\subfloat[$\gamma$ Cas Spectropolarimetry]{\label{fig:be-broad-gammacas}
\includegraphics[width=0.26\textwidth, angle=90]{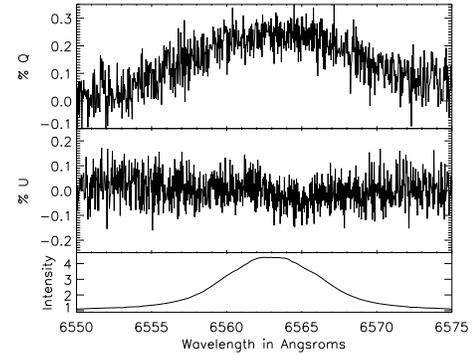}}
\quad
\subfloat[$\gamma$ Cas qu Plot]{\label{fig:be-broad-gammacasqu}
\includegraphics[width=0.26\textwidth, angle=90]{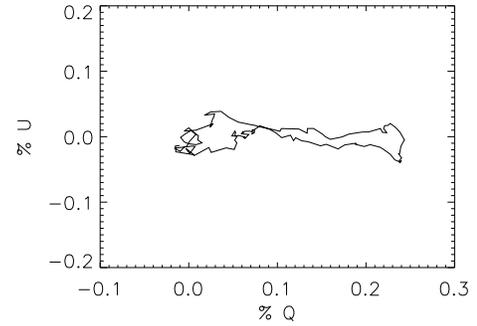}}
\caption[$\gamma$ Cas Broad-Effect Example]{{\bf a)} The average $\gamma$ Cas Spectropolarimetry at full resolution. All observations have been rotated to maximize Stokes q in line-center then averaged {\bf b)} the corresponding qu plot - a linear extension. The +q orientation is an artifact of the arbitrary alignment.}  
\label{fig:bebroadex}
\end{figure}

	$\gamma$ Cas was monitored quite heavily over a year and can use this target as a clear example for the entire class. This star did not show very significant H$_\alpha$ line-profile or spectropolarimetric variations throughout the monitoring period. All the spectropolarimetric observations were taken without binning and aligned in qu-space to maximize Stokes q in the central 50 pixels. These polarized spectra all had similar morphology but slightly varying magnitudes because of the varying telescope polarization properties at the many different telescope pointings used. The resulting 22 spectra were then averaged to make a single polarized spectra with only 0.07\% rms error at full spectral resolution (no bin-by-flux was applied to the observations). Figure \ref{fig:be-broad-gammacas} shows the spectropolarimetry. The polarization signature is almost entirely in the +q spectrum. Though the spectra have been arbitrarily rotated to maximize +q, this figure shows that the signature is almost entirely linear in qu-space. In the qu-loop plot, this polarization change becomes a simple horizontal excursion, as shown in figure \ref{fig:be-broad-gammacasqu}. The polarization change is roughly 0.3\% and is morphologically broader than the line-profile itself. There is a very small width to the extension which comes from a very slight variation in the u spectrum on the red side of the emission line, as seen in the middle panel of figure \ref{fig:be-broad-gammacas}. A decrease in polarization of the same 0.2\% magnitude had been reported in commissioning runs with the William-Wehlau spectropolarimeter (Eversberg et al. 1998), though at lower spectral resolution.
	
	Since the signature is broad and is a single linear excursion in qu-space, this star fits well with the traditional ``depolarization"  effect from scattering theory. The change in polarization is smaller in magnitude than the known intrinsic continuum polarization. The initial polarization studies all found a similar depolarization in continuum polarizations of: 1.0\% McLean \& Brown 1978, 0.8\% McLean 1979, 0.5\% in Poeckert \& Marlborough 1977.

\subsection{Clear Detections}

	The other nine targets with broad detections show very similar ``depolarization" morphology. All of them show broad changes that extend to the wings of the line. However, a more detailed examination is necessary. The example of $\gamma$ Cas illustrates the shape of these broad signatures, but the overall magnitude and morphology varies between the targets. For example, MWC 143 has a very large 1.1\% deviation (0.9\% q, 0.6\% u) but Omi Pup is barely detectable at 0.1\%. However, the main point that must be emphasized is that, regardless of the presence or type of absorption overlying the H$_\alpha$ emission, there is a very clear morphological difference between these stars and Herbig Ae/Be stars. The  ``polarization-in-absorption" that we've seen in the Herbig Ae/Be systems is simply not present. Instead, there is a very broad signature apparent in 10 of 30 stars with many showing some additional morphological features. Another 5 show more complex signatures worth describing in detail.

	There are some morphological differences that are worth mentioning specifically. In 25 Ori, the change in Stokes q is ``flat-topped" across the entire line center. The star $\zeta$ Tau shows a noticeable asymmetry, being larger in the blue side of the line. There is a small, narrow asymmetric spectropolarimetric signature in the absorptive part of the $\psi$ Per ``disky" profile. On the blue side of the central absorption there is a relative decrease with respect to the broad signature and a relative increase on the red side of the absorption. These morphological effects show that there is more to be learned about circumstellar environment. If the depolarization mechanism is the correct model to use, there is clearly a need for more detailed models for systems. Other spectropolarimetric studies of these stars have been done as well. $\omega$ Ori was a non-detection in Oudmaijer et al. 1999 with a continuum polarization of 0.30\% and a non-detection in Vink et al. 2002 with a continuum of 0.27\%. It should be noted that these effects extend to the other hydrogen lines. For instance, Oudmaijer et al. 2005 report a broad 0.4\% change across Pa$_\beta$ at 1.28$\mu$m in $\zeta$ Tau. This is only moderately less than the signature observed in H$_\alpha$.

	Before comparing these results with other systems, some interferrometric results must be reviewed. $\zeta$ Tau is a very well-studied target. Quirrenbach et al. 1994 resolved the H$_\alpha$ emission region using interferrometry and created a maximum-entropy map of the circumstellar region. An elliptical Gaussian model fits the observations with an axial ratio of 0.30 and an angular FWHM of 3.55mas. The map shows a highly elongated structure that was interpreted as a near edge-on disk. In a later interferrometric and spectropolarimetric survey, Quirrenbach et al. 1997 report resolving H$_\alpha$ emission regions in $\gamma$ Cas, $\phi$ Per, $\eta$ Tau, $\zeta$ Tau, and $\beta$ CMi with angular sizes of 1.5 to 3.5mas. These measurements show that the H$_\alpha$ emission is quite extended and is much larger than the stellar-radii scale. All of the stars were point-sources in the nearby continuum interferrometric measurements. In light of these observations, the depolarization effect from extended and less-scattered H$_\alpha$ emission is extremely likely. However, the simple models used to date are not enough to fit the morphologies observed here completely.

	Other stars show smaller, more complex signatures. There were five stars with detected spectropolarimetric signatures of very small magnitude: 18 Gem, $\alpha$ Col, $\beta$ CMi, R Pup, and $\eta$ Tau. There were a number of 18 Gem and $\eta$ Tau observations that were non-detections, as seen in figures \ref{fig:18gem} and \ref{fig:swap-rebinetatau}, but the signal-to-noise was not as good. The 18 Gem signature is a small antisymmetric change in Stokes u across the center of the line, as seen in figure \ref{fig:be-broad-18gem}. The signature for $\eta$ Tau is a small drop in Stokes u across line-center, as seen in figure \ref{fig:be-broad-etata}. In $\alpha$ Col and $\beta$ CMI, the detection was antisymmetric, spanned the line and was stable over two nights observations at the same telescope pointing. Both nights of data are shown in figures \ref{fig:be-broad-alfco} and \ref{fig:be-broad-betcm}. R Pup showed a small drop in q across the emission blue side of the simple emission line. 
	
	All of these detections do not fit in to the ``broad" morphology of the typical Be depolarization signature. The detections in $\alpha$ Col and $\beta$ CMi are extremely small, having an amplitude of $<$0.1\% with antisymmetric components. However, they do not fit in to the scheme of the disk-scattering theory either. The signatures are not double-peaked and symmetric and are not significantly wider than the H$_\alpha$ line. All that can be conclude is that more detailed modeling and investigation is necessary.

\subsection{3 Pup}

	3 Pup (MWC 570) is an emission line star of spectral type A3Iab (Simbad) with a classic ``disky" H$_\alpha$ line. This star provides yet another comparison between the different star types as it has an A spectral type and a distinctly different spectropolarimetric result. There is good quality archival ESPaDOnS data from February 7th and 8th 2006 showing a very large complex H$_\alpha$ signature. The polarization change, shown in figure \ref{fig:swp-3pup-esp-iqu} is centered on the ``disky" emission line but spans the entire width of the line. The polarization in the two emissive peaks is more than 0.5\% increasing to over 2\% in the central absorption.

\onecolumn

\begin{figure}
\centering
\subfloat[10 CMa]{\label{fig:be-broad-10cma}
\includegraphics[width=0.21\textwidth, angle=90]{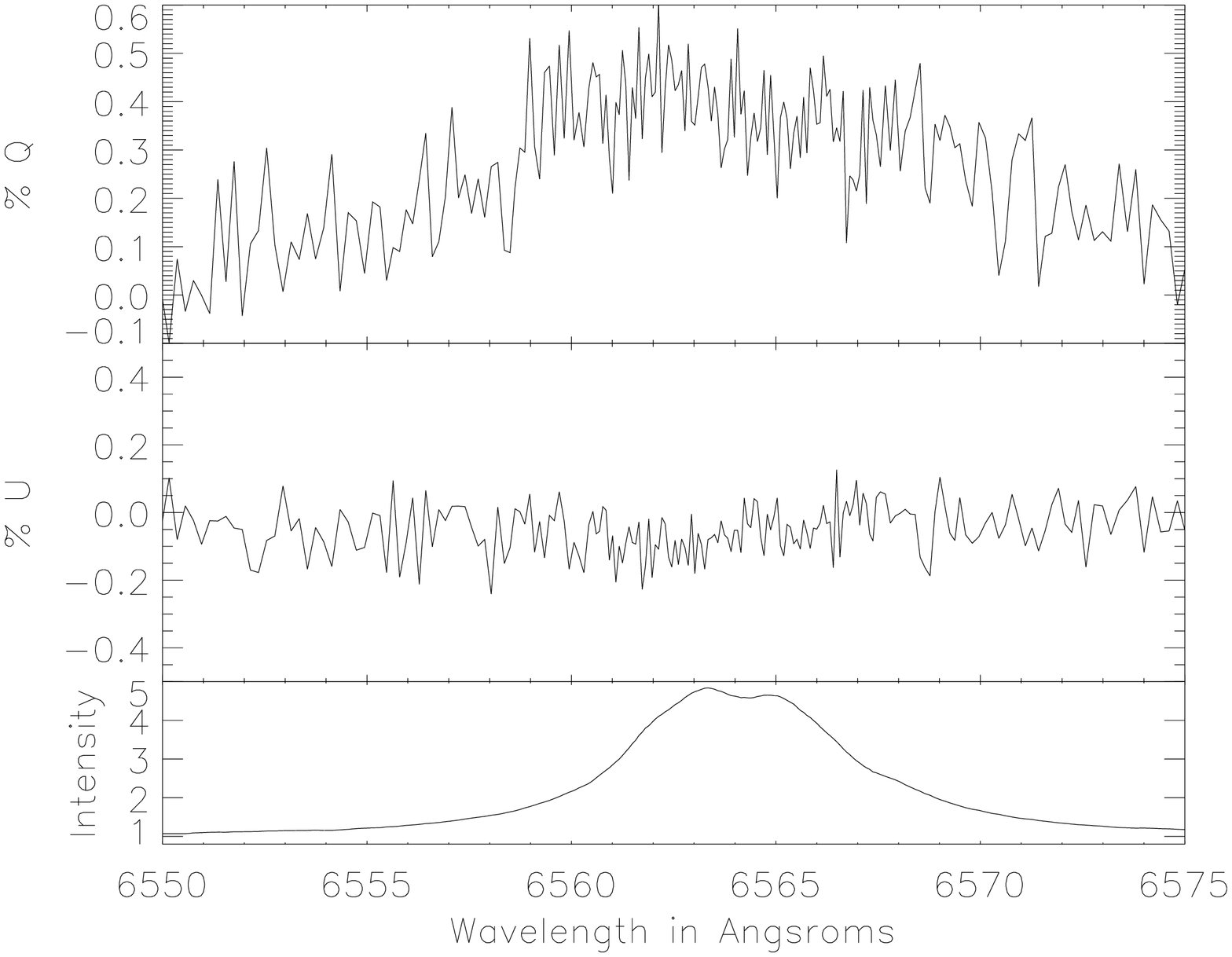}}
\quad
\subfloat[25 Ori]{\label{fig:be-broad-25ori}
\includegraphics[width=0.21\textwidth, angle=90]{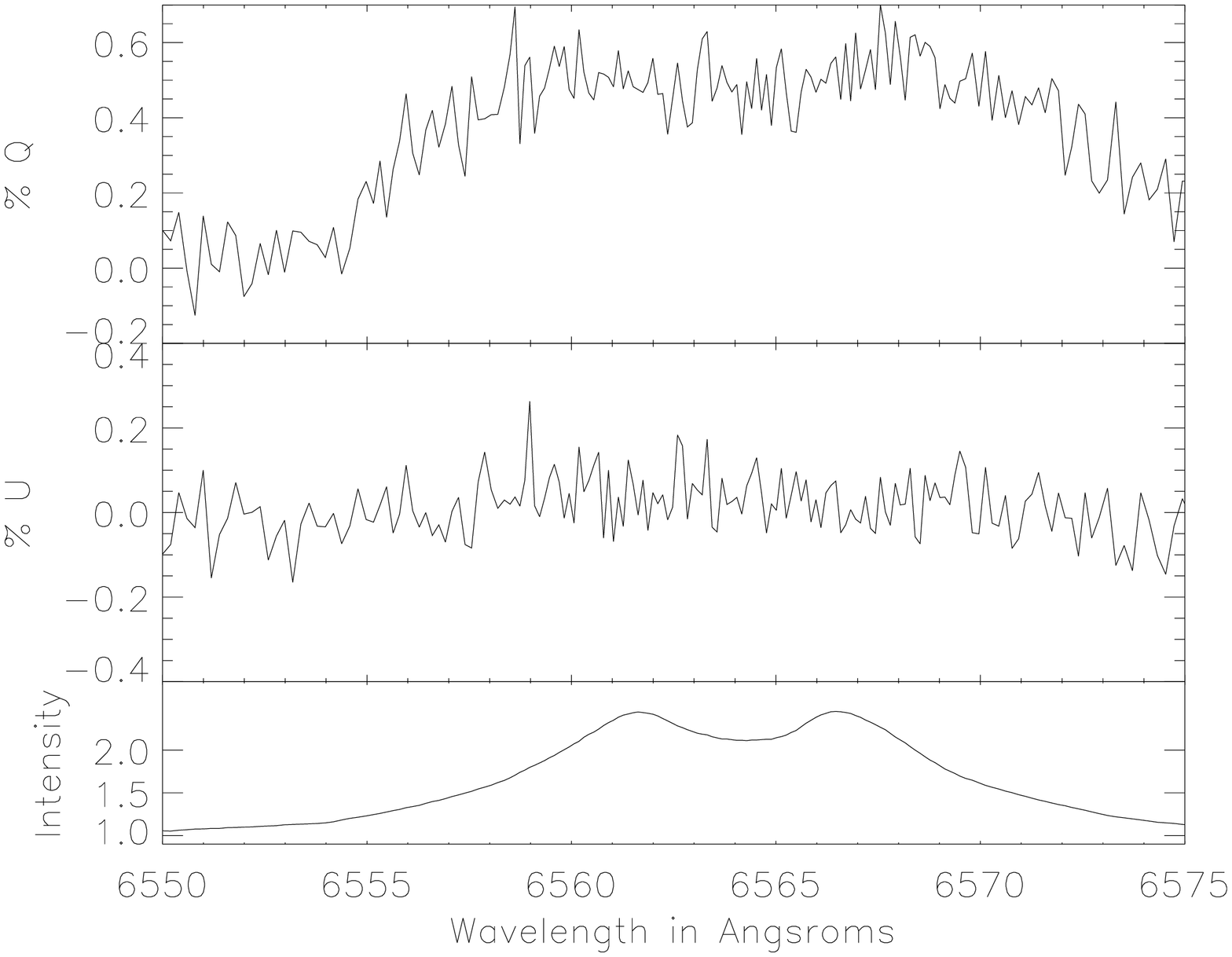}}
\quad
\subfloat[$\zeta$ Tau]{\label{fig:be-broad-zetat}
\includegraphics[width=0.21\textwidth, angle=90]{figs-swap-indiv-indivswap-rebin-zetatau.eps}}
\quad
\subfloat[$\psi$ Per]{\label{fig:be-broad-psiper}
\includegraphics[width=0.21\textwidth, angle=90]{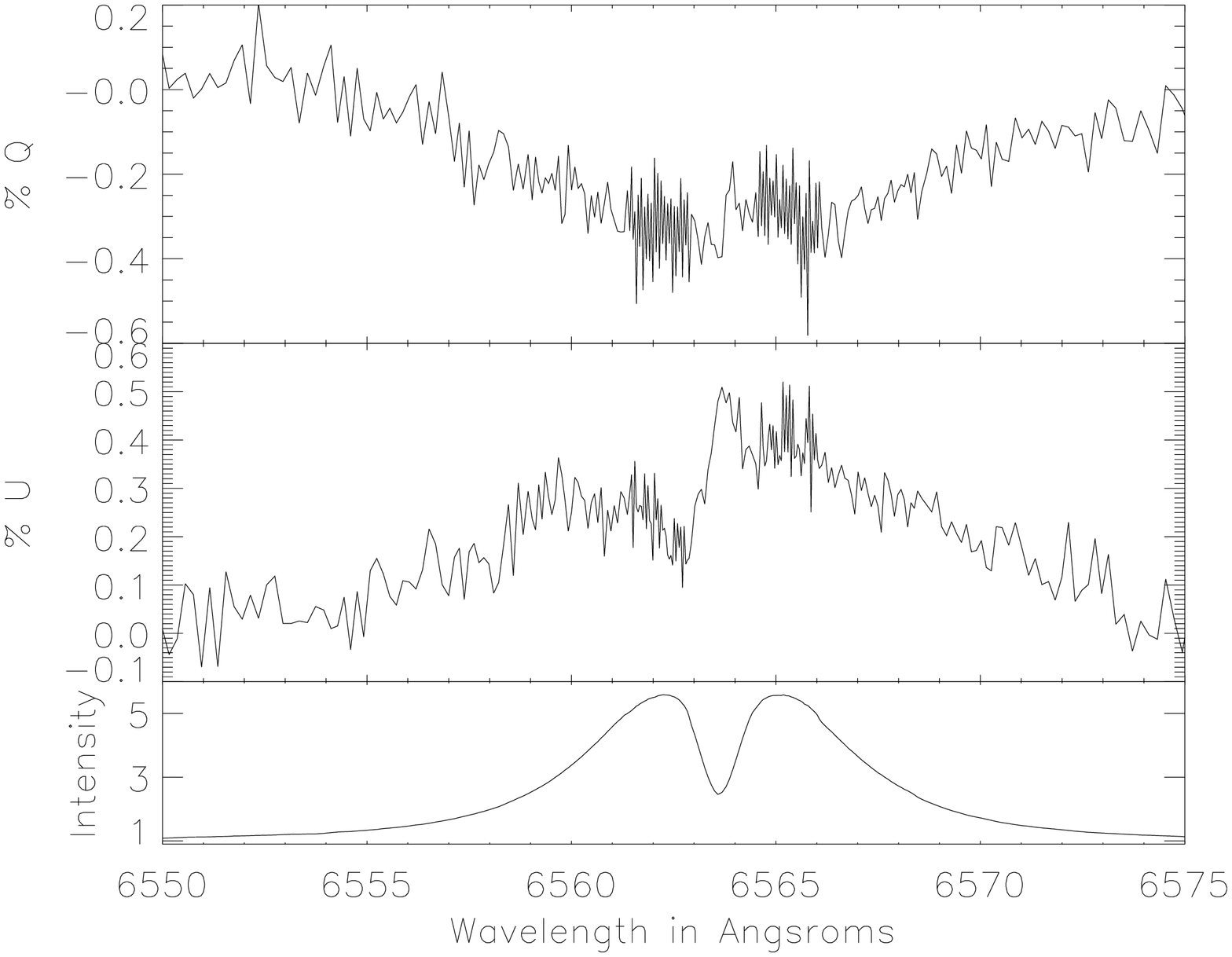}}
\quad
\subfloat[MWC 143]{\label{fig:be-broad-mwc143}
\includegraphics[width=0.21\textwidth, angle=90]{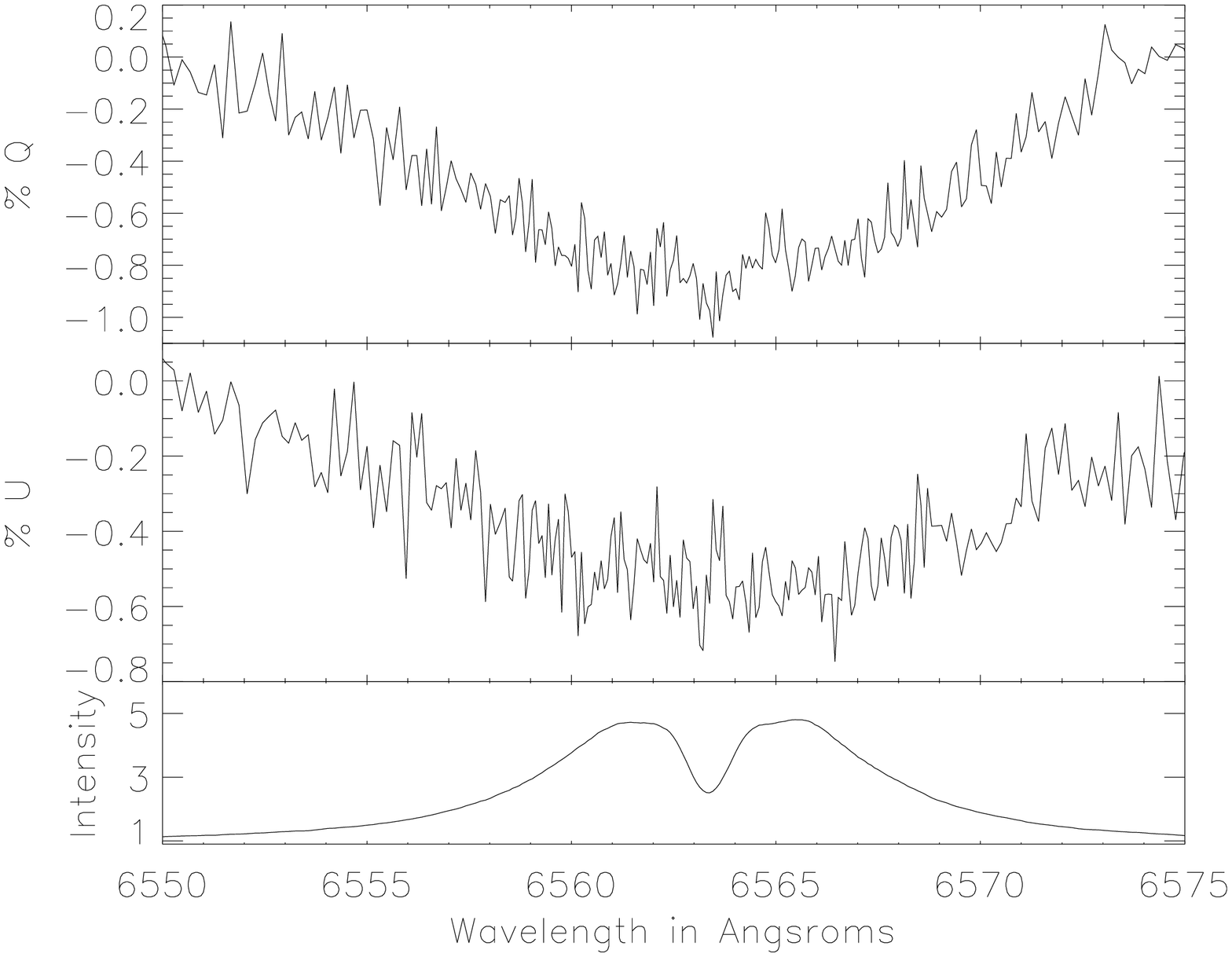}}
\quad
\subfloat[Omi Cas]{\label{fig:be-broad-omics}
\includegraphics[width=0.21\textwidth, angle=90]{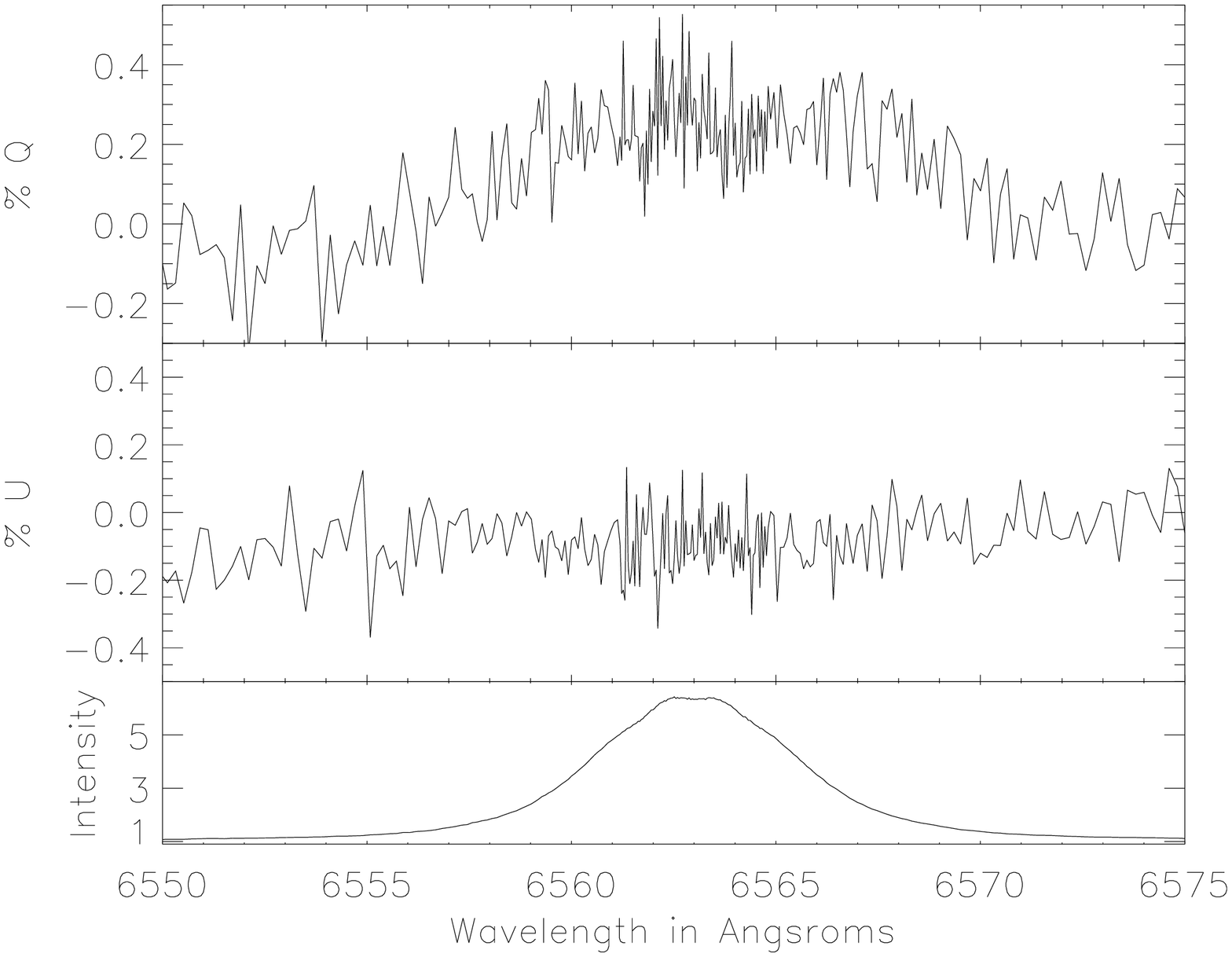}}
\quad
\subfloat[Omi Pup]{\label{fig:be-broad-ompup}
\includegraphics[width=0.21\textwidth, angle=90]{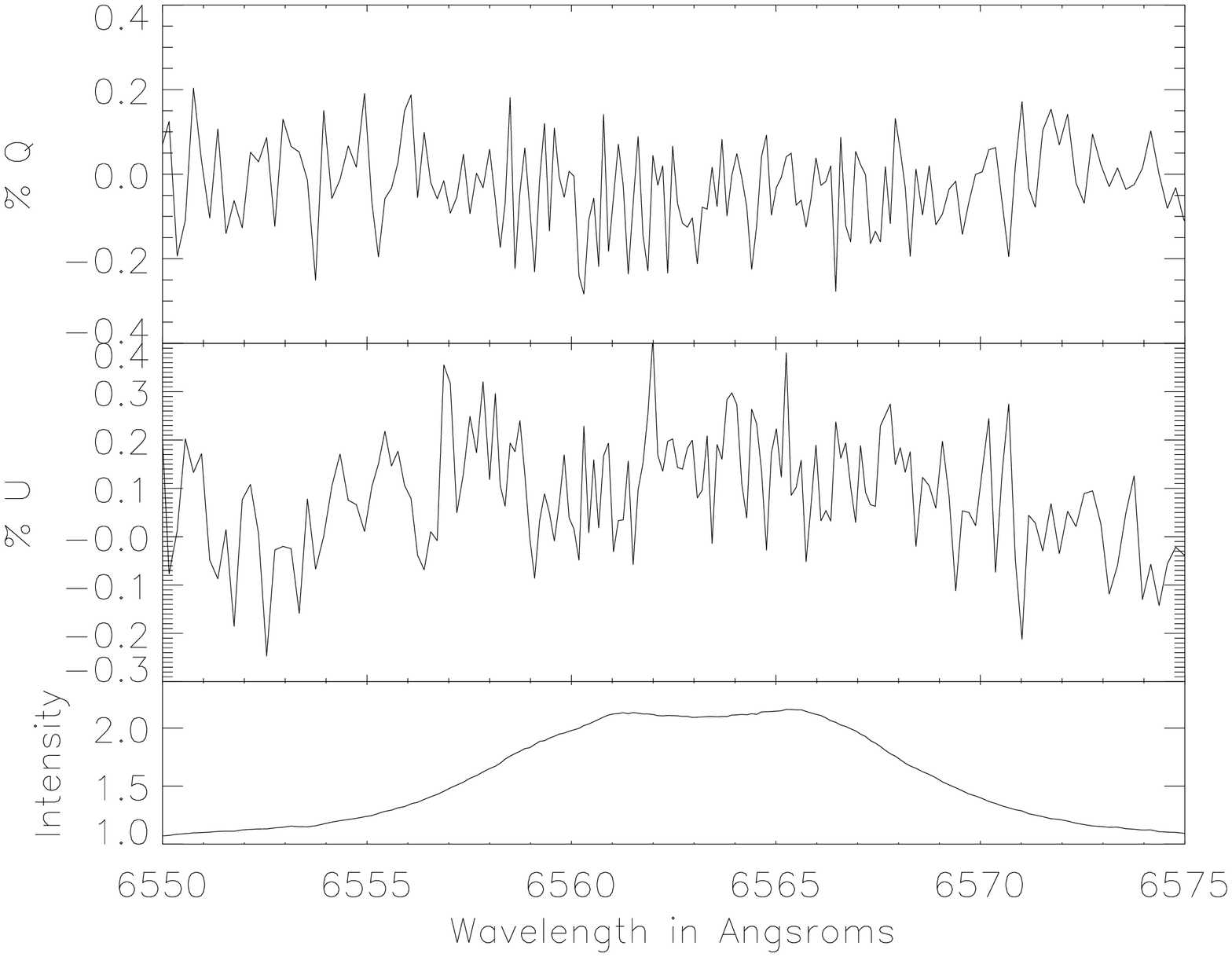}}
\quad
\subfloat[$\kappa$ Dra]{\label{fig:be-broad-kapdra}
\includegraphics[width=0.21\textwidth, angle=90]{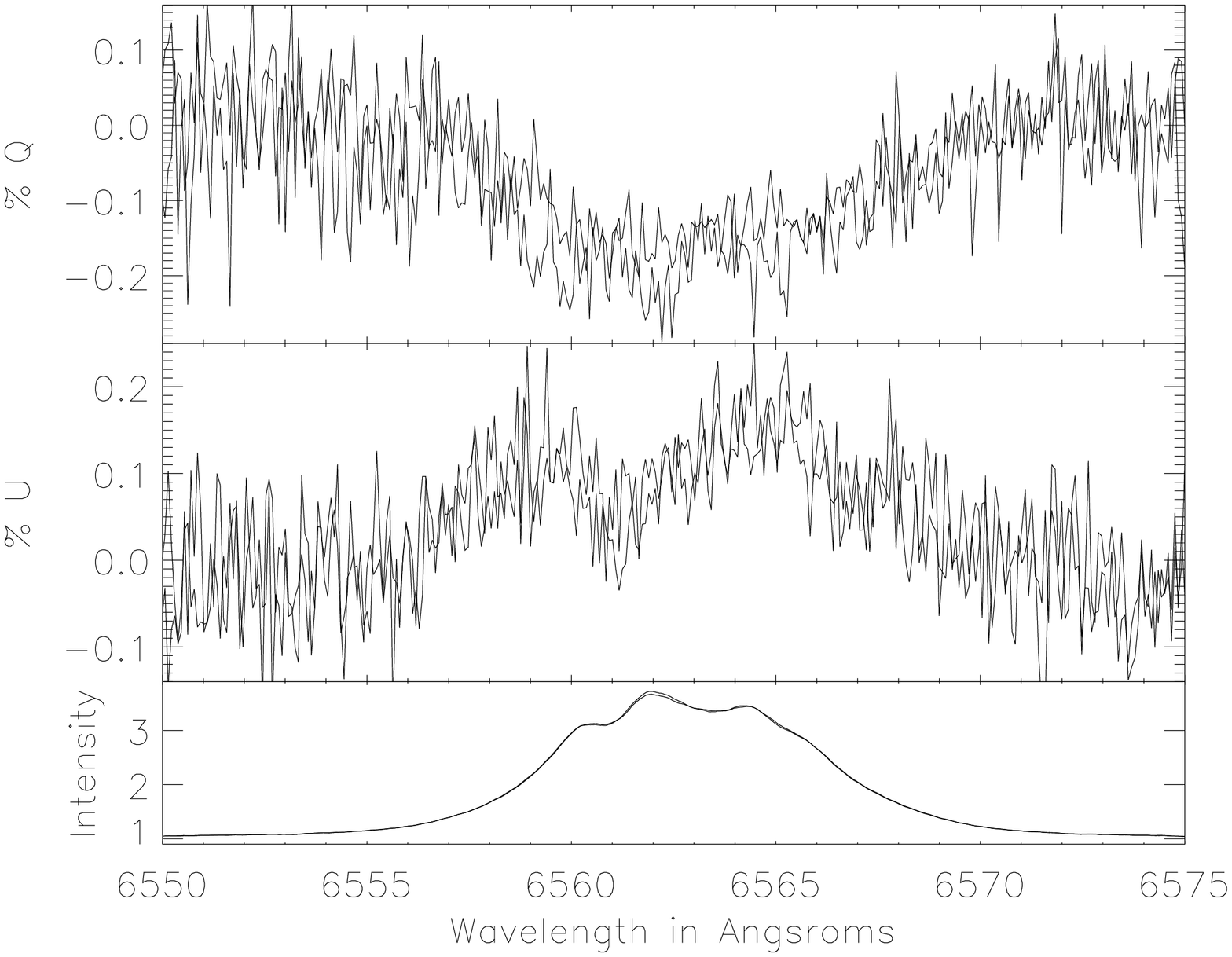}}
\quad
\subfloat[$\kappa$ CMa]{\label{fig:be-broad-kapcm}
\includegraphics[width=0.21\textwidth, angle=90]{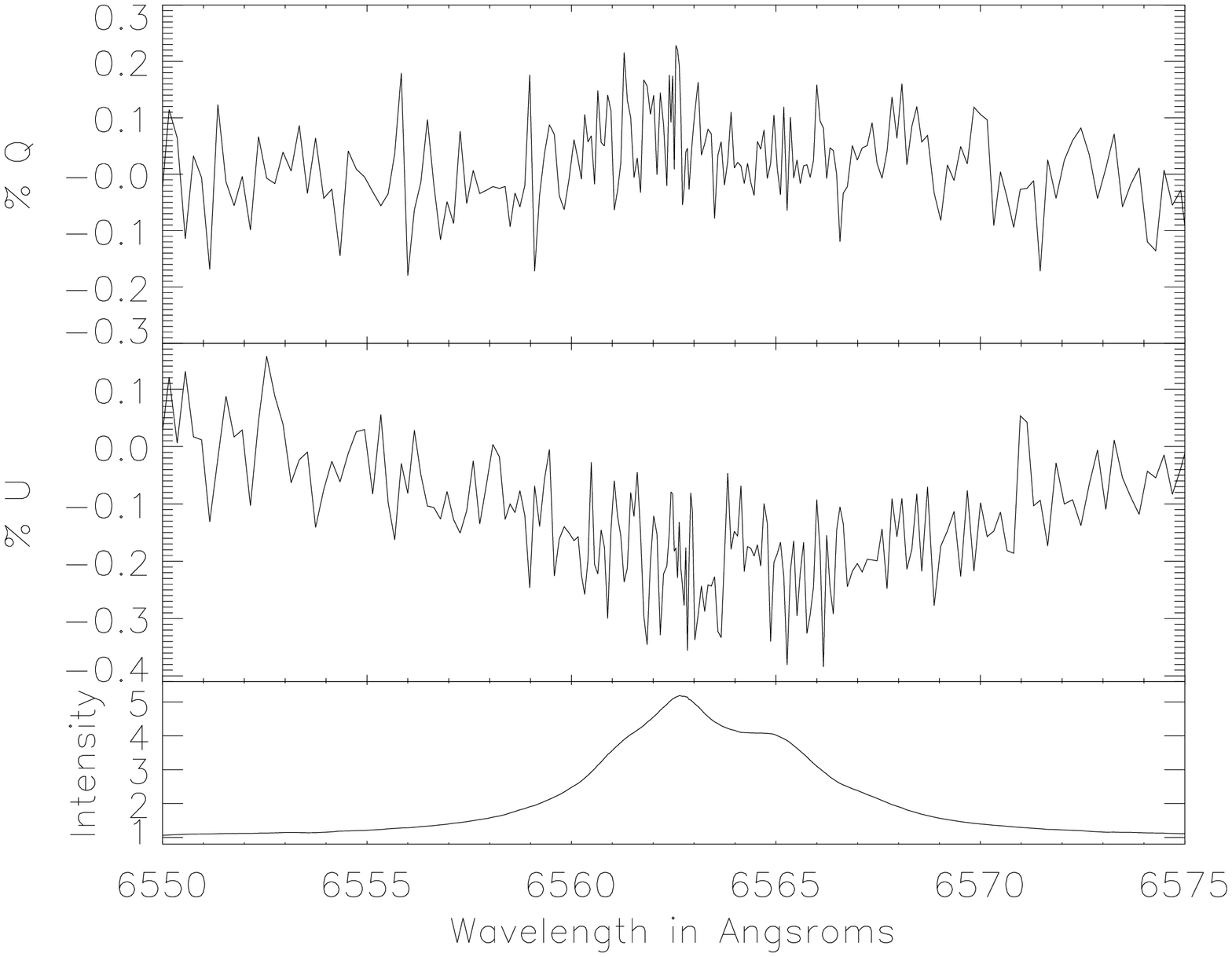}}
\caption[Broad Be Effects II]{Individual examples of HiVIS detections for 9 of the 10 stars with broad spectropolarimetric signatures. $\gamma$ Cas is shown in the previous figure. The individual detections are: {\bf a)}10 CMa  {\bf b)} 25 Ori.  {\bf c)} $\zeta$ Tau {\bf d)} $\psi$ Per {\bf e)} MWC 143  {\bf f)} Omi Cas.  {\bf g)} Omi Pup {\bf h)} $\kappa$ Dra {\bf i)} $\kappa$ CMa. The detections are all more broad than the H$_\alpha$ line itself. $\psi$ Per and $\kappa$ Dra show fairly strong morphological changes across the line and there is significant evidence for absorption. Omi Pup and 25 Ori are severely "flat-topped". $\zeta$ Tau is highly asymmetric with the strongest polarization change on the blue side of the emission line.}  
\label{fig:bebroad}
\end{figure}

\begin{figure}
\centering
\subfloat[18 Gem]{\label{fig:be-broad-18gem}
\includegraphics[width=0.35\textwidth, angle=90]{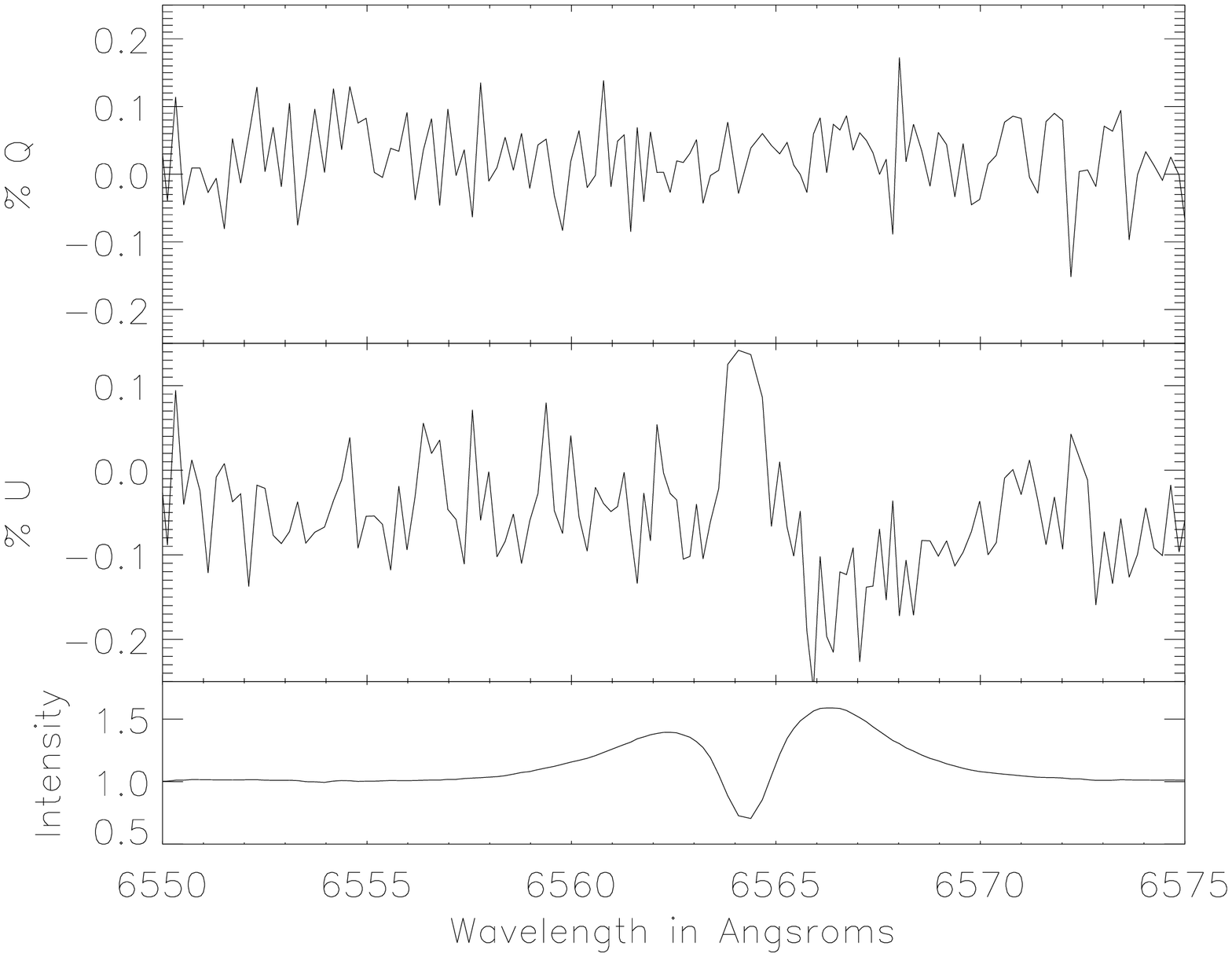}}
\quad
\subfloat[$\eta$ Tau]{\label{fig:be-broad-etata}
\includegraphics[width=0.35\textwidth, angle=90]{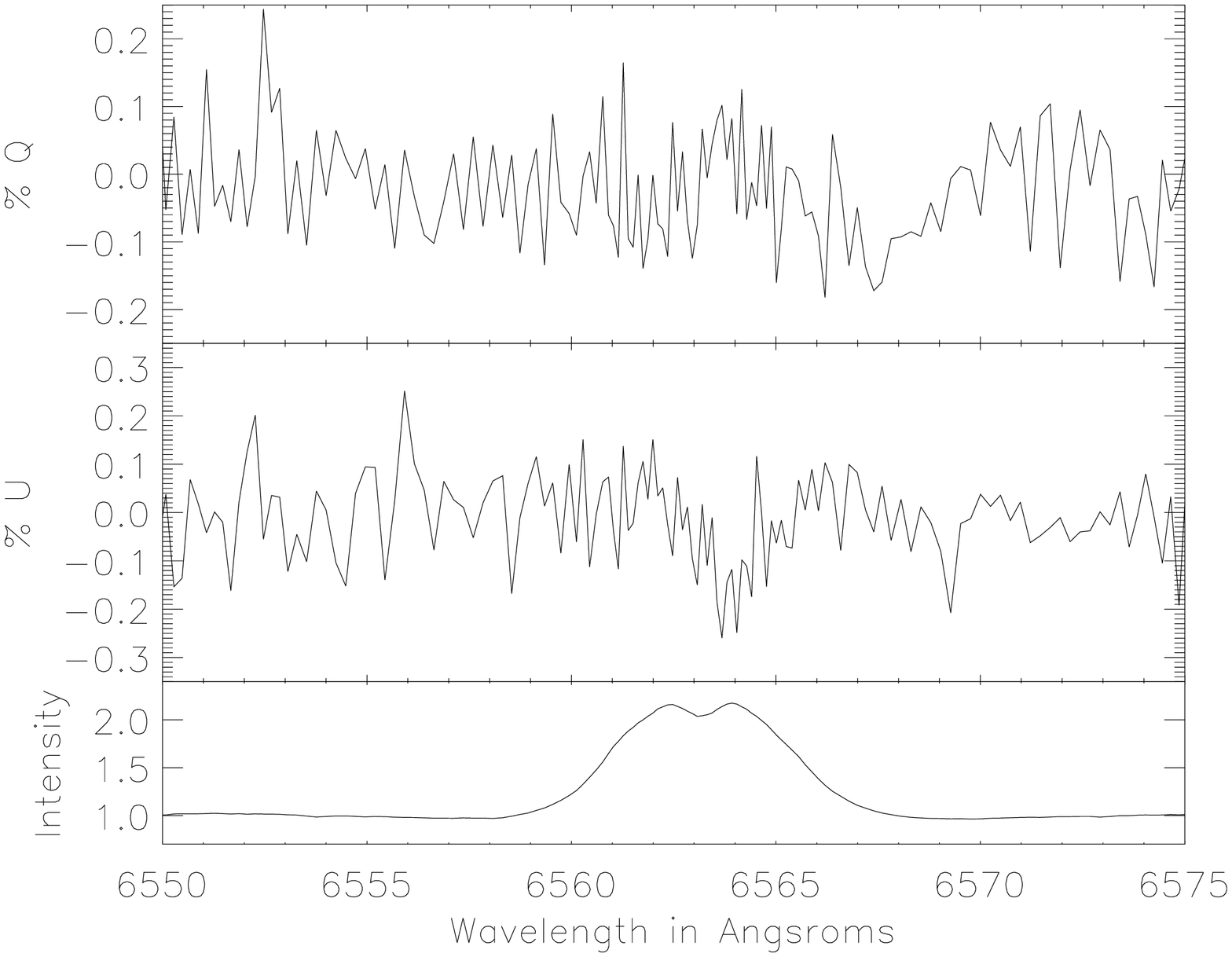}}
\quad
\subfloat[$\alpha$ Col]{\label{fig:be-broad-alfco}
\includegraphics[width=0.35\textwidth, angle=90]{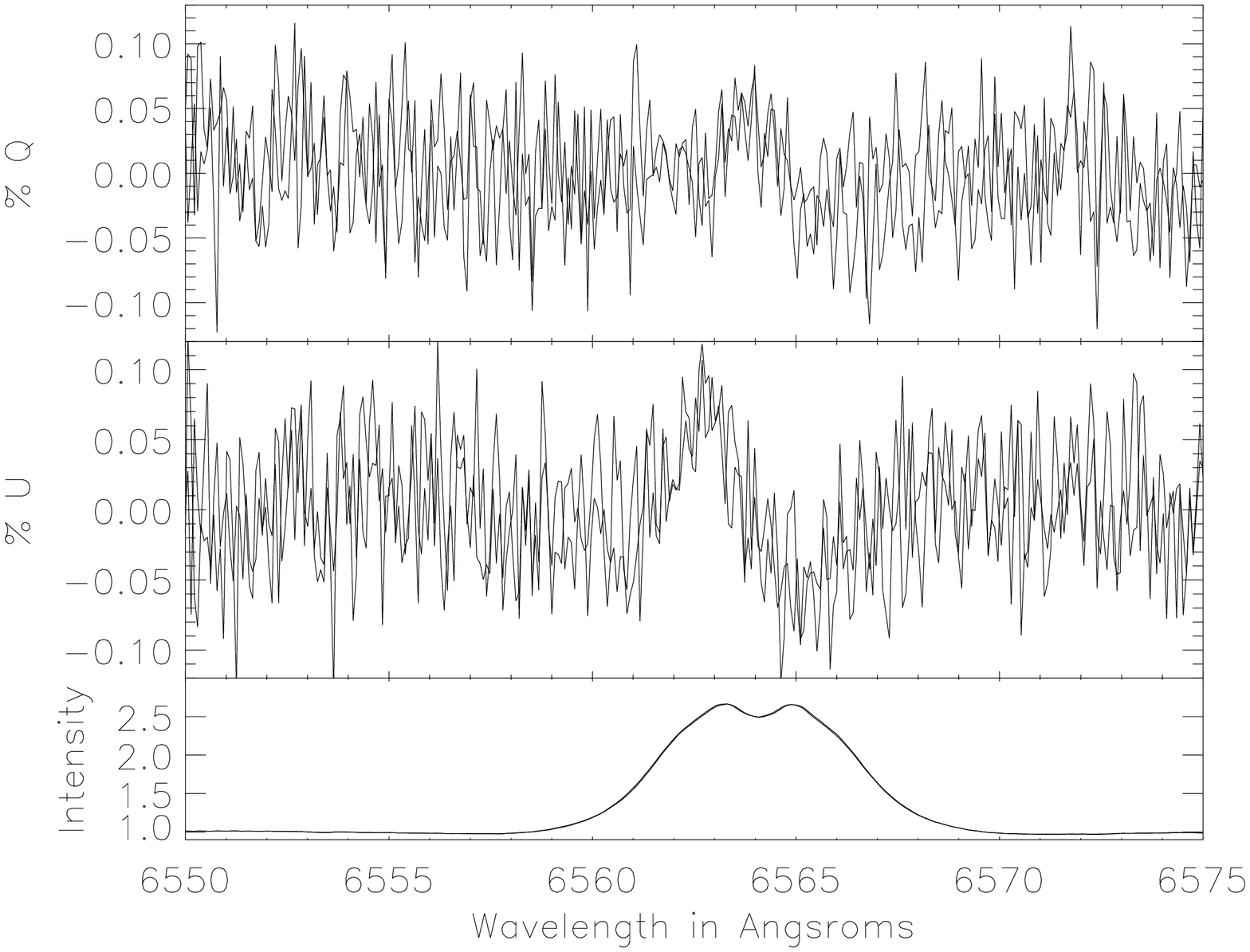}}
\quad
\subfloat[$\beta$ CMi]{\label{fig:be-broad-betcm}
\includegraphics[width=0.35\textwidth, angle=90]{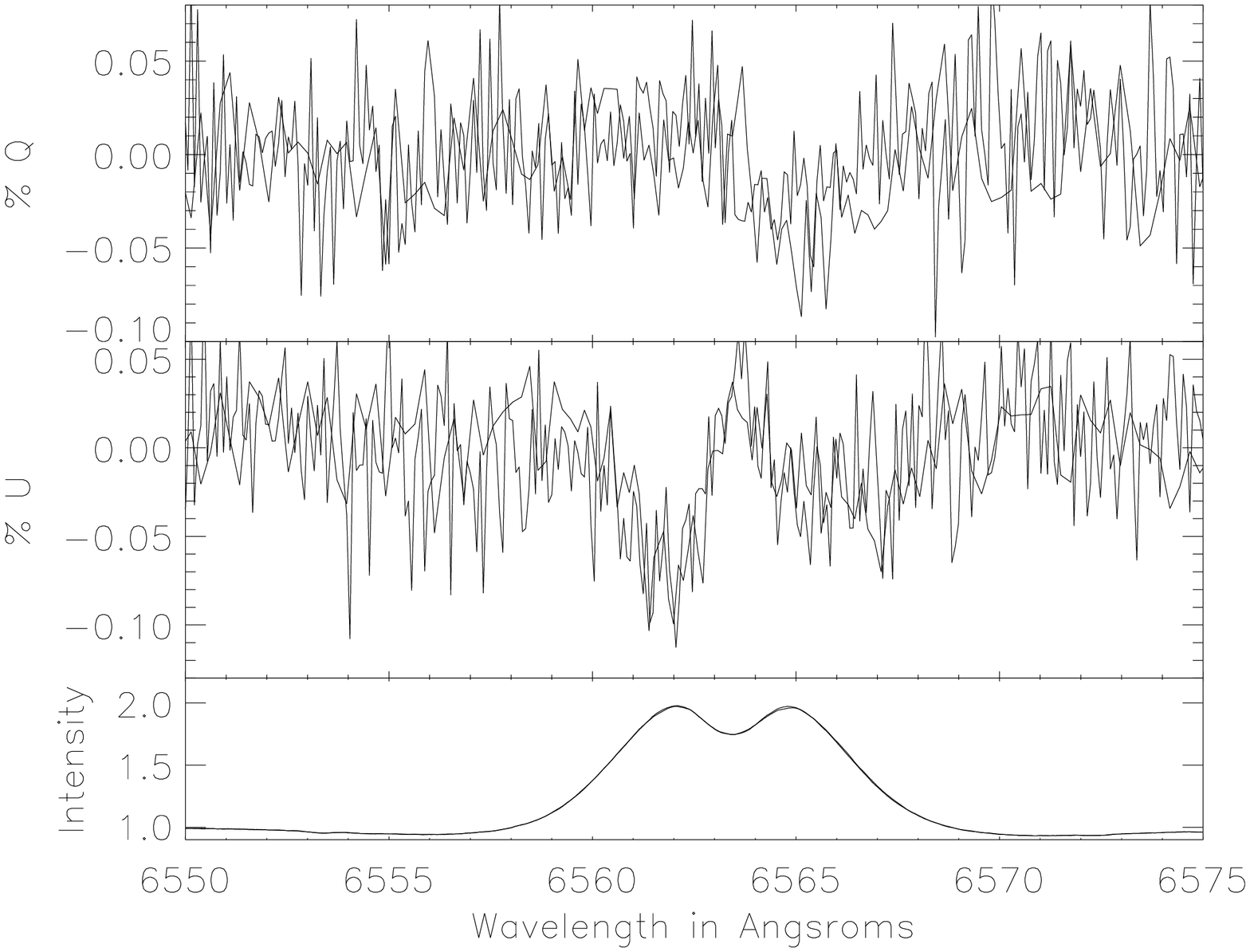}}
\quad
\subfloat[MWC 192]{\label{fig:be-broad-mwc192}
\includegraphics[width=0.35\textwidth, angle=90]{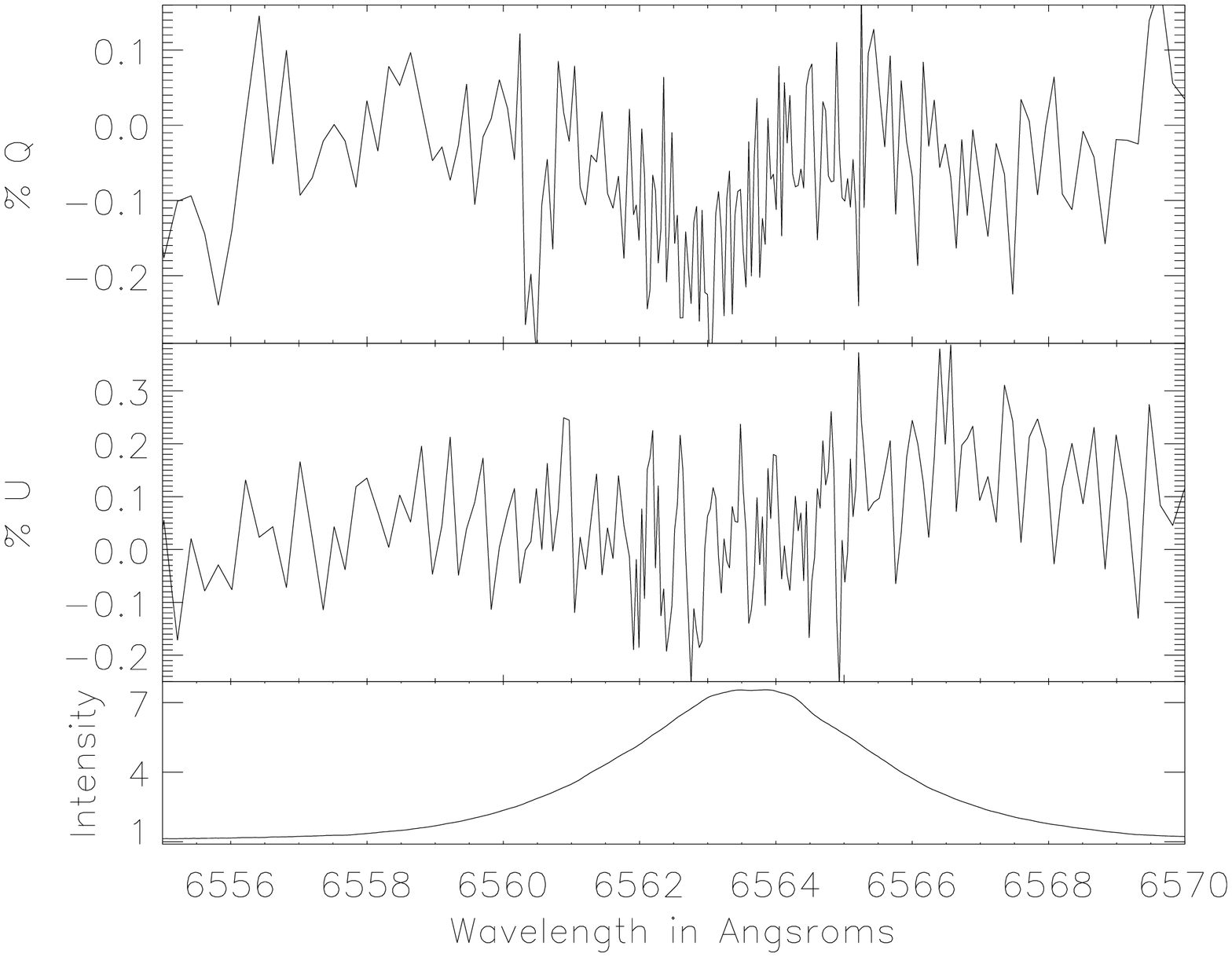}}
\caption[Other Be Effects ]{ The five smaller, more complex detections: {\bf a)}18 Gem  {\bf b)} $\eta$ Tau {\bf c)} $\alpha$ Col {\bf d)} $\beta$ CMi {\bf e)} R Pup. The signature in 18 Gem is reminiscent of several Herbig Ae/Be stars. Both $\alpha$ Col and $\beta$ CMi show two detections of similar signal-to-noise taken on two separate nights at the same pointing. The detections match beautifully despite the differing conditions.}   
\label{fig:otherbe}
\end{figure}

\begin{figure}
\centering
\subfloat[3 Pup ESPaDOnS Archive Spectropolarimetry]{\label{fig:swp-3pup-esp-iqu}
\includegraphics[width=0.35\textwidth, angle=90]{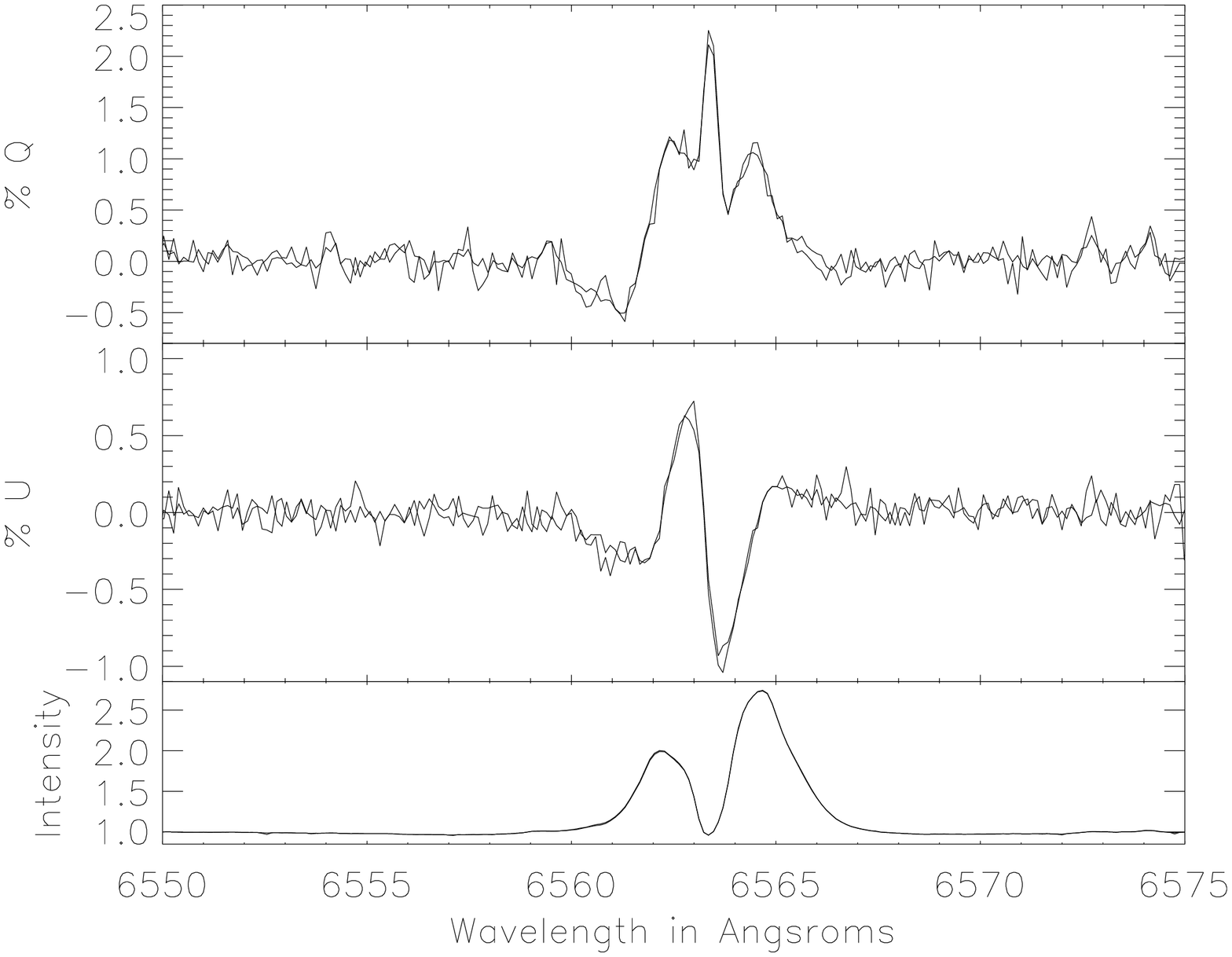}}
\quad
\subfloat[3 Pup ESPaDOnS Archive QU Plot]{\label{fig:swp-3pup-esp-qu}
\includegraphics[width=0.35\textwidth, angle=90]{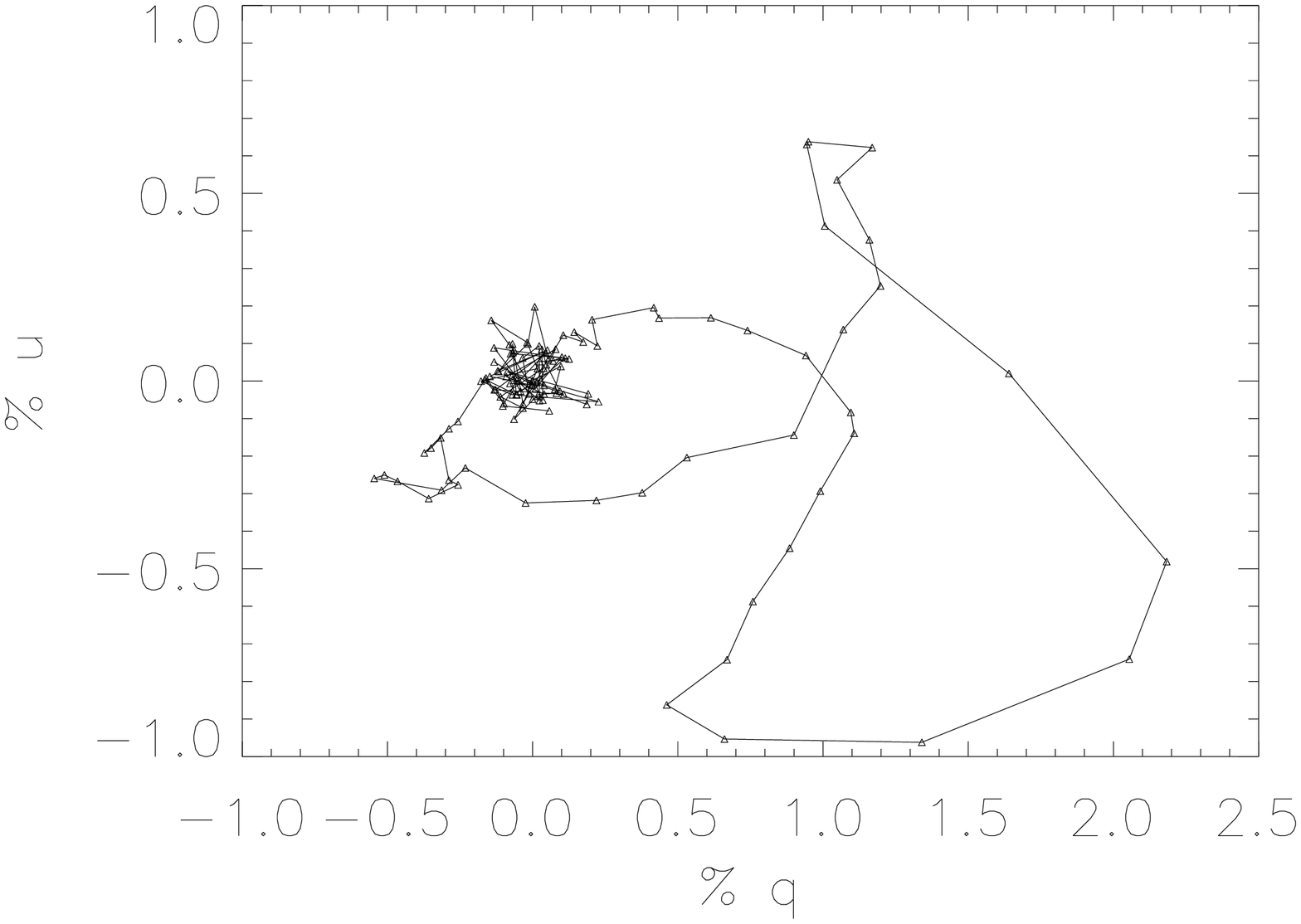}}
\quad
\subfloat[3 Pup HiVIS Spectropolarimetry]{\label{fig:swp-3pup-indiv}
\includegraphics[width=0.35\textwidth, angle=90]{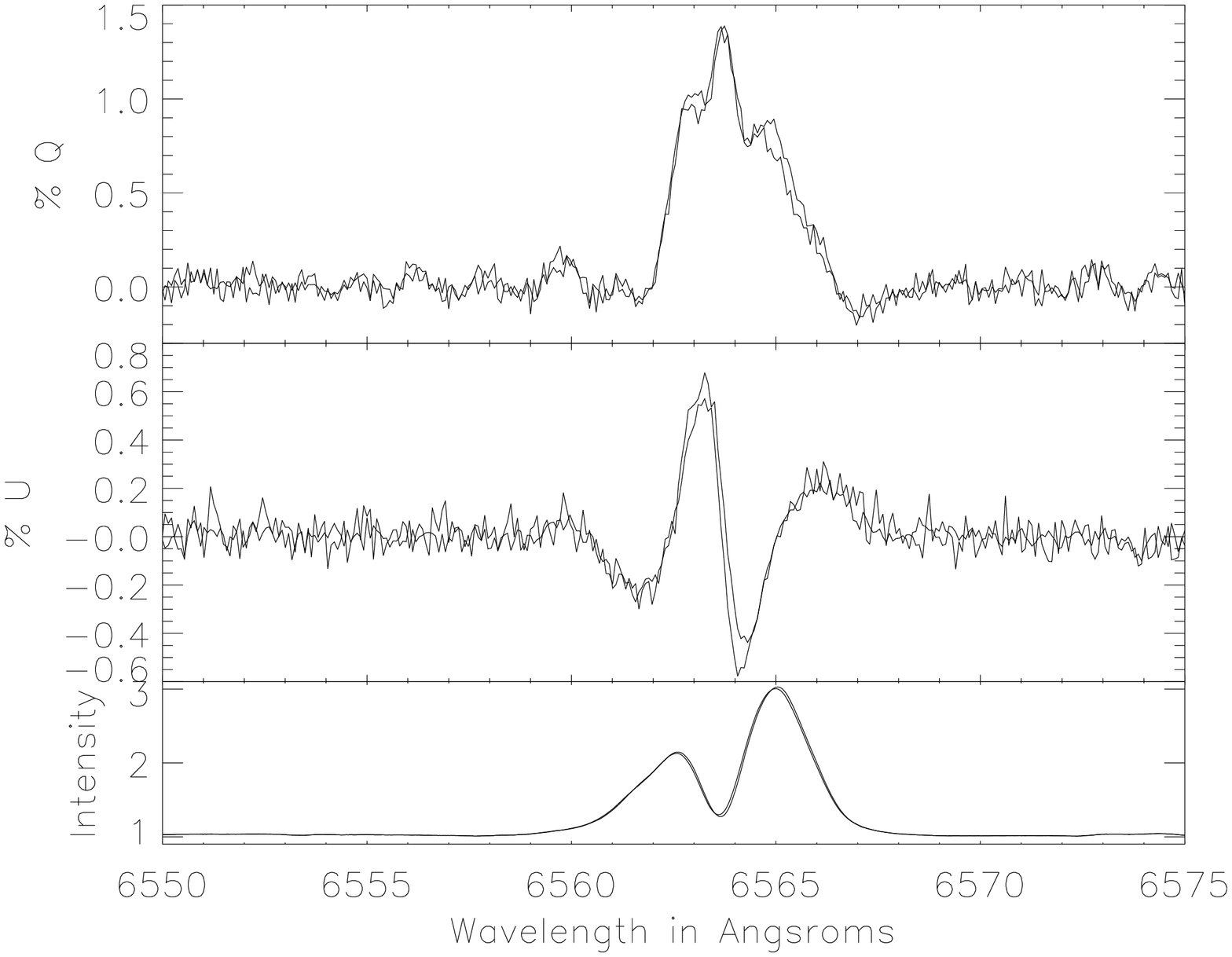}}
\quad
\subfloat[3 Pup HiVIS QU Plot]{\label{fig:swp-3pup-indiv-qu}
\includegraphics[width=0.35\textwidth, angle=90]{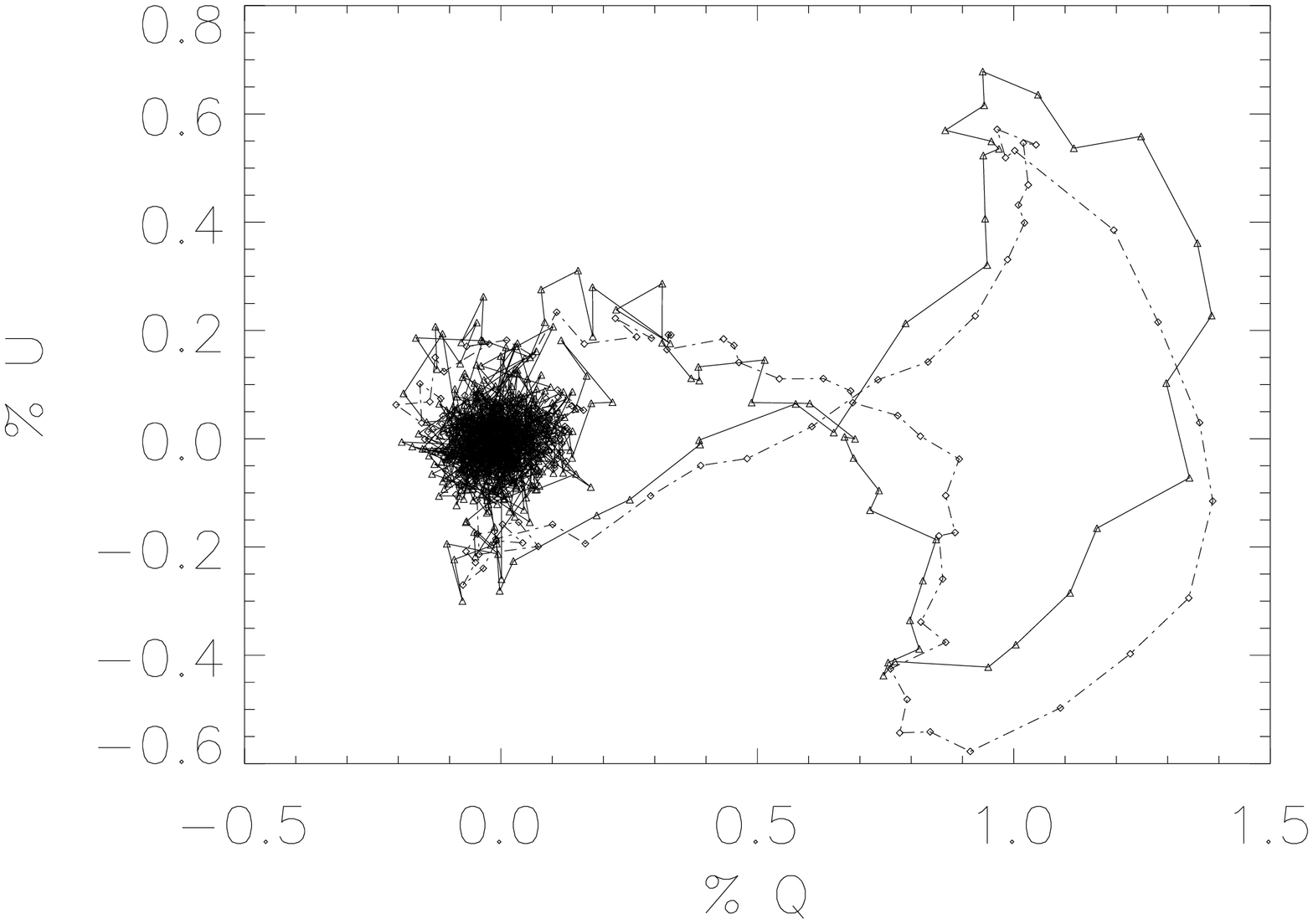}}
\caption[3 Pup Archive ESPaDOnS \& HiVIS Spectropolarimetry]{The spectropolarimetry from ESPaDOnS and HiVIS. {\bf a)} The ESPaDOnS archive data for 3 Pup on February 7th \& 8th, 2006 and {\bf b)} the corresponding QU plot for the higher S/N observation. Both observations match nearly perfectly and the qu-loops show the complicated structure of this signature. {\bf c)} Shows the HiVIS observations from late 2007 and {\bf d)} the two corresponding qu-loops. The observations were taken with only a small difference in pointing, but this difference is enough to rotate the qu-loops by roughly 10$^\circ$. Both have a nearly identical double-looped figure-8 form, but the loops themselves are slightly rotated with respect to each other.}
\label{fig:swp-3pup-esp}
\end{figure}

\begin{figure}
\centering
\subfloat[3 Pup Archive Iqu]{\label{fig:swp-3pup-esp-iqu-ss}
\includegraphics[width=0.5\textwidth, angle=90]{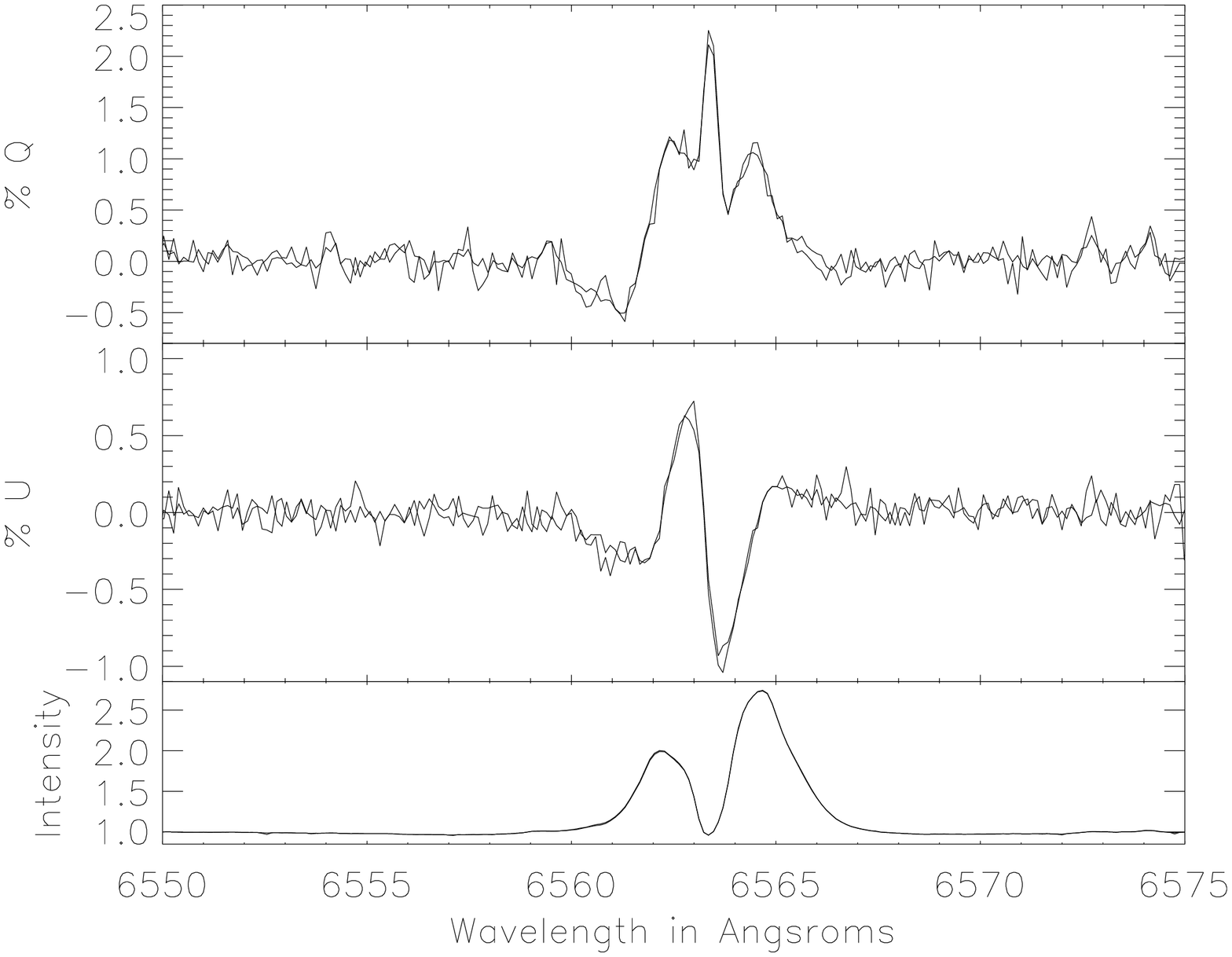}}
\quad
\subfloat[HD 58647]{\label{fig:swp-hd-iqu-ss}
\includegraphics[ width=0.35\textwidth, angle=90]{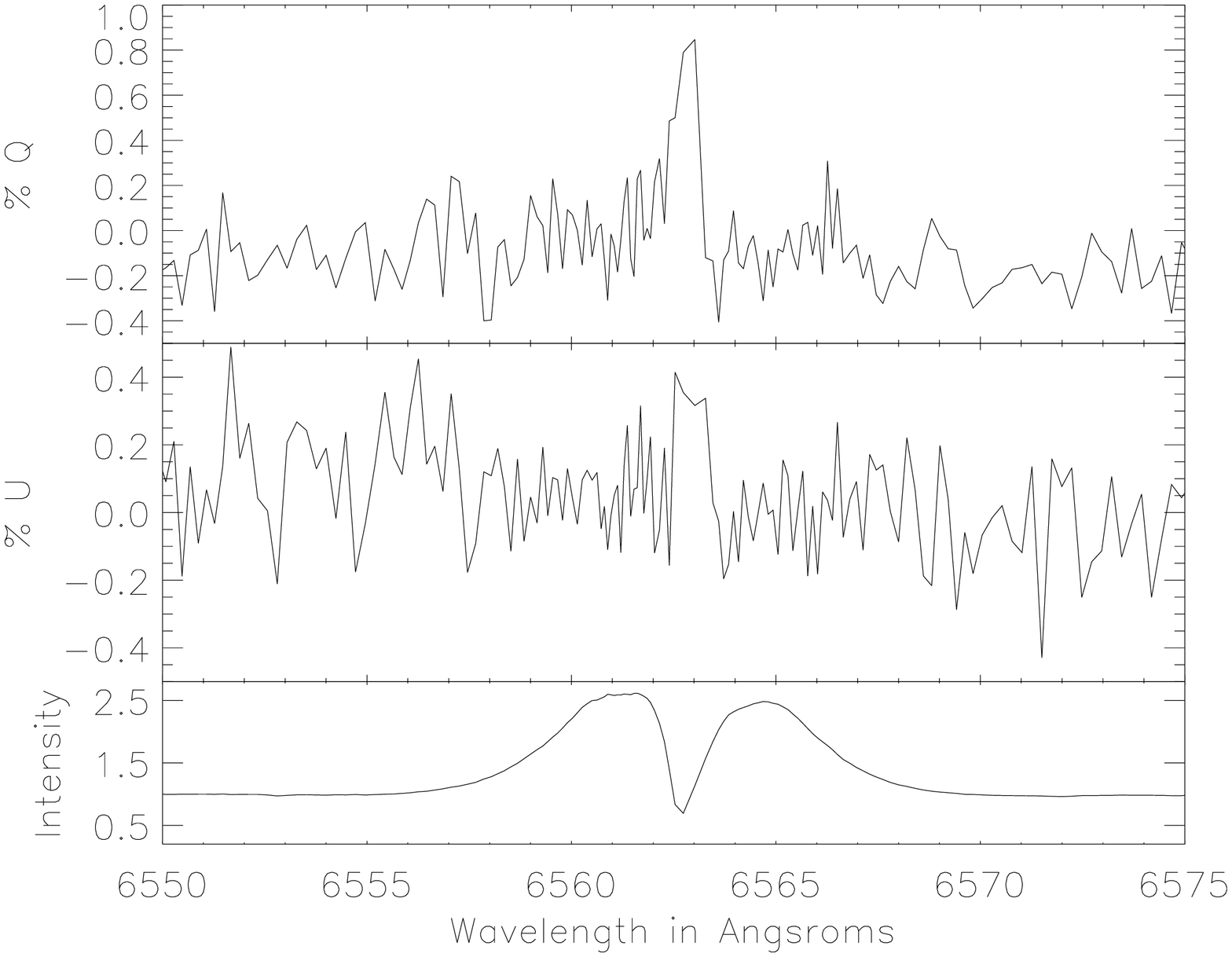}}
\quad
\subfloat[MWC 143]{\label{fig:be-broad-mwc143-ss}
\includegraphics[width=0.35\textwidth, angle=90]{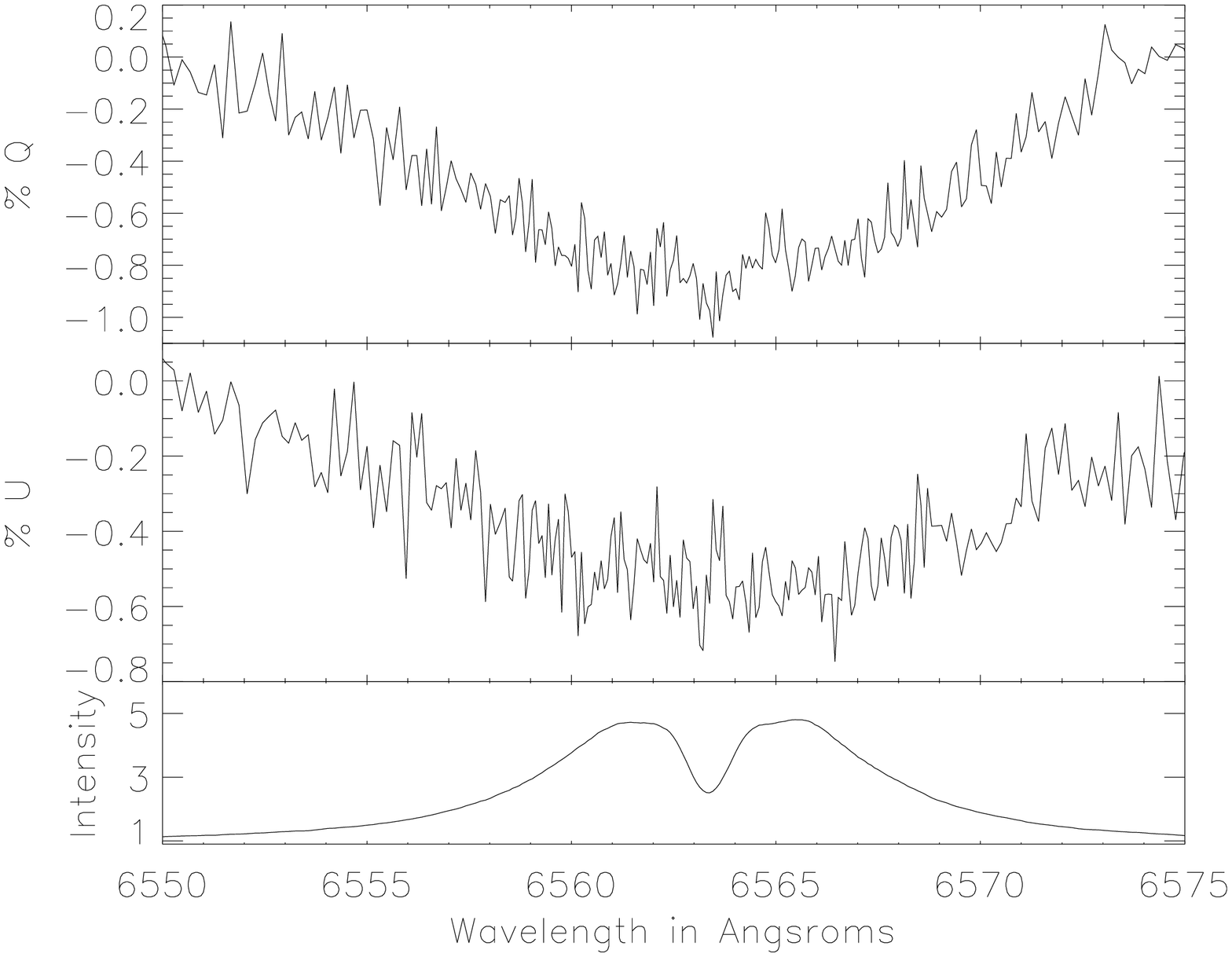}}
\caption[Side-by-Side Comparison]{A side-by-side comparison of the spectropolarimetric morphology of three separate stars with moderate-intensity ``disky" H$_\alpha$ line profiles. {\bf a)} The ESPaDOnS archive data for 3 Pup on February 7th \& 8th, 2006 and HiVIS observations of {\bf b)} HD 58647 and {\bf c)} MWC 143. 3 Pup is an A-type emission line star, HD 58647 is a Herbig Be star, and MWC 143 is a Be star. The morphology is completely different between the three stars though the H$_\alpha$ line profiles are quite similar. This comparison highlights an example of (a) an atypical observation (b) the Herbig Ae/Be ``polarization-in-absorption" and (c) a typical ``depolarization" in Be \& emission-line stars. }
\label{fig:swp-sidebyside}
\end{figure}

\twocolumn

	The qu-plot in figure \ref{fig:swp-3pup-esp-qu} shows a very complex structure. There are multiple components of narrow-wavelength range in Stokes q with a strongly antisymmetric Stokes u. In qu-space this translates into multiple loops. In the corresponding HiVIS observations, shown in figure \ref{fig:swp-3pup-indiv}, there is a similar overall morphology and strength, but with a slightly different exact shape. There has been no rotation applied to the HiVIS observations, but the alignment must be close. The qu-loops show the differences, regardless of rotation, much more clearly. The overall amplitude of the HiVIS observations is lower, and there is much more symmetry in the HiVIS loops. There double-loop structure is preserved, with a pattern somewhat resembling a figure-8. In the ESPaDOnS archive observations, the small third loop near (1.5,0.5) is quite clear whereas a simple cluster of points is only seen in the HiVIS loops.  However, the loss of the loop could be entirely caused by the bin-by-flux routine that is automatically applied to the HiVIS observations, while it was not applied to the ESPaDOnS data. The third loop at a q of 1.5\% is at the wavelengths of the central absorption, where HiVIS binning is roughly 5 pixles (since the threshold was 5x continuum).

	In the context of the scattering theories, this star is quite difficult to explain in any framework currently available. The polarization is very strong compared to most other observations. The form is also very complex, having strong deviations and asymmetries. The polarization is also strong across the entire width of the line: absorption, emission and wings. The overall intensity of the H$_\alpha$ line is quite low, being only 3 times continuum, and the central absorption is not extremely strong either, returning only to near continuum. This example provides yet another good example for the need for new, more detailed theories.

\subsection{Comparison Summary}

	The Be and emission line spectropolarimetry shows remarkably different spectropolarimetric profiles when compared to the Herbig Ae/Be stars even with the same shape and magnitude H$_\alpha$ line profile. The Be signatures are smaller in magnitude and typically span the entire line. The morphology of the Be spectropolarimetry does not seem to be tied to the absorption directly because the polarization effect is present across the entire line. There is some morphological change tied to the presence or strength of the absorption, but this has a smaller amplitude than in the Herbig Ae/Be systems. Systems with both strong and weak absorption in both strong and weak H$_\alpha$ lines showed the broad spectropolarimetric effects. However, Be stars with either symmetric unobscured emission lines or absorption lines did not show any detectable polarization changes. Herbig Ae/Be stars in contrast show a polarization that is much stronger and is directly tied to the absorptive component of the line. In windy systems with P-Cygni type line profiles the polarization can be over 2\% in the blue-shifted absorption trough. In many cases, especially AB Aurigae and MWC 480, the polarization at the emissive peak is identical to the continuum polarization within the 0.1\% noise. In both Be and HAe/Be stars, there are a wealth of spectropolarimetric morphologies even for the same line type. 
	
	A direct comparison between three different disky systems is shown in figure \ref{fig:swp-sidebyside}. Though all three stars have different spectral types, they all have mostly-symmetric H$_\alpha$ lines with central absorption features an line/continuum ratio's of 2.5-5. The star 3 Pup is not in either Be or HAeBe classes and shows a very strong, complex signature. HD 58647 is a Herbig Be star that shows a strong spectropolarimetric signature only in the very center of the central absorption and as mentioned in chapter 8, a very small antisymmetric signature that spans that absorption in some epochs. MWC 143 is a Be star that shows a broad clear spectropolarimetric signature that is wider than the emission line itself and shows no significant deviation from this broad spectropolarimetric trend in the central absorption.

	Comparison of these three stars highlights the range of effects seen in this survey. The detections show that, although the broad spectropolarimetric effect seems common (10/30) in Be and other B-type emission-line stars, it is certainly not common in the Herbig Ae/Be stars. One common thread in those Herbig Ae/Be stars was the consistent presence of spectropolarimetric signatures in and around absorptive components of the H$_\alpha$ line. In many of the stars, there was very large polarization changes in absorptive components while the wavelengths of strongest emission showed no polarization changes. The Be and emission-line stars do have their own morphological complexities that are worth investigating, and 5 of 30 detections that do not follow the broad morphological description. These more complex detections, in some cases, look very similar to some Herbig Ae/Be detections. Now that a very clear morphological difference has been established between Herbig Ae/Be stars and other emission-line stars, a new theory will be explored that may better fit the ``polarization-in-absorption" morphology of the Herbig Ae/Be stars.

\section{A New Theory: Optical Pumping}

	In light of the consistent detection of spectropolarimetric effects in and around the absorptive components of the Herbig Ae/Be star H$_\alpha$ lines, a new model that directly ties polarization to absorption was explored. In this new model, based on optical pumping, it is the absorptive process itself which induces the polarization. Analogous with scattered light being polarized with a dependence on scattering angle, absorbed light can be polarized with a dependence on angle as long as the gas is optically pumped. The absorbing hydrogen gas can be pumped by a strong anisotropic radiation source, namely the bright nearby star. This effect will be used to describe the resulting spectropolarimetric effects in this chapter. This chapter will follow the development of this theory as published in Kuhn et al. 2007 very closely. This chapter will be more pedagogical and provide more examples.

\subsection{Introduction to Optical Pumping}

	Optical pumping is an effect that can cause polarization during the absorption process. For a gas in a strong radiative environment, such as gas near a bright star, the populations in the atomic levels become unbalanced and can lead to polarization-dependent absorption. 
	
	There is one main argument showing that the absorptive effects can dominate scattering effects. The amount of highly polarized light scattered by an optically thin cloud at high scattering angles into the telescopes line of sight is small compared to the amount of light directly absorbed by the same optically thin material occulting the star. The ratio of absorbed flux to scattered flux is quite high - one cloud absorbs a given number of photons and scatters those into 4$\pi$ steradians. Only one line of sight corresponds to the telescope. This reason alone motivates the study of absorptive polarization effects because of the increased effect absorptive effects can have.

	There are four quantum numbers: (n,l,s,m). These are the principal number specifying the orbital radius (n), the azimuthal quantum number specifying the number of nodes in the z axis or the quantization axis (l), the spin number (s), and the magnetic quantum number (m), which is analagous to the quantized projection of the angular momentum onto the z axis.  Combining the orbit and spin angular momenta gives the total electronic angular momentum, j. The basic rules of quantum mechanics state that l can go from 0 to n-1 and m can go from -l to l.  So for say an n=3 state, l can be [0,1,2] and m can be [-2,-1,0,1,2].  The spin is always $\pm \frac{1}{2}$ for a single electron, but an atom can have any number of electrons, giving rise to even or odd combinations of spin numbers.  
	
	Hydrogen has only one electron so the spin is $\pm \frac{1}{2}$.  The fine structure is always {\bf J=L+S} or j=l$\pm \frac{1}{2}$.  Say l=2, then j=$\frac{5}{2}$ or $\frac{3}{2}$.  The magnetic quantum number, m, determines the spatial structure of the orbital, but does not play a role in the zeeman splitting, just the j values.  For the hydrogen, the levels n=1,2,3 have energies of -13.6, -3.4, -1.5 eV.  The H$_\alpha$ line is the n=3-2 transition, the first Balmer line, with an energy of $\delta E=1.9eV$.  The n=3 orbital has l=0,1,2 and one electron at spin $\frac{1}{2}$ so it has possible states of l=0, j=$\frac{1}{2}$, l=1 j=$\frac{1}{2}$ or j=$\frac{3}{2}$, l=2, j=$\frac{3}{2}$ or j=$\frac{5}{2}$.  The n=2 orbital has l=0 or 1 and available states of l=0, j=$\frac{1}{2}$ or l=1, j=$\frac{1}{2}$ or j=$\frac{3}{2}$.  When these states quantize along the z axis to make fine structure, the magnetic m number comes into play since m can range from -j to +j in steps of 1.  For the j=$\frac{5}{2}$ state, there are six levels, 5/2, 3/2, 1/2, -1/2, -3/2, and -5/2.  Since there's no magnetic field, these substates are indistinguishable in energy.  The selection rules allow transitions of $\delta l=\pm 1$ and $\delta m=\pm 1,0$.
	 
	Imbalances between atomic sublevels is actually somewhat common in solar and stellar astronomy. Resonant polarized scattering has been considered in the absence of magnetic fields by Warwick and Hyder 1965. In solar astronomy, it has been used in the presence of depolarizing magnetic fields through a mechanism known as the Hanle effect (cf. Hanle 1924, Stenflo 1994) and to describe scattering in stellar envelopes (cf. Ignace et al. 2004). From a quantum mechanical perspective, resonant scattering polarization is caused by unequal magnetic substate populations within the upper level of the resonant transition. This can be induced by anisotropy of the incident radiation and correlations between the upper sublevels which yields polarization of the scattered/re-radiated light (cf.  Stenflo 1998). 

	Polarized absorption results if the lower state magnetic atomic sublevels are unequally populated when the absorption occurs. This imbalance can be caused by a strong anisotropic radiation source, called an optical pump. Such an ``optically pumped'' gas has an opacity that will depend on the incident electric field direction. In general, unpolarized light incident on such an absorbing gas can emerge polarized simply by the absorption being stronger in one polarization state. Optical pumping has been demonstrated in the laboratory in Happer 1972 and has been discussed in the context of sensitive solar observations (cf. Trujillo Bueno and Landi Degl'Innocenti 1997) and in astronomical contexts by Yan \& Lazarian 2006, 2007, 2008.

	This mechanism can be illustrated by a simple classical example. Consider an atom in a low-density environment where there are no collisions to mix the sub-level imbalances induced by a light source. Suppose the lower level of a transition has a total angular momentum $j=1$ which is nominally degenerate with magnetic sublevels $m=\pm$1, and 0. Also suppose the upper level has $j=0$. In a cartesian coordinate system the electronic transitions or substates are associated with classical electronic oscillators aligned in the {\it x-y} plane (for $\Delta m=\pm$1) and in the $z$ direction (for $\Delta m=0$). A coordinate system is chosen with the pumping radiation incident on the gas in the $+z$ direction. In this case, only the $\Delta m=\pm 1$ transitions can be excited because the incident transverse electric field lies in the $x-y$ plane.  On the other hand the subsequent spontaneous downward transitions equally populate all lower magnetic sublevels. In equilibrium, only the $m=0$ substate will be significantly populated. Now consider a second unpolarized beam incident on the gas in the $+x$ direction. This beam has equal $y$ and $z$ electric field components since it is unpolarized. In this case, only the $z$ component of this electric field is absorbed and scattered by the pumped gas because only the $m=0$ electronic ground state is populated. Thus the emergent beam is 100\% linearly polarized in the $y$-direction solely because of the absorption. 
	
	In general a gas which is anisotropically excited, but is observed with light which is incident from a direction that differs from the optical pumping beam direction, will exhibit an emergent linear polarization with dominant electric field in a direction which is perpendicular to the plane of incidence with the pumping beam. This is the geometry expected from a star imbedded in a disk or outflowing wind where the occulting cloud is not directly along the line of sight to the center of the star. Optical pumping has been analyzed for an anisotropic stellar atmosphere in Trujillo Bueno \& Landi Degl'Innocenti (1997) and has been explored for circumstellar material in Kuhn et al. 2007.

\begin{table}[!h,!t,!b]
\begin{center}
\begin{footnotesize}
\caption{Flux-ratio Variables \label{tab-pumpvar}}
\begin{tabular}{ll}
\hline
\hline
Name                     &             \\
L                             &  Stellar luminosity          \\         
$\Omega_0$         &   Solid angle of detector seen from star           \\         
$\Omega_c$         &   Solid angle of cloud seen from star           \\         
$\frac{L \Omega_0}{4\pi}$ & Intensity of detected starlight, I       \\         
r$_c$                       &   Cloud radius       \\         
r$_s$                       &  Stellar radius (1)     \\         
d                               &  Star-Cloud distance       \\         
$\gamma$              &  Scattering angle, $\chi$ in chapter 7      \\         
2$\Theta$              &  Angular size of star seen by cloud (incidence angles)      \\         
f                               & Fraction of incident light scattered into 4$\pi$  \\
$\frac{f\Omega_0}{4\pi}$&Fraction of incident light scattered to detector \\
$(\frac{r_c}{r_s})^2$& Projected area of absorbing cloud   \\
(1-f)$(\frac{r_c}{r_s})^2 \frac{L \Omega_0}{4\pi}$ & Transmitted flux \\
\hline
\hline
\end{tabular}
\end{footnotesize}
\end{center}
\end{table}

	When the intervening cloud is optically thin, the polarization of the absorbed spectral feature can be larger than any scattered light spectral polarization signal. The ratio of polarized absorbed light to polarized scattered light will be computed to show in which regimes this absorbed signature is dominant. Table \ref{tab-pumpvar} lists the variables. Figure \ref{fig:pumpgeo} shows the geometry of a gas cloud near a star of luminosity $L$. Let the cloud and star have radii $r_c$ and $r_s$ and let the cloud be a distance $d$ from the star. For simplicity, set $r_s=1$. The solid angle subtended by the cloud and the observer's detector, as seen from the center of the star, are $\Omega_c$ and $\Omega_0$. The polarized scattered light signal is $Q_{sc}$ and the total continuum optical signal is $I_s$. When the cloud is not projected against  the disk of the star, the degree of polarization is:

\begin{equation}	
q_{sc}=\frac{Q_{sc}}{I_s} = {L\Omega_c/4\pi ~f \Omega_0/4\pi ~p_s\over L\Omega_0/4\pi} ~=~ p_s f {r_c^2\over 4 d^2}
\end{equation}

where $f$ is the fraction of light incident on the cloud which is scattered (assumed isotropically) and $p_s$ is the average intrinsic polarization of this scattered light. This equation describes the polarization of light scattered by a single cloud in terms of the amount of incident flux multiplied by the degree of polarization and amount of light scattered to the detector all normalized by the total stellar flux received at the detector. 

	When the cloud lies directly between the disk of the star and the observer, optical pumping can cause the absorbed light to be polarized. All the light removed from the beam appears as a polarized absorption feature in the spectrum. Since the absorbed light is scattered into all directions the absorption can polarize a much larger fraction of the light. This absorptive geometry can lead to a larger polarized optical signal. If $r_c < r_s$ then the relative transmitted polarized flux, $Q_{tr}$, normalized to the total, is:
	
\begin{equation}
q_{tr}=\frac{Q_{tr}}{I_s} = {(r_c/r_s)^2 L \Omega_0/4\pi~(1-f) p_a\over L\Omega_0/4\pi} = p_a(1- f){r_c^2 \over r_s^2}. 
\end{equation}

	Here $p_a$ is the absorptive (optically pumped) intrinsic polarization which is computed below. It will shown explicitly how this term is computed in the following sections. Typically this term is a small fraction of a percent and is no more than 1\%. The intrinsic scattered and absorbed polarization, $p_s$ and $p_a$, are due to population differences in, respectively, the upper and lower level magnetic substates, and to the anisotropy of the scattered or pumping radiation. The ratio of absorbed to scattered polarization is computed as:

\begin{equation}	
\frac{Q_{tr}}{Q_{sc}} = \frac{p_a}{p_s} \frac{(1-f)}{ f} 4d^2
\end{equation} 
 
 	The relation $r_s$=1 has been used. Since the cloud lies above the star, $d$ is always larger than $r_s$. In general, this alone illustrates the regimes where the absorptive effects dominate. In low optical depth cases, the fraction (1-f)/f is quite large. The further away the cloud is, the larger the ratio (though in absolute terms both scattered and absorbed fluxes decrease). It is shown below that the typical absorbed polarized fraction is less than 0.01 while scattered flux can approach 1 for clouds very close to the photosphere near the limb.
	
	As an example, the ratio for a cloud one stellar radii off the photosphere (d=2) absorbing near the limb will be computed at the maximum possible scattering angle. In this geometry, the cloud will be at a scattering angle of $\gamma=30^\circ$ and the star will subtend $2\Theta=60^\circ$. From chapter 7, if simple Rayleigh scattering is assumed, the degree of polarization can be calculated as the ratio Q/I from the scattering phase matrix:
	
\begin{equation}	
q ~=~ \frac{Q}{I} = \frac{sin^2\gamma}{(1+cos^2\gamma)} \approx 0.286
\end{equation} 
	
	For simplicity, we'll assume that the absorptive polarized fraction is 0.01 and that the optical depth is 0.1. In this case the ratio of absorbed to scattered polarization is high:
	
\begin{equation}	
\frac{Q_{tr}}{Q_{sc}} = \frac{p_a}{p_s} \frac{(1-f)}{ f} 4d^2 =  \frac{0.01}{0.286} \frac{(0.9)}{0.1} 16 \approx 5
\end{equation} 
	
	In this case, the optical-pumping polarization effect will dominate any scattered polarization. Now, this simple example ignores the entire spatial distribution of scattering and absorbing particles, but does provide a compelling reason to explore polarized absorption as an alternative to scattering polarization. For optically thin cases where the clouds are a few stellar radii away from the photosphere, this absorptive effect has just as much effect as the light scattered by the same cloud. 
	
	Our calculation is a very simplified version the general problem.  This example assumes that the ``cloud'' has a small line-of-sight velocity with respect to the star and have ignored doppler shifts or the spectral dependence of the stellar source. In realistic cases where the absorption is on top of an emission line, it is not stellar continuum radiation, but H$_\alpha$ emission that optically pumps the intervening gas. In general this enhances the optical pumping (since H$_\alpha$ can be significantly brighter than continuum) and complicates the geometry because the pumping source is a broad region with significant spectral dependence and doppler shifts induced by cloud motion. However, since this cloud only produces polarization at the specific wavelength where there is absorption, there is less difficulty explaining profiles where emission and continuum polarizations are identical. It is not the redistribution of polarized light, taking polarization from one wavelength and moving it to another, but the absorption itself which imprints the polarization. Thus the discussion here applies directly to the outflowing wind configuration of, for example, AB Aur. In this model, absorption causes polarization and the emission line can remain unpolarized. The absorptive polarization fraction from statistical equilibrium will be computed below followed by specific solutions.

\subsection{Statistical Equilibrium and H$_\alpha$}

In order to compute the intrinsic absorptive polarization amplitude $p_{a}$, the statistical equilibrium equation must be solved for the densities of the first 28 (LS coupled) hydrogen energy levels corresponding to all of the principal quantum numbers $n=1,2,3$ and $l=0,1,2$ sublevels. The ground state ($n=1$) consists of an $l=0,~j=1/2,~m=\pm 1/2$ doublet. The $n=2~(l=0,1)$ levels include two $j=1/2$ and one $j=3/2$ states for a total of 8 magnetic substates. Similarly there are 18 $n=3~(j=1/2,3/2,5/2)$ sublevels. A 28$\times$28 matrix statistical equilibrium equation is constructed from the radiative transition rates. Note that collisional transitions are unimportant in the density regimes of interest here. The solution for the individual sublevel densities is obtained by a singular value decomposition method.

First, one must obtain theoretical line strengths and Einstein $A$ and $B$ coefficients for all allowed dipole transitions between individual sublevels (Sobelman 1992). These coefficients describe fixed upper and lower magnetic substates so that stimulated emission and absorption $B$ coefficients for each transition are equal. The calculations have been checked against the fine structure $A$ coefficents and line strengths listed by NIST (Ralchenko et al. 2007).  As before, take the quantization axis to lie along the mean incident pumping radiation direction and take the opening half-coneangle as $\theta$. It is the angle-averaged mean radiation intensity that multiplies the Einstein $B$ coefficients in the rate equation. Next, derive distinct geometrical scaling factors for $\Delta m=0$ and $|\Delta m|=1$ transitions that follow from integrating Stenflo's (1994) eqns. 3.72 over the pumping radiation solid angle, $\Omega_s=\pi\theta^2$. For a limb-darkened illuminating stellar disk where $\mu =\cos\gamma$ and $I(\mu) = I_0(1-a+a\mu)$, obtain factors $C_{|\Delta m|}(\theta)$ that multiply the pumping blackbody background radiation terms in the rate equation. When the theoretical limb-darkening coefficients from Al-Naimiy (1978) are included, the coefficients can be written as:

\begin{scriptsize}
\begin{equation}
C_0(\theta ) = (1-a)(\frac{1}{2}-3\cos{\theta}/4+ \\ \cos^3{\theta}/4)~+~a(\frac{3}{8}(1-\cos^2{\theta})+3\cos^4{\theta}/16)
\end{equation}

\begin{equation}
C_1(\theta )=(1-a)(\frac{1}{2}-3\cos{\theta}/8-\cos^3{\theta}/8)~ \\ +~a(\frac{9}{16}-3\cos^2{\theta}/16-3\cos^4{\theta}/32)
\end{equation} 
\end{scriptsize}
 
Anisotropy of the illuminating stellar light source leads to a difference in the $m=\pm 3/2$ and $m=\pm 1/2$ populations within the $n=2, ~j=3/2$ level. This lower state anisotropy linearly polarizes the transmitted light from the star which passes through the cloud at a non-zero angle, $\gamma$, with respect to the $z$ (optical pumping) direction. Given the geometry defined by figure \ref{fig:pumpgeo}, one expects the emergent light to be partially polarized with electric field perpendicular to the plane of the figure. Most astronomical observations do not resolve the fine structure of the $n=2$ to $n=3$ H$_\alpha$ line but they will still exhibit a diluted linear polarization signal because only some of the transitions are insensitive to the pumping radiation anisotropy such as those which start and end on $j=1/2$ levels.

\subsection{Polarized Radiative Transfer and the Solutions}

The Stokes transfer equation simplifies in the geometry where the stellar photosphere is a background light source and the intervening cloud is optically thin. Take the positive Stokes $Q$ direction to correspond to light with electric field perpendicular to the plane of figure \ref{fig:pumpgeo}. The Stokes vector transfer equation (eq. 11.1 in Stenflo 1994) involves only $I$ and $Q$ intensity, and $\eta_I$ and $\eta_Q$ continuum-normalized opacity terms. Since the stellar photospheric background is the only source of photons observed through the cloud, there are no source or emissivity terms to consider in the polarized transfer equation. 

Typically, only small polarization ($Q << I$) is found in the stellar systems.  In this approximation, the transfer equation reveals that light transmitted through a diameter of the cloud will have polarization:

\begin{equation}
p_{a}=Q/I=\Delta Q/I~\approx 2r_c\eta_Q\kappa_{cont}
\end{equation}

where $\kappa_{cont}$ is the normalizing continuum opacity. Here $I$ is the continuum intensity and one can compute $\eta_{I,Q}\propto n_{I,Q}$ from the lower level population anisotropies and eqns. 3.74 (Stenflo 1994).  In this case the effective refractive index and opacity terms $n_{0,\pm}$ or H$_{0,\pm}$ (eqn 4.41) are computed from sums of the products of the $B$ coefficients and the sublevel populations, derived separately for the $\Delta m=\pm 1$ and $\Delta m=0$ transitions. The optical depth through the cloud is just $\tau = 2r_c \eta_I\kappa_{cont} \approx -\log{(1-f)}$ where $f$ is the fraction of H$_\alpha$ photons scattered, as defined above. For clouds with typical path lengths of roughly $ r_s$, the optically thin condition yields hydrogen densities that are effectively collisionless.

In the low density conditions of this disk/wind the statistical equilibrium rate equation depends on only stimulated emission, absorption, and spontaneous emission processes.  It is the difference in the angular dependence of the $\Delta m = 0$ versus $|\Delta m | = 1$ coupling to the anisotropic radiation field that leads to differences in the degenerate $n=2$ magnetic sublevel populations. Several qualitative features of the optical pumping solutions are intuitive. For intense pumping radiation fields the anisotropy yields a decreasing absorptive polarization -- in this regime the spontaneous emission is negligible and the net upward and downward transition rates between any two levels must be the same. In this case the equilibrium level populations are independent of the upward/downward transition rates. Consequently there can be no  sublevel density differences and the absorptive polarization must vanish. Similarly for weak radiation fields it is the spontaneous emission that defines transition rates and this term in the rate equation does not depend on the radiation anisotropy.  Thus at low and high radiation backgrounds the absorptive polarization approaches zero.

The absorptive polarization near H$_\alpha$ is also a function of the ratio of the H$_\alpha$ to $L_\beta$ (and $L_\alpha$) pumping flux. The ratio of the $n=2,~j=3/2$ to $j=1/2$ populations affects the net H$_\alpha$ polarization, since only the $j=3/2$ states are sensitive to radiation anisotropy. In the limit of no background H$_\alpha$ flux (and in the absence of collisions) all the hydrogen ends up in the ``sterile'' $n=2~l=0$ state because it has no decay route and there is no radiative absorption to depopulate it. Conversely, as the pumping H$_\alpha$ flux increases the equilibrium $n=2~j=3/2$ levels are populated and the induced linear polarization can increase.

This results in a rather weak temperature dependence of the optical pumping source on the absorptive polarization. Even though a decreasing source temperature decreases the pumping radiation, this effect on the polarization is compensated by the increasing ratio of H$_\alpha$ to $L_\beta$ flux. For a photospheric temperature between 30,000 to 3,000K, a change in pumping flux by nearly 4 orders of magnitude, the absorptive polarization is nearly constant. For example, a cloud located 1.1 stellar radii from a 30,000K star ($\theta = 65\degr$) that is $\gamma =65\degr$ from the line-of-sight to the center of the star, yields linear polarization of 0.53\%. If the star has a temperature of 3000K the polarization only decreases to 0.48\%.

For fixed radiation mean intensity, as the pumping half-cone angle ($\theta$) increases the anisotropy decreases and the polarization decreases. This effect continues until $\theta =90\degr$, at which point there is no difference between the dipole coupling of radiation to $\Delta m=0$ and $|\Delta m|=1$ transitions. The actual level population difference scales with $\theta$ as the difference between the derived geometric radiation coupling coefficients $C_0(\theta )$ and $C_1 (\theta )$ defined above. It is straightforward to see that this varies like $\theta^2$ at small angles. Thus the linear polarization $decreases$ for small cone angles (even though the anisotropy increases). This is because the pumping radiation intensity declines rapidly with $\sin{\theta}=1/d$ (where $d$ is in units of the stellar radius) as the cloud-star distance increases. Figure \ref{fig:pumpdist} shows how the linear polarization varies for a cloud seen at the projected edge of a star, as a function of the cloud-star distance, $d$. At small distances from the star the limb darkening function yields significant anisotropy that keeps the polarization from going to zero.

The absorptive polarization depends strongly on the angle between the absorbed beam and the mean pumping direction. As this angle ($\gamma$) increases the polarization perpendicular to the plane increases. For example, with a $T=10,000K$ source temperature and a $65\degr$ cone angle the polarization increases almost linearly from $0$ to $0.5\%$ as $\gamma$  varies from $0$ to $65\degr$.

\subsection{Specific Wind \& Disk Solutions}

With this new theory, detailed forward solutions for a wide range of circumstellar conditions can be developed. Here, the general form of the wavelength dependent linear polarization expected from intervening wind or disk structures will be computed. Consider two geometries; a radial outflow and orbiting material illuminated (optically pumped) by the stellar blackbody. By calculating polarization for clouds at a fixed distance from the photosphere, the polarization imprinted by a ring of material, much like the examples in chapter 7 can be traced. The geometry is essentially that defined in figure \ref{fig:pumpgeo} but allow that the wind or disk's symmetry axis is inclined at an angle $b$ with respect to the line-of-sight to the star.  Choose the common plane containing the line-of-sight direction and the symmetry axis to be oriented vertically. Then, as seen from the observer, the intervening clouds, a ring of material at a fixed distance, projects onto a horizontally oriented arc across the stellar disk. The ring of material is both illuminated by the star and absorbs the stellar light. Each point on this arc contributes absorptive polarization at a distinct velocity with respect to the observer.

\onecolumn

\begin{figure}
\centering
\subfloat[Geometry]{\label{fig:pumpgeo}  
\includegraphics[width=0.45\textwidth]{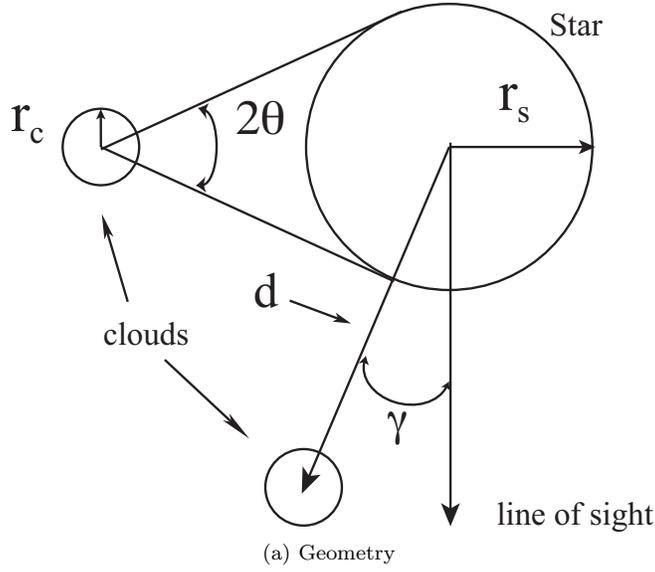}}
\caption[Optical Pumping Geometry]{The cloud geometries that can yield H$_\alpha$ polarization. There are two clouds, one occulting the limb of the star and another at a scattering angle of 90$^\circ$. The angular extent of the star is 2$\Theta$, the cloud is at a distance d from the center of the star and the scattering angle is $\gamma$. The cloud and star radii are r$_c$ and r$_s$ respectively.}
\label{fig:pumpgeo}
\end{figure}
 
\begin{figure}
\centering
\subfloat[Distance Dependence]{\label{fig:pumpdist}
\includegraphics[width=0.4\textwidth]{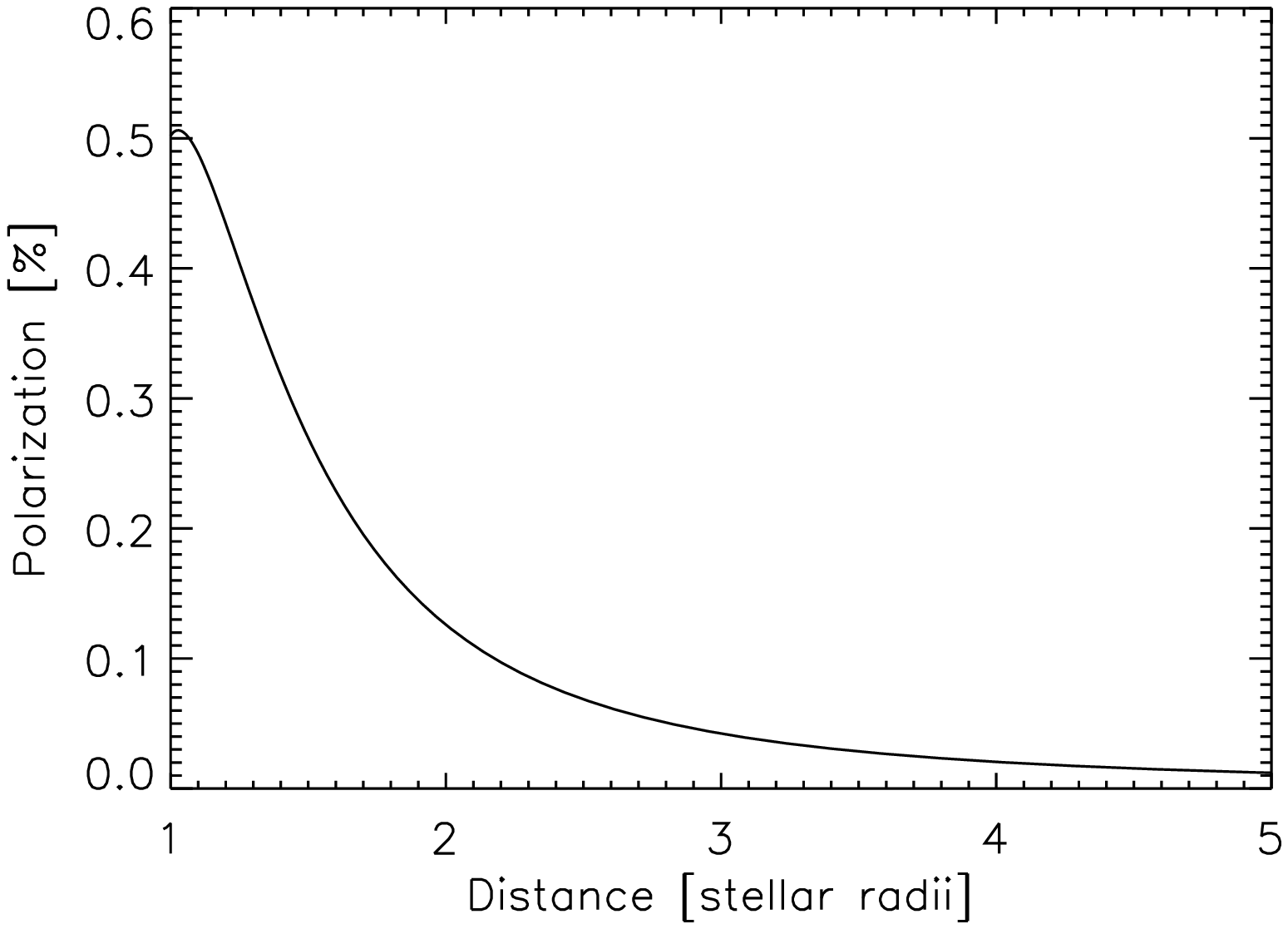}}
\quad
\subfloat[Polarization Calculations]{\label{fig:pumpcalc}
\includegraphics[width=0.4\textwidth]{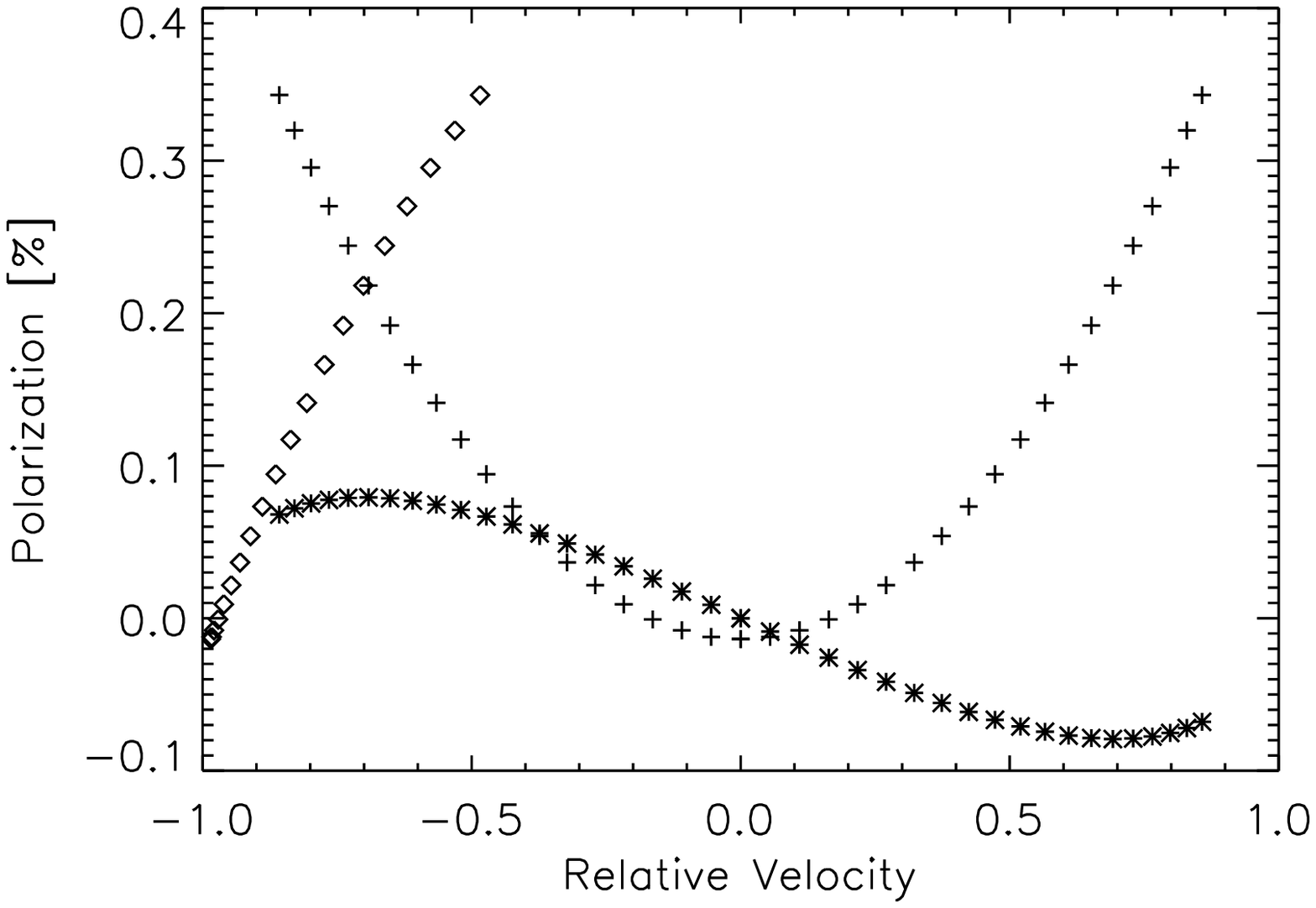}}
\caption[Optical Pumping Calculations]{{\bf a)} Polarization as a function of distance, $d$, between cloud and star assuming a T=10,000K A star, limb darkening, and with the intervening cloud at the projected edge of the stellar disk at the indicated radial distance from the star center.  {\bf b)}  Stokes $Q/I$ (plus symbols) and $U/I$ (star symbols) polarization versus relative velocity of a rotating hydrogen disk. The disk inclination angle is $b=80^o$, $T=10000$K, and $d=1.1 r_s$. The $Q/I$ polarization for a comparable outflowing wind is plotted with ``diamond'' symbols against units of the relative maximum outflow velocity. The H$_\alpha$ line center occurs at zero in the horizontal axis units and the polarization signal in each case is zero outside of the plotted points for that geometry. }
\label{fig:pumping}
\end{figure}

\twocolumn

We take positive Stokes $Q$ along the vertical sky direction. Positive $U$ corresponds to an electric field that is $45\degr$ to this. It follows that the absorptive polarization of a symmetric radial wind should yield only $+Q$ polarization at all velocities. Every cloud producing non-vertical polarization on one side of the photosphere will have another cloud on the other side with mirrored geometry, canceling any non-vertical polarization. Figure \ref{fig:pumpdist} shows the degree of polarization calculated for a cloud projected on the limb, the geometry of maximum polarization, assuming star with a  simple black-body spectrum with a temperature of 10,000K. This degree of polarization is fixed by the quantum-mechanical properties of hydrogen gas, the intensity of the radiation and the geometry. The velocity of the cloud, in contrast to the scattering models, only influences the wavelength of the absorption and does not produce other doppler effects.

The optical-pumping effect in disky systems has symmetric and anti-symmetric components as seen in figure  \ref{fig:pumpcalc}. In contrast to scattering theory, even a homogeneous rotating disk geometry must exhibit non-zero $q$ and $u$ line profiles because of the angular difference between the line of sight and radial directions in every point along the orbit. The zero line-of-sight velocity condition for orbiting material occurs along the vertical symmetry line in the center of the H$_\alpha$ line. For any non-zero angle between source and line-of-sight, there is polarization. The characteristic symmetric Stokes $q$ and antisymmetric Stokes $u$ profiles for such a disk, plotted as stars and diamonds in figure \ref{fig:pumpcalc}, show the degree of polarization rising towards larger velocities. These are plotted against units of the maximum rotation or outflow velocity of the disk or wind. The polarization is maximum when the material projects against the limb (largest asymmetry) on both the approaching and receding sides of the disk. The absorptive polarization outside of these plotted points must be zero, but spectral smoothing due to the finite spectral resolution of real measurements will yield some polarization outside of the absorptive region. The exact spectropolarimetric morphology caused by this inclined ring of material would be a double-peaked Stokes $q$ and anti-symmetric $u$. In qu-space, this would correspond to two separate excursions, one toward (+q,+u) and then another toward (+q,-u). However, the extensions would not be orthogonal because of the different amplitudes of the $q$ and $u$ signatures. The extensions would also not be exactly linear because of the differing curvature of the signatures.

In figure \ref{fig:pumpcalc} we've plotted the Stokes $q$ polarization across the velocity distribution of an outflow. The ring of material is assumed to have a spatially constant radial velocity and the velocity distribution has been normalized by the maximum projected outflow velocity. The outflow has it's maximum projected velocity (blue-shift) closest the line of sight. This occurs at the left-most portion of the curve, with zero polarization and a relative velocity of -1. The cloud occulting the limb is at the minimum relative velocity, in this case at a velocity near -0.5 and a polarization of roughly 0.35\%. Stokes $u$ must be zero, as can be seen in the anti-symmetry of Stokes $u$ for the disk calculation. Clouds that had the same magnitude velocity with opposite signs in the disk case will now have the same projected velocity in the wind case. This would leave only a Stokes $q$ signature that would correspond to a linear extension in qu-space.

Though neither of these calculations accounts for extended H$_\alpha$ source regions, a non-black-body stellar source, or other complications, the illustration shows that the main optical-pumping effects are easily interpreted. The polarization is less than 1\% of the continuum flux, polarization is only present where there is absorption, and the polarization is greatest where the absorbing material projects with maximum angular deviation between pumping source and line-of-sight directions. 

Though this simple calculation is only illustrative, the essential characteristics of optical pumping in a wind-type geometry can be inferred. The polarization will be highest for the parts of the H$_\alpha$ line occulting the source with greatest asymmetries. In stars with with extended H$_\alpha$ emission, the maximum polarization occurs at the wavelengths of highest angular difference between the incident radiation and line-of-sight. The degree of polarization is fixed by the quantum-mechanics - this effect will not produce extremely large polarizations and has a characteristic degree of polarization around 1\% continuum or less. Although most real stellar disks and outflows are not expected to be homogeneous in density or velocity, in general a non-zero $u$ polarization is expected (and observed) which, of course, depends on the rotation angle between the instrument $Q$-axis and the star as well as any geometrical complexities. The zero polarization level is also ill-defined in the observations since a wavelength independent constant has been removed from all observations and interstellar, telescope, and intrinsic polarization are difficult to disentangle with the HiVIS observations. Nevertheless the polarization amplitude and simple form of many Herbig Ae/Be detections is consistent with the ``diamond" symbols in figure \ref{fig:pumpcalc}. The polarization is only in the blue-shifted, absorbed component of the H$_\alpha$ line, and there is a characteristic amplitude of less than 1\% continuum.

\subsection{Optical Pumping Summary}

This new model shows how optical pumping can describe spectropolarimetric effects with the magnitude and form of the absorptive linear polarization features in some Herbig Ae/Be systems. The variation of Stokes $q$ and $u$ near absorptive components of the H$_\alpha$ line originates in this model from the intervening gas cloud projected against the disk of the star and the emission region. The polarization features contain information on the kinematics and inclination of the cloud. One immediate conclusion from these calculations is that the absorptive region is near the surface of the star, most likely within 2 stellar radii in order to have a polarization amplitude of about 0.5\%. The main properties of this model is that polarization is present only where there is absorption. The polarization is maximal where there is greatest asymmetry between the line-of-sight and the source direction. The quantum-mechanical properties of hydrogen put the amplitude of the polarization effects at typically a fraction of a percent (continuum). For constrained geometries it may be possible to devise an inversion technique to extract more specific information on non-homogeneous cloud structures in the innermost regions of these stellar systems. No attempt has been made to include extended H$_\alpha$ emission regions or other circumstellar complications, but the main framework for polarization-in-absorption is complete.

\section{Final Thesis Summary and Conclusions}

	This thesis presented many components of stellar spectropolarimetry in a very thorough, comprehensive and complete study. A new high-resolution spectropolarimeter was built and thoroughly calibrated. Custom processing and analysis packages were written and implemented. A massive amount of observations were performed on many Herbig Ae/Be, Be, and emission-line targets to compile a spectropolarimetric survey far larger in scope than any performed to date. The current spectropolarimetric theories were outlined and compared with the Herbig Ae/Be and Be/Emission-line survey results. Weakness of current theory to explain the Herbig Ae/Be observations were explained and an entirely new theory was outlined. The new theory, based on optical pumping, has the potential to explain the polarization-in-absorption seen in the Herbig Ae/Be systems. 
	
	In chapters 2 through 4, a discussion of the instrument, reduction and calibration was presented. With signatures of less than 0.1\% amplitude, there is a great need for precise and repeatable measurement and analysis. In order to accomplish this, many methods of testing and verifying were used. 
	
	A dedicated reduction script was developed for this instrument. Many methods of calculating and correcting the polarized spectra were explored with most of the methods giving identical results. This software package was cross-checked against another package, Libre-Esprit, in use on the ESPaDOnS spectropolarimeter giving similar results.
	
	Polarization calibration optics were installed on the spectrograph and response tests were performed. The spectrograph optics were shown to be quite stable. Polarized flat field measurements showed that the spectropolarimeter could measure linear polarization very accurately over the full useful wavelength range of the polarizer. The spectrograph induces a small amount of polarization in unpolarized light and this polarization is wavelength-dependent. This is automatically corrected by the reduction software.
	
	The polarization properties of the entire optical system was explored using unpolarized standard stars and highly polarized twilight. The polarization properties of the telescope are quite severe and complicated.  The effect of on detected spectropolarimetric signatures can be quite significant. However, the telescope can not induce any effects across a single spectral line. In effect, the telescope can reduce the overall magnitude of a spectropolarimetric effect, or rotate the plane of polarization. The telescope cannot create a spectropolarimetric signature or modify the morphology of the signature. Thus, any spectropolarimetric signature detected is a useful lower limit on the magnitude of the polarization change across a line.  
	
	In order to verify and validate the instrument and the results of the survey, a large cross-examination campaign was performed and presented in chapter 5. Nearly simultaneous observations of a few stars were performed with HiVIS and the ESPaDOnS spectropolarimeters. The results agreed quite well and illustrated the telescope polarization properties well. Archival ESPaDOnS observations were also used to validate the HiVIS measurements. Literature results were compiled from three other lower resolution instruments: ISIS, HPol and WWS. The results again agreed quite well for many stars observed with HiVIS. 
		
	The observing campaign was outlined in chapter 6. The spectroscopic observations and properties of the Herbig Ae/Be stars were discussed. The H$_\alpha$ line of these stars is typically strong and quite variable. The emission lines can be a few to thirty times continuum with many different absorptive components on top of the emission. Spectroscopic measurements for these stars over many time-scales were presented and discussed. A few short-term variability studies were reported illustrating the dynamic circumstellar environment in these stars. In addition to the HiVIS observations, many previous studies from the literature were discussed in a case-by-case basis to show stellar properties such as disk, wind, or jet phenomena or specific constraints on circumstellar material such as polar down flows or small-volume, high-temperature components in the circumstellar environment.	
	
	Chapter 7 outlined the current theory of spectropolarimetric line profiles, based on disk-scattering effects. The current theory predicts symmetric, broad, double-peaked spectropolarimetric signatures in the orbiting thin-disk case. In the case of outward radial motion, the disk-scattering theory predicts a shifting of the spectropolarimetric profiles toward the red side of the spectral line. Other spectropolarimetric effects, such as the depolarization effect and Zeeman splitting were discussed. The depolarization theory is essentially a dilution of continuum polarization by unpolarized emission. This effect produces broad signatures that are roughly inversely proportional to the amount of emission. Magnetic fields produce a more complex spectropolarimetric signature and other studies have shown that the field strengths of these stars is far too small to produce any measurable spectropolarimetric effect.
	
	In chapter 8, the spectropolarimetric survey of the Herbig Ae/Be stars was presented. The survey showed that roughly two thirds of the windy-type stars showed polarization effects in the blue-shifted absorptive components of the H$_\alpha$ line. Half of the disky-type stars also showed this polarization-in-absorption effect in the line-center absorptive component. The magnitude of the effect was typically 0.1\% to 2\% with several stars showing complex morphologies and complex geometries in qu-space. Only one star showed a spectropolarimetric effect across the entire H$_\alpha$ line. These observations are difficult to explain using current scattering theories. The depolarization effect causes a broad, monotonic change in polarization that is roughly inversely proportional to emission and is a linear extension in qu-space. The disk-scattering effect shows broad, symmetric morphology for orbiting material and a shift toward red wavelengths for out-flowing material. Both of these effects do not fit the observed morphology of polarization-in-absorption on the blue-shifted side of the windy-stars or the polarization-in-central-absorption-only for the disky-stars. The amplitude of the observed effects is also a consideration. Scattering theory has no natural amplitude, but most of the detections fall in the 0.3\% to 1.5\% range. 
	
	In order to solidify the case for exploring new alternative theories, a smaller survey of 30 Be and emission-line stars was performed and presented in chapter 9. These Be and emission-line stars had a very different type of spectropolarimetric morphology. In 10 of 30 stars there was a broad, monotonic spectropolarimetric signature present across the entire H$_\alpha$ line. This type of signature is typical of the depolarization effect. However, of these 10 broad detections, many showed more complex morphologies than a simple depolarization. In another 5 of 30 stars there were antisymmetric or other more complex spectropolarimetric signatures. This smaller survey is in good agreement with the collection of previous work done to date on these stars. The broad morphology is reproduced, but the higher resolution, higher signal-to-noise and larger sample of this survey shows that there are still some uncertainties about the exact form of the spectropolarimetric effects. 
	
	The inability of the scattering theories to fit the ``polarization-in-absorption" seen in the Herbig Ae/Be stars provided the motivation for developing a new theory for spectropolarimetric line profiles. This new theory is based on optical pumping - a method for producing polarization simply from absorption. This new theory is outlined in chapter 10. An anisotropic radiation source can induce an imbalance in the sub-states of an atomic transition. For instance, the H$_\alpha$ line is an n=3 to n=2 energy-level transition with many allowed sub-states in both levels. If the hydrogen gas is optically pumped by the stellar radiation, this non-isotropic gas will have a polarization-dependent absorption. In the circumstellar environment, radiative transitions dominate collisions and this effect can be much larger than scattering effects. When this theory is implemented in a simple thin-disk case, just as for scattering theory, there is only polarization where there is absorption and the spectropolarimetric effect is greatest at the wavelengths corresponding to the greatest angle between the line of sight and the absorbing material. Although detailed models to predict spectropolarimetric line profiles in the complicated environments have not yet been explored, this new theory has a very clear link between spectrpolarimetric effects and absorption. This new model shows promise in explaining the large collection of Herbig Ae/Be observations.

\section{Acknowledgments} This project involved many late nights on lonely mountain tops.  I have to give much thanks to those that helped me keep my sanity and stay focused. I dedicate this to those people. I'm glad I was supported by so many and I wasn't discouraged by those who think building instruments is simply engineering.

Don Mickey was very helpful and insightful. He participated in many useful discussions about polarimetry and optical design and offered much insight and encouragement. Jeff Kuhn was a great advisor. Though he should be hopelessly buried by work for his hundred other projects, somehow he managed to always find time and energy for this. I don't know how he does it.

This program was partially supported by the NSF AST-0123390 grant, the University of Hawaii and the AirForce Research Labs (AFRL). Some of this research used the facilities of the Canadian Astronomy Data Centre operated by the National Research Council of Canada with the support of the Canadian Space Agency. The archive was very helpful by providing the archival ESPaDOnS data.

This program also made use of observations obtained at the Canada-France-Hawaii Telescope (CFHT) which is operated by the National Research Council of Canada, the Institut National des Sciences de l'Univers of the Centre National de la Recherche Scientifique of France, and the University of Hawaii. These observations were reduced with the dedicated software package Libre-Esprit made available by J. -F. Donati. The Simbad data base operated by CDS, Strasbourg, France was very useful for compiling stellar properties.

\end{document}